\documentstyle[12pt,psfig]{article}
\renewcommand{\thepage}{\roman{page}}
\begin{document}

\newcommand{\be}{\begin{equation}}
\newcommand{\ee}{\end{equation}}
\newcommand{\bea}{\begin{eqnarray}}
\newcommand{\eea}{\end{eqnarray}}

\baselineskip=16pt
\def\boxx{\Box}
\def\mn{{\mu\nu}}
\def\e#1{{\hat e}_{(#1)}}
\def\t#1{{\hat \theta}^{(#1)}}
\def\p#1{{\partial}_{#1}}
\def\d{{\rm d}}
\def\R{{\bf R}}
\def\g{{\sqrt{-g}}}
\def\tr{\mathop{\rm Tr}\nolimits}
\def\lie{\pounds}
\def\bh{{\bar h}}
\def\x{{\bf x}}
\def\y{{\bf y}}
\def\k{{\bf k}}

\thispagestyle{plain}
\setcounter{page}{0}

\title{Lecture Notes on General Relativity}

\author{Sean M. Carroll \\ Institute for Theoretical Physics \\
University of California \\ Santa Barbara, CA 93106 \\
{\tt carroll@itp.ucsb.edu}}

\date{December 1997}

\maketitle

\begin{abstract}
These notes represent approximately one semester's worth of lectures
on introductory general relativity for beginning graduate students
in physics.  Topics include manifolds, Riemannian geometry,
Einstein's equations, and three applications: gravitational radiation,
black holes, and cosmology.
Individual chapters, and potentially updated versions, can be found at
{\tt http://itp.ucsb.edu/{\~{}}carroll/notes/}.
\end{abstract}

\vfill

\noindent{{NSF-ITP/97-147}\hfill{gr-qc/9712019}}

\eject

\section*{Table of Contents}

\begin{enumerate}
\setcounter{enumi}{-1}

\item{Introduction}

table of contents --- preface --- bibliography

\item{Special Relativity and Flat Spacetime}

the spacetime interval --- the metric --- Lorentz transformations ---
  spacetime diagrams --- vectors --- the tangent space ---
  dual vectors --- tensors --- tensor products --- the Levi-Civita
  tensor --- index manipulation --- electromagnetism ---
  differential forms --- Hodge duality --- worldlines --- proper time ---
  energy-momentum vector --- energy-momentum tensor --- perfect fluids
  --- energy-momentum conservation

\item{Manifolds}

examples --- non-examples --- maps --- continuity --- the chain rule ---
  open sets --- charts and atlases --- manifolds --- examples of charts ---
  differentiation --- vectors as derivatives --- coordinate bases ---
  the tensor transformation law --- partial derivatives are not tensors --- 
  the metric again --- canonical form of the metric --- Riemann normal 
  coordinates --- tensor densities --- volume forms and integration

\item{Curvature}

covariant derivatives and connections --- connection coefficients
  --- transformation properties --- the Christoffel connection --- 
  structures on manifolds --- parallel transport --- the parallel 
  propagator --- geodesics --- affine parameters --- the exponential
  map --- the Riemann curvature tensor --- symmetries of the Riemann
  tensor --- the Bianchi identity --- Ricci and Einstein tensors ---
  Weyl tensor --- simple examples --- geodesic deviation --- tetrads
  and non-coordinate bases --- the spin connection --- Maurer-Cartan
  structure equations --- fiber bundles and gauge transformations

\item{Gravitation}

the Principle of Equivalence --- gravitational redshift ---
  gravitation as spacetime curvature --- the Newtonian limit ---
  physics in curved spacetime --- Einstein's equations --- the Hilbert
  action --- the energy-momentum tensor again --- the Weak Energy
  Condition --- alternative theories --- the initial value problem ---
  gauge invariance and harmonic gauge --- domains of dependence ---
  causality

\item{More Geometry}

pullbacks and pushforwards --- diffeomorphisms --- integral
  curves --- Lie derivatives --- the energy-momentum tensor one
  more time --- isometries and Killing vectors

\item{Weak Fields and Gravitational Radiation}

the weak-field limit defined --- gauge transformations --- linearized
  Einstein equations --- gravitational plane waves --- transverse
  traceless gauge --- polarizations --- gravitational radiation by
  sources --- energy loss

\item{The Schwarzschild Solution and Black Holes}

spherical symmetry --- the Schwarzschild metric ---
  Birkhoff's theorem --- geodesics of Schwarzschild ---
  Newtonian vs. relativistic orbits --- perihelion precession ---
  the event horizon --- black holes --- Kruskal coordinates ---
  formation of black holes --- Penrose diagrams --- conformal infinity ---
  no hair --- charged black holes --- cosmic censorship --- extremal
  black holes --- rotating black holes --- Killing tensors --- the
  Penrose process --- irreducible mass --- black hole thermodynamics

\item{Cosmology}

homogeneity and isotropy --- the Robertson-Walker metric ---
  forms of energy and momentum --- Friedmann equations ---
  cosmological parameters --- evolution of the scale factor ---
  redshift --- Hubble's law

\end{enumerate}

\eject

\section*{Preface}

These lectures represent an introductory graduate course in general
relativity, both its foundations and applications.  They are a lightly
edited version of notes I handed out while teaching Physics
8.962, the graduate course in GR at MIT, during the Spring of 1996.
Although they are appropriately called ``lecture notes'', the level
of detail is fairly high, either including all necessary steps or
leaving gaps that can readily be filled in by the reader.  Nevertheless,
there are various ways in which these notes differ from a textbook;
most importantly, they are not organized into short sections that
can be approached in various orders, but are meant to be gone through from
start to finish.  A special effort has been made to maintain a 
conversational tone, in an attempt to go slightly beyond the bare
results themselves and into the context in which they belong.

The primary question facing any introductory treatment of general
relativity is the level of mathematical rigor at which to operate.
There is no uniquely proper solution, as different students will
respond with different levels of understanding and enthusiasm to
different approaches.  Recognizing this, I have tried to provide
something for everyone.  The lectures do not shy away from
detailed formalism (as for example in the introduction to manifolds),
but also attempt to include concrete examples and informal
discussion of the concepts under consideration.  

As these are advertised as lecture notes rather than an original
text, at times I have shamelessly stolen from various existing books
on the subject (especially those by Schutz, Wald, Weinberg, and
Misner, Thorne and Wheeler).  My philosophy was never to try to seek
originality for its own sake; however, originality sometimes 
crept in just because I thought I could be more clear than existing
treatments.  None of the substance of the material in these notes is 
new; the only reason for reading them is if an individual reader
finds the explanations here easier to understand than those elsewhere.

Time constraints during the actual semester prevented me from
covering some topics in the depth which they deserved, an obvious
example being the treatment of cosmology.  If the time and motivation
come to pass, I may expand and revise the existing notes; updated
versions will be available at 
{\tt http://itp.ucsb.edu/{\~{}}carroll/notes/}.  Of course I will
appreciate having my attention drawn to any typographical or
scientific errors, as well as suggestions for improvement of
all sorts.

Numerous people have contributed greatly both to my own understanding
of general relativity and to these notes in particular --- too many
to acknowledge with any hope of completeness.  
Special thanks are due to Ted Pyne, who learned
the subject along with me, taught me a great deal,
and collaborated on a predecessor to this
course which we taught as a seminar in the astronomy department at
Harvard.  Nick Warner taught the graduate course at MIT which I
took before ever teaching it, and his notes were (as comparison will
reveal) an important influence on these.  George Field offered a
great deal of advice and encouragement as I learned the subject and
struggled to teach it.  Tam\'as Hauer struggled along with me as the 
teaching assistant for 8.962, and was an invaluable help.  All of
the students in 8.962 deserve thanks for tolerating my idiosyncrasies
and prodding me to ever higher levels of precision.

During the course of writing these notes I was supported by
U.S. Dept.~of Energy contract no. DE-AC02-76ER03069 and National
Science Foundation grants PHY/92-06867 and PHY/94-07195.

\eject

\section*{Bibliography}

\noindent
The {\it typical} level of difficulty (especially mathematical) of the books
is indicated by a number of asterisks, one meaning mostly introductory
and three being advanced.  The asterisks are
normalized to these lecture notes, which would be given [**].  The 
first four books were frequently consulted in the preparation of these
notes, the next seven are other relativity texts which I have found
to be useful, and the last four are mathematical background references.

\begin{itemize}

\item B.F. Schutz, {\sl A First Course in General Relativity} (Cambridge,
1985) [*].  This is a very nice introductory text.  Especially useful
if, for example, you aren't quite clear on what the energy-momentum
tensor really means.

\item S. Weinberg, {\sl Gravitation and Cosmology} (Wiley, 1972) [**].
A really good book at what it does, especially strong on astrophysics, 
cosmology, and experimental tests.  However, it takes
an unusual non-geometric approach to the material, and 
doesn't discuss black holes.

\item C. Misner, K. Thorne and J. Wheeler, {\sl Gravitation}
(Freeman, 1973) [**].  A heavy book, in various senses.  Most things
you want to know are in here, although you might have to work hard
to get to them (perhaps learning something unexpected in the process).

\item R. Wald, {\sl General Relativity} (Chicago, 1984) [***].  Thorough
discussions of a number of advanced topics, including black holes,
global structure, and spinors.  The approach is more mathematically 
demanding than the previous books, and the basics are covered pretty quickly.

\item E. Taylor and J. Wheeler, {\sl Spacetime Physics} (Freeman, 1992)
[*].  A good introduction to special relativity.

\item R. D'Inverno, {\sl Introducing Einstein's Relativity} (Oxford, 1992)
[**].  A book I haven't looked at very carefully, but it seems as if all the
right topics are covered without noticeable ideological distortion.

\item A.P. Lightman, W.H. Press, R.H. Price, and S.A. Teukolsky,
{\sl Problem Book in Relativity and Gravitation} (Princeton, 1975) [**].
A sizeable collection of problems in all areas of GR, with fully
worked solutions, making it all the more difficult for instructors
to invent problems the students can't easily find the answers to.

\item N. Straumann, {\sl General Relativity and Relativistic Astrophysics}
(Springer-Verlag, 1984) [***].  A fairly high-level book, which starts out 
with a good deal of abstract geometry and goes on to detailed discussions
of stellar structure and other astrophysical topics.

\item F. de Felice and C. Clarke, {\sl Relativity on Curved Manifolds}
(Cambridge, 1990) [***].  A mathematical approach, but with an excellent
emphasis on physically measurable quantities.

\item S. Hawking and G. Ellis, {\sl The Large-Scale Structure of
Space-Time} (Cambridge, 1973) [***].  An advanced book which emphasizes
global techniques and singularity theorems.

\item R. Sachs and H. Wu, {\sl General Relativity for Mathematicians}
(Springer-Verlag, 1977) [***].  Just what the title says, although the
typically dry mathematics prose style is here enlivened by frequent
opinionated asides about both physics and mathematics (and the state
of the world).

\item B. Schutz, {\sl Geometrical Methods of Mathematical Physics}
(Cambridge, 1980) [**].  Another good book by Schutz, this one covering some
mathematical points that are left out of the GR book (but at a very
accessible level).  Included are
discussions of Lie derivatives, differential forms, and applications to
physics other than GR.

\item V. Guillemin and A. Pollack, {\sl Differential Topology}
(Prentice-Hall, 1974) [**].  An entertaining survey of manifolds, topology,
differential forms, and integration theory.

\item C. Nash and S. Sen, {\sl Topology and Geometry for Physicists}
(Academic Press, 1983) [***].  Includes homotopy, homology, fiber bundles
and Morse theory, with applications to physics; somewhat concise.

\item F.W. Warner, {\sl Foundations of Differentiable Manifolds and
Lie Groups} (Springer-Verlag, 1983) [***].  The standard text in the field,
includes basic topics such as manifolds and tensor fields as well as
more advanced subjects.

\end{itemize}

\eject
\thispagestyle{plain}
\renewcommand{\thepage}{\arabic{page}}
\setcounter{page}{1}
\setcounter{section}{0}
\setcounter{equation}{0}

\noindent{December 1997\hfill{\sl Lecture Notes on General Relativity}
\hfill{Sean M.~Carroll}}

\vskip .2in

\section{Special Relativity and Flat Spacetime}

We will begin with a whirlwind tour of special relativity (SR) and life in 
flat spacetime.  The point will be both to recall what SR is all about, and 
to introduce tensors and related concepts that will be crucial
later on, without the extra complications of curvature on top of everything
else.  Therefore, for this section we will always be working in flat
spacetime, and furthermore we will only use orthonormal (Cartesian-like)
coordinates.  Needless to say it is possible to do SR in any coordinate
system you like, but it turns out that introducing the necessary tools
for doing so would take us halfway to curved spaces anyway, so we will
put that off for a while.

It is often said that special relativity is a theory of 4-dimensional
spacetime: three of space, one of time.  But of course, the pre-SR world
of Newtonian mechanics featured three spatial dimensions and a time
parameter.  Nevertheless, there was not much temptation to consider these
as different aspects of a single 4-dimensional spacetime.  Why not?

\begin{figure}[h]
  \centerline{
  \psfig{figure=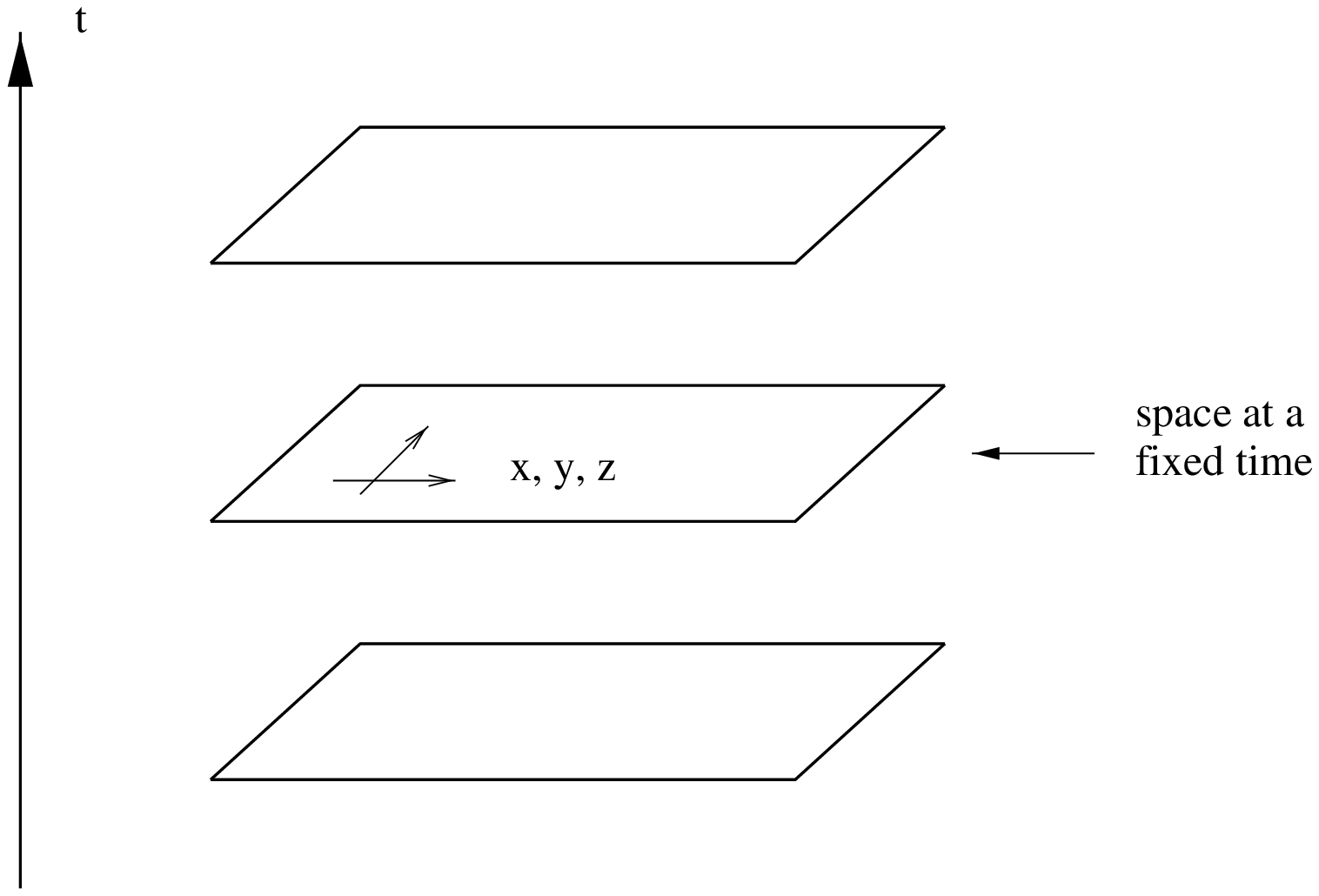,angle=-90,height=8cm}}
\end{figure}

Consider a garden-variety 2-dimensional plane.  It is typically 
convenient to label the points on such a plane by introducing coordinates,
for example by defining orthogonal $x$ and $y$ axes and projecting each
point onto these axes in the usual way.  However, it is clear that most
of the interesting geometrical facts about the plane are independent of
our choice of coordinates.  As a simple example, we can consider the
distance between two points, given by
\be
  s^2 = (\Delta x)^2 + (\Delta y)^2\ .\label{1.1}
\ee
In a different Cartesian coordinate system, defined by $x'$ and $y'$ axes
which are rotated with respect to the originals, the formula for the 
distance is unaltered:
\be
  s^2 = (\Delta x')^2 + (\Delta y')^2\ .\label{1.2}
\ee
We therefore say that the distance is invariant under such changes of
coordinates.

\begin{figure}[h]
  \centerline{
  \psfig{figure=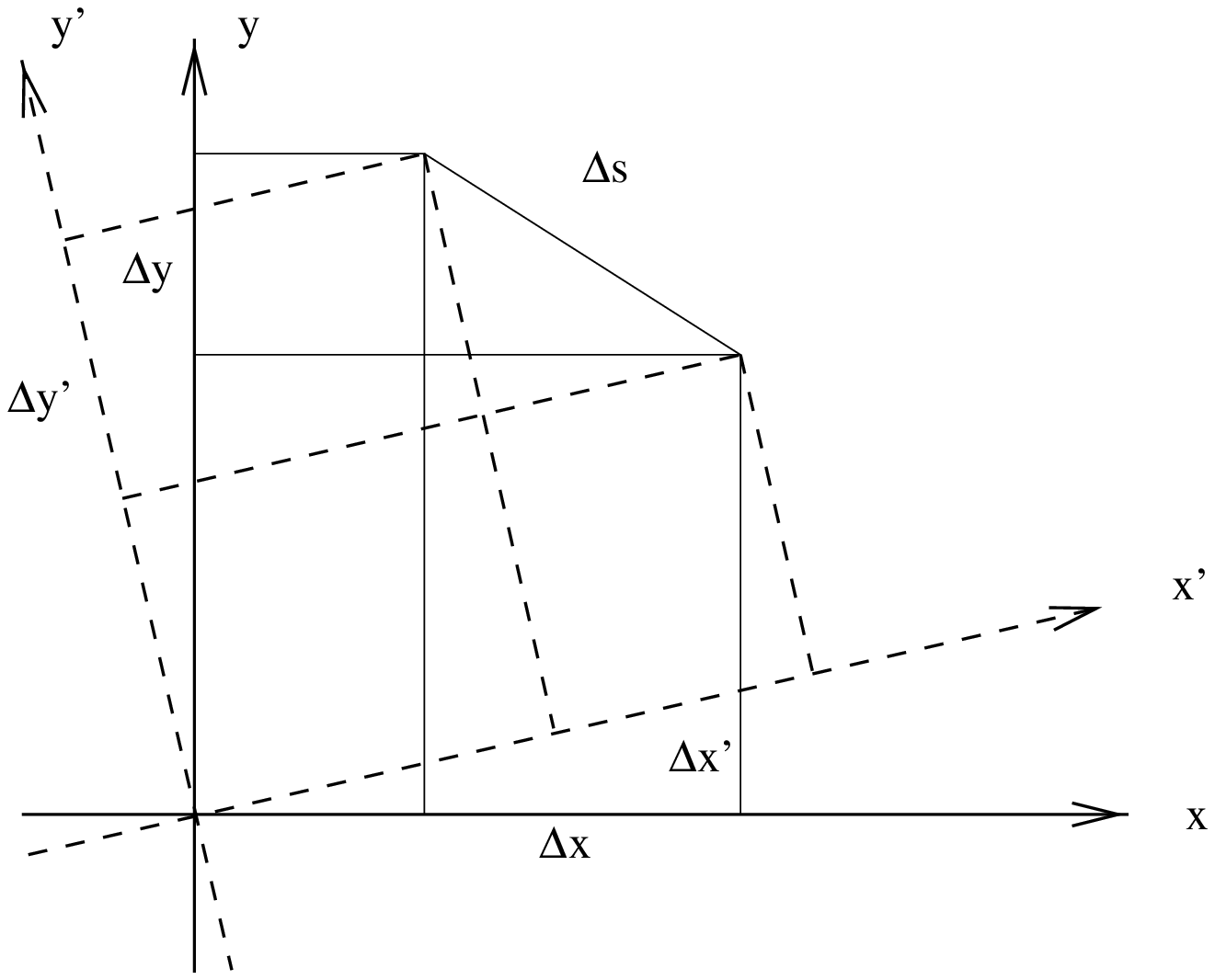,angle=-90,height=8cm}}
\end{figure}

This is why it is useful to think of the plane as 2-dimensional: although
we use two distinct numbers to label each point, the numbers are not the
essence of the geometry, since we can rotate axes into each other while
leaving distances and so forth unchanged.  In Newtonian physics this is
not the case with space and time; there is no useful notion of rotating
space and time into each other.  Rather, the notion of ``all of space at
a single moment in time'' has a meaning independent of coordinates.

Such is not the case in SR.  Let us consider coordinates $(t,x,y,z)$ on
spacetime, set up in the following way.  The spatial coordinates $(x,y,z)$
comprise a standard Cartesian system, constructed for example by welding
together rigid rods which meet at right angles.  The rods must be moving
freely, unaccelerated.  The time coordinate is
defined by a set of clocks which are not moving with respect to the
spatial coordinates.  (Since this is a thought experiment, we imagine that
the rods are infinitely long and there is one clock at every point in space.)
The clocks are synchronized in the following sense: if you travel from one
point in space to any other in a straight line at constant speed, the time
difference between the clocks at the ends of your journey is the same as
if you had made the same trip, at the same speed, in the other direction.
The coordinate system thus constructed is an {\bf inertial frame}.

An {\bf event} is defined as a single moment in space and time, characterized
uniquely by $(t,x,y,z)$.
Then, without any motivation for the moment, let us introduce the 
{\bf spacetime interval} between two events:
\be
  s^2 = -(c\Delta t)^2 + (\Delta x)^2 + (\Delta y)^2 + (\Delta z)^2\ .
  \label{1.3}
\ee
(Notice that it can be positive, negative, or zero even for two nonidentical
points.) Here, $c$ is some fixed conversion factor between space and time; 
that is, a fixed velocity.  Of course it will turn out to be the speed of 
light; the important thing, however, is not that photons happen to travel at 
that speed, but that there exists a $c$ such that the spacetime interval is
invariant under changes of coordinates.  In other words, if we set up a new
inertial frame $(t',x',y',z')$ by repeating our earlier procedure, but allowing
for an offset in initial position, angle, and velocity between the new rods and
the old, the interval is unchanged:
\be
  s^2 = -(c\Delta t')^2 + (\Delta x')^2 + (\Delta y')^2 + (\Delta z')^2\ .
  \label{1.4}
\ee
This is why it makes sense to think of SR as a theory of 4-dimensional
spacetime, known as {\bf Minkowski space}.  (This is a special case
of a 4-dimensional manifold, which we will deal with in detail later.)
As we shall see, the coordinate transformations which we have
implicitly defined do, in a sense, rotate space and time into each other.
There is no absolute notion of ``simultaneous events''; whether two things
occur at the same time depends on the coordinates used.  Therefore the
division of Minkowski space into space and time is a choice we make for 
our own purposes, not something intrinsic to the situation.

Almost all of the ``paradoxes'' associated with SR result from a stubborn
persistence of the Newtonian notions of a unique time coordinate and the
existence of ``space at a single moment in time.''
By thinking in terms of spacetime rather than space and time together,
these paradoxes tend to disappear.

Let's introduce some convenient notation.  Coordinates on spacetime will
be denoted by letters with Greek superscript indices running from $0$ to
$3$, with $0$ generally denoting the time coordinate.  Thus,
\be
  x^\mu :\quad \matrix{x^0 = ct\cr x^1 = x\cr x^2 = y\cr x^3 = z\cr}
  \label{1.5}
\ee
(Don't start thinking of the superscripts as exponents.)  Furthermore, for
the sake of simplicity we will choose units in which
\be
  c = 1 \ ;\label{1.6}
\ee
we will therefore leave out factors of $c$ in all subsequent formulae.
Empirically we know that $c$ is the speed of light, $3\times 10^8$ meters
per second; thus, we are working in units where 1 second equals $3\times 10^8$ 
meters.  Sometimes it will be useful to refer to the space and time 
components of $x^\mu$ separately, so we will use Latin superscripts to
stand for the space components alone:
\be
  x^i :\quad \matrix{x^1 = x\cr x^2 = y\cr x^3 = z\cr}\label{1.7}
\ee

It is also convenient to write the spacetime interval in a more compact
form.  We therefore introduce a $4\times 4$ matrix, the
{\bf metric}, which we write using two lower indices: 
\be
  \eta_{\mn} = \left(\matrix{-1 &0&0&0\cr 0&1&0&0\cr 
  0&0&1&0 \cr 0&0&0&1\cr}\right)\ .\label{1.8}
\ee
(Some references, especially field theory books, define the metric with
the opposite sign, so be careful.)  We then have the nice formula
\be
  s^2 = \eta_\mn \Delta x^\mu \Delta x^\nu\ . \label{1.9}
\ee
Notice that we use the {\bf summation convention}, in which indices which
appear both as superscripts and subscripts are summed over.  The content
of (1.9) is therefore just the same as (1.3).

Now we can consider coordinate transformations in spacetime at a somewhat
more abstract level than before.  What kind of transformations leave the
interval (1.9) invariant?  One simple variety are the translations, which
merely shift the coordinates:
\be
  x^\mu \rightarrow x^{\mu'} = x^\mu + a^\mu\ ,\label{1.10}
\ee
where $a^\mu$ is a set of four fixed numbers.  (Notice that we put the
prime on the index, not on the $x$.)  Translations leave the differences
$\Delta x^\mu$ unchanged, so it is not remarkable that the interval is
unchanged.  The only other kind of linear transformation is to multiply
$x^\mu$ by a (spacetime-independent) matrix:
\be
  x^{\mu'} = \Lambda^{\mu'}{}_\nu x^\nu\ , \label{1.11}
\ee
or, in more conventional matrix notation,
\be
  x' = \Lambda x\ .\label{1.12}
\ee
These transformations do not leave the differences $\Delta x^\mu$ unchanged, 
but multiply them also by the matrix $\Lambda$.  What kind of matrices will 
leave the interval invariant?  Sticking with the matrix notation, what we
would like is 
\bea
  s^2 = (\Delta x)^{\rm T} \eta (\Delta x)
  & = & (\Delta x')^{\rm T} \eta (\Delta x')\nonumber \\
  & = & (\Delta x)^{\rm T} \Lambda^{\rm T} \eta \Lambda (\Delta x)\ ,
  \label{1.13}
\eea
and therefore
\be
  \eta = \Lambda^{\rm T} \eta \Lambda \ ,\label{1.14}
\ee
or
\be
  \eta_{\rho\sigma} = \Lambda^{\mu'}{}_{\rho}\Lambda^{\nu'}{}_{\sigma}
  \eta_{\mu'\nu'}\ . \label{1.15}
\ee
We want to find the matrices $\Lambda^{\mu'}{}_\nu$ such that the components
of the matrix $\eta_{\mu'\nu'}$ are the same as those of $\eta_{\rho\sigma}$;
that is what it means for the interval to be invariant under these 
transformations.  

The matrices which satisfy (1.14) are known as the {\bf Lorentz 
transformations};  the set of them forms a group under matrix multiplication,
known as the {\bf Lorentz group}.  There is a close analogy between this
group and O(3), the rotation group in three-dimensional space.  The
rotation group can be thought of as $3\times 3$ matrices $R$ which satisfy
\be
  {\bf 1} = R^{\rm T}{\bf 1} R\ ,\label{1.16}
\ee
where {\bf 1} is the $3\times 3$ identity matrix.  The similarity with (1.14)
should be clear; the only difference is the minus sign in the first term
of the metric $\eta$, signifying the timelike direction.  The Lorentz
group is therefore often referred to as O(3,1).  (The $3\times 3$ identity
matrix is simply the metric for ordinary flat space.  Such a metric, in
which all of the eigenvalues are positive, is called {\bf Euclidean},
while those such as (1.8) which feature a single minus sign are called
{\bf Lorentzian}.)

Lorentz transformations fall into a number of categories.  First there
are the conventional {\bf rotations}, such as a rotation in the $x$-$y$
plane:
\be
  \Lambda^{\mu'}{}_\nu = \left(\matrix{ 1&0&0&0\cr
  0& \cos\theta & \sin\theta &0\cr 0& -\sin\theta & \cos\theta &0\cr
  0&0&0&1\cr}\right)\ .\label{1.17}
\ee
The rotation angle $\theta$ is a periodic variable with period $2\pi$.
There are also {\bf boosts}, which may be thought of as ``rotations
between space and time directions.''  An example is given by
\be
  \Lambda^{\mu'}{}_\nu = \left(\matrix{ \cosh\phi&-\sinh\phi &0&0\cr
  -\sinh\phi&\cosh\phi &0&0\cr 0&0&1&0\cr
  0&0&0&1\cr}\right)\ .\label{1.18}
\ee
The boost parameter $\phi$, unlike the rotation angle, is defined 
from $-\infty$ to $\infty$.  There are also discrete transformations 
which reverse the time direction or
one or more of the spatial directions.  (When these are excluded we
have the proper Lorentz group, SO(3,1).)  A general transformation
can be obtained by multiplying the individual transformations; the
explicit expression for this six-parameter matrix (three boosts,
three rotations) is not sufficiently pretty or useful to bother writing
down.  In general Lorentz transformations will not commute, so the 
Lorentz group is non-abelian.  The set of both translations and
Lorentz transformations is a ten-parameter non-abelian group,
the {\bf Poincar\'e group}.

You should not be surprised to learn that the boosts correspond to
changing coordinates by moving to a frame which travels at a constant
velocity, but let's see it more explicitly.  For the transformation
given by (1.18), the transformed coordinates $t'$ and $x'$ will be given 
by
\bea
  t' &=& t\cosh\phi - x \sinh\phi \nonumber \\
  x' &=& -t \sinh\phi + x\cosh\phi\ .\label{1.19}
\eea
From this we see that the point defined by $x'=0$ is moving; it has
a velocity
\be
  v = {x\over t} = {{\sinh\phi}\over{\cosh\phi}} = \tanh\phi\ .
  \label{1.20}
\ee
To translate into more pedestrian notation, we can replace
$\phi = \tanh^{-1}v$ to obtain
\bea
  t' &=& \gamma(t-vx)\nonumber \\
  x' &=& \gamma(x-vt)\label{1.21}
\eea
where $\gamma =1/\sqrt{1-v^2}$.  So indeed, our abstract approach has
recovered the conventional expressions for Lorentz transformations.
Applying these formulae leads to time dilation, length contraction,
and so forth.

An extremely useful tool is the {\bf spacetime diagram}, so let's 
consider Minkowski space from this point of view.  We can begin by
portraying the initial $t$ and $x$ axes at (what are conventionally
thought of as) right angles, and suppressing the $y$ and $z$ axes.
Then according to (1.19), under a boost in the $x$-$t$ plane 
the $x'$ axis ($t' = 0$) is given by
$t = x \tanh\phi$, while the $t'$ axis ($x' = 0$) is given by
$t = x/\tanh\phi$.  We therefore see that the space and time axes
are rotated into each other, although they scissor
together instead of remaining orthogonal in the traditional Euclidean
sense.  (As we shall see, the axes do in fact remain orthogonal
in the Lorentzian sense.)
This should come as no surprise, since if spacetime behaved
just like a four-dimensional version of space the world would be a 
very different place.

\begin{figure}
  \centerline{
  \psfig{figure=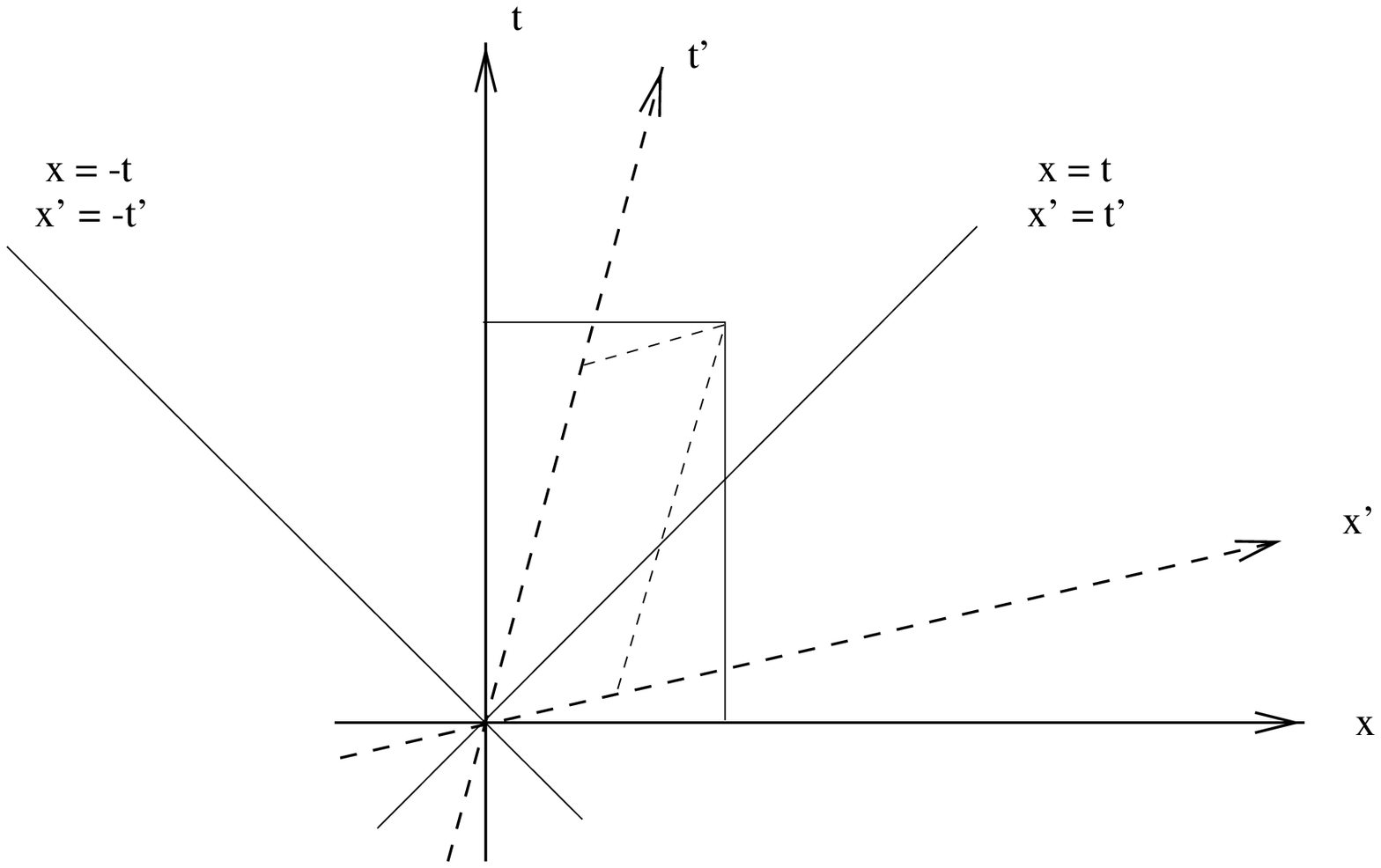,angle=-90,height=8cm}}
\end{figure}

It is also enlightening to consider the paths corresponding to travel
at the speed $c=1$.  These are given in the original coordinate system
by $x=\pm t$.  In the new system, a moment's thought reveals that the
paths defined by $x' = \pm t'$ are precisely the same as those defined
by $x=\pm t$; these trajectories are left invariant under Lorentz
transformations.  Of course we know that light travels at this speed;
we have therefore found that the speed of light is the same in any
inertial frame.  A set of points which are all connected to a single
event by straight lines moving at the speed of light is called a 
{\bf light cone}; this entire set is invariant under Lorentz
transformations.  Light cones are naturally divided into future
and past; the set of all points inside the future and past light
cones of a point $p$ are called {\bf timelike separated} from $p$,
while those outside the light cones are {\bf spacelike separated}
and those on the cones are {\bf lightlike} or {\bf null separated}
from $p$.  Referring back to (1.3), we see that the interval between
timelike separated points is negative, between spacelike separated
points is positive, and between null separated points is zero. 
(The interval is defined to be $s^2$, not the square root of this
quantity.)  Notice the distinction between this situation and that
in the Newtonian world; here, it is impossible to say (in a 
coordinate-independent way) whether a point that is spacelike separated
from $p$ is in the future of $p$, the past of $p$, or ``at the same 
time''.

To probe the structure of Minkowski space in more detail, it is 
necessary to introduce the concepts of vectors and tensors.  We will
start with vectors, which should be familiar.  Of course, in
spacetime vectors are four-dimensional, and are often referred to
as {\bf four-vectors}.  This turns out to make quite a bit of difference;
for example, there is no such thing as a cross product between two
four-vectors.

Beyond the simple fact of dimensionality, the most important thing to
emphasize is that each vector is located at a given point in spacetime.
You may be used to thinking of vectors as stretching from one point
to another in space, and even of ``free'' vectors which you can slide 
carelessly from point to point.  These are not useful concepts in
relativity.  Rather, to each point $p$ in spacetime we associate the
set of all possible vectors located at that point; this set is known
as the {\bf tangent space} at $p$, or $T_p$.  The name is inspired by
thinking of the set of vectors attached to a point on a simple curved
two-dimensional space as comprising a plane which is tangent to the
point.  But inspiration aside, it is important to think of these vectors
as being located at a single point, rather than stretching from one point
to another.  (Although this won't stop us from drawing them as arrows
on spacetime diagrams.)

\begin{figure}[h]
  \centerline{
  \psfig{figure=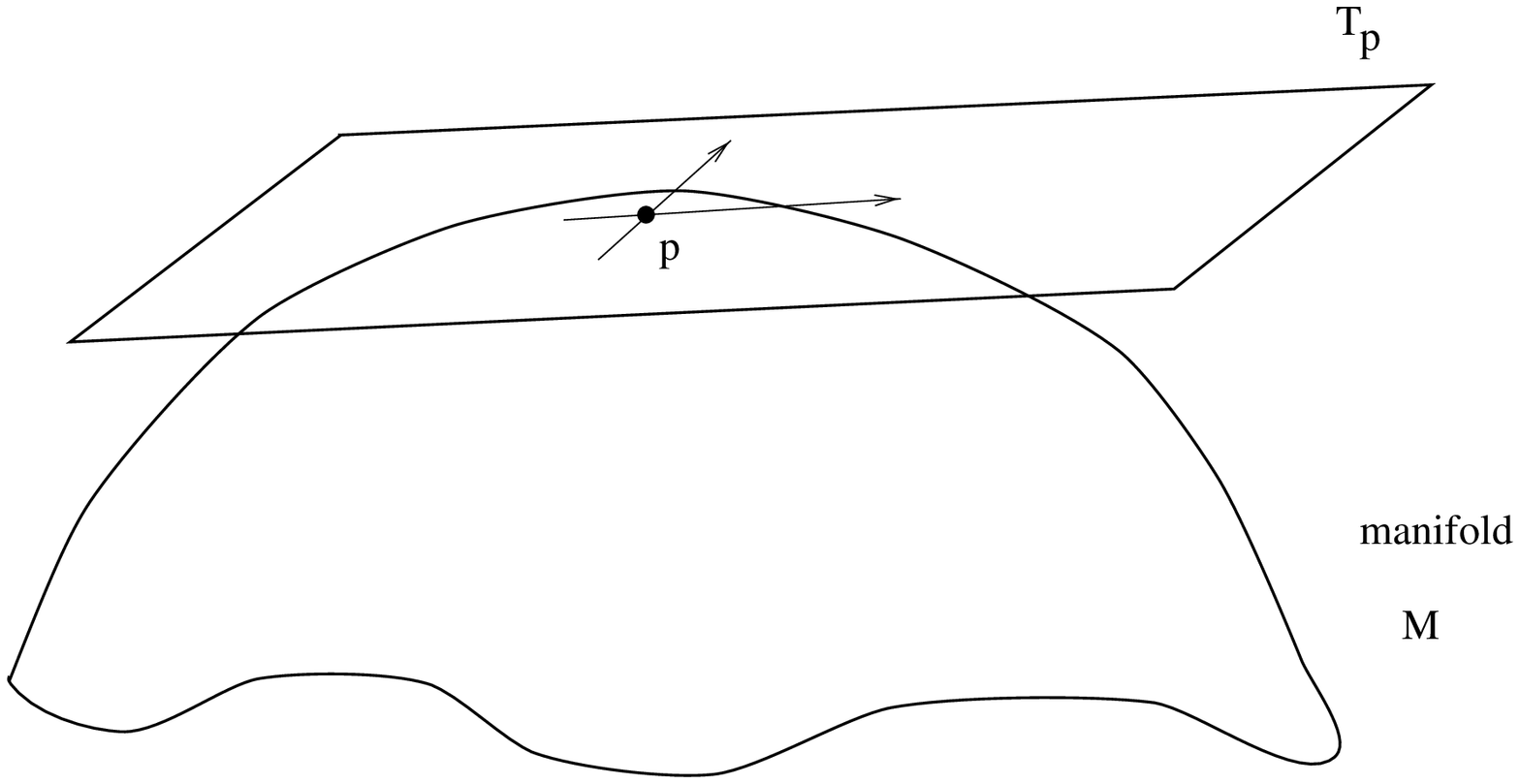,angle=-90,height=8cm}}
\end{figure}

Later we will relate the tangent space at each point to things we can
construct from the spacetime itself.  For right now, just think of $T_p$
as an abstract vector space for each point in spacetime.  A {\bf (real) vector 
space} is a collection of objects (``vectors'') which, roughly speaking, can 
be added together and multiplied by real numbers in a linear way.  Thus,
for any two vectors $V$ and $W$ and real numbers $a$ and $b$, we have
\be
  (a+b)(V+W) = aV+bV+aW+bW\ .\label{1.22}
\ee
Every vector space has an origin, {\it i.e.} a zero vector which functions
as an identity element under vector addition.  In many vector spaces there
are additional operations such as taking an inner (dot) product, but this
is extra structure over and above the elementary concept of a vector space.

A vector is a perfectly well-defined geometric object, as is a {\bf vector
field}, defined as a set of vectors with exactly one at each point in 
spacetime.  (The set of all the tangent spaces of a manifold $M$ is 
called the {\bf tangent bundle}, $T(M)$.)
Nevertheless it is often useful for concrete purposes to
decompose vectors into components with respect to some set of basis
vectors.  A {\bf basis} is any set of vectors which both spans the
vector space (any vector is a linear combination of basis vectors) and
is linearly independent (no vector in the basis is a linear combination of
other basis vectors).  For any given vector space, there will be an 
infinite number of legitimate bases, but each basis will consist of the
same number of vectors, known as the dimension of the space.  (For
a tangent space associated with a point in Minkowski space, the dimension
is of course four.)

Let us imagine that at each tangent space we set up a basis of four
vectors $\e\mu$, with $\mu\in\{0,1,2,3\}$ as usual.  In fact let us
say that each basis is adapted to the coordinates $x^\mu$; that is, the
basis vector $\e1$ is what we would normally think of pointing along
the $x$-axis, etc.  It is by no means necessary that we choose a basis
which is adapted to any coordinate system at all, although it is often
convenient.  (We really could be more precise here, but later
on we will repeat the discussion at an excruciating level of precision,
so some sloppiness now is forgivable.)  Then any abstract vector
$A$ can be written as a linear combination of basis vectors:
\be
  A = A^\mu \e\mu\ .\label{1.23}
\ee
The coefficients $A^\mu$ are the {\bf components} of the vector $A$.
More often than not we will forget the basis entirely and refer somewhat
loosely to ``the vector $A^\mu$'', but keep in mind that this is 
shorthand.  The real vector is an abstract geometrical entity, while the
components are just the coefficients of the basis vectors in some 
convenient basis.  (Since we will usually suppress the explicit basis
vectors, the indices will usually label components of vectors and 
tensors.  This is why there are parentheses around the indices on the
basis vectors, to remind us that this is a collection of vectors, not
components of a single vector.)

A standard example of a vector in spacetime is the tangent vector to
a curve.  A parameterized curve or path through spacetime is specified by
the coordinates as a function of the parameter, {\it e.g.} $x^\mu(\lambda)$.
The tangent vector $V(\lambda)$ has components
\be
  V^\mu = {{dx^\mu}\over{d\lambda}}\ .\label{1.24}
\ee
The entire vector is thus $V=V^\mu\e\mu$.  Under a Lorentz transformation
the coordinates $x^\mu$ change according to (1.11), while the 
parameterization $\lambda$ is unaltered; we can therefore deduce that
the components of the tangent vector must change as
\be
  V^\mu \rightarrow V^{\mu'} = \Lambda^{\mu'}{}_\nu V^\nu\ .\label{1.25}
\ee
However, the vector itself (as opposed to its components in some
coordinate system) is invariant under Lorentz transformations.  We can
use this fact to derive the transformation properties of the basis
vectors.  Let us refer to the set of basis vectors in the transformed
coordinate system as $\e{\nu'}$.
Since the vector is invariant, we have
\bea
  V = V^\mu\e\mu &=& V^{\nu'}\e{\nu'}\nonumber \\
  &=& \Lambda^{\nu'}{}_\mu V^\mu\e{\nu'}\ .\label{1.26}
\eea
But this relation must hold no matter what the numerical values of
the components $V^\mu$ are.  Therefore we can say
\be
  \e\mu = \Lambda^{\nu'}{}_\mu\e{\nu'}\ .\label{1.27}
\ee
To get the new basis $\e{\nu'}$ in terms of the old one $\e\mu$ we
should multiply by the inverse of the Lorentz transformation 
$\Lambda^{\nu'}{}_\mu$.  But the inverse of a Lorentz transformation
from the unprimed to the primed coordinates is also a Lorentz
transformation, this time from the primed to the unprimed systems.
We will therefore introduce a somewhat subtle notation, by writing
using the same symbol for both matrices, just with primed and unprimed
indices adjusted.  That is,
\be
  (\Lambda^{-1})^{\nu'}{}_\mu = \Lambda_{\nu'}{}^\mu\ ,\label{1.28}
\ee
or
\be
  \Lambda_{\nu'}{}^\mu \Lambda^{\sigma'}{}_\mu = \delta^{\sigma'}_{\nu'}
  \ ,\quad \Lambda_{\nu'}{}^\mu \Lambda^{\nu'}{}_\rho = 
  \delta^{\mu}_{\rho}\ ,\label{1.29}
\ee
where $\delta^{\mu}_{\rho}$ is the traditional Kronecker delta symbol
in four dimensions.  (Note that Schutz uses a different convention,
always arranging the two indices northwest/southeast; the important
thing is where the primes go.)
From (1.27) we then obtain the transformation rule for basis vectors:
\be
  \e{\nu'} = \Lambda_{\nu'}{}^\mu\e\mu\ .\label{1.30}
\ee
Therefore the set of basis vectors transforms via the inverse Lorentz
transformation of the coordinates or vector components.

It is worth pausing a moment to take all this in.  We introduced
coordinates labeled by upper indices, which transformed in a certain
way under Lorentz transformations.  We then considered vector components
which also were written with upper indices, which made sense since they
transformed in the same way as the coordinate functions.  (In a fixed
coordinate system, each of the four coordinates $x^\mu$ can be thought
of as a function on spacetime, as can each of the four components of
a vector field.)  The basis vectors associated with the coordinate
system transformed via the inverse matrix, and were labeled by a 
lower index.  This notation ensured that the invariant object constructed
by summing over the components and basis vectors was left unchanged by
the transformation, just as we would wish.  It's probably not giving too
much away to say that this will continue to be the case for more
complicated objects with multiple indices (tensors). 

Once we have set up a vector space, there is an associated vector space
(of equal dimension) which
we can immediately define, known as the {\bf dual vector space}.
The dual space is usually denoted by an asterisk, so that the dual 
space to the tangent space $T_p$ is called the {\bf cotangent space}
and denoted $T^*_p$.  The dual space is
the space of all linear maps from the original vector space to the real
numbers; in math lingo, if $\omega\in T_p^*$ is a dual vector, then it acts
as a map such that:
\be
  \omega(aV+bW) = a\omega(V) + b\omega(W) \in {\bf R}\ ,\label{1.31}
\ee
where $V$, $W$ are vectors and $a$, $b$ are real numbers.  The nice 
thing about these maps is that they form a vector space themselves; thus,
if $\omega$ and $\eta$ are dual vectors, we have
\be
  (a\omega+b\eta)(V) = a\omega(V) + b\eta(V)\ .\label{1.32}
\ee
To make this construction somewhat more concrete, we can introduce a
set of basis dual vectors $\t\nu$ by demanding
\be
  \t\nu(\e\mu) = \delta^\nu_\mu\ .\label{1.33}
\ee
Then every dual vector can be written in terms of its components, which
we label with lower indices:
\be
  \omega = \omega_\mu\t\mu\ .\label{1.34}
\ee
In perfect analogy with vectors, we will usually simply write
$\omega_\mu$ to stand for the entire dual vector.  In fact, you will
sometime see elements of $T_p$ (what we have called vectors) referred
to as {\bf contravariant vectors}, and elements of $T_p^*$ (what we
have called dual vectors) referred to as {\bf covariant vectors}.  Actually,
if you just refer to ordinary vectors as vectors with upper indices
and dual vectors as vectors with lower indices, nobody should be 
offended.  Another name for dual vectors is {\bf one-forms}, a somewhat
mysterious designation which will become clearer soon.

The component notation leads to a simple way of writing the action of
a dual vector on a vector:
\bea
  \omega(V) &=& \omega_\mu V^\nu\t\mu(\e\nu)\nonumber \\
  &=& \omega_\mu V^\nu \delta^\mu_\nu\nonumber \\
  &=& \omega_\mu V^\mu \in {\bf R} \ . \label{1.35}
\eea
This is why it is rarely necessary to write the basis vectors (and
dual vectors) explicitly; the components do all of the work.
The form of (1.35) also suggests that we can think of vectors as
linear maps on dual vectors, by defining 
\be
  V(\omega) \equiv \omega(V) = \omega_\mu V^\mu\ .\label{1.36}
\ee
Therefore, the dual space to the dual vector space is the original
vector space itself.

Of course in spacetime we will be interested not in a single 
vector space, but in fields of vectors and dual vectors. 
(The set of all cotangent spaces over $M$ is the
{\bf cotangent bundle}, $T^*(M)$.)  In that
case the action of a dual vector field on a vector field is not a single
number, but a {\bf scalar} (or just ``function'') on spacetime.
A scalar is a quantity without indices, which is unchanged under
Lorentz transformations.

We can use the same arguments that we earlier used for vectors to
derive the transformation properties of dual vectors.  The answers
are, for the components,
\be
  \omega_{\mu'} = \Lambda_{\mu'}{}^\nu\omega_\nu\ ,\label{1.37}
\ee
and for basis dual vectors,
\be
  \t{\rho'} = \Lambda^{\rho'}{}_\sigma \t\sigma\ .\label{1.38}
\ee
This is just what we would expect from index placement; the components
of a dual vector transform under the inverse transformation of those
of a vector.  Note that this ensures that the scalar (1.35) is
invariant under Lorentz transformations, just as it should be.

Let's consider some examples of dual vectors, first in other contexts
and then in Minkowski space.  Imagine the space of $n$-component
column vectors, for some integer $n$.  Then the dual space is that of
$n$-component row vectors, and the action is ordinary matrix
multiplication:
\bea
  V &=&\left(\matrix{V^1 \cr V^2 \cr \cdot\cr \cdot \cr \cdot \cr
  V^{n}\cr}\right)\ ,\quad
  \omega = \left(\omega_1\  \omega_2\  \cdots\ \omega_{n}\right)\ ,\nonumber \\
  \omega(V) &=&\left(\omega_1\ \omega_2\ \cdots\ \omega_{n}\right)
  \left(\matrix{V^1 \cr V^2 \cr \cdot\cr \cdot \cr \cdot\cr
  V^{n}\cr}\right) = \omega_i V^i\ . \label{1.39}
\eea
Another familiar example occurs in quantum mechanics, where vectors
in the Hilbert space are represented by kets, $|\psi\rangle$.  In this
case the dual space is the space of bras, $\langle\phi |$, and the
action gives the number $\langle \phi |\psi\rangle$.  (This is a
complex number in quantum mechanics, but the idea is precisely the 
same.)

In spacetime the simplest example of a dual vector is the {\bf gradient}
of a scalar function, the set of partial derivatives with respect to the
spacetime coordinates, which we denote by ``d'':
\be
  {\rm d}\phi = {{\p{}\phi}\over{\p{}x^\mu}} \t\mu\ .\label{1.40}
\ee
The conventional chain rule used to transform partial derivatives
amounts in this case to the transformation rule of components of dual
vectors:
\bea
  {{\p{}\phi}\over{\p{}x^{\mu'}}} &=&
  {{\p{}x^\mu}\over{\p{}x^{\mu'}}}{{\p{}\phi}\over{\p{}x^\mu}}\nonumber \\
  &=&\Lambda_{\mu'}{}^\mu {{\p{}\phi}\over{\p{}x^\mu}}\ ,
  \label{1.41}
\eea
where we have used (1.11) and (1.28) to relate the Lorentz transformation
to the coordinates.  The fact that the gradient is a dual vector leads
to the following shorthand notations for partial derivatives:
\be
  {{\p{}\phi}\over{\p{}x^\mu}}=\p\mu\phi = \phi,{}_\mu
  \ .\label{1.42}
\ee
(Very roughly speaking, ``$x^\mu$ has an upper index, but when it is in 
the denominator of a derivative it implies a lower index on the resulting 
object.'') I'm not a big fan of the comma notation, but we will use $\p\mu$ 
all the time.  Note that the gradient does in fact act in a natural way
on the example we gave above of a vector, the tangent vector to a curve.
The result is ordinary derivative of the function along the curve:
\be
  \p\mu\phi {{\partial x^\mu}\over{\partial \lambda}}
  = {{d\phi}\over{d\lambda}}\ .\label{1.43}
\ee

As a final note on dual vectors, there is a way to represent them as
pictures which is consistent with the picture of vectors as arrows.
See the discussion in Schutz, or in MTW (where it is taken to dizzying
extremes).

A straightforward generalization of vectors and dual vectors is the notion
of a {\bf tensor}.  Just as a dual vector is a linear map from vectors
to {\bf R}, a tensor $T$ of type (or rank) $(k,l)$ is a multilinear map 
from a collection of dual vectors and vectors to {\bf R}:
\bea
  T:~ &T^*_p\times\cdots\times T^*_p\times T_p\times\cdots
  \times T_p \rightarrow {\bf R}\nonumber \\
  & (k~{\rm times})\qquad\quad (l~{\rm times}) \label{1.44}
\eea
Here, ``$\times$'' denotes the Cartesian product, so that for example
$T_p\times T_p$ is the space of ordered pairs of vectors.  
Multilinearity means that the tensor acts linearly in each
of its arguments; for instance, for a tensor of type $(1,1)$, we have
\be
  T(a\omega+b\eta, cV+dW) = acT(\omega,V)+adT(\omega,W)+bcT(\eta,V)
  +bdT(\eta,W)\ .\label{1.45}
\ee
From this point of view, a scalar is a type $(0,0)$ tensor, a vector
is a type $(1,0)$ tensor, and a dual vector is a type $(0,1)$ tensor.

The space of all tensors of a fixed type $(k,l)$ forms a vector space;
they can be added together and multiplied by real numbers.  To construct
a basis for this space, we need to define a new operation known as the
{\bf tensor product}, denoted by $\otimes$.  If $T$ is a $(k,l)$ tensor
and $S$ is a $(m,n)$ tensor, we define a $(k+m,l+n)$ tensor
$T\otimes S$ by 
\bea
  \lefteqn{T\otimes S(\omega^{(1)},\ldots ,\omega^{(k)},\ldots 
  ,\omega^{(k+m)},V^{(1)},\ldots ,V^{(l)},\ldots ,V^{(l+n)})} \nonumber \\ 
  &  = 
  T(\omega^{(1)},\ldots ,\omega^{(k)},V^{(1)},\ldots ,V^{(l)})
  S(\omega^{(k+1)},\ldots ,\omega^{(k+m)},V^{(l+1)},\ldots ,V^{(l+n)})
  \ .\label{1.46}
\eea
(Note that the $\omega^{(i)}$ and $V^{(i)}$ are distinct
dual vectors and vectors, not components thereof.)
In other words, first act $T$ on the appropriate set of dual vectors and
vectors, and then act $S$ on the remainder, and then multiply the answers.
Note that, in general, $T\otimes S \neq S\otimes T$.

It is now straightforward to construct a basis for the space of all
$(k,l)$ tensors, by taking tensor products of basis vectors and dual
vectors; this basis will consist of all tensors of the form
\be
  \e{\mu_1}\otimes\cdots\otimes\e{\mu_k}\otimes
  \t{\nu_1}\otimes\cdots\otimes\t{\nu_l}\ .\label{1.47}
\ee
In a 4-dimensional spacetime there will be $4^{k+l}$ basis tensors
in all.  In component notation we then write our arbitrary tensor as
\be
  T = T^{\mu_1 \cdots \mu_k}{}_{\nu_1\cdots\nu_l}
  \e{\mu_1}\otimes\cdots\otimes\e{\mu_k}\otimes
  \t{\nu_1}\otimes\cdots\otimes\t{\nu_l}\ .\label{1.48}
\ee
Alternatively, we could define the components by acting the tensor
on basis vectors and dual vectors:
\be
  T^{\mu_1 \cdots \mu_k}{}_{\nu_1\cdots\nu_l} = 
  T(\t{\mu_1},\ldots, \t{\mu_k},\e{\nu_1},\ldots,\e{\nu_l})
  \ .\label{1.49}
\ee
You can check for yourself, using (1.33) and so forth, that these
equations all hang together properly.

As with vectors, we will usually take the shortcut of denoting the 
tensor $T$ by its components $T^{\mu_1 \cdots \mu_k}{}_{\nu_1\cdots\nu_l}$.
The action of the tensors on a set of vectors and dual vectors follows
the pattern established in (1.35):
\be
  T(\omega^{(1)},\ldots ,\omega^{(k)},V^{(1)},\ldots ,V^{(l)}) =
  T^{\mu_1 \cdots \mu_k}{}_{\nu_1\cdots\nu_l} 
  \omega^{(1)}_{\mu_1}\cdots\omega^{(k)}_{\mu_k}V^{(1)\nu_1}\cdots 
  V^{(l)\nu_l}\ .\label{1.50}
\ee
The order of the indices is obviously important, since the tensor
need not act in the same way on its various arguments.
Finally, the transformation of tensor components under Lorentz
transformations can be derived by applying what we already know
about the transformation of basis vectors and dual vectors.  The answer
is just what you would expect from index placement,
\be
  T^{\mu_1' \cdots \mu_k'}{}_{\nu_1'\cdots\nu_l'} = 
  \Lambda^{\mu_1'}{}_{\mu_1}\cdots\Lambda^{\mu_k'}{}_{\mu_k}
  \Lambda_{\nu_1'}{}^{\nu_1}\cdots\Lambda_{\nu_l'}{}^{\nu_l}
  T^{\mu_1 \cdots \mu_k}{}_{\nu_1\cdots\nu_l} \ .\label{1.51}
\ee
Thus, each upper index gets transformed like a vector, and each
lower index gets transformed like a dual vector.

Although we have defined tensors as linear maps from sets of vectors
and tangent vectors to {\bf R}, there is nothing that forces us to
act on a full collection of arguments.  Thus, a $(1,1)$ tensor also
acts as a map from vectors to vectors:
\be
  T^\mu{}_\nu :~ V^\nu\rightarrow T^\mu{}_\nu V^\nu\ .\label{1.52}
\ee
You can check for yourself that $T^\mu{}_\nu V^\nu$ is a vector
({\it i.e.} obeys the vector transformation law).  Similarly, we
can act one tensor on (all or part of) another tensor to obtain
a third tensor.  For example,
\be
  U^\mu{}_\nu = T^{\mu\rho}{}_\sigma S^{\sigma}{}_{\rho\nu}\label{1.53}
\ee
is a perfectly good $(1,1)$ tensor.

You may be concerned that this introduction to tensors has been somewhat
too brief, given the esoteric nature of the material.  In fact, the 
notion of tensors does not require a great deal of effort to master;
it's just a matter of keeping the indices straight, and the rules for
manipulating them are very natural.  Indeed, a number of books like to
{\it define} tensors as collections of numbers transforming according to
(1.51).  While this is operationally useful, it tends to obscure the
deeper meaning of tensors as geometrical entities with a life independent
of any chosen coordinate system.  There is, however, one subtlety which
we have glossed over.  The notions of dual vectors and tensors and bases
and linear maps belong to the realm of linear algebra, and are
appropriate whenever we have an abstract vector space at hand.  In the
case of interest to us we have not just a vector space, but a vector
space at each point in spacetime.  More often than not we are interested
in tensor fields, which can be thought of as tensor-valued functions on
spacetime.  Fortunately, none of the manipulations we defined above
really care whether we are dealing with a single vector space or a
collection of vector spaces, one for each event.  We will be able to
get away with simply calling things functions of $x^\mu$ when
appropriate.  However, you should keep straight the logical independence
of the notions we have introduced and their specific application to
spacetime and relativity.

Now let's turn to some examples of tensors.  First we consider the
previous example of column vectors and their duals, row vectors.  In
this system a $(1,1)$ tensor is simply a matrix, $M^i{}_j$.  Its action on
a pair $(\omega,V)$ is given by usual matrix multiplication:
\be
  M(\omega,V) = \left(\omega_1\ \omega_2\ \cdots\ \omega_{n}\right)
  \left(\matrix{M^1{}_1 & M^1{}_2 & \cdots & M^1{}_n \cr
  M^2{}_1 & M^2{}_2 & \cdots & M^2{}_n \cr
  \cdot &\cdot & \cdots &\cdot \cr \cdot &\cdot & \cdots &\cdot \cr 
  \cdot &\cdot & \cdots &\cdot \cr 
  M^n{}_1 & M^n{}_2 & \cdots & M^n{}_n \cr}\right)
  \left(\matrix{V^1 \cr V^2 \cr \cdot\cr \cdot \cr \cdot\cr
  V^{n}\cr}\right) = \omega_i M^i{}_j V^j\ .\label{1.54}
\ee
If you like, feel free to think of tensors as ``matrices with an
arbitrary number of indices.''

In spacetime, we have already seen some examples of tensors without
calling them that.  The most familiar example of a $(0,2)$ tensor is
the metric, $\eta_\mn$.  The action of the metric on two vectors is
so useful that it gets its own name, the {\bf inner product} (or
dot product):
\be
  \eta(V,W) = \eta_\mn V^\mu W^\nu = V\cdot W\ .\label{1.55}
\ee
Just as with the conventional Euclidean dot product, we will refer
to two vectors whose dot product vanishes as {\bf orthogonal}.
Since the dot product is a scalar, it is left invariant under
Lorentz transformations; therefore the basis vectors of any Cartesian
inertial frame, which are chosen to be orthogonal by definition, are
still orthogonal after a Lorentz transformation (despite the ``scissoring
together'' we noticed earlier).  The {\bf norm} of a vector is defined
to be inner product of the vector with itself; unlike in Euclidean
space, this number is not positive definite:
\[
  {\rm ~if~~}\eta_\mn V^\mu V^\nu
  {\rm ~~is~}\left\{\matrix{<0\ ,\  V^\mu {\rm ~is~timelike}\hfill\cr
  =0\ ,\ V^\mu {\rm ~is~lightlike~or~null}\hfill\cr
  >0\ ,\ V^\mu {\rm ~is~spacelike}\ .\hfill\cr}\right. 
\]
(A vector can have zero norm without being the zero vector.)
You will notice that the terminology is the same as that which we
earlier used to classify the relationship between two points in
spacetime; it's no accident, of course, and we will go into more
detail later.

Another tensor is the Kronecker delta $\delta^\mu_\nu$, of type $(1,1)$,
which you already know the components of.  Related to this and
the metric is the {\bf inverse metric} $\eta^\mn$, a type $(2,0)$
tensor defined as the inverse of the metric:
\be
  \eta^\mn\eta_{\nu\rho} = \eta_{\rho\nu}\eta^{\nu\mu}
  = \delta^\rho_\mu\ .\label{1.56}
\ee
In fact, as you can check, the inverse metric has exactly the same
components as the metric itself.  (This is only true in flat space in
Cartesian coordinates, and will fail to hold in more general situations.)
There is also the {\bf Levi-Civita tensor}, a $(0,4)$ tensor:
\be
  \epsilon_{\mu\nu\rho\sigma}=\left\{\matrix{+1 {\rm ~if~}\mu\nu\rho\sigma
  {\rm ~is~an~even~permutation~of~0123}\hfill\cr 
  -1 {\rm ~if~}\mu\nu\rho\sigma {\rm ~is~an~odd~permutation~of~0123}\hfill\cr
  0{\rm ~otherwise}\ .\hfill\cr}\right.
  \label{1.57}
\ee
Here, a ``permutation of 0123'' is an ordering of the numbers 0, 1,
2, 3 which can be obtained by starting with 0123 and exchanging two of the
digits; an even permutation is obtained by an even number of such exchanges,
and an odd permutation is obtained by an odd number.  Thus, for example,
$\epsilon_{0321}=-1$.

It is a remarkable property of the above tensors -- the metric, the inverse
metric, the Kronecker delta, and the Levi-Civita tensor -- that, even though
they all transform according to the tensor transformation law (1.51), their
components remain unchanged in {\it any} Cartesian coordinate system in
flat spacetime.  In some sense this makes them bad examples of tensors, since
most tensors do not have this property.  In fact, even these tensors do not
have this property once we go to more general coordinate systems, with the
single exception of the Kronecker delta.  This tensor has exactly the same
components in any coordinate system in any spacetime.  This makes sense
from the definition of a tensor as a linear map; the Kronecker tensor can
be thought of as the identity map from vectors to vectors (or from dual
vectors to dual vectors), which clearly must have the same components
regardless of coordinate system.  The other tensors (the metric, its inverse,
and the Levi-Civita tensor) characterize the structure of spacetime, and
all depend on the metric.  We shall therefore have to treat them more
carefully when we drop our assumption of flat spacetime.

A more typical example of a tensor is the {\bf electromagnetic field
strength tensor}.  We all know that the electromagnetic fields are made up
of the electric field vector $E_i$ and the magnetic field vector $B_i$.
(Remember that we use Latin indices for spacelike components 1,2,3.)
Actually these are only ``vectors'' under rotations in space, not under
the full Lorentz group.  In fact they are components of a $(0,2)$
tensor $F_{\mu\nu}$, defined by
\be
  F_{\mu\nu} = \left(\matrix{0&-E_1 &-E_2 & -E_3\cr E_1 & 0 & B_3 & -B_2\cr
  E_2 & -B_3 & 0 & B_1\cr E_3 & B_2 & -B_1 & 0 \cr}\right)
  =-F_{\nu\mu}\ .\label{1.58}
\ee
From this point of view it is easy to transform the electromagnetic
fields in one reference frame to those in another, by application of (1.51).
The unifying power of the tensor formalism is evident: rather than a
collection of two vectors whose relationship and transformation
properties are rather mysterious, we have a single tensor field to
describe all of electromagnetism.  (On the other hand, don't get 
carried away; sometimes it's more convenient to work in a single
coordinate system using the electric and magnetic field vectors.)

With some examples in hand we can now be a little more systematic 
about some properties of tensors.  First consider the operation of
{\bf contraction}, which turns a $(k,l)$ tensor into a $(k-1,l-1)$
tensor.  Contraction proceeds by summing over one upper and one lower
index:
\be
  S^{\mu\rho}{}_\sigma = T^{\mu\nu\rho}{}_{\sigma\nu}\ .\label{1.59}
\ee
You can check that the result is a well-defined tensor.  Of course it
is only permissible to contract an upper index with a lower index (as
opposed to two indices of the same type).  Note also that the order
of the indices matters, so that you can get different tensors by
contracting in different ways; thus,
\be
  T^{\mu\nu\rho}{}_{\sigma\nu}\neq T^{\mu\rho\nu}{}_{\sigma\nu}
  \label{1.60}
\ee
in general.

The metric and inverse metric can be used to {\bf raise and lower
indices} on tensors.  That is, given a tensor 
$T^{\alpha\beta}{}_{\gamma\delta}$, we can use the metric to define new
tensors which we choose to denote by the same letter $T$:
\bea
  T^{\alpha\beta\mu}{}_\delta &=&\eta^{\mu\gamma}
  T^{\alpha\beta}{}_{\gamma\delta}\ , \nonumber \\
  T_\mu{}^\beta{}_{\gamma\delta} &=&\eta_{\mu\alpha}
  T^{\alpha\beta}{}_{\gamma\delta}\ , \nonumber \\
  T_{\mu\nu}{}^{\rho\sigma} &=&\eta_{\mu\alpha}\eta_{\nu\beta}
  \eta^{\rho\gamma}\eta^{\sigma\delta}
  T^{\alpha\beta}{}_{\gamma\delta}\ ,  \label{1.61}
\eea
and so forth.  Notice that raising and lowering does not change the
position of an index relative to other indices, and also that ``free''
indices (which are not summed over) must be the same on both sides
of an equation, while ``dummy'' indices (which are summed over) only
appear on one side.  As an example, we can turn vectors and dual vectors
into each other by raising and lowering indices: 
\bea
  V_\mu &=&\eta_\mn V^\nu\nonumber \\
  \omega^\mu &=&\eta^\mn \omega_\nu\ . \label{1.62}
\eea
This explains why the gradient in three-dimensional flat Euclidean space
is usually thought of as an ordinary vector, even though we have seen
that it arises as a dual vector; in Euclidean space (where the metric
is diagonal with all entries $+1$) a dual vector is turned into a vector
with precisely the same components when we raise its index.  You may then
wonder why we have belabored the distinction at all.  One simple reason,
of course, is that in a Lorentzian spacetime the components are not
equal:
\be
  \omega^\mu = (-\omega_0, \omega_1, \omega_2, \omega_3)\ .\label{1.63}
\ee
In a curved spacetime, where the form of the metric is generally more
complicated, the difference is rather more dramatic.  But there
is a deeper reason, namely that tensors generally have a ``natural''
definition which is independent of the metric.  Even though we will
always have a metric available, it is helpful to be aware of the logical
status of each mathematical object we introduce.  The gradient, and
its action on vectors, is perfectly well defined regardless of any metric,
whereas the ``gradient with upper indices'' is not.  (As an example, we
will eventually want to take variations of functionals with respect to
the metric, and will therefore have to know exactly how the functional 
depends on the metric, something that is easily obscured by the index 
notation.)

Continuing our compilation of tensor jargon, we refer to a tensor as
{\bf symmetric} in any of its indices if it is unchanged under exchange
of those indices.  Thus, if 
\be
  S_{\mu\nu\rho} = S_{\nu\mu\rho}\ ,\label{1.64}
\ee
we say that $S_{\mu\nu\rho}$ is symmetric in its first two indices, while
if
\be
  S_{\mu\nu\rho} = S_{\mu\rho\nu} = S_{\rho\mu\nu} = S_{\nu\mu\rho}
  = S_{\nu\rho\mu} = S_{\rho\nu\mu} \ ,\label{1.65}
\ee
we say that $S_{\mu\nu\rho}$ is symmetric in all three of its indices.
Similarly, a tensor is {\bf antisymmetric} (or ``skew-symmetric'') in
any of its indices if it changes sign when those indices are exchanged;
thus,
\be
  A_{\mu\nu\rho} = -A_{\rho\nu\mu}\label{1.66}
\ee
means that $A_{\mu\nu\rho}$ is antisymmetric in its first and third
indices (or just ``antisymmetric in $\mu$ and $\rho$'').  If a tensor
is (anti-) symmetric in all of its indices, we refer to it as simply
(anti-) symmetric (sometimes with the redundant modifier ``completely'').  
As examples, the metric $\eta_{\mn}$ and the inverse
metric $\eta^{\mu\nu}$ are symmetric, while the Levi-Civita tensor
$\epsilon_{\mu\nu\rho\sigma}$ and the electromagnetic field strength
tensor $F_{\mu\nu}$ are antisymmetric.  (Check for yourself that if you
raise or lower a set of indices which are symmetric or antisymmetric,
they remain that way.)  Notice that it makes no sense to exchange upper
and lower indices with each other, so don't succumb to the temptation
to think of the Kronecker delta $\delta^\alpha_\beta$ as symmetric.
On the other hand, the fact that lowering an index on $\delta^\alpha_\beta$
gives a symmetric tensor (in fact, the metric) means that the order of
indices doesn't really matter, which is why we don't keep track index
placement for this one tensor.

Given any tensor, we can {\bf symmetrize} (or antisymmetrize) any
number of its upper or lower indices.  To symmetrize, we take the
sum of all permutations of the relevant indices and divide by the 
number of terms:
\be
  T_{(\mu_1\mu_2\cdots\mu_n)\rho}{}^\sigma = {1\over {n!}}
  \left(T_{\mu_1\mu_2\cdots\mu_n\rho}{}^\sigma +
  {\rm sum~over~permutations~of~indices~}\mu_1\cdots\mu_n \right)
  \ ,\label{1.67}
\ee
while antisymmetrization comes from the alternating sum:
\be
  T_{[\mu_1\mu_2\cdots\mu_n]\rho}{}^\sigma = {1\over {n!}}
  \left(T_{\mu_1\mu_2\cdots\mu_n\rho}{}^\sigma +
  {\rm alternating~sum~over~permutations~of~indices~ }
  \mu_1\cdots\mu_n \right)\ .\label{1.68}
\ee
By ``alternating sum'' we mean that permutations which are the result
of an odd number of exchanges are given a minus sign, thus: 
\be
  T_{[\mu\nu\rho]\sigma} = {1\over 6}\left(T_{\mu\nu\rho\sigma} 
  - T _{\mu\rho\nu\sigma} + T_{\rho\mu\nu\sigma} - T_{\nu\mu\rho\sigma} 
  + T_{\nu\rho\mu\sigma} - T_{\rho\nu\mu\sigma} 
  \right)\ .\label{1.69}
\ee
Notice that round/square brackets denote symmetrization/antisymmetrization.
Furthermore, we may sometimes want to (anti-) symmetrize indices which
are not next to each other, in which case we use vertical bars to denote
indices not included in the sum:
\be
  T_{(\mu |\nu |\rho)} = {1\over 2}\left(T_{\mu\nu\rho}
  + T_{\rho\nu\mu}\right)\ .\label{1.70}
\ee
Finally, some people use a convention in which the factor of $1/n!$
is omitted.  The one used here is a good one, since (for example) a
symmetric tensor satisfies
\be
  S_{\mu_1 \cdots \mu_n} = S_{(\mu_1 \cdots \mu_n)} \ ,\label{1.71}
\ee
and likewise for antisymmetric tensors.

We have been very careful so far to distinguish clearly between things
that are always true (on a manifold with arbitrary metric) and things
which are only true in Minkowski space in Cartesian coordinates.  One
of the most important distinctions arises with {\bf partial derivatives}.
If we are working in flat spacetime with Cartesian coordinates, 
then the partial derivative of a $(k,l)$ tensor is a
$(k,l+1)$ tensor; that is,
\be
  T_\alpha{}^\mu{}_\nu = \partial_\alpha R^\mu{}_\nu \label{1.72}
\ee
transforms properly under Lorentz transformations.  However, this will
no longer be true in more general spacetimes, and we will have to
define a ``covariant derivative'' to take the place of the partial
derivative.  Nevertheless, we can still use the fact that partial
derivatives give us tensor in this special case, as long as we keep
our wits about us.  (The one exception to this warning is the partial
derivative of a scalar, $\partial_\alpha\phi$, which is a perfectly
good tensor [the gradient] in any spacetime.)

We have now accumulated enough tensor know-how to illustrate some of
these concepts using actual physics.  Specifically, we will examine
{\bf Maxwell's equations} of electrodynamics.  In $19^{\rm th}$-century
notation, these are 
\bea
  \nabla\times{\bf B} - \p{t}{\bf E} &=&4\pi{\bf J}\nonumber \\
  \nabla\cdot{\bf E} &=&4\pi\rho\nonumber \\
  \nabla\times{\bf E} + \p{t}{\bf B} &=&0\nonumber \\
  \nabla\cdot{\bf B} &=&0\ . \label{1.73}
\eea
Here, {\bf E} and {\bf B} are the electric and magnetic field
3-vectors, {\bf J} is the current, $\rho$ is the charge density, and
$\nabla\times$ and $\nabla\cdot$ are the conventional curl and divergence.
These equations are invariant under Lorentz transformations, of course;
that's how the whole business got started.  But they don't look obviously
invariant; our tensor notation can fix that.  Let's begin by writing
these equations in just a slightly different notation,
\bea
  \epsilon^{ijk}\p{j}B_k - \p0 E^i &=&
  4\pi J^i\nonumber \\
  \p{i}E^i &=&4\pi J^0\nonumber \\
  \epsilon^{ijk}\p{j}E_k + \p0 B^i &=&0\nonumber \\
  \p{i}B^i &=&0\ . \label{1.74}
\eea
In these expressions, spatial indices have been raised and lowered
with abandon, without any attempt to keep straight where the metric
appears.  This is because $\delta_{ij}$ is the metric on flat 3-space,
with $\delta^{ij}$ its inverse (they are equal as matrices).  We can
therefore raise and lower indices at will, since the components don't
change.  Meanwhile,
the three-dimensional Levi-Civita tensor $\epsilon^{ijk}$ is defined
just as the four-dimensional one, although with one fewer index.  We
have replaced the charge density by $J^0$; this is legitimate because
the density and current together form the {\bf current 4-vector},
$J^\mu = (\rho,J^1,J^2,J^3)$.

From these expressions, and the definition (1.58) of the field strength
tensor $F_{\mu\nu}$, it is easy to get a completely tensorial 
$20^{\rm th}$-century version of Maxwell's equations.  Begin by noting
that we can express the field strength with upper indices as
\bea
  F^{0i} &=&E^i\nonumber \\  F^{ij} &=&\epsilon^{ijk}B_k\ . \label{1.75}
\eea
(To check this, note for example that $F^{01} = \eta^{00}\eta^{11}
F_{01}$ and $F^{12} = \epsilon^{123}B_3$.)  Then the first two 
equations in (1.74) become
\bea
  \p{j}F^{ij} - \p0 F^{0i} & =& 4\pi J^i\nonumber \\
  \p{i}F^{0i} &=&4\pi J^0\ . \label{1.76}
\eea
Using the antisymmetry of $F^{\mn}$, we see that these may be combined 
into the single tensor equation
\be
  \p\mu F^{\nu\mu} = 4\pi J^\nu\ .\label{1.77}
\ee
A similar line of reasoning, which is left as an exercise to you,
reveals that the third and fourth equations in (1.74) can be written
\be
  \p{[\mu} F_{\nu\lambda]}= 0\ .\label{1.78}
\ee
The four traditional Maxwell equations are thus replaced by two, 
thus demonstrating the economy of tensor notation.  More importantly,
however, both sides of equations (1.77) and (1.78) manifestly transform
as tensors; therefore, if they are true in one inertial frame, they
must be true in any Lorentz-transformed frame.  This is why tensors
are so useful in relativity --- we often want to express relationships
without recourse to any reference frame, and it is necessary that the
quantities on each side of an equation transform in the same way under
change of coordinates.  As a matter of jargon, we will sometimes refer
to quantities which are written in terms of tensors as {\bf covariant}
(which has nothing to do with ``covariant'' as opposed to 
``contravariant'').  Thus, we say that (1.77) and (1.78) together serve
as the covariant form of Maxwell's equations, while (1.73) or (1.74)
are non-covariant.

Let us now introduce a special class of tensors, known as 
{\bf differential forms} (or just ``forms'').  A differential
$p$-form is a $(0,p)$ tensor which is completely antisymmetric.
Thus, scalars are automatically 0-forms, and dual vectors are automatically
one-forms (thus explaining this terminology from a while back).  We also 
have the 2-form $F_{\mu\nu}$ and the 4-form $\epsilon_{\mu\nu\rho\sigma}$.
The space of all $p$-forms is denoted $\Lambda^p$, and the space of all
$p$-form fields over a manifold $M$ is denoted $\Lambda^p(M)$.
A semi-straightforward exercise in combinatorics reveals that the number
of linearly independent $p$-forms on an $n$-dimensional vector space is 
$n!/(p!(n-p)!)$.  So at a point on a 4-dimensional spacetime there is one 
linearly independent 0-form, 
four 1-forms, six 2-forms, four 3-forms, and one 4-form.  There
are no $p$-forms for $p>n$, since all of the components will automatically
be zero by antisymmetry.

Why should we care about differential forms?  This is a hard question to
answer without some more work, but the basic idea is that forms can
be both differentiated and integrated, without the help of any 
additional geometric structure.  We will delay integration theory until
later, but see how to differentiate forms shortly.

Given a $p$-form $A$ and a $q$-form $B$, we can form a $(p+q)$-form
known as the {\bf wedge product} $A\wedge B$ by taking the antisymmetrized 
tensor product:
\be
  (A\wedge B)_{\mu_1 \cdots \mu_{p+q}} = {{(p+q)!}\over{p!\ q!}}
  A_{[\mu_1\cdots \mu_p} B_{\mu_{p+1}\cdots\mu_{p+q}]}\ .\label{1.79}
\ee
Thus, for example, the wedge product of two 1-forms is
\be
  (A\wedge B)_{\mu\nu} = 2A_{[\mu}B_{\nu]} = A_\mu B_\nu
  - A_\nu B_\mu\ .\label{1.80}
\ee
Note that
\be
  A\wedge B = (-1)^{pq} B\wedge A\ ,\label{1.81}
\ee
so you can alter the order of a wedge product if you are careful
with signs.

The {\bf exterior derivative} ``d'' allows us to differentiate
$p$-form fields to obtain $(p+1)$-form fields.  It is defined as an
appropriately normalized antisymmetric partial derivative:
\be
  ({\rm d}A)_{\mu_1\cdots\mu_{p+1}} = (p+1)\p{[\mu_1}
  A_{\mu_2\cdots \mu_{p+1}]}\ .\label{1.82}
\ee
The simplest example is the gradient, which is the exterior derivative
of a 1-form:
\be
  (\d \phi)_\mu = \p\mu \phi\ .\label{1.83}
\ee
The reason why the exterior derivative deserves special attention
is that {\it it is a tensor}, even in curved spacetimes, unlike its
cousin the partial derivative.  Since we haven't studied curved
spaces yet, we cannot prove this, but (1.82) defines an honest tensor
no matter what the metric and coordinates are.

Another interesting fact about exterior differentiation is that, for
any form $A$,
\be
  \d(\d A) =0\ ,\label{1.84}
\ee
which is often written $\d^2 =0$.  This identity is a consequence of
the definition of $\d$ and the fact that partial derivatives commute,
$\p\alpha\p\beta = \p\beta\p\alpha$ (acting on anything).  This leads
us to the following mathematical aside, just for fun.
We define a $p$-form $A$ to be {\bf closed} if $\d A=0$, and {\bf exact}
if $A = \d B$ for some $(p-1)$-form $B$.  Obviously, all exact forms are
closed, but the converse is not necessarily true.  On a manifold $M$, 
closed $p$-forms comprise a vector space $Z^p(M)$, and exact forms
comprise a vector space $B^p(M)$.  Define a new vector space as the
closed forms modulo the exact forms:
\be
  H^p(M) = {{Z^p(M)}\over{B^p(M)}}\ .\label{1.85}
\ee
This is known as the $p$th de~Rham cohomology vector space, and depends
only on the topology of the manifold $M$.  (Minkowski space is topologically
equivalent to {\bf R}$^4$, which is uninteresting, so that all of the
$H^p(M)$ vanish for $p>0$; for $p=0$ we have $H^0(M)={\bf R}$.  
Therefore in Minkowski space all closed forms 
are exact except for zero-forms; zero-forms can't be exact since
there are no $-1$-forms for them to be the exterior derivative of.)
It is striking that information about the topology can be extracted in 
this way, which essentially involves the solutions to differential 
equations.  The dimension $b_p$ of the space $H^p(M)$ is called the $p$th
Betti number of $M$, and the Euler characteristic is given by the
alternating sum
\be
  \chi(M) = \sum_{p=0}^{n} (-1)^p b_p\ .\label{1.86}
\ee
Cohomology theory is the basis for much of modern differential topology.

Moving back to reality, the final operation on differential forms we will
introduce is {\bf Hodge duality}.  We define the ``Hodge star operator''
on an $n$-dimensional manifold as a map from $p$-forms to $(n-p)$-forms,
\be
  (*A)_{\mu_1\cdots\mu_{n-p}} = {1\over {p!}} 
  \epsilon^{\nu_1\cdots\nu_p}{}_{\mu_1\cdots\mu_{n-p}}
  A_{\nu_1\cdots\nu_p}\ ,\label{1.87}
\ee
mapping $A$ to ``$A$ dual''.  Unlike our other operations
on forms, the Hodge dual does depend on the metric of the manifold
(which should be obvious, since we had to raise some indices on the
Levi-Civita tensor in order to define (1.87)).  Applying the Hodge star
twice returns either plus or minus the original form:
\be
  **A = (-1)^{s+p(n-p)}A\ ,\label{1.88}
\ee
where $s$ is the number of minus signs in the eigenvalues of the
metric (for Minkowski space, $s=1$).  

Two facts on the Hodge dual:  First, ``duality'' in the sense of Hodge is 
different than the relationship between vectors and dual vectors, although 
both can be thought of as the space of linear maps from the original space 
to {\bf R}.  Notice that the dimensionality of the space of $(n-p)$-forms
is equal to that of the space of $p$-forms, so this has at least a chance
of being true.  In the case of forms, the linear map defined by an 
$(n-p)$-form acting on a $p$-form is given by the dual of the wedge
product of the two forms.  Thus, if $A^{(n-p)}$ is an $(n-p)$-form and 
$B^{(p)}$ is a $p$-form at some point in spacetime, we have
\be
  *(A^{(n-p)}\wedge B^{(p)}) \in {\bf R}\ .\label{1.89}
\ee
The second fact concerns differential forms in 3-dimensional Euclidean
space.  The Hodge dual of the wedge product of two 1-forms gives another
1-form:
\be
  *(U\wedge V)_i = \epsilon_i{}^{jk}U_j V_k\ .\label{1.90}
\ee
(All of the prefactors cancel.)  Since 1-forms in Euclidean space are
just like vectors, we have a map from two vectors to a single vector.
You should convince yourself that this is just the conventional
cross product, and that the appearance of the Levi-Civita tensor
explains why the cross product changes sign under parity (interchange
of two coordinates, or equivalently basis vectors).  This is why the
cross product only exists in three dimensions --- because only in three
dimensions do we have an interesting map from two dual vectors to a
third dual vector.  If you wanted to you could define a map from
$n-1$ one-forms to a single one-form, but I'm not sure it would be
of any use.

Electrodynamics provides an especially compelling example of the use
of differential forms.  From the definition of the exterior derivative,
it is clear that equation (1.78) can be concisely expressed as closure
of the two-form $F_{\mu\nu}$:
\be
  \d F = 0\ .\label{1.91}
\ee
Does this mean that $F$ is also exact?  Yes; as we've noted, Minkowski
space is topologically trivial, so all closed forms are exact.  There
must therefore be a one-form $A_\mu$ such that
\be
  F = \d A\ .\label{1.92}
\ee
This one-form is the familiar {\bf vector potential} of electromagnetism,
with the $0$ component given by the scalar potential, $A_0 = \phi$. 
If one starts from the view that the $A_\mu$ is the fundamental field
of electromagnetism, then (1.91) follows as an identity (as opposed to a
dynamical law, an equation of motion).  Gauge invariance is expressed
by the observation that the theory is invariant under $A \rightarrow
A +\d \lambda$ for some scalar (zero-form) $\lambda$, and this is also 
immediate from the relation (1.92). The
other one of Maxwell's equations, (1.77), can be expressed as an equation
between three-forms:
\be
  \d(*F) = 4\pi(* J)\ ,\label{1.93}
\ee
where the current one-form $J$ is just the current four-vector with
index lowered.  Filling in the details is left for you to do.

As an intriguing aside, Hodge duality is the basis for one of the hottest
topics in theoretical physics today.  It's hard not to notice that the
equations (1.91) and (1.93) look very similar.  Indeed, if we set $J_\mu=0$,
the equations are invariant under the ``duality transformations''
\bea
  F &\rightarrow \ast F\ ,\nonumber \\
  \ast F &\rightarrow -F\ .  \label{1.94}
\eea
We therefore say that the vacuum Maxwell's equations are duality
invariant, while the invariance is spoiled in the presence of charges.
We might imagine that magnetic as well as electric monopoles existed in
nature; then we could add a magnetic current term $4\pi(*J_M)$ to the
right hand side of (1.91), and the equations would be invariant under duality
transformations plus the additional replacement $J \leftrightarrow J_M$.
(Of course a nonzero right hand side to (1.91) is inconsistent with $F=\d A$,
so this idea only works if $A_\mu$ is not a fundamental variable.)
Long ago Dirac considered the idea of magnetic monopoles and showed that
a necessary condition for their existence is that the fundamental
monopole charge be inversely proportional to the fundamental electric
charge.  Now, the fundamental electric charge is a small number;
electrodynamics is ``weakly coupled'', which is why perturbation theory
is so remarkably successful in quantum electrodynamics (QED).  But
Dirac's condition on magnetic charges implies that a duality 
transformation takes a theory of weakly coupled electric charges to
a theory of strongly coupled magnetic monopoles (and vice-versa).
Unfortunately monopoles don't exist (as far as we know), so these
ideas aren't directly applicable to electromagnetism; but there are
some theories (such as supersymmetric non-abelian gauge theories) for
which it has been long conjectured that some sort of duality symmetry
may exist.  If it did, we would have the opportunity to analyze a 
theory which looked strongly coupled (and therefore hard to solve) by
looking at the weakly coupled dual version.  Recently work by Seiberg 
and Witten and others has provided very strong evidence that this is
exactly what happens in certain theories.  The hope is that these
techniques will allow us to explore various phenomena which we know
exist in strongly coupled quantum field theories, such as confinement
of quarks in hadrons.

We've now gone over essentially everything there is to know about the
care and feeding of tensors.  In the next section we will look more
carefully at the rigorous definitions of manifolds and tensors, but
the basic mechanics have been pretty well covered.  Before jumping
to more abstract mathematics, let's review how
physics works in Minkowski spacetime.

Start with the worldline of a single particle.  This is specified
by a map ${\bf R} \rightarrow M$, where $M$ is the manifold representing
spacetime; we usually think of the path as a parameterized curve
$x^\mu(\lambda)$.  As mentioned earlier, the tangent vector to this
path is $dx^\mu/d\lambda$ (note that it depends on the parameterization).  
An object of primary interest is the norm
of the tangent vector, which serves to characterize the path; if the
tangent vector is timelike/null/spacelike at some parameter value
$\lambda$, we say that the path is timelike/null/spacelike at that
point.  This explains why the same words are used to classify vectors
in the tangent space and intervals between two points --- because a
straight line connecting, say, two timelike separated points will 
itself be timelike at every point along the path.

\begin{figure}
  \centerline{
  \psfig{figure=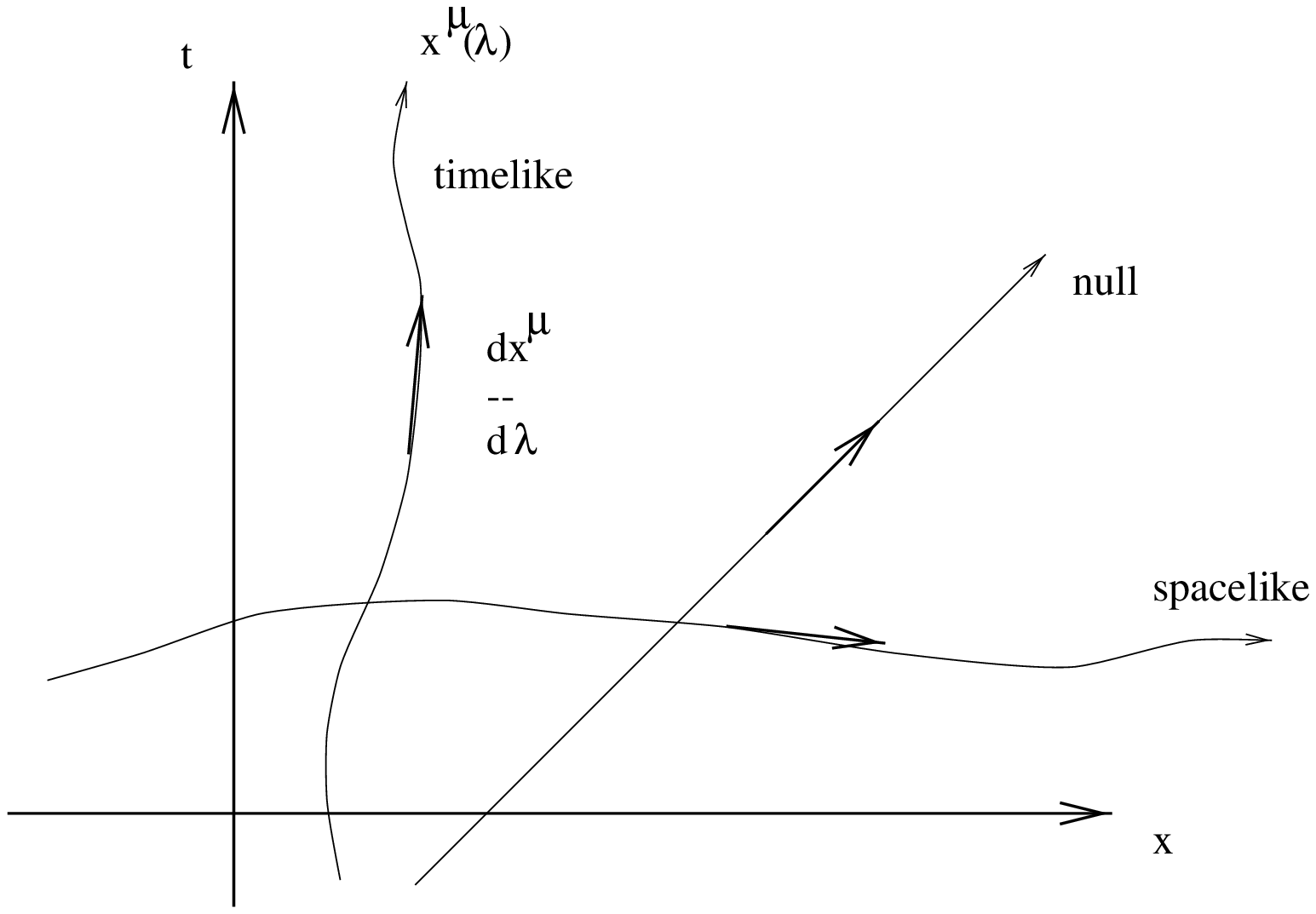,angle=-90,height=8cm}}
\end{figure}

Nevertheless, it's important to be aware of the sleight of hand which is 
being pulled here.  The metric, as a $(0,2)$ tensor, is a machine which 
acts on two vectors (or two copies of the same vector) to produce a number.
It is therefore very natural to classify tangent vectors according to
the sign of their norm.  But the interval between two points isn't
something quite so natural; it depends on a specific choice of path
(a ``straight line'') which connects the points, and this choice in
turn depends on the fact that spacetime is flat (which allows a 
unique choice of straight line between the points).  A more natural
object is the {\bf line element}, or infinitesimal interval:
\be
  ds^2 = \eta_\mn dx^\mu dx^\nu\ .\label{1.95}
\ee
From this definition it is tempting to take the square root and
integrate along a path to obtain a finite interval.  But since $ds^2$
need not be positive, we define different procedures for different
cases.  For spacelike paths we define the {\bf path length}
\be
  \Delta s = \int \sqrt{\eta_\mn{{dx^\mu}\over{d\lambda}}
  {{dx^\nu}\over{d\lambda}}}\ d\lambda\ ,\label{1.96}
\ee
where the integral is taken over the path.  For null paths the interval
is zero, so no extra formula is required.  For timelike paths we define
the {\bf proper time}
\be
  \Delta \tau = \int \sqrt{-\eta_\mn{{dx^\mu}\over{d\lambda}}
  {{dx^\nu}\over{d\lambda}}}\ d\lambda\ ,\label{1.97}
\ee
which will be positive.  Of course we may consider paths that are
timelike in some places and spacelike in others, but fortunately
it is seldom necessary since the paths of physical particles never
change their character (massive particles move on timelike paths,
massless particles move on null paths).  Furthermore, the phrase
``proper time'' is especially appropriate, since $\tau$ {\it actually
measures the time elapsed on a physical clock carried along the path}.
This point of view makes the ``twin paradox'' and similar puzzles
very clear; two worldlines, not necessarily straight, which intersect at 
two different events in spacetime will have proper times measured by the
integral (1.97) along the appropriate paths, and these two numbers will
in general be different even if the people travelling along them were
born at the same time.

Let's move from the consideration of paths in general to the paths
of massive particles (which will always be timelike).  Since the proper 
time is measured by a clock travelling on a timelike worldline,
it is convenient to use $\tau$ as the parameter along the path.
That is, we use (1.97) to compute $\tau(\lambda)$, which (if $\lambda$
is a good parameter in the first place) we can invert to obtain
$\lambda(\tau)$, after which we can think of the path as
$x^\mu(\tau)$.  The tangent vector in this parameterization is known
as the {\bf four-velocity}, $U^\mu$:
\be
  U^\mu = {{dx^\mu}\over{d\tau}}\ .\label{1.98}
\ee
Since $d\tau^2 = -\eta_\mn dx^\mu dx^\nu$, the four-velocity is
automatically normalized:
\be
  \eta_\mn U^\mu U^\nu = -1\ .\label{1.99}
\ee
(It will always be negative, since we are only defining it for timelike
trajectories.  You could define an analogous vector for spacelike
paths as well; null paths give some extra problems since the norm is
zero.)  In the rest frame of a particle, its four-velocity has
components $U^\mu = (1,0,0,0)$.

A related vector is the {\bf energy-momentum four-vector}, defined by
\be
  p^\mu = m U^\mu\ ,\label{1.100}
\ee
where $m$ is the mass of the particle.  The mass is a fixed quantity
independent of inertial frame; what you may be used to thinking of as
the ``rest mass.''  It turns out to be much more convenient to take
this as the mass once and for all, rather than thinking of mass as
depending on velocity.  The {\bf energy} of a particle is simply
$p^0$, the timelike component of its energy-momentum vector. 
Since it's only one component of a four-vector, it is not invariant
under Lorentz transformations; that's to be expected, however, since
the energy of a particle at rest is not the same as that of the same
particle in motion.  In the
particle's rest frame we have $p^0 =m$; recalling that we have set
$c=1$, we find that we have found the equation that made Einstein
a celebrity, $E=mc^2$.  (The field equations of general relativity are
actually much more important than this one, but ``$R_\mn - {1\over 2}
Rg_\mn = 8\pi G T_\mn$'' doesn't elicit the visceral reaction that
you get from ``$E=mc^2$''.)  In a moving frame we can find the components
of $p^\mu$ by performing a Lorentz transformation; for a particle 
moving with (three-) velocity $v$ along the $x$ axis we have
\be
  p^\mu = (\gamma m, v\gamma m, 0 ,0)\ ,\label{1.101}
\ee
where $\gamma = 1/\sqrt{1-v^2}$.  For small $v$, this gives
$p^0 = m+{1\over 2}mv^2$ (what we usually think of as rest energy
plus kinetic energy) and $p^1 = mv$ (what we usually think of as
[Newtonian] momentum).  So the energy-momentum vector lives up to its 
name.

The centerpiece of pre-relativity physics is Newton's 2nd Law,
or ${\bf f}=m{\bf a} = d{\bf p}/dt$.  An analogous equation should hold
in SR, and the requirement that it be tensorial leads us directly
to introduce a force four-vector $f^\mu$ satisfying
\be
  f^\mu = m{{d^2}\over{d\tau^2}}x^\mu(\tau) = {{d}\over{d\tau}}
  p^\mu(\tau)\ .\label{1.102}
\ee
The simplest example of a force in Newtonian physics is the force
due to gravity.  In relativity, however, gravity is not described
by a force, but rather by the curvature of spacetime itself.  Instead,
let us consider electromagnetism.  The three-dimensional Lorentz
force is given by ${\bf f} = q({\bf E} + {\bf v}\times{\bf B})$,
where $q$ is the charge on the particle.  We would like a tensorial
generalization of this equation.  There turns out to be a unique answer:
\be
  f^\mu = qU^\lambda F_\lambda{}^\mu\ .\label{1.103}
\ee
You can check for yourself that this reduces to the Newtonian version 
in the limit of small velocities.  Notice how the requirement that
the equation be tensorial, which is one way of guaranteeing Lorentz
invariance, severely restricted the possible expressions we could
get.  This is an example of a very general phenomenon, in which 
a small number of an apparently endless variety of possible physical laws 
are picked out by the demands of symmetry.

Although $p^\mu$ provides a complete description of the energy and
momentum of a particle, for extended systems it is necessary to go
further and define the {\bf energy-momentum tensor} (sometimes called
the stress-energy tensor), $T^\mn$.  This is a symmetric $(2,0)$ tensor 
which tells us all we need to know about the energy-like aspects of a 
system: energy density, pressure, stress, and so
forth.  A general definition of $T^\mn$ is ``the flux of four-momentum
$p^\mu$ across a surface of constant $x^\nu$''.  To make this more
concrete, let's consider the very general category of matter which
may be characterized as a {\bf fluid} --- a continuum of matter
described by macroscopic quantities  such as temperature, pressure,
entropy, viscosity, etc.  In fact this definition is so general that
it is of little use.  In general relativity essentially all interesting
types of matter can be thought of as {\bf perfect fluids}, from
stars to electromagnetic fields to the entire universe.  Schutz
defines a perfect fluid to be one with no heat conduction and no
viscosity, while Weinberg defines it as a fluid which looks isotropic
in its rest frame; these two viewpoints turn out to be equivalent.
Operationally, you should think of a perfect fluid as one which may
be completely characterized by its pressure and density.

To understand perfect fluids, let's start with the even simpler example 
of {\bf dust}.  Dust is defined as a collection of particles at rest
with respect to each other, or alternatively as a perfect fluid with
zero pressure.  Since the particles all have an equal velocity in
any fixed inertial frame, we can imagine a ``four-velocity field''
$U^\mu(x)$ defined all over spacetime.  (Indeed, its components are
the same at each point.)  Define the {\bf number-flux four-vector}
to be
\be
  N^\mu = n U^\mu\ ,\label{1.104}
\ee
where $n$ is the number density of the particles as measured in their
rest frame.  Then $N^0$ is the number density of particles as measured
in any other frame, while $N^i$ is the flux of particles in the $x^i$
direction.  Let's now imagine that each of the particles have the same
mass $m$.  Then in the rest frame the energy density of the dust is
given by
\be
  \rho = nm\ .\label{1.105}
\ee
By definition, the energy density completely specifies the dust.  But
$\rho$ only measures the energy density in the rest frame; what about
other frames?  We notice that both $n$ and $m$ are $0$-components of
four-vectors in their rest frame; specifically, $N^\mu = (n,0,0,0)$
and $p^\mu = (m,0,0,0)$.  Therefore $\rho$ is the $\mu = 0$, 
$\nu =0$ component of the tensor $p\otimes N$ as measured in its
rest frame.  We are therefore led to define the energy-momentum tensor
for dust:
\be
  T^\mn_{\rm dust} = p^\mu N^\nu = nm U^\mu U^\nu = \rho U^\mu U^\nu
  \ ,\label{1.106}
\ee
where $\rho$ is defined as the energy density in the rest frame.

Having mastered dust, more general perfect fluids are not much more
complicated.  Remember that ``perfect'' can be taken to mean
``isotropic in its rest frame.''  This in turn means that $T^\mn$
is diagonal --- there is no net flux of any component of momentum
in an orthogonal direction.  Furthermore, the nonzero spacelike 
components must all be equal, $T^{11} = T^{22}=T^{33}$.  The only
two independent numbers are therefore $T^{00}$ and one of the
$T^{ii}$; we can choose to call the first of these the energy density
$\rho$, and the second the pressure $p$.  (Sorry that it's the same
letter as the momentum.)  The energy-momentum tensor of a perfect
fluid therefore takes the following form in its rest frame:
\be
  T^\mn = \left(\matrix{\rho&0&0&0\cr 0&p&0&0\cr 0&0&p&0\cr
  0&0&0&p\cr}\right)\ .\label{1.107}
\ee
We would like, of course, a formula which is good in any frame.
For dust we had $T^\mn = \rho U^\mu U^\nu$, so we might begin by
guessing $(\rho+p)U^\mu U^\nu$, which gives
\be
  \left(\matrix{\rho+p &0&0&0\cr 0&0&0&0\cr 0&0&0&0\cr
  0&0&0&0\cr}\right)\ .\label{1.108}
\ee
To get the answer we want we must therefore add
\be
  \left(\matrix{-p&0&0&0\cr 0&p&0&0\cr 0&0&p&0\cr
  0&0&0&p\cr}\right)\ .\label{1.109}
\ee
Fortunately, this has an obvious covariant generalization, namely
$p\eta^\mn$.  Thus, the general form of the energy-momentum 
tensor for a perfect fluid is
\be
  T^\mn = (\rho+p) U^\mu U^\nu + p\eta^\mn\ .\label{1.110}
\ee
This is an important formula for applications such as stellar
structure and cosmology.

As further examples, let's consider the energy-momentum tensors
of electromagnetism and scalar field theory.  Without any
explanation at all, these are given by
\be
  T^\mn_{\rm e+m} = {-1\over{4\pi}}(F^{\mu\lambda}F^\nu{}_\lambda
  -{1\over 4}\eta^{\mu\nu} F^{\lambda\sigma}F_{\lambda\sigma})
  \ ,\label{1.111}
\ee
and
\be
  T^\mn_{\rm scalar} = \eta^{\mu\lambda}\eta^{\nu\sigma}
  \p\lambda\phi\p\sigma\phi - {1\over 2}\eta^\mn (\eta^{\lambda\sigma}
  \p\lambda\phi\p\sigma\phi + m^2\phi^2)\ .\label{1.112}
\ee
You can check for yourself that, for example, $T^{00}$ in each case
is equal to what you would expect the energy density to be.

Besides being symmetric, $T^\mn$ has the even more important property
of being {\it conserved}.  In this context, conservation is
expressed as the vanishing of the ``divergence'':
\be
  \p\mu T^\mn =0\ .\label{1.113}
\ee  
This is a set of four equations, one for each value of $\nu$.  The
$\nu =0$ equation corresponds to conservation of energy,  while
$\p\mu T^{\mu k}=0$ expresses conservation of the $k^{\rm th}$
component of the momentum.  We are not going to prove this in
general; the proof follows for any individual source of matter from
the equations of motion obeyed by that kind of matter.  In fact, one
way to define $T^\mn$ would be ``a $(2,0)$ tensor with units of
energy per volume, which is conserved.''  You can prove conservation
of the energy-momentum tensor for electromagnetism, for example, by
taking the divergence of (1.111) and using Maxwell's equations
as previously discussed.

A final aside: we have already mentioned that in general relativity
gravitation does not count as a ``force.''  As a related point, the
gravitational field also does not have an energy-momentum tensor.
In fact it is very hard to come up with a sensible local expression
for the energy of a gravitational field; a number of suggestions
have been made, but they all have their drawbacks.  Although there
is no ``correct'' answer, it is an important issue from the point of
view of asking seemingly reasonable questions such as ``What is the
energy emitted per second from a binary pulsar as the result of
gravitational radiation?''

\eject

\thispagestyle{plain}

\setcounter{equation}{0}

\noindent{December 1997 \hfill {\sl Lecture Notes on General Relativity}
\hfill{Sean M.~Carroll}}

\vskip .2in

\setcounter{section}{1}
\section{Manifolds}

After the invention of special relativity, Einstein tried for a 
number of years to invent a Lorentz-invariant theory of gravity,
without success.  His eventual breakthrough was to replace Minkowski 
spacetime with a curved spacetime, where the curvature was created by 
(and reacted back on) energy and momentum.  Before we explore how this
happens, we have to learn a bit about the mathematics of curved 
spaces.  First we will take a look at manifolds in general, and 
then in the next section study curvature.  In the interest of 
generality we will usually work in $n$ dimensions, although you are
permitted to take $n=4$ if you like.

A manifold (or sometimes ``differentiable manifold'') is one of the
most fundamental concepts in mathematics and physics.  We are all
aware of the properties of $n$-dimensional Euclidean space, $\R^n$,
the set of $n$-tuples $(x^1,\ldots,x^n)$.
The notion of a manifold captures the idea of a space which may be
curved and have a complicated topology, but in local regions looks
just like $\R^n$.  (Here by ``looks like'' we do not mean that the
metric is the same, but only basic notions of analysis like open sets,
functions, and coordinates.)  The entire manifold is constructed by
smoothly sewing together these local regions.  Examples of manifolds
include:
\begin{itemize}
  \item $\R^n$ itself, including the line ($\R$), the plane
  ($\R^2$), and so on.  This should be obvious, since $\R^n$ looks 
  like $\R^n$ not only locally but globally.
  \item The $n$-sphere, $S^n$.  This can be defined as the locus
  of all points some fixed distance from the origin in $\R^{n+1}$.
  The circle is of course $S^1$, and the two-sphere $S^2$ will be one
  of our favorite examples of a manifold.


  \item The $n$-torus $T^n$ results from taking an 
  $n$-dimensional cube and identifying opposite sides.  Thus $T^2$ is
  the traditional surface of a doughnut.

\begin{figure}[h]
  \centerline{
  \psfig{figure=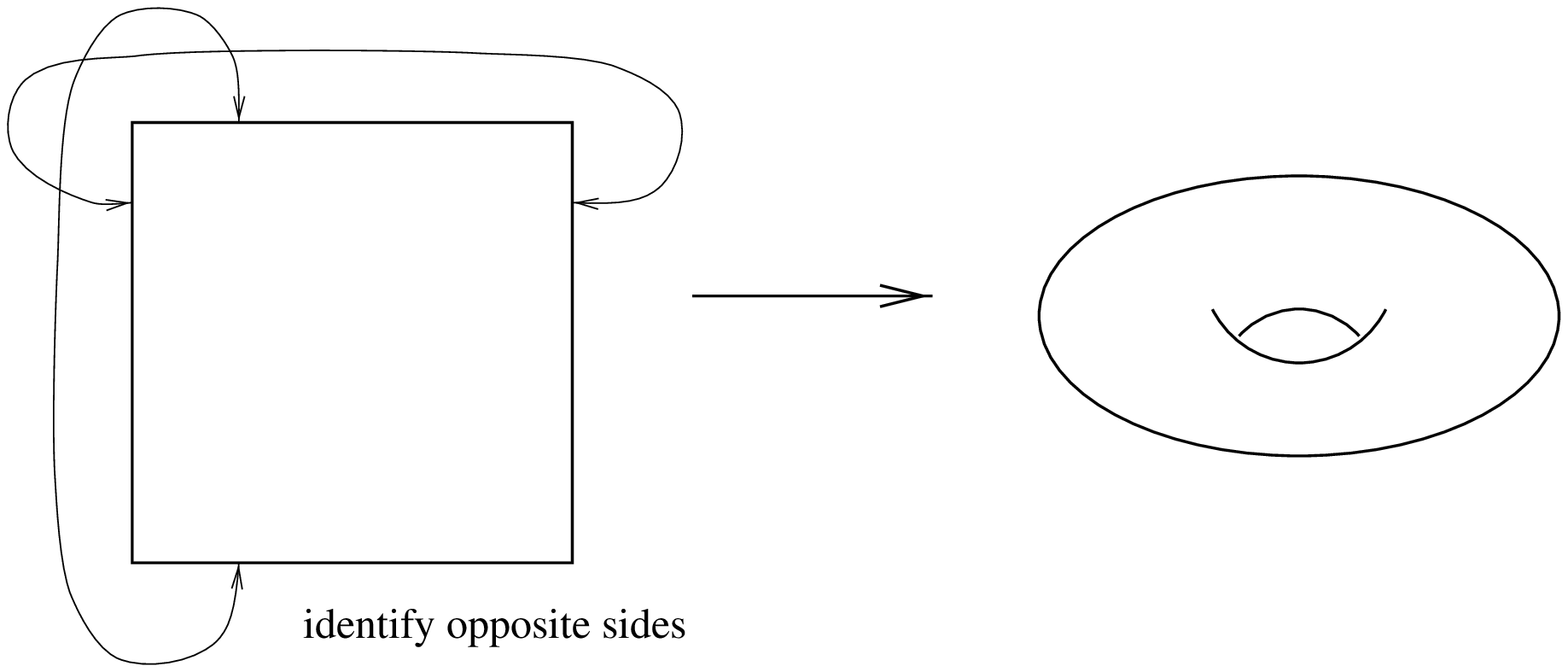,angle=-90,height=5cm}}
\end{figure}

  \item A Riemann surface of genus $g$ is essentially a
  two-torus with $g$ holes instead of just one.  $S^2$ may be thought
  of as a Riemann surface of genus zero.  For those of you who know 
  what the words mean, every ``compact orientable boundaryless''
  two-dimensional manifold is a Riemann surface of some genus.

\begin{figure}[h]
  \centerline{
  \psfig{figure=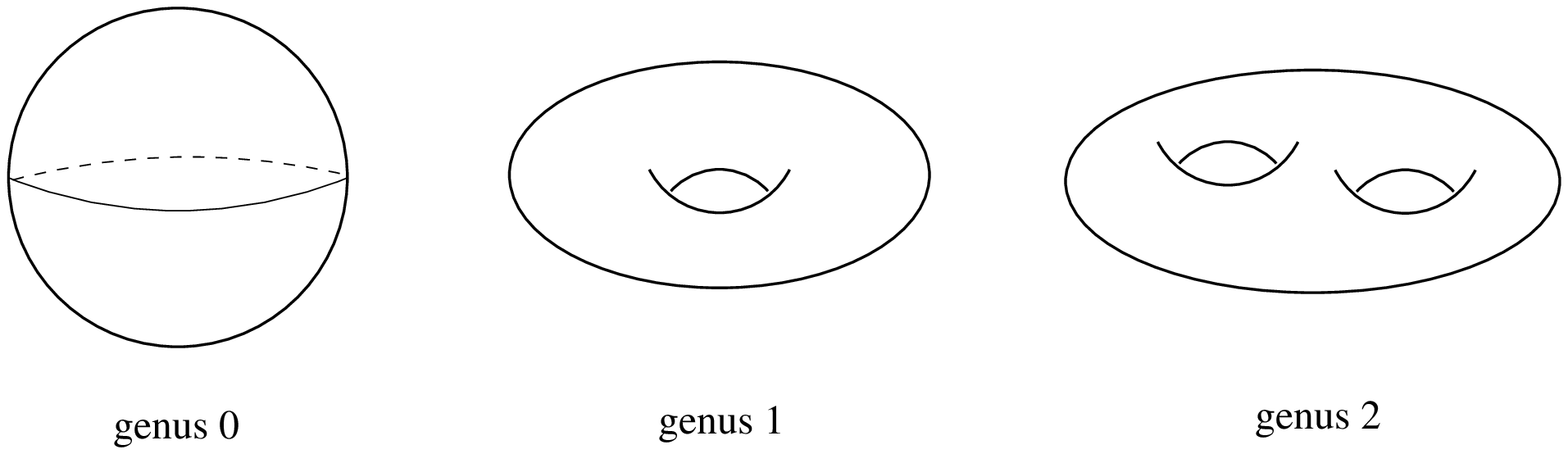,angle=-90,height=5cm}}
\end{figure}

  \item More abstractly, a set of continuous transformations
  such as rotations in $\R^n$ forms a manifold.  Lie groups are manifolds
  which also have a group structure.

  \item The direct product of two manifolds is a manifold.
  That is, given manifolds $M$ and $M'$ of dimension $n$ and $n'$, we can 
  construct a manifold $M\times M'$, of dimension $n+n'$, consisting of 
  ordered pairs $(p,p')$ for all $p\in M$ and $p'\in M'$.

\end{itemize}

With all of these examples, the notion of a manifold may seem vacuous;
what isn't a manifold?  There are plenty of things which are not
manifolds, because somewhere they do not look locally like $\R^n$.
Examples include a one-dimensional line running into a two-dimensional
plane, and two cones stuck together at their vertices.  (A single cone
is okay; you can imagine smoothing out the vertex.)

\begin{figure}[h]
  \centerline{
  \psfig{figure=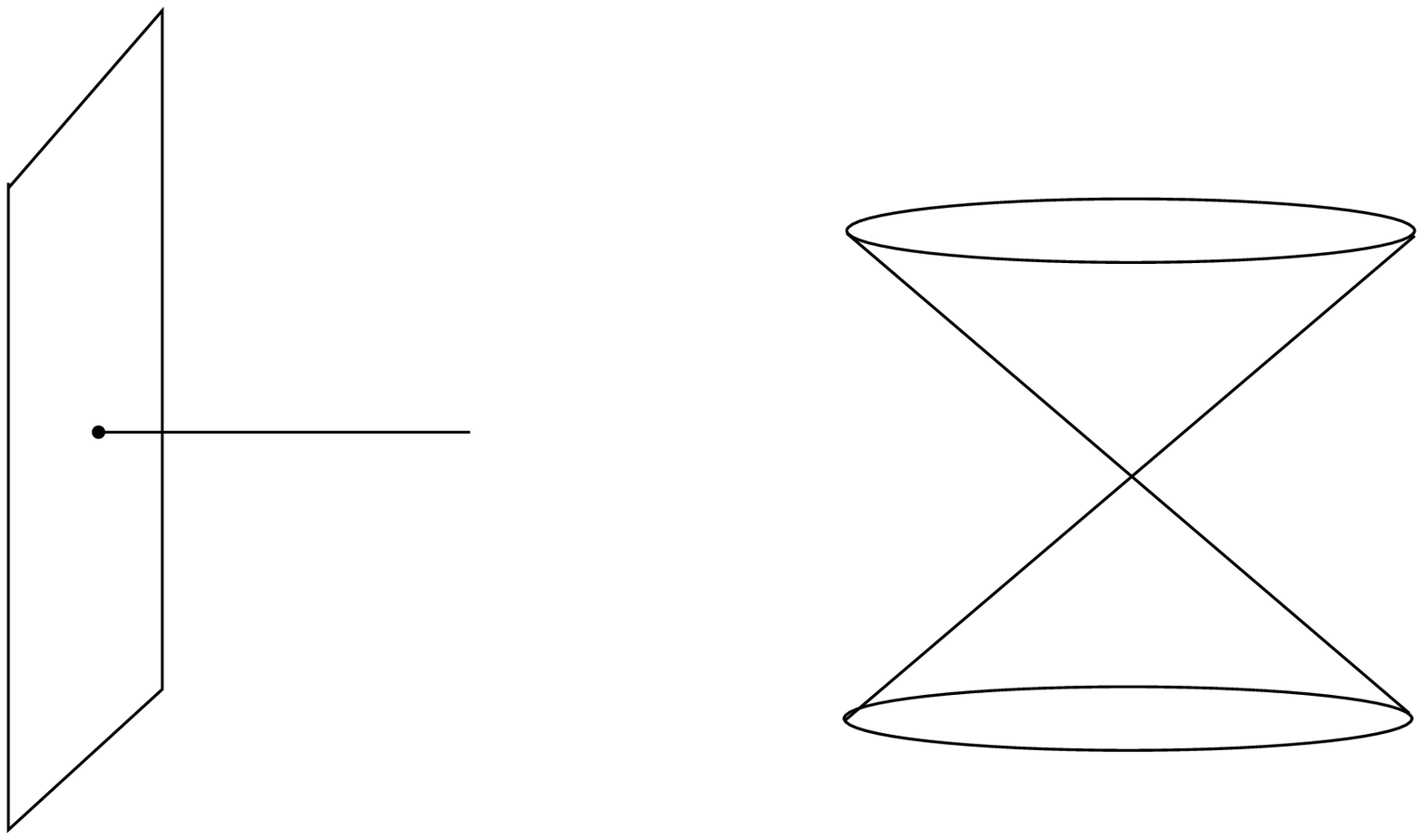,angle=-90,height=5cm}}
\end{figure}

We will now approach the rigorous definition of this simple idea, which
requires a number of preliminary definitions.  Many of them are pretty
clear anyway, but it's nice to be complete.

The most elementary notion is that of a {\bf map} between two sets.
(We assume you know what a set is.)  Given two sets $M$ and $N$, a
map $\phi: M\rightarrow N$ is a relationship which assigns, to each
element of $M$, exactly one element of $N$.  A map is therefore just a
simple generalization of a function.  The canonical picture of a
map looks like this:

\begin{figure}[h]
  \centerline{
  \psfig{figure=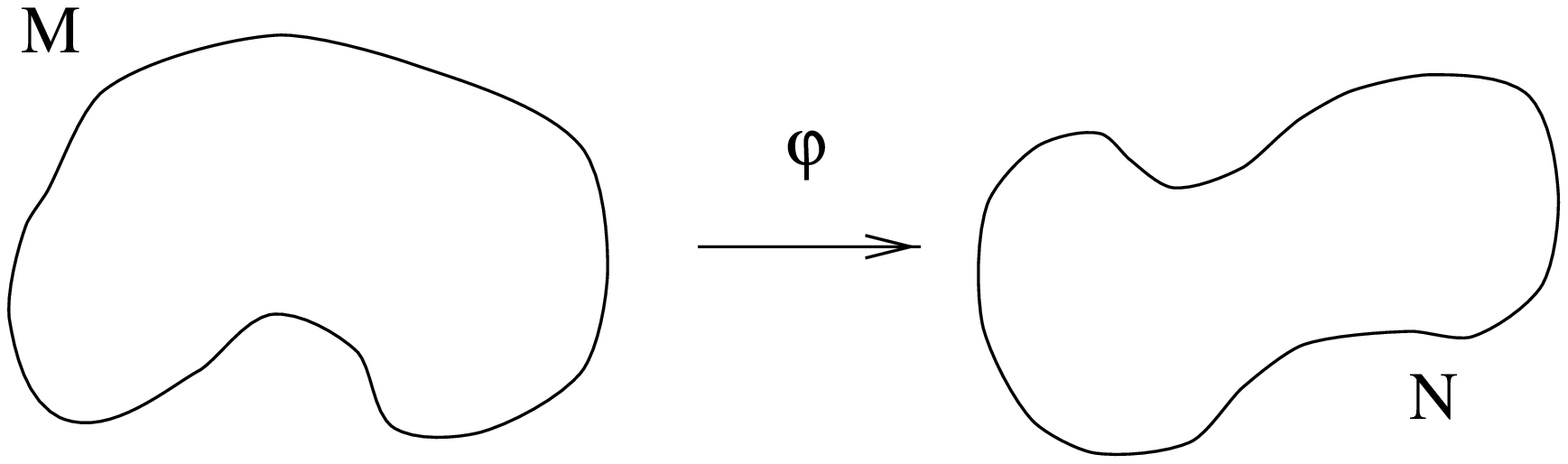,angle=-90,height=4cm}}
\end{figure}

Given two maps $\phi: A\rightarrow B$ and $\psi:B\rightarrow C$,
we define the {\bf composition} $\psi\circ\phi: A\rightarrow C$
by the operation $(\psi\circ\phi)(a)=\psi(\phi(a))$.  So $a\in A$,
$\phi(a)\in B$, and thus $(\psi\circ\phi)(a)\in C$.  The order in
which the maps are written makes sense, since the one on the right
acts first.  In pictures:

\begin{figure}[h]
  \centerline{
  \psfig{figure=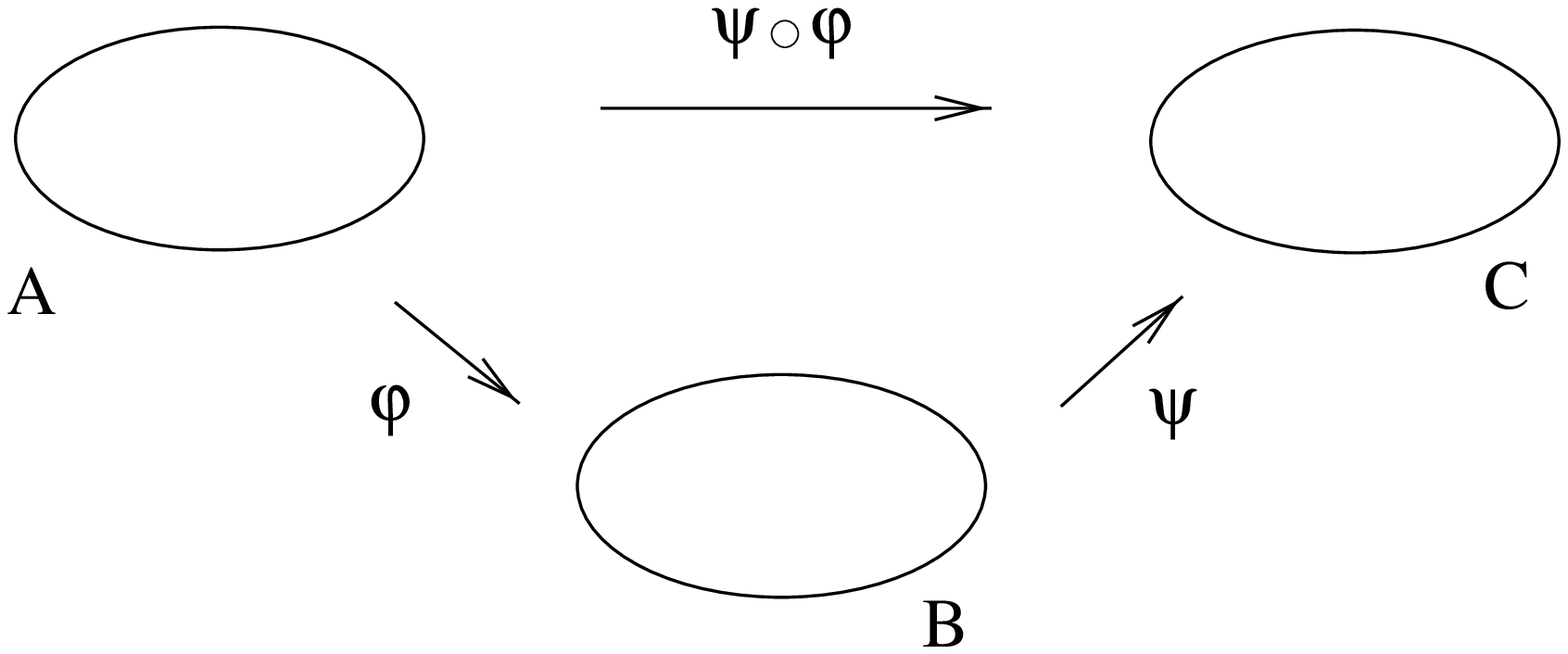,angle=-90,height=6cm}}
\end{figure}

A map $\phi$ is called {\bf one-to-one} (or ``injective'') if each
element of $N$ has at most one element of $M$ mapped into it, and
{\bf onto} (or ``surjective'') if each element of $N$ has at least one 
element of $M$ mapped into it.  (If you think about it, a better name for
``one-to-one'' would be ``two-to-two''.)  Consider a function
$\phi: \R\rightarrow\R$.  Then $\phi(x)=e^x$ is one-to-one, but not onto;
$\phi(x)=x^3-x$ is onto, but not one-to-one; $\phi(x)=x^3$ is both; and
$\phi(x)=x^2$ is neither.  

\begin{figure}
  \centerline{
  \psfig{figure=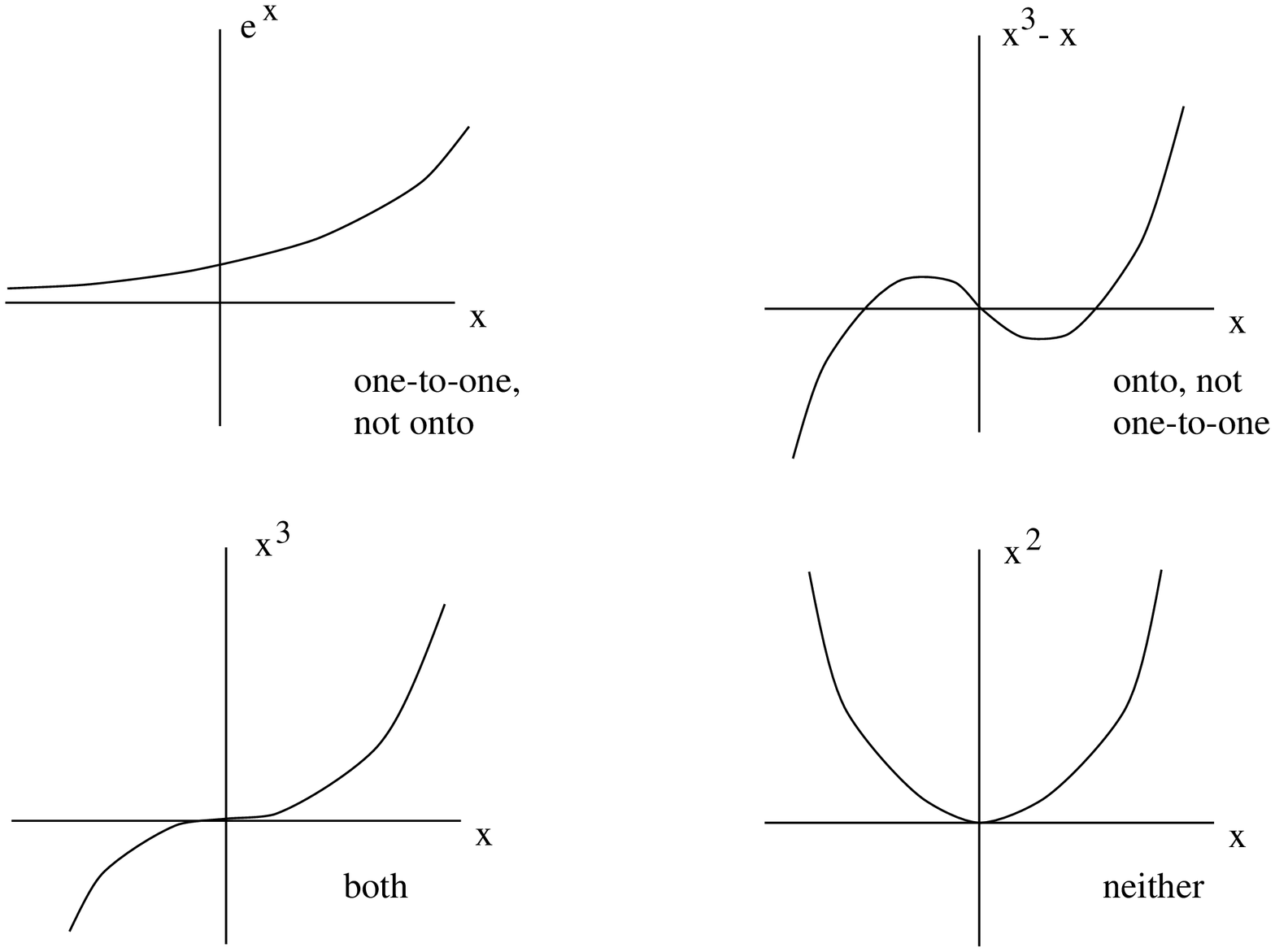,angle=-90,height=10cm}}
\end{figure}

The set $M$ is known as the {\bf domain}
of the map $\phi$, and the set of points in $N$ which $M$ gets mapped into
is called the {\bf image} of $\phi$.  For some subset $U\subset N$,
the set of elements of $M$ which get mapped to $U$ is called the
{\bf preimage} of $U$ under $\phi$, or $\phi^{-1}(U)$.  A map which
is both one-to-one and onto is known as {\bf invertible} (or 
``bijective'').  In this case we can define the {\bf inverse map}
$\phi^{-1}:N\rightarrow M$ by $(\phi^{-1}\circ\phi)(a)=a$.  (Note that
the same symbol $\phi^{-1}$ is used for both the preimage and the inverse
map, even though the former is always defined and the latter is
only defined in some special cases.)  Thus:

\begin{figure}[h]
  \centerline{
  \psfig{figure=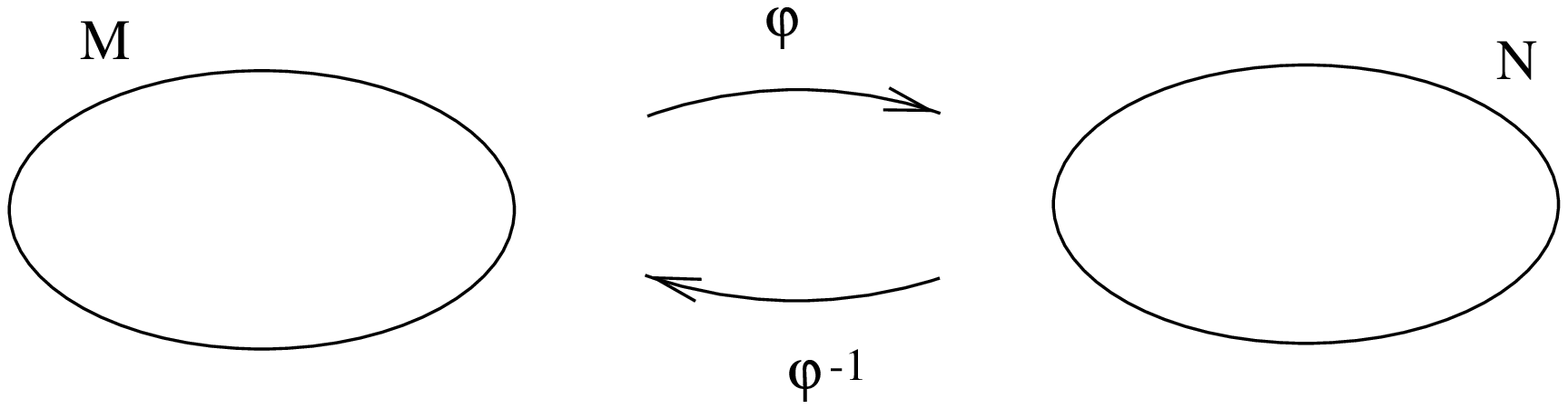,angle=-90,height=4cm}}
\end{figure}

The notion of {\bf continuity} of a map between topological spaces (and 
thus manifolds) is actually a very subtle one, the precise formulation of
which we won't really
need.  However the intuitive notions of continuity and differentiability
of maps $\phi:\R^m\rightarrow\R^n$ between Euclidean spaces are useful.
A map from $\R^m$ to $\R^n$ takes an $m$-tuple $(x^1,x^2,\ldots,x^m)$
to an $n$-tuple $(y^1,y^2,\ldots,y^n)$, and can therefore
be thought of as a collection of $n$ functions $\phi^i$ of $m$ variables:
\be
  \matrix{y^1 =\phi^1(x^1,x^2,\ldots,x^m)\cr 
  y^2=\phi^2(x^1,x^2,\ldots,x^m)\cr
  \cdot\cr\cdot\cr\cdot\cr y^n=\phi^n(x^1,x^2,\ldots,x^m)\ .\cr}
  \label{2.1}
\ee
We will refer to any one of these functions as $C^p$ if it is continuous 
and $p$-times differentiable, and refer to the entire map 
$\phi:\R^m\rightarrow\R^n$ as $C^p$ if each of its component functions
are at least $C^p$.  Thus a $C^0$ map is continuous but not 
necessarily differentiable,
while a $C^\infty$ map is continuous and can be differentiated as many
times as you like.  $C^\infty$ maps are sometimes called {\bf smooth}.
We will call two sets $M$ and $N$ {\bf diffeomorphic} if there exists
a $C^\infty$ map $\phi:M\rightarrow N$ with a $C^\infty$ inverse 
$\phi^{-1}:N\rightarrow M$; the map $\phi$ is then called a diffeomorphism.

Aside: The notion of two spaces being diffeomorphic only applies to
manifolds, where a notion of differentiability is inherited from the
fact that the space resembles $\R^n$ locally.  But ``continuity'' of maps
between topological spaces (not necessarily manifolds) can be defined,
and we say that two such spaces are ``homeomorphic,'' which means 
``topologically equivalent to,'' if there is a continuous map 
between them with a continuous inverse.  It is therefore conceivable that
spaces exist which are homeomorphic but not diffeomorphic; topologically
the same but with distinct ``differentiable structures.''  In 1964 Milnor 
showed that $S^7$ had 28 different differentiable structures; 
it turns out that for $n<7$ there is only one differentiable 
structure on $S^n$, while for $n>7$ the number grows very large.  
$\R^4$ has infinitely many differentiable structures.

One piece of conventional calculus that we will need later is the
{\bf chain rule}.  Let us imagine that we have maps $f:\R^m\rightarrow
\R^n$ and $g:\R^n\rightarrow \R^l$, and therefore the composition
$(g\circ f):\R^m\rightarrow\R^l$.  

\begin{figure}[h]
  \centerline{
  \psfig{figure=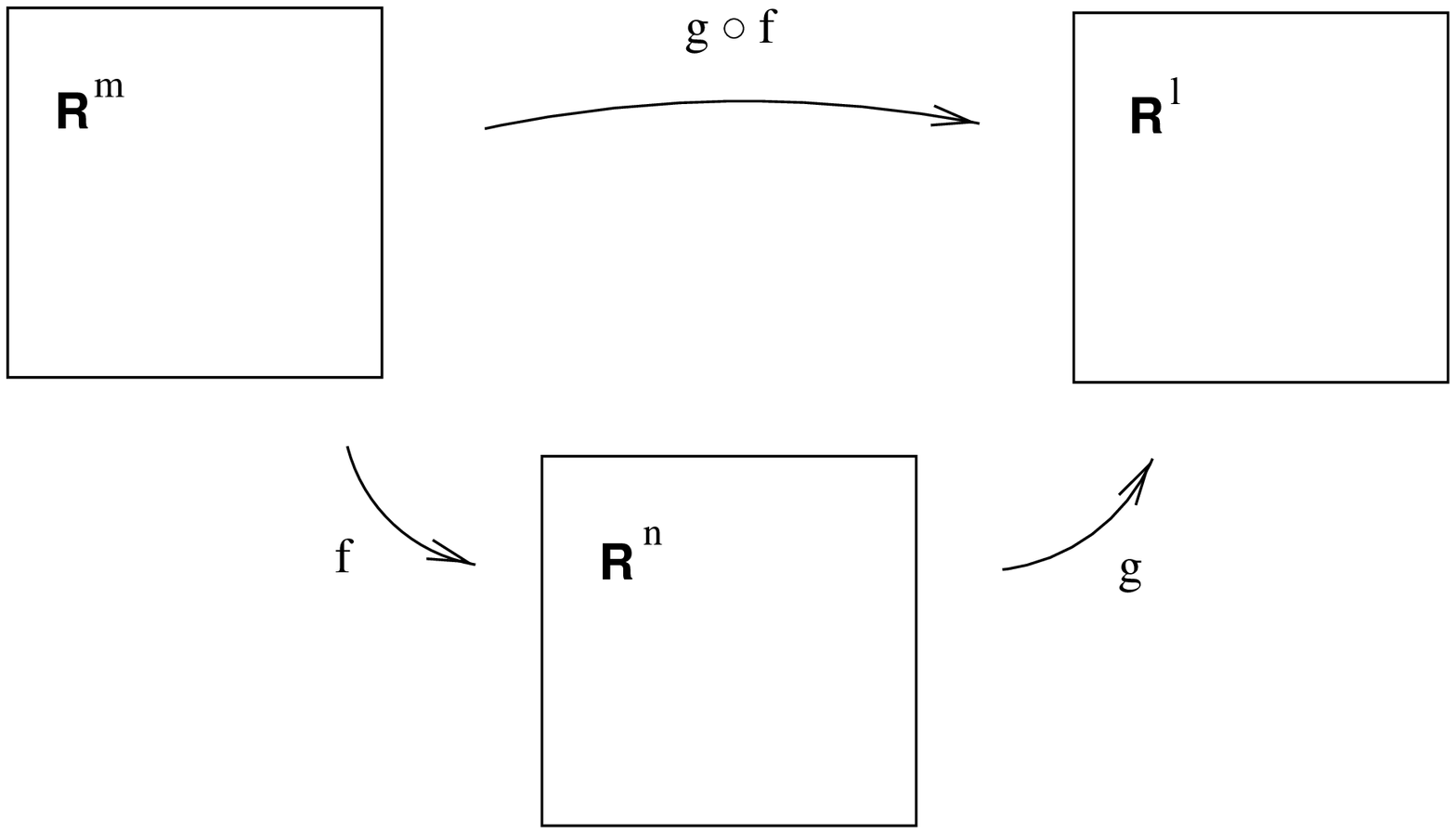,angle=-90,height=5cm}}
\end{figure}

\noindent We can represent each space in terms of coordinates:
$x^a$ on $\R^m$, $y^b$ on $\R^n$, and $z^c$ on $\R^l$, where the
indices range over the appropriate values. 
The chain rule relates the partial derivatives of the 
composition to the partial derivatives of the individual maps:
\be
  {{\partial}\over{\partial x^a}}(g\circ f)^c =
  \sum_b {{\partial f^b}\over{\partial x^a}}{{\partial g^c}
  \over{\partial y^b}}\ .\label{2.2}
\ee
This is usually abbreviated to
\be
  {{\partial}\over{\partial x^a}} = \sum_b {{\partial y^b}\over
  {\partial x^a}}{{\partial}\over{\partial y^b}}\ .\label{2.3}
\ee
There is nothing illegal or immoral about using this form of
the chain rule, but you should be able to visualize the maps that
underlie the construction.  Recall that when $m=n$ the 
determinant of the matrix
$\partial y^b/\partial x^a$ is called the {\bf Jacobian} of the
map, and the map is invertible whenever the
Jacobian is nonzero.

These basic definitions were presumably familiar to you, even if only 
vaguely remembered.  We will now put them to use in the rigorous definition
of a manifold.  Unfortunately, a somewhat baroque procedure is required
to formalize this relatively intuitive notion.  We will first have to
define the notion of an open set, on which we can put coordinate systems,
and then sew the open sets together in an appropriate way.

Start with the notion of an {\bf open ball}, which is the set of all
points $x$ in $\R^n$ such that $|x-y|<r$ for some fixed $y\in \R^n$ and
$r\in \R$, where $|x-y|=[\sum_i(x^i-y^i)^2]^{1/2}$.  Note that this is
a strict inequality --- the open ball is the interior of an $n$-sphere
of radius $r$ centered at $y$.

\begin{figure}[h]
  \centerline{
  \psfig{figure=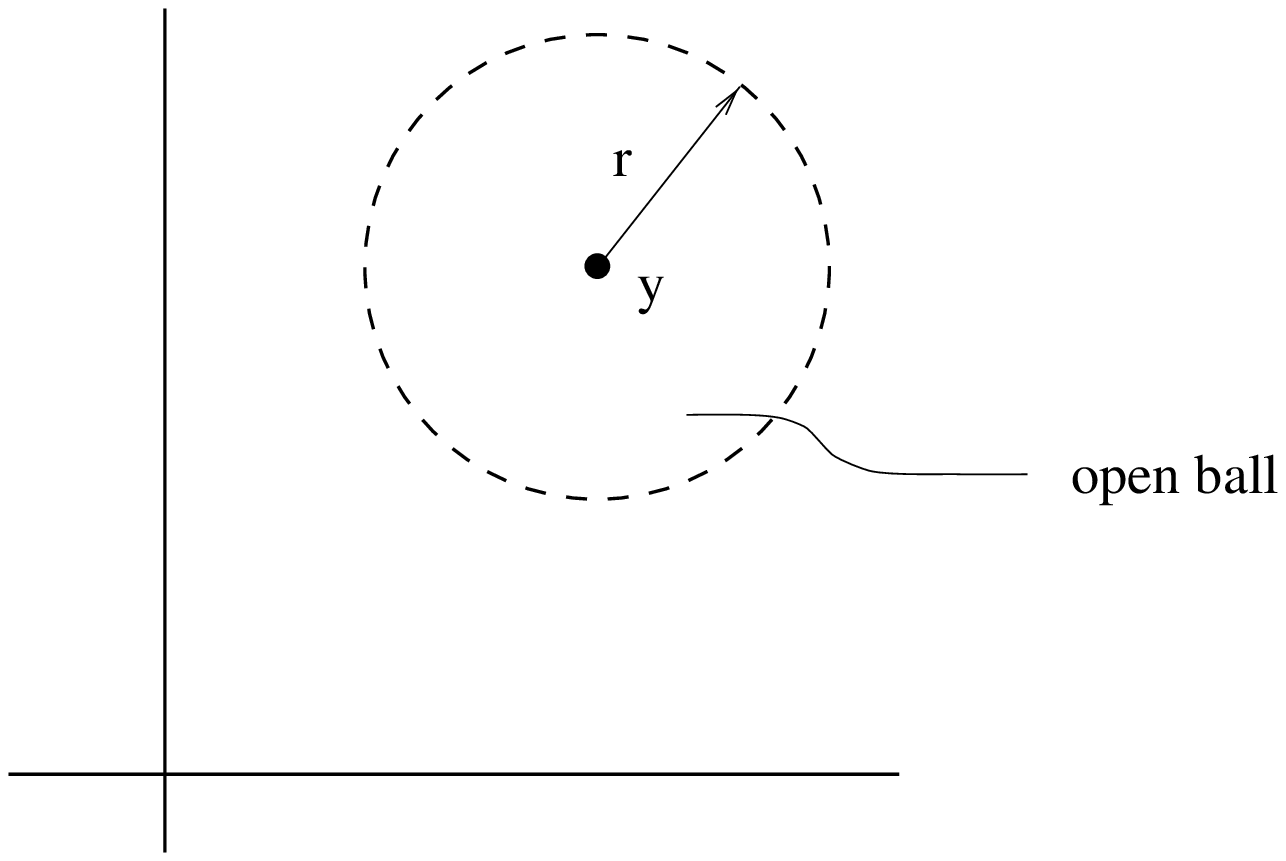,angle=-90,height=6cm}}
\end{figure}

\noindent An {\bf open set} in $\R^n$ is a set constructed from an arbitrary
(maybe infinite) union of open balls.  In other words, $V\subset \R^n$ is
open if, for any $y\in V$, there is an open ball centered at $y$ which
is completely inside $V$.  Roughly speaking, an open set is the interior
of some $(n-1)$-dimensional closed surface (or the union of several such
interiors).  By defining a notion of open sets, we have equipped $\R^n$
with a topology --- in this case, the ``standard metric topology.''

A {\bf chart} or {\bf coordinate system} consists of a subset $U$ of a 
set $M$, along with a one-to-one map $\phi:U\rightarrow\R^n$, such that the 
image $\phi(U)$ is open in $\R$.  (Any map is onto its image, so the
map $\phi:U\rightarrow \phi(U)$ is invertible.)  We then can say that
$U$ is an open set in $M$.  (We have thus induced a topology on $M$,
although we will not explore this.)

\begin{figure}[h]
  \centerline{
  \psfig{figure=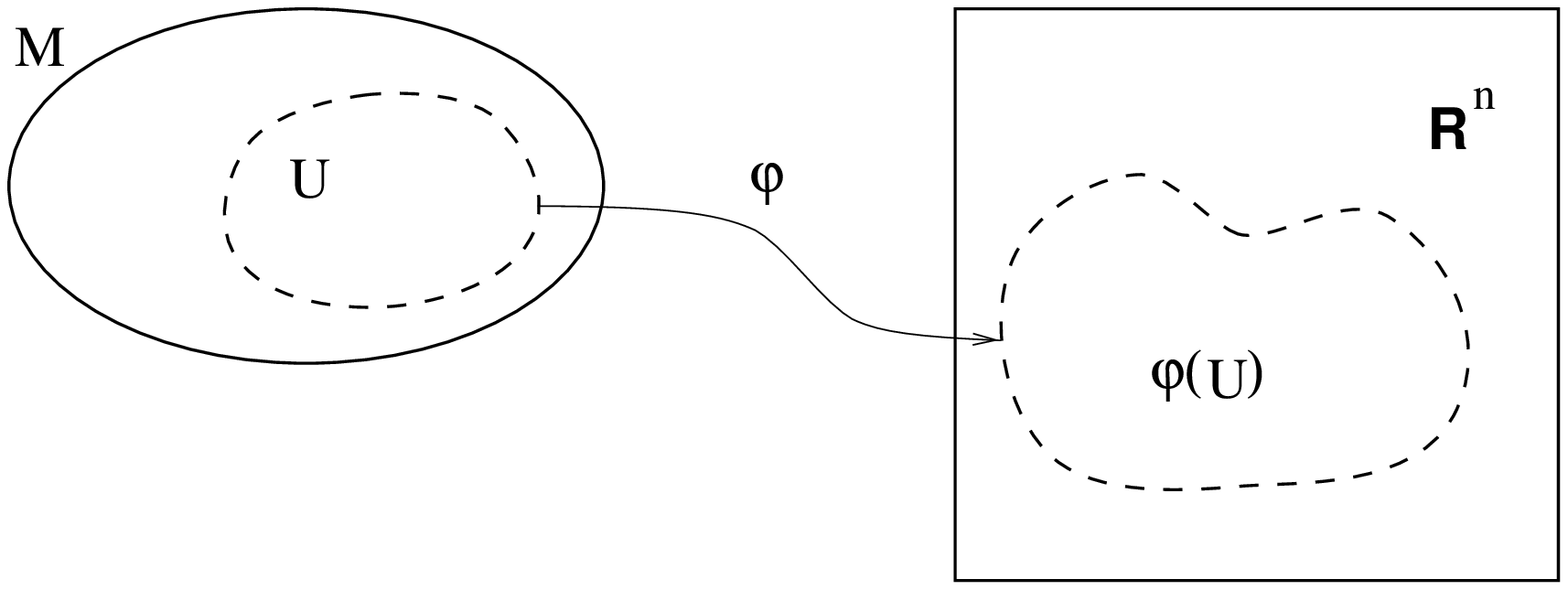,angle=-90,height=5cm}}
\end{figure}

\noindent A {\bf $C^\infty$ atlas} is an indexed collection of charts 
$\{(U_\alpha,\phi_\alpha)\}$ which satisfies two conditions:
\begin{enumerate}
  \item The union of the $U_\alpha$ is equal to $M$; that is, the
   $U_\alpha$ cover $M$.
  \item The charts are smoothly sewn together.  More precisely, if
   two charts overlap, $U_\alpha\cap U_\beta\neq\emptyset$, then the
   map $(\phi_\alpha\circ\phi_\beta^{-1})$ takes points in 
   $\phi_\beta(U_\alpha\cap U_\beta)\subset\R^n$ {\it onto}
   $\phi_\alpha(U_\alpha\cap U_\beta)\subset\R^n$, and all of these
   maps must be $C^\infty$ where they are defined.  This should be clearer 
   in pictures:
\end{enumerate}

\begin{figure}[h]
  \centerline{
  \psfig{figure=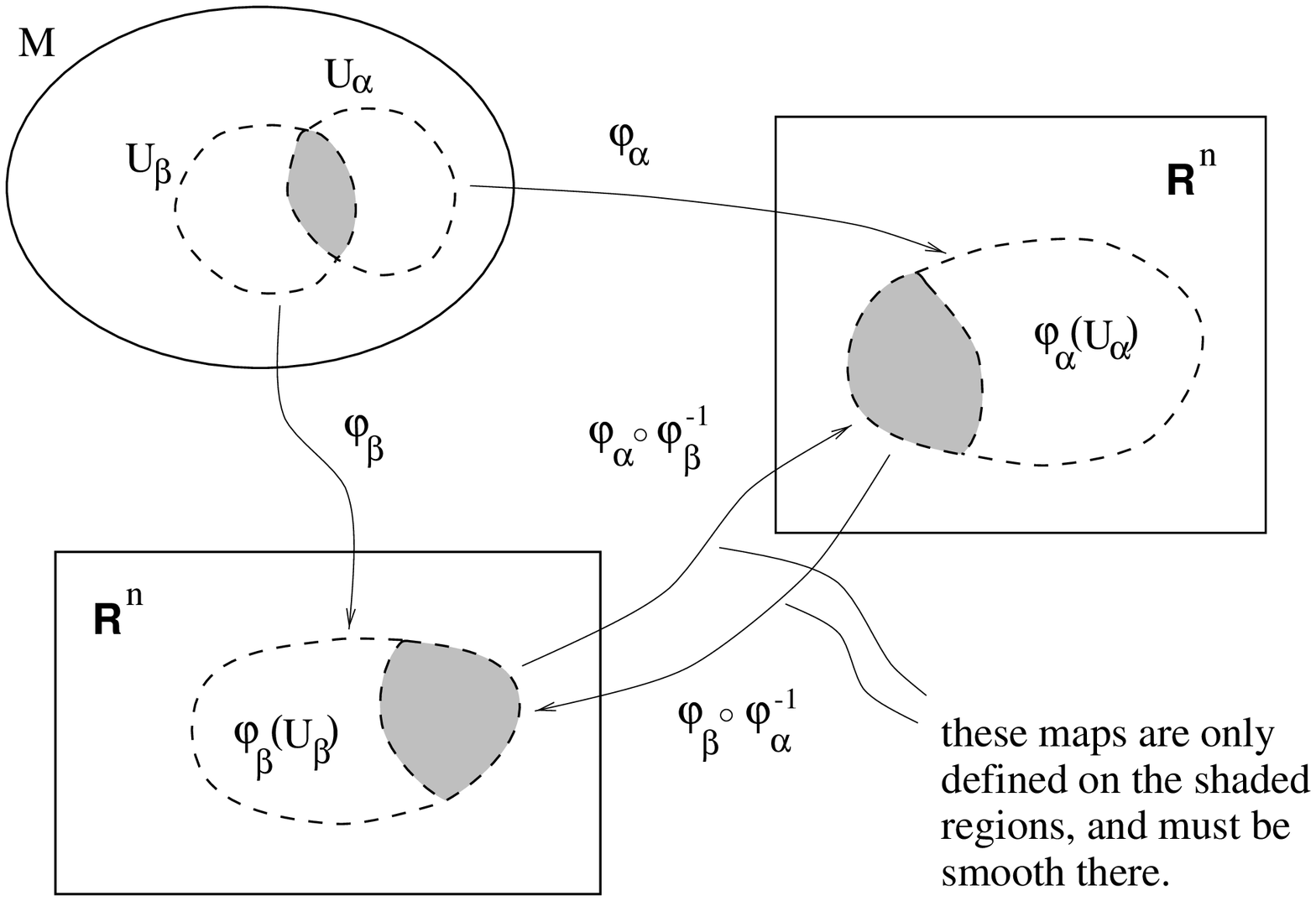,angle=-90,height=8cm}}
\end{figure}

\noindent So a chart is what we normally think of as a coordinate system on
some open set, and an atlas is a system of charts which are smoothly
related on their overlaps.

At long last, then: a $C^\infty$ $n$-dimensional {\bf manifold} (or 
$n$-manifold for short) is simply a set $M$ along with a ``maximal atlas'',
one that contains every possible compatible chart.  (We can also replace 
$C^\infty$ by $C^p$ in all the above definitions.  For our purposes the degree
of differentiability of a manifold is not crucial; we will always assume
that any manifold is as differentiable as necessary for the application
under consideration.)  The requirement that the 
atlas be maximal is so that two equivalent spaces equipped with
different atlases don't count as different manifolds.
This definition captures in
formal terms our notion of a set that looks locally like $\R^n$.
Of course we will rarely have to make use of the full power of the
definition, but precision is its own reward.  

One thing that is nice about our definition is that it does not rely
on an embedding of the manifold in some higher-dimensional Euclidean
space.  In fact any $n$-dimensional manifold can be embedded in
$\R^{2n}$ (``Whitney's embedding theorem''), 
and sometimes we will make use of this fact (such as
in our definition of the sphere above).  But it's important to 
recognize that the manifold has an individual existence independent
of any embedding.  We have no reason to believe, for example, that
four-dimensional spacetime is stuck in some larger space.  (Actually
a number of people, string theorists and so forth, believe that our
four-dimensional world is part of a ten- or eleven-dimensional
spacetime, but as far as GR
is concerned the 4-dimensional view is perfectly adequate.)

Why was it necessary to be so finicky about charts and their overlaps,
rather than just covering every manifold with a single chart?  Because
most manifolds cannot be covered with just one chart.  Consider the
simplest example, $S^1$.  There is a conventional coordinate system,
$\theta: S^1\rightarrow\R$, where $\theta=0$ at the top of the circle
and wraps around to $2\pi$.  However, in the definition of a chart we
have required that the image $\theta(S^1)$ be open in $\R$.  If we 
include either $\theta=0$ or $\theta=2\pi$, we have a closed interval
rather than an open one; if we exclude both points, we haven't covered
the whole circle.  So we need at least two charts, as shown.

\begin{figure}[h]
  \centerline{
  \psfig{figure=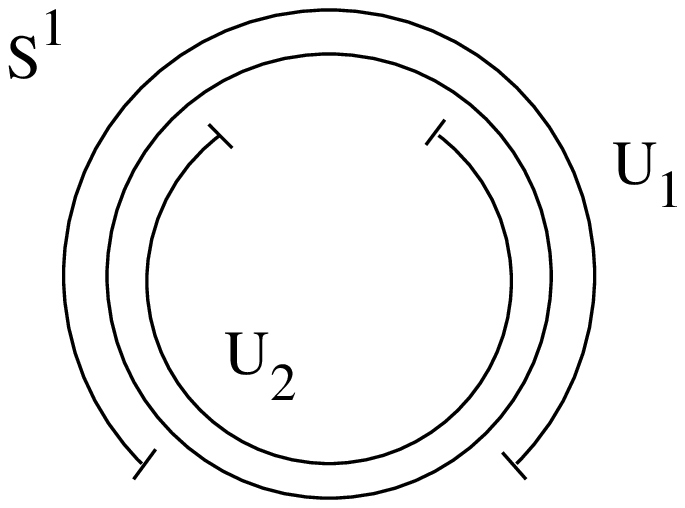,angle=-90,height=5cm}}
\end{figure}

A somewhat more complicated example is provided by $S^2$, where once
again no single chart will cover the manifold.  A Mercator projection,
traditionally used for world maps, misses both the North and South poles
(as well as the International Date Line, which involves the same problem 
with $\theta$ that we found for $S^1$.)  Let's take $S^2$ to be the set
of points in $\R^3$ defined by $(x^1)^2 +(x^2)^2 +(x^3)^2 =1$.  We
can construct a chart from an open set $U_1$, defined to be the sphere 
minus the north pole, via ``stereographic projection'':

\begin{figure}[h]
  \centerline{
  \psfig{figure=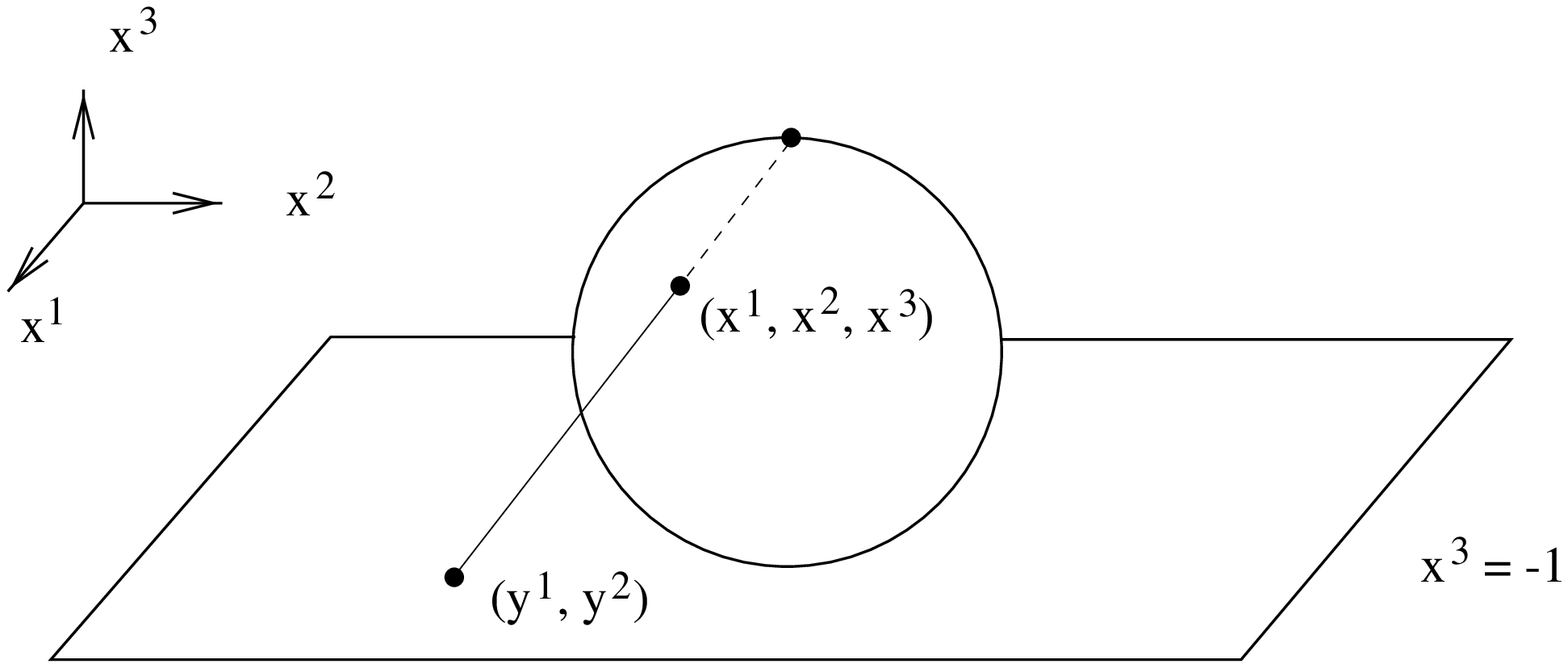,angle=-90,height=5cm}}
\end{figure}

\noindent Thus, we draw a straight line from the north pole to the plane
defined by $x^3 = -1$, and assign to the point on $S^2$ intercepted by
the line the Cartesian coordinates $(y^1,y^2)$ of the appropriate point
on the plane.  Explicitly, the map is given by
\be
  \phi_1(x^1,x^2,x^3) \equiv (y^1,y^2) = \left(
  {{2x^1}\over{1-x^3}}\ ,\ {{2x^2}\over{1-x^3}}\right)\ .\label{2.4}
\ee
You are encouraged to check this for yourself.  Another chart
$(U_2,\phi_2)$ is obtained by projecting from the south pole to the
plane defined by $x^3 = +1$.  The resulting coordinates cover the 
sphere minus the south pole, and are given by
\be
  \phi_2(x^1,x^2,x^3) \equiv (z^1,z^2) = \left(
  {{2x^1}\over{1+x^3}}\ ,\ {{2x^2}\over{1+x^3}}\right)\ .\label{2.5}
\ee
Together, these two charts cover the entire manifold, and they overlap
in the region $-1<x^3<+1$.  Another thing you can check is that the
composition $\phi_2\circ\phi_1^{-1}$ is given by
\be
  z^i = {{4y^i}\over{[(y^1)^2 +(y^2)^2]}}\ ,\label{2.6}
\ee
and is $C^\infty$ in the region of overlap.  As long as we restrict
our attention to this region, (2.6) is just what we normally think
of as a change of coordinates.

We therefore see the necessity of charts and atlases: many manifolds
cannot be covered with a single coordinate system.  (Although some
can, even ones with nontrivial topology.  Can you think of a single
good coordinate system that covers the cylinder, $S^1\times\R$?)
Nevertheless, it is very often most convenient to work with a single
chart, and just keep track of the set of points which aren't included.

The fact that manifolds look locally like $\R^n$, which is manifested
by the construction of coordinate charts, introduces the possibility
of analysis on manifolds, including operations such as differentiation
and integration.  Consider two manifolds $M$ and $N$ of dimensions
$m$ and $n$, with coordinate charts $\phi$ on $M$ and $\psi$ on $N$.
Imagine we have a function $f:M\rightarrow N$,

\begin{figure}[h]
  \centerline{
  \psfig{figure=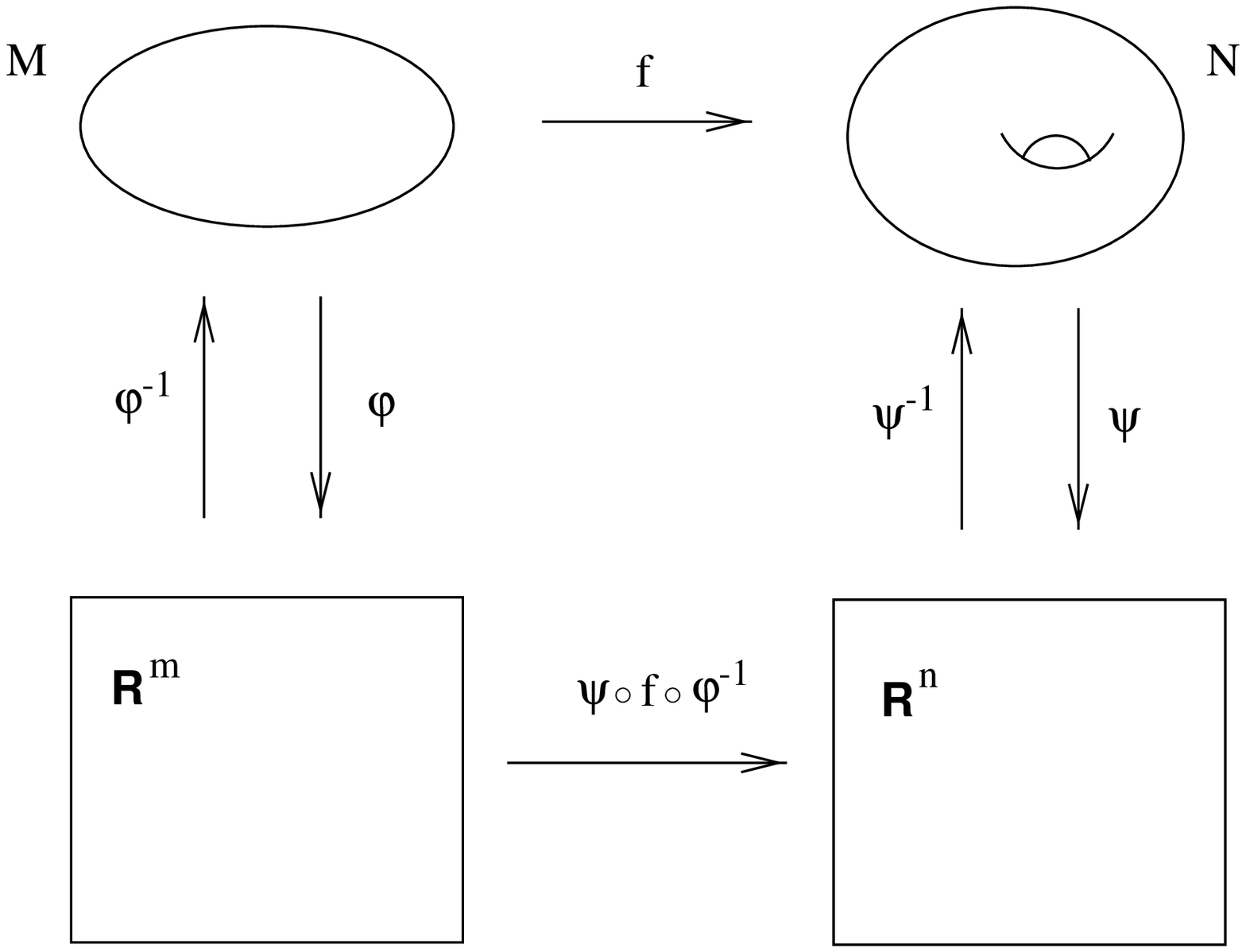,angle=-90,height=7cm}}
\end{figure}

\noindent Just thinking of $M$ and $N$ as sets, we cannot nonchalantly
differentiate the map $f$, since we don't know what such an operation
means.  But the coordinate charts allow us to construct the map
$(\psi\circ f\circ\phi^{-1}):\R^m\rightarrow\R^n$.  (Feel free to
insert the words ``where the maps are defined'' wherever appropriate,
here and later on.)  This is just
a map between Euclidean spaces, and all of the concepts of advanced
calculus apply.  For example $f$, thought of as an $N$-valued
function on $M$, can be differentiated to obtain ${\partial f}/
{\partial x^\mu}$, where the $x^\mu$ represent $\R^m$.
The point is that this notation is a shortcut, and what is really
going on is
\be
  {{\partial f}\over{\p{} x^\mu}} \equiv {{\partial}\over{\p{} x^\mu}}
  (\psi\circ f\circ\phi^{-1})(x^{\mu})\ .\label{2.7}
\ee
It would be far too unwieldy (not to mention pedantic) to write out
the coordinate maps explicitly in every case.  The shorthand notation
of the left-hand-side will be sufficient for most purposes.

Having constructed this groundwork, we can now proceed to introduce
various kinds of structure on manifolds.  We begin with vectors and
tangent spaces.  In our discussion of special relativity we were
intentionally vague about the definition of vectors and their 
relationship to the spacetime.  One point that was stressed was the
notion of a tangent space --- the set of all vectors at a single
point in spacetime.  The reason for this emphasis was to remove from
your minds the idea that a vector stretches from one point on the
manifold to another, but instead is just an object associated with
a single point.  What is temporarily lost by adopting this view is
a way to make sense of statements like ``the vector points in the
$x$ direction'' --- if the tangent space is merely an abstract vector
space associated with each point, it's hard to know what this should
mean.  Now it's time to fix the problem.

Let's imagine that we wanted to construct the tangent space at a
point $p$ in a manifold $M$, using only things that are intrinsic
to $M$ (no embeddings in higher-dimensional spaces etc.).  One first
guess might be to use our intuitive knowledge that  there are
objects called ``tangent vectors to curves'' which
belong in the tangent space.  We might therefore consider the set
of all parameterized curves through $p$ --- that is,
the space of all (nondegenerate) maps $\gamma:\R\rightarrow M$ such
that $p$ is in the image of $\gamma$.  The temptation is to define
the tangent space as simply the space of all tangent vectors to
these curves at the point $p$.  But this is obviously cheating; the
tangent space $T_p$ is supposed to be the space of vectors at $p$,
and before we have defined this we don't have an independent notion
of what ``the tangent vector to a curve'' is supposed to mean.  In
some coordinate system $x^\mu$ any curve through $p$ defines an
element of $\R^n$ specified by the $n$ real numbers
$dx^\mu/d\lambda$ (where $\lambda$ is the parameter along the curve),
but this map is clearly coordinate-dependent, which is not what we
want.  

Nevertheless we are on the right track, we just have to make things
independent of coordinates.  To this end we define ${\cal F}$ to be
the space of all smooth functions on $M$ (that is, $C^\infty$ maps
$f:M\rightarrow \R$).  Then we notice that each curve through $p$
defines an operator on this space, the directional derivative, which
maps $f\rightarrow df/d\lambda$ (at $p$).  We will make the following
claim: {\it the tangent space $T_p$ can be identified with the space
of directional derivative operators along curves through $p$}.  To
establish this idea we must demonstrate two things: first, that the
space of directional derivatives is a vector space, and second that
it is the vector space we want (it has the same dimensionality as $M$, 
yields a natural idea of a vector pointing along a certain direction, 
and so on).

The first claim, that directional derivatives form a vector space,
seems straightforward enough.  Imagine two operators ${d\over{d\lambda}}$
and ${d\over{d\eta}}$ representing derivatives along two curves through $p$.
There is no problem adding these and scaling by real numbers, to
obtain a new operator $a{d\over{d\lambda}}+ b{d\over{d\eta}}$.  It is
not immediately obvious, however, that the space closes; {\it i.e.},
that the resulting operator is itself a derivative operator.  A good
derivative operator is one that acts linearly on functions, and obeys
the conventional Leibniz (product) rule on products of functions.
Our new operator is manifestly linear, so we need to verify that it
obeys the Leibniz rule.  We have
\bea
  \left(a{d\over{d\lambda}}+ b{d\over{d\eta}}\right)(fg)
  & = &  af{{dg}\over{d\lambda}} + ag{{df}\over{d\lambda}} +
  bf{{dg}\over{d\eta}} + bg{{df}\over{d\eta}} \nonumber \\
  & = &  \left(a{{df}\over{d\lambda}}+ b{{df}\over{d\eta}}\right)g +
  \left(a{{dg}\over{d\lambda}}+ b{{dg}\over{d\eta}}\right)f\ .
  \label{2.8}
\eea
As we had hoped, the product rule is satisfied, and the set of 
directional derivatives is therefore a vector space.

Is it the vector space that we would like to identify with the tangent
space?  The easiest way to become convinced is to find a basis for
the space.  Consider again a coordinate chart with coordinates $x^\mu$.
Then there is an obvious set of $n$ directional derivatives at $p$,
namely the partial derivatives $\p\mu$ at $p$.

\begin{figure}[h]
  \centerline{
  \psfig{figure=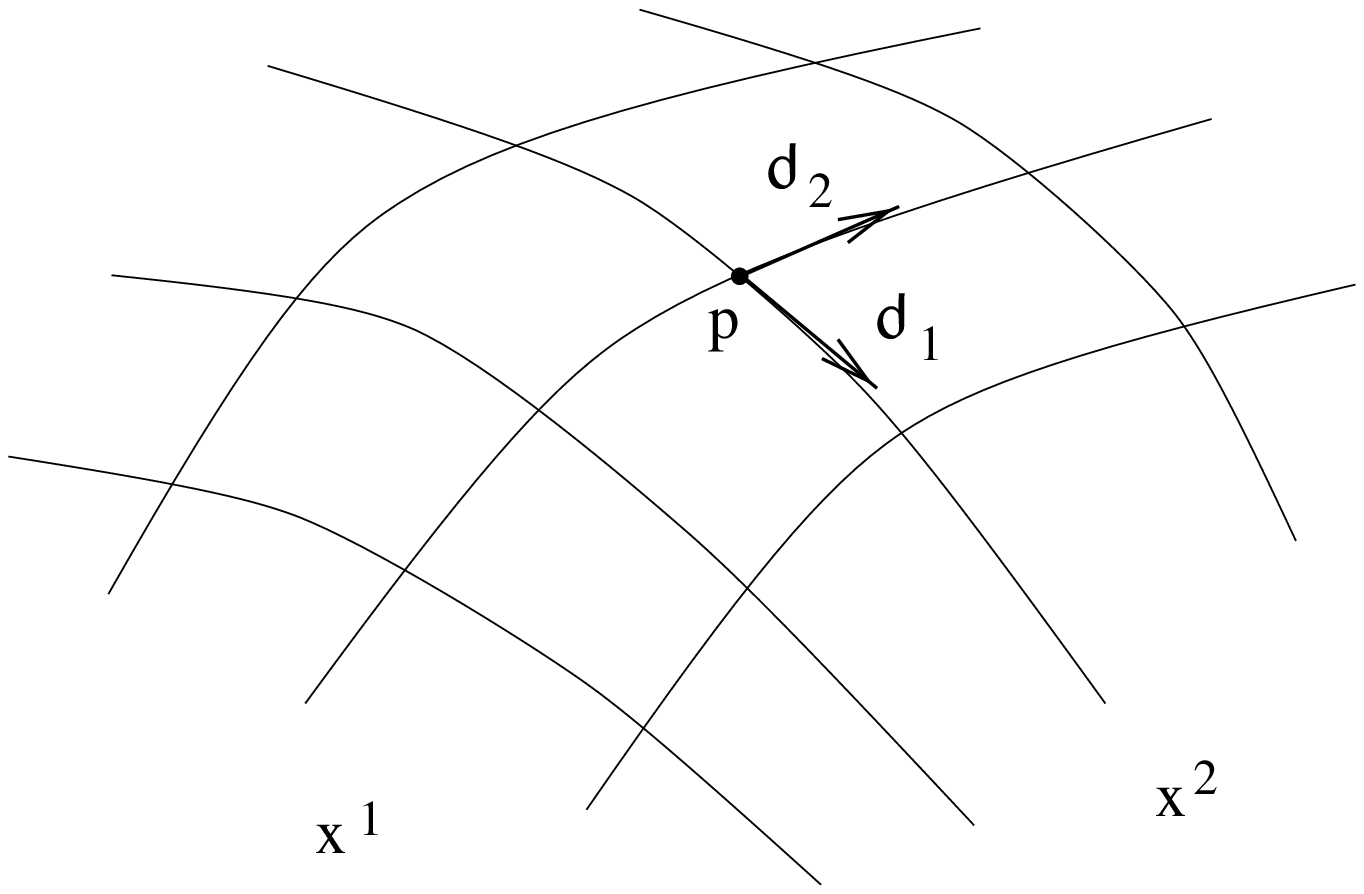,angle=-90,height=5cm}}
\end{figure}

\noindent We are now going to claim that the partial derivative
operators $\{\p\mu\}$ at $p$ form a basis for the tangent space $T_p$.
(It follows immediately that $T_p$ is $n$-dimensional, since that is
the number of basis vectors.)  To see this we will show that any 
directional derivative can be decomposed into a sum of real numbers
times partial derivatives.  This is in fact just the familiar 
expression for the components of a tangent vector, but it's nice to
see it from the big-machinery approach.  Consider an $n$-manifold $M$,
a coordinate chart $\phi:M\rightarrow \R^n$, a curve 
$\gamma:\R\rightarrow M$, and a function $f:M\rightarrow\R$.
This leads to the following tangle of maps:

\begin{figure}[h]
  \centerline{
  \psfig{figure=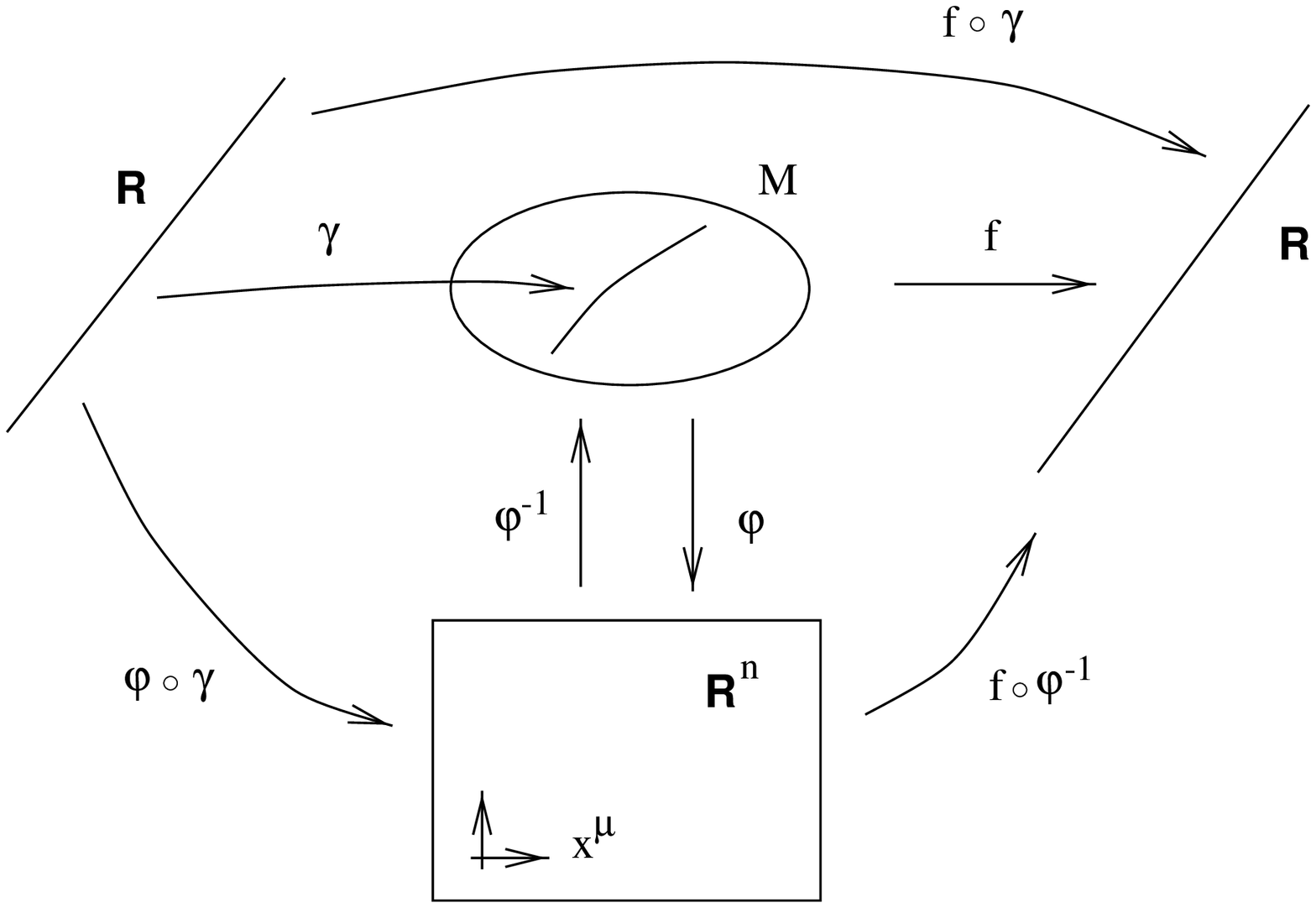,angle=-90,height=8cm}}
\end{figure}

\noindent If $\lambda$ is the parameter along $\gamma$, we want to
expand the vector/operator ${{d}\over{d\lambda}}$ in terms of the
partials $\p\mu$.  Using the chain rule (2.2), we have
\bea
  {d\over{d\lambda}}f &=&  {d\over{d\lambda}}(f\circ\gamma)\nonumber \\
  &=& {d\over{d\lambda}}[(f\circ\phi^{-1})\circ(\phi\circ\gamma)]\nonumber \\
  &=& {{d(\phi\circ\gamma)^\mu}\over{d\lambda}}
  {{\partial(f\circ\phi^{-1})}\over{\partial x^\mu}}\nonumber \\
  &=&  {{dx^\mu}\over{d\lambda}}\p\mu f\ . \label{2.9}
\eea
The first line simply takes the informal expression on the left hand
side and rewrites it as an honest derivative of the function
$(f\circ\gamma):\R\rightarrow\R$.  The second line just comes from the
definition of the inverse map $\phi^{-1}$ (and associativity of the
operation of composition).  The third line is the formal chain rule
(2.2), and the last line is a return to the informal notation of
the start.  Since the function $f$ was arbitrary, we have
\be
  {d\over{d\lambda}} = {{dx^\mu}\over{d\lambda}}\p\mu\ .\label{2.10}
\ee
Thus, the partials $\{\p\mu\}$ do indeed represent a good basis for the
vector space of directional derivatives, which we can therefore
safely identify with the tangent space.

Of course, the vector represented by ${d\over{d\lambda}}$ is one
we already know; it's the tangent vector to the curve with parameter
$\lambda$.  Thus (2.10) can be thought of as a restatement of (1.24),
where we claimed the that components of the tangent vector were 
simply $dx^\mu/d\lambda$.  The only difference is that we are working
on an arbitrary manifold, and we have specified our basis vectors
to be $\e\mu=\p\mu$.

This particular basis ($\e\mu=\p\mu$) is known as a {\bf coordinate
basis} for $T_p$; it is the formalization of the notion of setting up
the basis vectors to point along the coordinate axes.  
There is no reason why we are limited to coordinate
bases when we consider tangent vectors; it is sometimes more convenient,
for example, to use orthonormal bases of some sort.  However, the
coordinate basis is very simple and natural, and we will use it 
almost exclusively throughout the course.

One of the advantages of the rather abstract point of view we have
taken toward vectors is that the transformation law is
immediate.  Since the basis vectors are $\e\mu=\p\mu$, the basis
vectors in some new coordinate system $x^{\mu'}$ are given by
the chain rule (2.3) as
\be
  \p{\mu'} = {{\partial x^\mu}\over{\partial x^{\mu'}}}\p\mu\ .
  \label{2.11}
\ee
We can get the transformation law for vector components by the same
technique used in flat space, demanding the the vector $V=V^\mu\p\mu$
be unchanged by a change of basis.  We have
\bea
  V^\mu\p\mu &=&  V^{\mu'}\p{\mu'}\nonumber \\
  &=&  V^{\mu'}{{\partial x^\mu}\over{\partial x^{\mu'}}}\p\mu\ ,
  \label{2.12}
\eea
and hence (since the matrix $\partial x^{\mu'}/\partial x^\mu$ is the 
inverse of the matrix $\partial x^{\mu}/\partial x^{\mu'}$),
\be
  V^{\mu'} = {{\partial x^{\mu'}}\over{\partial x^{\mu}}}V^\mu
  \ .\label{2.13}
\ee
Since the basis vectors are usually not written explicitly, the rule
(2.13) for transforming components is what we call the ``vector
transformation law.''  We notice that it is compatible with the 
transformation of vector components in special relativity under
Lorentz transformations, $V^{\mu'} = \Lambda^{\mu'}{}_\mu V^\mu$,
since a Lorentz transformation is a special kind of coordinate
transformation, with $x^{\mu'} = \Lambda^{\mu'}{}_\mu x^\mu$.  But
(2.13) is much more general, as it encompasses the behavior of vectors
under arbitrary changes of coordinates (and therefore bases), not just 
linear transformations.  As usual, we are trying to emphasize a
somewhat subtle ontological distinction --- tensor components do not
change when we change coordinates, they change when we change the
basis in the tangent space, but we have decided to use the coordinates
to define our basis.  Therefore a change of coordinates induces a
change of basis:

\begin{figure}[h]
  \centerline{
  \psfig{figure=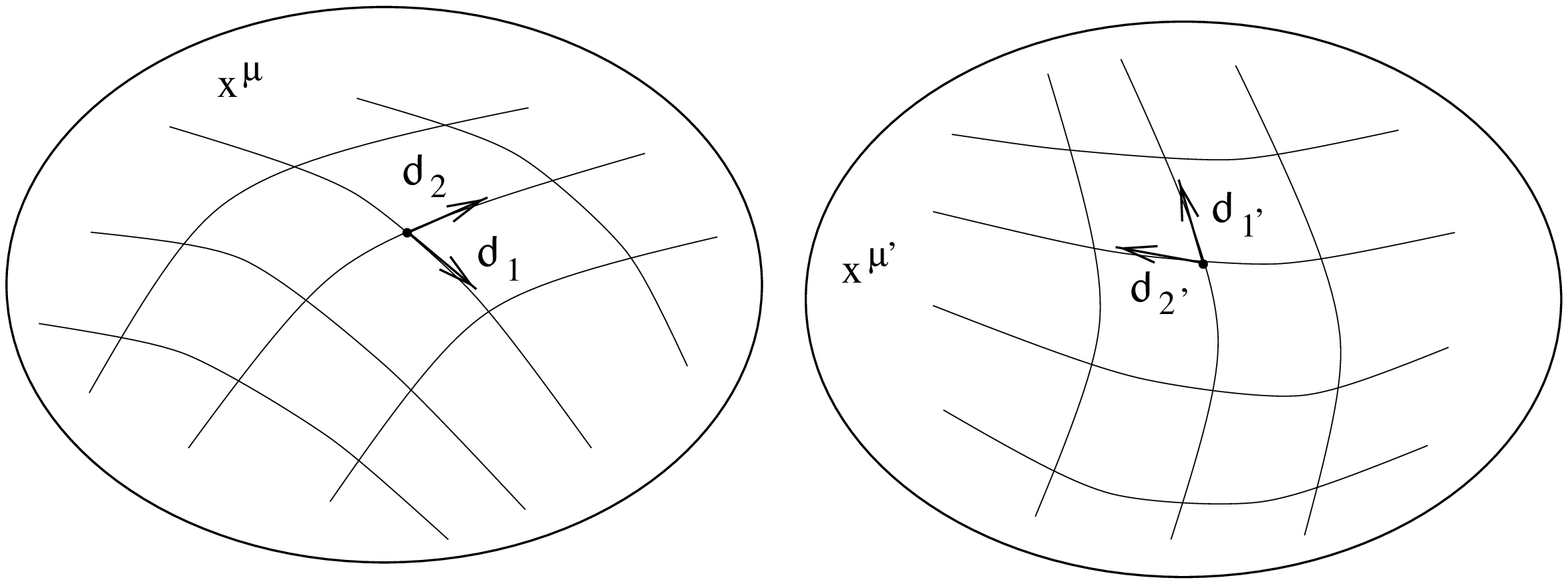,angle=-90,height=6cm}}
\end{figure}

Having explored the world of vectors, we continue to retrace the steps
we took in flat space, and now consider dual vectors (one-forms).
Once again the cotangent space $T^*_p$ is the set of linear maps
$\omega:T_p\rightarrow\R$.  The canonical example of a one-form is
the gradient of a function $f$, denoted $\d f$.  Its action on a
vector ${d\over{d\lambda}}$ is exactly the directional derivative of
the function:
\be
  \d f\left({d\over{d\lambda}}\right)={{df}\over{d\lambda}}\ .\label{2.14}
\ee
It's tempting to think, ``why shouldn't the function $f$ itself be 
considered the one-form, and $df/d\lambda$ its action?''  The point
is that a one-form, like a vector, exists only at the point it
is defined, and does not depend on information at other points on $M$.
If you know a function in some neighborhood of a point you can take
its derivative, but not just from knowing its value at the point;
the gradient, on the other hand, encodes precisely the
information necessary to take the directional derivative along any curve
through $p$, fulfilling its role as a dual vector.

Just as the partial derivatives along coordinate axes provide a natural
basis for the tangent space, the gradients of the coordinate functions
$x^\mu$ provide a natural basis for the cotangent space.  Recall that in
flat space we constructed a basis for $T^*_p$ by demanding that
$\t\mu(\e\nu)=\delta^\mu_\nu$.  Continuing the same philosophy on an
arbitrary manifold, we find that (2.14) leads to
\be
  \d x^\mu(\p\nu) = {{\partial x^\mu}\over{\partial x^\nu}} 
  =\delta^\mu_\nu\ .\label{2.15}
\ee
Therefore the gradients $\{\d x^\mu\}$ are an appropriate set of
basis one-forms; an arbitrary one-form is expanded into components
as $\omega = \omega_\mu\,\d x^\mu$.

The transformation properties of basis dual vectors and components
follow from what is by now the usual procedure.  We obtain, for
basis one-forms,
\be
  \d x^{\mu'} = {{\partial x^{\mu'}}\over{\partial x^{\mu}}}\,\d x^\mu
  \  ,\label{2.16}
\ee
and for components,
\be
  \omega_{\mu'} = {{\partial x^{\mu}}\over{\partial x^{\mu'}}}\omega_\mu
  \ .\label{2.17}
\ee
We will usually write the components $\omega_\mu$ when we speak about a
one-form $\omega$.

The transformation law for general tensors follows this same pattern of
replacing the Lorentz transformation matrix used in flat space with a
matrix representing more general coordinate transformations.
A $(k,l)$ tensor $T$ can be expanded
\be
  T = T^{\mu_1 \cdots \mu_k}{}_{\nu_1\cdots\nu_l}
  \p{\mu_1}\otimes\cdots\otimes\p{\mu_k}\otimes
  \d x^{\nu_1}\otimes\cdots\otimes\d x^{\nu_l}\ ,\label{2.18}
\ee
and under a coordinate transformation the components change according
to
\be
  T^{\mu_1' \cdots \mu_k'}{}_{\nu_1'\cdots\nu_l'} = 
  {{\partial x^{\mu_1'}}\over{\partial x^{\mu_1}}}\cdots
  {{\partial x^{\mu_k'}}\over{\partial x^{\mu_k}}}
  {{\partial x^{\nu_1}}\over{\partial x^{\nu_1'}}}\cdots
  {{\partial x^{\nu_l}}\over{\partial x^{\nu_l'}}}
  T^{\mu_1 \cdots \mu_k}{}_{\nu_1\cdots\nu_l} \ .\label{2.19}
\ee
This tensor transformation law is straightforward to remember, since
there really isn't anything else it could be, given the placement
of indices.  However, it is often easier to transform a tensor by
taking the identity of basis vectors and one-forms as partial
derivatives and gradients at face value, and simply substituting in the
coordinate transformation.  As an example consider a symmetric $(0, 2)$ 
tensor $S$ on a 2-dimensional manifold, whose components in a coordinate 
system 
$(x^1=x, x^2=y)$ are given by
\be
  S_\mn = \left(\matrix{x&0\cr 0&1\cr}\right)\ .\label{2.20}
\ee
This can be written equivalently as
\bea
  S &=&  S_\mn (\d x^\mu \otimes \d x^\nu)\nonumber \\
  &=&  x(\d x)^2 + (\d y)^2\ , \label{2.21}
\eea
where in the last line the tensor product symbols are suppressed
for brevity.  Now consider new coordinates
\bea
  x' &=&  x^{1/3}\nonumber \\ y' &=&  e^{x+y}\ . \label{2.22}
\eea
This leads directly to
\bea
  x &=&  (x')^3\nonumber \\ y &=&  \ln(y') - (x')^3\nonumber \\
  \d x &=&  3(x')^2 \,\d x'\nonumber \\ \d y &=&  {1\over{y'}}\,\d y' 
  - 3(x')^2\,\d x'\ .
  \label{2.23}
\eea
We need only plug these expressions directly into (2.21) to obtain
(remembering that tensor products don't commute, so $\d x'\,\d y' \neq
\d y' \,\d x'$):
\be
  S= 9(x')^4[1+(x')^3](\d x')^2 -3{{(x')^2}\over{y'}}(\d x' \,\d y'
  +\d y' \,\d x') + {1\over{(y')^2}}(\d y')^2\ ,\label{2.24}
\ee
or
\be
  S_{\mu'\nu'} = \left(\matrix{9(x')^4[1+(x')^3]&-3{{(x')^2}\over{y'}}\cr
  -3{{(x')^2}\over{y'}}&{1\over{(y')^2}}\cr}\right)\ .\label{2.25}
\ee
Notice that it is still symmetric.  We did not use the transformation
law (2.19) directly, but doing so would have yielded the same result,
as you can check.

For the most part the various tensor operations we defined in flat
space are unaltered in a more general setting: contraction,
symmetrization, etc.  There are three important exceptions:
partial derivatives, the metric, and the Levi-Civita
tensor.  Let's look at the partial derivative first.

The unfortunate fact is that the partial derivative of a tensor is
not, in general, a new tensor.  The gradient, which is the partial
derivative of a scalar, is an honest $(0,1)$ tensor, as we have 
seen.  But the partial derivative of higher-rank tensors is not
tensorial, as we can see by considering
the partial derivative of a one-form, $\p\mu W_\nu$, and changing to
a new coordinate system:
\bea
  {{\partial}\over{\partial x^{\mu'}}}W_{\nu'} &=& 
  {{\partial x^{\mu}}\over{\partial x^{\mu'}}}
  {{\partial}\over{\partial x^{\mu}}}\left({{\partial x^{\nu}}
  \over{\partial x^{\nu'}}}W_\nu\right)\nonumber \\
  &=& {{\partial x^{\mu}}\over{\partial x^{\mu'}}}
  {{\partial x^{\nu}}\over{\partial x^{\nu'}}}
  \left({{\partial}\over{\partial x^{\mu}}}W_\nu\right)
  + W_\nu{{\partial x^{\mu}}\over{\partial x^{\mu'}}}
  {{\partial}\over{\partial x^\mu}}
  {{\partial x^{\nu}}\over{\partial x^{\nu'}}}\ .
   \label{2.26}
\eea
The second term in the last line should not be there if $\p\mu W_\nu$
were to transform as a $(0,2)$ tensor.  As you can see, it arises because 
the derivative of the transformation matrix does not vanish, as it did 
for Lorentz transformations in flat space.  

On the other hand, the exterior derivative operator $\d$ does form
an antisymmetric $(0,p+1)$ tensor when acted on a $p$-form.  For $p=1$ we 
can see this from (2.26); the offending non-tensorial term can be written
\be
  W_\nu{{\partial x^{\mu}}\over{\partial x^{\mu'}}}
  {{\partial}\over{\partial x^\mu}}
  {{\partial x^{\nu}}\over{\partial x^{\nu'}}} =
  W_\nu{{\partial^2 x^{\nu}}\over{\partial x^{\mu'}
  \partial x^{\nu'}}}\ .\label{2.27}
\ee
This expression is symmetric in $\mu'$ and $\nu'$, since partial
derivatives commute.  But the exterior derivative is defined to be
the antisymmetrized partial derivative, so this term vanishes (the
antisymmetric part of a symmetric expression is zero).  We are then
left with the correct tensor transformation law; extension to arbitrary
$p$ is straightforward.  So the exterior derivative is a legitimate
tensor operator; it is not, however, an adequate substitute for the
partial derivative, since it is only defined on forms.  In the next
section we will define a covariant derivative, which can be thought
of as the extension of the partial derivative to arbitrary manifolds.

The metric tensor is such an important object in curved space that
it is given a new symbol, $g_\mn$ (while $\eta_\mn$ is reserved
specifically for the Minkowski metric).  There are few restrictions
on the components of $g_\mn$, other than that it be a symmetric
$(0,2)$ tensor.  It is usually taken to be non-degenerate, meaning that
the determinant $g=|g_\mn|$ doesn't vanish.  This allows us to define
the inverse metric $g^\mn$ via
\be
  g^\mn g_{\nu\sigma} = \delta^\mu_\sigma\ .\label{2.28}
\ee
The symmetry of $g_\mn$ implies that $g^\mn$ is also symmetric.
Just as in special relativity, the metric and its inverse may be 
used to raise and lower indices on tensors.

It will take several weeks to fully appreciate the role of the
metric in all of its glory, but for purposes of inspiration we can
list the various uses to which $g_\mn$ will be put: (1) the metric
supplies a notion of ``past'' and ``future''; (2) the metric
allows the computation of path length and proper time; (3) the metric
determines the ``shortest distance'' between two points (and therefore
the motion of test particles); (4) the metric replaces the Newtonian
gravitational field $\phi$; (5) the metric provides a notion of
locally inertial frames and therefore a sense of ``no rotation'';
(6) the metric determines causality, by defining the speed of light
faster than which no signal can travel; (7) the metric replaces the
traditional Euclidean three-dimensional dot product of Newtonian
mechanics; and so on.  Obviously these ideas are not all completely
independent, but we get some sense of the importance of this tensor.

In our discussion of path lengths in special relativity we (somewhat
handwavingly) introduced the line element as $ds^2 = \eta_\mn 
dx^\mu dx^\nu$, which was used to get the length of a path.  Of course
now that we know that $\d x^\mu$ is really a basis dual vector, it 
becomes natural to use the terms ``metric'' and ``line element''
interchangeably, and write
\be
  ds^2 = g_\mn \,\d x^\mu \,\d x^\nu\ .\label{2.29}
\ee
(To be perfectly consistent we should write this as ``$g$'', and sometimes
will, but more often than not $g$ is used for the determinant 
$|g_\mn|$.)
For example, we know that the Euclidean line element in a 
three-dimensional space with Cartesian coordinates is
\be
  ds^2 = (\d x)^2 + (\d y)^2 + (\d z)^2\ .\label{2.30}
\ee
We can now change to any coordinate system we choose.  For example,
in spherical coordinates we have
\bea
  x &=&  r\sin\theta \cos\phi\nonumber \\
  y &=&  r\sin\theta \sin\phi\nonumber \\ z &=&  r\cos\theta\ ,  \label{2.31}
\eea
which leads directly to
\be
  ds^2 = \d r^2 + r^2 \,\d\theta^2 + r^2\sin^2\theta\,\d \phi^2\ .\label{2.32}
\ee
Obviously the components of the metric look different than those in
Cartesian coordinates, but all of the properties of the space remain
unaltered.

Perhaps this is a good time to note that most references are not
sufficiently picky to distinguish between
``$dx$'', the informal notion of an infinitesimal displacement, and
``$\d x$'', the rigorous notion of a basis one-form given by the
gradient of a coordinate function.  In fact our notation ``$ds^2$''
does not refer to the exterior derivative of anything, or the square
of anything; it's just conventional shorthand for the metric tensor.
On the other hand, ``$(\d x)^2$'' refers specifically to the $(0,2)$
tensor $\d x\otimes\d x$.

A good example of a
space with curvature is the two-sphere, which can be thought of
as the locus of points in $\R^3$ at distance 1 from the origin.  The 
metric in the $(\theta, \phi)$ coordinate system comes from setting $r=1$
and $\d r=0$ in (2.32):
\be
  ds^2 = \d\theta^2 + \sin^2\theta\,\d \phi^2\ .\label{2.33}
\ee
This is completely consistent with the interpretation of $ds$ as an
infinitesimal length, as illustrated in the figure.

\begin{figure}
  \centerline{
  \psfig{figure=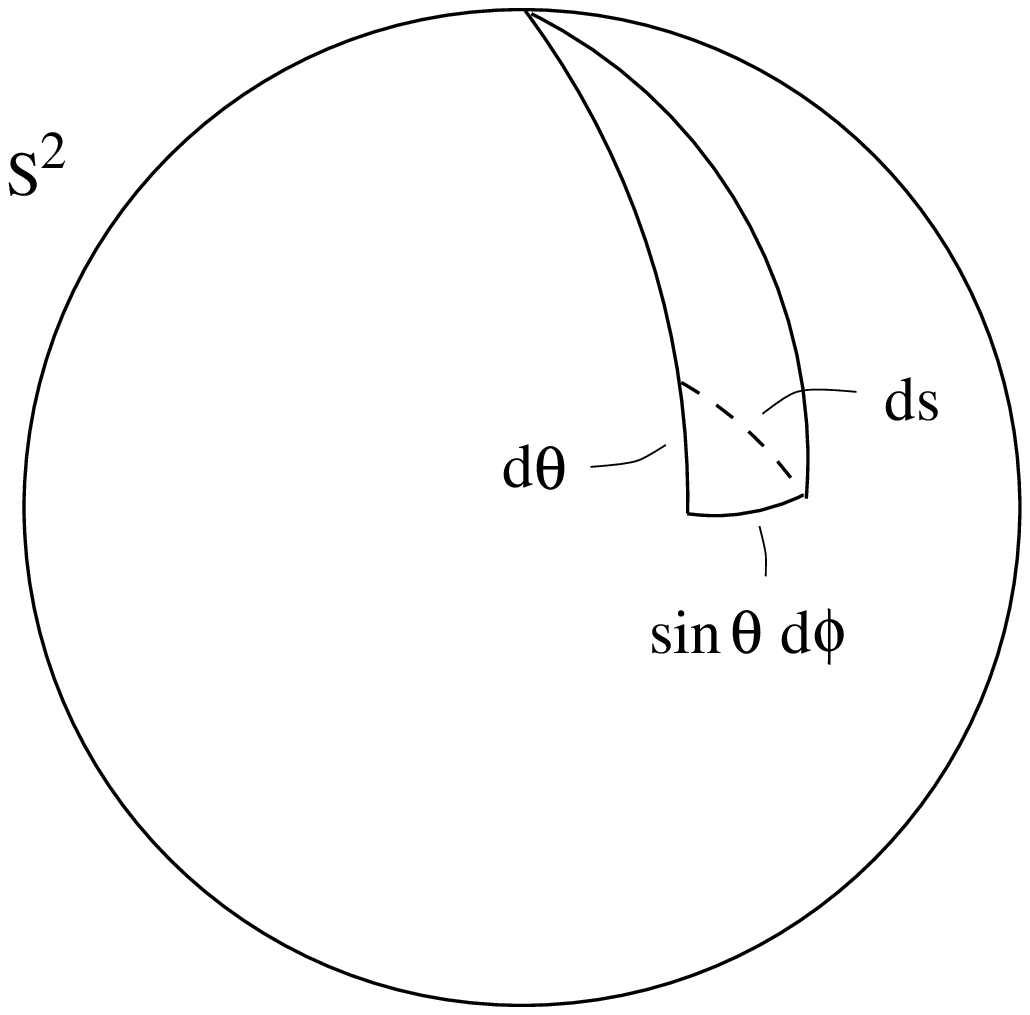,angle=-90,height=6cm}}
\end{figure}

As we shall see, the metric tensor contains all the information we need
to describe the curvature of the manifold (at least in Riemannian
geometry; we will actually indicate somewhat more general approaches).
In Minkowski space we can choose coordinates in which the components of
the metric are constant; but it should be clear that the existence of 
curvature is more subtle than having the metric depend on the coordinates,
since in the example above we showed how the metric in flat Euclidean
space in spherical coordinates is a function of $r$ and $\theta$.  Later,
we shall see that constancy of the metric components is sufficient
for a space to be flat, and in fact there always exists a coordinate 
system on any flat space in which the metric is constant.  But we might
not want to work in such a coordinate system, and we might not even 
know how to find it; therefore we will want a more precise characterization
of the curvature, which will be introduced down the road.

A useful characterization of the metric is obtained by putting $g_\mn$
into its {\bf canonical form}.  In this form the metric components become
\be
  g_\mn = {\rm ~diag~}(-1,-1,\ldots,-1,+1,+1,\ldots,+1,
  0,0,\ldots,0)\ ,\label{2.34}
\ee
where ``diag'' means a diagonal matrix with the given elements.  If
$n$ is the dimension of the manifold, $s$ is the number of $+1$'s in
the canonical form, and $t$ is the number of $-1$'s, then
$s-t$ is the {\bf signature}
of the metric (the difference in the number of minus and plus signs),
and $s+t$ is the {\bf rank} of the metric (the number of nonzero
eigenvalues).  If a metric is continuous, the rank and signature of
the metric tensor field are the same at every point, and if the metric
is nondegenerate the rank is equal to the dimension $n$.  We will
always deal with continuous, nondegenerate metrics.  If all of the
signs are positive ($t=0$) the metric is called {\bf Euclidean} or
{\bf Riemannian} (or just ``positive definite''), while if there is a 
single minus ($t=1$) it is called {\bf Lorentzian} or 
{\bf pseudo-Riemannian}, and any metric with some $+1$'s and some 
$-1$'s is called ``indefinite.''  (So the word ``Euclidean''
sometimes means that the space is flat, and sometimes doesn't, but
always means that the canonical form is strictly positive; the
terminology is unfortunate but standard.)  The spacetimes of interest
in general relativity have Lorentzian metrics.

We haven't yet demonstrated that it is always possible to but the metric
into canonical form.  In fact it is always
possible to do so at some point $p\in M$, but in general it will only
be possible at that single point, not in any neighborhood of $p$.
Actually we can do slightly better than this; it turns out that at
any point $p$ there exists a coordinate system in which $g_\mn$ takes
its canonical form and the first derivatives $\p\sigma g_\mn$ all vanish 
(while the second derivatives $\p\rho \p\sigma g_\mn$ cannot be made
to all vanish).  Such coordinates are known as {\bf Riemann normal
coordinates}, and  the associated basis vectors constitute
a {\bf local Lorentz frame}.  Notice
that in Riemann normal coordinates (or RNC's) the metric at $p$ looks 
like that of flat space ``to first order.''  This is the rigorous
notion of the idea that ``small enough regions of spacetime look like
flat (Minkowski) space.''  (Also, there is no difficulty in simultaneously
constructing sets of basis vectors at every point in $M$ such that the
metric takes its canonical form; the problem is that in general this
will not be a {\it coordinate} basis, and there will be no way to
make it into one.)

We won't consider the detailed proof of this statement; it can be
found in Schutz, pp.~158-160, where it goes by the name of the 
``local flatness theorem.''  (He also calls local Lorentz frames
``momentarily comoving reference frames,'' or MCRF's.)  It is useful
to see a sketch of the proof, however, for the specific case of a
Lorentzian metric in four dimensions.  The idea is to consider the
transformation law for the metric
\be
  g_{\mu'\nu'} = {{\partial x^\mu}\over{\partial x^{\mu'}}}
  {{\partial x^\nu}\over{\partial x^{\nu'}}} g_\mn\ ,\label{2.35}
\ee
and expand both sides in Taylor series in the sought-after
coordinates $x^{\mu'}$.  The expansion of the old coordinates $x^\mu$
looks like
\be
  x^\mu = \left({{\partial x^\mu}\over{\partial x^{\mu'}}}\right)_p
  x^{\mu'} + {1\over 2} \left({{\partial^2 x^\mu}\over
  {\partial x^{\mu_1'}\partial x^{\mu_2'}}}\right)_p 
  x^{\mu_1'}x^{\mu_2'} + {1\over 6} \left({{\partial^3 x^\mu}\over
  {\partial x^{\mu_1'}\partial x^{\mu_2'}\partial x^{\mu_3'}}}\right)_p 
  x^{\mu_1'}x^{\mu_2'}x^{\mu_3'} +\cdots\ ,\label{2.36}
\ee
with the other expansions proceeding along the same lines.  (For
simplicity we have set $x^\mu(p)=x^{\mu'}(p)=0$.)  Then, using some
extremely schematic notation, the expansion of (2.35) to second order is
\bea
  \lefteqn{\left(g'\right)_p + \left(\partial' g'\right)_p x' +
  \left(\partial'\partial' g'\right)_p x' x'}\nonumber \\
   &=&  
  \left({{\partial x}\over{\partial x'}}{{\partial x}\over{\partial x'}}
  g\right)_p + \left({{\partial x}\over{\partial x'}}
  {{\partial^2 x}\over{\partial x'\partial x'}}g +
  {{\partial x}\over{\partial x'}}{{\partial x}\over{\partial x'}}
  \partial' g\right)_p x' \nonumber \\
  & & \quad + \left({{\partial x}\over{\partial x'}}
  {{\partial^3 x}\over{\partial x'\partial x'\partial x'}}g +
  {{\partial^2 x}\over{\partial x'\partial x'}}
  {{\partial^2 x}\over{\partial x'\partial x'}}g +
  {{\partial x}\over{\partial x'}}
  {{\partial^2 x}\over{\partial x'\partial x'}}\partial' g +
  {{\partial x}\over{\partial x'}}{{\partial x}\over{\partial x'}}
  \partial'\partial' g\right)_p x' x'\ .  \label{2.37}
\eea
We can set terms of equal order in $x'$ on each side equal to each other.
Therefore, the components $g_{\mu'\nu'}(p)$, 10 numbers in all (to
describe a symmetric two-index tensor), are determined by the
matrix $(\partial x^\mu/\partial x^{\mu'})_p$.  This is a $4\times 4$
matrix with no constraints; thus, 16 numbers we are free to choose.
Clearly this is enough freedom to put the 10 numbers of $g_{\mu'\nu'}(p)$
into canonical form, at least as far as having enough degrees of
freedom is concerned.  (In fact there are some limitations --- if you go
through the procedure carefully, you find for example that you cannot
change the signature and rank.)  The six remaining degrees of freedom can
be interpreted as exactly the six parameters of the Lorentz group;
we know that these leave the canonical form unchanged.  At first
order we have the derivatives $\p{\sigma'}g_{\mu'\nu'}(p)$, four
derivatives of ten components for a total of 40 numbers.  But looking
at the right hand side of (2.37) we see that we now have the additional
freedom to choose $(\partial^2 x^\mu/\partial x^{\mu'_1}
\partial x^{\mu_2'})_p$.  In this set of numbers there are 10 
independent choices of the indices $\mu_1'$ and $\mu_2'$ (it's symmetric, 
since partial derivatives commute) and four choices of $\mu$, for
a total of 40 degrees of freedom.  This is precisely the amount of
choice we need to determine all of the first derivatives of the metric,
which we can therefore set to zero.  At second order, however, we are
concerned with $\p{\rho'}\p{\sigma'}g_{\mu'\nu'}(p)$; this is symmetric
in $\rho'$ and $\sigma'$ as well as $\mu'$ and $\nu'$, for a total of
$10\times 10=100$ numbers.  Our ability to make additional choices
is contained in $(\partial^3 x^\mu/\partial x^{\mu'_1}\partial x^{\mu'_2}
\partial x^{\mu_3'})_p$.  This is symmetric in the three lower indices,
which gives 20 possibilities, times four for the upper index gives us
80 degrees of freedom --- 20 fewer than we require to set the second
derivatives of the metric to zero.  So in fact we cannot make the second
derivatives vanish; the deviation from flatness must therefore be
measured by the 20 coordinate-independent degrees of freedom representing
the second derivatives of the 
metric tensor field.  We will see later how this comes about, when we
characterize curvature using the Riemann tensor, which will turn out
to have 20 independent components.

The final change we have to make to our tensor knowledge now that
we have dropped the assumption of flat space has to do with the
Levi-Civita tensor, $\epsilon_{\mu_1\mu_2\cdots\mu_n}$.  Remember
that the flat-space version of this object, which we will now 
denote by $\tilde\epsilon_{\mu_1\mu_2\cdots\mu_n}$, was defined 
as
\be
  \tilde\epsilon_{\mu_1\mu_2\cdots\mu_n}=\left\{\matrix{
  +1 {\rm ~if~}\mu_1\mu_2\cdots\mu_n
  {\rm ~is~an~even~permutation~of~}01\cdots (n-1)\ ,\hfill\cr 
  -1 {\rm ~if~}\mu_1\mu_2\cdots\mu_n 
  {\rm ~is~an~odd~permutation~of~}01\cdots (n-1)\ ,\hfill\cr
  0{\rm ~otherwise}\ .\hfill\cr}\right.
  \label{2.38}
\ee
We will now define the {\bf Levi-Civita symbol} to be exactly this
$\tilde\epsilon_{\mu_1\mu_2\cdots\mu_n}$ --- that is, an object
with $n$ indices which has the components specified above {\it in
any coordinate system}.  This is called a ``symbol,'' of course,
because it is not a tensor; it is defined not to change under
coordinate transformations.  We can relate its behavior to that of
an ordinary tensor by first noting that, given some $n\times n$ matrix
$M^\mu{}_{\mu'}$, the determinant $|M|$ obeys
\be
  \tilde\epsilon_{\mu_1'\mu_2'\cdots\mu_n'} |M| =
  \tilde\epsilon_{\mu_1\mu_2\cdots\mu_n} M^{\mu_1}{}_{\mu_1'}
  M^{\mu_2}{}_{\mu_2'}\cdots M^{\mu_n}{}_{\mu_n'}\ .\label{2.39}
\ee
This is just a true fact about the determinant which you can find in
a sufficiently enlightened linear algebra book.  If follows that,
setting $M^\mu{}_{\mu'}=\partial x^\mu/\partial x^{\mu'}$, we have
\be
  \tilde\epsilon_{\mu_1'\mu_2'\cdots\mu_n'} =
  \left|{{\partial x^{\mu'}}\over{\partial x^\mu}}\right|
  \tilde\epsilon_{\mu_1\mu_2\cdots\mu_n}
  {{\partial x^{\mu_1}}\over{\partial x^{\mu_1'}}}
  {{\partial x^{\mu_2}}\over{\partial x^{\mu_2'}}}\cdots
  {{\partial x^{\mu_n}}\over{\partial x^{\mu_n'}}}\ .\label{2.40}
\ee
This is close to the tensor transformation law, except for the 
determinant out front.  Objects which transform in this way are
known as {\bf tensor densities}.  Another example is given by
the determinant of the metric, $g=|g_\mn|$.  It's easy to check
(by taking the determinant of both sides of (2.35)) that under a
coordinate transformation we get
\be
  g(x^{\mu'}) = \left|{{\partial x^{\mu'}}\over{\partial x^\mu}}
  \right|^{-2} g(x^\mu)\ .\label{2.41}
\ee
Therefore $g$ is also not a tensor; it transforms in a way similar
to the Levi-Civita symbol, except that the
Jacobian is raised to the $-2$ power.  The power to which the
Jacobian is raised is known as the {\bf weight} of the tensor
density; the Levi-Civita symbol is a density of weight $1$, while
$g$ is a (scalar) density of weight $-2$.

However, we don't like tensor densities, we like tensors.  There
is a simple way to convert a density into an honest tensor ---
multiply by $|g|^{w/2}$, where $w$ is the weight of the density
(the absolute value signs are there because $g<0$ for Lorentz
metrics).  The result will transform according to the tensor
transformation law.  Therefore, for example, we can define the 
Levi-Civita tensor as
\be
  \epsilon_{\mu_1\mu_2\cdots\mu_n}= \sqrt{|g|}\,
  \tilde\epsilon_{\mu_1\mu_2\cdots\mu_n}\ .\label{2.42}
\ee
It is this tensor
which is used in the definition of the Hodge dual, (1.87), which
is otherwise unchanged when generalized to arbitrary manifolds.
Since this is a real tensor, we can raise indices, etc.  Sometimes
people define a version of the Levi-Civita symbol with upper indices,
$\tilde\epsilon^{\mu_1\mu_2\cdots\mu_n}$, whose components are
numerically equal to the symbol with lower indices.  This turns out
to be a density of weight $-1$, and is related to the tensor with
upper indices by
\be
  \epsilon^{\mu_1\mu_2\cdots\mu_n} = {\rm ~sgn}(g){1\over{\sqrt{|g|}}}
  \,\tilde\epsilon^{\mu_1\mu_2\cdots\mu_n}\ .\label{2.43}
\ee

As an aside, we should come clean and admit that, even with the factor
of $\sqrt{|g|}$, the Levi-Civita tensor is in some sense not a true
tensor, because on some manifolds it cannot be globally defined.  Those
on which it can be defined are called {\bf orientable}, and we will deal
exclusively with orientable manifolds in this course.  An example of
a non-orientable manifold is the M\"obius strip; see Schutz's
{\sl Geometrical Methods in Mathematical Physics} (or a similar text) 
for a discussion.

One final appearance of tensor densities is in integration on manifolds.
We will not do this subject justice, but at least a casual glance is
necessary.  You have probably been exposed to the fact that in ordinary
calculus on $\R^n$ the volume
element $d^nx$ picks up a factor of the Jacobian under
change of coordinates:
\be
  d^nx' = \left|{{\partial x^{\mu'}}\over{\partial x^\mu}}
  \right| d^nx\ .\label{2.44}
\ee
There is actually a beautiful explanation of this formula from the
point of view of differential forms, which arises from the following fact:
{\it on an $n$-dimensional manifold, the integrand is properly understood
as an $n$-form}.  The naive volume element $d^nx$ is itself a density
rather than an $n$-form, but there is no difficulty in using it
to construct a real $n$-form.
To see how this works, we should make the identification
\be
  d^nx \leftrightarrow \d x^0\wedge \cdots \wedge\,\d x^{n-1}
  \ .\label{2.45}
\ee
The expression on the right hand side can be misleading, because it looks 
like a tensor (an $n$-form, actually) but is really a density.  
Certainly if we have two
functions $f$ and $g$ on $M$, then $\d f$ and $\d g$ are one-forms, 
and $\d f\wedge \,\d g$ is a two-form.  But we would like to interpret
the right hand side of (2.45) as a coordinate-dependent object which, 
in the $x^\mu$ coordinate system, acts like $\d x^0\wedge \cdots 
\wedge\,\d x^{n-1}$.  This sounds tricky, but in fact it's just an
ambiguity of notation, and in practice we will just use the 
shorthand notation ``$d^nx$''.

To justify this song and dance, let's see how (2.45) changes
under coordinate transformations.  First notice that the 
definition of the wedge product allows us to write
\be
  \d x^0\wedge \cdots \wedge\,\d x^{n-1} = {1\over {n!}}
  \tilde\epsilon_{\mu_1\cdots\mu_n}
  \,\d x^{\mu_1}\wedge \cdots \wedge\,\d x^{\mu_n}\ ,\label{2.46}
\ee
since both the wedge product and the Levi-Civita symbol are completely
antisymmetric.  Under a coordinate transformation 
$\tilde\epsilon_{\mu_1\cdots\mu_n}$ stays the same while the one-forms
change according to (2.16), leading to
\bea
  \tilde\epsilon_{\mu_1\cdots\mu_n}
  \,\d x^{\mu_1}\wedge \cdots \wedge \,\d x^{\mu_n} 
  &=& \tilde\epsilon_{\mu_1\cdots\mu_n}
  {{\partial x^{\mu_1}}\over{\partial x^{\mu_1'}}}\cdots
  {{\partial x^{\mu_n}}\over{\partial x^{\mu_n'}}}
  \,\d x^{\mu_1'}\wedge \cdots \wedge \,\d x^{\mu_n'}\nonumber \\
  &=&  \left|{{\partial x^{\mu}}\over{\partial x^{\mu'}}}\right|
  \tilde\epsilon_{\mu_1'\cdots\mu_n'}
  \,\d x^{\mu_1'}\wedge \cdots \wedge \,\d x^{\mu_n'}\ .  \label{2.47}
\eea
Multiplying by the Jacobian on both sides recovers (2.44).

It is clear that the naive volume element $d^nx$ transforms as a 
density, not a tensor, but it is straightforward to construct an
invariant volume element by multiplying by $\sqrt{|g|}$:
\be
  \sqrt{|g'|}\,\d x^{0'}\wedge \cdots \wedge\,\d x^{(n-1)'} 
  = \sqrt{|g|}\,\d x^0\wedge \cdots \wedge\,\d x^{n-1}\ ,\label{2.48}
\ee
which is of course just $(n!)^{-1}\epsilon_{\mu_1\cdots\mu_n}
\,\d x^{\mu_1}\wedge \cdots \wedge \,\d x^{\mu_n}$.
In the interest of simplicity we will usually write the volume 
element as $\sqrt{|g|}\,d^nx$, rather than as the explicit wedge
product $\sqrt{|g|}\,\d x^0\wedge \cdots \wedge\,\d x^{n-1}$; it will 
be enough to keep in mind that it's supposed to be an $n$-form.

As a final aside to finish this section, let's consider one of the
most elegant and powerful theorems of differential geometry: Stokes's
theorem.  This theorem is the generalization of the fundamental 
theorem of calculus, $\int^a_b dx = a-b$.  Imagine that we have an
$n$-manifold $M$ with boundary $\partial M$, and an $(n-1)$-form
$\omega$ on $M$.  (We haven't discussed manifolds with boundaries,
but the idea is obvious; $M$ could for instance be the interior of
an $(n-1)$-dimensional closed surface $\partial M$.)  Then $\d \omega$
is an $n$-form, which can be integrated over $M$, while $\omega$ itself
can be integrated over $\partial M$.  Stokes's theorem is then
\be
  \int_M \d \omega = \int_{\partial M}\omega\ .\label{2.49}
\ee
You can convince yourself that different special cases of this
theorem include not only the fundamental theorem of calculus, but
also the theorems of Green, Gauss, and Stokes, familiar from vector
calculus in three dimensions.

\eject

\thispagestyle{plain}

\setcounter{equation}{0}

\noindent{December 1997 \hfill {\sl Lecture Notes on General Relativity}
\hfill{Sean M.~Carroll}}

\vskip .2in

\setcounter{section}{2}
\section{Curvature}

In our discussion of manifolds, it became clear that there were
various notions we could talk about as soon as the manifold was
defined; we could define functions, take their derivatives, consider
parameterized paths, set up tensors, and so on.  Other concepts, such
as the volume of a region or the length of a path, required some additional
piece of structure, namely the introduction of a metric.  It would be
natural to think of the notion of ``curvature'', which we have already
used informally, is something that depends on the metric.  Actually this
turns out to be not quite true, or at least incomplete.  In fact there
is one additional structure we need to introduce --- a ``connection'' 
--- which is characterized by the curvature.  We will show how the 
existence of a metric implies a certain connection, whose curvature
may be thought of as that of the metric.

The connection becomes necessary when we attempt to address the
problem of the partial derivative not being a good tensor operator.
What we would like is a covariant derivative; that is, an operator
which reduces to the partial derivative in flat space with Cartesian
coordinates, but transforms as a tensor on an arbitrary manifold.
It is conventional to spend a certain amount of time motivating the
introduction of a covariant derivative, but in fact the need is
obvious; equations such as $\p\mu T^{\mn}=0$ are going to have to 
be generalized to curved space somehow.  So let's agree that a 
covariant derivative would be a good thing to have, and go about setting
it up.

In flat space in Cartesian coordinates, the partial derivative operator
$\p\mu$ is a map from $(k,l)$ tensor fields to $(k,l+1)$ tensor fields,
which acts linearly on its arguments and obeys the Leibniz rule on
tensor products.  All of this continues to be true in the more general
situation we would now like to consider, but the map provided by the
partial derivative depends on the coordinate system used.
We would therefore like to define a {\bf covariant derivative} operator
$\nabla$ to perform the functions of the partial derivative, but 
in a way independent of coordinates.  We therefore require that 
$\nabla$ be a map from $(k,l)$ tensor fields to $(k,l+1)$ tensor fields
which has these two properties:
\begin{enumerate}
  \item linearity: ~$\nabla(T+S) = \nabla T + \nabla S$ ;
  \item Leibniz (product) rule: ~$\nabla(T\otimes S) = 
(\nabla T)\otimes S + T\otimes (\nabla S)$ .
\end{enumerate}

If $\nabla$ is going to obey the Leibniz rule, it can always be written
as the partial derivative plus some linear transformation.  That is,
to take the covariant derivative we first take the partial derivative,
and then apply a correction to make the result covariant.  (We aren't
going to prove this reasonable-sounding statement, but Wald goes into
detail if you are interested.)  Let's consider what this means for the
covariant derivative of a vector $V^\nu$.  It means that, for each
direction $\mu$, the covariant derivative $\nabla_\mu$ will be given
by the partial derivative $\partial_\mu$ plus a correction specified
by a matrix $(\Gamma_\mu)^\rho{}_\sigma$ (an $n\times n$ matrix, where
$n$ is the dimensionality of the manifold, for each $\mu$).  In fact
the parentheses are usually dropped and we write these matrices,
known as the {\bf connection coefficients}, with haphazard index
placement as $\Gamma^\rho_{\mu\sigma}$.  We therefore have
\be
  \nabla_\mu V^\nu = \partial_\mu V^\nu + \Gamma^\nu_{\mu\lambda}
  V^\lambda\ .\label{3.1}
\ee
Notice that in the second term the index originally on $V$ has moved
to the $\Gamma$, and a new index is summed over.  If this is the
expression for the covariant derivative of a vector in terms of the 
partial derivative, we should be able to determine the transformation
properties of $\Gamma^\nu_{\mu\lambda}$ by demanding that the left
hand side be a $(1,1)$ tensor.  That is, we want the transformation
law to be
\be
  \nabla_{\mu'}V^{\nu'} = {{\partial x^\mu}\over{\partial x^{\mu'}}}
  {{\partial x^{\nu'}}\over{\partial x^{\nu}}}\nabla_{\mu}V^{\nu}
  \ .\label{3.2}
\ee
Let's look at the left side first; we can expand it using (3.1) and
then transform the parts that we understand:
\bea
  \nabla_{\mu'}V^{\nu'} &=&\partial_{\mu'} V^{\nu'} 
  + \Gamma^{\nu'}_{\mu'\lambda'}V^{\lambda'}\cr
  &=& {{\partial x^\mu}\over{\partial x^{\mu'}}}
  {{\partial x^{\nu'}}\over{\partial x^{\nu}}}\partial_{\mu} V^{\nu}
  + {{\partial x^\mu}\over{\partial x^{\mu'}}} V^\nu
  {{\partial}\over{\partial x^{\mu}}}
  {{\partial x^{\nu'}}\over{\partial x^{\nu}}}
  +  \Gamma^{\nu'}_{\mu'\lambda'}{{\partial x^{\lambda'}}\over
  {\partial x^{\lambda}}}V^{\lambda}\ . \label{3.3}
\eea
The right side, meanwhile, can likewise be expanded:
\be
  {{\partial x^\mu}\over{\partial x^{\mu'}}}
  {{\partial x^{\nu'}}\over{\partial x^{\nu}}}\nabla_{\mu}V^{\nu}
  = {{\partial x^\mu}\over{\partial x^{\mu'}}}
  {{\partial x^{\nu'}}\over{\partial x^{\nu}}}\partial_{\mu}V^{\nu}
  + {{\partial x^\mu}\over{\partial x^{\mu'}}}
  {{\partial x^{\nu'}}\over{\partial x^{\nu}}}
  \Gamma^\nu_{\mu\lambda}V^{\lambda}\ .\label{3.4}
\ee
These last two expressions are to be equated; the first terms in each
are identical and therefore cancel, so we have
\be
  \Gamma^{\nu'}_{\mu'\lambda'}{{\partial x^{\lambda'}}\over
  {\partial x^{\lambda}}}V^{\lambda} + 
  {{\partial x^\mu}\over{\partial x^{\mu'}}} V^\lambda
  {{\partial}\over{\partial x^{\mu}}}
  {{\partial x^{\nu'}}\over{\partial x^{\lambda}}}
  = {{\partial x^\mu}\over{\partial x^{\mu'}}}
  {{\partial x^{\nu'}}\over{\partial x^{\nu}}}
  \Gamma^\nu_{\mu\lambda}V^{\lambda}\ ,\label{3.5}
\ee
where we have changed a dummy index
from $\nu$ to $\lambda$.  This equation must be true for any vector
$V^\lambda$, so we can eliminate that on both sides.  Then the 
connection coefficients in the primed coordinates may be isolated by
multiplying by $\partial x^{\lambda}/\partial x^{\lambda'}$.  The result
is
\be
  \Gamma^{\nu'}_{\mu'\lambda'} = {{\partial x^\mu}\over{\partial x^{\mu'}}}
  {{\partial x^\lambda}\over{\partial x^{\lambda'}}}
  {{\partial x^{\nu'}}\over{\partial x^{\nu}}} \Gamma^\nu_{\mu\lambda}
  - {{\partial x^\mu}\over{\partial x^{\mu'}}} 
  {{\partial x^\lambda}\over{\partial x^{\lambda'}}}
  {{\partial^2 x^{\nu'}}\over{\partial x^{\mu}\partial x^{\lambda}}}\ .
  \label{3.6}
\ee
This is not, of course, the tensor transformation law; the second term
on the right spoils it.  That's okay, because {\it the connection
coefficients are not the components of a tensor}.  They are purposefully
constructed to be non-tensorial, but in such a way that the combination
(3.1) transforms as a tensor --- the extra terms in the transformation
of the partials and the $\Gamma$'s exactly cancel.  This is why we
are not so careful about index placement on the connection coefficients;
they are not a tensor, and therefore you should try not to raise and
lower their indices.

What about the covariant derivatives of other sorts of tensors?
By similar reasoning to that used for vectors, the covariant 
derivative of a one-form can also be expressed as a partial
derivative plus some linear transformation.  But there is no reason
as yet that the matrices representing this transformation should be
related to the coefficients $\Gamma^\nu_{\mu\lambda}$.  In general
we could write something like
\be
  \nabla_\mu \omega_\nu = \partial_\mu \omega_\nu + 
  \widetilde{\Gamma}^\lambda_{\mu\nu}
  \omega_\lambda\ ,\label{3.7}
\ee
where $\widetilde{\Gamma}^\lambda_{\mu\nu}$ is a new set of matrices
for each $\mu$.  (Pay attention to where all of the various indices go.)
It is straightforward to derive that the transformation properties
of $\widetilde{\Gamma}$ must be the same as those of $\Gamma$, but
otherwise no relationship has been established.  To do so, we need to
introduce two new properties that we would like our covariant derivative
to have (in addition to the two above):
\begin{enumerate}
\setcounter{enumi}{2}
\item commutes with contractions: ~$\nabla_\mu(T^\lambda{}_{\lambda\rho})
=(\nabla T)_\mu{}^\lambda{}_{\lambda\rho}$ ,
\item reduces to the partial derivative on scalars: ~$\nabla_\mu\phi
=\p\mu\phi$ .
\end{enumerate}

\noindent There is no way to ``derive'' these properties; we are simply
demanding that they be true as part of the definition of a covariant
derivative.

Let's see what these new properties imply.  Given some one-form field
$\omega_\mu$ and vector field $V^\mu$, we can take the covariant
derivative of the scalar defined by $\omega_\lambda V^\lambda$ to
get
\bea
  \nabla_\mu(\omega_\lambda V^\lambda) &=& 
  (\nabla_\mu \omega_\lambda)V^\lambda + \omega_\lambda
  (\nabla_\mu V^\lambda)\cr
  &=& (\p\mu\omega_\lambda)V^\lambda + 
  \widetilde{\Gamma}^\sigma_{\mu\lambda}\omega_\sigma V^\lambda
  +\omega_\lambda(\p\mu V^\lambda) + 
  \omega_\lambda\Gamma^\lambda_{\mu\rho}V^\rho\ . \label{3.8}
\eea
But since $\omega_\lambda V^\lambda$ is a scalar, this must also
be given by the partial derivative:
\bea
  \nabla_\mu(\omega_\lambda V^\lambda) &=& \partial_\mu
  (\omega_\lambda V^\lambda) \cr &=& 
  (\partial_\mu \omega_\lambda)V^\lambda + \omega_\lambda
  (\partial_\mu V^\lambda)\ . \label{3.9}
\eea
This can only be true if the terms in (3.8) with connection
coefficients cancel each other; that is, rearranging dummy indices,
we must have
\be
  0 = \widetilde{\Gamma}^\sigma_{\mu\lambda}\omega_\sigma V^\lambda
  + {\Gamma}^\sigma_{\mu\lambda}\omega_\sigma V^\lambda\ .\label{3.10}
\ee
But both $\omega_\sigma$ and $V^\lambda$ are completely arbitrary,
so
\be
  \widetilde{\Gamma}^\sigma_{\mu\lambda} = - \Gamma^\sigma_{\mu\lambda}
  \ .\label{3.11}
\ee
The two extra conditions we have imposed therefore allow us to express
the covariant derivative of a one-form using the same connection
coefficients as were used for the vector, but now with a minus sign
(and indices matched up somewhat differently):
\be
  \nabla_\mu\omega_\nu = \partial_\mu\omega_\nu
  -\Gamma^\lambda_{\mu\nu}\omega_\lambda\ .\label{3.12}
\ee

It should come as no surprise that the connection coefficients
encode all of the information necessary to take the covariant
derivative of a tensor of arbitrary rank.  The formula is quite
straightforward; for each upper index you introduce a term with
a single $+\Gamma$, and for each lower index a term with a single
$-\Gamma$:
\bea
  \nabla_\sigma T^{\mu_1 \mu_2 \cdots \mu_k}{}_{\nu_1
  \nu_2 \cdots \nu_l} &=&  \partial_\sigma T^{\mu_1 \mu_2 \cdots 
  \mu_k}{}_{\nu_1 \nu_2 \cdots \nu_l} \cr
  &&  +\Gamma^{\mu_1}_{\sigma\lambda}\, T^{\lambda \mu_2 \cdots 
  \mu_k}{}_{\nu_1 \nu_2 \cdots \nu_l} 
  +\Gamma^{\mu_2}_{\sigma\lambda}\, T^{\mu_1 \lambda \cdots 
  \mu_k}{}_{\nu_1 \nu_2 \cdots \nu_l} +\cdots\cr
  && -\Gamma^\lambda_{\sigma\nu_1}T^{\mu_1 \mu_2 \cdots 
  \mu_k}{}_{\lambda \nu_2 \cdots \nu_l}
  -\Gamma^\lambda_{\sigma\nu_2}T^{\mu_1 \mu_2 \cdots \mu_k}{}_{\nu_1
  \lambda \cdots \nu_l} - \cdots \ . \label{3.13}
\eea
This is the general expression for the covariant derivative.
You can check it yourself; it comes from the set of axioms we have
established, and the usual requirements that tensors of various 
sorts be coordinate-independent entities.
Sometimes an alternative notation is used; just as commas are
used for partial derivatives, semicolons are used for covariant
ones:
\be
  \nabla_\sigma T^{\mu_1 \mu_2 \cdots \mu_k}{}_{\nu_1
  \nu_2 \cdots \nu_l} \equiv T^{\mu_1 \mu_2 \cdots \mu_k}{}_{\nu_1
  \nu_2 \cdots \nu_l ;\sigma}\ .\label{3.14}
\ee
Once again, I'm not a big fan of this notation.

To define a covariant derivative, then, we need to put a
``connection'' on our manifold, which is specified in some
coordinate system by a set of coefficients $\Gamma^\lambda_\mn$
($n^3=64$ independent components in $n=4$ dimensions) 
which transform according to (3.6).
(The name ``connection'' comes from the fact that it is used to
transport vectors from one tangent space to another, as we will
soon see.)  There are evidently a large number of connections
we could define on any manifold, and each of them implies a
distinct notion of covariant differentiation.  In general relativity 
this freedom is not a big concern, because it turns out that every
metric defines a unique connection, which is the one used in GR.
Let's see how that works.

The first thing to notice is that the difference of two connections
is a $(1,2)$ tensor.  If we have two sets of connection coefficients,
$\Gamma^\lambda_\mn$ and $\widehat\Gamma^\lambda_\mn$, their difference
$S_{\mn}{}^\lambda = \Gamma^\lambda_\mn-\widehat\Gamma^\lambda_\mn$ (notice
index placement) transforms as
\bea
  S_{\mu'\nu'}{}^{\lambda'} &=& \Gamma^{\lambda'}_{\mu'\nu'}
  -\widehat\Gamma^{\lambda'}_{\mu'\nu'}\cr
  &=&{{\partial x^\mu}\over{\partial x^{\mu'}}}
  {{\partial x^\nu}\over{\partial x^{\nu'}}}
  {{\partial x^{\lambda'}}\over{\partial x^{\lambda}}} 
  \Gamma^\lambda_{\mu\nu} - {{\partial x^\mu}\over{\partial x^{\mu'}}} 
  {{\partial x^\nu}\over{\partial x^{\nu'}}}
  {{\partial^2 x^{\lambda'}}\over{\partial x^{\mu}\partial x^{\nu}}}
  - {{\partial x^\mu}\over{\partial x^{\mu'}}}
  {{\partial x^\nu}\over{\partial x^{\nu'}}}
  {{\partial x^{\lambda'}}\over{\partial x^{\lambda}}} 
  \widehat\Gamma^\lambda_{\mu\nu}
  + {{\partial x^\mu}\over{\partial x^{\mu'}}} 
  {{\partial x^\nu}\over{\partial x^{\nu'}}}
  {{\partial^2 x^{\lambda'}}\over{\partial x^{\mu}\partial x^{\nu}}}\cr
  &=& {{\partial x^\mu}\over{\partial x^{\mu'}}}
  {{\partial x^\nu}\over{\partial x^{\nu'}}}
  {{\partial x^{\lambda'}}\over{\partial x^{\lambda}}} 
  (\Gamma^\lambda_{\mu\nu}-\widehat\Gamma^\lambda_{\mu\nu})\cr
  &=& {{\partial x^\mu}\over{\partial x^{\mu'}}}
  {{\partial x^\nu}\over{\partial x^{\nu'}}}
  {{\partial x^{\lambda'}}\over{\partial x^{\lambda}}}
  S_{\mu\nu}{}^{\lambda}\ . \label{3.15}
\eea
This is just the tensor transormation law, so $S_{\mn}{}^\lambda$ is 
indeed a tensor.  This implies that any set of connections can be
expressed as some fiducial connection plus a tensorial correction.

Next notice that, given a connection specified by $\Gamma^\lambda_\mn$,
we can immediately form another connection simply by
permuting the lower indices.  That is, the set of coefficients
$\Gamma^\lambda_{\nu\mu}$ will also transform according to (3.6)
(since the partial derivatives appearing in the last term can be
commuted), so they determine a distinct connection.  There is thus
a tensor we can associate with any given connection, known as the
{\bf torsion tensor}, defined by
\be
  T_{\mu\nu}{}^{\lambda} = \Gamma^\lambda_\mn - \Gamma^\lambda_{\nu\mu}
  = 2\Gamma^\lambda_{[\mu\nu]}\ .\label{3.16}
\ee
It is clear that the torsion is antisymmetric its lower indices, and
a connection which is symmetric in its lower indices is known as
``torsion-free.''

We can now define a unique connection on a manifold with a metric
$g_\mn$ by introducing two additional properties:
\begin{itemize}
\item torsion-free: $\Gamma^\lambda_{\mu\nu}=
\Gamma^\lambda_{(\mu\nu)}$.
\item metric compatibility: $\nabla_\rho g_\mn=0$.
\end{itemize}

\noindent A connection is metric compatible if the covariant derivative
of the metric with respect to that connection is everywhere zero. 
This implies a couple of nice properties.  First, it's easy to show
that the inverse metric also has zero covariant derivative,
\be
  \nabla_\rho g^\mn = 0\ .\label{3.17}
\ee
Second, a metric-compatible covariant derivative commutes with
raising and lowering of indices.  Thus, for some vector field
$V^\lambda$,
\be
  g_{\mu\lambda}\nabla_\rho V^\lambda = \nabla_\rho
  (g_{\mu\lambda} V^\lambda) = \nabla_\rho V_\mu\ .\label{3.18}
\ee
With non-metric-compatible connections one must be very careful about
index placement when taking a covariant derivative.

Our claim is therefore that there is exactly one torsion-free
connection on a given manifold which is compatible with some given
metric on that manifold.  We do not want to make these two requirements
part of the definition of a covariant derivative; they simply 
single out one of the many possible ones.

We can demonstrate both existence and uniqueness by deriving a
manifestly unique expression for the connection coefficients in terms
of the metric.
To accomplish this, we expand out the equation of metric 
compatibility for three different permutations of the indices:
\bea
  \nabla_\rho g_\mn &=& \p\rho g_\mn - \Gamma^\lambda_{\rho\mu}
  g_{\lambda\nu} - \Gamma^\lambda_{\rho\nu}g_{\mu\lambda} = 0\cr
  \nabla_\mu g_{\nu\rho} &=& \p\mu g_{\nu\rho} -\Gamma^\lambda_{\mu\nu}
  g_{\lambda\rho} - \Gamma^\lambda_{\mu\rho}g_{\nu\lambda} = 0\cr
  \nabla_\nu g_{\rho\mu} &=& \p\nu g_{\rho\mu} -\Gamma^\lambda_{\nu\rho}
  g_{\lambda\mu} - \Gamma^\lambda_{\nu\mu} g_{\rho\lambda} = 0\ .
  \label{3.19}
\eea
We subtract the second and third of these from the first, and use
the symmetry of the connection to obtain
\be
  \p\rho g_\mn - \p\mu g_{\nu\rho} - \p\nu g_{\rho\mu}
  +2\Gamma^\lambda_\mn g_{\lambda\rho} = 0\ .\label{3.20}
\ee
It is straightforward to solve this for the connection by multiplying
by $g^{\sigma\rho}$.  The result is
\be
  \Gamma^\sigma_\mn = {1\over 2} g^{\sigma\rho}(\p\mu g_{\nu\rho} + 
  \p\nu g_{\rho\mu} - \p\rho g_\mn)\ .\label{3.21}
\ee
This is one of the most important formulas in this subject; commit
it to memory.  Of course, we have only proved that if a 
metric-compatible and torsion-free connection exists, it must be of
the form (3.21); you can check for yourself (for those of you without
enough tedious computation in your lives) that the right hand side
of (3.21) transforms like a connection.  

This connection we have derived from the metric is the one on which
conventional general relativity is based (although we will keep an open
mind for a while longer).  It is known by different names:
sometimes the {\bf Christoffel} connection, sometimes the 
{\bf Levi-Civita} connection, sometimes the {\bf Riemannian} connection.
The associated connection coefficients are sometimes called 
{\bf Christoffel symbols} and written as $\left\{{}^{\,\,\sigma}_{\mu\nu}
\right\}$; we will sometimes call them Christoffel symbols, but we
won't use the funny notation.  The study of manifolds with metrics 
and their associated connections is called ``Riemannian geometry.''
As far as I can tell the study of more general connections can be 
traced back to Cartan, but I've never heard it called ``Cartanian
geometry.''

Before putting our covariant derivatives to work, we should mention
some miscellaneous properties.  First, let's emphasize again that the
connection does not {\it have} to be constructed from the metric.  In
ordinary flat space there is an implicit connection we use all the
time --- the Christoffel connection constructed from the flat metric.
But we could, if we chose, use a different connection, while keeping
the metric flat.  Also notice that the coefficients of the Christoffel
connection in flat space will vanish in Cartesian coordinates, but not
in curvilinear coordinate systems.  Consider for example the plane in
polar coordinates, with metric
\be
  ds^2 = \d r^2 + r^2\d\theta^2\ .\label{3.22}
\ee
The nonzero components of the inverse metric are readily found to be
$g^{rr}=1$ and $g^{\theta\theta}=r^{-2}$.  (Notice that we use
$r$ and $\theta$ as indices in an obvious notation.)  We can compute
a typical connection coefficient:
\bea
  \Gamma^r_{rr} &=& {1\over 2} g^{r\rho}(\p{r} g_{r\rho} + 
  \p{r} g_{\rho r} - \p\rho g_{rr})\cr
  &=& {1\over 2} g^{rr}(\p{r} g_{rr} + 
  \p{r} g_{rr} - \p{r} g_{rr})\cr
  && + {1\over 2} g^{r\theta}(\p{r} g_{r\theta} + 
  \p{r} g_{\theta r} - \p\theta g_{rr})\cr
  &=& {1\over 2}(1)(0+0-0) + {1\over 2}(0)(0+0-0)\cr
  &=&0\ . \label{3.23}
\eea
Sadly, it vanishes.  But not all of them do:
\bea
  \Gamma^r_{\theta\theta} &=& {1\over 2} g^{r\rho}
  (\p{\theta} g_{\theta\rho} +  \p{\theta} g_{\rho \theta} 
  - \p\rho g_{\theta\theta})\cr
  &=& {1\over 2}g^{rr}
  (\p{\theta} g_{\theta r} +  \p{\theta} g_{r \theta} 
  - \p{r} g_{\theta\theta})\cr
  &=& {1\over 2}(1)(0+0-2r)\cr
  &=& -r\ . \label{3.24}
\eea
Continuing to turn the crank, we eventually find
\bea
  \Gamma^r_{\theta r} &=& \Gamma^r_{r\theta} = 0\cr
  \Gamma^\theta_{rr} &=& 0\cr
  \Gamma^\theta_{r\theta} &=& \Gamma^\theta_{\theta r} = {1\over r}\cr
  \Gamma^\theta_{\theta\theta} &=& 0\ . \label{3.25}
\eea
The existence of nonvanishing connection coefficients in curvilinear
coordinate systems is the ultimate cause of the formulas for the
divergence and so on that you find in books on electricity and
magnetism.

Contrariwise, even in a curved space it is still possible to make
the Christoffel symbols vanish at any one point.  This is just
because, as we saw in the last section, we can always make the first 
derivative of the metric vanish at a point; so by (3.21) the connection
coefficients derived from this metric will also vanish.  Of course
this can only be established at a point, not in some neighborhood
of the point.

Another useful property is that the formula for the divergence
of a vector (with respect to the Christoffel connection) has a
simplified form.  The covariant divergence of $V^\mu$ is given by
\be
  \nabla_\mu V^\mu = \p\mu V^\mu +\Gamma^\mu_{\mu\lambda}V^\lambda
  \ .\label{3.26}
\ee
It's easy to show (see pp.~106-108 of Weinberg) that the Christoffel
connection satisfies
\be
  \Gamma^\mu_{\mu\lambda}= {1\over{\sqrt{|g|}}}\p\lambda
  \sqrt{|g|}\ ,\label{3.27}
\ee
and we therefore obtain
\be
  \nabla_\mu V^\mu = {1\over{\sqrt{|g|}}}\p\mu(\sqrt{|g|}V^\mu)
  \ .\label{3.28}
\ee
There are also formulas for the divergences of higher-rank tensors,
but they are generally not such a great simplification.

As the last factoid we should mention about connections, let us 
emphasize (once more) that the exterior derivative is a well-defined
tensor in the absence of any connection.  The reason this needs to
be emphasized is that, if you happen to be using a symmetric 
(torsion-free) connection, the exterior derivative (defined to be the
antisymmetrized partial derivative) happens to be equal to the
antisymmetrized covariant derivative:
\bea
  \nabla_{[\mu}\omega_{\nu]} &=& \p{[\mu}\omega_{\nu]}
  -\Gamma^\lambda_{[\mn]}\omega_\lambda \cr
  &=& \p{[\mu}\omega_{\nu]}\ . \label{3.29}
\eea
This has led some misfortunate souls to fret about the ``ambiguity''
of the exterior derivative in spaces with torsion, where the above
simplification does not occur.  There is no ambiguity: the exterior
derivative does not involve the connection, no matter what connection
you happen to be using, and therefore the torsion never enters the
formula for the exterior derivative of anything.

Before moving on, let's review the process by which we have been
adding structures to our mathematical constructs.  We started with
the basic notion of a set, which you were presumed to know (informally,
if not rigorously).  We introduced the concept of open subsets of
our set; this is equivalent
to introducing a topology, and promoted the set to a topological
space.  Then by demanding that each open set look like a region of
$\R^n$ (with $n$ the same for each set) and that the coordinate
charts be smoothly sewn together, the topological space became a
manifold.  A manifold is simultaneously a very flexible and powerful
structure, and comes equipped naturally with a tangent bundle,
tensor bundles of various ranks, the
ability to take exterior derivatives, and so forth.  We then proceeded
to put a metric on the manifold, resulting in a manifold with metric
(or sometimes ``Riemannian manifold'').
Independently of the metric we found we could introduce a connection,
allowing us to take covariant derivatives.  Once we have a metric,
however,
there is automatically a unique torsion-free metric-compatible
connection.  (In principle there is nothing to stop us from introducing
more than one connection, or more than one metric, on any given
manifold.)  The situation is thus as portrayed in the diagram on
the next page.

\eject

\begin{figure}
  \centerline{
  \psfig{figure=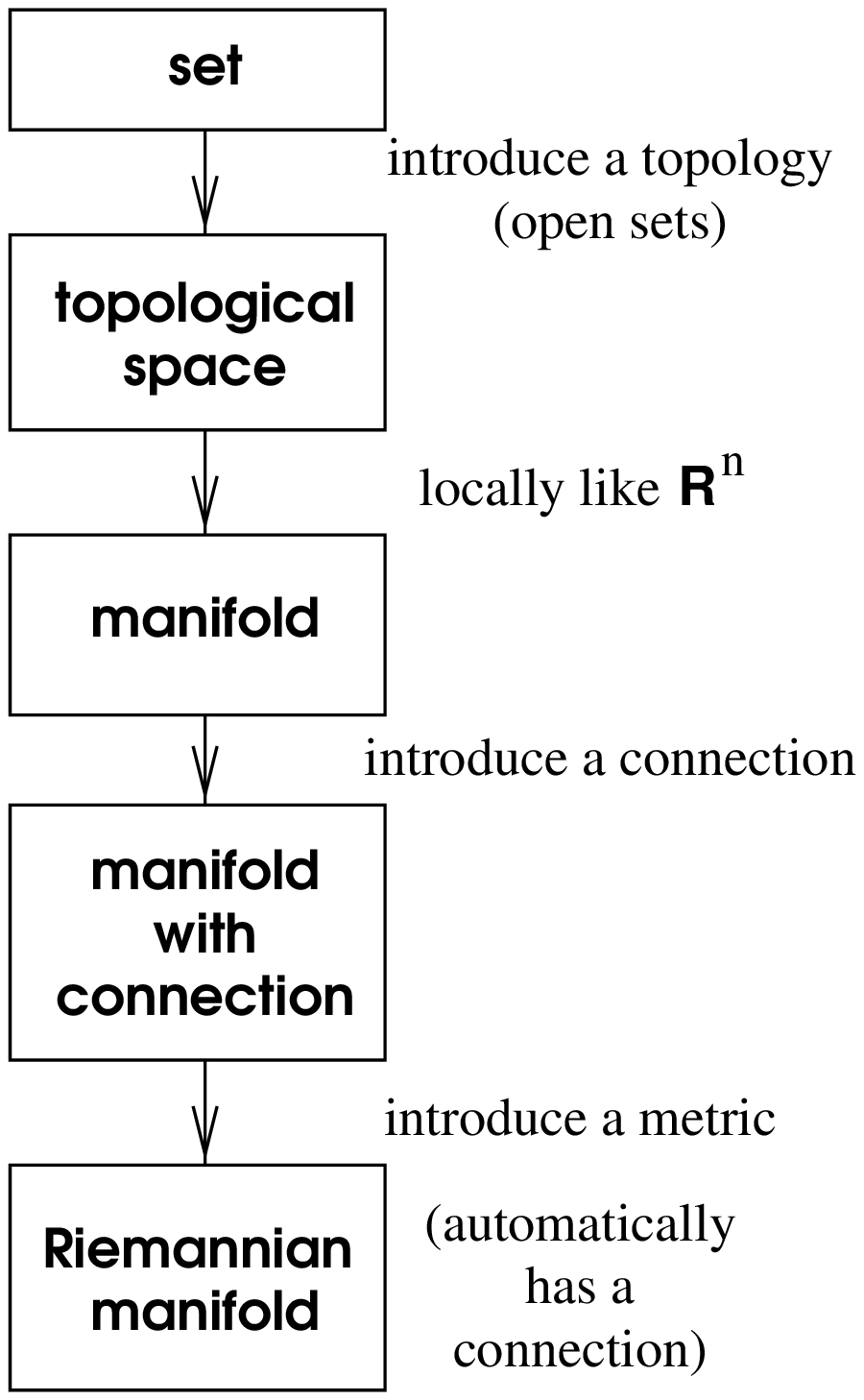,angle=0,height=9cm}}
\end{figure}

Having set up the machinery of connections, the first thing we will
do is discuss parallel transport.  Recall that in flat space it was
unnecessary to be very careful about the fact that vectors were
elements of tangent spaces defined at individual points; it is
actually very natural to compare vectors at different points (where
by ``compare'' we mean add, subtract, take the dot product, etc.).
The reason why it is natural is because it makes sense, in flat space,
to ``move a vector from one point to another while keeping it constant.''
Then once we get the vector from one point to another we can do the
usual operations allowed in a vector space.

\begin{figure}[h]
  \centerline{
  \psfig{figure=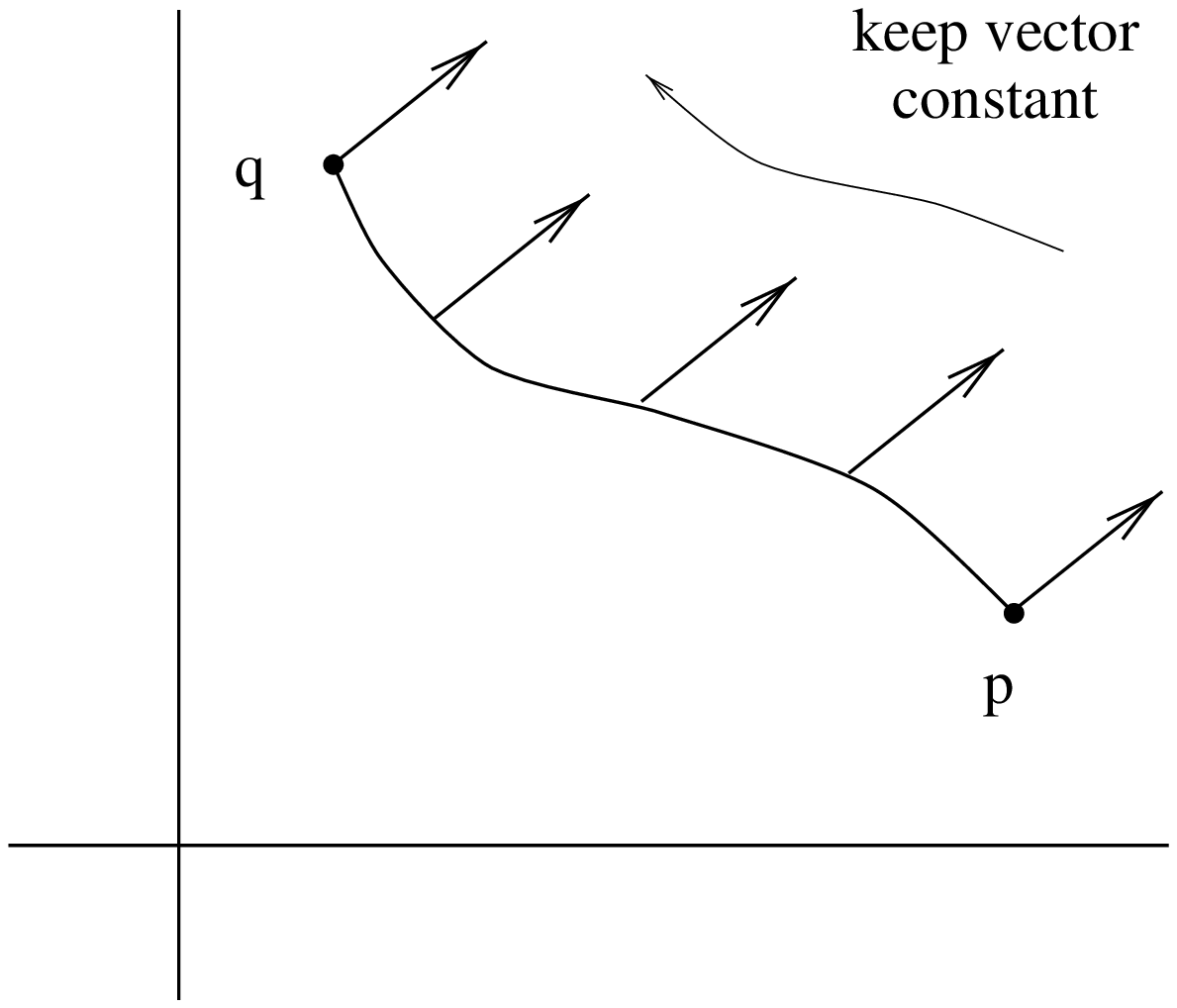,angle=0,height=5.5cm}}
\end{figure}

The concept of moving a vector along a path, keeping constant all
the while, is known as parallel transport.  As we shall see, parallel
transport is defined whenever we have a connection; the intuitive
manipulation of vectors in flat space makes implicit use of the
Christoffel connection on this space.  The crucial difference between
flat and curved spaces is that, in a curved space, {\it the result
of parallel transporting a vector from one point to another will
depend on the path taken between the points}.  Without yet assembling
the complete mechanism of parallel transport, we can use our 
intuition about the two-sphere to see that this is the case. Start
with a vector on the equator, pointing along a line of constant
longitude.  Parallel transport it up to the north pole along a line
of longitude in the
obvious way.  Then take the original vector, parallel transport it
along the equator by an angle $\theta$, and then move it up to the
north pole as before.
It is clear that the vector, parallel transported along two paths,
arrived at the same destination with two different values (rotated
by $\theta$).

\begin{figure}[h]
  \centerline{
  \psfig{figure=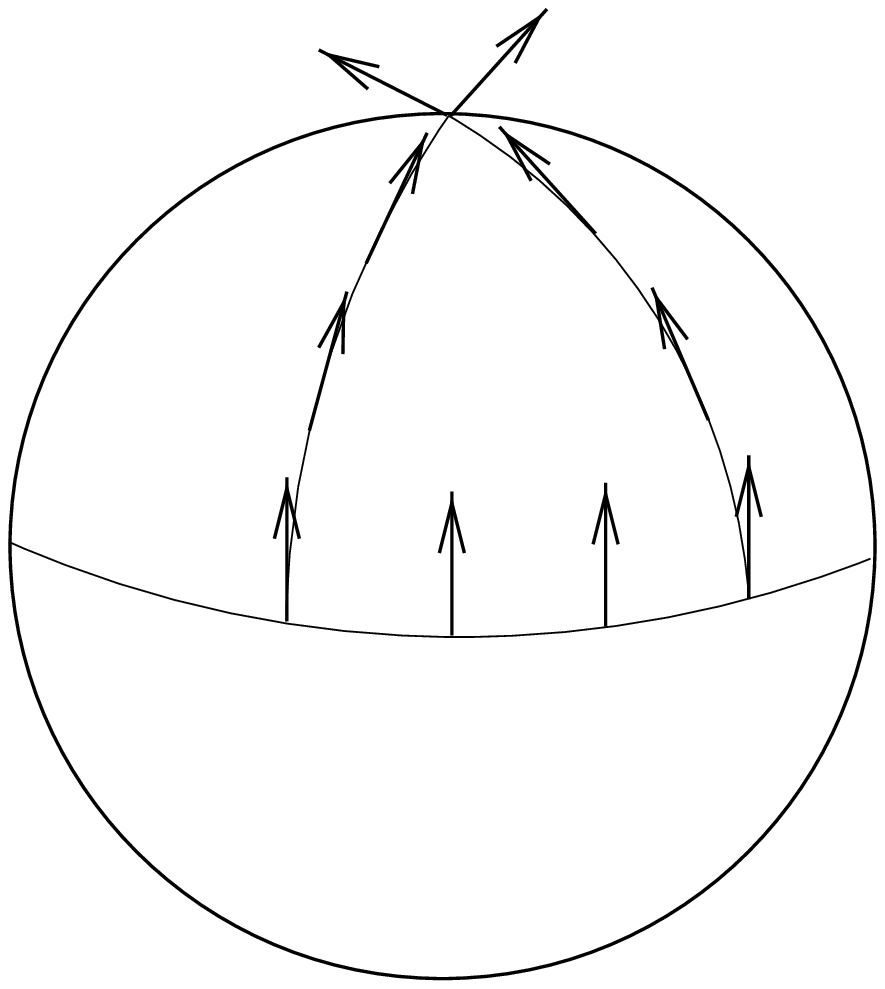,angle=0,height=5cm}}
\end{figure}

It therefore appears as if there is no natural way to uniquely move
a vector from one tangent space to another; we can always parallel
transport it, but the result depends on the path, and there is no
natural choice of which path to take.  Unlike some of the problems we
have encountered, {\it there is no solution to this one} --- we
simply must learn to live with the fact that two vectors can only
be compared in a natural way if they are elements of the same tangent
space.  For example, two particles passing by each other have a
well-defined relative velocity (which cannot be greater than the 
speed of light).  But two particles at different points on a curved
manifold do not have any well-defined notion of relative velocity ---
the concept simply makes no sense.  Of course, in certain special
situations it is still useful to talk as if it did make sense, but it
is necessary to understand that occasional usefulness is not a 
substitute for rigorous definition.  In cosmology, for example, the
light from distant galaxies is redshifted with respect to the frequencies
we would observe from a nearby stationary source.  Since this 
phenomenon bears such a close resemblance to the conventional Doppler
effect due to relative motion, it is very tempting to say that the
galaxies are ``receding away from us'' at a speed defined by their
redshift.  At a rigorous level this is nonsense, what Wittgenstein would
call a ``grammatical mistake'' --- the galaxies are not receding, since
the notion of their velocity with respect to us is not well-defined.
What is actually happening is that the metric of spacetime between
us and the galaxies has changed (the universe has expanded) along the
path of the photon from here to there, leading to an increase in the
wavelength of the light.  As an example of how you can go wrong,
naive application of the Doppler formula to the redshift of galaxies
implies that some of them are receding faster than light, in apparent
contradiction with relativity.  The resolution of this apparent paradox
is simply that the very notion of their recession should not be 
taken literally.

Enough about what we cannot do; let's see what we can.  Parallel
transport is supposed to be the curved-space generalization of the
concept of ``keeping the vector constant'' as we move it along
a path; similarly for a tensor of arbitrary rank.  Given a 
curve $x^\mu(\lambda)$, the requirement of constancy of a tensor $T$
along this curve in flat space is simply ${{dT}\over{d\lambda}} = 
{{dx^\mu}\over{d\lambda}}{{\partial T}\over{\partial x^\mu}}=0$.
We therefore define the covariant derivative along the path to be
given by an operator
\be
  {{D}\over {d\lambda}} = {{dx^\mu}\over{d\lambda}}\nabla_\mu
  \ .\label{3.30}
\ee
We then define {\bf parallel transport} of the tensor $T$ along
the path $x^\mu(\lambda)$ to be the requirement that, along the
path,
\be
  \left({{D}\over{d\lambda}}T\right)^{\mu_1 \mu_2 \cdots 
  \mu_k}{}_{\nu_1 \nu_2 \cdots \nu_l} \equiv {{dx^\sigma}\over
  {d\lambda}}\nabla_\sigma T^{\mu_1 \mu_2 \cdots 
  \mu_k}{}_{\nu_1 \nu_2 \cdots \nu_l} = 0\ .\label{3.31}
\ee
This is a well-defined tensor equation, since both the tangent vector
$dx^\mu/d\lambda$ and the covariant derivative $\nabla T$ are tensors.
This is known as the {\bf equation of parallel transport}.  For
a vector it takes the form
\be
  {{d}\over{d\lambda}} V^\mu
  + \Gamma^\mu_{\sigma\rho}{{dx^\sigma}\over{d\lambda}}V^\rho = 0\ .
  \label{3.32}
\ee
We can look at the parallel transport equation as a first-order
differential equation defining an initial-value problem: given a tensor
at some point along the path, there will be a unique continuation of
the tensor to other points along the path such that the continuation
solves (3.31).  We say that such a tensor is parallel transported.

The notion of parallel transport is obviously dependent on the 
connection, and different connections lead to different answers.
If the connection is metric-compatible, the metric is always
parallel transported with respect to it:
\be
  {{D}\over{d\lambda}}g_\mn = {{dx^\sigma}\over{d\lambda}}
  \nabla_\sigma g_\mn =0\ .\label{3.33}
\ee
It follows that the inner product of two parallel-transported
vectors is preserved.  That is, if $V^\mu$ and $W^\nu$ are
parallel-transported along a curve $x^\sigma(\lambda)$, we have
\bea
  {{D}\over{d\lambda}}(g_\mn V^\mu W^\nu) &=&
  \left({{D}\over{d\lambda}}g_\mn\right)V^\mu W^\nu +
  g_\mn \left({{D}\over{d\lambda}} V^\mu\right)W^\nu +
  g_\mn V^\mu\left({{D}\over{d\lambda}} W^\nu\right)\cr
  &=& 0\ . \label{3.34}
\eea
This means that parallel transport with respect to a metric-compatible
connection preserves the norm of vectors, the sense of orthogonality,
and so on.

One thing they don't usually tell you in GR books is that you can
write down an explicit and general solution to the parallel transport
equation, although it's somewhat formal.  First notice that for some
path $\gamma :\lambda \rightarrow x^\sigma(\lambda)$, solving the parallel
transport equation for a vector $V^\mu$ amounts to finding a matrix
$P^\mu{}_\rho(\lambda,\lambda_0)$ which relates the vector at its
initial value $V^\mu(\lambda_0)$ to its value somewhere later down the
path:
\be
  V^\mu(\lambda) = P^\mu{}_\rho(\lambda,\lambda_0)V^\rho(\lambda_0)
  \ .\label{3.35}
\ee
Of course the matrix $P^\mu{}_\rho(\lambda,\lambda_0)$, known as the
{\bf parallel propagator}, depends on the
path $\gamma$ (although it's hard to find a notation which indicates
this without making $\gamma$ look like an index).  If we define
\be
  A^\mu{}_\rho(\lambda) = -\Gamma^\mu_{\sigma\rho} 
  {{dx^\sigma}\over{d \lambda}}\ ,\label{3.36}
\ee
where the quantities on the right hand side are evaluated at 
$x^\nu(\lambda)$, then the parallel transport equation becomes
\be
  {{d}\over{d\lambda}}V^\mu = A^\mu{}_\rho V^\rho\ .\label{3.37}
\ee
Since the parallel propagator must work for any vector, substituting
(3.35) into (3.37) shows that $P^\mu{}_\rho(\lambda,\lambda_0)$ also obeys
this equation:
\be
  {{d}\over{d\lambda}}P^\mu{}_\rho(\lambda,\lambda_0) = 
  A^\mu{}_\sigma(\lambda) P^\sigma{}_\rho(\lambda,\lambda_0)
  \ .\label{3.38}
\ee
To solve this equation, first integrate both sides:
\be
  P^\mu{}_\rho(\lambda,\lambda_0)=\delta^\mu_\rho
  +\int^\lambda_{\lambda_0} A^\mu{}_\sigma(\eta) 
  P^\sigma{}_\rho(\eta,\lambda_0)\, d\eta\ .\label{3.39}
\ee
The Kronecker delta, it is easy to see, provides the correct
normalization for $\lambda=\lambda_0$.

We can solve (3.39) by iteration, taking the right hand side and
plugging it into itself repeatedly, giving
\be
  P^\mu{}_\rho(\lambda,\lambda_0)=\delta^\mu_\rho
  +\int^\lambda_{\lambda_0} A^\mu{}_\rho(\eta) \, d\eta
  +\int^\lambda_{\lambda_0} \int^\eta_{\lambda_0}
  A^\mu{}_\sigma(\eta) A^\sigma{}_\rho(\eta')\, d\eta' d\eta
  +\cdots\ .\label{3.40}
\ee
The $n$th term in this series is an integral over an $n$-dimensional
right triangle, or $n$-simplex.

\eject

\[
  \int^\lambda_{\lambda_0} A(\eta_1) \, d\eta_1 \qquad
  \int^\lambda_{\lambda_0} \int^{\eta_2}_{\lambda_0}
  A(\eta_2) A(\eta_1)\, d\eta_1 d\eta_2 \qquad
  \int^\lambda_{\lambda_0} \int^{\eta_3}_{\lambda_0}\int^{\eta_2}_{\lambda_0}
  A(\eta_3) A(\eta_2) A(\eta_1)\, d^3\eta 
\]

\begin{figure}[h]
  \centerline{
  \psfig{figure=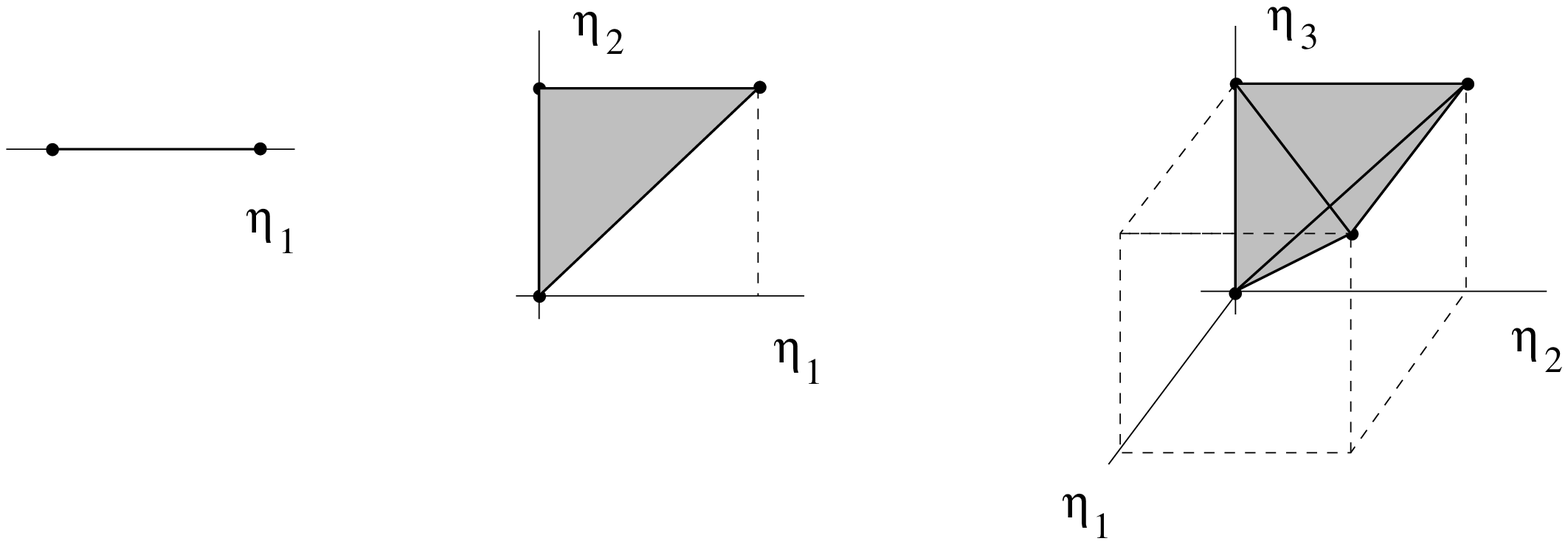,angle=0,height=5cm}}
\end{figure}

It would simplify things if we could consider such an integral to
be over an $n$-cube instead of an $n$-simplex; is there some way
to do this?  There are $n!$ such
simplices in each cube, so we would have to multiply by $1/n!$ to
compensate for this extra volume.  But we also want to get the
integrand right; using matrix notation, the integrand at $n$th order
is $A(\eta_n)A(\eta_{n-1})\cdots A(\eta_1)$, but with the special
property that $\eta_n\geq \eta_{n-1}\geq \cdots \geq \eta_1$.
We therefore define the {\bf path-ordering symbol}, ${\cal P}$,
to ensure that this condition holds.  In other words, the expression
\be
  {\cal P}[A(\eta_n)A(\eta_{n-1})\cdots A(\eta_1)]\label{3.41}
\ee
stands for the product of the $n$ matrices $A(\eta_i)$, ordered in
such a way that the largest value of $\eta_i$ is on the left, and
each subsequent value of $\eta_i$ is less than or equal to the 
previous one.  We then can express the $n$th-order term in (3.40) as
\bea
  \lefteqn{\int^\lambda_{\lambda_0}\int^{\eta_n}_{\lambda_0}\cdots
  \int^{\eta_2}_{\lambda_0} A(\eta_n) A(\eta_{n-1})\cdots
  A(\eta_1)\, d^n\eta} \cr
  &=& {1\over{n!}}\int^\lambda_{\lambda_0}
  \int^\lambda_{\lambda_0}\cdots\int^\lambda_{\lambda_0}
  {\cal P}[A(\eta_n) A(\eta_{n-1})\cdots A(\eta_1)]\, d^n\eta\ .
  \label{3.42}
\eea
This expression contains no substantive statement about the matrices
$A(\eta_i)$; it is just notation.  But we can now write
(3.40) in matrix form as
\be
  P(\lambda,\lambda_0) = {\bf 1} + \sum^\infty_{n=1}{1\over {n!}}
  \int^\lambda_{\lambda_0} {\cal P}[A(\eta_n) A(\eta_{n-1})\cdots 
  A(\eta_1)]\, d^n\eta\ .\label{3.43}
\ee
This formula is just the series expression for an exponential; we
therefore say that the parallel propagator is given by the path-ordered
exponential
\be
  P(\lambda,\lambda_0) = {\cal P}\exp\left(\int^\lambda_{\lambda_0}
  A(\eta)\, d\eta\right)\ ,\label{3.44}
\ee
where once again this is just notation; the path-ordered exponential
is defined to be the right hand side of (3.43).  We can write it more
explicitly as
\be
  P^\mu{}_\nu(\lambda,\lambda_0) ={\cal P}\exp\left(-
  \int^\lambda_{\lambda_0}\Gamma^\mu_{\sigma\nu}{{dx^\sigma}\over
  {d\eta}}\, d\eta\right)\ .\label{3.45}
\ee
It's nice to have an explicit formula, even if it is rather abstract.
The same kind of expression appears in quantum field theory as
``Dyson's Formula,'' where it arises because the Schr\"odinger
equation for the time-evolution operator has the same form as (3.38).

As an aside, an especially interesting example of the parallel
propagator occurs when the path is a loop, starting and ending at the
same point.  Then if the connection is metric-compatible, the
resulting matrix will just be a Lorentz transformation on the tangent
space at the point.  This transformation is known as the ``holonomy''
of the loop.  If you know the holonomy of every possible loop, that 
turns out to be equivalent to knowing the metric.  This fact has let
Ashtekar and his collaborators to examine general relativity in
the ``loop representation,'' where the fundamental variables are 
holonomies rather than the explicit metric.  They have made some
progress towards quantizing the theory in this approach, although the
jury is still out about how much further progress can be made.

With parallel transport understood, the next logical step is to
discuss geodesics.  A geodesic is the curved-space generalization
of the notion of a ``straight line'' in Euclidean space.  We all
know what a straight line is: it's the path of shortest distance
between two points.  But there is an equally good definition ---
a straight line is a path which parallel transports its own
tangent vector.  On a manifold with an arbitrary (not necessarily
Christoffel) connection, these two concepts do not quite coincide,
and we should discuss them separately.

We'll take the second definition first, since it is computationally
much more straightforward.  The tangent vector to a path $x^\mu(\lambda)$
is $dx^\mu/d\lambda$.  The condition that it be parallel transported 
is thus
\be
  {{D}\over{d\lambda}}{{dx^\mu}\over{d\lambda}}=0\ ,\label{3.46}
\ee
or alternatively
\be
  {{d^2x^\mu}\over{d\lambda^2}}+\Gamma^\mu_{\rho\sigma}
  {{dx^\rho}\over{d\lambda}}{{dx^\sigma}\over{d\lambda}}=0\ .
  \label{3.47}
\ee
This is the {\bf geodesic equation}, another one which you should
memorize.  We can easily see that it reproduces the usual notion
of straight lines if the connection coefficients are the Christoffel
symbols in Euclidean space; in that case we can choose Cartesian
coordinates in which $\Gamma^\mu_{\rho\sigma}=0$, and the geodesic
equation is just $d^2x^\mu/d\lambda^2=0$, which is the equation for
a straight line.

That was embarrassingly simple; let's turn to the more nontrivial case
of the shortest distance definition.  As we know, there are various
subtleties involved in the definition of distance in a Lorentzian
spacetime; for null paths the distance is zero, for timelike paths
it's more convenient to use the proper time, etc.  So in the name of
simplicity let's do the calculation just for a timelike path ---
the resulting equation will turn
out to be good for any path, so we are not losing
any generality.  We therefore consider the proper time functional,
\be
  \tau = \int \left(-g_\mn {{dx^\mu}\over{d\lambda}}
  {{dx^\nu}\over{d\lambda}}\right)^{1/2}\, d\lambda\ ,\label{3.48}
\ee
where the integral is over the path.  To search for shortest-distance
paths, we will do the usual calculus of variations treatment to seek
extrema of this functional.  (In fact they will turn out to be curves
of {\it maximum} proper time.)

We want to consider the change in the proper time under
infinitesimal variations of the path,
\bea
  x^\mu &\rightarrow & x^\mu+\delta x^\mu\cr
  g_\mn &\rightarrow & g_\mn + \delta x^\sigma\partial_\sigma g_\mn
  \ . \label{3.49}
\eea
(The second line comes from Taylor expansion in curved spacetime, which
as you can see uses the partial derivative,
not the covariant derivative.)  Plugging this into (3.48), we get
\bea
  \tau + \delta\tau &=& \int\left(-g_\mn {{dx^\mu}\over{d\lambda}}
  {{dx^\nu}\over{d\lambda}} - \p\sigma g_\mn {{dx^\mu}\over{d\lambda}}
  {{dx^\nu}\over{d\lambda}}\delta x^\sigma 
  -2 g_\mn {{dx^\mu}\over{d\lambda}}{{d(\delta x^\nu)}\over{d\lambda}}
  \right)^{1/2}\, d\lambda\cr
  &=& \int\left(-g_\mn {{dx^\mu}\over{d\lambda}}
  {{dx^\nu}\over{d\lambda}}\right)^{1/2}
  \left[1+\left(-g_\mn {{dx^\mu}\over{d\lambda}}
  {{dx^\nu}\over{d\lambda}}\right)^{-1}\right.\cr
  && \qquad\qquad\left.
  \times\left(-\p\sigma g_\mn {{dx^\mu}\over{d\lambda}}
  {{dx^\nu}\over{d\lambda}}\delta x^\sigma 
  -2 g_\mn {{dx^\mu}\over{d\lambda}}{{d(\delta x^\nu)}\over{d\lambda}}
  \right)\right]^{1/2}\, d\lambda\ . \label{3.50}
\eea
Since $\delta x^\sigma$ is assumed to be small, we can expand the
square root of the expression in square brackets to find
\be
  \delta\tau = \int\left(-g_\mn {{dx^\mu}\over{d\lambda}}
  {{dx^\nu}\over{d\lambda}}\right)^{-1/2}
  \left(-{1\over 2}\p\sigma g_\mn {{dx^\mu}\over{d\lambda}}
  {{dx^\nu}\over{d\lambda}}\delta x^\sigma 
  - g_\mn {{dx^\mu}\over{d\lambda}}{{d(\delta x^\nu)}\over{d\lambda}}
  \right)\, d\lambda\ .\label{3.51}
\ee
It is helpful at this point to change the parameterization of our
curve from $\lambda$, which was arbitrary, to the proper time $\tau$
itself, using
\be
  d\lambda = \left(-g_\mn {{dx^\mu}\over{d\lambda}}
  {{dx^\nu}\over{d\lambda}}\right)^{-1/2}\, d\tau\ .\label{3.52}
\ee
We plug this into (3.51) (note: we plug it in for every appearance
of $d\lambda$) to obtain
\bea
  \delta\tau &=&\int \left[-{1\over 2}\p\sigma g_\mn 
  {{dx^\mu}\over{d\tau}} {{dx^\nu}\over{d\tau}}\delta x^\sigma 
  - g_\mn {{dx^\mu}\over{d\tau}}{{d(\delta x^\nu)}\over{d\tau}}
  \right]\, d\tau\cr
  &=& \int \left[-{1\over 2}\p\sigma g_\mn 
  {{dx^\mu}\over{d\tau}} {{dx^\nu}\over{d\tau}}
  +{{d}\over{d\tau}}\left(g_{\mu\sigma} {{dx^\mu}\over{d\tau}}\right)
  \right]\delta x^\sigma\, d\tau\ , \label{3.53}
\eea
where in the last line we have integrated by parts, avoiding possible
boundary contributions by demanding that the variation $\delta x^\sigma$
vanish at the endpoints of the path.  Since we are searching for
stationary points, we want $\delta \tau$ to vanish for any variation;
this implies
\be
  -{1\over 2}\p\sigma g_\mn {{dx^\mu}\over{d\tau}} {{dx^\nu}\over{d\tau}}
  + {{dx^\mu}\over{d\tau}} {{dx^\nu}\over{d\tau}} \p\nu g_{\mu\sigma}
  +g_{\mu\sigma}{{d^2x^\mu}\over{d\tau^2}} = 0\ ,\label{3.54}
\ee
where we have used $d g_{\mu\sigma}/d\tau=(dx^\nu/d\tau)\p\nu 
g_{\mu\sigma}$.  Some shuffling of dummy indices reveals 
\be
  g_{\mu\sigma}{{d^2x^\mu}\over{d\tau^2}} +{1\over 2}\left(
  -\p\sigma g_{\mn} + \p\nu g_{\mu\sigma} + \p\mu g_{\nu\sigma}
  \right){{dx^\mu}\over{d\tau}} {{dx^\nu}\over{d\tau}} =0\ ,\label{3.55}
\ee
and multiplying by the inverse metric finally leads to
\be
  {{d^2x^\rho}\over{d\tau^2}} +{1\over 2}g^{\rho\sigma}\left(
  \p\mu g_{\nu\sigma} + \p\nu g_{\sigma\mu}-\p\sigma g_{\mn} 
  \right){{dx^\mu}\over{d\tau}} {{dx^\nu}\over{d\tau}} =0\ .\label{3.56}
\ee
We see that this is precisely the geodesic equation (3.32), but
with the specific choice of Christoffel connection (3.21).  Thus,
on a manifold with metric, extremals of the length functional are
curves which parallel transport their tangent vector with respect
to the Christoffel connection associated with that metric.  It doesn't
matter if there is any other connection defined on the same manifold.
Of course, in GR the Christoffel connection is the only one which
is used, so the two notions are the same.

The primary usefulness of geodesics in general relativity is that
they are the paths followed by unaccelerated particles.  In fact,
the geodesic equation can be thought of as the generalization of
Newton's law ${\bf f}=m{\bf a}$ for the case ${\bf f}=0$.  It is
also possible to introduce forces by adding terms to the right hand
side; in fact, looking back to the expression (1.103) for the 
Lorentz force in special relativity, it is tempting to guess that
the equation of motion for a particle of mass $m$ and charge $q$
in general relativity should be
\be
  {{d^2x^\mu}\over{d\tau^2}}+\Gamma^\mu_{\rho\sigma}
  {{dx^\rho}\over{d\tau}}{{dx^\sigma}\over{d\tau}}=
  {q\over m}F^\mu{}_\nu{{dx^\nu}\over{d\tau}}\ .\label{3.57}
\ee
We will talk about this more later, but in fact your guess would
be correct.

Having boldly derived these expressions, we should say some more 
careful words about the parameterization of a geodesic path.
When we presented the geodesic equation as the requirement that
the tangent vector be parallel transported, (3.47), we parameterized
our path with some parameter $\lambda$, whereas when we found 
the formula (3.56) for the extremal of the spacetime interval we wound
up with a very specific parameterization, the proper time.  Of course
from the form of (3.56) it is clear that a transformation 
\be
  \tau \rightarrow \lambda = a\tau +b \ ,\label{3.58}
\ee
for some constants $a$ and $b$, leaves the equation invariant.  Any
parameter related to the proper time in this way is called an
{\bf affine parameter}, and is just as good as the proper time
for parameterizing a geodesic.  What was hidden in our derivation
of (3.47) was that {\it the demand that the tangent vector be parallel
transported actually constrains the parameterization of the curve},
specifically to one related to the proper time by (3.58).  In other
words, if you start at some point and with some initial direction,
and then construct a curve by beginning to walk in that direction
and keeping your tangent vector parallel transported, you will not
only define a path in the manifold but also (up to linear transformations)
define the parameter along the path.

Of course, there is nothing to stop you from using any other 
parameterization you like, but then (3.47) will not be satisfied.
More generally you will satisfy an equation of the form
\be
  {{d^2x^\mu}\over{d\alpha^2}}+\Gamma^\mu_{\rho\sigma}
  {{dx^\rho}\over{d\alpha}}{{dx^\sigma}\over{d\alpha}}=
  f(\alpha){{dx^\mu}\over{d\alpha}}\ ,\label{3.59}
\ee
for some parameter $\alpha$ and some function $f(\alpha)$.
Conversely, if (3.59) is satisfied along a curve you can always find
an affine parameter $\lambda(\alpha)$ for which the geodesic equation
(3.47) will be satisfied.

An important property of geodesics in a spacetime with Lorentzian
metric is that the character (timelike/null/spacelike) of the
geodesic (relative to a metric-compatible connection) never changes.  
This is simply because parallel transport preserves inner products,
and the character is determined by the inner product of the tangent
vector with itself.  This is why we were consistent to consider
purely timelike paths when we derived (3.56); for spacelike paths
we would have derived the same equation, since the only difference
is an overall minus sign in the final answer.  There are also null
geodesics, which satisfy the same equation, except that the proper
time cannot be used as a parameter (some set of allowed parameters
will exist, related to each other by linear transformations).  You
can derive this fact either from the simple requirement that the
tangent vector be parallel transported, or by extending the variation
of (3.48) to include all non-spacelike paths.

Let's now explain the earlier remark that timelike geodesics are
maxima of the proper time.  The reason we know this is true is
that, given any timelike curve (geodesic or not), we can approximate
it to arbitrary accuracy by a null curve.  To do this all we have
to do is to consider ``jagged'' null curves which follow the
timelike one:

\eject

\begin{figure}[h]
  \centerline{
  \psfig{figure=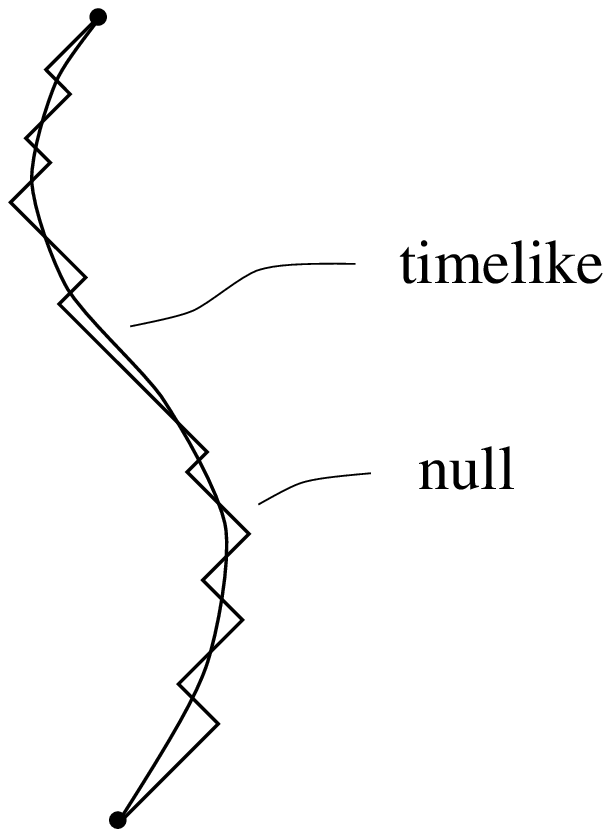,angle=0,height=7cm}}
\end{figure}

\noindent As we increase the number of sharp corners, the null curve
comes closer and closer to the timelike curve while still having
zero path length.  Timelike geodesics cannot therefore be curves
of minimum proper time, since they are always infinitesimally close
to curves of zero proper time; in fact they maximize the proper time.
(This is how you can remember which twin in the twin paradox ages
more --- the one who stays home is basically on a geodesic, and
therefore experiences more proper time.)  Of course even this is
being a little cavalier; actually every time we say ``maximize''
or ``minimize'' we should add the modifier ``locally.''  It is often
the case that between two points on a manifold there is more than
one geodesic.  For instance, on $S^2$ we can draw a great circle 
through any two points, and imagine travelling between them either 
the short way or the long way around.
One of these is obviously longer than the other, although
both are stationary points of the length functional.

The final fact about geodesics before we move on to curvature proper
is their use in mapping the tangent space at a point $p$ to a local
neighborhood of $p$.  To do this we notice that any geodesic 
$x^\mu(\lambda)$ which passes through $p$ can be specified by its
behavior at $p$; let us choose the parameter value to be 
$\lambda(p)=0$, and the tangent vector at $p$ to be 
\be
  {{d x^\mu}\over{d\lambda}}(\lambda=0)=k^\mu\ ,\label{3.60}
\ee
for $k^\mu$ some vector at $p$ (some element of $T_p$).  Then
there will be a unique point on the manifold $M$ which lies on
this geodesic where the
parameter has the value $\lambda=1$.  We define the {\bf exponential
map} at $p$, $\exp_p :T_p\rightarrow M$, via
\be
  \exp_p(k^\mu) = x^\nu(\lambda = 1)\ ,\label{3.61}
\ee
where $x^\nu(\lambda)$ solves the geodesic equation subject to (3.60).
\begin{figure}
  \centerline{
  \psfig{figure=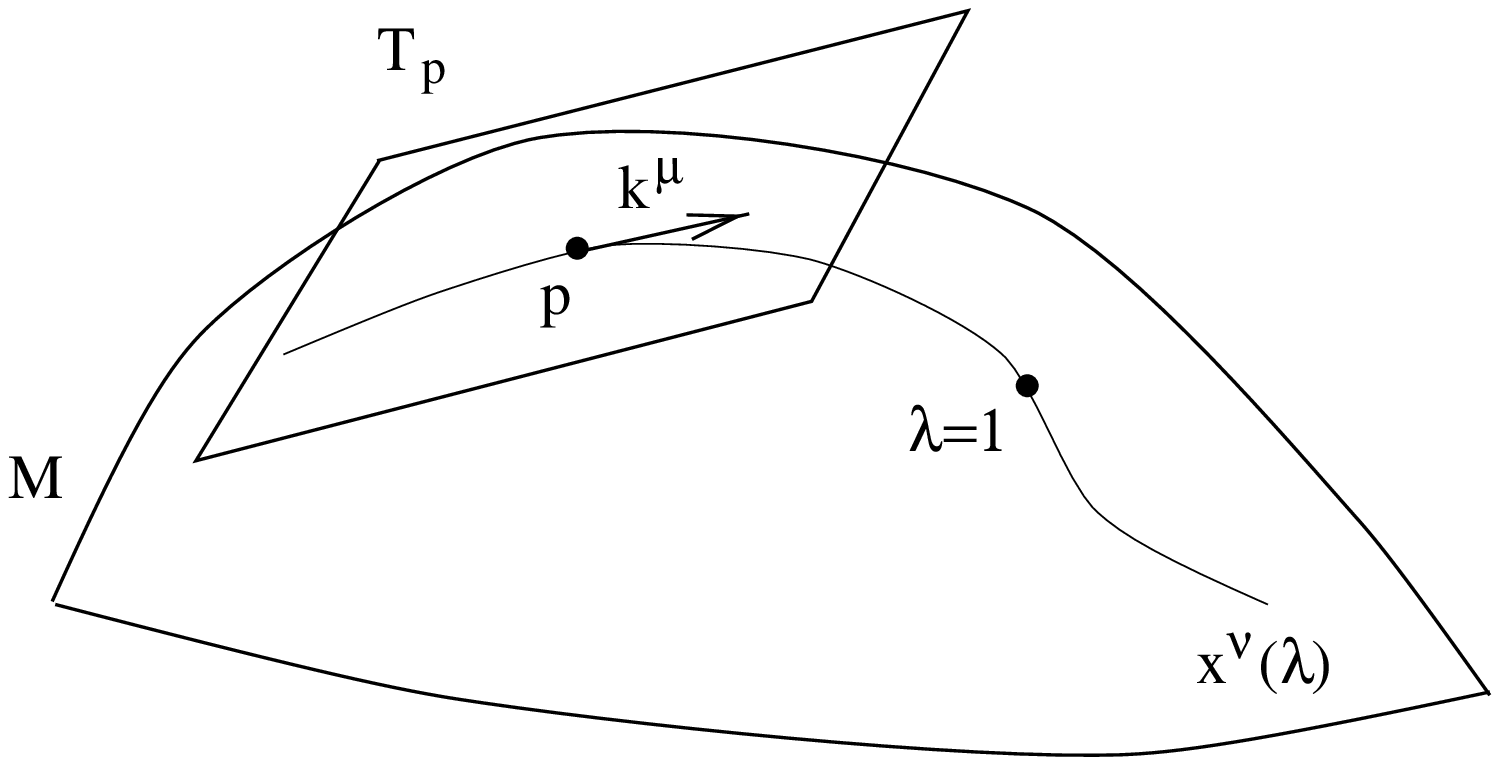,angle=0,height=5cm}}
\end{figure}
For some set of tangent vectors $k^\mu$ near the zero vector, 
this map will be well-defined, and in fact invertible.  Thus in the
neighborhood of $p$ given by the range of the map on this set of
tangent vectors, the the tangent vectors themselves define a coordinate
system on the manifold.  In this coordinate system, any geodesic
through $p$ is expressed trivially as
\be
  x^\mu(\lambda) = \lambda k^\mu\ ,\label{3.62}
\ee
for some appropriate vector $k^\mu$.

We won't go into detail about the properties of the exponential map,
since in fact we won't be using it much, but it's important to emphasize
that the range of the map is not necessarily the whole manifold, and the
domain is not necessarily the whole tangent space.  The range can fail
to be all of $M$ simply because there can be two points which are not
connected by any geodesic.  (In a Euclidean signature metric this is 
impossible, but not in a Lorentzian spacetime.)  The domain can fail
to be all of $T_p$ because a geodesic may run into a singularity, which
we think of as ``the edge of the manifold.''  Manifolds which have
such singularities are known as {\bf geodesically incomplete}.  This
is not merely a problem for careful mathematicians; in fact the
``singularity theorems'' of Hawking and Penrose state that, for
reasonable matter content (no negative energies), spacetimes in
general relativity are almost guaranteed to be geodesically incomplete.
As examples, the two most useful spacetimes in GR --- the Schwarzschild
solution describing black holes and the Friedmann-Robertson-Walker
solutions describing homogeneous, isotropic cosmologies --- both feature
important singularities.

Having set up the machinery of parallel transport and covariant
derivatives, we are at last prepared to discuss curvature proper.
The curvature is quantified by the Riemann tensor, which is derived
from the connection.  The idea behind this measure of curvature
is that we know what we mean by ``flatness'' of a connection ---
the conventional (and usually implicit) Christoffel connection
associated with a Euclidean or Minkowskian metric has a number of
properties which can be thought of as different manifestations 
of flatness.  These include the fact that parallel transport around
a closed loop leaves a vector unchanged, that covariant derivatives
of tensors commute, and that initially parallel geodesics remain
parallel.  As we shall see, the Riemann tensor arises when we 
study how any of these properties are altered in more general
contexts.

We have already argued, using the two-sphere as an example, that parallel
transport of a vector around a closed loop in a curved space will lead 
to a transformation of the vector.  The resulting transformation 
depends on the total curvature enclosed by the loop; it would be more
useful to have a local description of the curvature at each point,
which is what the Riemann tensor is supposed to provide.
One conventional way to introduce the Riemann tensor, therefore,
is to consider parallel transport around an 
infinitesimal loop.  We are not going to do that here, but take a 
more direct route.  (Most of the presentations in the literature are
either sloppy, or correct but very difficult to follow.)  Nevertheless,
even without working through the details, it is possible to see what
form the answer should take.  Imagine that we parallel transport a
vector $V^\sigma$ around a closed loop defined by two vectors $A^\nu$ and 
$B^\mu$:

\begin{figure}[h]
  \centerline{
  \psfig{figure=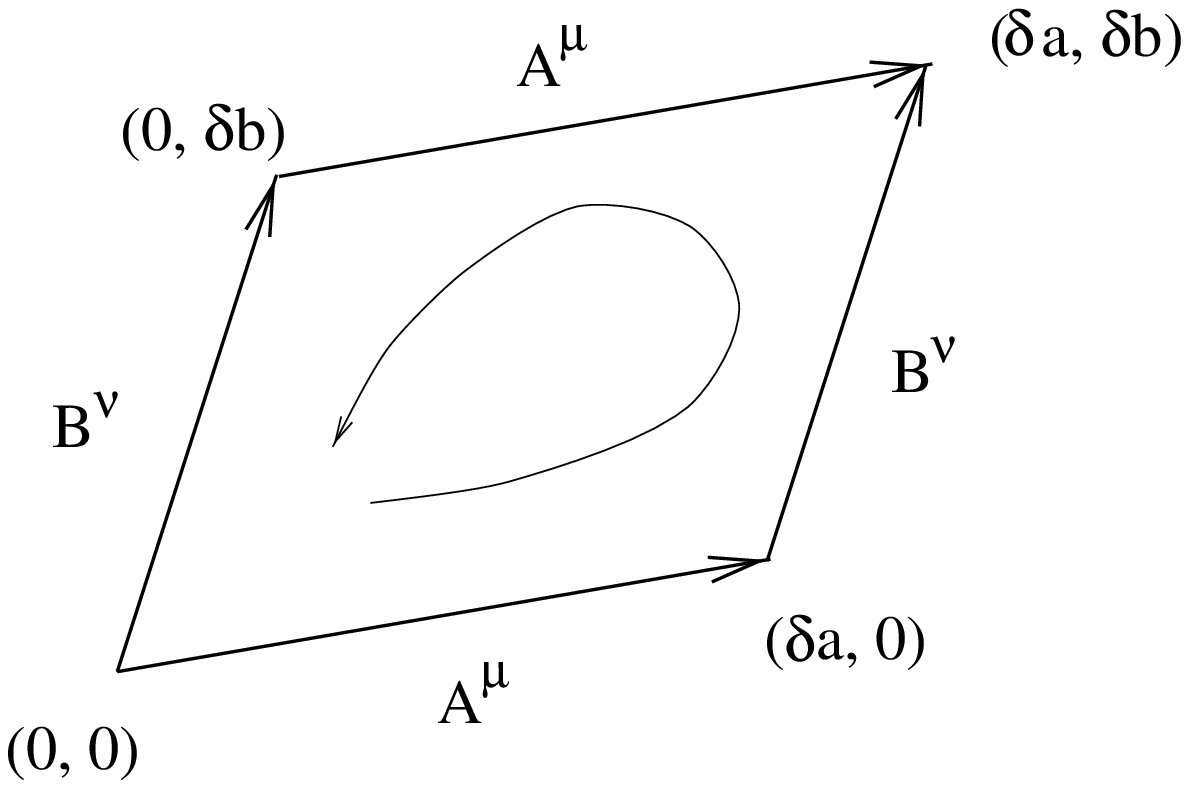,angle=0,height=5cm}}
\end{figure}

\noindent The (infinitesimal) lengths of the sides of the loop are $\delta 
a$ and $\delta b$, respectively.  Now, we know the action of parallel
transport is independent of coordinates, so there should be some
tensor which tells us how the vector changes when it comes back to 
its starting point; it will be a linear transformation on a vector,
and therefore involve one upper and one lower index.  But it will
also depend on the two vectors $A$ and $B$ which define the loop;
therefore there should be two additional lower indices to contract with
$A^\nu$ and $B^\mu$.  Furthermore, the tensor should be antisymmetric
in these two indices, since interchanging the vectors corresponds to
traversing the loop in the opposite direction, and should give the
inverse of the original answer.  (This is consistent with the
fact that the transformation should vanish if $A$ and $B$
are the same vector.)  We therefore expect that the expression for
the change $\delta V^\rho$ experienced by this vector when parallel
transported around the loop should be of the form
\be
  \delta V^\rho = (\delta a) (\delta b) A^\nu B^\mu
  R^\rho{}_{\sigma \mu\nu} V^\sigma\ ,\label{3.63}
\ee
where $R^\rho{}_{\sigma \mu\nu}$ is a $(1,3)$ tensor known as the
{\bf Riemann tensor} (or simply ``curvature tensor'').  It is
antisymmetric in the last two indices:
\be
  R^\rho{}_{\sigma \mu\nu}=-R^\rho{}_{\sigma \nu\mu}\ .\label{3.64}
\ee
(Of course, if (3.63) is taken as a definition of the Riemann tensor,
there is a convention that needs to be chosen for the ordering of
the indices.  There is no agreement at all on what this convention
should be, so be careful.)

Knowing what we do about parallel transport, we could very carefully
perform the necessary manipulations to see what happens to the
vector under this operation, and the result would be a formula for 
the curvature tensor in terms of the connection coefficients.  It is
much quicker, however, to consider a related operation, the 
commutator of two covariant derivatives.  The relationship between
this and parallel transport around a loop should be evident; the
covariant derivative of a tensor in a certain direction measures
how much the tensor changes relative to what it would have been if
it had been parallel transported (since the covariant derivative of
a tensor in a direction along which it is parallel transported is
zero).  The commutator of two covariant derivatives, then, measures
the difference between parallel transporting the tensor first one
way and then the other, versus the opposite ordering.

\begin{figure}[h]
  \centerline{
  \psfig{figure=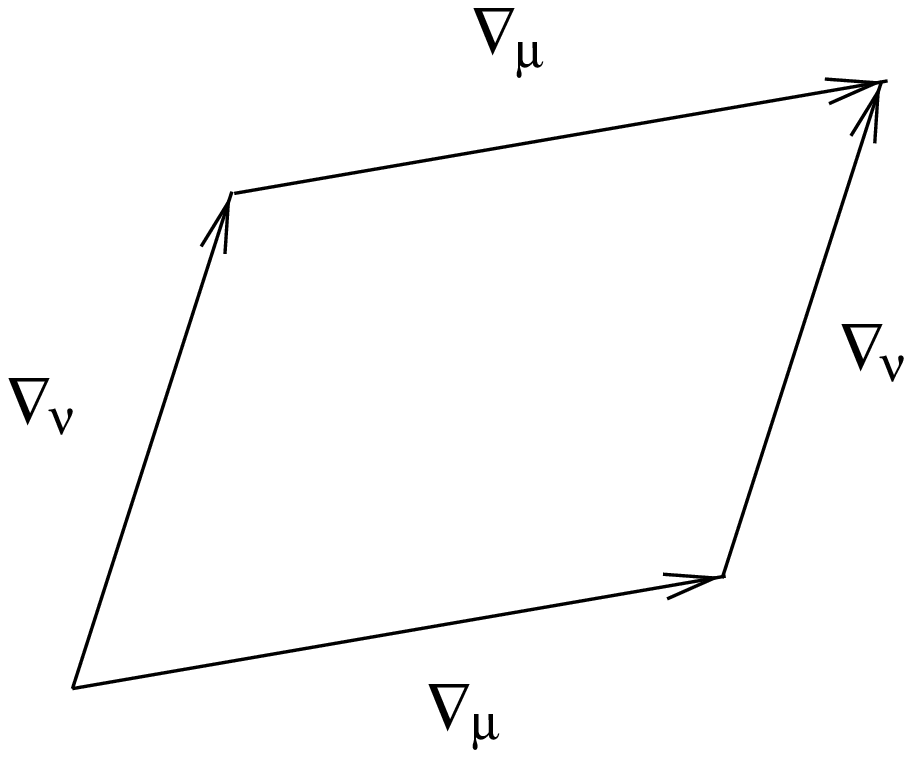,angle=0,height=5cm}}
\end{figure}

The actual computation is very straightforward.  Considering a
vector field $V^\rho$, we take
\bea
  [\nabla_\mu,\nabla_\nu]V^\rho &=& \nabla_\mu\nabla_\nu
  V^\rho - \nabla_\nu\nabla_\mu V^\rho \cr
  &=&\p\mu(\nabla_\nu V^\rho) -\Gamma^\lambda_{\mn} \nabla_\lambda
  V^\rho + \Gamma^\rho_{\mu\sigma} \nabla_\nu V^\sigma
  - (\mu \leftrightarrow \nu)\cr
  &=& \p\mu \p\nu V^\rho + (\p\mu \Gamma^\rho_{\nu\sigma})V^\sigma
  +\Gamma^\rho_{\nu\sigma}\p\mu V^\sigma - \Gamma^\lambda_{\mn}
  \p\lambda V^\rho - \Gamma^\lambda_\mn \Gamma^\rho_{\lambda\sigma}
  V^\sigma \cr
  &&\qquad +\Gamma^\rho_{\mu\sigma}\p\nu V^\sigma + \Gamma^\rho_{\mu\sigma}
  \Gamma^\sigma_{\nu\lambda}V^\lambda - (\mu\leftrightarrow \nu )\cr
  &=& (\p\mu\Gamma^\rho_{\nu\sigma}-\p\nu\Gamma^\rho_{\mu\sigma}
  +\Gamma^\rho_{\mu\lambda}\Gamma^\lambda_{\nu\sigma}
  -\Gamma^\rho_{\nu\lambda}\Gamma^\lambda_{\mu\sigma})V^\sigma 
  - 2\Gamma^\lambda_{[\mn]}\nabla_\lambda V^\rho \ . \label{3.65}
\eea
In the last step we have relabeled some dummy indices and eliminated
some terms that cancel when antisymmetrized.  We recognize that the
last term is simply the torsion tensor,
and that the left hand side is manifestly a tensor; therefore the
expression in parentheses must be a tensor
itself.  We write
\be
  [\nabla_\mu,\nabla_\nu]V^\rho = R^\rho{}_{\sigma\mn}V^\sigma
  - T_{\mn}{}^\lambda\nabla_\lambda V^\rho\ ,\label{3.66}
\ee
where the Riemann tensor is identified as
\be
  R^\rho{}_{\sigma\mn}=\p\mu\Gamma^\rho_{\nu\sigma}- \p\nu
  \Gamma^\rho_{\mu\sigma}+\Gamma^\rho_{\mu\lambda}
  \Gamma^\lambda_{\nu\sigma} -\Gamma^\rho_{\nu\lambda}
  \Gamma^\lambda_{\mu\sigma}\ .\label{3.67}
\ee
There are a number of things to notice about the derivation of
this expression:
\begin{itemize}
\item Of course we have not demonstrated that (3.67) is actually 
the same tensor that appeared in (3.63), but in fact it's true (see Wald for
a believable if tortuous demonstration).  
\item It is perhaps surprising
that the commutator $[\nabla_\mu,\nabla_\nu]$, which appears to be a
differential operator, has an action on vector fields which (in the
absence of torsion, at any rate) is a simple multiplicative 
transformation.  The Riemann tensor measures that part of the
commutator of covariant derivatives which is proportional to the
vector field, while the torsion tensor measures the part which is 
proportional to the covariant derivative of the vector field; the
second derivative doesn't enter at all.
\item Notice that the expression (3.67) is constructed
from non-tensorial elements; you can check that the transformation
laws all work out to make this particular combination a legitimate
tensor.
\item The antisymmetry of $R^\rho{}_{\sigma\mn}$ in
its last two indices is immediate from this formula and its derivation.
\item We constructed the curvature tensor completely from the
connection (no mention of the metric was made).  We were sufficiently
careful that the above expression is true for any connection, whether
or not it is metric compatible or torsion free.
\item Using what are by now our usual methods, the action
of $[\nabla_\rho,\nabla_\sigma]$ can be computed on a tensor of arbitrary
rank.  The answer is
\bea
  [\nabla_\rho,\nabla_\sigma]X^{\mu_1\cdots 
  \mu_k}{}_{\nu_1\cdots\nu_l} &=& 
  {} -T_{\rho\sigma}{}^\lambda\nabla_\lambda
  X^{\mu_1\cdots \mu_k}{}_{\nu_1\cdots\nu_l} \cr &&\quad
  +R^{\mu_1}{}_{\lambda\rho\sigma} X^{\lambda \mu_2\cdots \mu_k}{}_{\nu_1
  \cdots \nu_l}+R^{\mu_2}{}_{\lambda\rho\sigma} X^{\mu_1\lambda\cdots 
  \mu_k}{}_{\nu_1 \cdots \nu_l} +\cdots \cr &&\quad
  -R^{\lambda}{}_{\nu_1\rho\sigma} X^{\mu_1\cdots \mu_k}{}_{\lambda\nu_2
  \cdots \nu_l} - R^{\lambda}{}_{\nu_2\rho\sigma} X^{\mu_1\cdots 
  \mu_k}{}_{\nu_1\lambda\cdots \nu_l} - \cdots \ . \label{3.68}
\eea
\end{itemize}

A useful notion is that
of the commutator of two vector fields $X$ and $Y$, which is a third 
vector field with components 
\be
  [X,Y]^\mu = X^\lambda\p\lambda Y^\mu - Y^\lambda\p\lambda X^\mu\ .
  \label{3.69}
\ee
Both the torsion tensor and the Riemann tensor, thought of as
multilinear maps, have elegant expressions in terms of the
commutator.  Thinking of the torsion as a map from two vector fields to
a third vector field, we have
\be
  T(X,Y) = \nabla_X Y - \nabla_Y X - [X,Y]\ ,\label{3.70}
\ee
and thinking of the Riemann tensor as a map from three vector fields
to a fourth one, we have
\be
  R(X,Y)Z = \nabla_X\nabla_Y Z-\nabla_Y\nabla_X Z 
  - \nabla_{[X,Y]}Z\ .\label{3.71}
\ee
In these expressions, the notation $\nabla_X$ refers to the covariant
derivative along the vector field $X$; in components, $\nabla_X = 
X^\mu\nabla_\mu$.  Note that the two vectors $X$ and $Y$ in (3.71)
correspond to the two antisymmetric indices in the component form
of the Riemann tensor.  The last term in (3.71), involving the
commutator $[X,Y]$, vanishes when $X$ and $Y$ are taken to be the
coordinate basis vector fields (since $[\p\mu ,\p\nu]=0$), which
is why this term did not arise when we originally took the commutator
of two covariant derivatives.  We will not use this notation
extensively, but you might see it in the literature, so you should
be able to decode it.

Having defined the curvature tensor as something which characterizes
the connection, let us now admit that in GR we are most concerned with
the Christoffel connection.  In this case the connection is derived 
from the metric, and the associated curvature may be thought of as 
that of the metric itself.  This identification allows us to finally
make sense of our informal notion that spaces for which the metric
looks Euclidean or Minkowskian are flat.  In fact it works both ways:
if the components of the metric are constant in some coordinate system,
the Riemann tensor will vanish, while if the Riemann tensor vanishes
we can always construct a coordinate system in which the metric components
are constant.

The first of these is easy to show.  If we are in some coordinate system
such that $\p\sigma g_\mn=0$ (everywhere, not just at a point), then 
$\Gamma^\rho_\mn = 0$ and $\p\sigma\Gamma^\rho_\mn = 0$; thus
$R^\rho{}_{\sigma\mn}=0$ by (3.67).  But this is a tensor equation, and
if it is true in one coordinate system it must be true in any coordinate
system.  Therefore, the statement that the Riemann tensor vanishes
is a necessary condition for it to be possible to find coordinates in
which the components of $g_\mn$ are constant everywhere.

It is also a sufficient condition, although we have to work harder to
show it.  Start by choosing Riemann normal coordinates at some point
$p$, so that $g_\mn = \eta_\mn$ at $p$.  (Here we are using $\eta_\mn$
in a generalized sense, as a matrix with either $+1$ or $-1$ for each
diagonal element and zeroes elsewhere.  The actual arrangement of
the $+1$'s and $-1$'s depends on the canonical form of the metric, but
is irrelevant for the present argument.)  Denote the basis vectors at
$p$ by $\e\mu$, with components $\e\mu^\sigma$.  Then by construction
we have
\be
  g_{\sigma\rho}\e\mu^\sigma \e\nu^\rho (p) =\eta_\mn\ .\label{3.72}
\ee
Now let us parallel transport the entire set of basis vectors from
$p$ to another point $q$; the vanishing of the Riemann tensor ensures
that the result will be independent of the path taken between $p$
and $q$.  Since parallel transport with respect to a metric compatible
connection preserves inner products, we must have
\be
  g_{\sigma\rho}\e\mu^\sigma \e\nu^\rho (q) =\eta_\mn\ .\label{3.73}
\ee
We therefore have specified a set of vector fields which
everywhere define a basis in which the metric components are constant.
This is completely unimpressive; it can be done on any manifold,
regardless of what the curvature is.  What we would like to show
is that this is a coordinate basis (which can only be true
if the curvature vanishes).
We know that if the $\e\mu$'s are a coordinate basis, their
commutator will vanish:
\be
  [\e\mu,\e\nu] = 0\ .\label{3.74}
\ee
What we would really like is the converse: that if the commutator
vanishes we can find coordinates $y^\mu$ such that $\e\mu = {{\partial}
\over{\partial y^\mu}}$.  In fact this is a true result, known as
{\bf Frobenius's Theorem}.  It's something of a mess to prove, involving
a good deal more mathematical apparatus than we have bothered to set
up.  Let's just take it for granted (skeptics can consult Schutz's
{\sl Geometrical Methods} book).  Thus, we would like to demonstrate
(3.74) for the vector fields we have set up.  Let's use the expression
(3.70) for the torsion:
\be
  [\e\mu,\e\nu] = \nabla_{\e\mu} \e\nu - \nabla_{\e\nu}\e\mu
  - T(\e\mu,\e\nu)\ .\label{3.75}
\ee
The torsion vanishes by hypothesis.  The covariant derivatives will
also vanish, given the method by which we constructed our vector fields;
they were made by parallel transporting along arbitrary paths.  If the
fields are parallel transported along arbitrary paths, they are
certainly parallel transported along the vectors $\e\mu$, and therefore
their covariant derivatives in the direction of these vectors will
vanish.  Thus (3.70) implies that the commutator vanishes, and therefore
that we can find a coordinate system $y^\mu$ for which these vector
fields are the partial derivatives.  In this coordinate system the
metric will have components $\eta_\mn$, as desired.

The Riemann tensor, with four indices, naively has $n^4$ independent
components in an $n$-dimensional space.  In fact the antisymmetry
property (3.64) means that there are only $n(n-1)/2$ independent values
these last two indices can take on, leaving us with $n^3(n-1)/2$
independent components.  When we consider the Christoffel connection,
however, there are a number of other symmetries that reduce the 
independent components further.  Let's consider these now.

The simplest way to derive these additional symmetries is to examine
the Riemann tensor with all lower indices,
\be
  R_{\rho\sigma\mn} = g_{\rho\lambda}R^\lambda{}_{\sigma\mn}\ .
  \label{3.76}
\ee
Let us further consider the components of this tensor in Riemann
normal coordinates established at a point $p$.  Then the Christoffel
symbols themselves will vanish, although their derivatives will not.
We therefore have
\bea
  R_{\rho\sigma\mn} &=& g_{\rho\lambda}
  (\p\mu\Gamma^\lambda_{\nu\sigma}- \p\nu
  \Gamma^\lambda_{\mu\sigma})\cr
  &=& {1\over 2}g_{\rho\lambda}g^{\lambda\tau}(
  \p\mu\p\nu g_{\sigma\tau} + \p\mu\p\sigma g_{\tau\nu}
  -\p\mu\p\tau g_{\nu\sigma} - \p\nu\p\mu g_{\sigma\tau} 
  - \p\nu\p\sigma g_{\tau\mu}+\p\nu\p\tau g_{\mu\sigma})\cr
  &=&{1\over 2}(\p\mu\p\sigma g_{\rho\nu}
  -\p\mu\p\rho g_{\nu\sigma} - \p\nu\p\sigma g_{\rho\mu} 
  +\p\nu\p\rho g_{\mu\sigma})\ . \label{3.77}
\eea
In the second line we have used $\partial_\mu g^{\lambda\tau}=0$
in RNC's, and in the third line the fact that partials commute.
From this expression we can notice immediately two properties
of $R_{\rho\sigma\mn}$; it is antisymmetric in its first two
indices,
\be
  R_{\rho\sigma\mn}=-R_{\sigma\rho\mn}\ ,\label{3.78}
\ee
and it is invariant under interchange of the first pair of
indices with the second:
\be
  R_{\rho\sigma\mn}= R_{\mn\rho\sigma}\ .\label{3.79}
\ee
With a little more work, which we leave to your imagination, 
we can see that the sum of cyclic
permutations of the last three indices vanishes:
\be
  R_{\rho\sigma\mn} + R_{\rho\mn\sigma} + R_{\rho\nu\sigma\mu}
  =0 \ .\label{3.80}
\ee
This last property is equivalent to the vanishing of the antisymmetric
part of the last three indices:
\be
  R_{\rho[\sigma\mn]} =0 \ .\label{3.81}
\ee
All of these properties have been derived in a special coordinate
system, but they are all tensor equations; therefore they will be
true in any coordinates.  Not all of them are independent; with some 
effort, you can show that (3.64), (3.78) 
and (3.81) together imply (3.79).  The logical
interdependence of the equations is usually less important than
the simple fact that they are true.  

Given these relationships between the different components of the
Riemann tensor, how many independent quantities remain?  Let's
begin with the facts that $R_{\rho\sigma\mn}$ is antisymmetric
in the first two indices, antisymmetric in the last two indices,
and symmetric under interchange of these two pairs.  This means that
we can think of it as a symmetric matrix $R_{[\rho\sigma][\mn]}$,
where the pairs $\rho\sigma$ and $\mu\nu$ are thought of as individual
indices.  An $m\times m$ symmetric matrix has $m(m+1)/2$ independent
components, while an $n\times n$ antisymmetric matrix has $n(n-1)/2$
independent components.  We therefore have
\be
  {1\over 2}\left[{1\over 2}n(n-1)\right]\left[{1\over 2}n(n-1)
  +1\right] = {1\over 8}(n^4-2n^3+3n^2-2n)\label{3.82}
\ee
independent components.  We still have to deal with the additional
symmetry (3.81).  An immediate consequence of (3.81) is that the
totally antisymmetric part of the Riemann tensor vanishes,
\be
  R_{[\rho\sigma\mn]} =0 \ .\label{3.83}
\ee
In fact, this equation plus the other symmetries (3.64), (3.78)
and (3.79) are enough to imply (3.81), as can be easily shown
by expanding (3.83) and messing with the resulting terms.
Therefore imposing the additional constraint of (3.83) is equivalent
to imposing (3.81), once the other symmetries have been accounted
for.  How many independent restrictions does this represent?  
Let us imagine decomposing
\be
  R_{\rho\sigma\mn}=X_{\rho\sigma\mn}+R_{[\rho\sigma\mn]}\ .
  \label{3.84}
\ee
It is easy to see that any totally antisymmetric 4-index tensor
is automatically antisymmetric in its first and last indices, and
symmetric under interchange of the two pairs.  Therefore these
properties are independent restrictions on $X_{\rho\sigma\mn}$,
unrelated to the requirement (3.83).  Now a
totally antisymmetric 4-index tensor has $n(n-1)(n-2)(n-3)/4!$
terms, and therefore (3.83) reduces the number of independent 
components by this amount.  We are left with
\be
  {1\over 8}(n^4-2n^3+3n^2-2n)-{1\over {24}}n(n-1)(n-2)(n-3) =
  {1\over{12}}n^2(n^2-1)\label{3.85}
\ee
independent components of the Riemann tensor.

In four dimensions, therefore, the Riemann tensor has 20 independent
components.  (In one dimension it has none.)  These twenty functions
are precisely the 20 degrees of freedom in the second derivatives
of the metric which we could not set to zero by a clever choice of
coordinates.  This should reinforce your confidence that the 
Riemann tensor is an appropriate measure of curvature.

In addition to the algebraic symmetries of the Riemann tensor (which
constrain the number of independent components at any point), there
is a differential identity which it obeys (which constrains its 
relative values at different points).  Consider the covariant derivative
of the Riemann tensor, evaluated in Riemann normal coordinates:
\bea
  \nabla_\lambda R_{\rho\sigma\mn}&=&\p\lambda
  R_{\rho\sigma\mn}\cr
  &=& {1\over 2}\p\lambda(\p\mu\p\sigma g_{\rho\nu}
  -\p\mu\p\rho g_{\nu\sigma} - \p\nu\p\sigma g_{\rho\mu} 
  +\p\nu\p\rho g_{\mu\sigma})\ . \label{3.86}
\eea
We would like to consider the sum of cyclic permutations of the
first three indices:
\bea
  \lefteqn{\nabla_\lambda R_{\rho\sigma\mn} +
  \nabla_\rho R_{\sigma\lambda\mn}+\nabla_\sigma R_{\lambda\rho\mn}} \cr
  &=& {1\over 2}
  (\p\lambda\p\mu\p\sigma g_{\rho\nu} -\p\lambda\p\mu\p\rho g_{\nu\sigma} 
  -\p\lambda\p\nu\p\sigma g_{\rho\mu}+\p\lambda\p\nu\p\rho g_{\mu\sigma}\cr
  &&+\p\rho\p\mu\p\lambda g_{\sigma\nu} -\p\rho\p\mu\p\sigma g_{\nu\lambda} 
  -\p\rho\p\nu\p\lambda g_{\sigma\mu}+\p\rho\p\nu\p\sigma g_{\mu\lambda}\cr
  &&+\p\sigma\p\mu\p\rho g_{\lambda\nu} -\p\sigma\p\mu\p\lambda g_{\nu\rho} 
  -\p\sigma\p\nu\p\rho g_{\lambda\mu}+\p\sigma\p\nu\p\lambda g_{\mu\rho})
  \cr & =& 0\ .\label{3.87}
\eea
Once again, since this is an equation between tensors it is true in any
coordinate system, even though we derived it in a particular one.
We recognize by now that the antisymmetry $R_{\rho\sigma\mn}=-
R_{\sigma\rho\mn}$ allows us to write this result as
\be
  \nabla_{[\lambda}R_{\rho\sigma]\mn}=0\ .\label{3.88}
\ee
This is known as the {\bf Bianchi identity}.  (Notice that for a
general connection there would be additional terms involving the
torsion tensor.)  It is closely related 
to the Jacobi identity, since (as you can show) it basically expresses
\be
  [[\nabla_\lambda,\nabla_\rho],\nabla_\sigma]
  +[[\nabla_\rho,\nabla_\sigma],\nabla_\lambda]
  +[[\nabla_\sigma,\nabla_\lambda],\nabla_\rho]=0\ .\label{3.89}
\ee

It is frequently useful to consider contractions of the Riemann
tensor.  Even without the metric, we can form a contraction known
as the {\bf Ricci tensor}:
\be
  R_{\mn} = R^\lambda{}_{\mu\lambda\nu}\ .\label{3.90}
\ee
Notice that, for the curvature tensor formed from an arbitrary
(not necessarily Christoffel) connection, there are a number
of independent contractions to take. Our primary concern is with the 
Christoffel connection, for which (3.90) is the only independent contraction
(modulo conventions for the sign, which of course change from 
place to place).  The Ricci tensor associated with the Christoffel
connection is symmetric,
\be
  R_{\mn} = R_{\nu\mu}\ ,\label{3.91}
\ee
as a consequence of the various symmetries of the Riemann tensor.
Using the metric, we can take a further contraction
to form the {\bf Ricci scalar}:
\be
  R = R^\mu{}_\mu = g^\mn R_\mn\ .\label{3.92}
\ee

An especially useful form of the Bianchi identity comes from
contracting twice on (3.87):
\bea
  0&=& g^{\nu\sigma}g^{\mu\lambda}(\nabla_\lambda R_{\rho\sigma\mn} 
  +\nabla_\rho R_{\sigma\lambda\mn}+\nabla_\sigma R_{\lambda\rho\mn})\cr
  &=&\nabla^\mu R_{\rho\mu}-\nabla_\rho R + \nabla^\nu R_{\rho\nu}\ ,
  \label{3.93}
\eea
or
\be
  \nabla^\mu R_{\rho\mu} = {1\over 2}\nabla_\rho R\ .\label{3.94}
\ee
(Notice that, unlike the partial derivative, it makes sense to raise
an index on the covariant derivative, due to metric compatibility.)
If we define the {\bf Einstein tensor} as
\be
  G_{\mu\nu} = R_\mn -{1\over 2} R g_\mn\ ,\label{3.95}
\ee
then we see that the twice-contracted Bianchi identity (3.94)
is equivalent to
\be
  \nabla^\mu G_{\mn} = 0\ .\label{3.96}
\ee
The Einstein tensor, which is symmetric due to the symmetry of the
Ricci tensor and the metric, will be of great importance in general
relativity.

The Ricci tensor and the Ricci scalar contain information about
``traces'' of the Riemann tensor.  It is sometimes useful to consider
separately those pieces of the Riemann tensor which the Ricci
tensor doesn't tell us about.  We therefore invent the {\bf Weyl
tensor}, which is basically the Riemann tensor with all of its
contractions removed.  It is given in $n$ dimensions by
\be
  C_{\rho\sigma\mn} = R_{\rho\sigma\mn} - {2\over{(n-2)}}
  \left(g_{\rho[\mu}R_{\nu]\sigma} - g_{\sigma[\mu}R_{\nu]\rho}
  \right) +{2\over{(n-1)(n-2)}}R g_{\rho[\mu}g_{\nu]\sigma}\ .\label{3.97}
\ee
This messy formula is designed so that all possible contractions of
$C_{\rho\sigma\mn}$ vanish, while it retains the symmetries of the
Riemann tensor:
\bea
  C_{\rho\sigma\mn} &=& C_{[\rho\sigma][\mn]}\ ,\cr
  C_{\rho\sigma\mn} &=& C_{\mn\rho\sigma}\ ,\cr
  C_{\rho[\sigma\mn]} &=&0\ . \label{3.98}
\eea
The Weyl tensor is only defined in three or more dimensions, and
in three dimensions it vanishes identically.  For $n\geq 4$ it
satisfies a version of the Bianchi identity,
\be
  \nabla^\rho C_{\rho\sigma\mn} = -2{{(n-3)}\over{(n-2)}}
  \left(\nabla_{[\mu}R_{\nu]\sigma} + {1\over{2(n-1)}}
  g_{\sigma[\nu}\nabla_{\mu]}R\right)\ .\label{3.99}
\ee
One of the most important properties of the Weyl tensor is that
it is invariant under {\bf conformal transformations}.  This means that
if you compute $C_{\rho\sigma\mn}$ for some metric $g_{\mn}$, and
then compute it again for a metric given by $\Omega^2 (x)g_{\mn}$, 
where $\Omega(x)$ is an arbitrary nonvanishing function of
spacetime, you get the same answer.  For this reason it is often
known as the ``conformal tensor.''

After this large amount of formalism, it might be time to step back
and think about what curvature means for some simple examples.
First notice that, according to (3.85), in 1, 2, 3 and 4 dimensions
there are 0, 1, 6 and 20 components of the curvature tensor,
respectively.  (Everything we say about the curvature in these
examples refers to the curvature associated with the Christoffel
connection, and therefore the metric.)
This means that one-dimensional manifolds (such as $S^1$) are never
curved; the intuition you have that tells you that a circle is
curved comes from thinking of it embedded in a certain flat
two-dimensional plane.  (There is something called ``extrinsic
curvature,'' which characterizes the way something is embedded
in a higher dimensional space.  Our notion of curvature is ``intrinsic,''
and has nothing to do with such embeddings.)

The distinction between intrinsic and extrinsic curvature is also
important in two dimensions, where the curvature has one independent
component.  (In fact, all of the information about the curvature is 
contained in the single component of the Ricci scalar.)  Consider
a cylinder, $\R\times S^1$.
\begin{figure}
  \centerline{
  \psfig{figure=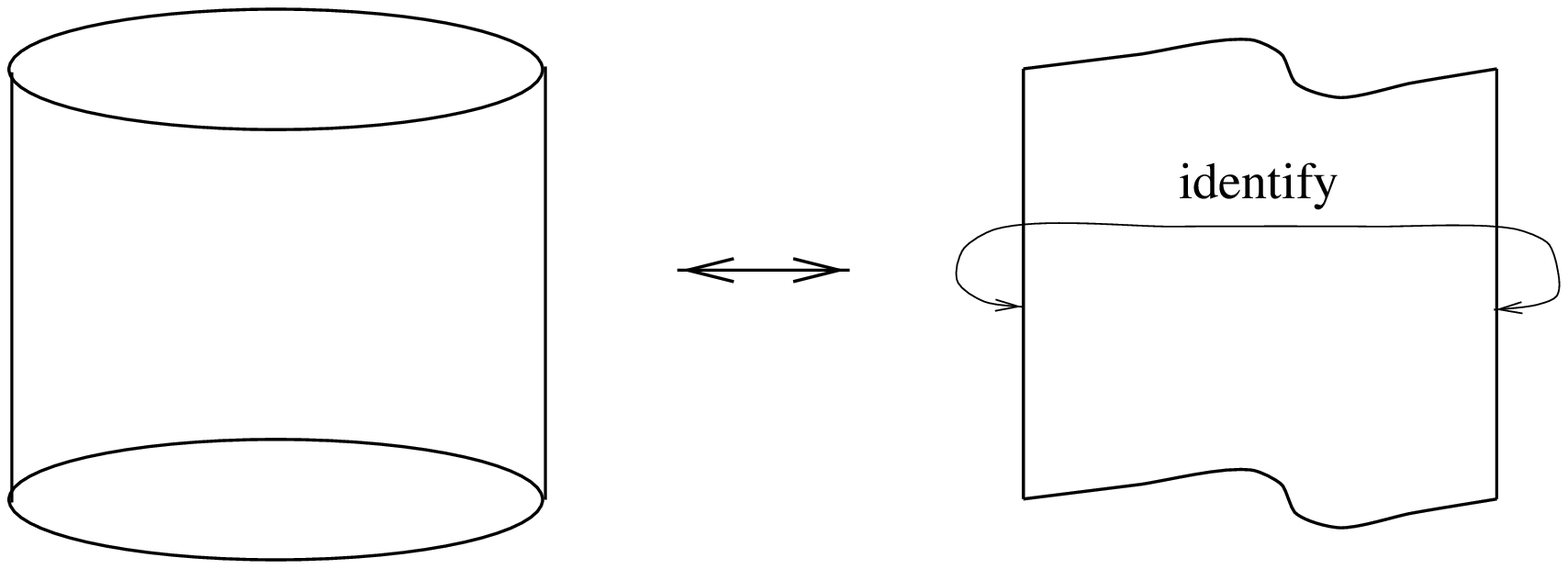,angle=0,height=5cm}}
\end{figure}
Although this looks curved from our point of view, it
should be clear that we can put a metric on the cylinder whose
components are constant in an appropriate coordinate system ---
simply unroll it and use the induced metric from the plane.  In this
metric, the cylinder is flat.  (There is also nothing to stop us from
introducing a different metric in which the cylinder is not flat, but
the point we are trying to emphasize is that it can be made flat in
some metric.)  The same story holds for the torus:

\begin{figure}[h]
  \centerline{
  \psfig{figure=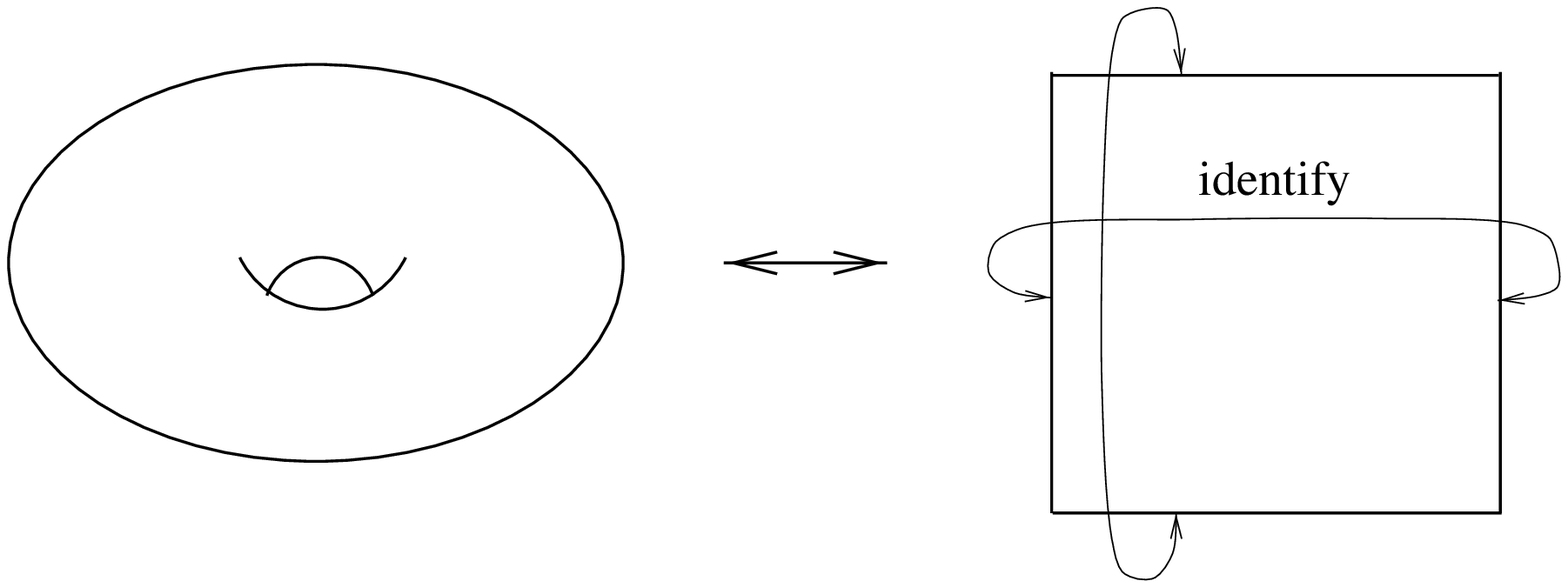,angle=0,height=5cm}}
\end{figure}

\noindent We can think of the torus as a square region of the plane
with opposite sides identified (in other words, $S^1\times S^1$),
from which it is clear that it can have a flat metric even though
it looks curved from the embedded point of view.

A cone is an example of a two-dimensional manifold with nonzero
curvature at exactly one point.  We can see this also by unrolling
it; the cone is equivalent to the plane with a ``deficit angle''
removed and opposite sides identified:

\eject

\begin{figure}[h]
  \centerline{
  \psfig{figure=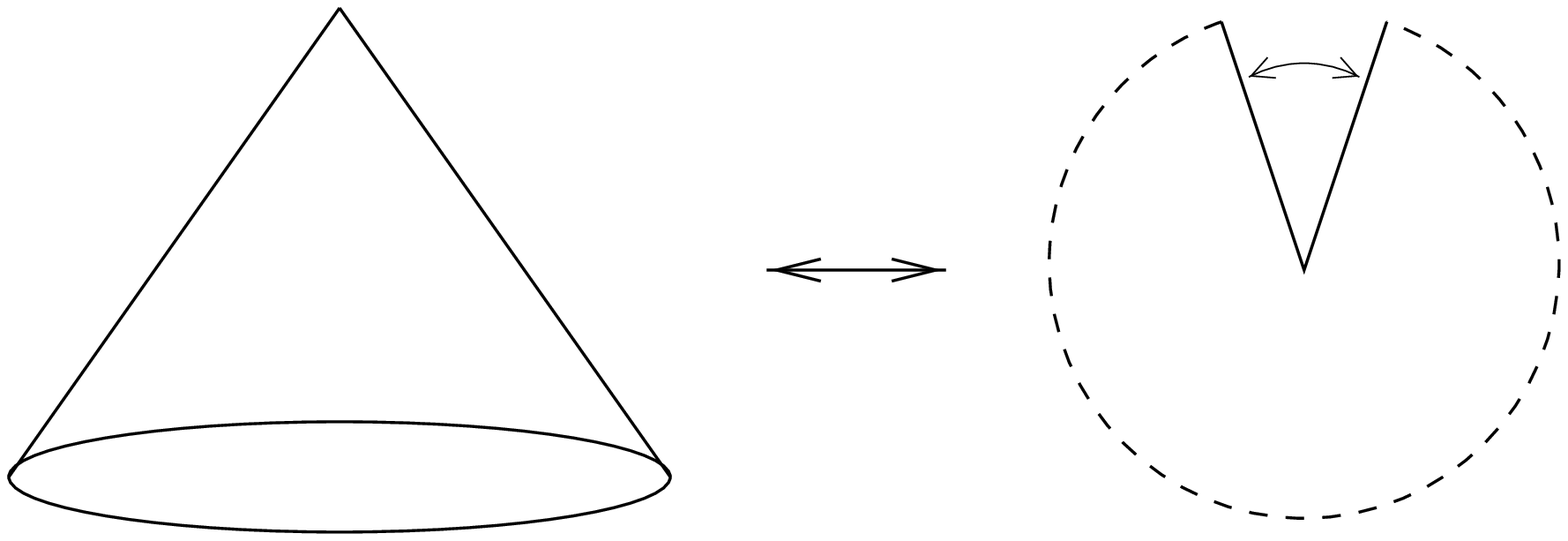,angle=0,height=5cm}}
\end{figure}

\noindent In the metric inherited from this description as part of the 
flat plane, the cone is flat everywhere but at its vertex.  This can
be seen by considering parallel transport of a vector around various
loops; if a loop does not enclose the vertex, there will be no overall
transformation, whereas a loop that does enclose the vertex (say, just
one time) will lead to a rotation by an angle which is just the
deficit angle.

\begin{figure}[h]
  \centerline{
  \psfig{figure=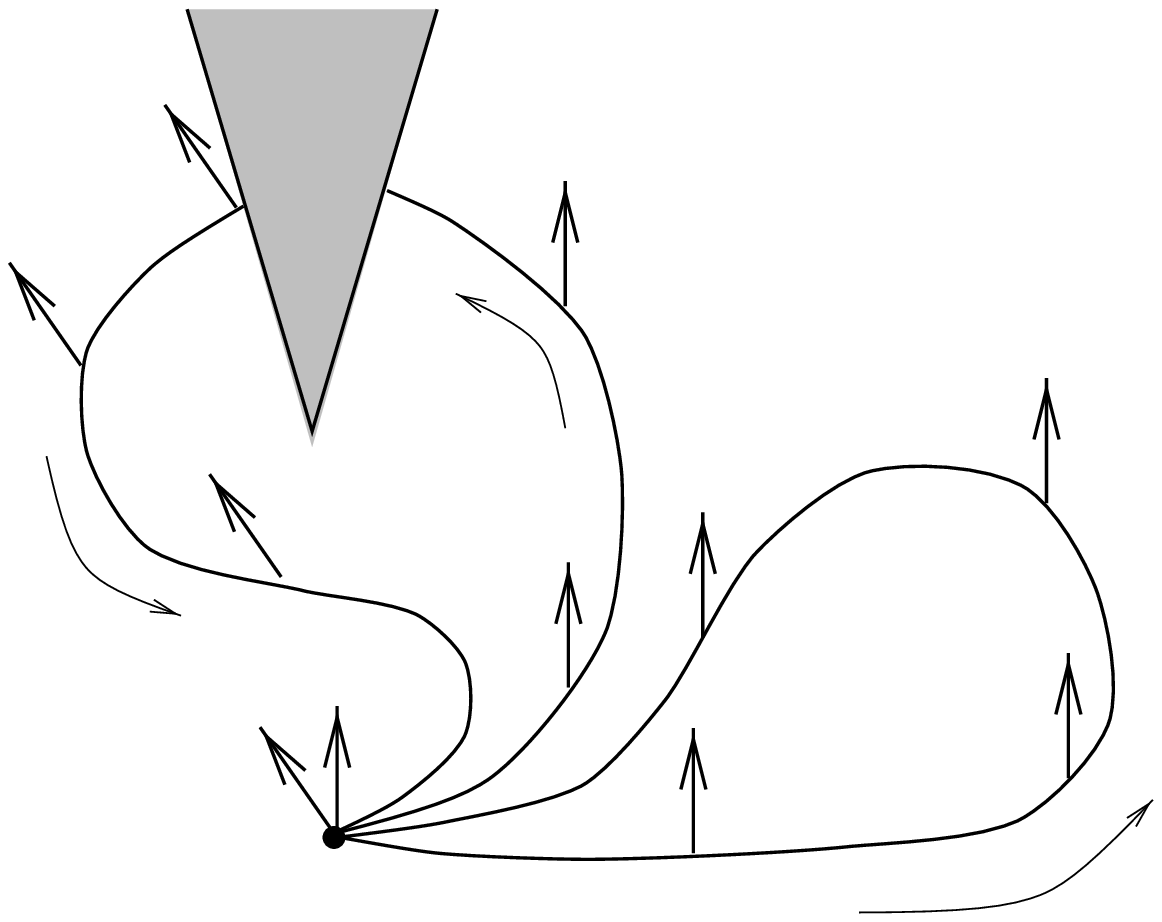,angle=0,height=7cm}}
\end{figure}

Our favorite example is of course the two-sphere, with metric
\be
  ds^2 = a^2(\d\theta^2 + \sin^2\theta ~ \d\phi^2)\ ,\label{3.100}
\ee
where $a$ is the radius of the sphere (thought of as embedded in
$\R^3$).  Without going through the details, the nonzero connection
coefficients are
\bea
  \Gamma^\theta_{\phi\phi} &=& -\sin\theta \cos\theta\cr
  \Gamma^\phi_{\theta\phi} = \Gamma^\phi_{\phi\theta} &=&
  \cot\theta\ . \label{3.101}
\eea
Let's compute a promising component of the Riemann tensor:
\bea
  R^\theta{}_{\phi\theta\phi} &=& \p\theta
  \Gamma^\theta_{\phi\phi} - \p\phi \Gamma^\theta_{\theta\phi}
  +\Gamma^\theta_{\theta\lambda}\Gamma^\lambda_{\phi\phi}
  -\Gamma^\theta_{\phi\lambda}\Gamma^\lambda_{\theta\phi}\cr
  &=& (\sin^2\theta - \cos^2\theta) -(0) + (0) - (-\sin\theta 
  \cos\theta)(\cot\theta)\cr
  &=& \sin^2\theta\ . \label{3.102}
\eea
(The notation is obviously imperfect, since the Greek letter $\lambda$
is a dummy index which is summed over, while the Greek letters
$\theta$ and $\phi$ represent specific coordinates.)  Lowering an
index, we have
\bea
  R_{\theta\phi\theta\phi} &=& g_{\theta\lambda}
  R^\lambda{}_{\phi\theta\phi}\cr
  &=&g_{\theta\theta}R^\theta{}_{\phi\theta\phi}\cr
  &=& a^2\sin^2\theta\ . \label{3.103}
\eea
It is easy to check that all of the components of the Riemann tensor
either vanish or are related to this one by symmetry.  We can go on
to compute the Ricci tensor via $R_{\mn}=g^{\alpha\beta}R_{\alpha\mu
\beta\nu}$.  We obtain
\bea
  R_{\theta\theta} &=& g^{\phi\phi}R_{\phi\theta\phi\theta}
  = 1\cr
  R_{\theta\phi} &=& R_{\phi\theta} = 0\cr
  R_{\phi\phi} &=& g^{\theta\theta}R_{\theta\phi\theta\phi}
  = \sin^2\theta\ . \label{3.104}
\eea
The Ricci scalar is similarly straightforward:
\be
  R = g^{\theta\theta}R_{\theta\theta}+ g^{\phi\phi}R_{\phi\phi}
  = {2\over{a^2}}\ .\label{3.105}
\ee
Therefore the Ricci scalar, which for a two-dimensional manifold
completely characterizes the curvature, is a constant over this
two-sphere.  This is a reflection of the fact that the manifold is
``maximally symmetric,'' a concept we will define more precisely later
(although it means what you think it should).  In any number of
dimensions the curvature of a maximally symmetric space satisfies
(for some constant $a$)
\be
  R_{\rho\sigma\mu\nu} = a^{-2}(g_{\rho\mu}g_{\sigma\nu}
  - g_{\rho\nu}g_{\sigma\mu})\ ,\label{3.106}
\ee
which you may check is satisfied by this example.

Notice that the Ricci scalar is not only constant for the two-sphere,
it is manifestly positive.  We say that the sphere is ``positively
curved'' (of course a convention or two came into play, but fortunately
our conventions conspired so that spaces which everyone agrees to call
positively curved actually have a positive Ricci scalar).  
From the point of view of someone living on a manifold which is
embedded in a higher-dimensional Euclidean space, 
if they are sitting at a point of positive curvature the
space curves away from them in the same way in any direction, while
in a negatively curved space it curves away in opposite directions.
Negatively curved spaces are therefore saddle-like.

\begin{figure}
  \centerline{
  \psfig{figure=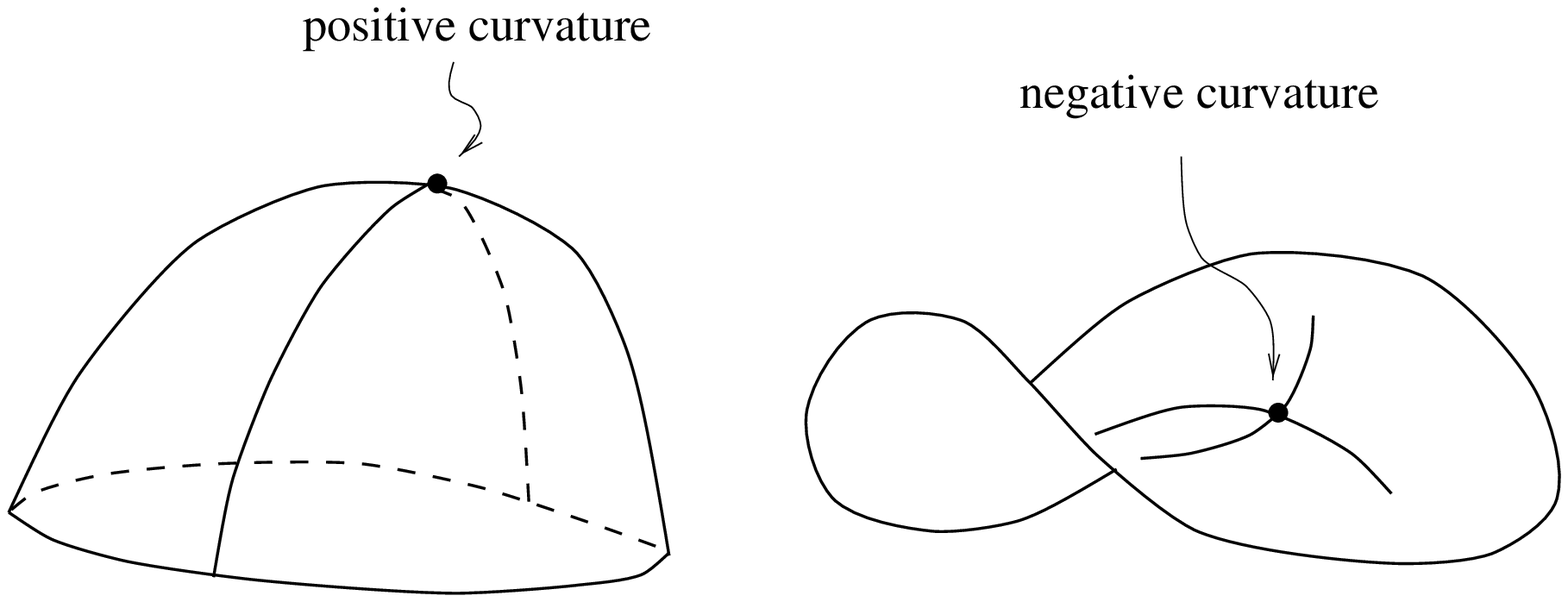,angle=0,height=6cm}}
\end{figure}

Enough fun with examples.  There is one more topic we have to cover
before introducing general relativity itself: geodesic deviation.
You have undoubtedly heard that the defining property of Euclidean (flat)
geometry is the parallel postulate: initially parallel lines remain
parallel forever.  Of course in a curved space this is not true; on
a sphere, certainly, initially parallel geodesics will eventually
cross.  We would like to quantify this behavior for an arbitrary
curved space.

The problem is that the notion of ``parallel'' does not extend
naturally from flat to curved spaces.  Instead what we will do is
to construct a one-parameter family of geodesics, $\gamma_s(t)$.
That is, for each $s\in\R$, $\gamma_s$ is a geodesic parameterized
by the affine parameter $t$.
The collection of these curves defines a smooth two-dimensional
surface (embedded in a manifold $M$ of arbitrary dimensionality).  The
coordinates on this surface may be chosen to be $s$ and $t$, provided
we have chosen a family of geodesics which do not cross.  The entire
surface is the set of points $x^\mu(s,t)\in M$.  We have two natural
vector fields: the tangent vectors to the geodesics,
\be
  T^\mu = {{\partial x^\mu}\over{\partial t}}\ ,\label{3.107}
\ee
and the ``deviation vectors''
\be
  S^\mu = {{\partial x^\mu}\over{\partial s}}\ .\label{3.108}
\ee
This name derives from the informal notion that $S^\mu$ points
from one geodesic towards the neighboring ones.

\begin{figure}
  \centerline{
  \psfig{figure=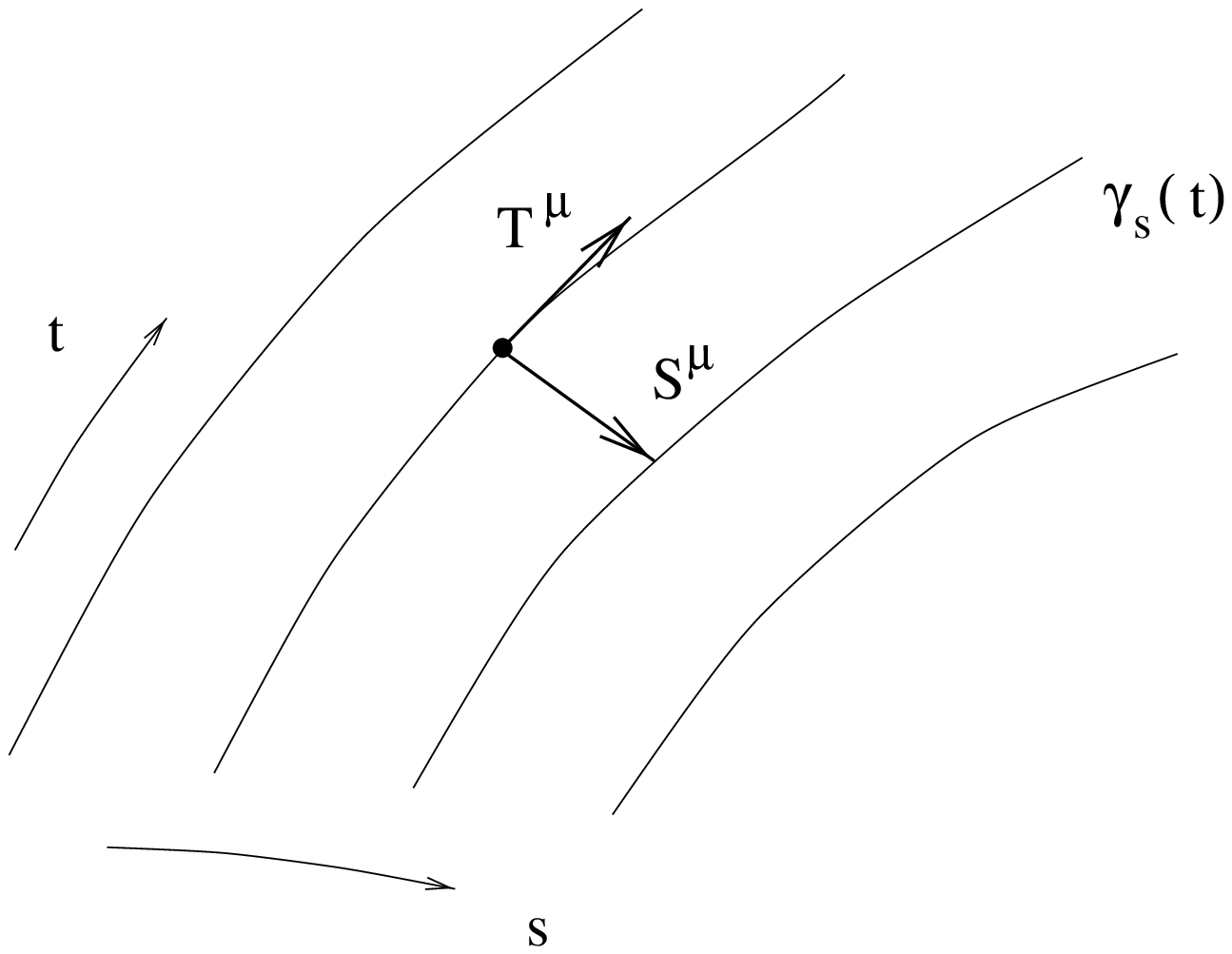,angle=0,height=7cm}}
\end{figure}

The idea that $S^\mu$ points from one geodesic to the next inspires
us to define the ``relative velocity of geodesics,''
\be
  V^\mu = (\nabla_TS)^\mu = T^\rho\nabla_\rho S^\mu\ ,\label{3.109}
\ee
and the ``relative acceleration of geodesics,''
\be
  a^\mu = (\nabla_T V)^\mu =T^\rho\nabla_\rho V^\mu\ .\label{3.110}
\ee
You should take the names with a grain of salt, but these vectors
are certainly well-defined.

Since $S$ and $T$ are basis vectors adapted to a coordinate system,
their commutator vanishes:
\[
  [S,T]=0\ .
\]
We would like to consider the conventional case where the torsion
vanishes, so from (3.70) we then have
\be
  S^\rho\nabla_\rho T^\mu = T^\rho\nabla_\rho S^\mu \ .\label{3.111}
\ee
With this in mind, let's compute the acceleration:
\bea
  a^\mu &=& T^\rho\nabla_\rho(T^\sigma\nabla_\sigma S^\mu)\cr
  &=& T^\rho\nabla_\rho (S^\sigma\nabla_\sigma T^\mu)\cr
  &=&(T^\rho\nabla_\rho S^\sigma)(\nabla_\sigma T^\mu) +
  T^\rho S^\sigma\nabla_\rho\nabla_\sigma T^\mu\cr
  &=&(S^\rho\nabla_\rho T^\sigma)(\nabla_\sigma T^\mu) +
  T^\rho S^\sigma(\nabla_\sigma\nabla_\rho T^\mu
  +R^\mu{}_{\nu\rho\sigma}T^\nu)\cr
  &=& (S^\rho\nabla_\rho T^\sigma)(\nabla_\sigma T^\mu) +
  S^\sigma\nabla_\sigma(T^\rho\nabla_\rho T^\mu)
  -(S^\sigma\nabla_\sigma T^\rho)\nabla_\rho T^\mu
  +R^\mu{}_{\nu\rho\sigma}T^\nu T^\rho S^\sigma\cr
  &=& R^\mu{}_{\nu\rho\sigma}T^\nu T^\rho S^\sigma\ .
  \label{3.112}
\eea
Let's think about this line by line.  The first line is the definition
of $a^\mu$, and the second line comes directly from (3.111).  The
third line is simply the Leibniz rule.  The fourth line replaces a
double covariant derivative by the derivatives in the opposite order
plus the Riemann tensor.  In the fifth line we use Leibniz again (in
the opposite order from usual), and then we cancel two identical terms 
and notice that the term involving $T^\rho\nabla_\rho T^\mu$ vanishes
because $T^\mu$ is the tangent vector to a geodesic.  The result,
\be
  a^\mu = {{D^2}\over{dt^2}}S^\mu = R^\mu{}_{\nu\rho\sigma}T^\nu 
  T^\rho S^\sigma\ ,\label{3.113}
\ee
is known as the {\bf geodesic deviation equation}.  It expresses
something that we might have expected: the relative acceleration
between two neighboring geodesics is proportional to the curvature.

Physically, of course, the acceleration of neighboring geodesics 
is interpreted as a manifestation of gravitational tidal forces.
This reminds us that we are very close to doing physics by now.

There is one last piece of formalism which it would be nice to cover
before we move on to gravitation proper.  What we will do is to consider
once again (although much more concisely) the formalism of connections
and curvature, but this time we will use sets of basis vectors in the
tangent space which are {\it not} derived from any coordinate system.
It will turn out that this slight change in emphasis reveals a different
point of view on the connection and curvature, one in which the
relationship to gauge theories in particle physics is much more
transparent.  In fact the concepts to be introduced are very
straightforward, but the subject is a notational nightmare, so it
looks more difficult than it really is.

Up until now we have been taking advantage of the fact that a natural
basis for the tangent space $T_p$ at a point $p$ is given by the 
partial derivatives with respect to the coordinates
at that point, $\e\mu = \p\mu$.  Similarly, a
basis for the cotangent space $T^*_p$ is given by the gradients of
the coordinate functions, $\t\mu = \d x^\mu$.  There is nothing to
stop us, however, from setting up any bases we like.  Let us therefore
imagine that at each point in the manifold we introduce a set of
basis vectors $\e{a}$ (indexed by a Latin letter rather than Greek, to
remind us that they are not related to any coordinate system).  We will
choose these basis vectors to be ``orthonormal'', in a sense which is
appropriate to the signature of the manifold we are working on.  That is,
if the canonical form of the metric is written $\eta_{ab}$, we demand
that the inner product of our basis vectors be
\be
  g(\e{a},\e{b}) = \eta_{ab}\ ,\label{3.114}
\ee
where $g(~,~)$ is the usual metric tensor.  Thus, in a Lorentzian 
spacetime $\eta_{ab}$ represents the Minkowski metric, while in a 
space with positive-definite metric it would represent the Euclidean 
metric.  The set of vectors comprising an orthonormal basis 
is sometimes known as a {\bf tetrad} (from Greek {\it tetras}, 
``a group of four'') or {\bf vielbein} (from the German
for ``many legs'').  In different numbers of dimensions it 
occasionally becomes a {\it vierbein} (four), {\it dreibein} (three), 
{\it zweibein} (two), and so on.  (Just as we cannot in general find
coordinate charts which cover the entire manifold, we will often not
be able to find a single set of smooth basis vector fields which are
defined everywhere.  As usual, we can overcome this problem by working
in different patches and making sure things are well-behaved on the
overlaps.)

The point of having a basis is that any vector can be expressed as a
linear combination of basis vectors.  Specifically, we can express our
old basis vectors $\e\mu = \p\mu$ in terms of the new ones:
\be
  \e\mu = e^a_\mu\e{a}\ .\label{3.115}
\ee
The components $e^a_\mu$ form an $n\times n$ invertible matrix.  (In
accord with our usual practice of blurring the distinction between
objects and their components, we will refer to the $e^a_\mu$ as
the tetrad or vielbein, and often in the plural as ``vielbeins.'')  
We denote their inverse by switching indices
to obtain $e^\mu_a$, which satisfy
\be
  e^\mu_a e^a_\nu=\delta^\mu_\nu\ ,\qquad
  e^a_\mu e^\mu_b = \delta^a_b\ .\label{3.116}
\ee
These serve as the components of the vectors $\e{a}$ in the coordinate
basis:
\be
  \e{a} = e^\mu_a \e\mu\ .\label{3.117}
\ee
In terms of the inverse vielbeins, (3.114) becomes
\be
  g_{\mu\nu} e^\mu_a e^\nu_b = \eta_{ab}\ ,\label{3.118}
\ee
or equivalently
\be
  g_{\mu\nu} = e_\mu^a e_\nu^b \eta_{ab}\ .\label{3.119}
\ee
This last equation sometimes leads people to say that the vielbeins
are the ``square root'' of the metric.

We can similarly set up an orthonormal basis of one-forms in
$T^*_p$, which we denote $\t{a}$.  They may be chosen to be compatible
with the basis vectors, in the sense that
\be
  \t{a}(\e{b}) = \delta^a_b\ .\label{3.120}
\ee
It is an immediate consequence of this that the orthonormal one-forms
are related to their coordinate-based cousins $\t\mu = \d x^\mu$ by
\be
  \t\mu = e^\mu_a \t{a}\label{3.121}
\ee
and
\be
  \t{a} = e^a_\mu \t\mu\ .\label{3.122}
\ee
The vielbeins $e^a_\mu$ thus serve double duty as the components of the
coordinate basis vectors in terms of the orthonormal basis vectors, and
as components of the orthonormal basis one-forms in terms of the
coordinate basis one-forms; while the inverse vielbeins serve as the
components of the orthonormal basis vectors in terms of the coordinate
basis, and as components of the coordinate basis one-forms in terms of
the orthonormal basis.

Any other vector can be expressed in terms of its components in the
orthonormal basis.  If a vector $V$ is written in the coordinate
basis as $V^\mu\e\mu$ and in the orthonormal basis as $V^a\e{a}$,
the sets of components will be related by
\be
  V^a = e^a_\mu V^\mu\ .\label{3.123}
\ee
So the vielbeins allow us to ``switch from Latin to Greek indices
and back.''
The nice property of tensors, that there is usually only one
sensible thing to do based on index placement, is of great help here.
We can go on to refer to multi-index tensors in either basis, or even
in terms of mixed components:
\be
  V^a{}_b = e^a_\mu V^\mu{}_b = e^\nu_b V^a{}_\nu =
  e^a_\mu e^\nu_b V^\mu{}_\nu \ .\label{3.124}
\ee
Looking back at (3.118), we see that the components of the metric tensor
in the orthonormal basis are just those of the flat metric,
$\eta_{ab}$.  (For this reason the Greek indices are sometimes referred
to as ``curved'' and the Latin ones as ``flat.'')  In fact we can
go so far as to raise and lower the Latin indices using the flat metric
and its inverse $\eta^{ab}$.  You can check for yourself that everything
works okay ({\it e.g.}, that the lowering an index with the metric
commutes with changing from orthonormal to coordinate bases).
 
By introducing a new set of basis vectors and one-forms, we 
necessitate a return to our favorite topic of transformation properties.
We've been careful all along to emphasize that the tensor transformation
law was only an indirect outcome of a coordinate transformation; the 
real issue was a change of basis.  Now that we have non-coordinate
bases, these bases can be changed independently of the coordinates.
The only restriction is that the orthonormality property (3.114) be
preserved.  But we know what kind of transformations preserve the 
flat metric --- in a Euclidean signature metric they are orthogonal
transformations, while in a Lorentzian signature metric they are
Lorentz transformations.  We therefore consider changes of basis of
the form
\be
  \e{a}\rightarrow \e{a'} = \Lambda_{a'}{}^a(x) \e{a}\ ,\label{3.125}
\ee
where the matrices $\Lambda_{a'}{}^a(x)$ represent position-dependent
transformations which (at each point) leave the canonical form of the
metric unaltered:
\be
  \Lambda_{a'}{}^a\Lambda_{b'}{}^b\eta_{ab} = \eta_{a'b'}\ .
  \label{3.126}
\ee
In fact these matrices correspond to what in flat space we called
the inverse Lorentz transformations (which operate on basis vectors);
as before we also have ordinary Lorentz transformations $\Lambda^{a'}{}_a$,
which transform the basis one-forms.  As far as components are concerned,
as before we transform upper indices with $\Lambda^{a'}{}_a$ and lower
indices with $\Lambda_{a'}{}^a$.

So we now have the freedom to perform a Lorentz transformation (or
an ordinary Euclidean rotation, depending on the signature) at every
point in space.  These transformations are therefore called {\bf local
Lorentz transformations}, or LLT's.  We still have our usual freedom to
make changes in coordinates, which are called {\bf general coordinate
transformations}, or GCT's.  Both can happen at the same time, resulting
in a mixed tensor transformation law:
\be
  T^{a'\mu'}{}_{b'\nu'} =  \Lambda^{a'}{}_a {{\partial x^{\mu'}}\over
  {\partial x^\mu}} \Lambda_{b'}{}^b {{\partial x^{\nu}}\over
  {\partial x^{\nu'}}}T^{a\mu}{}_{b\nu}\ .\label{3.127}
\ee

Translating what we know about tensors into non-coordinate bases is
for the most part merely a matter of sticking vielbeins in the right
places.  The crucial exception comes when we begin to differentiate
things.  In our ordinary formalism, the covariant derivative of
a tensor is given by its partial derivative plus correction terms, one
for each index, involving the tensor and the connection coefficients.
The same procedure will continue to be true for the non-coordinate
basis, but we replace the ordinary connection coefficients 
$\Gamma^\lambda_\mn$ by the {\bf spin connection}, 
denoted $\omega_\mu{}^a{}_b$.  Each Latin index gets
a factor of the spin connection in the usual way:
\be
  \nabla_\mu X^a{}_b=\partial_\mu X^a{}_b+\omega_\mu{}^a{}_cX^c{}_b
  -\omega_\mu{}^c{}_b X^a{}_c\ .\label{3.128}
\ee
(The name ``spin connection'' comes from the fact that this can be
used to take covariant derivatives of spinors, which is actually
impossible using the conventional connection coefficients.)  In the
presence of mixed Latin and Greek indices we get terms of both kinds.

The usual demand that a tensor be independent of the way it is
written allows us to derive a relationship between the spin connection,
the vielbeins, and the $\Gamma^\nu_{\mu\lambda}$'s.
Consider the covariant derivative of a vector $X$,
first in a purely coordinate basis:
\bea
  \nabla X&=& (\nabla_\mu X^\nu)\d x^\mu\otimes 
  \partial_\nu\cr &=&(\partial_\mu X^\nu +\Gamma^\nu_{\mu\lambda}
  X^\lambda)\d x^\mu\otimes \partial_\nu\ . \label{3.129}
\eea
Now find the same object in a mixed basis, and convert into the
coordinate basis:
\bea
  \nabla X&=& (\nabla_\mu X^a)\d x^\mu\otimes 
  \e{a}\cr &=&(\partial_\mu X^a +\omega_\mu{}^a{}_b X^b)\d x^\mu
  \otimes \e{a}\cr
  &=&(\partial_\mu(e^a_\nu X^\nu)+\omega_\mu{}^a{}_b e^b_\lambda 
  X^\lambda)\d x^\mu \otimes(e^\sigma_a \partial_\sigma)\cr
  &=&e^\sigma_a(e^a_\nu \partial_\mu X^\nu +X^\nu \partial_\mu 
  e^a_\nu +\omega_\mu{}^a{}_b e^b_\lambda X^\lambda)\d x^\mu \otimes
  \partial_\sigma\cr
  &=&(\partial_\mu X^\nu+e^\nu_a \partial_\mu e^a_\lambda X^\lambda
  +e^\nu_a e^b_\lambda \omega_\mu{}^a{}_b X^\lambda)\d x^\mu \otimes
  \partial_\nu \ . \label{3.130}
\eea
Comparison with (3.129) reveals
\be
  \Gamma^\nu_{\mu\lambda}=e^\nu_a \partial_\mu e^a_\lambda
  +e^\nu_a e^b_\lambda \omega_\mu{}^a{}_b\ ,\label{3.131}
\ee
or equivalently
\be
  \omega_\mu{}^a{}_b = e^a_\nu e^\lambda_b \Gamma^\nu_{\mu\lambda}
  - e^\lambda_b\p\mu e^a_\lambda\ .\label{3.132}
\ee
A bit of manipulation allows us to write this relation as the
vanishing of the covariant derivative of the vielbein, 
\be
  \nabla_\mu e_\nu^a=0\ ,\label{3.133}
\ee
which is sometimes known as the ``tetrad postulate.''
Note that this is always true; we did not need to assume anything
about the connection in order to derive it.  Specifically, we did not
need to assume that the connection was metric compatible or torsion
free. 

Since the connection may be thought of as something we need to fix
up the transformation law of the covariant derivative, it should
come as no surprise that the spin connection does not itself obey
the tensor transformation law.  Actually, under GCT's the one lower
Greek index does transform in the right way, as a one-form.  But
under LLT's the spin connection transforms inhomogeneously, as
\be
  \omega_\mu{}^{a'}{}_{b'} = \Lambda^{a'}{}_a\Lambda_{b'}{}^b
  \omega_\mu{}^a{}_b - \Lambda_{b'}{}^c\p\mu\Lambda^{a'}{}_c
  \ .\label{3.134}
\ee
You are encouraged to check for yourself that this results in the
proper transformation of the covariant derivative.

So far we have done nothing but empty formalism, translating things
we already knew into a new notation.  But the work we are doing does
buy us two things.  The first, which we already alluded to, is the
ability to describe spinor fields on spacetime and take their
covariant derivatives; we won't explore this further right now.
The second is a change in viewpoint, in which we can think of 
various tensors as tensor-valued differential forms.  For 
example, an object like
$X_\mu{}^a$, which we think of as a $(1,1)$ tensor written with mixed
indices, can also be thought of as a ``vector-valued one-form.''
It has one lower Greek index, so we think of it as a one-form, but
for each value of the lower index it is a vector.  Similarly a
tensor $A_{\mu\nu}{}^{a}{}_b$, antisymmetric in $\mu$ and $\nu$,
can be thought of as a ``$(1,1)$-tensor-valued two-form.''  Thus, any
tensor with some number of antisymmetric lower Greek indices and some
number of Latin indices can be thought of as a differential form, but
taking values in the tensor bundle.  (Ordinary differential forms are 
simply scalar-valued forms.)  The usefulness of this
viewpoint comes when we consider exterior derivatives.  If we want
to think of $X_\mu{}^a$ as a vector-valued one-form, we are tempted to
take its exterior derivative:
\be
  (\d X)_{\mn}{}^a = \p\mu X_\nu{}^a - \p\nu X_\mu{}^a\ .\label{3.135}
\ee
It is easy to check that this object transforms like a two-form (that
is, according to the transformation law for $(0,2)$ tensors) under
GCT's, but not as a vector under LLT's (the Lorentz transformations
depend on position, which introduces an inhomogeneous term into the
transformation law).  But we can fix this by judicious use of the
spin connection, which can be thought of as a one-form.  (Not a
tensor-valued one-form, due to the nontensorial transformation law
(3.134).)  Thus, the object
\be
  (\d X)_{\mn}{}^a +(\omega\wedge X)_{\mn}{}^a
  = \p\mu X_\nu{}^a - \p\nu X_\mu{}^a
  +\omega_\mu{}^a{}_b X_\nu{}^b - \omega_\nu{}^a{}_b X_\mu{}^b 
  \ ,\label{3.136}
\ee
as you can verify at home, transforms as a proper tensor.

An immediate application of this formalism is to the expressions
for the torsion and curvature, the two tensors which characterize
any given connection.  The torsion, with two antisymmetric lower
indices, can be thought of as a vector-valued two-form $T_{\mn}{}^a$.
The curvature, which is always antisymmetric in its last two 
indices, is a $(1,1)$-tensor-valued two-form, $R^a{}_{b\mn}$.
Using our freedom to suppress indices on differential forms, we
can write the defining relations for these two tensors as
\be
  T^a = \d e^a + \omega^a{}_b\wedge e^b\label{3.137}
\ee
and
\be
  R^a{}_b = \d \omega^a{}_b + \omega^a{}_c\wedge\omega^c{}_b\ .
  \label{3.138}
\ee
These are known as the {\bf Maurer-Cartan structure
equations}.  They are equivalent to the usual definitions; let's go
through the exercise of showing this for the torsion, and you can
check the curvature for yourself.  We have
\bea
  T_\mn{}^\lambda &=& e^\lambda_a T_\mn{}^a\cr
  &=& e^\lambda_a(\p\mu e_\nu{}^a - \p\nu e_\mu{}^a
  +\omega_\mu{}^a{}_b e_\nu{}^b - \omega_\nu{}^a{}_b e_\mu{}^b)\cr
  &=& \Gamma^\lambda_{\mn} - \Gamma^\lambda_{\nu\mu}\ ,
  \label{3.139}
\eea
which is just the original definition we gave.  Here we have used
(3.131), the expression for the $\Gamma^\lambda_{\mn}$'s in terms
of the vielbeins and spin connection.  We can also express identities
obeyed by these tensors as
\be
  \d T^a + \omega^a{}_b\wedge T^b = R^a{}_b\wedge e^b\label{3.140}
\ee
and
\be
  \d R^a{}_b + \omega^a{}_c\wedge R^c{}_b - R^a{}_c\wedge
  \omega^c{}_b=0\ .\label{3.141}
\ee
The first of these is the generalization of $R^\rho{}_{[\sigma\mn]}=0$,
while the second is the Bianchi identity $\nabla_{[\lambda|}
R^\rho{}_{\sigma |\mn]}=0$.  (Sometimes both equations are called
Bianchi identities.)

The form of these expressions leads to an almost irresistible
temptation to define a ``covariant-exterior derivative'', which
acts on a tensor-valued form by taking the ordinary exterior
derivative and then adding appropriate terms with the spin 
connection, one for each Latin index.  Although we won't do that
here, it is okay to give in to this temptation, and in fact the
right hand side of (3.137) and the
left hand sides of (3.140) and (3.141) can be thought
of as just such covariant-exterior derivatives.  But be careful, 
since (3.138) cannot; you can't take any sort of covariant derivative
of the spin connection, since it's not a tensor.

So far our equations have been true for general connections; let's
see what we get for the Christoffel connection.  The torsion-free
requirement is just that (3.137) vanish; this does not lead immediately
to any simple statement about the coefficients of the spin connection.
Metric compatibility is expressed as the vanishing of the covariant
derivative of the metric: $\nabla g=0$.  We can see what this leads
to when we express the metric in the orthonormal basis, where its
components are simply $\eta_{ab}$:
\bea
  \nabla_\mu \eta_{ab}&=&\partial_\mu \eta_{ab}
  -\omega_\mu{}^c{}_a \eta_{cb}-\omega_\mu{}^c{}_b \eta_{ac}\cr
  &=&-\omega_{\mu ab}-\omega_{\mu ba}\ . \label{3.142}
\eea
Then setting this equal to zero implies
\be
  \omega_{\mu ab}=-\omega_{\mu ba}\ .\label{3.143}
\ee
Thus, metric compatibility is equivalent to the antisymmetry of the
spin connection in its Latin indices.  (As before, such a statement
is only sensible if both indices are either upstairs or downstairs.)
These two conditions together allow us to express the spin connection 
in terms of the vielbeins.  There is an explicit formula which expresses
this solution, but in practice it is easier to simply solve the
torsion-free condition
\be
  \omega^{ab}\wedge e_b = -\d e^a\ ,\label{3.144}
\ee
using the asymmetry of the spin connection, to find the 
individual components.

We now have the means to compare the formalism of connections and
curvature in Riemannian geometry to that of gauge theories in 
particle physics.  (This is an aside, which is hopefully comprehensible
to everybody, but not an essential ingredient of the course.)
In both situations, the fields of interest live
in vector spaces which are assigned to each point in spacetime.
In Riemannian geometry the vector spaces include the tangent space,
the cotangent space, and the higher tensor spaces constructed from
these.  In gauge theories, on the other hand, we are concerned with
``internal'' vector spaces.  The distinction is that the tangent
space and its relatives are intimately associated with the manifold
itself, and were naturally defined once the manifold was set up;
an internal vector space can be of any dimension we like, and has to
be defined as an independent addition to the manifold.  In math lingo,
the union of the base manifold with the internal vector spaces (defined
at each point) is a {\bf fiber bundle}, and each copy of the vector
space is called the ``fiber'' (in perfect accord with our definition
of the tangent bundle).

Besides the base manifold (for us, spacetime) and the fibers, the other
important ingredient in the definition of a fiber bundle is the
``structure group,'' a Lie group which acts on the fibers to describe 
how they are sewn together on overlapping coordinate patches.  Without
going into details, the structure group for the tangent bundle in a
four-dimensional spacetime is generally GL$(4,\R)$, the group of
real invertible $4\times 4$ matrices; if we have a Lorentzian metric,
this may be reduced to the Lorentz group SO$(3,1)$.  Now imagine that
we introduce an internal three-dimensional vector space, and sew the
fibers together with ordinary rotations; the structure group of
this new bundle is then SO$(3)$.  A field that lives in this 
bundle might be denoted $\phi^A(x^\mu)$, where $A$ runs from one
to three; it is a three-vector (an internal one, unrelated to
spacetime) for each point on the manifold.  We have freedom to choose
the basis in the fibers in any way we wish; this means that ``physical
quantities'' should be left invariant under local SO(3) transformations
such as 
\be
  \phi^A(x^\mu)\rightarrow \phi^{A'}(x^\mu)=
  O^{A'}{}_A(x^\mu)\phi^A(x^\mu)\ ,\label{3.145}
\ee
where $O^{A'}{}_A(x^\mu)$ is a matrix in SO(3) which depends on
spacetime.  Such transformations are known as {\bf gauge transformations},
and theories invariant under them are called ``gauge theories.''

For the most part it is not hard to arrange things such that physical
quantities are invariant under gauge transformations.  The one
difficulty arises when we consider partial derivatives, $\p\mu \phi^A$.
Because the matrix $O^{A'}{}_A(x^\mu)$ depends on spacetime, it will 
contribute an unwanted term to the transformation of the partial derivative.
By now you should be able to guess the solution: introduce a connection
to correct for the inhomogeneous term in the transformation law.  We
therefore define a connection on the fiber bundle to be an object
$A_\mu{}^A{}_B$, with two ``group indices'' and one spacetime index.
Under GCT's it transforms as a one-form, while under gauge transformations
it transforms as
\be
  A_\mu{}^{A'}{}_{B'} = O^{A'}{}_A O_{B'}{}^B A_\mu{}^A{}_B
  - O_{B'}{}^C\p\mu O^{A'}{}_C\ .\label{3.146}
\ee
(Beware: our conventions are so drastically different from those
in the particle physics literature that I won't even try to get
them straight.)  With this transformation law, the ``gauge covariant
derivative''
\be
  D_\mu \phi^A = \p\mu\phi^A + A_\mu{}^A{}_B\phi^B\label{3.147}
\ee
transforms ``tensorially'' under gauge transformations, as you are
welcome to check.  (In ordinary electromagnetism the connection is
just the conventional vector potential.  No indices are necessary,
because the structure group U(1) is one-dimensional.)

It is clear that this notion of a connection on an internal fiber
bundle is very closely related to the connection on the tangent bundle,
especially in the orthonormal-frame picture we have been discussing.
The transformation law (3.146), for example, is exactly the same as
the transformation law (3.134) for the spin connection.  We can also
define a curvature or ``field strength'' tensor which is a two-form,
\be
  F^A{}_B = \d A^A{}_B + A^A{}_C\wedge A^C{}_B\ ,\label{3.148}
\ee    
in exact correspondence with (3.138).  We can parallel transport things
along paths, and there is a construction analogous to the
parallel propagator; the trace of the matrix obtained by parallel
transporting a vector around a closed curve is called a ``Wilson
loop.''

We could go on in the development of the relationship between the
tangent bundle and internal vector bundles, but time is short and we
have other fish to fry.  Let us instead finish by emphasizing the
important {\it difference} between the two constructions.  The
difference stems from the fact that the tangent bundle is closely
related to the base manifold, while other fiber bundles are tacked
on after the fact.  It makes sense to say that a vector in the
tangent space at $p$ ``points along a path'' through $p$; but this
makes no sense for an internal vector bundle.  There is therefore
no analogue of the coordinate basis for an internal space ---
partial derivatives along curves have nothing to do with internal
vectors.  It follows in turn that there is nothing like the vielbeins,
which relate orthonormal bases to coordinate bases.  The torsion tensor,
in particular, is only defined for a connection on the tangent bundle,
not for any gauge theory connections; it can be thought of as the
covariant exterior derivative of the vielbein, and no such construction
is available on an internal bundle.  You should appreciate the 
relationship between the different uses of the notion of a connection,
without getting carried away.

\eject
\thispagestyle{plain}

\setcounter{equation}{0}

\noindent{December 1997 \hfill {\sl Lecture Notes on General Relativity}
\hfill{Sean M.~Carroll}}

\vskip .2in

\setcounter{section}{3}
\section{Gravitation}

Having paid our mathematical dues, we are now prepared to examine
the physics of gravitation as described by general relativity.  This
subject falls naturally into two pieces:  how the curvature of
spacetime acts on matter to manifest itself as ``gravity'', and 
how energy and momentum influence spacetime to create curvature.
In either case it would be legitimate to start at the top, by stating
outright the laws governing physics in curved spacetime and working
out their consequences.  Instead, we will try to be a little more
motivational, starting with basic physical principles and attempting
to argue that these lead naturally to an almost unique physical 
theory.

The most basic of these physical principles is the Principle of
Equivalence, which comes in a variety of forms.  The earliest form
dates from Galileo and Newton, and is known as the {\bf Weak
Equivalence Principle}, or WEP.  The WEP states that the ``inertial
mass'' and ``gravitational mass'' of any object are equal.  To see
what this means, think about Newton's Second Law.  This relates
the force exerted on an object to the acceleration it undergoes,
setting them proportional to each other with the constant of 
proportionality being the inertial mass $m_i$:
\be
  {\bf f} = m_i {\bf a}\ .\label{4.1}
\ee
The inertial mass clearly has a universal character, related to the
resistance you feel when you try to push on the object; it is the
same constant no matter what kind of force is being exerted.  We also
have the law of gravitation, which states that the gravitational
force exerted on an object is proportional to the gradient of a scalar
field $\Phi$, known as the gravitational potential.  The constant of
proportionality in this case is called the gravitational mass $m_g$:
\be
  {\bf f}_g = -m_g \nabla\Phi\ .\label{4.2}
\ee
On the face of it, $m_g$ has a very different character than $m_i$;
it is a quantity specific to the gravitational force.  If you like, it
is the ``gravitational charge'' of the body.  Nevertheless, Galileo
long ago showed (apocryphally by dropping weights off of the Leaning
Tower of Pisa, actually by rolling balls down inclined planes) that
the response of matter to gravitation was universal --- every object
falls at the same rate in a gravitational field, independent of
the composition of the object.  In Newtonian mechanics this 
translates into the WEP, which is simply
\be
  m_i = m_g\label{4.3}
\ee
for any object.  An immediate consequence is that the behavior of
freely-falling test particles is universal, independent of their mass
(or any other qualities they may have); in fact we have
\be
  {\bf a}= - \nabla\Phi\ .\label{4.4}
\ee

The universality of gravitation, as implied by the WEP, can be stated
in another, more popular, form.  Imagine that we consider a physicist
in a tightly sealed box, unable to observe the outside world, who is
doing experiments involving the motion of test particles, for example
to measure the local gravitational field.  Of course she would obtain
different answers if the box were sitting on the moon or
on Jupiter than she would on the Earth.  But the answers would also
be different if the box were accelerating at a constant velocity;
this would change the acceleration of the freely-falling particles
with respect to the box.  The WEP implies that there is no way to
disentangle the effects of a gravitational field from those of being
in a uniformly accelerating frame, simply by observing the behavior
of freely-falling particles.  This follows from the universality of
gravitation; it would be possible to distinguish between uniform
acceleration and an electromagnetic field, by observing the behavior
of particles with different charges.  But with gravity it is 
impossible, since the ``charge'' is necessarily proportional to the
(inertial) mass.

To be careful, we should limit our claims about the impossibility of
distinguishing gravity from uniform acceleration by restricting our
attention to ``small enough regions of spacetime.''  If the sealed
box were sufficiently big, the gravitational field would change from
place to place in an observable way, while the effect of acceleration
is always in the same direction.  In a rocket ship or elevator, the
particles always fall straight down:

\begin{figure}[h]
  \centerline{
  \psfig{figure=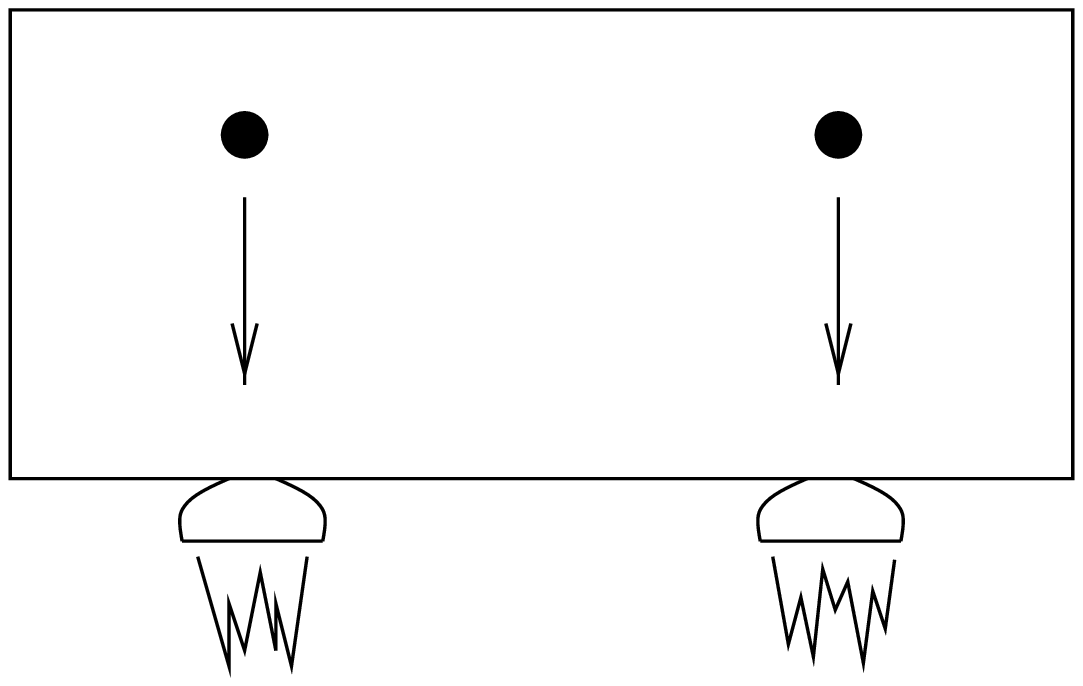,angle=0,height=6cm}}
\end{figure}

\noindent In a very big box in a gravitational field, however, the
particles will move toward the center of the Earth (for example), which
might be a different direction in different regions:

\eject

\begin{figure}
  \centerline{
  \psfig{figure=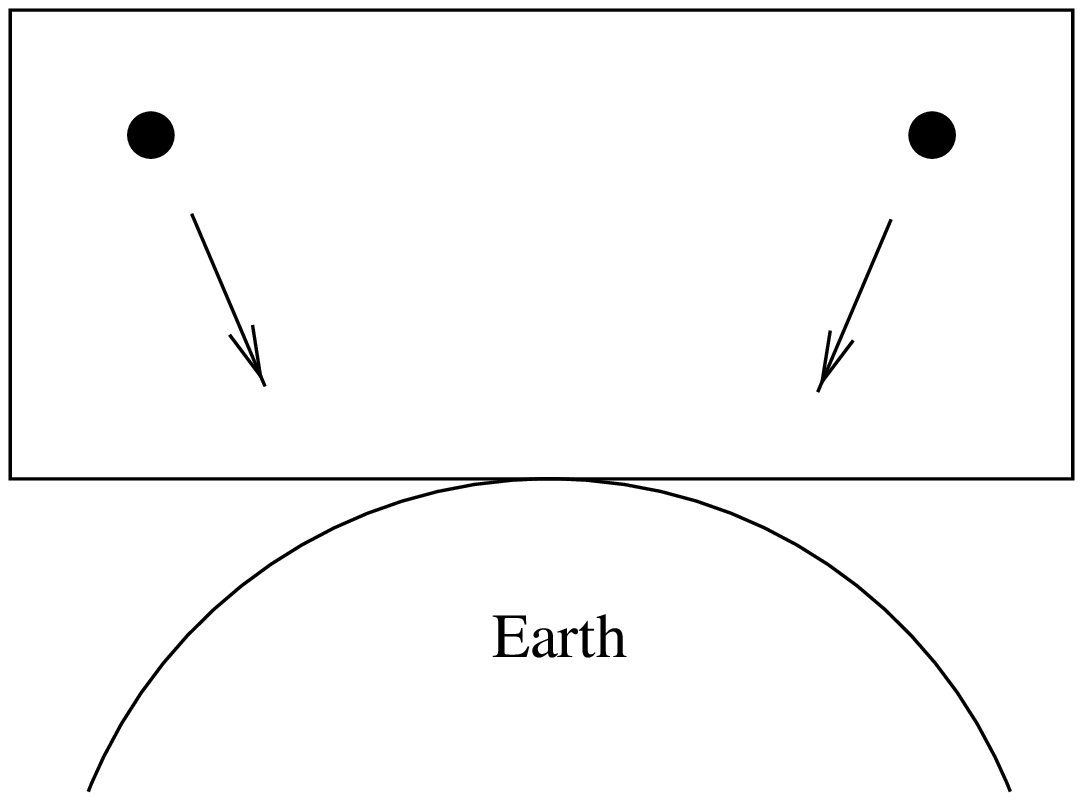,angle=0,height=6cm}}
\end{figure}

\noindent The WEP can therefore be stated as ``the laws of freely-falling
particles are the same in a gravitational field and a uniformly
accelerated frame, in small enough regions of spacetime.''  In larger
regions of spacetime there will be inhomogeneities in the gravitational
field, which will lead to tidal forces which can be detected.

After the advent of special relativity, the concept of mass lost
some of its uniqueness, as it became clear that mass was simply a
manifestation of energy and momentum ($E=mc^2$ and all that).  It
was therefore natural for Einstein to think about generalizing the
WEP to something more inclusive.  His idea was simply that there
should be no way whatsoever for the physicist in the box to distinguish
between uniform acceleration and an external gravitational field, no
matter what experiments she did (not only by dropping test particles).
This reasonable extrapolation became what is now known as the 
{\bf Einstein Equivalence Principle}, or EEP: ``In small enough
regions of spacetime, the laws of physics reduce to those of special
relativity; it is impossible to detect the existence of a gravitational
field.''

In fact, it is hard to imagine theories which respect the WEP but
violate the EEP.  Consider a hydrogen atom, a bound state of a proton
and an electron.  Its mass is actually less than the sum of the masses
of the proton and electron considered individually, because there is
a negative binding energy --- you have to put energy into the atom
to separate the proton and electron.  According to the WEP, the 
gravitational mass of the hydrogen atom is therefore less than the
sum of the masses of its constituents; the gravitational field couples
to electromagnetism (which holds the atom together) in exactly the
right way to make the gravitational mass come out right.  This means
that not only must gravity couple to rest mass universally, but to
all forms of energy and momentum --- which is practically the claim of 
the EEP.  It is possible to come up with counterexamples, however; for
example, we could imagine a theory of gravity in which freely falling 
particles began to rotate as they moved through a gravitational field.  
Then they could fall along the same paths as they would in an accelerated 
frame (thereby satisfying the WEP), but you could nevertheless detect the
existence of the gravitational field (in violation of the EEP).
Such theories seem contrived, but there is no law of nature which
forbids them.

Sometimes a distinction is drawn between ``gravitational laws of
physics'' and ``non-gravitational laws of physics,'' and the EEP
is defined to apply only to the latter.  Then one defines the
``Strong Equivalence Principle'' (SEP) to include all of the laws of
physics, gravitational and otherwise.  I don't find this a particularly
useful distinction, and won't belabor it.  For our purposes, the
EEP (or simply ``the principle of equivalence'') includes all of the
laws of physics.

It is the EEP which implies (or at least suggests) that we should
attribute the action of gravity to the curvature of spacetime.  
Remember that in special relativity a prominent role is played by
inertial frames --- while it was not possible to single out some
frame of reference as uniquely ``at rest'', it was possible to single
out a family of frames which were ``unaccelerated'' (inertial).
The acceleration of a charged particle in an electromagnetic field
was therefore uniquely defined with respect to these frames.  The
EEP, on the other hand, implies that gravity is inescapable --- there
is no such thing as a ``gravitationally neutral object'' with respect 
to which we can measure the acceleration due to gravity.  It follows
that ``the acceleration due to gravity'' is not something which can
be reliably defined, and therefore is of little use.  

Instead, it makes more sense to {\it define} ``unaccelerated'' as
``freely falling,'' and that is what we shall do.  This point of 
view is the origin of the idea that gravity is not a ``force'' ---
a force is something which leads to acceleration, and our definition
of zero acceleration is ``moving freely in the presence of whatever
gravitational field happens to be around.''

This seemingly innocuous step has profound implications for the nature
of spacetime.  In SR, we had a procedure for starting at some point
and constructing an inertial frame which stretched throughout spacetime,
by joining together rigid rods and attaching clocks to them.  But, again
due to inhomogeneities in the gravitational field, this is no longer 
possible.  If we start in some freely-falling state and build a large
structure out of rigid rods, at some distance away freely-falling
objects will look like they are ``accelerating'' with respect to this
reference frame, as shown in the figure on the next page.

\eject

\begin{figure}
  \centerline{
  \psfig{figure=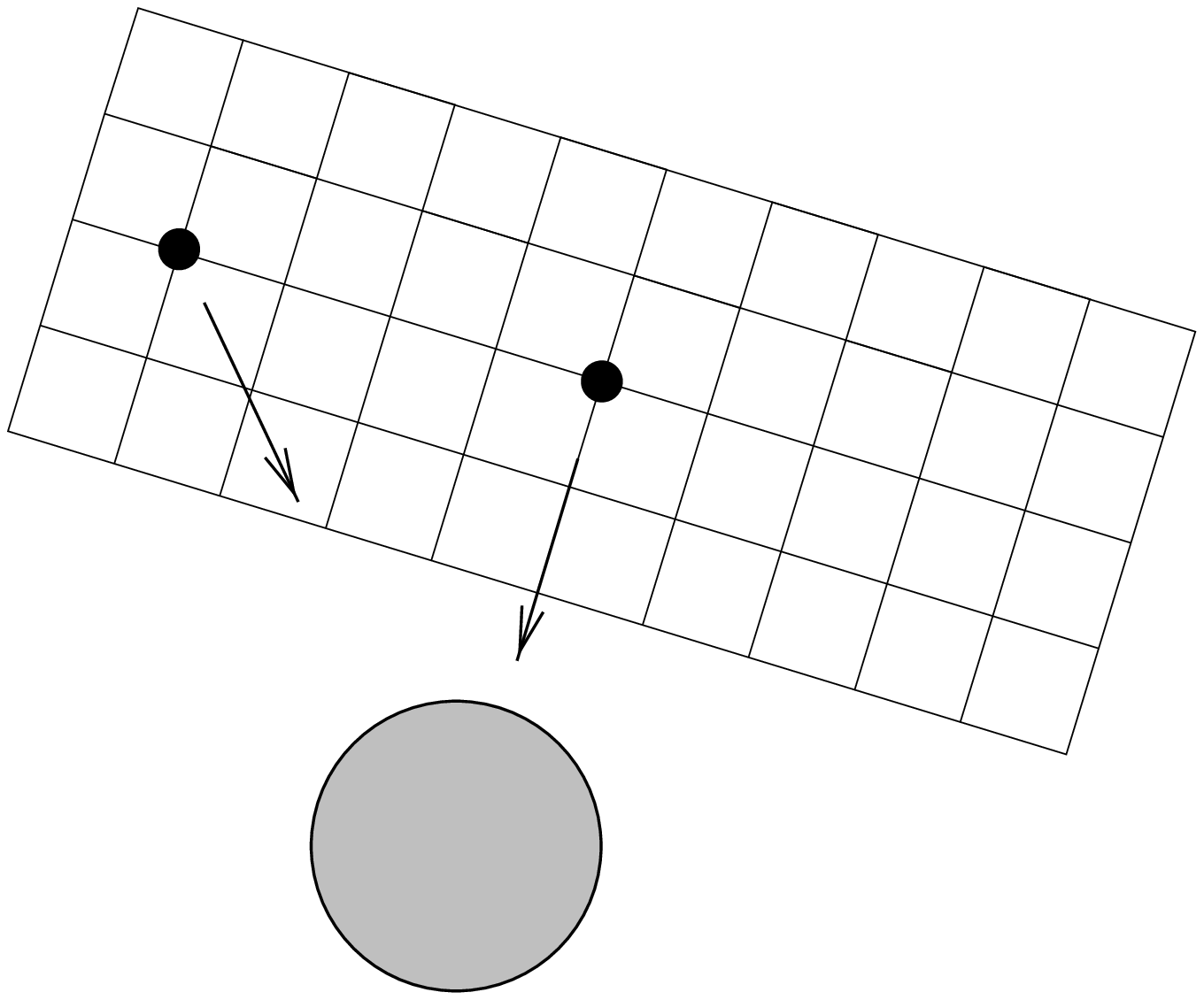,angle=0,height=6cm}}
\end{figure}

The solution is to retain the notion of inertial frames, but to 
discard the hope that they can be uniquely extended throughout space
and time.  Instead we can define {\bf locally inertial frames}, those
which follow the motion of freely falling particles in small enough
regions of spacetime.  (Every time we say ``small enough regions'',
purists should imagine a limiting procedure in which we take the appropriate
spacetime volume to zero.)  This is the best we can do, but it forces
us to give up a good deal.  For example, we can no longer speak with
confidence about the relative velocity of far away objects, since the
inertial reference frames appropriate to those objects are independent
of those appropriate to us.

So far we have been talking strictly about physics, without jumping
to the conclusion that spacetime should be described as a curved manifold.
It should be clear, however, why such a conclusion is appropriate.
The idea that the laws of special relativity should be obeyed in
sufficiently small regions of spacetime, and further that local inertial 
frames can be established in such regions, corresponds to our ability
to construct Riemann normal coordinates at any one point on a manifold ---
coordinates in which the metric takes its canonical form and the Christoffel 
symbols vanish.  The impossibility of comparing velocities (vectors) at 
widely separated regions corresponds to the path-dependence of 
parallel transport on a curved manifold.  These considerations were
enough to give Einstein the idea that gravity was a manifestation of
spacetime curvature.  But in fact we can be even more persuasive.
(It is impossible to ``prove'' that gravity should be thought of as
spacetime curvature, since scientific hypotheses can only be falsified,
never verified [and not even really falsified, as Thomas Kuhn has famously
argued].  But there is nothing to be dissatisfied with about convincing
plausibility arguments, if they lead to empirically successful theories.)

Let's consider one of the celebrated predictions of the EEP, the
gravitational redshift.  Consider two boxes, a distance $z$ apart,
moving (far away from any matter, so we assume in the absence of any
gravitational field) with some constant acceleration $a$.  At
time $t_0$ the trailing box emits a photon of wavelength $\lambda_0$.

\eject

\begin{figure}
  \centerline{
  \psfig{figure=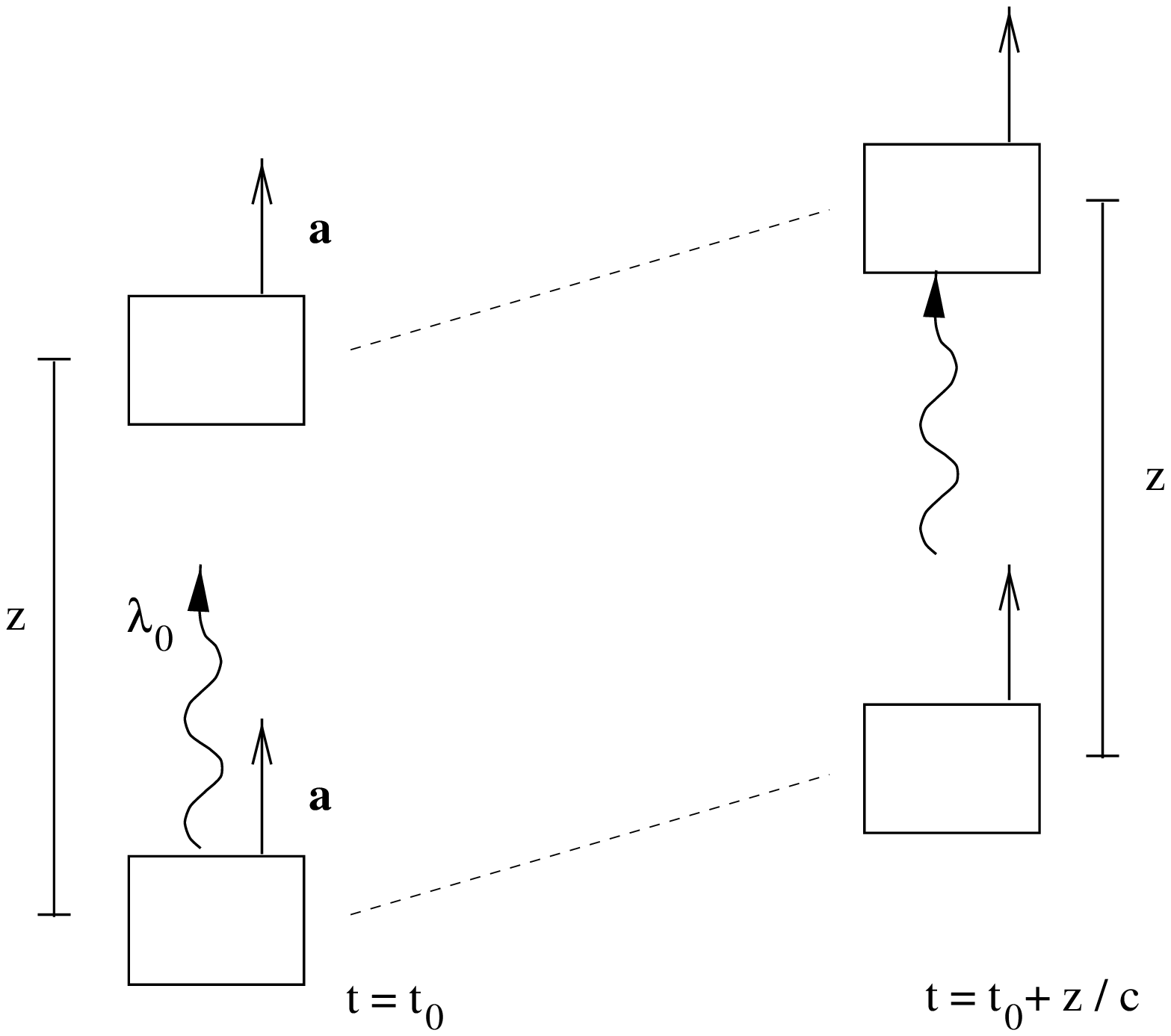,angle=0,height=7cm}}
\end{figure}

The boxes remain a constant distance apart, so the photon reaches
the leading box after a time $\Delta t = z/c$ in the reference frame
of the boxes.  In this time the boxes will have picked up an additional
velocity $\Delta v = a\Delta t = az/c$.  Therefore, the photon reaching 
the lead box will be redshifted by the conventional Doppler effect by
an amount
\be
  {{\Delta \lambda}\over {\lambda_0}}={{\Delta v}\over c} =
  {{az}\over{c^2}}\ .\label{4.5}
\ee
(We assume $\Delta v/c$ is small, so we only work to first order.)
According to the EEP, the same thing should happen in a uniform
gravitational field.  So we imagine a tower of height $z$ sitting
on the surface of a planet, with $a_g$ the strength of the gravitational
field (what Newton would have called the ``acceleration due to gravity'').

\begin{figure}[h]
  \centerline{
  \psfig{figure=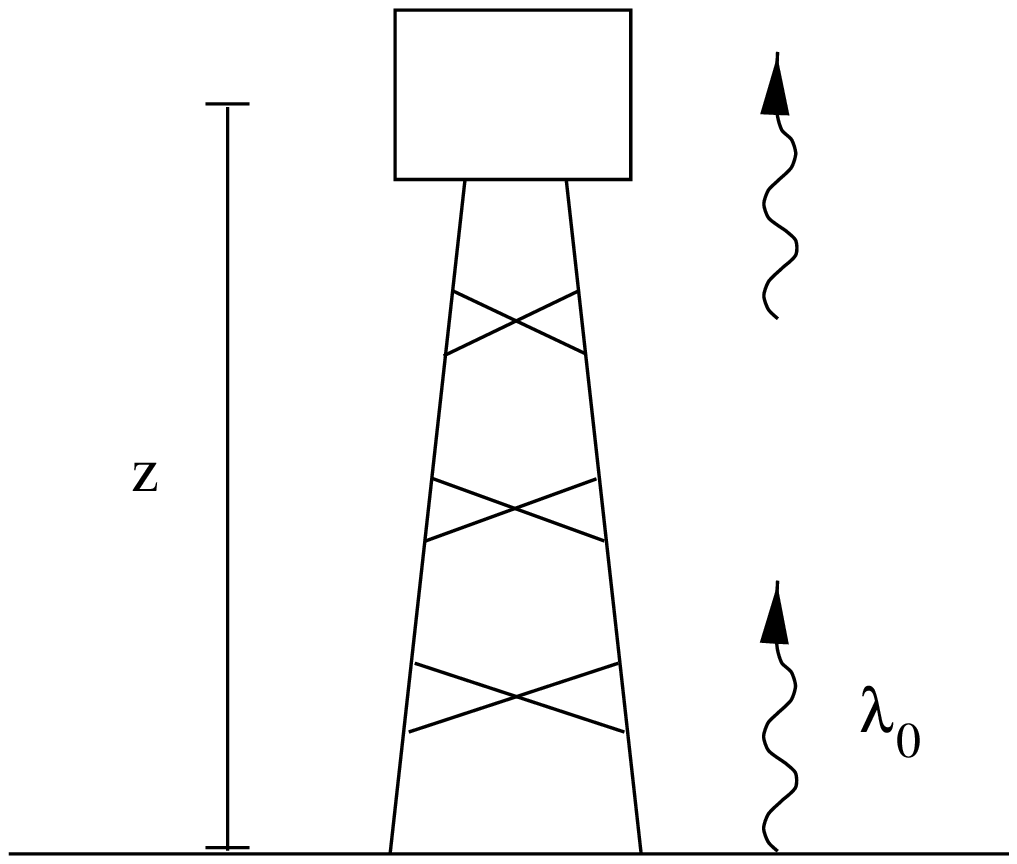,angle=0,height=6cm}}
\end{figure}

\noindent This situation is supposed to be indistinguishable from the 
previous one, from the point of view of an observer in a box at the top of 
the tower (able to detect the emitted photon, but otherwise unable
to look outside the box).  Therefore, a photon emitted from the
ground with wavelength $\lambda_0$ should be redshifted by an amount
\be
  {{\Delta \lambda}\over {\lambda_0}}={{a_gz}\over{c^2}}\ .\label{4.6}
\ee
This is the famous gravitational redshift.  Notice that it is a
direct consequence of the EEP, not of the details of general 
relativity.  It has been verified experimentally, first by Pound
and Rebka in 1960.  They used the M\"ossbauer effect to measure the
change in frequency in $\gamma$-rays as they traveled from the ground to
the top of Jefferson Labs at Harvard.

The formula for the redshift is more often stated in terms of the
Newtonian potential $\Phi$, where ${\bf a}_g = \nabla\Phi$.
(The sign is changed with respect to the usual convention, since
we are thinking of ${\bf a}_g$ as the acceleration of the reference
frame, not of a particle with respect to this reference frame.)
A non-constant gradient of $\Phi$ is like a time-varying
acceleration, and the equivalent net velocity is given by integrating
over the time between emission and absorption of the photon.  We
then have
\bea
  {{\Delta \lambda}\over {\lambda_0}} &=& 
  {1\over {c}}\int \nabla\Phi\ dt \cr
  &=&  {1\over {c^2}}\int \p{z}\Phi\ dz \cr
  &=&  \Delta\Phi\ , \label{4.7}
\eea
where $\Delta\Phi$ is the total change in the gravitational potential,
and we have once again set $c=1$.  This simple formula for the
gravitational redshift continues to be true in more general
circumstances.  Of course, by using the Newtonian potential at all,
we are restricting our domain of validity to weak gravitational
fields, but that is usually completely justified for observable
effects.

The gravitational redshift leads to another argument that we should
consider spacetime as curved.  Consider the same experimental setup
that we had before, now portrayed on the spacetime diagram on the
next page.

\begin{figure}
  \centerline{
  \psfig{figure=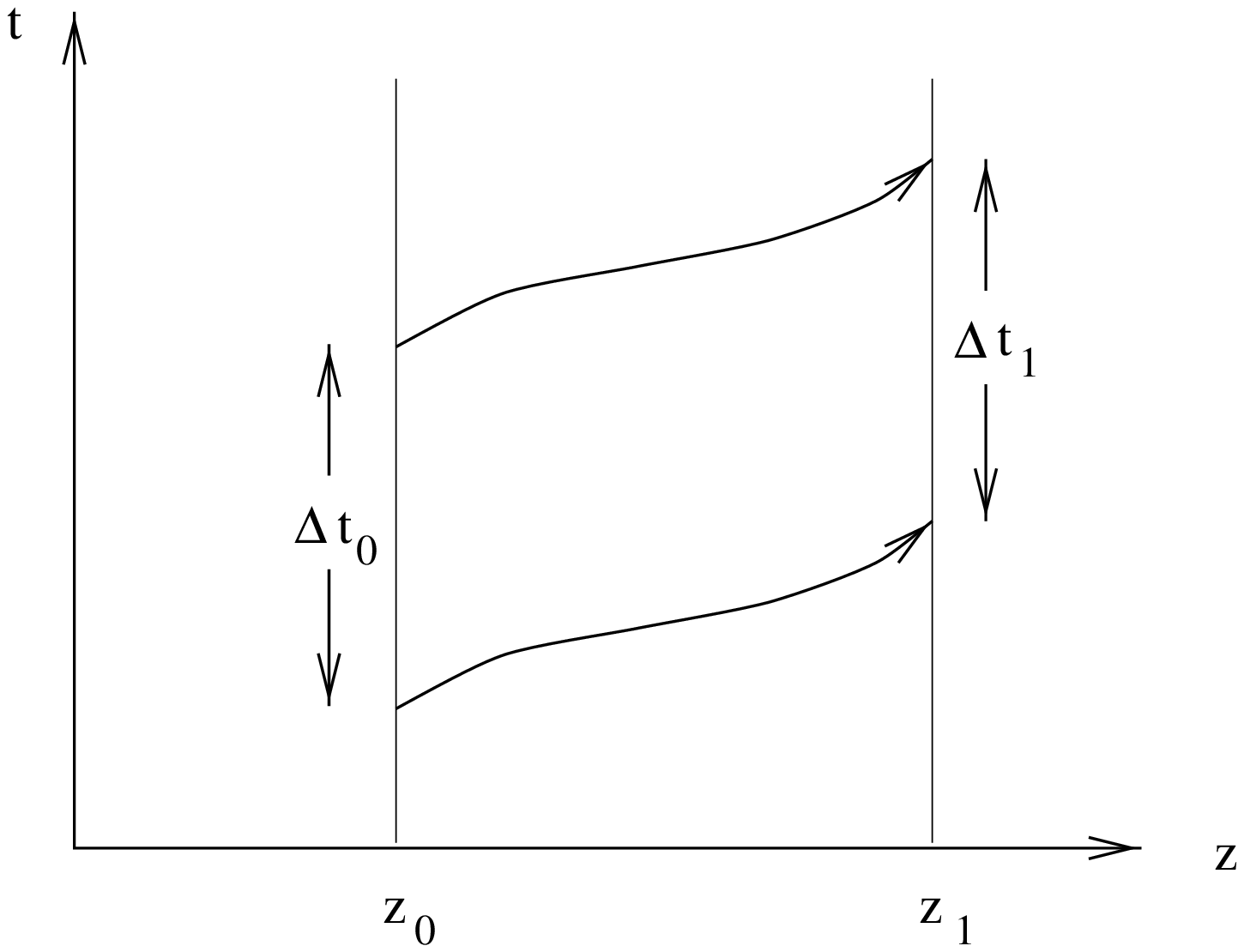,angle=0,height=6cm}}
\end{figure}

The physicist on the ground emits a beam of light with wavelength
$\lambda_0$ from a height $z_0$, which travels to the top of the
tower at height $z_1$.  The time between when the beginning of any
single wavelength of the light is emitted and the end of that same 
wavelength is emitted is $\Delta t_0 = \lambda_0/c$, and the same time 
interval for the absorption is $\Delta t_1=\lambda_1/c$.  Since we imagine 
that the gravitational field is not varying with time, the paths through
spacetime followed by the leading and trailing edge of the single
wave must be precisely congruent.  (They are represented by some
generic curved paths, since we do not pretend that we know just what
the paths will be.)  Simple geometry tells us that the times
$\Delta t_0$ and $\Delta t_1$ must be the same.  But of course they
are not; the gravitational redshift implies that $\Delta t_1 >
\Delta t_0$.  (Which we can interpret as ``the clock on the tower
appears to run more quickly.'')  The fault lies with ``simple
geometry''; a better description of what happens is to imagine that
spacetime is curved.

All of this should constitute more than enough motivation for our
claim that, in the presence of gravity, spacetime should be thought
of as a curved manifold.  Let us now take this to be true and begin
to set up how physics works in a curved spacetime.  The principle
of equivalence tells us that the laws of physics, in small enough
regions of spacetime, look like those of special relativity.  We
interpret this in the language of manifolds as the statement that
these laws, when written in Riemannian normal coordinates $x^\mu$
based at some point $p$, are described by equations which take the
same form as they would in flat space.  The simplest example is
that of freely-falling (unaccelerated) particles.  In flat space
such particles move in straight lines; in equations, this is 
expressed as the vanishing of the second derivative of the parameterized
path $x^\mu(\lambda)$:
\be
  {{d^2 x^\mu}\over{d\lambda^2}} = 0\ .\label{4.8}
\ee
According to the EEP, exactly this equation should hold in
curved space, as long as the coordinates $x^\mu$ are RNC's.  What
about some other coordinate system?  As it stands, (4.8) is
not an equation between tensors.  However, there is a unique 
tensorial equation which reduces to (4.8) when the Christoffel
symbols vanish; it is
\be
  {{d^2 x^\mu}\over{d\lambda^2}}+\Gamma^\mu_{\rho\sigma}
  {{d x^\rho}\over{d\lambda}}{{d x^\sigma}\over{d\lambda}} = 0\ .
  \label{4.9}
\ee
Of course, this is simply the geodesic equation.  In general relativity,
therefore, free particles move along geodesics; we have mentioned this
before, but now you know why it is true.

As far as free particles go, we have argued that curvature of
spacetime is necessary to describe gravity; we have not yet shown that
it is sufficient.  To do so, we can show how the usual results of
Newtonian gravity fit into the picture.  We define the ``Newtonian
limit'' by three requirements: the particles are moving slowly
(with respect to the speed of light), the gravitational field is
weak (can be considered a perturbation of flat space), and the field
is also static (unchanging with time).  Let us see what these
assumptions do to the geodesic equation, taking the proper time
$\tau$ as an affine parameter.  ``Moving slowly'' means that
\be
  {{dx^i}\over{d\tau}}<<{{dt}\over{d\tau}}\ ,\label{4.10}
\ee
so the geodesic equation becomes
\be
  {{d^2 x^\mu}\over{d\tau^2}}+\Gamma^\mu_{00}
  \left({{d t}\over{d\tau}}\right)^2 = 0\ .
  \label{4.11}
\ee
Since the field is static, the relevant Christoffel symbols 
$\Gamma^\mu_{00}$ simplify:
\bea
  \Gamma^\mu_{00}&=& {1\over 2} g^{\mu\lambda}
  (\p0 g_{\lambda 0} + \p0 g_{0 \lambda} - \p\lambda g_{00})\cr
  &=&  - {1\over 2} g^{\mu\lambda}\p\lambda g_{00}\ . \label{4.12}
\eea
Finally, the weakness of the gravitational field allows us to
decompose the metric into the Minkowski form plus a small
perturbation:
\be
  g_\mn = \eta_\mn + h_\mn\ ,\qquad |h_\mn |<<1\ .\label{4.13}
\ee
(We are working in Cartesian coordinates, so $\eta_\mn$ is the
canonical form of the metric.  The ``smallness condition'' on
the metric perturbation $h_\mn$ doesn't really make sense in
other coordinates.)  From the definition of the inverse metric,
$g^\mn g_{\nu\sigma}=\delta^\mu_\sigma$, we find that to first
order in $h$,
\be
  g^\mn = \eta^\mn - h^\mn\ ,\label{4.14}
\ee
where $h^\mn = \eta^{\mu\rho}\eta^{\nu\sigma}h_{\rho\sigma}$. In 
fact, we can use the Minkowski metric to raise and lower indices on
an object of any definite order in $h$, since the corrections would
only contribute at higher orders.

Putting it all together, we find
\be
  \Gamma^\mu_{00}= - {1\over 2} \eta^{\mu\lambda}\p\lambda h_{00}
  \ .\label{4.15}
\ee
The geodesic equation (4.11) is therefore
\be
  {{d^2 x^\mu}\over{d\tau^2}}={1\over 2} \eta^{\mu\lambda}
  \p\lambda h_{00} \left({{d t}\over{d\tau}}\right)^2\ .\label{4.16}
\ee
Using $\p0 h_{00}=0$, the $\mu=0$ component of this is just
\be
  {{d^2 t}\over{d\tau^2}}=0\ .\label{4.17}
\ee
That is, ${{dt}\over{d\tau}}$ is constant.  To examine the spacelike
components of (4.16), recall that the spacelike components of 
$\eta^\mn$ are just those of a $3\times 3$ identity matrix.  We
therefore have
\be
  {{d^2 x^i}\over{d\tau^2}}={1\over 2}\left({{d t}\over{d\tau}}
  \right)^2 \p{i} h_{00} \ .\label{4.18}
\ee
Dividing both sides by $\left({{d t}\over{d\tau}}\right)^2$ has the
effect of converting the derivative on the left-hand side
from $\tau$ to $t$, leaving us with
\be
  {{d^2 x^i}\over{d t^2}}={1\over 2}\p{i} h_{00} \ .\label{4.19}
\ee
This begins to look a great deal like Newton's theory of gravitation.
In fact, if we compare this equation to (4.4), we find that they
are the same once we identify
\be
  h_{00} = -2\Phi\ ,\label{4.20}
\ee
or in other words
\be
  g_{00} = -(1+2\Phi)\ .\label{4.21}
\ee
Therefore, we have shown that the curvature of spacetime is indeed
sufficient to describe gravity in the Newtonian limit, as long as
the metric takes the form (4.21).  It remains, of course, to find
field equations for the metric which imply that this is the form
taken, and that for a single gravitating body we recover the 
Newtonian formula
\be
  \Phi = -{{GM}\over r}\ ,\label{4.22}
\ee
but that will come soon enough.

Our next task is to show how the remaining laws of physics, beyond 
those governing freely-falling particles, adapt to the curvature
of spacetime.  The procedure essentially follows the paradigm
established in arguing that free particles move along geodesics.
Take a law of physics in flat space, traditionally written in terms
of partial derivatives and the flat metric.  According to the 
equivalence principle this law will hold in the presence of gravity,
as long as we are in Riemannian normal coordinates.  Translate the
law into a relationship between tensors; for example, change
partial derivatives to covariant ones.  In RNC's this version of the 
law will reduce to the flat-space one,
but tensors are coordinate-independent objects, so the tensorial
version must hold in any coordinate system.

This procedure is sometimes given a name, the {\bf Principle of
Covariance}.  I'm not sure that it deserves its own name, since
it's really a consequence of the EEP plus the requirement that 
the laws of physics be independent of coordinates.  (The requirement
that laws of physics be independent of coordinates is essentially
impossible to even imagine being untrue.  Given some experiment,
if one person uses one coordinate system to predict a result and
another one uses a different coordinate system, they had better agree.)
Another name is the ``comma-goes-to-semicolon rule'', since at
a typographical level the thing you have to do is replace partial
derivatives (commas) with covariant ones (semicolons).

We have already implicitly used the principle of covariance (or
whatever you want to call it) in deriving the statement that free
particles move along geodesics.  For the most part, it is very simple
to apply it to interesting cases.  Consider for example the formula
for conservation of energy in flat spacetime, $\p\mu T^\mn =0$.
The adaptation to curved spacetime is immediate:
\be
  \nabla_\mu T^\mn =0\ .\label{4.23}
\ee
This equation expresses the conservation of energy in the presence
of a gravitational field.

Unfortunately, life is not always so easy.  Consider Maxwell's 
equations in special relativity, where it would seem that the principle
of covariance can be applied in a straightforward way.  The
inhomogeneous equation $\p\mu F^{\nu\mu} = 4\pi J^\nu$ becomes
\be
  \nabla_\mu F^{\nu\mu} = 4\pi J^\nu\ ,\label{4.24}
\ee
and the homogeneous one $\p{[\mu} F_{\nu\lambda]}= 0$ becomes
\be
  \nabla_{[\mu} F_{\nu\lambda]}= 0\ .\label{4.25}
\ee
On the other hand, we could also write Maxwell's equations in
flat space in terms of differential forms as
\be
  \d(*F) = 4\pi(* J)\ ,\label{4.26}
\ee
and
\be
  \d F = 0\ .\label{4.27}
\ee
These are already in perfectly tensorial form, since we have shown
that the exterior derivative is a well-defined tensor operator regardless
of what the connection is.  We therefore begin to worry a little bit;
what is the guarantee that the process of writing a law of physics in
tensorial form gives a unique answer?  In fact, as we have mentioned
earlier, the differential forms versions of Maxwell's equations should
be taken as fundamental.  Nevertheless, in this case it happens to make no 
difference, since in the absence of torsion (4.26) is identical to (4.24), 
and (4.27) is identical to (4.25); the symmetric part of the connection
doesn't contribute.  Similarly, the definition of the field strength tensor 
in terms of the potential $A_\mu$ can be written either as
\be
  F_{\mn} = \nabla_\mu A_\nu - \nabla_\nu A_\mu\ ,\label{4.28}
\ee
or equally well as
\be
  F=\d A\ .\label{4.29}
\ee

The worry about uniqueness is a real one, however.  Imagine that
two vector fields $X^\mu$ and $Y^\nu$ obey a law in flat space
given by
\be
  Y^\mu\p\mu \p\nu X^\nu = 0\ .\label{4.30}
\ee
The problem in writing this as a tensor equation should be clear:
the partial derivatives can be commuted, but covariant derivatives
cannot.  If we simply replace the partials in (4.30) by covariant
derivatives, we get a different answer than we would if we had
first exchanged the order of the derivatives (leaving the equation
in flat space invariant) and then replaced them.  The difference is
given by
\be
  Y^\mu\nabla_\mu \nabla_\nu X^\nu-Y^\mu\nabla_\nu \nabla_\mu X^\nu 
  = -R_{\mn}Y^\mu X^\nu\ .\label{4.31}
\ee
The prescription for generalizing laws from flat to curved 
spacetimes does not guide us in choosing the order of the
derivatives, and therefore is ambiguous about whether a term such
as that in (4.31) should appear in the presence of gravity.
(The problem of ordering covariant derivatives is similar to the
problem of operator-ordering ambiguities in quantum mechanics.)

In the literature you can find various prescriptions for dealing with
ambiguities such as this, most of which are sensible pieces of advice such 
as remembering to preserve gauge invariance for electromagnetism.  But
deep down the real answer is that there is no way to resolve these
problems by pure thought alone; the fact is that there may be more
than one way to adapt a law of physics to curved space, and ultimately
only experiment can decide between the alternatives.

In fact, let us be honest about the principle of equivalence: it serves
as a useful guideline, but it does not deserve to be treated as a
fundamental principle of nature.  From the modern point of view, we
do not expect the EEP to be rigorously true.  Consider the following
alternative version of (4.24):
\be
  \nabla_\mu[(1+\alpha R)F^{\nu\mu}] = 4\pi J^\nu\ ,\label{4.32}
\ee
where $R$ is the Ricci scalar and $\alpha$ is some coupling constant.
If this equation correctly described electrodynamics in curved
spacetime, it would be possible to measure $R$ even in an arbitrarily
small region, by doing experiments with charged particles.  The 
equivalence principle therefore demands that $\alpha=0$.  But
otherwise this is a perfectly respectable equation, consistent with
charge conservation and other desirable features of electromagnetism,
which reduces to the usual equation in flat space.  Indeed, in a 
world governed by quantum mechanics we expect all possible couplings
between different fields (such as gravity and electromagnetism) that
are consistent with the symmetries of the theory (in this case,
gauge invariance).  So why is it reasonable to set $\alpha=0$? The
real reason is one of scales.  Notice that the Ricci tensor involves
second derivatives of the metric, which is dimensionless, so $R$
has dimensions of (length)$^{-2}$ (with $c=1$).  Therefore $\alpha$ must 
have dimensions of (length)$^{2}$.  But since the coupling represented by
$\alpha$ is of gravitational origin, the only reasonable expectation
for the relevant length scale is
\be
  \alpha \sim l_P^2\ ,\label{4.33}
\ee
where $l_P$ is the Planck length
\be
  l_P = \left({{G\hbar}\over{c^3}}\right)^{1/2}=1.6\times
  10^{-33}{\rm ~cm}\ ,\label{4.34}
\ee
where $\hbar$ is of course Planck's constant.  So the length scale
corresponding to this coupling is extremely small, and for any
conceivable experiment we expect the typical scale of variation for
the gravitational field to be much larger.  Therefore the reason why
this equivalence-principle-violating term can be safely ignored is 
simply because $\alpha R$ is probably a fantastically small number, far
out of the reach of any experiment.  On the other hand, we might as
well keep an open mind, since our expectations are not always borne
out by observation.

Having established how physical laws govern the behavior of
fields and objects in a curved spacetime, we can complete the
establishment of general relativity proper by introducing Einstein's
field equations, which govern how the metric responds to energy
and momentum.  We will actually do this in two ways: first by an
informal argument close to what Einstein himself was thinking,
and then by starting with an action and deriving the 
corresponding equations of motion.

The informal argument begins with the realization that we would 
like to find an equation which supersedes the Poisson equation for
the Newtonian potential:
\be
  \nabla^2\Phi = 4\pi G\rho\ ,\label{4.35}
\ee
where $\nabla^2 = \delta^{ij}\p{i}\p{j}$ is the Laplacian in 
space and $\rho$ is the mass density.  (The explicit form of
$\Phi$ given in (4.22) is one solution of (4.35),
for the case of a pointlike mass distribution.)  What characteristics
should our sought-after equation possess?  On the left-hand side
of (4.35) we have a second-order differential operator acting on the
gravitational potential, and on the right-hand side a measure of
the mass distribution.  A relativistic generalization should take
the form of an equation between tensors.  We know what the tensor
generalization of the mass density is; it's the energy-momentum
tensor $T_\mn$.  The gravitational potential, meanwhile, should
get replaced by the metric tensor.  We might therefore guess
that our new equation will have $T_\mn$ set proportional to some
tensor which is second-order in derivatives of the metric.  In
fact, using (4.21) for the metric in the Newtonian limit and
$T_{00}=\rho$, we see that in this limit we are looking for an
equation that predicts
\be
  \nabla^2 h_{00} = -8\pi G T_{00}\ ,\label{4.36}
\ee
but of course we want it to be completely tensorial.

The left-hand side of (4.36) does not obviously generalize to
a tensor.  The first choice might be to act the D'Alembertian
$\boxx = \nabla^\mu \nabla_\mu$ on the metric $g_\mn$, but this
is automatically zero by metric compatibility.  Fortunately, there
is an obvious quantity which is not zero and is constructed from
second derivatives (and first derivatives) of the metric: the
Riemann tensor $R^\rho{}_{\sigma\mn}$.  It doesn't have the right
number of indices, but we can contract it to form the Ricci tensor
$R_{\mn}$, which does (and is symmetric to boot).  It is therefore
reasonable to guess that the gravitational field equations are
\be
  R_{\mn} = \kappa T_{\mn}\ ,\label{4.37}
\ee
for some constant $\kappa$.  In fact, Einstein did suggest this
equation at one point.  There is a problem, unfortunately, with
conservation of energy.  According to the Principle of Equivalence,
the statement of energy-momentum conservation in curved spacetime
should be
\be
  \nabla^\mu T_{\mn} =0\ ,\label{4.38}
\ee
which would then imply
\be
  \nabla^\mu R_{\mn} =0\ .\label{4.39}
\ee
This is certainly not true in an arbitrary geometry; we have seen
from the Bianchi identity (3.94) that
\be
  \nabla^\mu R_{\mn} ={1\over 2}\nabla_\nu R\ .\label{4.40}
\ee
But our proposed field equation implies that $R=\kappa g^{\mn}
T_{\mn} = \kappa T$, so taking these together we have
\be
  \nabla_\mu T=0\ .\label{4.41}
\ee
The covariant derivative of a scalar is just the partial derivative,
so (4.41) is telling us that $T$ is constant throughout spacetime.
This is highly implausible, since $T=0$ in vacuum while $T>0$ in
matter.  We have to try harder.

(Actually we are cheating slightly, in taking the equation
$\nabla^\mu T_{\mn} =0$ so seriously.  If as we said, the equivalence
principle is only an approximate guide, we could imagine that there are
nonzero terms on the right-hand side involving the curvature tensor.  
Later we will be more precise and argue that they are strictly zero.)

Of course we don't have to try much harder, since we already know
of a symmetric $(0,2)$ tensor, constructed from the Ricci tensor,
which is automatically conserved: the Einstein tensor
\be
  G_\mn = R_\mn - {1\over 2} R g_\mn\ ,\label{4.42}
\ee
which always obeys $\nabla^\mu G_\mn =0$.  We are therefore led to
propose
\be
  G_\mn = \kappa T_\mn\label{4.43}
\ee
as a field equation for the metric.  This equation satisfies all
of the obvious requirements; the right-hand side is a covariant
expression of the energy and momentum density in the form of a
symmetric and conserved $(0,2)$ tensor, while the left-hand side
is a symmetric and conserved $(0,2)$ tensor constructed from the metric
and its first and second derivatives.  It only remains to see whether
it actually reproduces gravity as we know it.  

To answer this, note that contracting both sides of (4.43) yields (in
four dimensions)
\be
  R = -\kappa T\ ,\label{4.44}
\ee
and using this we can rewrite (4.43) as
\be
  R_\mn = \kappa(T_\mn -{1\over 2}T g_\mn)\ .\label{4.45}
\ee
This is the same equation, just written slightly differently.  We would 
like to see if it predicts Newtonian gravity in the weak-field, 
time-independent, slowly-moving-particles limit.  In this limit the 
rest energy $\rho=T_{00}$ will be much larger than the other terms
in $T_\mn$, so we want to focus on the $\mu=0$, $\nu=0$ 
component of (4.45).  In the weak-field limit, we write (in accordance
with (4.13) and (4.14))
\bea
  g_{00} &=&  -1 +h_{00}\ ,\cr
  g^{00} &=&  -1 -h_{00}\ . \label{4.46}
\eea
The trace of the energy-momentum tensor, to lowest nontrivial
order, is
\be
  T = g^{00}T_{00} = -T_{00}\ .\label{4.47}
\ee
Plugging this into (4.45), we get
\be
  R_{00} = {1\over 2} \kappa T_{00}\ .\label{4.48}
\ee
This is an equation relating derivatives of the metric to the
energy density.  To find the explicit expression in terms of the
metric, we need to evaluate $R_{00} = R^\lambda{}_{0\lambda 0}$.
In fact we only need $R^i{}_{0i0}$, since $R^0{}_{000}=0$.  We
have
\be
  R^i{}_{0j0} = \p{j}\Gamma^i_{00} - \p{0}\Gamma^i_{j0}
  +\Gamma^i_{j\lambda}\Gamma^\lambda_{00}
  -\Gamma^i_{0\lambda}\Gamma^\lambda_{j0}\ .\label{4.49}
\ee
The second term here is a time derivative, which vanishes for
static fields.  The third and fourth terms are of the form $(\Gamma)^2$,
and since $\Gamma$ is first-order in the metric perturbation these
contribute only at second order, and can be neglected.  We are left
with $R^i{}_{0j0} = \p{j}\Gamma^i_{00}$.  From this we get
\bea
  R_{00} &=&  R^i{}_{0i0}\cr
  &=& \p{i}\left({1\over 2} g^{i\lambda}
  (\p0 g_{ \lambda 0} + \p0 g_{0\lambda} - \p\lambda g_{00})\right)\cr
  &=&  -{1\over 2}\eta^{ij}\p{i}\p{j} h_{00}\cr
  &=&  -{1\over 2}\nabla^2h_{00}\ . \label{4.50}
\eea
Comparing to (4.48), we see that the $00$ component of (4.43) in the
Newtonian limit predicts
\be
  \nabla^2 h_{00} = -\kappa T_{00}\ .\label{4.51}
\ee
But this is exactly (4.36), if we set $\kappa = 8\pi G$.

So our guess seems to have worked out.  With the normalization fixed
by comparison with the Newtonian limit, we can present {\bf Einstein's
equations} for general relativity:
\be
  R_{\mn} -{1\over 2}R g_\mn = 8\pi G T_{\mn}\ .\label{4.52}
\ee
These tell us how the curvature of spacetime reacts to the presence
of energy-momentum.  Einstein, you may have heard, thought that the
left-hand side was nice and geometrical, while the right-hand side
was somewhat less compelling.

Einstein's equations may be thought of as second-order differential
equations for the metric tensor field $g_\mn$.  There are ten 
independent equations (since both sides are symmetric two-index
tensors), which seems to be exactly right for the ten unknown functions
of the metric components.  However, the Bianchi identity $\nabla^\mu
G_\mn=0$ represents four constraints on the functions $R_{\mn}$, so
there are only six truly independent equations in (4.52).  In fact
this is appropriate, since if a metric is a solution to Einstein's
equation in one coordinate system $x^\mu$ it should also be a 
solution in any other coordinate system $x^{\mu'}$.  This means that
there are four unphysical degrees of freedom in $g_\mn$ (represented
by the four functions $x^{\mu'}(x^\mu)$), and we should expect that
Einstein's equations only constrain the six coordinate-independent
degrees of freedom.

As differential equations, these are
extremely complicated; the Ricci scalar and tensor are contractions 
of the Riemann tensor, which involves derivatives and products 
of the Christoffel symbols, which in turn involve the inverse metric
and derivatives of the metric.  Furthermore, the energy-momentum
tensor $T_\mn$ will generally involve the metric as well.  The
equations are also nonlinear, so that two known solutions cannot
be superposed to find a third.  It is therefore very difficult to
solve Einstein's equations in any sort of generality, and it is
usually necessary to make some simplifying assumptions.  Even
in vacuum, where we set the energy-momentum tensor to zero, the
resulting equations (from (4.45))
\be
  R_\mn=0\label{4.53}
\ee
can be very difficult to solve.  The most popular sort of 
simplifying assumption is that the metric has a significant
degree of symmetry, and we will talk later on about how symmetries
of the metric make life easier.

The nonlinearity of general relativity is worth remarking on.  In
Newtonian gravity the potential due to two point masses is simply
the sum of the potentials for each mass, but clearly this does not
carry over to general relativity (outside the weak-field limit).
There is a physical reason for this, namely that in GR the
gravitational field couples to itself.  This can be thought of as
a consequence of the equivalence principle --- if gravitation did
not couple to itself, a ``gravitational atom'' (two particles
bound by their mutual gravitational attraction) would have a 
different inertial mass (due to the negative binding energy) than
gravitational mass.  From a particle physics point of view this
can be expressed in terms of Feynman diagrams.  The electromagnetic
interaction between two electrons can be thought of as due to 
exchange of a virtual photon:

\begin{figure}[h]
  \centerline{
  \psfig{figure=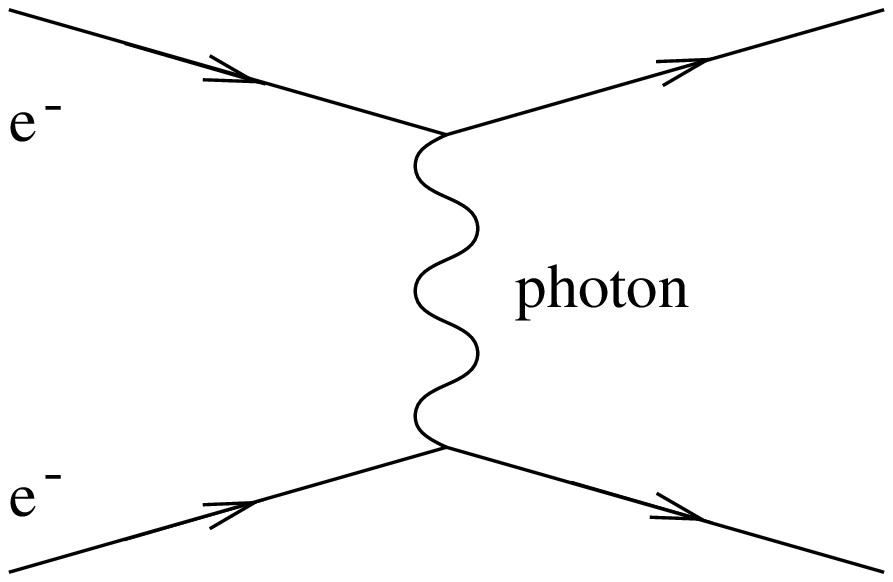,angle=0,height=4cm}}
\end{figure}

\noindent But there is no diagram in which two photons exchange
another photon between themselves; electromagnetism is linear.
The gravitational interaction, meanwhile, can be thought of as due
to exchange of a virtual graviton (a quantized perturbation of the
metric).  The nonlinearity manifests itself as the fact that both
electrons and gravitons (and anything else) can exchange virtual
gravitons, and therefore exert a gravitational force:

\begin{figure}[h]
  \centerline{
  \psfig{figure=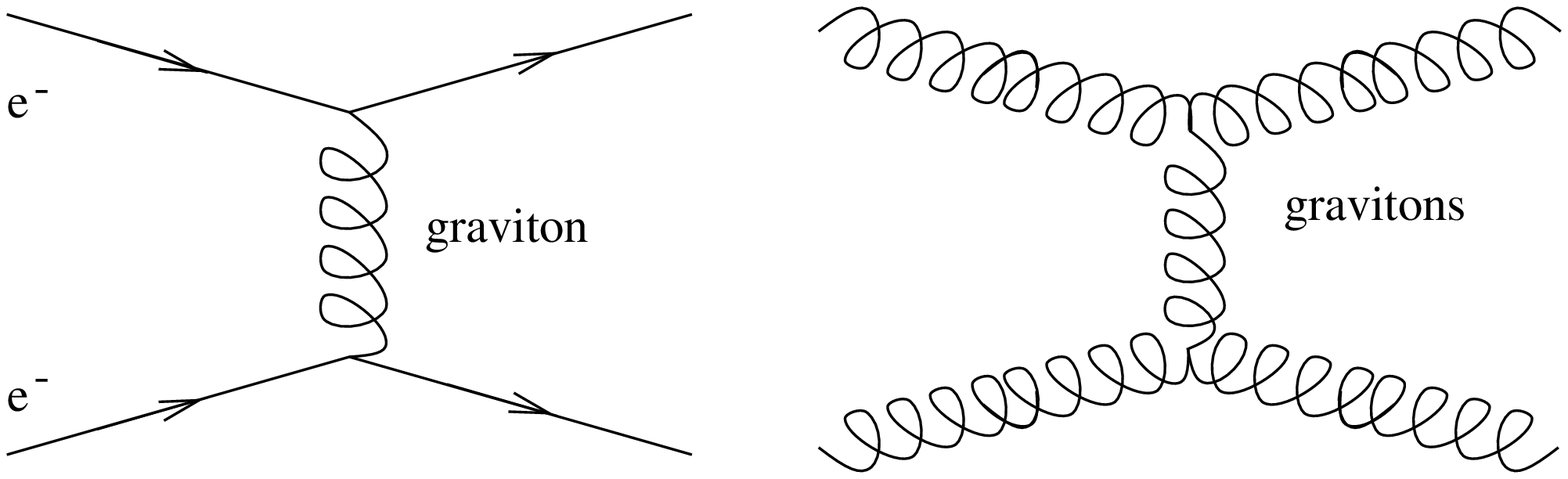,angle=0,height=4cm}}
\end{figure}

\noindent There is nothing profound about this feature of gravity;
it is shared by most gauge theories, such as quantum chromodynamics,
the theory of the strong interactions.  (Electromagnetism is 
actually the exception; the linearity can be traced to the fact 
that the relevant gauge group, U(1), is abelian.)  But it does
represent a departure from the Newtonian theory.  (Of course this
quantum mechanical language of Feynman diagrams is somewhat 
inappropriate for GR, which has not [yet] been successfully quantized,
but the diagrams are just a convenient shorthand for remembering
what interactions exist in the theory.)
 
To increase your confidence that Einstein's equations as we have
derived them are indeed the correct field equations for the metric,
let's see how they can be derived from a more modern viewpoint,
starting from an action principle.  (In fact the equations were
first derived by Hilbert, not Einstein, and Hilbert did it using
the action principle.  But he had been inspired by Einstein's 
previous papers on the subject, and Einstein himself derived the
equations independently, so they are rightly named after Einstein.
The action, however, is rightly called the Hilbert action.)
The action should be the integral over spacetime of a Lagrange
density (``Lagrangian'' for short, although strictly speaking
the Lagrangian is the integral over space of the Lagrange density):
\be
  S_H=\int d^nx {\cal L}_H\ .\label{4.54}
\ee
The Lagrange density is a tensor density, which can be written as
$\g$ times a scalar.  What scalars can we make out of the metric?
Since we know that the metric can be set equal to its canonical form
and its first derivatives set to zero at any one point, any nontrivial
scalar must involve at least second derivatives of the metric.
The Riemann tensor is of course made from second derivatives of the
metric, and we argued earlier that the only independent scalar we
could construct from the Riemann tensor was the Ricci scalar
$R$.  What we did not show, but is nevertheless true, is that any
nontrivial tensor made from the metric and its first and second
derivatives can be expressed in terms of the metric and the Riemann
tensor.  Therefore, the only independent scalar constructed from
the metric, which is no higher than second order in its derivatives,
is the Ricci scalar.  Hilbert figured that this was therefore the
simplest possible choice for a Lagrangian, and proposed 
\be
  {\cal L}_H = \g R\ .\label{4.55}
\ee
The equations of motion should come from varying the action
with respect to the metric.  In fact let us consider variations
with respect to the inverse metric $g^\mn$, which are slightly
easier but give an equivalent set of equations.  Using
$R=g^\mn R_\mn$, in general we will have
\bea
  \delta S &=&  \int d^nx\left[\g g^\mn \delta R_\mn + \g R_\mn \delta 
  g^\mn+ R\delta\g\right]\cr
  &=& (\delta S)_1 +(\delta S)_2 +(\delta S)_3 \ . \label{4.56}
\eea
The second term $(\delta S)_2$ is already in the form of some
expression times $\delta g^\mn$; let's examine the others more
closely.

Recall that the Ricci tensor is the contraction of the Riemann tensor,
which is given by
\be
  R^\rho{}_{\mu\lambda\nu} = \p\lambda \Gamma^\lambda_{\nu\mu}
  +\Gamma^\rho_{\lambda\sigma}\Gamma^\sigma_{\nu\mu}
  - (\lambda \leftrightarrow \nu)\ .\label{4.57}
\ee
The variation of this with respect the metric can be found first
varying the connection with respect to the metric, and then 
substituting into this expression.  Let us however consider arbitrary
variations of the connection, by replacing
\be
  \Gamma^\rho_{\nu\mu}\rightarrow \Gamma^\rho_{\nu\mu}+
  \delta\Gamma^\rho_{\nu\mu}\ .\label{4.58}
\ee
The variation $\delta\Gamma^\rho_{\nu\mu}$ is the difference of
two connections, and therefore is itself a tensor.  We can thus
take its covariant derivative,
\be
  \nabla_\lambda(\delta\Gamma^\rho_{\nu\mu})=
  \p\lambda(\delta\Gamma^\rho_{\nu\mu})
  +\Gamma^\rho_{\lambda\sigma}\delta\Gamma^\sigma_{\nu\mu}
  -\Gamma^\sigma_{\lambda\nu}\delta\Gamma^\rho_{\sigma\mu}
  -\Gamma^\sigma_{\lambda\mu}\delta\Gamma^\rho_{\nu\sigma}\ .
  \label{4.59}
\ee
Given this expression (and a small amount of labor) it is easy
to show that
\be
  \delta R^\rho{}_{\mu\lambda\nu}= 
  \nabla_\lambda(\delta\Gamma^\rho_{\nu\mu})
  -\nabla_\nu(\delta\Gamma^\rho_{\lambda\mu})\ .\label{4.60}
\ee
You can check this yourself.  Therefore, the contribution of 
the first term in (4.56) to $\delta S$ can be written
\bea
  (\delta S)_1 &=& 
  \int d^nx \g ~g^\mn \left[\nabla_\lambda(
  \delta\Gamma^\lambda_{\nu\mu})
  -\nabla_\nu(\delta\Gamma^\lambda_{\lambda\mu})\right]\cr
  &=& \int d^nx \g ~ {\nabla_\sigma}\left[g^{\mu\sigma}(\delta
  \Gamma^\lambda_{\lambda\mu}) - g^{\mn}(\delta
  \Gamma^\sigma_{\mu\nu})\right]\ , \label{4.61}
\eea
where we have used metric compatibility and relabeled some dummy
indices.  But now we have the integral with respect to the natural
volume element of the covariant divergence of a vector; by Stokes's
theorem, this is equal to a boundary contribution at infinity which
we can set to zero by making the variation vanish at infinity.
(We haven't actually shown that Stokes's theorem, as mentioned
earlier in terms of differential forms, can be thought of this way,
but you can easily convince yourself it's true.)  Therefore this
term contributes nothing to the total variation.

To make sense of the $(\delta S)_3$ term we need to use the following
fact, true for any matrix $M$:
\be
  \tr(\ln M) = \ln(\det M)\ .\label{4.62}
\ee
Here, $\ln M$ is defined by $\exp(\ln M)=M$.  (For numbers this
is obvious, for matrices it's a little less straightforward.)
The variation of this identity yields
\be
  \tr(M^{-1} \delta M) = {1\over{\det M}}\delta(\det M)\ .\label{4.63}
\ee
Here we have used the cyclic property of the trace to allow us to
ignore the fact that $M^{-1}$ and $\delta M$ may not commute.  Now we
would like to apply this to the inverse metric, $M = g^\mn$.  Then
$\det M=g^{-1}$ (where $g=\det g_{\mn}$), and
\be
  \delta(g^{-1})={1\over g}g_\mn \delta g^\mn\ .\label{4.64}
\ee
Now we can just plug in:
\bea
  \delta\g &=&  \delta[(-g^{-1})^{-1/2}]\cr
  &=& -{1\over 2}(-g^{-1})^{-3/2}\delta(-g^{-1})\cr
  &=&  -{1\over 2}\g g_\mn \delta g^\mn\ . \label{4.65}
\eea
Hearkening back to (4.56), and remembering that $(\delta S)_1$ does
not contribute, we find
\be
  \delta S = \int d^nx \g ~\left[R_{\mn} -{1\over 2} Rg_\mn\right]
  \delta g^\mn\ .\label{4.66}
\ee
This should vanish for arbitrary variations, so we are led to 
Einstein's equations in vacuum:
\be
  {1\over{\g}}{{\delta S}\over{\delta g^\mn}}
  =R_{\mn} -{1\over 2} Rg_\mn =0\ .\label{4.67}
\ee

The fact that this simple action leads to the same vacuum field
equations as we had previously arrived at by more informal
arguments certainly reassures us that we are doing something
right.  What we would really like, however, is to get the non-vacuum
field equations as well.  That means we consider an action of
the form
\be
  S={{1}\over{8\pi G}}S_H+S_M\ ,\label{4.68}
\ee
where $S_M$ is the action for matter, and we have presciently normalized
the gravitational action (although the proper normalization is somewhat 
convention-dependent).  Following through the same procedure as
above leads to 
\be
  {1\over{\g}}{{\delta S}\over{\delta g^\mn}}
  ={{1}\over{8\pi G}}\left(R_{\mn} -{1\over 2} Rg_\mn\right)
  +{1\over{\g}} {{\delta S_M}\over{\delta g^\mn}}=0\ ,\label{4.69}
\ee
and we recover Einstein's equations if we can set
\be
  T_\mn = -{1\over{\g}}{{\delta S_M}\over{\delta g^\mn}}\ .\label{4.70}
\ee
What makes us think that we can make such an identification?  
In fact (4.70) turns out to be the best way to define a symmetric
energy-momentum tensor.  The tricky part is to show that it is 
conserved, which is in fact automatically true, 
but which we will not justify until the next section.  

We say that (4.70) provides the ``best'' definition of the energy-momentum
tensor because it is not the only one you will find.  In flat Minkowski 
space, there is an alternative definition which is sometimes given in books
on electromagnetism or field theory.  In this context energy-momentum
conservation arises as a consequence of symmetry of the Lagrangian
under spacetime translations.  {\it Noether's theorem}
states that every symmetry of a Lagrangian implies the existence
of a conservation law; invariance under the four spacetime translations 
leads to a tensor $S^\mn$ which obeys $\p\mu S^\mn=0$ (four relations,
one for each value of $\nu$).  The details can be found in Wald or
in any number of field theory books.  Applying Noether's procedure
to a Lagrangian which depends on some fields $\psi^i$ and their
first derivatives $\p\mu\psi^i$, we obtain
\be
  S^\mn={{\delta{\cal L}}\over{\delta(\p\mu\psi^i)}}\partial^\nu\psi^i
  -\eta^\mn{\cal L}\ ,\label{4.71}
\ee
where a sum over $i$ is implied.  You can check that this tensor
is conserved by virtue of the equations of motion of the matter
fields.  $S^\mn$ often goes by the name ``canonical
energy-momentum tensor''; however, there are a number of reasons
why it is more convenient for us to use (4.70).  First and foremost,
(4.70) is in fact what appears on the right hand side of 
Einstein's equations when they are derived from an action, and it
is not always possible to generalize (4.71) to curved spacetime.  
But even in flat space (4.70) has its advantages; it is
manifestly symmetric, and also guaranteed to be gauge invariant,
neither of which is true for (4.71).  We will therefore stick with
(4.70) as the definition of the energy-momentum tensor.

Sometimes it is useful to think about Einstein's equations without
specifying the theory of matter from which $T_\mn$ is derived.  
This leaves us with a great deal of arbitrariness; consider for
example the question ``What metrics obey Einstein's equations?''
In the absence of some constraints on $T_\mn$, the answer is ``any
metric at all''; simply take the metric of your choice, compute the
Einstein tensor $G_\mn$ for this metric, and then demand that
$T_\mn$ be equal to $G_\mn$.  (It will automatically be conserved,
by the Bianchi identity.)  Our real concern is with the existence
of solutions to Einstein's equations in the presence of ``realistic''
sources of energy and momentum, whatever that means.  The most
common property that is demanded of $T_\mn$ is that it represent
positive energy densities --- no negative masses are allowed.  In
a locally inertial frame this requirement can be stated as
$\rho = T_{00} \geq 0$.  To turn this into a coordinate-independent
statement, we ask that
\be
  T_\mn V^\mu V^\nu \geq 0\ ,\qquad{\rm ~for~all~timelike~vectors~}
  V^\mu\ .\label{4.72}
\ee
This is known as the {\bf Weak Energy Condition}, or WEC.  It seems 
like a fairly
reasonable requirement, and many of the important theorems about 
solutions to general relativity (such as the singularity theorems
of Hawking and Penrose) rely on this condition or something very
close to it.  Unfortunately it is not set in stone; indeed, it is
straightforward to invent otherwise respectable classical field
theories which violate the WEC, and almost impossible to invent a
quantum field theory which obeys it.  Nevertheless, it is legitimate
to assume that the WEC holds in all but the most extreme conditions.
(There are also stronger energy conditions, but they are even less
true than the WEC, and we won't dwell on them.)

We have now justified Einstein's equations in two different ways:
as the natural covariant generalization of Poisson's equation for
the Newtonian gravitational potential, and as the result of varying
the simplest possible action we could invent for the metric.  The
rest of the course will be an exploration of the consequences of
these equations, but before we start on that road let us briefly
explore ways in which the equations could be modified.  There are
an uncountable number of such ways, but we will consider four
different possibilities: the introduction of a cosmological constant,
higher-order terms in the action, gravitational scalar fields, and
a nonvanishing torsion tensor.

The first possibility is the cosmological constant; George Gamow
has quoted Einstein as calling this the biggest mistake of his
life.  Recall that in our search for the simplest possible action
for gravity we noted that any nontrivial scalar had to be of at
least second order in derivatives of the metric; at lower order all
we can create is a constant.  Although a constant does not by itself
lead to very interesting dynamics, it has an important effect if we
add it to the conventional Hilbert action.  We therefore consider
an action given by
\be
  S=\int d^nx \g(R-2\Lambda)\ ,\label{4.73}
\ee
where $\Lambda$ is some constant.  The resulting field equations 
are 
\be
  R_\mn -{1\over 2}R g_\mn +\Lambda g_\mn =0\ ,\label{4.74}
\ee
and of course there would be an energy-momentum tensor on the
right hand side if we had included an action for matter.  $\Lambda$
is the cosmological constant; it was originally introduced by 
Einstein after it became clear that there were no solutions to his
equations representing a static cosmology (a universe unchanging
with time on large scales) with a nonzero matter content.  If the
cosmological constant is tuned just right, it is possible to find
a static solution, but it is unstable to small perturbations.
Furthermore, once Hubble demonstrated that the universe is expanding,
it became less important to find static solutions, and Einstein
rejected his suggestion.  Like Rasputin, however, the cosmological
constant has proven difficult to kill off.  If we like we can move
the additional term in (4.74) to the right hand side, and think
of it as a kind of energy-momentum tensor, with $T_\mn = -
\Lambda g_\mn$ (it is automatically conserved by metric compatibility).
Then $\Lambda$ can be interpreted as the ``energy density of the
vacuum,'' a source of energy and momentum that is present even in
the absence of matter fields.  This interpretation is important because
quantum field theory predicts that the vacuum should have some sort
of energy and momentum.  In ordinary quantum mechanics, an harmonic
oscillator with frequency $\omega$ and minimum classical energy
$E_0=0$ upon quantization has a ground state with energy
$E_0={1\over 2}\hbar\omega$.  A quantized field can be thought of
as a collection of an infinite number of harmonic oscillators, and
each mode contributes to the ground state energy.  The result is
of course infinite, and must be appropriately regularized, for 
example by introducing a cutoff at high frequencies.  The final
vacuum energy, which is the regularized sum of the energies of
the ground state oscillations of all the fields of the theory, has
no good reason to be zero and in fact would be expected to have
a natural scale
\be
  \Lambda \sim m_P^4\ ,\label{4.75}
\ee
where the Planck mass $m_P$ is approximately $10^{19}$~GeV, or
$10^{-5}$ grams.  Observations of the universe on large scales 
allow us to constrain the actual value of $\Lambda$, which turns
out to be smaller than (4.75) by at least a factor of $10^{120}$.
This is the largest known discrepancy between theoretical estimate
and observational constraint in physics, and convinces many people
that the ``cosmological constant problem'' is one of the most
important unsolved problems today.  On the other hand the 
observations do not tell us that $\Lambda$ is strictly zero, and
in fact allow values that can have important consequences for the
evolution of the universe.  This mistake of Einstein's therefore
continues to bedevil both physicists, who would like to understand
why it is so small, and astronomers, who would like to determine
whether it is really small enough to be ignored.

A somewhat less intriguing generalization of the Hilbert action would
be to include scalars of more than second order in derivatives of
the metric.  We could imagine an action of the form
\be
  S = \int d^nx \g (R + \alpha_1 R^2 + \alpha_2 R_\mn R^\mn
  +\alpha_3 g^\mn \nabla_\mu R \nabla_\nu R +\cdots)\ ,\label{4.76}
\ee
where the $\alpha$'s are coupling constants and the dots represent
every other scalar we can make from the curvature tensor, its 
contractions, and its derivatives.  Traditionally, such terms have
been neglected on the reasonable grounds that they merely complicate
a theory which is already both aesthetically pleasing and empirically
successful.  However, there are at least three more substantive reasons 
for this neglect.  First, as we shall see below, Einstein's equations
lead to a well-posed initial value problem for the metric, in which
``coordinates'' and ``momenta'' specified at an initial time can be
used to predict future evolution.  With higher-derivative terms, we
would require not only those data, but also some number of derivatives
of the momenta.  Second, the main source of dissatisfaction with 
general relativity on the part of particle physicists is that it cannot
be renormalized (as far as we know), and Lagrangians with higher 
derivatives tend generally to make theories less renormalizable rather
than more.  Third, by the same arguments we used above when speaking
about the limitations of the principle of equivalence, the extra terms
in (4.76) should be suppressed (by powers of the Planck mass to some
power) relative to the usual Hilbert term, and therefore would not be
expected to be of any practical importance to the low-energy world.
None of these reasons are completely persuasive, and indeed people
continue to consider such theories, but for the most part these models
do not attract a great deal of attention.

A set of models which does attract attention are known as 
{\bf scalar-tensor theories} of gravity, since they involve both
the metric tensor $g_\mn$ and a fundamental scalar field, $\lambda$.
The action can be written
\be
  S = \int d^nx \g \left[f(\lambda)R+{1\over 2}g^\mn(\p\mu\lambda)
  (\p\nu\lambda) -V(\lambda)\right]\ ,\label{4.77}
\ee
where $f(\lambda)$ and $V(\lambda)$ are functions which define the
theory.  Recall from (4.68) that the coefficient of the Ricci scalar
in conventional GR is proportional to the inverse of Newton's constant
$G$.  In scalar-tensor theories, then, where this coefficient is replaced
by some function of a field which can vary throughout spacetime,
the ``strength'' of gravity (as measured by the local value of Newton's
constant) will be different from place to place and time to time.
In fact the most famous scalar-tensor theory, invented by Brans and
Dicke and now named after them, was inspired by a suggestion of 
Dirac's that the gravitational constant varies with time.  Dirac had
noticed that there were some interesting numerical coincidences one
could discover by taking combinations of cosmological numbers such as the
Hubble constant $H_0$ (a measure of the expansion rate of the universe)
and typical particle-physics parameters such as the mass of the pion,
$m_\pi$.  For example,
\be
  {{m_\pi^3}\over{H_0}}\sim {{cG}\over{\hbar^2}}\ .\label{4.78}
\ee
If we assume for the moment that this relation is not simply an
accident, we are faced with the problem that the Hubble ``constant''
actually changes with time (in most cosmological models), while the
other quantities conventionally do not.  Dirac therefore proposed that
in fact $G$ varied with time, in such a way as to maintain (4.78);
satisfying this proposal was the motivation of Brans and Dicke.
These days, experimental test of general relativity are sufficiently
precise that we can state with confidence that, if Brans-Dicke theory
is correct, the predicted change in $G$ over space and time must be
very small, much slower than that necessary to satisfy Dirac's
hypothesis.  (See Weinberg for details on Brans-Dicke theory and
experimental tests.)
Nevertheless there is still a great deal of work being
done on other kinds of scalar-tensor theories, which turn out to be
vital in superstring theory and may have important consequences in 
the very early universe.

As a final alternative to general relativity, we should mention the
possibility that the connection really is not derived from the metric,
but in fact has an independent existence as a fundamental field.
We will leave it as an exercise for you to show that it is possible
to consider the conventional action for general relativity but treat
it as a function of both the metric $g_\mn$ and a torsion-free connection
$\Gamma^\lambda_{\rho\sigma}$,
and the equations of motion derived from varying such an action with
respect to the connection imply that $\Gamma^\lambda_{\rho\sigma}$ is
actually the Christoffel connection associated with $g_\mn$.  We could
drop the demand that the connection be torsion-free, in which case the
torsion tensor could lead to additional propagating degrees of freedom.
Without going into details, the basic reason why such theories do not
receive much attention is simply because the torsion is itself a tensor;
there is nothing to distinguish it from other, ``non-gravitational''
tensor fields.  Thus, we do not really lose any generality by considering
theories of torsion-free connections (which lead to GR) plus any number
of tensor fields, which we can name what we like.

With the possibility in mind that one of these alternatives (or, more
likely, something we have not yet thought of) is actually realized in
nature, for the rest of the course we will work under the assumption
that general relativity as based on Einstein's equations or the Hilbert
action is the correct theory, and work out its consequences.  These
consequences, of course, are constituted by the solutions to Einstein's
equations for various sources of energy and momentum, and the behavior
of test particles in these solutions.  Before considering specific
solutions in detail, lets look more abstractly at the initial-value
problem in general relativity.

In classical Newtonian mechanics, the behavior of a single particle 
is of course governed by ${\bf f} = m{\bf a}$.  If the particle is
moving under the influence of some potential energy field $\Phi(x)$,
then the force is ${\bf f} = -\nabla \Phi$, and the particle obeys
\be
  m{{d^2x^i}\over{dt^2}}=-\p{i}\Phi\ .\label{4.79}
\ee
This is a second-order differential equation for $x^i(t)$, which we
can recast as a system of two coupled first-order equations by 
introducing the momentum ${\bf p}$:
\bea
  {{dp^i}\over{dt}}&=& -\p{i}\Phi\cr
  {{dx^i}\over{dt}} &=&  {1\over m}p^i\ . \label{4.80}
\eea
The initial-value problem is simply the procedure of specifying a
``state'' $(x^i,p^i)$ which serves as a boundary condition with which
(4.80) can be uniquely solved.  You may think of (4.80) as allowing you,
once you are given the coordinates and momenta at some time $t$, to
evolve them forward an infinitesimal amount to a time $t+\delta t$,
and iterate this procedure to obtain the entire solution.

We would like to formulate the analogous problem in general relativity.
Einstein's equations $G_\mn = 8\pi G T_\mn$ are of course covariant;
they don't single out a preferred notion of ``time'' through which
a state can evolve.  Nevertheless, we can by hand pick a spacelike 
hypersurface (or ``slice'') $\Sigma$, specify initial data on that 
hypersurface, and see if we can evolve uniquely from it to a 
hypersurface in the future.  (``Hyper'' because a constant-time 
slice in four dimensions will be three-dimensional, whereas
``surfaces'' are conventionally two-dimensional.)
This process does violence to the manifest covariance of the theory,
but if we are careful we should wind up with a formulation that is
equivalent to solving Einstein's equations all at once throughout
spacetime.

\begin{figure}
  \centerline{
  \psfig{figure=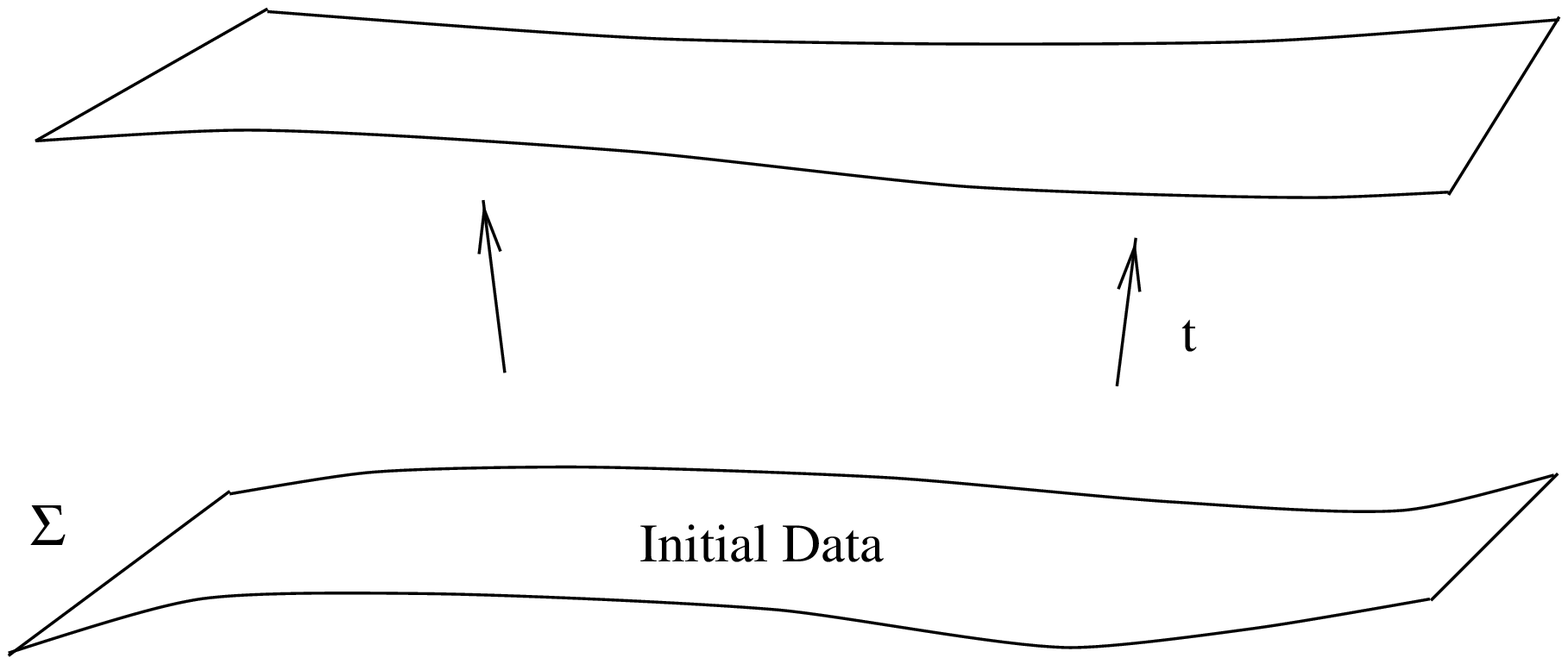,angle=0,height=5cm}}
\end{figure}

Since the metric is the fundamental variable, our first guess is that
we should consider the values $g_\mn |_\Sigma$ of the metric on our
hypersurface to be the ``coordinates'' and the time derivatives 
$\p{t}g_\mn |_\Sigma$ (with respect to some specified time coordinate)
to be the ``momenta'', which together specify the state.  (There
will also be coordinates and momenta for the matter fields, which we
will not consider explicitly.)  In fact the equations $G_\mn = 8\pi 
G T_\mn$ do involve second derivatives of the metric with respect to 
time (since the connection involves first derivatives of the metric and
the Einstein tensor involves first derivatives of the connection), 
so we seem to be on the right track.  However, the Bianchi identity 
tells us that $\nabla_\mu G^\mn=0$.  We can rewrite this equation as
\be
  \p0 G^{0\nu}=-\p{i}G^{i\nu}-\Gamma^\mu_{\mu\lambda}G^{\lambda\nu}
  -\Gamma^\nu_{\mu\lambda}G^{\mu\lambda}\ .\label{4.81}
\ee
A close look at the right hand side reveals that there are no
{\it third}-order time derivatives; therefore there cannot be any
on the left hand side.  Thus, although $G^\mn$ as a whole involves
second-order time derivatives of the metric, the specific components
$G^{0\nu}$ do not.  Of the ten independent components in Einstein's
equations, the four represented by
\be
  G^{0\nu}=8\pi GT^{0\nu}\label{4.82}
\ee
cannot be used to evolve the initial data $(g_\mn,\p{t}g_\mn)_\Sigma$.
Rather, they serve as {\bf constraints} on this initial data; we are
not free to specify any combination of the metric and its time 
derivatives on the hypersurface $\Sigma$, since they must obey the 
relations (4.82).  The remaining equations,
\be
  G^{ij}=8\pi GT^{ij}\label{4.83}
\ee
are the dynamical evolution equations for the metric.  Of course,
these are only six equations for the ten unknown functions 
$g_\mn(x^\sigma)$, so the solution will inevitably involve a fourfold
ambiguity.  This is simply the freedom that we have already mentioned,
to choose the four coordinate functions throughout spacetime.

It is a straightforward but unenlightening exercise to sift through
(4.83) to find that not all second time derivatives of the metric 
appear.  In fact we find that $\p{t}^2g_{ij}$ appears in (4.83), but
not $\p{t}^2g_{0\nu}$.  Therefore a ``state'' in general relativity
will consist of a specification of the spacelike components of the
metric $g_{ij}|_\Sigma$ and their first time derivatives 
$\p{t}g_{ij}|_\Sigma$ on the hypersurface $\Sigma$, from which we can
determine the future evolution using (4.83), up to an unavoidable
ambiguity in fixing the remaining components $g_{0\nu}$.  The
situation is precisely analogous to that in electromagnetism, where
we know that no amount of initial data can suffice to determine the
evolution uniquely since there will always be the freedom to perform
a gauge transformation $A_\mu \rightarrow A_\mu +\p\mu\lambda$.
In general relativity, then, coordinate transformations play a role
reminiscent of gauge transformations in electromagnetism, in that
they introduce ambiguity into the time evolution.

One way to cope with this problem is to simply ``choose a gauge.''
In electromagnetism this means to place a condition on the vector
potential $A_\mu$, which will restrict our freedom to perform gauge
transformations.  For example we can choose Lorentz gauge, in which
$\nabla_\mu A^\mu=0$, or temporal gauge, in which $A_0=0$.  We can
do a similar thing in general relativity, by fixing our coordinate
system.  A popular choice is {\bf harmonic gauge} (also known as
Lorentz gauge and a host of other names), in which
\be
  \boxx x^\mu =0\ .\label{4.84}
\ee
Here $\boxx=\nabla^\mu\nabla_\mu$ is the covariant D'Alembertian, 
and it is crucial to realize when we take the covariant derivative
that the four functions $x^\mu$ are just functions, not components
of a vector.  This condition is therefore simply
\bea
  0&=&  \boxx x^\mu\cr
  &=&  g^{\rho\sigma}\p\rho\p\sigma x^\mu - g^{\rho\sigma}
  \Gamma^\lambda_{\rho\sigma}\p\lambda x^\mu\cr
  &=&  -g^{\rho\sigma}\Gamma^\lambda_{\rho\sigma}\ . \label{4.85}
\eea
In flat space, of course, Cartesian coordinates (in which
$\Gamma^\lambda_{\rho\sigma}=0$) are harmonic coordinates.  (As a
general principle, any function $f$ which satisfies $\boxx f=0$
is called an ``harmonic function.'')

To see that this choice of coordinates successfully fixes our gauge
freedom, let's rewrite the condition (4.84) in a somewhat simpler form.
We have
\be
  g^{\rho\sigma}\Gamma^\mu_{\rho\sigma}={1\over 2}g^{\rho\sigma}
  g^\mn(\p\rho g_{\sigma\nu}+\p\sigma g_{\nu\rho} -\p\nu
  g_{\rho\sigma})\ ,\label{4.86}
\ee
from the definition of the Christoffel symbols.  Meanwhile, from
$\p\rho(g^\mn g_{\sigma\nu})=\p\rho\delta^\mu_\sigma=0$ we have
\be
  g^\mn\p\rho g_{\sigma\nu} = - g_{\sigma\nu}\p\rho g^\mn\ .\label{4.87}
\ee
Also, from our previous exploration of the variation of the determinant
of the metric (4.65), we have
\be
  {1\over 2}g_{\rho\sigma}\p\nu g^{\rho\sigma} =
  -{1\over{\g}}\,\p\nu \g\ .\label{4.88}
\ee
Putting it all together, we find that (in general),
\be
  g^{\rho\sigma}\Gamma^\mu_{\rho\sigma}={1\over{\g}}\,\p\lambda
  (\g g^{\lambda\mu})\ .\label{4.89}
\ee
The harmonic gauge condition (4.85) therefore is equivalent to
\be
  \p\lambda(\g g^{\lambda\mu})=0\ .\label{4.90}
\ee
Taking the partial derivative of this with respect to $t=x^0$ 
yields
\be
  {{\partial^2}\over{\partial t^2}}(\g g^{0\nu})=
  -{{\partial}\over{\partial x^i}}\left[{{\partial}\over
  {\partial t}}(\g g^{i\nu})\right]\ .\label{4.91}
\ee
This condition represents a second-order differential equation
for the previously unconstrained metric components $g^{0\nu}$, in
terms of the given initial data.  We have therefore succeeded in
fixing our gauge freedom, in that we can now solve for the evolution
of the entire metric in harmonic coordinates.  (At least locally;
we have been glossing over the fact our gauge choice may not be 
well-defined globally, and we would have to resort to working in
patches as usual.  The same problem appears in gauge theories in
particle physics.)  Note that we still have some freedom remaining;
our gauge condition (4.84) restricts how the coordinates stretch from
our initial hypersurface $\Sigma$ throughout spacetime, but we can 
still choose coordinates $x^i$ on $\Sigma$ however we like.  This 
corresponds to the fact that making a coordinate transformation
$x^\mu \rightarrow x^\mu +\delta^\mu$, with $\boxx \delta^\mu=0$,
does not violate the harmonic gauge condition.

We therefore have a well-defined initial value problem for general
relativity; a state is specified by the spacelike components of the
metric and their time derivatives on a spacelike hypersurface $\Sigma$;
given these, the spacelike components (4.83) of Einstein's equations
allow us to evolve the metric forward in time, up to an ambiguity
in coordinate choice which may be resolved by choice of gauge.
We must keep in mind that the initial data are not arbitrary, but
must obey the constraints (4.82).  (Once we impose the constraints on
some spacelike hypersurface, the equations of motion guarantee that they
remain satisfied, as you can check.)  The constraints serve a useful
purpose, of guaranteeing that the result remains spacetime covariant
after we have split our manifold into ``space'' and ``time.''
Specifically, the $G^{i0}=8\pi GT^{i0}$ constraint implies that
the evolution is independent of our choice of coordinates on
$\Sigma$, while $G^{00}=8\pi GT^{00}$ enforces invariance under
different ways of slicing spacetime into spacelike hypersurfaces.

\begin{figure}
  \centerline{
  \psfig{figure=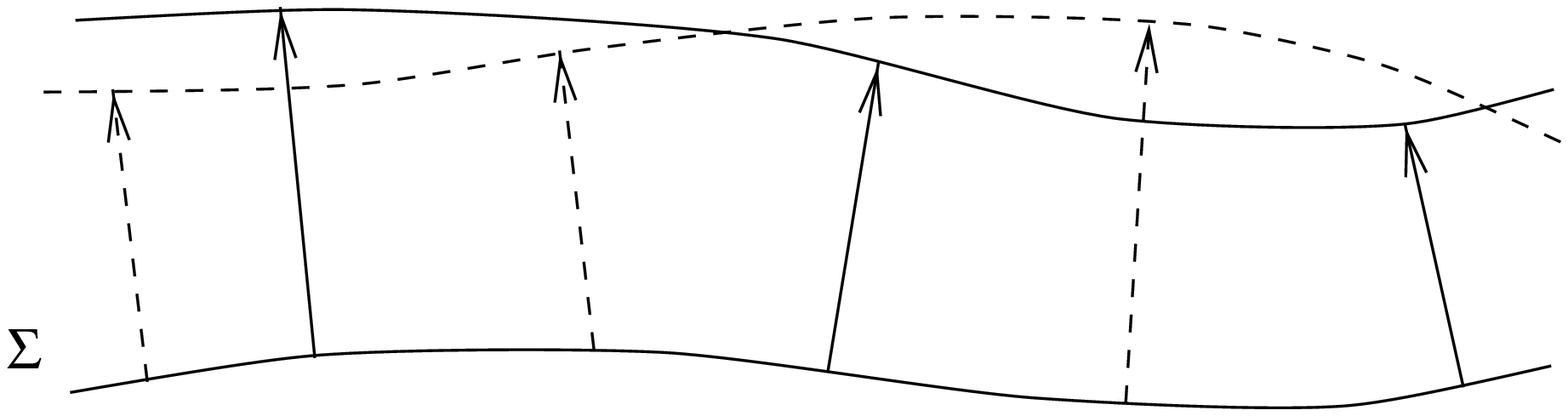,angle=0,height=4cm}}
\end{figure}

Once we have seen how to cast Einstein's equations as an initial
value problem, one issue of crucial importance is the existence of
solutions to the problem.  That is, once we have specified a spacelike
hypersurface with initial data, to what extent can we be guaranteed
that a unique spacetime will be determined?  Although one can do a 
great deal of hard work to answer this question with some precision,
it is fairly simple to get a handle on the ways in which a well-defined
solution can fail to exist, which we now consider.

It is simplest to first consider the problem of evolving matter fields
on a fixed background spacetime, rather than the evolution of the
metric itself.  We therefore consider a spacelike hypersurface $\Sigma$
in some manifold $M$ with fixed metric $g_\mn$, and furthermore look
at some connected subset $S$ in $\Sigma$.  Our guiding principle will
be that no signals can travel faster than the speed of light; therefore
``information'' will only flow along timelike or null trajectories
(not necessarily geodesics).  We define the {\bf future domain of
dependence} of $S$, denoted $D^+(S)$, as the set of all points $p$ such
that {\it every} past-moving, timelike or null, inextendible curve through
$p$ must intersect $S$.  (``Inextendible'' just means that the curve
goes on forever, not ending at some finite point.)  We interpret this
definition in such a way that $S$ itself is a subset of $D^+(S)$.  (Of
course a rigorous formulation does not require additional interpretation
over and above the definitions, but we are not being as rigorous
as we could be right now.)  Similarly, we 
define the past domain of dependence $D^-(S)$ in the same way, but
with ``past-moving'' replaced by ``future-moving.''  Generally
speaking, some points in $M$ will be in one of the domains of dependence,
and some will be outside; we define the boundary of $D^+(S)$ to be
the {\bf future Cauchy horizon} $H^+(S)$, and likewise the boundary of 
$D^-(S)$ to be the past Cauchy horizon $H^-(S)$.  You can convince
yourself that they are both null surfaces.

\begin{figure}[h]
  \centerline{
  \psfig{figure=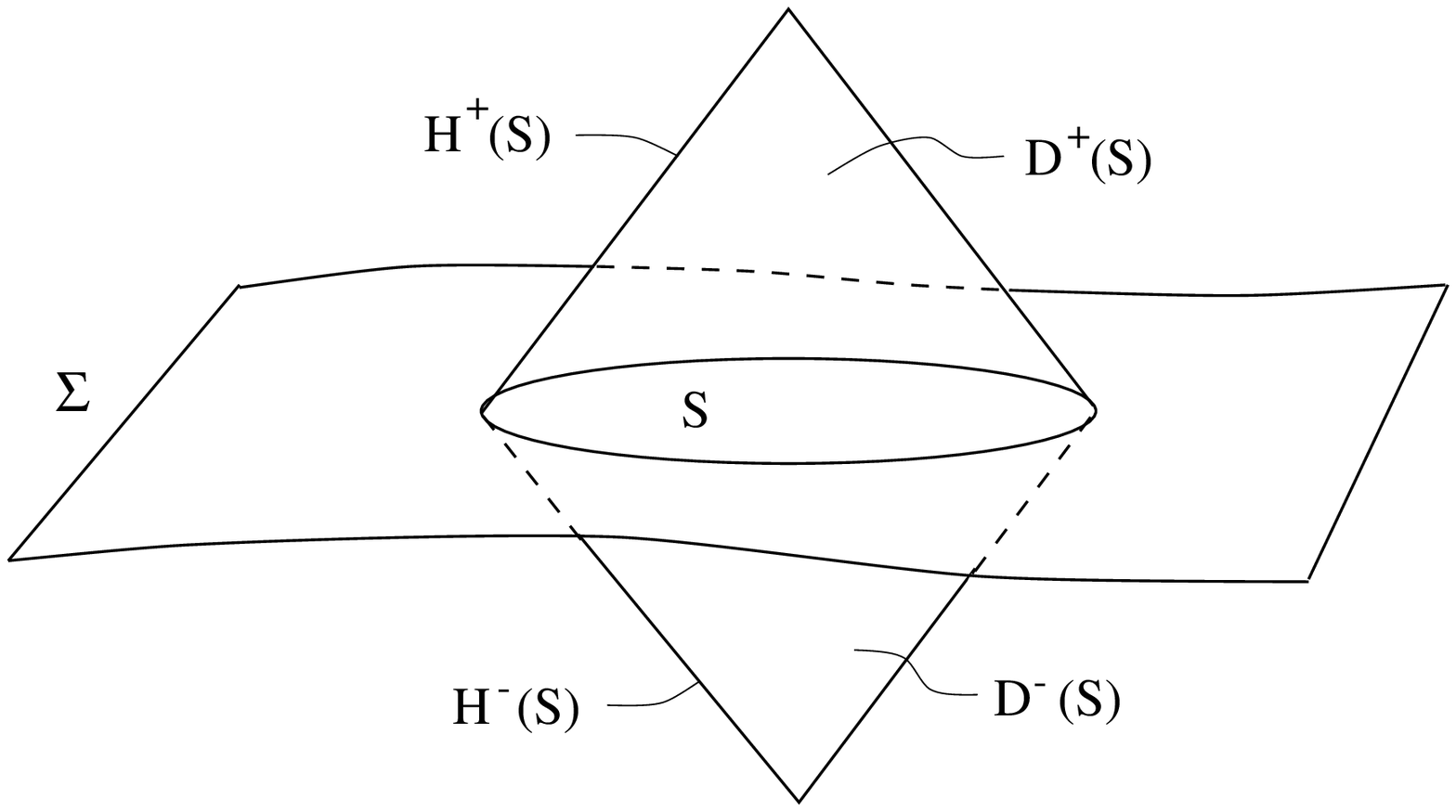,angle=0,height=6cm}}
\end{figure}

The usefulness of these definitions should be apparent; if nothing
moves faster than light, than signals cannot propagate outside the
light cone of any point $p$.  Therefore, if every curve which
remains inside this light cone must intersect $S$, then information
specified on $S$ should be sufficient to predict what the situation
is at $p$.  (That is, initial data for matter fields given on $S$
can be used to solve for the value of the fields at $p$.)  The set
of all points for which we can predict what happens by knowing 
what happens on $S$ is simply the union $D^+(S)\cup D^-(S)$.

We can easily extend these ideas from the subset $S$ to the entire
hypersurface $\Sigma$.  The important point is that $D^+(\Sigma)\cup 
D^-(\Sigma)$ might fail to be all of $M$, even if $\Sigma$ itself
seems like a perfectly respectable hypersurface that extends 
throughout space.  There are a number of ways in which this can
happen.  One possibility is that we have just chosen a ``bad''
hypersurface (although it is hard to give a general prescription for
when a hypersurface is bad in this sense).  Consider Minkowski space,
and a spacelike hypersurface $\Sigma$ which remains to the past of
the light cone of some point.

\begin{figure}[h]
  \centerline{
  \psfig{figure=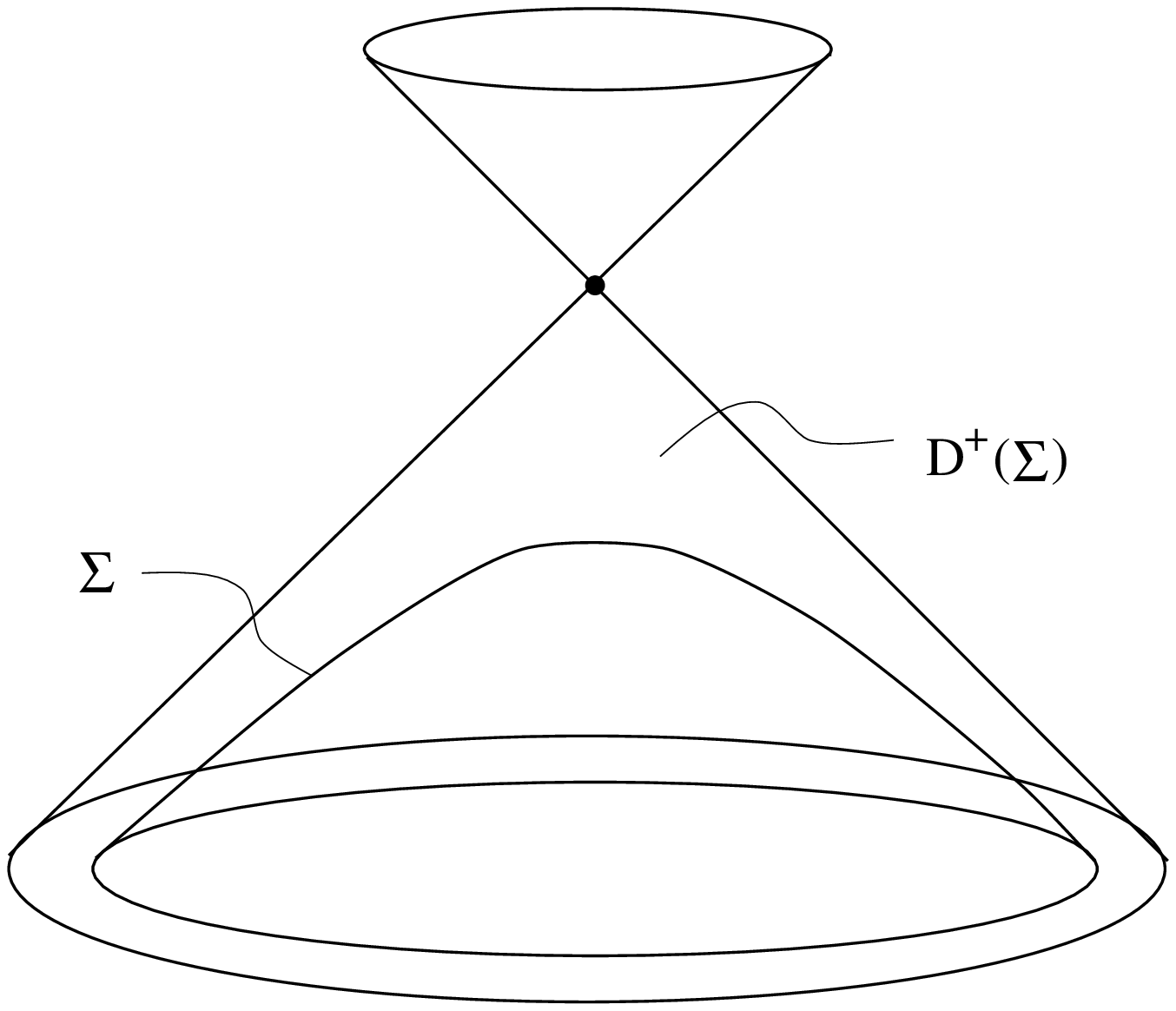,angle=0,height=6cm}}
\end{figure}

\noindent In this case $\Sigma$ is a nice spacelike surface, but
it is clear that $D^+(\Sigma)$ ends at the light cone, and we cannot
use information on $\Sigma$ to predict what happens throughout 
Minkowski space.  Of course, there are other surfaces we could have
picked for which the domain of dependence would have been the entire
manifold, so this doesn't worry us too much.

A somewhat more nontrivial example is known as {\bf Misner space}.
This is a two-dimensional spacetime with the topology of $\R^1\times
S^1$, and a metric for which the light cones progressively tilt as
you go forward in time.
\begin{figure}
  \centerline{
  \psfig{figure=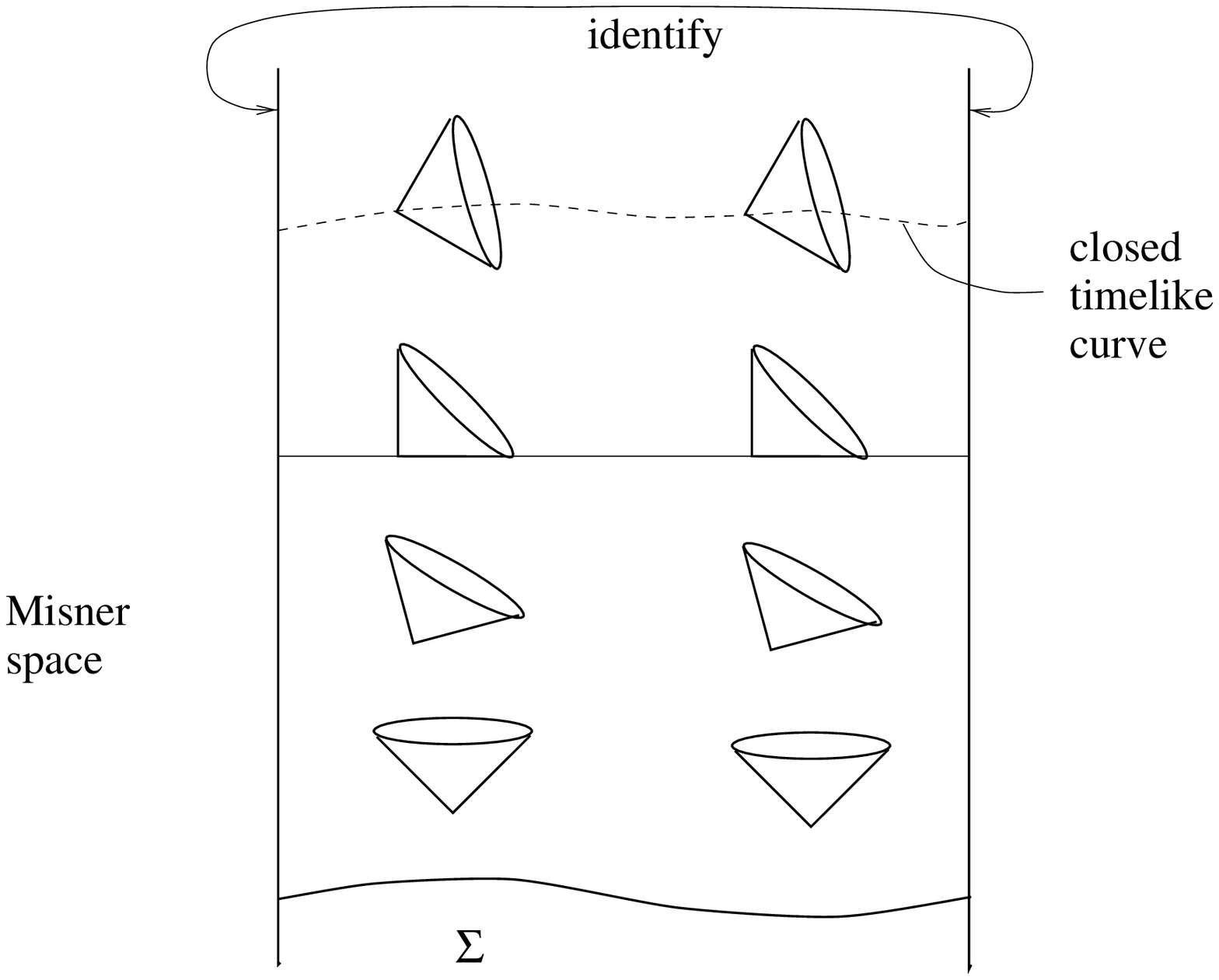,angle=0,height=8cm}}
\end{figure}
Past a certain point, it is possible to travel on a timelike
trajectory which wraps around the $S^1$ and comes back to itself; this
is known as a {\bf closed timelike curve}.  If we had specified a
surface $\Sigma$ to this past of this point, then none of the points
in the region containing closed timelike curves are in the domain of
dependence of $\Sigma$, since the closed timelike curves themselves
do not intersect $\Sigma$.  This is obviously a worse problem than
the previous one, since a well-defined initial value problem does not
seem to exist in this spacetime.  (Actually problems like this are
the subject of some current research interest, so I won't claim that
the issue is settled.)

A final example is provided by the existence of singularities, points
which are not in the manifold even though they can be reached by
travelling along a geodesic for a finite distance.  Typically these
occur when the curvature becomes infinite at some point; if this
happens, the point can no longer be said to be part of the spacetime.
Such an occurrence can lead to the emergence of a Cauchy horizon ---
a point $p$ which is in the future of a singularity cannot be in the
domain of dependence of a hypersurface to the past of the singularity,
because there will be curves from $p$ which simply end at the 
singularity.

\begin{figure}[h]
  \centerline{
  \psfig{figure=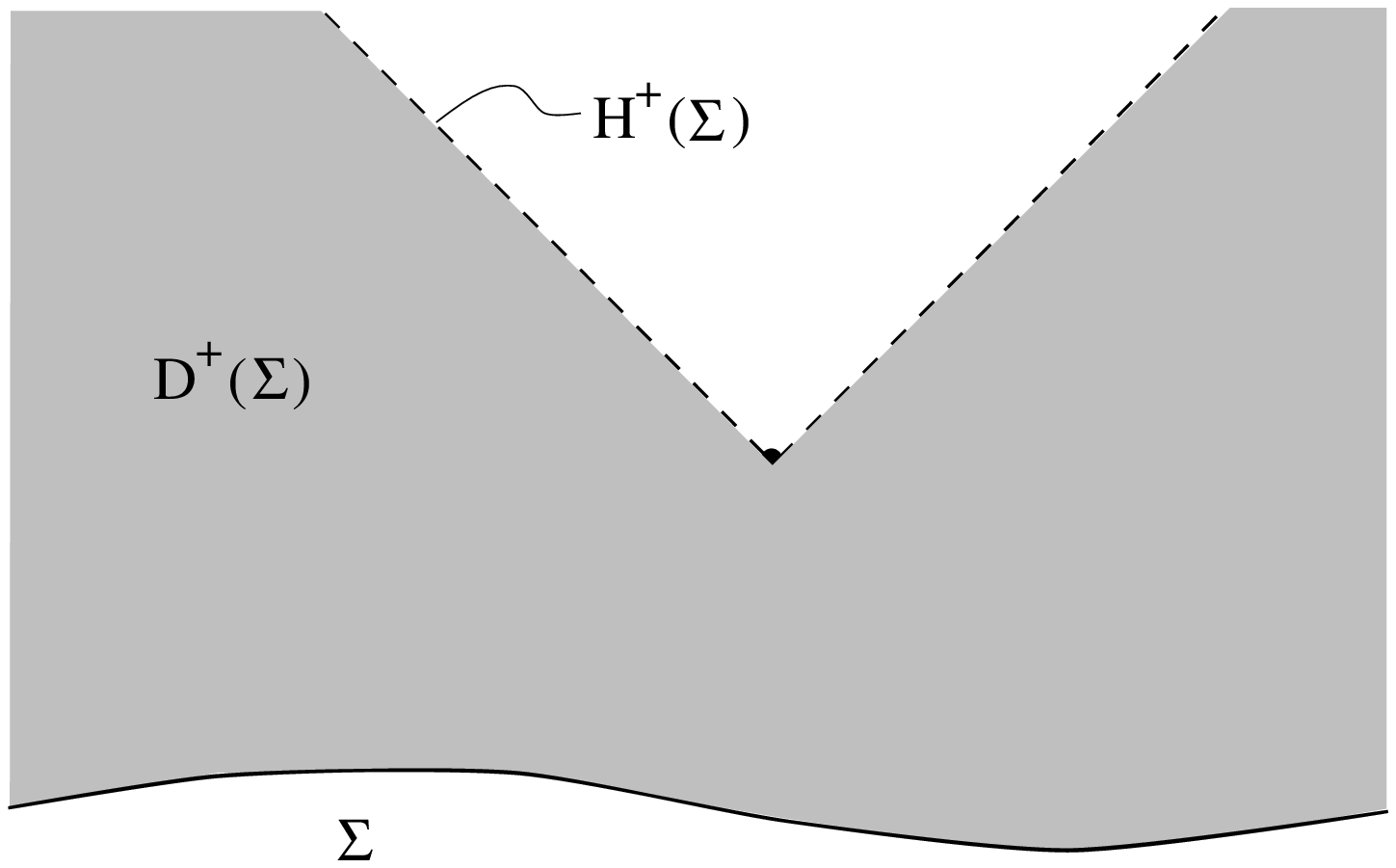,angle=0,height=6cm}}
\end{figure}

All of these obstacles can also arise in the initial value problem
for GR, when we try to evolve the metric itself from initial data.
However, they are of different degrees of troublesomeness.  The 
possibility of picking a ``bad'' initial hypersurface does not arise
very often, especially since most solutions are found globally (by
solving Einstein's equations throughout spacetime).  The one 
situation in which you have to be careful is in numerical solution
of Einstein's equations, where a bad choice of hypersurface can lead
to numerical difficulties even if in principle a complete solution
exists.  Closed timelike curves seem to be something that GR
works hard to avoid --- there are certainly solutions which contain
them, but evolution from generic initial data does not usually produce
them.  Singularities, on the other hand, are practically unavoidable.
The simple fact that the gravitational force is always attractive
tends to pull matter together, increasing the curvature, and generally
leading to some sort of singularity.  This is something which we 
apparently must learn to live with, although there is some hope that
a well-defined theory of quantum gravity will eliminate the 
singularities of classical GR.

\eject
\thispagestyle{plain}

\setcounter{equation}{0}

\noindent{December 1997 \hfill {\sl Lecture Notes on General Relativity}
\hfill{Sean M.~Carroll}}

\vskip .2in

\setcounter{section}{4}
\section{More Geometry}

With an understanding of how the laws of physics adapt to curved
spacetime, it is undeniably tempting to start in on applications.  
However, a few extra mathematical techniques will simplify our task a 
great deal, so we will pause briefly to explore the geometry of manifolds
some more.

When we discussed manifolds in section 2, we introduced maps
between two different manifolds and how maps could be composed.  We
now turn to the use of such maps in carrying along tensor fields 
from one manifold to another.  We therefore consider two manifolds
$M$ and $N$, possibly of different dimension, with coordinate systems
$x^\mu$ and $y^\alpha$, respectively.  We imagine that we have a
map $\phi:M\rightarrow N$ and a function $f:N\rightarrow \R$.

\begin{figure}[h]
  \centerline{
  \psfig{figure=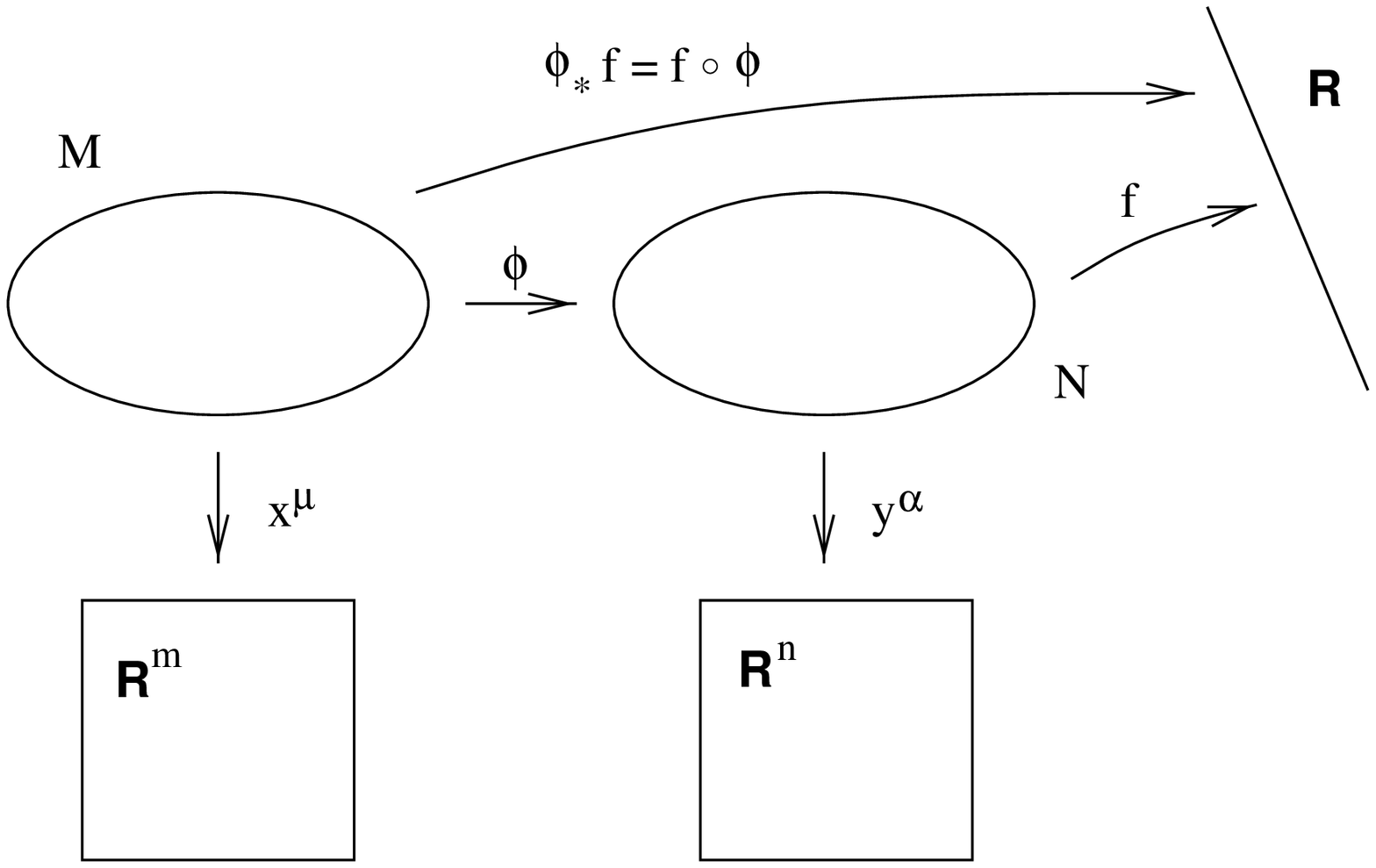,angle=0,height=7cm}}
\end{figure}

\noindent It is obvious that we can compose $\phi$ with $f$ to 
construct a map $(f\circ\phi):M\rightarrow\R$, which is simply
a function on $M$.  Such a construction is sufficiently useful that
it gets its own name; we define the {\bf pullback} of $f$ by $\phi$,
denoted $\phi_*f$, by
\be
  \phi_* f = (f\circ\phi)\ .\label{5.1}
\ee
The name makes sense, since we think of $\phi_*$ as ``pulling back''
the function $f$ from $N$ to $M$.

We can pull functions back, but we cannot push them forward.  If
we have a function $g:M\rightarrow\R$, there is no way we can compose
$g$ with $\phi$ to create a function on $N$; the arrows don't fit 
together correctly.  But recall that a vector can be thought of
as a derivative operator that maps smooth functions to real numbers.
This allows us to define the {\bf pushforward} of a vector; if $V(p)$ is 
a vector at a point $p$ on $M$, we define the pushforward vector $\phi^*V$ 
at the point $\phi(p)$ on $N$ by giving its action on functions on $N$:
\be
  (\phi^*V)(f) = V(\phi_*f)\ .\label{5.2}
\ee
So to push forward a vector field we say ``the action of
$\phi^*V$ on any function is simply the action of $V$ on the pullback
of that function.''

This is a little abstract, and it would be nice to have a more concrete 
description.  We know that a basis for vectors on $M$ is given by
the set of partial derivatives $\p\mu={{\partial}\over{\partial x^\mu}}$, 
and a basis on $N$ is given by the set of partial derivatives $\p\alpha=
{{\partial}\over{\partial y^\alpha}}$.  Therefore we would like to relate
the components of $V=V^\mu\p\mu$ to those of $(\phi^*V)=(\phi^*V)^\alpha
\p\alpha$.  We can find the sought-after relation by applying the
pushed-forward vector to a test function and using the chain
rule (2.3):
\bea
  (\phi^*V)^\alpha\p\alpha f &=&  V^\mu\p\mu (\phi_* f)\cr
  &=&  V^\mu\p\mu (f\circ\phi)\cr
  &=& V^\mu {{\partial y^\alpha}\over{\partial x^\mu}}\p\alpha f\ .
  \label{5.3}
\eea
This simple formula makes it irresistible to think of the pushforward
operation $\phi^*$ as a matrix operator, $(\phi^*V)^\alpha = 
(\phi^*)^\alpha{}_\mu V^\mu$, with the matrix being given by
\be
  (\phi^*)^\alpha{}_\mu = {{\partial y^\alpha}\over{\partial x^\mu}}
  \ .\label{5.4}
\ee
The behavior of a vector under a pushforward thus bears an 
unmistakable resemblance to the vector transformation law under
change of coordinates.  In fact it is a generalization, since when
$M$ and $N$ are the same manifold the constructions are (as we shall
discuss) identical; but
don't be fooled, since in general $\mu$ and $\alpha$ have different
allowed values, and there is no reason for the matrix ${{\partial y^\alpha}
/{\partial x^\mu}}$ to be invertible.

It is a rewarding exercise to convince yourself that, although you can
push vectors forward from $M$ to $N$ (given a map $\phi:
M\rightarrow N$), you cannot in general pull them back --- just keep trying 
to invent an appropriate construction until the futility of the attempt
becomes clear.  Since one-forms are dual to vectors, you should 
not be surprised to hear that one-forms can be pulled back (but not
in general pushed forward).  To do this, remember that one-forms are linear 
maps from vectors to the real numbers.  The pullback $\phi_*\omega$ of a
one-form $\omega$ on $N$ can therefore be defined by its action on
a vector $V$ on $M$, by equating it with the action of $\omega$
itself on the pushforward of $V$:
\be
  (\phi_*\omega)(V)=\omega(\phi^*V)\ .\label{5.5}
\ee
Once again, there is a simple matrix description of the pullback 
operator on forms, $(\phi_*\omega)_\mu =(\phi_*)_\mu{}^\alpha 
\omega_\alpha$, which we can derive using the chain rule.  It is
given by
\be
  (\phi_*)_\mu{}^\alpha = {{\partial y^\alpha}\over{\partial x^\mu}}
  \ .\label{5.6}
\ee
That is, it is the same matrix as the pushforward (5.4), but of course a 
different index is contracted when the matrix acts to pull back
one-forms.

There is a way of thinking about why pullbacks and pushforwards work
on some objects but not others, which may or may not be helpful.
If we denote the set of smooth functions on $M$ by ${\cal F}(M)$,
then a vector $V(p)$ at a point $p$ on $M$ ({\it i.e.}, an element of 
the tangent space $T_pM$) can be thought of as an operator from 
${\cal F}(M)$ to $\R$.  But we already know that the pullback operator
on functions maps ${\cal F}(N)$ to ${\cal F}(M)$ (just as $\phi$
itself maps $M$ to $N$, but in the opposite direction).  Therefore
we can define the pushforward $\phi_*$ acting on vectors simply by
composing maps, as we first defined the pullback of functions:

\begin{figure}[h]
  \centerline{
  \psfig{figure=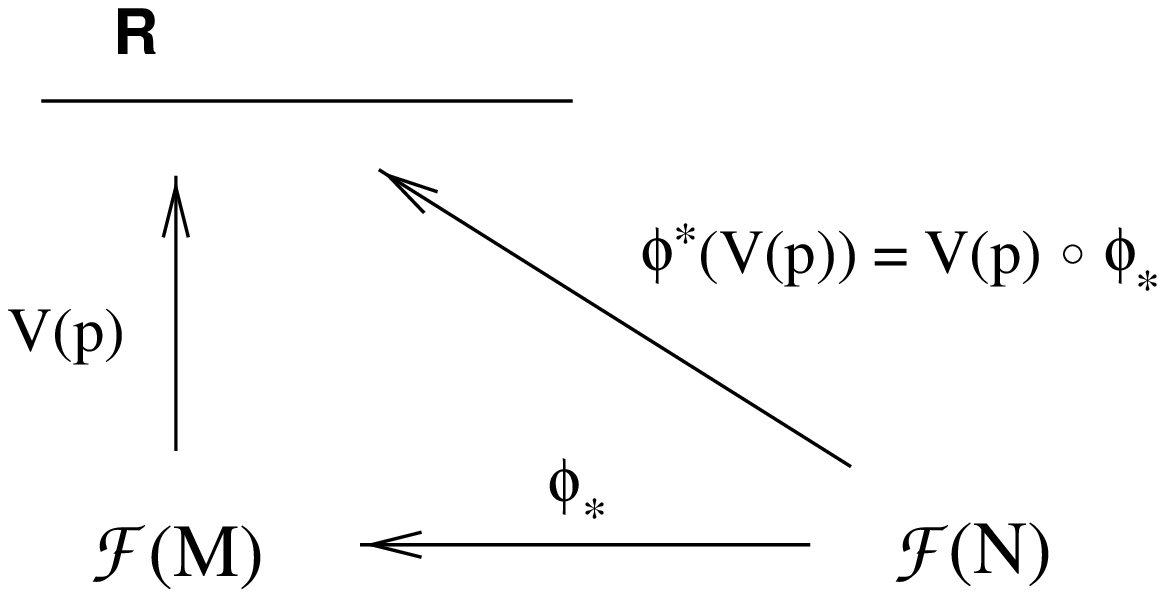,angle=0,height=4cm}}
\end{figure}

\noindent Similarly, if $T_qN$ is the tangent space at a point $q$ on $N$,
then a one-form $\omega$ at $q$ 
({\it i.e.}, an element of the cotangent space $T_q^*N$) can be thought of 
as an operator from $T_qN$ to $\R$.  Since the pushforward
$\phi^*$ maps $T_pM$ to $T_{\phi(p)}N$, the pullback $\phi_*$
of a one-form can also be thought of as mere composition of maps:

\begin{figure}[h]
  \centerline{
  \psfig{figure=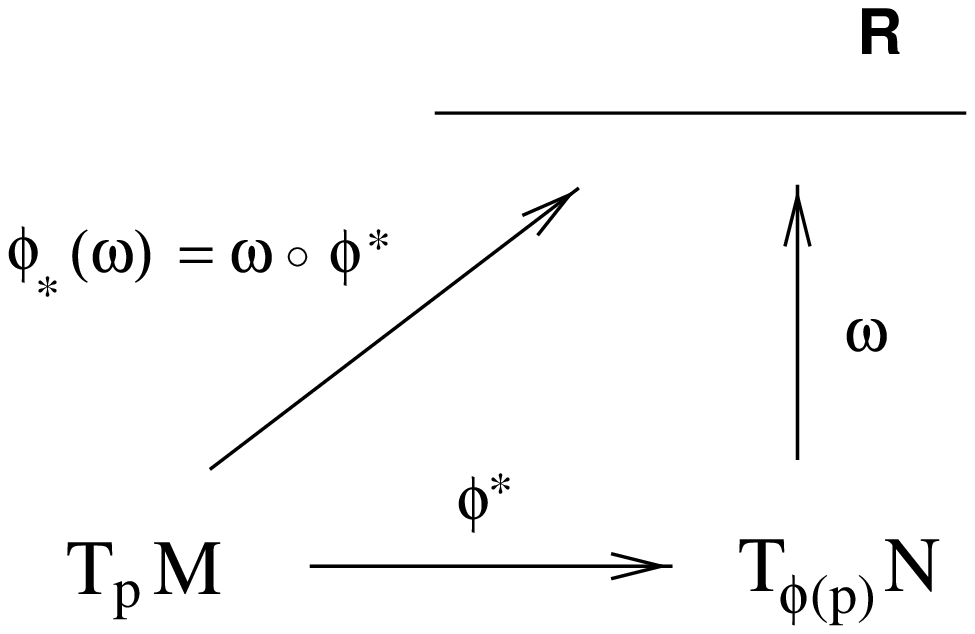,angle=0,height=4cm}}
\end{figure}

\noindent If this is not helpful, don't worry about it.  But do keep
straight what exists and what doesn't; the actual concepts are simple,
it's just remembering which map goes what way that leads to confusion.

You will recall further that a $(0,l)$ tensor --- one with $l$ lower
indices and no upper ones --- is a linear map from the direct product
of $l$ vectors to $\R$.  We can therefore pull back not only one-forms,
but tensors with an arbitrary number of lower indices.  The definition
is simply the action of the original tensor on the pushed-forward
vectors:
\be
  (\phi_* T)(V^{(1)}, V^{(2)},\ldots ,V^{(l)})=T(\phi^*V^{(1)}, 
  \phi^*V^{(2)},\ldots ,\phi^*V^{(l)})\ ,\label{5.7}
\ee
where $T_{\alpha_1 \cdots \alpha_l}$ is a $(0,l)$ tensor on $N$.  We can 
similarly push forward any $(k,0)$ tensor $S^{\mu_1 \cdots \mu_k}$
by acting it on pulled-back one-forms:
\be
  (\phi^* S)(\omega^{(1)}, \omega^{(2)},\ldots ,\omega^{(k)})=
  S(\phi_*\omega^{(1)}, \phi_*\omega^{(2)},\ldots ,\phi_*\omega^{(k)})
  \ .\label{5.8}
\ee
Fortunately, the matrix representations of the pushforward (5.4) and
pullback (5.6) extend to the higher-rank tensors simply by assigning
one matrix to each index; thus, for the pullback of a $(0,l)$ tensor,
we have
\be
  (\phi_* T)_{\mu_1 \cdots \mu_l} = {{\partial y^{\alpha_1}}
  \over{\partial x^{\mu_1}}}\cdots{{\partial y^{\alpha_l}}
  \over{\partial x^{\mu_l}}}T_{\alpha_1 \cdots \alpha_l}\ ,\label{5.9}
\ee
while for the pushforward of a $(k,0)$ tensor we have
\be
  (\phi^* S)^{\alpha_1 \cdots \alpha_k} = {{\partial y^{\alpha_1}}
  \over{\partial x^{\mu_1}}}\cdots{{\partial y^{\alpha_k}}
  \over{\partial x^{\mu_k}}}S^{\mu_1 \cdots \mu_k}\ .\label{5.10}
\ee
Our complete picture is therefore:

\begin{figure}[h]
  \centerline{
  \psfig{figure=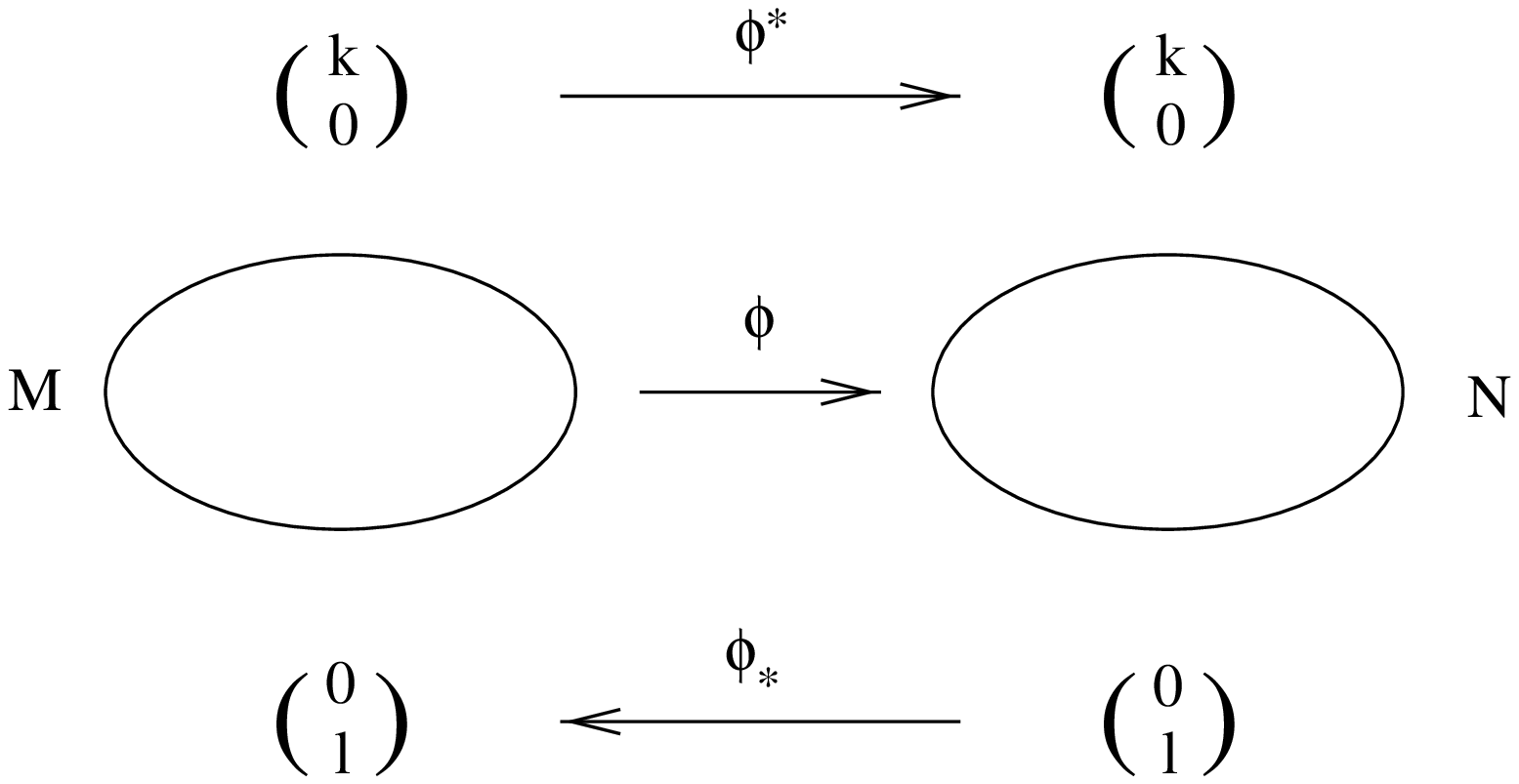,angle=0,height=5cm}}
\end{figure}

\noindent Note that tensors with both upper and lower indices can
generally be neither pushed forward nor pulled back.

This machinery becomes somewhat less imposing once we see it at work
in a simple example.  One common occurrence of a map between two
manifolds is when $M$ is actually a submanifold of $N$; then there is
an obvious map from $M$ to $N$ which just takes an element of $M$ to
the ``same'' element of $N$.  Consider our usual example, the two-sphere
embedded in $\R^3$, as the locus of points a unit distance from the 
origin.  If we put coordinates $x^\mu=(\theta,\phi)$ on $M=S^2$ and 
$y^\alpha=(x,y,z)$ on $N=\R^3$, the map $\phi:M\rightarrow N$ is given by
\be
  \phi(\theta,\phi)=(\sin\theta \cos\phi,\sin\theta \sin\phi,
  \cos\theta)\ .\label{5.11}
\ee
In the past we have considered the metric $ds^2=\d x^2+\d y^2+ \d z^2$
on $\R^3$, and said that it induces a metric $\d\theta^2 +\sin^2\theta
~\d\phi^2$ on $S^2$, just by substituting (5.11) into this flat metric
on $\R^3$.  We didn't really justify such a statement at the
time, but now we can do so.  (Of course it would be easier if we
worked in spherical coordinates on $\R^3$, but doing it the hard way
is more illustrative.)  The matrix of partial derivatives is given
by
\be
  {{\partial y^{\alpha}}\over{\partial x^{\mu}}}=
  \left(\matrix{\cos\theta \cos\phi &\cos\theta \sin\phi &-\sin\theta\cr
  -\sin\theta \sin\phi &\sin\theta \cos\phi & 0\cr}\right)\ .\label{5.12}
\ee
The metric on $S^2$ is obtained by simply pulling back the metric from
$\R^3$,
\bea
  (\phi^* g)_\mn &=&  {{\partial y^{\alpha}}
  \over{\partial x^{\mu}}}{{\partial y^{\beta}}
  \over{\partial x^{\nu}}}g_{\alpha\beta}\cr
  &=& \left(\matrix{1&0\cr 0& \sin^2\theta\cr}\right)\ ,\label{5.13}
\eea
as you can easily check.  Once again, the answer is the same as you
would get by naive substitution, but now we know why.

We have been careful to emphasize that a map $\phi:M\rightarrow N$ can
be used to push certain things forward and pull other things back.
The reason why it generally doesn't work both ways can be traced to the
fact that $\phi$ might not be invertible.  If $\phi$ is invertible
(and both $\phi$ and $\phi^{-1}$ are smooth, which we always implicitly 
assume), then it defines a diffeomorphism between $M$ and $N$.  In this 
case $M$ and $N$ are the same abstract manifold.  The beauty
of diffeomorphisms is that we can use both $\phi$ and $\phi^{-1}$ to
move tensors from $M$ to $N$; this will allow us to define the 
pushforward and pullback of arbitrary tensors.  Specifically, for a
$(k,l)$ tensor field $T^{\mu_1 \cdots \mu_k}{}_{\nu_1 \cdots \mu_l}$
on $M$, we define the pushforward by
\be
  (\phi^*T)(\omega^{(1)},\ldots ,\omega^{(k)},V^{(1)},\ldots ,V^{(l)})
  = T(\phi_*\omega^{(1)},\ldots ,\phi_*\omega^{(k)},
  [\phi^{-1}]^*V^{(1)},\ldots ,[\phi^{-1}]^*V^{(l)})\ ,\label{5.14}
\ee
where the $\omega^{(i)}$'s are one-forms on $N$ and the $V^{(i)}$'s
are vectors on $N$.  In components this becomes
\be
  (\phi^*T)^{\alpha_1 \cdots \alpha_k}{}_{\beta_1 \cdots \beta_l}
  = {{\partial y^{\alpha_1}}
  \over{\partial x^{\mu_1}}}\cdots{{\partial y^{\alpha_k}}
  \over{\partial x^{\mu_k}}}{{\partial x^{\nu_1}}
  \over{\partial y^{\beta_1}}}\cdots{{\partial x^{\nu_l}}
  \over{\partial y^{\beta_l}}}T^{\mu_1 \cdots \mu_k}{}_{\nu_1
  \cdots \nu_l}\ .\label{5.15}
\ee
The appearance of the inverse matrix $\partial x^\nu/\partial y^\beta$
is legitimate because $\phi$ is invertible.  Note that we could also
define the pullback in the obvious way, but there is no need to write
separate equations because the pullback $\phi_*$ is the same as the
pushforward via the inverse map, $[\phi^{-1}]^*$.

We are now in a position to explain the relationship between 
diffeomorphisms and coordinate transformations.  The relationship is
that they are two different ways of doing precisely the same thing.
If you like, diffeomorphisms are ``active coordinate transformations'',
while traditional coordinate transformations are ``passive.''  Consider
an $n$-dimensional manifold $M$ with coordinate functions
$x^\mu :M\rightarrow \R^n$.  To change coordinates we can either simply
introduce new functions $y^\mu :M\rightarrow \R^n$ (``keep the manifold
fixed, change the coordinate maps''), or we could just as well introduce
a diffeomorphism $\phi:M\rightarrow M$, after which the coordinates would
just be the pullbacks $(\phi_*x)^\mu:M\rightarrow \R^n$ (``move the 
points on the manifold, and then evaluate the coordinates of the new
points'').  In this sense, (5.15) really is the tensor transformation
law, just thought of from a different point of view.

\begin{figure}[h]
  \centerline{
  \psfig{figure=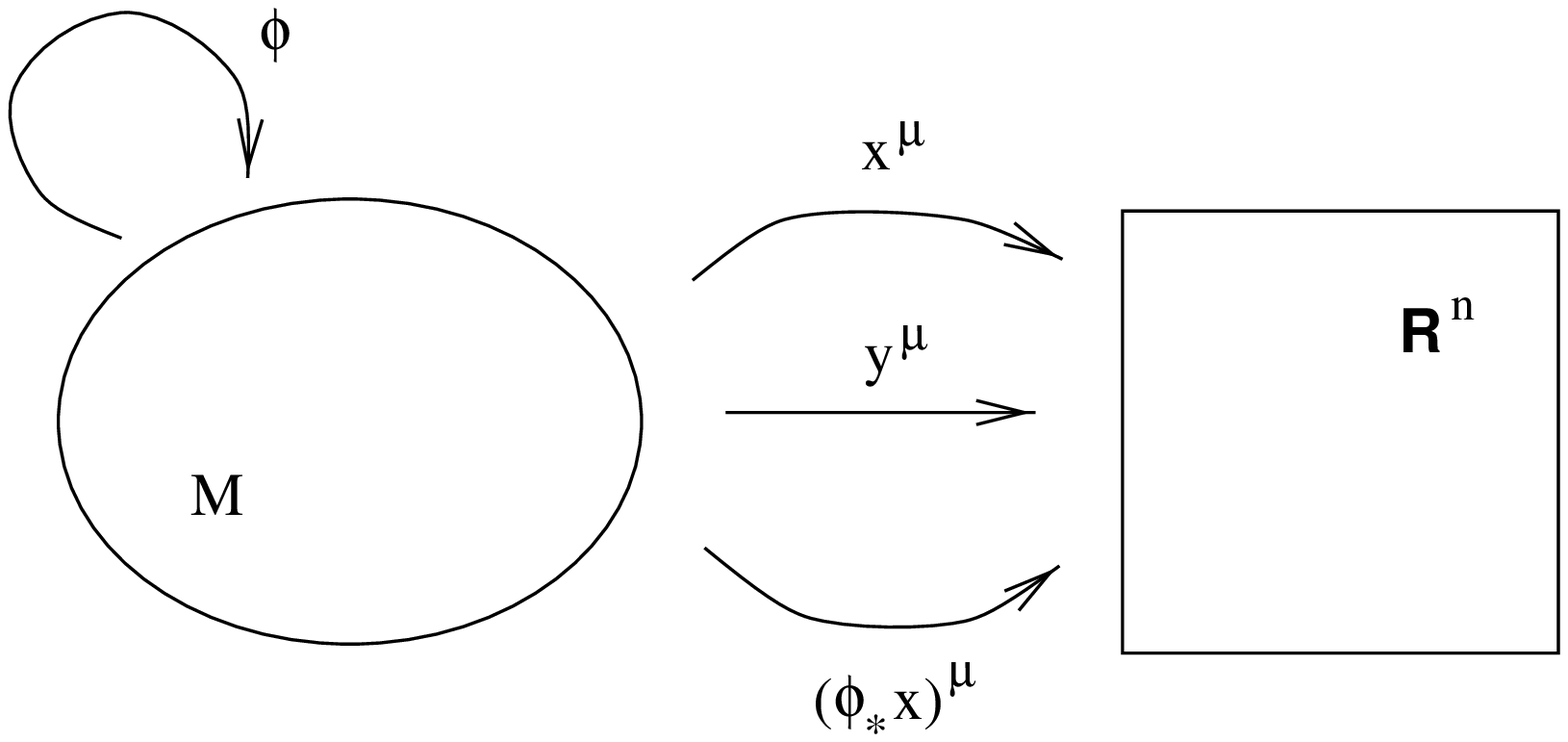,angle=0,height=5cm}}
\end{figure}

Since a diffeomorphism allows us to pull back and push forward arbitrary
tensors, it provides another way of comparing tensors at different 
points on a manifold.  Given a diffeomorphism $\phi:M\rightarrow M$ and
a tensor field $T^{\mu_1 \cdots \mu_k}{}_{\nu_1 \cdots \mu_l}(x)$, we
can form the difference between the value of the tensor at some point
$p$ and $\phi_*[T^{\mu_1 \cdots \mu_k}{}_{\nu_1 \cdots \mu_l}(\phi(p))]$,
its value at $\phi(p)$ pulled back to $p$.
This suggests that we could define another kind of derivative operator
on tensor fields, one which categorizes the rate of change of the
tensor as it changes under the diffeomorphism.  For that, however, a
single discrete diffeomorphism is insufficient; we require a one-parameter
family of diffeomorphisms, $\phi_t$.  This family can be thought of as
a smooth map $\R\times M\rightarrow M$, such that for each $t\in\R$
$\phi_t$ is a diffeomorphism and $\phi_s\circ\phi_t=\phi_{s+t}$.  Note
that this last condition implies that $\phi_0$ is the identity map.

One-parameter families of diffeomorphisms can be thought of as arising
from vector fields (and vice-versa).  If we consider what happens to
the point $p$ under the entire family $\phi_t$, it is clear that it
describes a curve in $M$; since the same thing will be true of every
point on $M$, these curves fill the manifold (although there can be
degeneracies where the diffeomorphisms have fixed points).  We can define
a vector field $V^\mu(x)$ to be the set of tangent vectors to each of
these curves at every point, evaluated at $t=0$.  An example on $S^2$
is provided by the diffeomorphism $\phi_t(\theta,\phi)=(\theta,\phi+t)$.

\begin{figure}
  \centerline{
  \psfig{figure=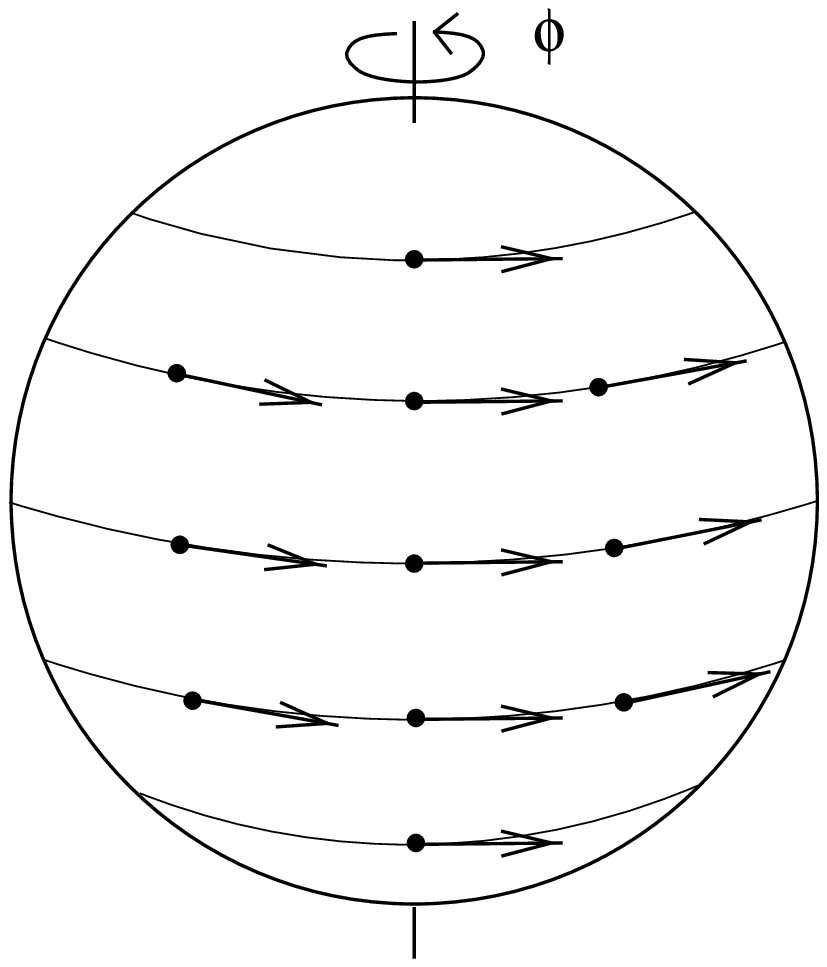,angle=0,height=5cm}}
\end{figure}

We can reverse the construction to define a one-parameter family of
diffeomorphisms from any vector field.  Given a vector field
$V^\mu(x)$, we define the {\bf integral curves} of the vector field
to be those curves $x^\mu(t)$ which solve
\be
  {{dx^\mu}\over {dt}}=V^\mu\ .\label{5.16}
\ee
Note that this familiar-looking equation is now to be interpreted
in the opposite sense from our usual way --- we are given the vectors,
from which we define the curves.  Solutions to (5.16) are guaranteed
to exist as long as we don't do anything silly like run into the
edge of our manifold; any standard differential geometry text will 
have the proof, which amounts to finding a clever coordinate system in
which the problem reduces to the fundamental theorem of ordinary
differential equations.  Our diffeomorphisms $\phi_t$ represent ``flow
down the integral curves,'' and the associated vector field is referred
to as the {\bf generator} of the diffeomorphism.  (Integral curves are
used all the time in elementary physics, just not given the name.
The ``lines of magnetic flux'' traced out by iron filings in the
presence of a magnet are simply the integral curves of the magnetic
field vector {\bf B}.)

Given a vector field $V^\mu(x)$, then, we have a family of 
diffeomorphisms parameterized by $t$, and we can ask how fast a tensor 
changes as we travel down the integral curves.  For each $t$ we can define
this change as
\be
  \Delta_t T^{\mu_1 \cdots \mu_k}{}_{\nu_1 \cdots \mu_l}(p)
  = \phi_{t*}[T^{\mu_1 \cdots \mu_k}{}_{\nu_1 \cdots \mu_l}(\phi_t(p))]
  - T^{\mu_1 \cdots \mu_k}{}_{\nu_1 \cdots \mu_l}(p)\ .\label{5.17}
\ee
Note that both terms on the right hand side are tensors at $p$.

\begin{figure}[h]
  \centerline{
  \psfig{figure=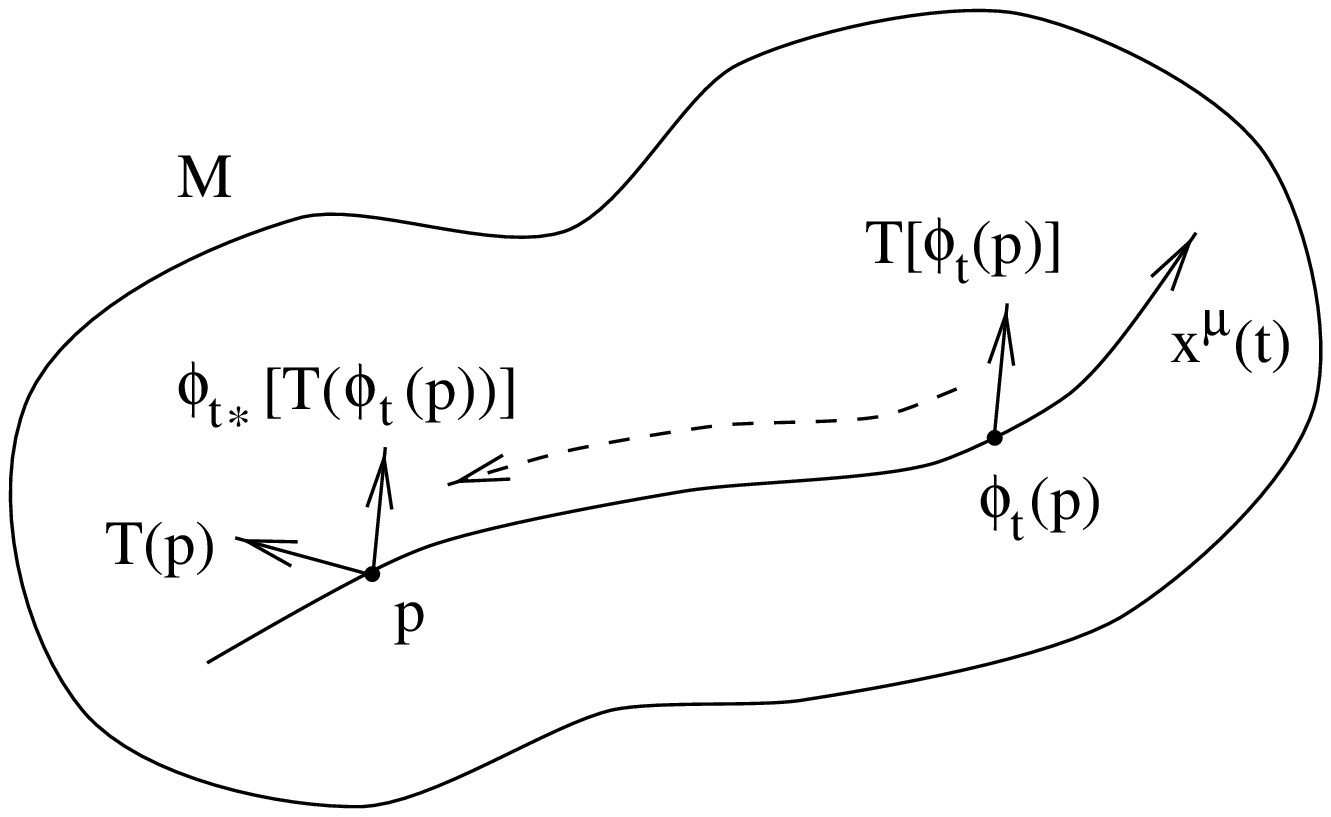,angle=0,height=6cm}}
\end{figure}

\noindent We then define the {\bf Lie derivative} of the tensor along the 
vector field as
\be
  \lie_VT^{\mu_1 \cdots \mu_k}{}_{\nu_1 \cdots \mu_l} =
  \lim_{t\rightarrow 0}\left({{\Delta_t 
  T^{\mu_1 \cdots \mu_k}{}_{\nu_1 \cdots \mu_l}}\over{t}}\right)\ .
  \label{5.18}
\ee
The Lie derivative is a map from $(k,l)$ tensor fields to $(k,l)$ tensor
fields, which is manifestly independent of coordinates.  Since the
definition essentially amounts to the conventional definition of an
ordinary derivative applied to the component functions of the tensor,
it should be clear that it is linear,
\be
  \lie_V(aT+bS) = a\lie_VT + b\lie_VS\ ,\label{5.19}
\ee
and obeys the Leibniz rule,
\be
  \lie_V(T\otimes S) = (\lie_VT)\otimes S+T\otimes(\lie_VS)\ ,\label{5.20}
\ee
where $S$ and $T$ are tensors and $a$ and $b$ are constants.  The Lie
derivative is in fact a more primitive notion than the covariant 
derivative, since it does not require specification of a connection
(although it does require a vector field, of course).  A moment's
reflection shows that it reduces to the ordinary derivative on
functions,
\be
  \lie_Vf = V(f) = V^\mu\p\mu f\ .\label{5.21}
\ee

To discuss the action of the Lie derivative on tensors in terms of
other operations we know, it is convenient to choose a coordinate
system adapted to our problem.  Specifically, we will work in
coordinates $x^\mu$ for which $x^1$ is the parameter along the
integral curves (and the other coordinates
are chosen any way we like).  Then the vector field takes the form
$V=\partial/\partial x^1$; that is, it has components $V^\mu =(1,0,0,
\ldots, 0)$.  The magic of this coordinate system is that a
diffeomorphism by $t$ amounts to a coordinate transformation from
$x^\mu$ to $y^\mu=(x^1+t,x^2,\ldots,x^n)$.  Thus, from (5.6) the
pullback matrix is simply
\be
  (\phi_{t*})_\mu{}^\nu = \delta^\nu_\mu\ ,\label{5.22}
\ee
and the components of the tensor pulled back from $\phi_t(p)$ to
$p$ are simply
\be
  \phi_{t*}[T^{\mu_1 \cdots \mu_k}{}_{\nu_1 \cdots \mu_l}(\phi_t(p))]
  =T^{\mu_1 \cdots \mu_k}{}_{\nu_1 \cdots \mu_l}(x^1+t,x^2,\ldots,x^n)
  \ .\label{5.23}
\ee
In this coordinate system, then, the Lie derivative becomes 
\be
  \lie_VT^{\mu_1 \cdots \mu_k}{}_{\nu_1 \cdots \mu_l} =
  {{\partial}\over{\partial x^1}}
  T^{\mu_1 \cdots \mu_k}{}_{\nu_1 \cdots \mu_l}\ ,\label{5.24}
\ee
and specifically the derivative of a vector field $U^\mu(x)$ is
\be
  \lie_VU^\mu = {{\partial U^\mu}\over{\partial x^1}}\ .\label{5.25}
\ee
Although this expression is clearly not covariant, we know that
the commutator $[V,U]$ is a well-defined tensor, and in this
coordinate system
\bea
  [V,U]^\mu &=&  V^\nu\p\nu U^\mu-U^\nu\p\nu V^\mu\cr
  &=& {{\partial U^\mu}\over{\partial x^1}}\ .\label{5.26}
\eea
Therefore the Lie derivative of $U$ with respect to $V$ has the
same components in this coordinate system as the commutator of
$V$ and $U$; but since both are vectors, they must be equal in
any coordinate system:
\be
  \lie_VU^\mu = [V,U]^\mu\ .\label{5.27}
\ee
As an immediate consequence, we have $\lie_VS=-\lie_WV$.  It is
because of (5.27) that the commutator is sometimes called the ``Lie
bracket.''

To derive the action of $\lie_V$ on a one-form $\omega_\mu$,
begin by considering the action on the scalar $\omega_\mu U^\mu$
for an arbitrary vector field $U^\mu$.  First use the fact that
the Lie derivative with respect to a vector field reduces to the action 
of the vector itself when applied to a scalar:
\bea
  \lie_V(\omega_\mu U^\mu) &=&  V(\omega_\mu U^\mu)\cr
  &=&  V^\nu\p\nu(\omega_\mu U^\mu)\cr
  &=&  V^\nu(\p\nu\omega_\mu)U^\mu + V^\nu\omega_\mu(\p\nu U^\mu)\ .
  \label{5.28}
\eea
Then use the Leibniz rule on the original scalar:
\bea
  \lie_V(\omega_\mu U^\mu) &=&  (\lie_V\omega)_\mu U^\mu
  +\omega_\mu (\lie_V U)^\mu \cr
  &=&  (\lie_V\omega)_\mu U^\mu + \omega_\mu V^\nu\p\nu U^\mu
  -\omega_\mu U^\nu\p\nu V^\mu\ .\label{5.29}
\eea
Setting these expressions equal to each other and requiring that 
equality hold for arbitrary $U^\mu$, we see that
\be
  \lie_V \omega_\mu = V^\nu\p\nu \omega_\mu + (\p\mu V^\nu)
  \omega_\nu\ ,\label{5.30}
\ee
which (like the definition of the commutator) is completely covariant,
although not manifestly so.

By a similar procedure we can define the Lie derivative of an
arbitrary tensor field.  The answer can be written
\bea
  \lie_V T^{\mu_1 \mu_2 \cdots \mu_k}{}_{\nu_1
  \nu_2 \cdots \nu_l} &=&   V^\sigma\partial_\sigma T^{\mu_1 \mu_2 \cdots 
  \mu_k}{}_{\nu_1 \nu_2 \cdots \nu_l} \cr
  && \quad -(\p\lambda V^{\mu_1}) T^{\lambda \mu_2 \cdots 
  \mu_k}{}_{\nu_1 \nu_2 \cdots \nu_l} 
  -(\p\lambda V^{\mu_2}) T^{\mu_1 \lambda \cdots 
  \mu_k}{}_{\nu_1 \nu_2 \cdots \nu_l} -\cdots\cr
  &&\quad +(\p{\nu_1}V^\lambda)T^{\mu_1 \mu_2 \cdots 
  \mu_k}{}_{\lambda \nu_2 \cdots \nu_l}
  +(\p{\nu_2}V^\lambda)T^{\mu_1 \mu_2 \cdots \mu_k}{}_{\nu_1
  \lambda \cdots \nu_l} + \cdots \ .\label{5.31}
\eea
Once again, this expression is covariant, despite appearances.  It
would undoubtedly be comforting, however, to have an equivalent
expression that looked manifestly tensorial.  In fact it turns out
that we can write
\bea
  \lie_V T^{\mu_1 \mu_2 \cdots \mu_k}{}_{\nu_1
  \nu_2 \cdots \nu_l} &=&  V^\sigma\nabla_\sigma T^{\mu_1 \mu_2 \cdots 
  \mu_k}{}_{\nu_1 \nu_2 \cdots \nu_l} \cr
  && \quad -(\nabla_\lambda V^{\mu_1}) T^{\lambda \mu_2 \cdots 
  \mu_k}{}_{\nu_1 \nu_2 \cdots \nu_l} 
  -(\nabla_\lambda V^{\mu_2}) T^{\mu_1 \lambda \cdots 
  \mu_k}{}_{\nu_1 \nu_2 \cdots \nu_l} -\cdots\cr
  &&\quad +(\nabla_{\nu_1}V^\lambda)T^{\mu_1 \mu_2 \cdots 
  \mu_k}{}_{\lambda \nu_2 \cdots \nu_l}
  +(\nabla_{\nu_2}V^\lambda)T^{\mu_1 \mu_2 \cdots \mu_k}{}_{\nu_1
  \lambda \cdots \nu_l} + \cdots \ ,\label{5.32}
\eea
where $\nabla_\mu$ represents {\it any} symmetric (torsion-free)
covariant derivative (including, of course, one derived from a 
metric).  You can check that all of the terms which would involve
connection coefficients if we were to expand (5.32) would cancel,
leaving only (5.31).  Both versions of the formula for a Lie derivative
are useful at different times.  A particularly useful formula is for
the Lie derivative of the metric:
\bea
  \lie_V g_\mn &=&  V^\sigma\nabla_\sigma g_\mn
  +(\nabla_{\mu}V^\lambda)g_{\lambda\nu} +(\nabla_{\nu}V^\lambda)
  g_{\mu\lambda}\cr
  &=& \nabla_\mu V_\nu + \nabla_\nu V_\mu\cr
  &=&  2\nabla_{(\mu} V_{\nu)}\ ,\label{5.33}
\eea
where $\nabla_\mu$ is the covariant derivative derived from $g_\mn$. 

Let's put some of these ideas into the context of general relativity.
You will often hear it proclaimed that GR is a ``diffeomorphism
invariant'' theory.  What this means is that, if the universe is 
represented by a manifold $M$ with metric $g_\mn$ and matter fields
$\psi$, and $\phi:M\rightarrow M$ is a diffeomorphism, then the sets 
$(M,g_\mn,\psi)$ and $(M, \phi_*g_\mn,\phi_*\psi)$ represent the same
physical situation.  Since diffeomorphisms are just active coordinate
transformations, this is a highbrow way of saying that the theory is
coordinate invariant.  Although such a statement is true, it is a 
source of great misunderstanding, for the simple fact that it conveys
very little information.  Any semi-respectable theory of physics is
coordinate invariant, including those based on special relativity or
Newtonian mechanics; GR is not unique in this regard.  When people say 
that GR is diffeomorphism invariant, more likely than not they have one 
of two (closely related) concepts in mind: the theory is free of ``prior 
geometry'', and there is no preferred coordinate system for spacetime.
The first of these stems from the fact that the metric is a 
dynamical variable, and along with it the connection and volume
element and so forth.  Nothing is given to us ahead of time, unlike in
classical mechanics or SR.  As a consequence, there is no way to
simplify life by sticking to a specific coordinate system adapted to
some absolute elements of the geometry.  This state of affairs forces
us to be very careful; it is possible that two purportedly distinct
configurations (of matter and metric) in GR are actually ``the same'',
related by a diffeomorphism.  In a path integral approach to quantum
gravity, where we would like to sum over all possible configurations,
special care must be taken not to overcount by allowing physically
indistinguishable configurations to contribute more than once.  In
SR or Newtonian mechanics, meanwhile, the existence of a preferred 
set of coordinates saves us from such ambiguities.  The fact that GR
has no {\it preferred} coordinate system is often garbled into the
statement that it is coordinate invariant (or ``generally covariant'');
both things are true, but one has more content than the other.

On the other hand, the fact of diffeomorphism invariance can be put
to good use.  Recall that the complete action for gravity coupled to
a set of matter fields $\psi^i$ is given by a sum of the Hilbert action
for GR plus the matter action,
\be
  S={1\over{8\pi G}}S_H[g_\mn]+S_M[g_\mn,\psi^i]\ .\label{5.34}
\ee
The Hilbert action $S_H$ is diffeomorphism invariant when considered
in isolation, so the matter action $S_M$ must also be if the action as
a whole is to be invariant.  We can write the variation in $S_M$ under
a diffeomorphism as
\be
  \delta S_M=\int d^nx {{\delta S_M}\over{\delta g_\mn}}\delta g_\mn
  + \int d^nx {{\delta S_M}\over{\delta \psi^i}}\delta \psi^i\ .
  \label{5.35}
\ee
We are not considering arbitrary variations of the fields, only those
which result from a diffeomorphism.  Nevertheless, the matter equations
of motion tell us that the variation of $S_M$ with respect to $\psi^i$
will vanish for any variation (since the gravitational part of the action
doesn't involve the matter fields).  Hence, for a diffeomorphism
invariant theory the first term on the right hand side of (5.35) must
vanish.  If the diffeomorphism in generated by a vector field $V^\mu(x)$,
the infinitesimal change in the metric is simply given by its Lie
derivative along $V^\mu$; by (5.33) we have
\bea
  \delta g_\mn &=& \lie_V g_\mn\cr
  &=&  2\nabla_{(\mu} V_{\nu)}\ .\label{5.36}
\eea
Setting $\delta S_M=0$ then implies
\bea
  0 &=&  \int d^nx {{\delta S_M}\over{\delta g_\mn}}
  \nabla_{\mu} V_{\nu} \cr
  &=&  -\int d^nx\g V_\nu\nabla_\mu\left({1\over{\g}}{{\delta S_M}\over
  {\delta g_\mn}}\right)\ ,\label{5.37}
\eea
where we are able to drop the symmetrization of $\nabla_{(\mu} V_{\nu)}$
since $\delta S_M/\delta g_\mn$ is already symmetric.  Demanding that
(5.37) hold for diffeomorphisms generated by arbitrary vector fields
$V^\mu$, and using the definition (4.70) of the energy-momentum tensor,
we obtain precisely the law of energy-momentum conservation,
\be
  \nabla_\mu T^\mn=0\ .\label{5.38}
\ee
This is why we claimed earlier that the conservation of $T_\mn$ was
more than simply a consequence of the Principle of Equivalence; it is
much more secure than that, resting only on the diffeomorphism
invariance of the theory.

There is one more use to which we will put the machinery we have
set up in this section: symmetries of tensors.  We say that a
diffeomorphism $\phi$ is a {\bf symmetry} of some tensor
$T$ if the tensor is invariant after being pulled back under $\phi$:
\be
  \phi_*T = T\ .\label{5.39}
\ee
Although symmetries may be discrete, it is more common to have a
one-parameter family of symmetries $\phi_t$.  If the family is
generated by a vector field $V^\mu(x)$, then (5.39) amounts to
\be
  \lie_V T=0\ .\label{5.40}
\ee
By (5.25), one implication of a symmetry is that, if $T$ is
symmetric under some one-parameter family of diffeomorphisms, we can
always find a coordinate system in which the components of $T$ are
all independent of one of the coordinates (the integral curve
coordinate of the vector field).  The converse is also true; if 
all of the components are independent of one of the coordinates,
then the partial derivative vector field associated with that 
coordinate generates a symmetry of the tensor.

The most important symmetries are those of the metric, for which
$\phi_*g_\mn = g_\mn$.  A diffeomorphism of this type is called an
{\bf isometry}.  If a one-parameter family of isometries is 
generated by a vector field $V^\mu(x)$, then $V^\mu$ is known as
a {\bf Killing vector field}.  The condition that $V^\mu$ be a Killing
vector is thus
\be
  \lie_V g_\mn=0\ ,\label{5.41}
\ee
or from (5.33),
\be
  \nabla_{(\mu}V_{\nu)}=0\ .\label{5.42}
\ee
This last version is {\bf Killing's equation}.
If a spacetime has a Killing vector, then we know we can find a 
coordinate system in which the metric is independent of one of the
coordinates.  

By far the most useful fact about Killing vectors is that {\it Killing 
vectors imply conserved quantities associated with the motion of
free particles}.  If $x^\mu(\lambda)$ is a geodesic with tangent
vector $U^\mu=dx^\mu/d\lambda$, and $V^\mu$ is a Killing vector, then
\bea
  U^\nu\nabla_\nu(V_\mu U^\mu)&=&  U^\nu U^\mu \nabla_\nu V_\mu
  +V_\mu U^\nu \nabla_\nu U^\mu \cr
  &=& 0\ ,\label{5.43}
\eea
where the first term vanishes from Killing's equation and the second
from the fact that $x^\mu(\lambda)$ is a geodesic.  Thus, the
quantity $V_\mu U^\mu$ is conserved along the particle's worldline.
This can be understood physically: by definition the metric is
unchanging along the direction of the Killing vector.  Loosely speaking,
therefore, a free particle will not feel any ``forces''
in this direction, and the component of its momentum in that
direction will consequently be conserved.

Long ago we referred to the concept of a space with maximal symmetry, 
without offering a rigorous definition.  The rigorous
definition is that a {\bf maximally symmetric space} is one which
possesses the largest possible number of Killing vectors, which on
an $n$-dimensional manifold is $n(n+1)/2$.  We will not prove
this statement, but it is easy to understand at an informal level.
Consider the Euclidean space $\R^n$, where the isometries are well
known to us: translations and rotations.  In general there will
be $n$ translations, one for each direction we can move.  There will
also be $n(n-1)/2$ rotations; for each of $n$ dimensions there are
$n-1$ directions in which we can rotate it, but we must divide by
two to prevent overcounting (rotating $x$ into $y$ and rotating $y$ 
into $x$ are two versions of the same thing).  We therefore have
\be
  n+{{n(n-1)}\over 2}=  {{n(n+1)}\over2} \label{5.44}
\ee
independent Killing vectors.  The same kind of counting argument applies
to maximally symmetric spaces with curvature (such as spheres) or
a non-Euclidean signature (such as Minkowski space), although the 
details are marginally different.

Although it may or may not be simple to actually solve Killing's
equation in any given spacetime, it is frequently possible to 
write down some Killing vectors by inspection.  (Of course a ``generic''
metric has no Killing vectors at all, but to keep things simple we
often deal with metrics with high degrees of symmetry.)  For example
in $\R^2$ with metric $ds^2 = \d x^2+\d y^2$, independence of the
metric components with respect to $x$ and $y$ immediately yields two
Killing vectors: 
\bea
  X^\mu&=& (1,0)\ ,\cr
  Y^\mu&=& (0,1)\ .\label{5.45}
\eea
These clearly 
represent the two translations.  The one rotation would correspond to
the vector $R=\partial/\partial\theta$ if we were in polar coordinates;
in Cartesian coordinates this becomes
\be
  R^\mu = (-y,x)\ .\label{5.46}
\ee  
You can check for yourself that this actually does solve Killing's
equation.

Note that in $n\geq 2$ dimensions, there can be more Killing vectors
than dimensions.  This is because a set of Killing vector fields can
be linearly independent, even though at any one point on the manifold
the vectors at that point are linearly dependent.  It is trivial to
show (so you should do it yourself) that a linear combination
of Killing vectors with {\it constant} coefficients is still a Killing
vector (in which case the linear combination does not count as an
independent Killing vector), 
but this is not necessarily true with coefficients which 
vary over the manifold.  You will also show that the commutator of two
Killing vector fields is a Killing vector field; this is very useful
to know, but it may be the case that the commutator gives you a vector
field which is not linearly independent (or it may simply vanish).
The problem of finding all of the Killing vectors of a metric is therefore
somewhat tricky, as it is sometimes not clear when to stop looking.

\eject
\thispagestyle{plain}

\setcounter{equation}{0}

\noindent{December 1997 \hfill {\sl Lecture Notes on General Relativity}
\hfill{Sean M.~Carroll}}

\vskip .2in

\setcounter{section}{5}
\section{Weak Fields and Gravitational Radiation}

When we first derived Einstein's equations, we checked that we were
on the right track by considering the Newtonian limit.  This amounted
to the requirements that the gravitational field be weak, that it
be static (no time derivatives), and that test particles be moving
slowly.  In this section we will consider a less restrictive situation,
in which the field is still weak but it can vary with time, and there
are no restrictions on the motion of test particles.  This will 
allow us to discuss phenomena which are absent or ambiguous in the
Newtonian theory, such as gravitational radiation (where the field
varies with time) and the deflection of light (which involves
fast-moving particles).

The weakness of the gravitational field is once again expressed as our 
ability to decompose the metric into the flat Minkowski metric plus a
small perturbation,
\be
  g_\mn = \eta_\mn + h_\mn\ ,\qquad |h_\mn |<<1\ .\label{6.1}
\ee
We will restrict ourselves to coordinates in which $\eta_\mn$
takes its canonical form, $\eta_\mn = {\rm diag}(-1,+1,+1,+1)$.  The
assumption that $h_\mn$ is small allows us to ignore anything that
is higher than first order in this quantity, from which we immediately
obtain
\be
  g^\mn = \eta^\mn - h^\mn\ ,\label{6.2}
\ee
where $h^\mn = \eta^{\mu\rho}\eta^{\nu\sigma}h_{\rho\sigma}$. As before, 
we can raise and lower indices using $\eta^\mn$ and $\eta_\mn$, since
the corrections would be of higher order in the perturbation.
In fact, we can think of the linearized version of general relativity
(where effects of higher than first order in $h_\mn$ are neglected)
as describing a theory of a symmetric tensor field $h_\mn$
propagating on a flat background spacetime.  This theory is Lorentz
invariant in the sense of special relativity; under a Lorentz
transformation $x^{\mu'} = \Lambda^{\mu'}{}_\mu x^\mu$, the
flat metric $\eta_\mn$ is invariant, while the perturbation transforms
as 
\be
  h_{\mu'\nu'}=\Lambda_{\mu'}{}^\mu \Lambda_{\nu'}{}^\nu h_\mn\ .
  \label{6.3}
\ee
(Note that we could have considered small perturbations about some
other background spacetime besides Minkowski space.  In that case
the metric would have been written $g_\mn = g_\mn^{(0)}+h_\mn$, and
we would have derived a theory of a symmetric tensor propagating on
the curved space with metric $g_\mn^{(0)}$.  Such an approach is
necessary, for example, in cosmology.)

We want to find the equation of motion obeyed by the perturbations
$h_\mn$, which come by examining Einstein's equations to first order.
We begin with the Christoffel symbols, which are given by
\bea
  \Gamma^\rho_{\mn}&=& {1\over 2} g^{\rho\lambda}
  (\p\mu g_{\nu\lambda} + \p\nu g_{\lambda\mu} - \p\lambda g_{\mn})\cr
  &=& {1\over 2}\eta^{\rho\lambda}(\p\mu h_{\nu\lambda} 
  + \p\nu h_{\lambda\mu} - \p\lambda h_{\mn})\ . \label{6.4}
\eea
Since the connection coefficients are first order quantities, the
only contribution to the Riemann tensor will come from the derivatives
of the $\Gamma$'s, not the $\Gamma^2$ terms.  Lowering an index for
convenience, we obtain
\bea
  R_{\mn\rho\sigma} &=&  \eta_{\mu\lambda}\p\rho
  \Gamma^\lambda_{\nu\sigma} - \eta_{\mu\lambda}\p\sigma
  \Gamma^\lambda_{\nu\rho} \cr 
  &=& {1\over 2}(\p\rho\p\nu h_{\mu\sigma} +  \p\sigma\p\mu h_{\nu\rho}
  -\p\sigma\p\nu h_{\mu\rho}-\p\rho\p\mu h_{\nu\sigma})\ .
  \label{6.5}
\eea
The Ricci tensor comes from contracting over $\mu$ and $\rho$,
giving
\be
  R_\mn = {1\over 2}(\p\sigma\p\nu h^\sigma{}_\mu
  +\p\sigma\p\mu h^\sigma{}_\nu - \p\mu\p\nu h - \boxx h_\mn)\ ,
  \label{6.6}
\ee
which is manifestly symmetric in $\mu$ and $\nu$.  In this expression
we have defined the trace of the perturbation as $h=\eta^\mn h_\mn =
h^\mu{}_\mu$, and the D'Alembertian is simply the one from flat space,
$\boxx = -\p{t}^2+\p{x}^2+\p{y}^2+\p{z}^2$.  Contracting again to
obtain the Ricci scalar yields
\be
  R = \p\mu\p\nu h^\mn - \boxx h\ .\label{6.7}
\ee
Putting it all together we obtain the Einstein tensor:
\bea
  G_\mn &=&  R_\mn - {1\over 2}\eta_\mn R\cr
  &=& {1\over 2}(\p\sigma\p\nu h^\sigma{}_\mu
  +\p\sigma\p\mu h^\sigma{}_\nu - \p\mu\p\nu h - \boxx h_\mn 
  -\eta_\mn \p\mu\p\nu h^\mn + \eta_\mn \boxx h)\ . \label{6.8}
\eea
Consistent with our interpretation of the linearized theory as
one describing a symmetric tensor on a flat background, the linearized
Einstein tensor (6.8) can be derived by varying the following
Lagrangian with respect to $h_\mn$:
\be
  {\cal L} = {1\over 2}\left[(\p\mu h^\mn)(\p\nu h) - 
  (\p\mu h^{\rho\sigma})(\p\rho h^\mu{}_\sigma) + {1\over 2}
  \eta^\mn(\p\mu h^{\rho\sigma})(\p\nu h_{\rho\sigma})
  -{1\over 2}\eta^\mn(\p\mu h)(\p\nu h)\right]\ .\label{6.9}
\ee
I will spare you the details.

The linearized field equation is of course $G_\mn=8\pi GT_\mn$,
where $G_\mn$ is given by (6.8) and $T_\mn$ is the energy-momentum
tensor, calculated to zeroth order in $h_\mn$.  We do not include
higher-order corrections to the energy-momentum tensor because the
amount of energy and momentum must itself be small for the weak-field
limit to apply.  In other words, the lowest nonvanishing order in
$T_\mn$ is automatically of the same order of magnitude as the
perturbation.  Notice that the conservation law to lowest order is
simply $\p\mu T^\mn=0$.  We will most often be concerned
with the vacuum equations, which as usual are just $R_\mn=0$, where
$R_\mn$ is given by (6.6).

With the linearized field equations in hand, we are almost prepared
to set about solving them.  First, however, we should deal with the
thorny issue of gauge invariance.  This issue arises because the
demand that $g_\mn=\eta_\mn+h_\mn$ does not completely specify the
coordinate system on spacetime; there may be other coordinate systems
in which the metric can still be written as the Minkowski metric
plus a small perturbation, but the perturbation will be different.  
Thus, the decomposition of the metric into a flat background plus a
perturbation is not unique.

We can think about this from a highbrow point of view.  The notion
that the linearized theory can be thought of as one governing the
behavior of tensor fields on a flat background can be formalized in
terms of a ``background spacetime'' $M_b$, a ``physical spacetime''
$M_p$, and a diffeomorphism $\phi:M_b\rightarrow M_p$.  As manifolds
$M_b$ and $M_p$ are ``the same'' (since they are diffeomorphic), but
we imagine that they possess some different tensor fields;
on $M_b$ we have defined the flat Minkowski metric $\eta_\mn$, while
on $M_p$ we have some metric $g_{\alpha\beta}$ which obeys Einstein's
equations.  (We imagine that $M_b$ is equipped with coordinates
$x^\mu$ and $M_p$ is equipped with coordinates $y^\alpha$, although
these will not play a prominent role.)
The diffeomorphism $\phi$ allows us to move tensors back
and forth between the background and physical spacetimes.  Since we
would like to construct our linearized theory as one taking place 
on the flat background spacetime, we are interested in the pullback
$(\phi_*g)_\mn$ of the physical metric.  We can define the perturbation
as the difference between the pulled-back physical metric and the
flat one:
\be
  h_\mn = (\phi_*g)_\mn - \eta_\mn\ .\label{6.10}
\ee
From this definition, there is no reason for the components of $h_\mn$
to be small; however, if the gravitational fields on $M_p$ are weak,
then for {\it some} diffeomorphisms $\phi$ we will have $|h_\mn| << 1$.
We therefore limit our attention only to those diffeomorphisms for which
this is true.  Then the fact that $g_{\alpha\beta}$ obeys Einstein's
equations on the physical spacetime means that $h_\mn$ will obey
the linearized equations on the background spacetime (since $\phi$,
as a diffeomorphism, can be used to pull back Einstein's equations
themselves).

\begin{figure}[h]
  \centerline{
  \psfig{figure=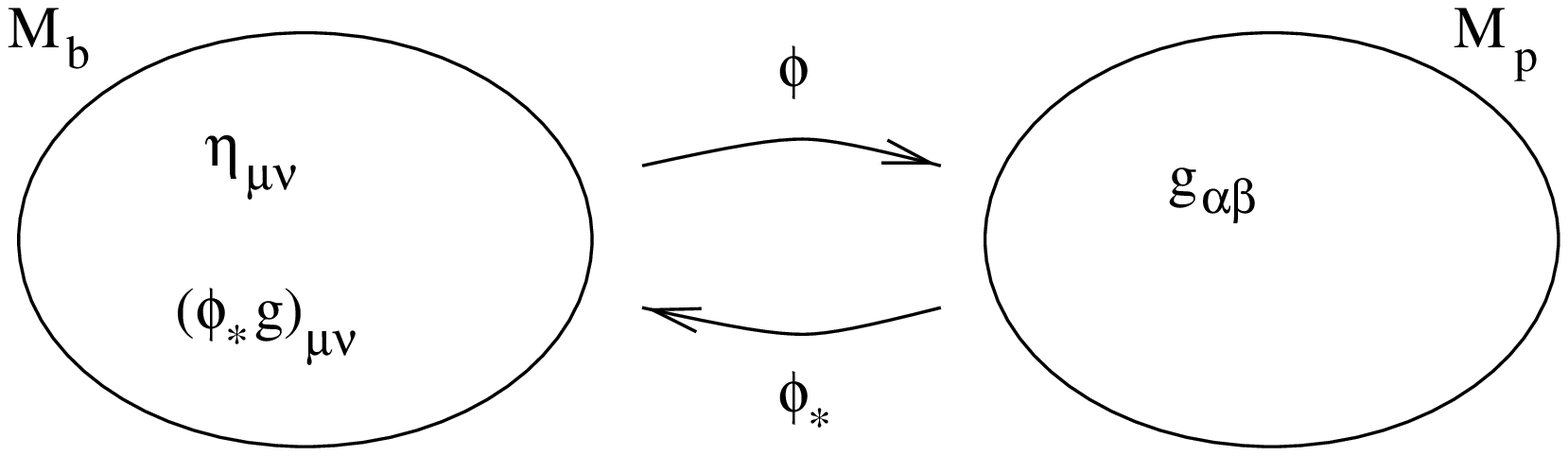,angle=0,height=4cm}}
\end{figure}

In this language, the issue of gauge invariance is simply the fact that
there are a large number of permissible diffeomorphisms between $M_b$
and $M_p$ (where ``permissible'' means that the perturbation is small).
Consider a vector field $\xi^\mu(x)$ on the background spacetime.
This vector field generates a one-parameter family of diffeomorphisms
$\psi_\epsilon:M_b\rightarrow M_b$.  For $\epsilon$ sufficiently small,
if $\phi$ is a diffeomorphism for which the perturbation defined 
by (6.10) is small than so will $(\phi\circ\psi_\epsilon)$ be, although
the perturbation will have a different value.  

\begin{figure}[h]
  \centerline{
  \psfig{figure=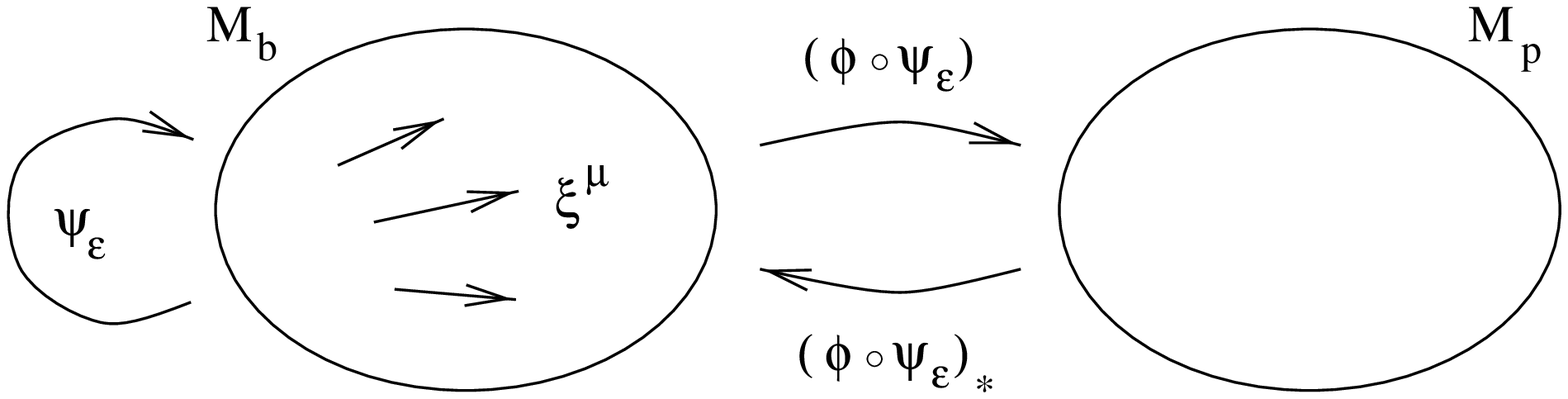,angle=0,height=4cm}}
\end{figure}

\noindent Specifically, we can 
define a family of perturbations parameterized by $\epsilon$:
\bea
  h_\mn^{(\epsilon)} &=&  [(\phi\circ\psi_\epsilon)_*g]_\mn 
  - \eta_\mn\cr
  &=& [\psi_{\epsilon *}(\phi_*g)]_\mn - \eta_\mn\ . \label{6.11}
\eea
The second equality is based on the fact that the pullback under a
composition is given by the composition of the pullbacks in the
opposite order, which follows from the fact that the pullback itself
moves things in the opposite direction from the original map.  Plugging
in the relation (6.10), we find
\bea
  h_\mn^{(\epsilon)} &=&  \psi_{\epsilon *}(h +\eta)_\mn
  -\eta_\mn \cr
  &=&  \psi_{\epsilon *}(h_\mn) +\psi_{\epsilon *}(\eta_\mn)-\eta_\mn
  \label{6.12}
\eea
(since the pullback of the sum of two tensors is the sum of the 
pullbacks).  Now we use our assumption that $\epsilon$ is small; in
this case $\psi_{\epsilon *}(h_\mn)$ will be equal to $h_\mn$ to
lowest order, while the other two terms give us a Lie derivative:
\bea
  h_\mn^{(\epsilon)} &=&  \psi_{\epsilon *}(h_\mn)
  +\epsilon\left[{{\psi_{\epsilon *}(\eta_\mn)-\eta_\mn}\over 
  \epsilon}\right] \cr
  &=& h_\mn + \epsilon \lie_\xi\eta_\mn \cr
  &=& h_\mn + 2\epsilon\partial_{(\mu}\xi_{\nu)}\ . \label{6.13}
\eea
The last equality follows from our previous computation of the Lie
derivative of the metric, (5.33), plus the fact that covariant 
derivatives are simply partial derivatives to lowest order.

The infinitesimal diffeomorphisms $\phi_\epsilon$ provide a
different representation of the same physical situation, while
maintaining our requirement that the perturbation be small.  Therefore,
the result (6.12) tells us what kind of metric perturbations denote
physically equivalent spacetimes --- those related to each other by
$2\epsilon\partial_{(\mu}\xi_{\nu)}$, for some vector $\xi^\mu$.
The invariance of our theory under such transformations is analogous
to traditional gauge invariance of electromagnetism under 
$A_\mu \rightarrow A_\mu + \p\mu\lambda$.  (The analogy is
different from the previous analogy we drew with electromagnetism,
relating local Lorentz transformations in the orthonormal-frame
formalism to changes of basis in an internal vector bundle.)  In
electromagnetism the invariance comes about because the field
strength $F_\mn = \p\mu A_\nu - \p\nu A_\mu$ is left unchanged by
gauge transformations; similarly, we find that the transformation (6.13)
changes the linearized Riemann tensor by
\bea
  \delta R_{\mn\rho\sigma} &=&  
  {1\over 2}(\p\rho\p\nu\p\mu\xi_\sigma +\p\rho\p\nu\p\sigma\xi_\mu 
  + \p\sigma\p\mu\p\nu\xi_\rho + \p\sigma\p\mu\p\rho\xi_\nu \cr
  & & \qquad - \p\sigma\p\nu\p\mu\xi_\rho - \p\sigma\p\nu\p\rho\xi_\mu 
  - \p\rho\p\mu\p\nu\xi_\sigma - \p\rho\p\mu\p\sigma\xi_\nu )\cr
  & =& 0\ . \label{6.14}
\eea
Our abstract derivation of the appropriate gauge transformation for
the metric perturbation is verified by the fact that it leaves the
curvature (and hence the physical spacetime) unchanged.

Gauge invariance can also be understood from the slightly more
lowbrow but considerably more direct route of infinitesimal coordinate
transformations.  Our diffeomorphism $\psi_\epsilon$ can be thought
of as changing coordinates from $x^\mu$ to $x^\mu -\epsilon\xi^\mu$.
(The minus sign, which is unconventional, comes from the fact that the
``new'' metric is pulled back from a small distance forward along the 
integral curves, which is equivalent to replacing the coordinates by
those a small distance backward along the curves.)  Following through
the usual rules for transforming tensors under coordinate transformations,
you can derive precisely (6.13) --- although you have to cheat somewhat
by equating components of tensors in two different coordinate systems.
See Schutz or Weinberg for an example.

When faced with a system that is invariant under some kind of gauge
transformations, our first instinct is to fix a gauge.  We have
already discussed the harmonic coordinate system, and will return to
it now in the context of the weak field limit.  Recall that this gauge
was specified by $\boxx x^\mu=0$, which we showed was equivalent to
\be
  g^\mn \Gamma^\rho_\mn =0\ .\label{6.15}
\ee
In the weak field limit this becomes
\be
  {1\over 2}\eta^\mn \eta^{\lambda\rho}(\p\mu h_{\nu\lambda} 
  +\p\nu h_{\lambda\mu} -\p\lambda h_\mn)=0\ ,\label{6.16}
\ee
or
\be
  \p\mu h^\mu{}_\lambda - {1\over 2}\p\lambda h = 0\ .\label{6.17}
\ee
This condition is also known as Lorentz gauge (or Einstein gauge or
Hilbert gauge or de~Donder gauge or Fock gauge).
As before, we still have some gauge freedom remaining, since we can
change our coordinates by (infinitesimal) harmonic functions.

In this gauge, the linearized Einstein equations $G_\mn = 8\pi GT_\mn$
simplify somewhat, to
\be
   \boxx h_\mn - {1\over 2}\eta_\mn \boxx h=-16\pi GT_\mn\ ,\label{6.18}
\ee
while the vacuum equations $R_\mn=0$ take on the elegant form
\be
  \boxx h_\mn=0\ ,\label{6.19}
\ee
which is simply the conventional relativistic wave equation.
Together, (6.19) and (6.17) determine the evolution of a disturbance
in the gravitational field in vacuum in the harmonic gauge.

It is often convenient to work with a slightly different description
of the metric perturbation.  We define the ``trace-reversed''
perturbation $\bar h_\mn$ by
\be
  \bar h_\mn= h_\mn - {1\over 2}\eta_\mn h\ .\label{6.20}
\ee
The name makes sense, since $\bar h^\mu{}_\mu=-h^\mu{}_\mu$.  (The
Einstein tensor is simply the trace-reversed Ricci tensor.)  In 
terms of $\bar h_\mn$ the harmonic gauge condition becomes
\be
  \p\mu \bar h^\mu{}_\lambda =0\ .\label{6.21}
\ee
The full field equations are
\be
  \boxx \bar h_\mn = -16\pi G T_\mn\ ,\label{6.22}
\ee
from which it follows immediately that the vacuum equations are
\be
  \boxx \bar h_\mn = 0\ .\label{6.23}
\ee

From (6.22) and our previous exploration of the Newtonian limit, it
is straightforward to derive the weak-field metric for a stationary
spherical source such as a planet or star.  Recall that previously we 
found that Einstein's
equations predicted that $h_{00}$ obeyed the Poisson equation (4.51)
in the weak-field limit, which implied
\be
  h_{00} = -2\Phi\ ,\label{6.24}
\ee 
where $\Phi$ is the conventional Newtonian potential, $\Phi=-GM/r$.
Let us now assume that the energy-momentum tensor of our source is
dominated by its rest energy density $\rho=T_{00}$.  (Such an
assumption is not generally necessary in the weak-field limit, but
will certainly hold for a planet or star, which is what we would
like to consider for the moment.)  Then the other components of
$T_\mn$ will be much smaller than $T_{00}$, and from (6.22) the same
must hold for $\bar h_\mn$. 
If $\bar h_{00}$ is much larger than $\bar h_{ij}$, we will have
\be
  h = -\bar h=-\eta^\mn \bar h_\mn = \bar h_{00}\ ,\label{6.25}
\ee
and then from (6.20) we immediately obtain
\be
  \bar h_{00} = 2 h_{00} =-4\Phi\ .\label{6.26}
\ee
The other components of $\bar h_{\mn}$ are negligible, from which
we can derive
\be
  h_{i0} = \bar h_{i0} - {1\over 2}\eta_{i0}\bar h = 0\ ,\label{6.27}
\ee
and
\be
  h_{ij}= \bar h_{ij} - {1\over 2}\eta_{ij}\bar h = -2\Phi\delta_{ij}\ .
  \label{6.28}
\ee
The metric for a star or planet in the weak-field limit is therefore
\be
  ds^2 = -(1+2\Phi)\d t^2 +(1-2\Phi)(\d x^2 +\d y^2 +\d z^2)\ .
  \label{6.29}
\ee

A somewhat less simplistic application of the weak-field limit
is to gravitational radiation.  Those of you familiar with the 
analogous problem in electromagnetism will notice that the 
procedure is almost precisely the same.  We begin by considering the 
linearized equations in vacuum (6.23).  Since the flat-space
D'Alembertian has the form $\boxx = -\p{t}^2 +\nabla^2$, the field
equation is in the form of a wave equation for $\bh_\mn$.  As all
good physicists know, the thing to do when faced with such an equation
is to write down complex-valued solutions, and then take the real
part at the very end of the day.  So we recognize that a particularly
useful set of solutions to this wave equation are the plane waves, given 
by
\be
  \bh_\mn = C_\mn e^{ik_\sigma x^\sigma}\ ,\label{6.30}
\ee
where $C_\mn$ is a constant, symmetric, $(0,2)$ tensor, and $k^\sigma$
is a constant vector known as the {\bf wave vector}.   
To check that it is a solution, we plug in:
\bea
  0&=& \boxx \bh_\mn\cr &=& \eta^{\rho\sigma}\p\rho\p\sigma
  \bh_\mn\cr &=&  \eta^{\rho\sigma}\p\rho (i k_\sigma\bh_\mn)\cr
  &=&  - \eta^{\rho\sigma}k_\rho k_\sigma\bh_\mn\cr
  &=&  -k_\sigma k^\sigma \bh_\mn\ . \label{6.31}
\eea
Since (for an interesting solution) not all of the components of
$h_\mn$ will be zero everywhere, we must have
\be
  k_\sigma k^\sigma=0\ .\label{6.32}
\ee
The plane wave (6.30) is therefore a solution to the linearized
equations if the wavevector is null; this is loosely translated into
the statement that gravitational waves propagate at the speed of light.
The timelike component of the wave vector is often referred to as 
the {\bf frequency} of the wave, and we write $k^\sigma = (\omega,
k^1,k^2,k^3)$.  (More generally, an observer moving with four-velocity
$U^\mu$ would observe the wave to have a frequency $\omega=-k_\mu U^\mu$.)
Then the condition that the wave vector be null becomes
\be
  \omega^2 = \delta_{ij}k^i k^j\ .\label{6.33}
\ee
Of course our wave is far from the most general 
solution; any (possibly infinite) number of distinct plane waves
can be added together and will still solve the linear equation (6.23).
Indeed, any solution can be written as such a superposition.

There are a number of free parameters to specify the wave: ten numbers 
for the coefficients $C_\mn$ and three for the null vector $k^\sigma$.
Much of these are the result of coordinate freedom and gauge freedom,
which we now set about eliminating.  We begin by imposing the 
harmonic gauge condition, (6.21).  This implies that
\bea
  0 &=&  \p\mu\bh^\mn \cr &=&  \p\mu(C^\mn e^{ik_\sigma x^\sigma})\cr
  &=& iC^\mn k_\mu e^{ik_\sigma x^\sigma}\ , \label{6.34}
\eea
which is only true if
\be
  k_\mu C^\mn=0\ .\label{6.35}
\ee
We say that the wave vector is orthogonal to $C^\mn$.  These are four
equations, which reduce the number of independent components of $C_\mn$
from ten to six.

Although we have now imposed the harmonic gauge condition, there is
still some coordinate freedom left.  Remember that any coordinate
transformation of the form 
\be
  x^\mu \rightarrow x^\mu + \zeta^\mu\label{6.36}
\ee
will leave the harmonic coordinate condition
\be
  \boxx x^\mu=0 \label{6.37}
\ee
satisfied as long as we have
\be
  \boxx \zeta^\mu=0\ .\label{6.38}
\ee
Of course, (6.38) is itself a wave equation for $\zeta^\mu$; once
we choose a solution, we will have used up all of our gauge freedom.
Let's choose the following solution:
\be
  \zeta_\mu = B_\mu e^{ik_\sigma x^\sigma}\ ,\label{6.39}
\ee
where $k_\sigma$ is the wave vector for our gravitational wave
and the $B_\mu$ are constant coefficients.

We now claim that this
remaining freedom allows us to convert from whatever coefficients
$C^{\rm (old)}_\mn$ that characterize our gravitational wave to a new
set $C^{\rm (new)}_\mn$, such that
\be
  C^{{\rm (new)}\mu}{}_\mu = 0\label{6.40}
\ee
and
\be
  C^{\rm (new)}_{0\nu}=0\ .\label{6.41}
\ee
(Actually this last condition is both a choice of gauge and a 
choice of Lorentz frame.  The choice of gauge sets $U^\mu 
C^{\rm (new)}_{\mu\nu}=0$ for some constant timelike vector $U^\mu$,
while the choice of frame makes $U^\mu$ point along the time axis.)
Let's see how this is possible by solving explicitly for the necessary
coefficients $B_\mu$.  Under the transformation (6.36),
the resulting change in our metric perturbation can be written
\be
  h^{\rm (new)}_\mn = h^{\rm (old)}_\mn -\p\mu \zeta_\nu
  -\p\nu \zeta_\mu\ ,\label{6.42}
\ee
which induces a change in the trace-reversed perturbation,
\bea
  \bh^{\rm (new)}_\mn &=&  h^{\rm (new)}_\mn -{1\over 2}
  \eta_\mn h^{\rm (new)}\cr & =& h^{\rm (old)}_\mn -\p\mu \zeta_\nu
  -\p\nu \zeta_\mu -{1\over 2}\eta_\mn(h^{\rm (old)}
  -2\p\lambda \zeta^\lambda)\cr
  &=& \bh^{\rm (old)}_\mn -\p\mu \zeta_\nu -\p\nu \zeta_\mu
  +\eta_\mn\p\lambda \zeta^\lambda\ . \label{6.43}
\eea
Using the specific forms (6.30) for the solution and (6.39) for the
transformation, we obtain
\be
  C^{\rm (new)}_\mn =C^{\rm (old)}_\mn - ik_\mu B_\nu -i k_\nu B_\mu
  +i\eta_\mn k_\lambda B^\lambda\ .\label{6.44}
\ee
Imposing (6.40) therefore means
\be
  0= C^{{\rm (old)}\mu}{}_\mu +2ik_\lambda B^\lambda\ ,\label{6.45}
\ee
or
\be
  k_\lambda B^\lambda = {{i}\over 2}C^{{\rm (old)}\mu}{}_\mu\ .
  \label{6.46}
\ee
Then we can impose (6.41), first for $\nu=0$:
\bea
  0 &=&  C^{\rm (old)}_{00}-2ik_0B_0-ik_\lambda B^\lambda\cr
  &=& C^{\rm (old)}_{00}-2ik_0B_0 +{1\over 2}C^{{\rm (old)}\mu}{}_\mu
  \ , \label{6.47}
\eea
or
\be
  B_0=-{{i}\over {2k_0}}\left(C^{\rm (old)}_{00}+{1\over 2}
  C^{{\rm (old)}\mu}{}_\mu\right)\ .\label{6.48}
\ee
Then impose (6.41) for $\nu=j$:
\bea
  0 &=&  C^{\rm (old)}_{0j}-ik_0 B_j -ik_jB_0\cr
  &=&  C^{\rm (old)}_{0j}-ik_0B_j -ik_j\left[
  {{-i}\over {2k_0}}\left(C^{\rm (old)}_{00}+{1\over 2}
  C^{{\rm (old)}\mu}{}_\mu\right)\right]\ , \label{6.49}
\eea
or 
\be
  B_j={{i}\over{2(k_0)^2}}\left[-2k_0C^{\rm (old)}_{0j}
  +k_j\left(C^{\rm (old)}_{00}+{1\over 2}
  C^{{\rm (old)}\mu}{}_\mu\right)\right]\ .\label{6.50}
\ee
To check that these choices are mutually consistent, we should plug
(6.48) and (6.50) back into (6.40), which I will leave to you.
Let us assume that we have performed this transformation, and refer
to the new components $C_\mn^{\rm (new)}$ simply as $C_\mn$.

Thus, we began with the ten independent numbers in the symmetric
matrix $C_\mn$.  Choosing harmonic gauge implied the four conditions
(6.35), which brought the number of independent components down to
six.  Using our remaining gauge freedom led to the one condition (6.40)
and the four conditions (6.41); but when $\nu=0$ (6.41) implies
(6.35), so we have a total of four additional constraints, which
brings us to two independent components.  We've used up all of our
possible freedom, so these two numbers represent the physical
information characterizing our plane wave in this gauge.  This can
be seen more explicitly by choosing our spatial coordinates such
that the wave is travelling in the $x^3$ direction; that is,
\be
  k^\mu = (\omega,0,0,k^3) = (\omega,0,0,\omega)\ ,\label{6.51}
\ee
where we know that $k^3=\omega$ because the wave vector is null.
In this case, $k^\mu C_\mn=0$ and $C_{0\nu}=0$ together imply
\be
  C_{3\nu}=0\ .\label{6.52}
\ee
The only nonzero components of $C_\mn$ are therefore $C_{11}$,
$C_{12}$, $C_{21}$, and $C_{22}$.  But $C_\mn$ is traceless and
symmetric, so in general we can write
\be
  C_\mn = \left(\matrix{0&0&0&0\cr 0&C_{11}&C_{12}&0\cr
  0&C_{12}&-C_{11}&0\cr 0&0&0&0\cr}\right)\ .\label{6.53}
\ee
Thus, for a plane wave in this gauge travelling in the $x^3$
direction, the two components $C_{11}$ and $C_{12}$ (along with the
frequency $\omega$) completely characterize the wave.

In using up all of our gauge freedom, we have gone to a subgauge of
the harmonic gauge known as the {\bf transverse traceless gauge}
(or sometimes ``radiation gauge'').  The name comes from the fact
that the metric perturbation is traceless and perpendicular to the
wave vector.  Of course, we have been working with the trace-reversed
perturbation $\bh_\mn$ rather than the perturbation $h_\mn$ itself;
but since $\bh_\mn$ is traceless (because $C_\mn$ is), and is equal to
the trace-reverse of $h_\mn$, in this gauge we have
\be
  \bh_\mn^{\rm TT} = h_\mn^{\rm TT}\qquad 
  {\rm (transverse~traceless~gauge)}\ .\label{6.54}
\ee
Therefore we can drop the bars over $h_\mn$, as long as we are in
this gauge.

One nice feature of the transverse traceless gauge is that if you
are given the components of a plane wave in some arbitrary gauge, you
can easily convert them into the transverse traceless components.
We first define a tensor $P_\mn$ which acts as a projection operator:
\be
  P_\mn = \eta_\mn - n_\mu n_\nu\ .\label{6.55}
\ee
You can check that this projects vectors onto hyperplanes orthogonal
to the unit vector $n_\mu$.  Here we take
$n_\mu$ to be a {\it spacelike} unit vector, which we choose
to lie along the direction of propagation of the wave:
\be
  n_0=0\ ,\qquad n_j = k_j/\omega\ .\label{6.56}
\ee
Then the transverse part of some perturbation $h_\mn$ is simply
the projection $P_\mu{}^\rho P_\nu{}^\sigma h_{\rho\sigma}$, and
the transverse traceless part is obtained by subtracting off the
trace:
\be
  h^{\rm TT}_\mn = P_\mu{}^\rho P_\nu{}^\sigma h_{\rho\sigma}
  -{1\over 2}P_\mn P^{\rho\sigma} h_{\rho\sigma}\ .\label{6.57}
\ee
For details appropriate to more general cases, see the discussion in
Misner, Thorne and Wheeler.

To get a feeling for the physical effects due to gravitational
waves, it is useful to consider the motion of test particles in
the presence of a wave.  It is certainly insufficient to solve
for the trajectory of a single particle, since that would only
tell us about the values of the coordinates along the world line.
(In fact, for any single particle we can find transverse traceless 
coordinates in which the particle appears stationary to first
order in $h_\mn$.)
To obtain a coordinate-independent measure of the wave's effects,
we consider the relative motion of nearby particles, as described
by the geodesic deviation equation.  If we consider some nearby
particles with four-velocities described by a single vector field
$U^\mu(x)$ and separation vector $S^\mu$, we have
\be
  {{D^2}\over{d\tau^2}}S^\mu = R^\mu{}_{\nu\rho\sigma}
  U^\nu U^\rho S^\sigma\ .\label{6.58}
\ee
We would like to compute the left-hand side to first order in
$h_\mn$.  If we take our test particles to be moving slowly
then we can express the four-velocity as a unit vector in the
time direction plus corrections of order $h_\mn$ and higher;
but we know that the Riemann tensor is already first order, so
the corrections to $U^\nu$ may be ignored, and we write
\be
  U^\nu = (1,0,0,0)\ .\label{6.59}
\ee
Therefore we only need to compute $R^\mu{}_{00\sigma}$, or
equivalently $R_{\mu 00\sigma}$.  From (6.5) we have
\be
  R_{\mu 00\sigma}={1\over 2}(\p0\p0 h_{\mu\sigma} + \p\sigma
  \p\mu h_{00} - \p\sigma\p0 h_{\mu 0} - \p\mu\p0 h_{\sigma 0})
  \ .\label{6.60}
\ee
But $h_{\mu 0}=0$, so
\be
  R_{\mu 00\sigma}={1\over 2}\p0\p0 h_{\mu\sigma}\ .\label{6.61}
\ee
Meanwhile, for our slowly-moving particles we have $\tau=x^0=t$
to lowest order, so the geodesic deviation equation becomes
\be
  {{\partial^2}\over{\partial t^2}}S^\mu = {1\over 2}S^\sigma
  {{\partial^2}\over{\partial t^2}} h^\mu{}_\sigma\ .\label{6.62}
\ee
For our wave travelling in the $x^3$ direction, this implies
that only $S^1$ and $S^2$ will be affected --- the test particles
are only disturbed in directions perpendicular to the wave
vector.  This is of course familiar from electromagnetism,
where the electric and magnetic fields in a plane wave are
perpendicular to the wave vector.

Our wave is characterized by the two numbers, which for future
convenience we will rename as $C_+ = C_{11}$ and $C_\times = 
C_{12}$.  Let's consider their effects separately, beginning
with the case $C_\times=0$.  Then we have 
\be
  {{\partial^2}\over{\partial t^2}}S^1 = {1\over 2} S^1
  {{\partial^2}\over{\partial t^2}}
  (C_+ e^{ik_\sigma x^\sigma})\label{6.63}
\ee
and
\be
  {{\partial^2}\over{\partial t^2}}S^2 = -{1\over 2} S^2
  {{\partial^2}\over{\partial t^2}} 
  (C_+ e^{ik_\sigma x^\sigma})\ .\label{6.64}
\ee
These can be immediately solved to yield, to lowest order,
\be
  S^1 = \left(1+{1\over 2}C_+ e^{ik_\sigma x^\sigma}
  \right)S^1(0)\label{6.65}
\ee
and 
\be
  S^2 = \left(1-{1\over 2}C_+ e^{ik_\sigma x^\sigma}
  \right)S^2(0)\ .\label{6.66}
\ee
Thus, particles initially separated in the $x^1$ direction will
oscillate back and forth in the $x^1$ direction, and likewise
for those with an initial $x^2$ separation.  That is, if we 
start with a ring of stationary particles in the $x$-$y$ plane,
as the wave passes they will bounce back and forth in the 
shape of a ``$+$'':

\begin{figure}[h]
  \centerline{
  \psfig{figure=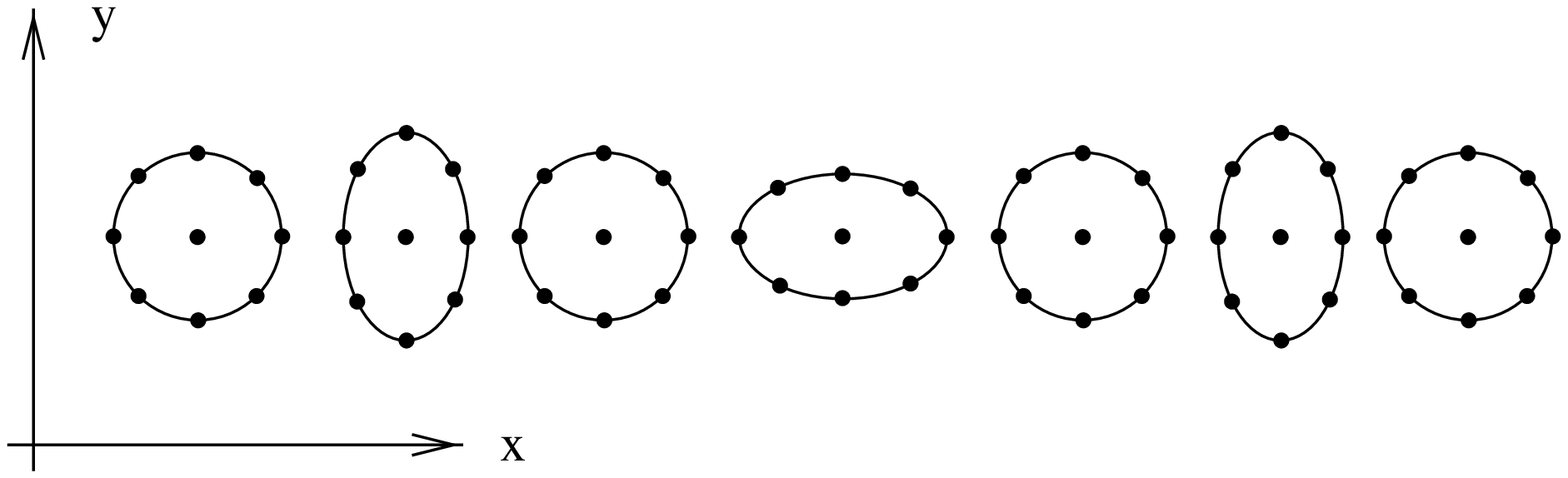,angle=0,height=4cm}}
\end{figure}

\noindent On the other hand, the equivalent analysis for the
case where $C_+=0$ but $C_\times\neq 0$ would yield the solution
\be
  S^1 = S^1(0)+{1\over 2}C_\times e^{ik_\sigma x^\sigma}
  S^2(0)\label{6.67}
\ee
and 
\be
  S^2 = S^2(0)+{1\over 2}C_\times e^{ik_\sigma x^\sigma}
  S^1(0)\ .\label{6.68}
\ee
In this case the circle of particles would bounce back and forth
in the shape of a ``$\times$'':

\begin{figure}[h]
  \centerline{
  \psfig{figure=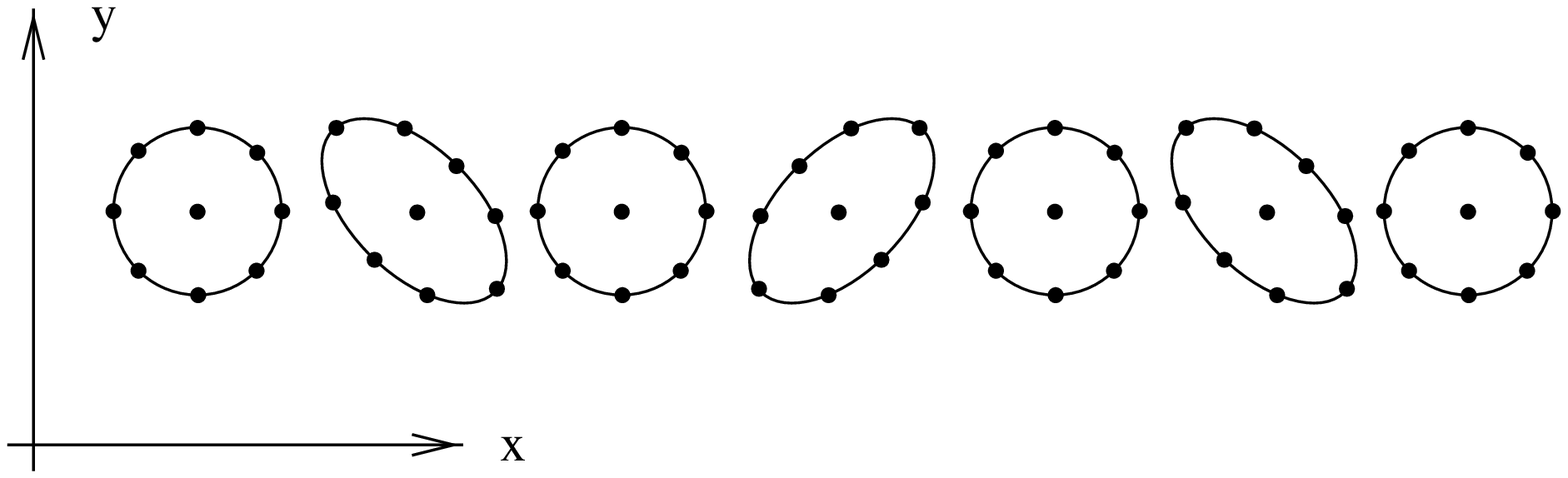,angle=0,height=4cm}}
\end{figure}

\noindent The notation $C_+$ and $C_\times$ should therefore be
clear.  These two quantities measure the two independent modes
of linear polarization of the gravitational wave.  If we liked
we could consider right- and left-handed circularly polarized 
modes by defining
\bea
  C_R &=&  {1\over {\sqrt2}}(C_+ +i C_\times)\ ,\cr
  C_L &=&  {1\over {\sqrt2}}(C_+ -i C_\times)\ . \label{6.69}
\eea
The effect of a pure $C_R$ wave would be to rotate the particles
in a right-handed sense,

\begin{figure}[h]
  \centerline{
  \psfig{figure=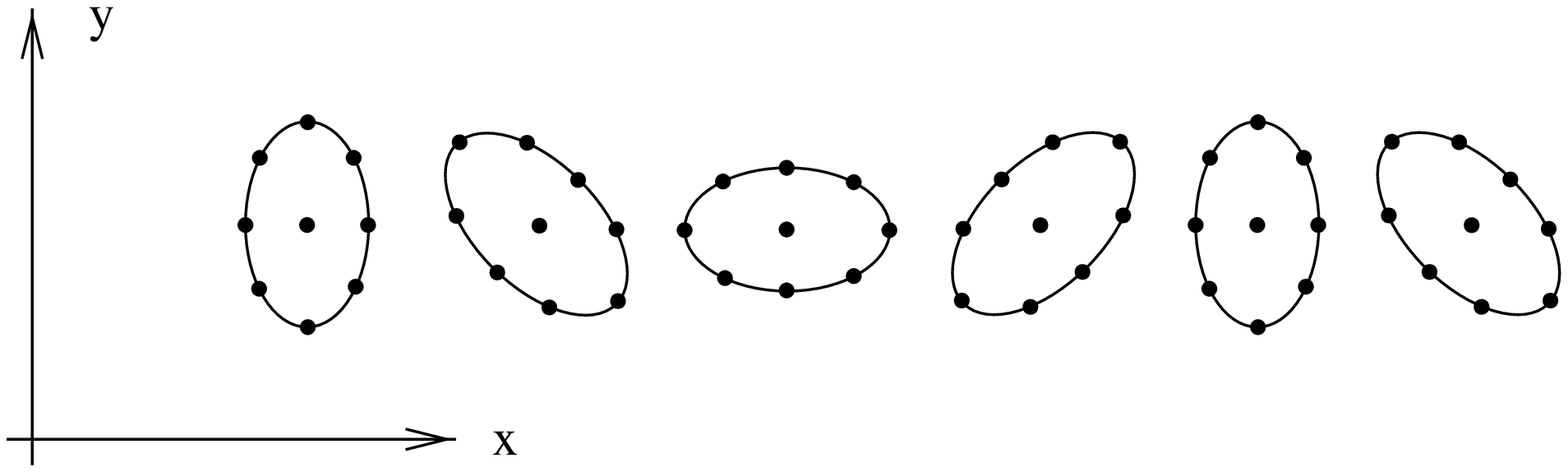,angle=0,height=4cm}}
\end{figure}

\noindent and similarly for the left-handed mode $C_L$.  (Note
that the individual particles do not travel around the ring; they
just move in little epicycles.)

We can relate the polarization states of classical gravitational waves
to the kinds of particles we would expect to find upon 
quantization.  The electromagnetic field has two independent
polarization states which are described by vectors in
the $x$-$y$ plane; equivalently, a single polarization mode
is invariant under a rotation by $360^\circ$ in this plane.
Upon quantization this theory yields the photon, a massless
spin-one particle.  The neutrino, on the other hand, is also a 
massless particle, described by a field which picks up a
minus sign under rotations by $360^\circ$; it is invariant 
under rotations of $720^\circ$, and we say it has 
spin-${1\over 2}$.  The general rule is that the spin $S$ is
related to the angle $\theta$ under which the polarization modes
are invariant by $S=360^\circ/\theta$.
The gravitational field, whose waves
propagate at the speed of light, should lead to massless particles
in the quantum theory.  Noticing that the polarization modes we
have described are invariant under rotations of $180^\circ$ in
the $x$-$y$ plane, we expect the associated particles
--- ``gravitons'' --- to be
spin-2.  We are a long way from detecting such particles (and it
would not be a surprise if we never detected them directly), but any
respectable quantum theory of gravity should predict their existence.

With plane-wave solutions to the linearized vacuum equations in our
possession, it remains to discuss the generation of gravitational 
radiation by sources.  For this purpose it is necessary to consider
the equations coupled to matter, 
\be
  \boxx \bh_\mn = -16\pi G T_\mn\ .\label{6.70}
\ee
The solution to such an equation can be obtained using a
Green's function, in precisely the same way as the analogous problem
in electromagnetism.  Here we will review the outline of the method.

The Green's function $G(x^\sigma - y^\sigma)$ for the D'Alembertian operator
$\boxx$ is the solution of the wave equation in the presence of a 
delta-function source:
\be
  {\boxx}_x G(x^\sigma - y^\sigma) = \delta^{(4)}(x^\sigma - y^\sigma)
  \ ,\label{6.71}
\ee
where ${\boxx}_x$ denotes the D'Alembertian with respect to the coordinates
$x^\sigma$.  The usefulness of such a function resides in the fact that
the general solution to an equation such as (6.70) can be written
\be
  \bh_\mn(x^\sigma) = -16\pi G \int G(x^\sigma - y^\sigma)T_\mn(y^\sigma)
  ~d^4y\ ,\label{6.72}
\ee
as can be verified immediately.  (Notice that no factors of $\g$ are
necessary, since our background is simply flat spacetime.)  The 
solutions to (6.71) have of course been worked out long ago, and
they can be thought of as either ``retarded'' or ``advanced,'' depending
on whether they represent waves travelling forward or backward in time.
Our interest is in the retarded Green's function, which represents the
accumulated effects of signals to the past of the point under 
consideration.  It is given by
\be
  G(x^\sigma - y^\sigma) = -{{1}\over{4\pi |\x-\y |}}\delta
  [|\x-\y | - (x^0-y^0)] ~\theta(x^0-y^0)\ .\label{6.73}
\ee
Here we have used boldface to denote the spatial vectors
$\x = (x^1,x^2,x^3)$ and $\y=(y^1,y^2,y^3)$, with norm 
$|\x-\y |=[\delta_{ij}(x^i-y^i)(x^j-y^j)]^{1/2}$.  The theta function
$\theta(x^0-y^0)$ equals $1$ when $x^0> y^0$, and zero otherwise.
The derivation of (6.73) would take us too far afield, but it can be
found in any standard text on electrodynamics or partial differential
equations in physics.

Upon plugging (6.73) into (6.72), we can use the delta function to
perform the integral over $y^0$, leaving us with
\be
  \bh_\mn(t,\x) =4G\int {{1}\over {|\x-\y |}}T_\mn(t-|\x-\y |,\y)
  ~d^3y\ ,\label{6.74}
\ee
where $t=x^0$.  The term ``retarded time'' is used to refer to the 
quantity
\be
  t_r = t-|\x-\y |\ .\label{6.75}
\ee
The interpretation of (6.74) should be clear: the disturbance in the
gravitational field at $(t,\x)$ is a sum of the influences from the energy 
and momentum sources at the point $(t_r,\x-\y)$ on the past
light cone.

\begin{figure}[h]
  \centerline{
  \psfig{figure=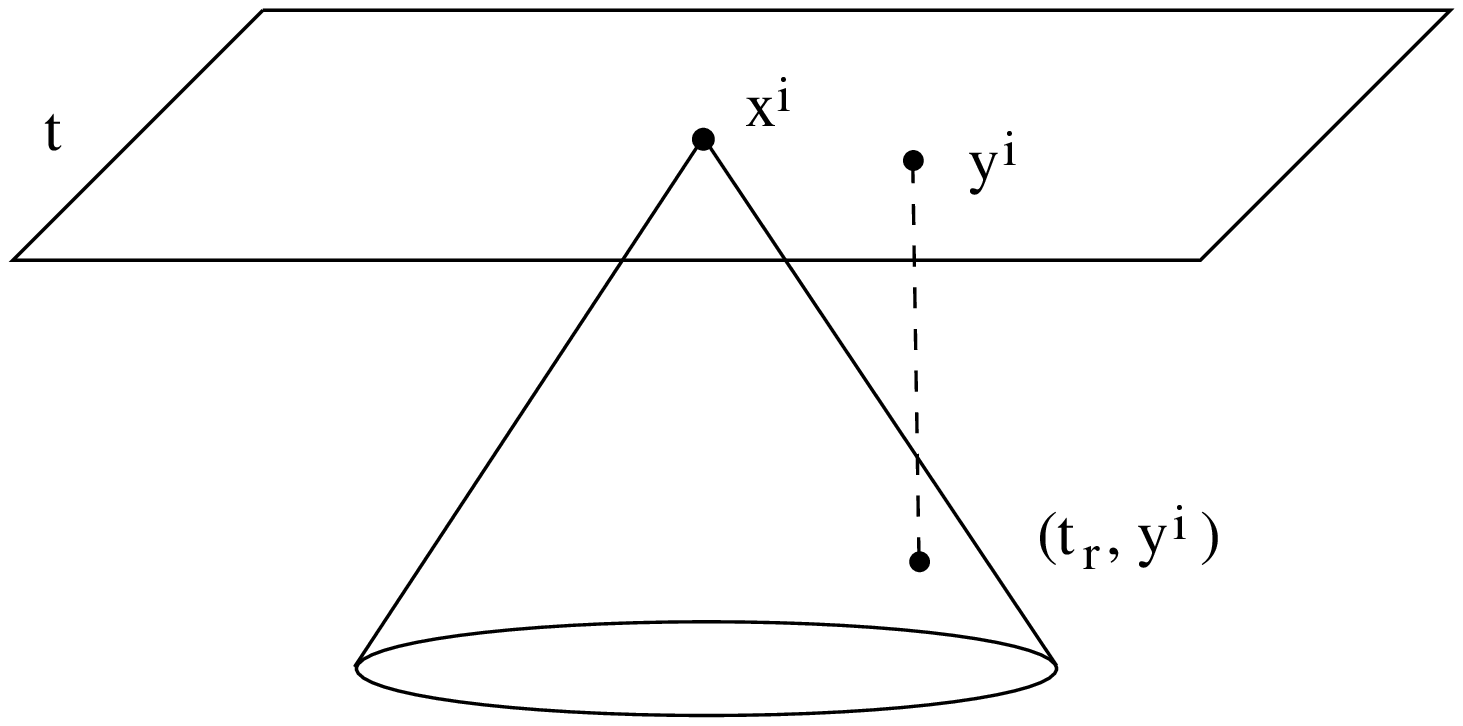,angle=0,height=6cm}}
\end{figure}

Let us take this general solution and consider the case where the
gravitational radiation is emitted by an isolated source, fairly far away,
comprised of nonrelativistic matter; these approximations will be made
more precise as we go on.  First we need to set up some conventions for
Fourier transforms, which always make life easier when dealing with
oscillatory phenomena.  Given a function of spacetime $\phi(t,\x)$, we
are interested in its Fourier transform (and inverse) with
respect to time alone,
\bea
  \widetilde\phi(\omega,\x) &=&  {{1}\over{\sqrt{2\pi}}}\int
  dt~e^{i\omega t}\phi(t,\x)\ ,\cr
  \phi(t,\x) &=&  {{1}\over{\sqrt{2\pi}}}\int d\omega~e^{-i\omega t}
  \widetilde\phi(\omega,\x)\ . \label{6.76}
\eea
Taking the transform of the metric perturbation, we obtain
\bea
  \widetilde\bh_\mn(\omega,\x) &=&  {{1}\over{\sqrt{2\pi}}}
  \int dt~e^{i\omega t}\bh_\mn(t,\x)\cr
  &=&  {{4G}\over{\sqrt{2\pi}}}\int dt~ d^3y~e^{i\omega t}~
  {{T_\mn(t-|\x-\y |,\y)}\over {|\x-\y |}}\cr
  &=&  {{4G}\over{\sqrt{2\pi}}}\int dt_r~d^3y~e^{i\omega t_r}
  e^{i\omega |\x-\y |}{{T_\mn(t_r,\y)}\over {|\x-\y |}}\cr
  &=&  4G \int d^3y~e^{i\omega |\x-\y |}{{\widetilde T_\mn(\omega,\y)}
  \over {|\x-\y |}}\ . \label{6.77}
\eea
In this sequence, the first equation is simply the definition of the
Fourier transform, the second line comes from the solution (6.74), the
third line is a change of variables from $t$ to $t_r$, and the fourth
line is once again the definition of the Fourier transform.

We now make the approximations that our source is isolated, far away, and 
slowly moving.  This means that we can consider the source to be centered
at a (spatial) distance $R$, with the different parts of the source at
distances $R+\delta R$ such that $\delta R << R$.  Since it is slowly
moving, most of the radiation emitted will be at frequencies $\omega$
sufficiently low that $\delta R<<\omega^{-1}$.  (Essentially, light 
traverses the source much faster than the components of the source itself 
do.)

\begin{figure}[h]
  \centerline{
  \psfig{figure=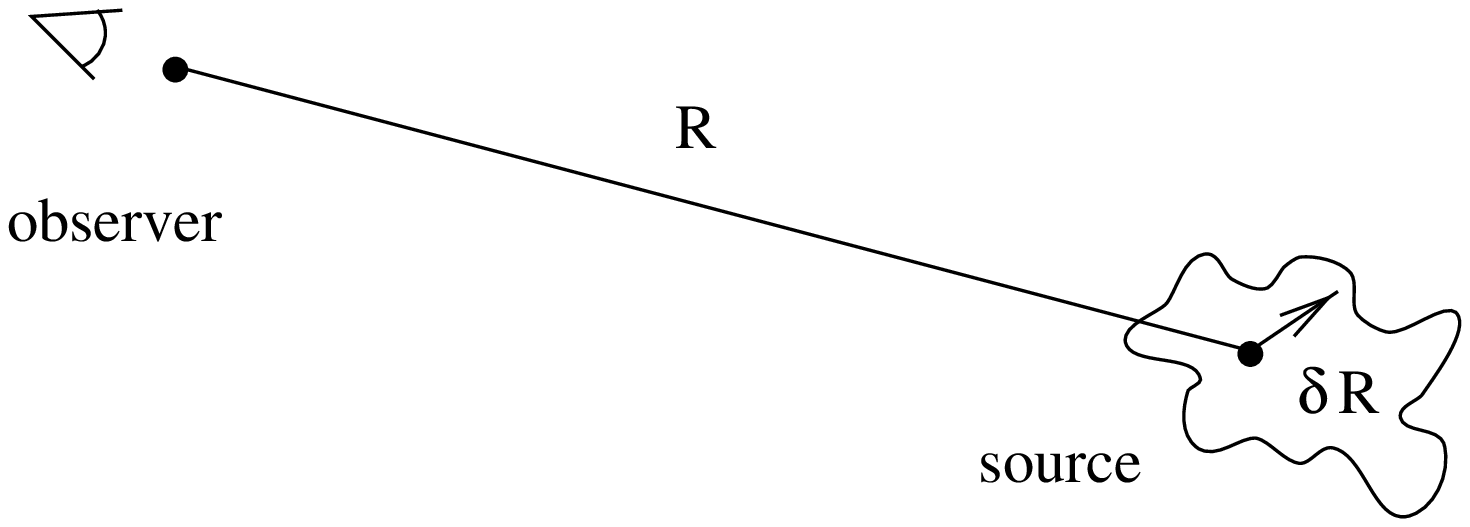,angle=0,height=5cm}}
\end{figure}

\noindent Under these approximations, the term 
$e^{i\omega |\x-\y |}/|\x-\y |$ can be replaced by $e^{i\omega R}/R$
and brought outside the integral.  This leaves us with
\be
  \widetilde\bh_\mn(\omega,\x) = 4G {{e^{i\omega R}}\over R}
  \int d^3y~\widetilde T_\mn(\omega,\y)\ .\label{6.78}
\ee

In fact there is no need to compute all of the components of 
$\widetilde\bh_\mn(\omega,\x)$, since the harmonic gauge condition
$\p\mu \bh^\mn(t,\x)=0$ in Fourier space implies
\be
  \widetilde\bh{}^{0\nu} = {i\over \omega}\p{i}\widetilde\bh{}^{i\nu}\ .
  \label{6.79}
\ee
We therefore only need to concern ourselves with the spacelike 
components of $\widetilde\bh_\mn(\omega,\x)$.  From (6.78) we
therefore want to take the integral of the spacelike components
of $\widetilde T_\mn(\omega,\y)$.  We begin by integrating by parts in 
reverse:
\be
  \int d^3y~\widetilde T^{ij}(\omega,\y)=\int \p{k}
  (y^i\widetilde T^{kj})~d^3y - \int y^i(\p{k}\widetilde T^{kj})~d^3y 
  \ .\label{6.80}
\ee
The first term is a surface integral which will vanish since the 
source is isolated, while the second can be related to 
$\widetilde T^{0j}$ by the Fourier-space version of $\p\mu T^\mn=0$:
\be
  -\p{k}\widetilde T^{k\mu}=i\omega \widetilde T^{0\mu}\ .
  \label{6.81}
\ee 
Thus,
\bea
  \int d^3y~\widetilde T^{ij}(\omega,\y)&=& 
  i\omega \int y^i \widetilde T^{0j}~d^3y \cr
  &=&  {{i\omega}\over 2}\int (y^i \widetilde T^{0j}
  +y^j \widetilde T^{0i})~d^3y \cr
  &=&  {{i\omega}\over 2}\int\left[\p{l}(y^i y^j \widetilde T^{0l})
  -y^i y^j (\p{l}\widetilde T^{0l})\right]~d^3y \cr
  &=&  -{{\omega^2}\over 2}\int y^i y^j \widetilde T^{00}~d^3y\ .
  \label{6.82}
\eea
The second line is justified since we know that the left hand side
is symmetric in $i$ and $j$, while the third and fourth lines are simply
repetitions of reverse integration by parts and conservation of $T^\mn$.
It is conventional to define the {\bf quadrupole moment tensor} of the 
energy density of the source,
\be
  q_{ij}(t) = 3\int y^i y^j T^{00}(t,\y)~d^3y\ ,\label{6.83}
\ee
a constant tensor on each surface of constant time.  In terms of the
Fourier transform of the quadrupole moment,
our solution takes on the compact form
\be
  \widetilde\bh_{ij}(\omega,\x) = -{{2G\omega^2}\over 3}
  {{e^{i\omega R}}\over R} \widetilde q_{ij}(\omega)\ ,\label{6.84}
\ee
or, transforming back to $t$,
\bea
  \bh_{ij}(t,\x) &=&  -{1\over{\sqrt{2\pi}}}{{2G}\over {3R}}
  \int d\omega~e^{-i\omega(t-R)}\omega^2\widetilde q_{ij}(\omega)\cr
  &=&  {1\over{\sqrt{2\pi}}}{{2G}\over {3R}}{{d^2}\over{dt^2}}
  \int d\omega~e^{-i\omega t_r}\widetilde q_{ij}(\omega)\cr
  &=&  {{2G}\over {3R}}{{d^2 q_{ij}}\over{dt^2}}(t_r) \ , \label{6.85}
\eea
where as before $t_r = t-R$.  

The gravitational wave produced by an isolated nonrelativistic object
is therefore proportional to the second derivative of the quadrupole
moment of the energy density at the point where the past light cone 
of the observer intersects the source.  In contrast, the leading
contribution to electromagnetic radiation comes from the changing
{\it dipole} moment of the charge density.  The difference can be
traced back to the universal nature of gravitation.  A changing dipole
moment corresponds to motion of the center of density --- charge density
in the case of electromagnetism, energy density in the case of 
gravitation.  While there is nothing to stop the center of charge of an 
object from oscillating, oscillation of the center of mass of an isolated
system violates conservation of momentum.  (You can shake a body up and
down, but you and the earth shake ever so slightly in the opposite
direction to compensate.)  The quadrupole moment, which
measures the shape of the system, is generally smaller than the dipole
moment, and for this reason (as well as the weak coupling of matter
to gravity) gravitational radiation is typically much weaker than
electromagnetic radiation.

It is always educational to take a general solution and apply it to
a specific case of interest.  One case of genuine interest is the
gravitational radiation emitted by a binary star (two stars in orbit
around each other).  For simplicity let us consider two stars of mass
$M$ in a circular orbit in the $x^1$-$x^2$ plane, at distance $r$ from 
their common center of mass.

\begin{figure}[h]
  \centerline{
  \psfig{figure=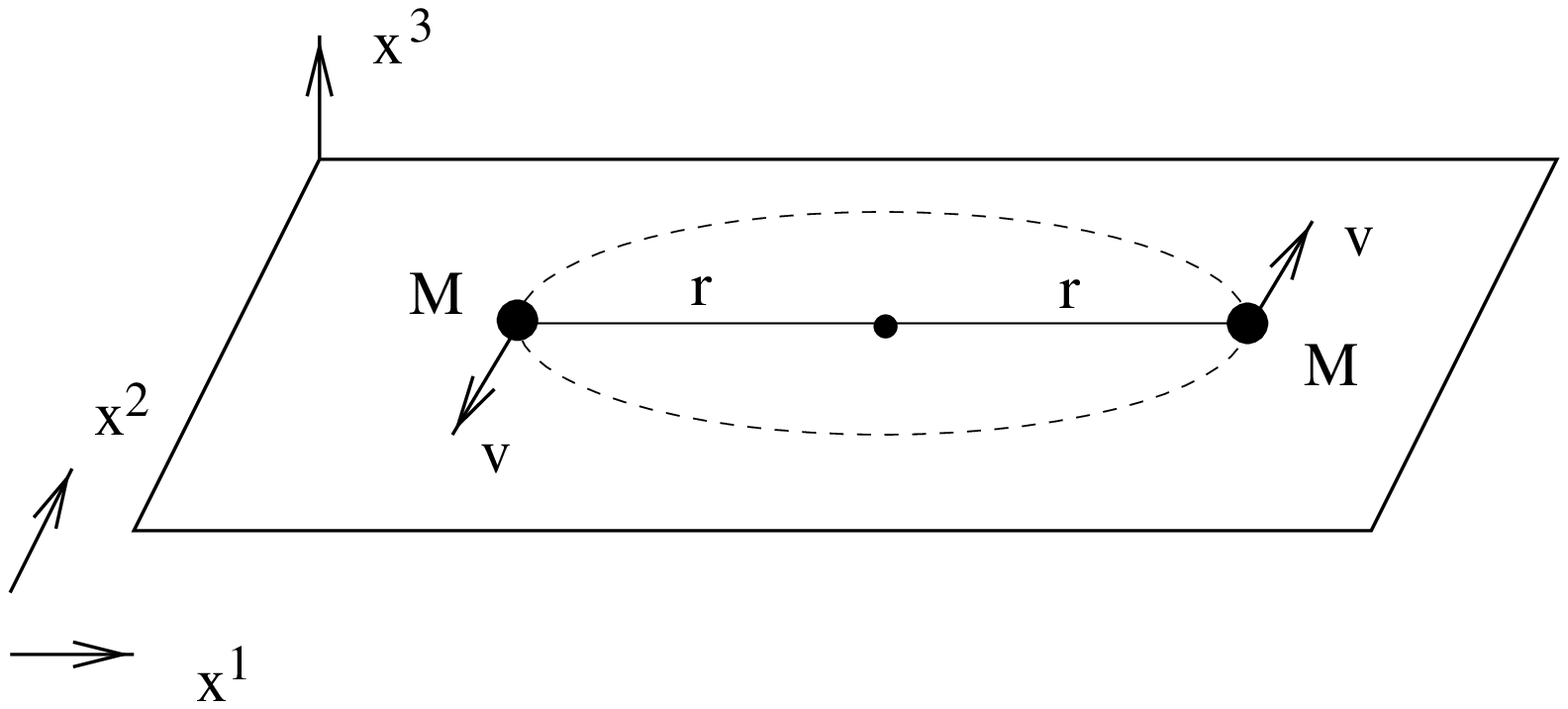,angle=0,height=5cm}}
\end{figure}

\noindent We will treat the motion of the stars in the Newtonian
approximation, where we can discuss their orbit just as Kepler would
have.  Circular orbits are most easily characterized by equating the
force due to gravity to the outward ``centrifugal'' force:
\be
  {{GM^2}\over{(2r)^2}} = {{Mv^2}\over r}\ ,\label{6.86}
\ee
which gives us
\be
  v=\left( {{GM}\over{4r}}\right)^{1/2}\ .\label{6.87}
\ee
The time it takes to complete a single orbit is simply
\be
  T = {{2\pi r}\over v}\ ,\label{6.88}
\ee
but more useful to us is the angular frequency of the orbit,
\be
  \Omega = {{2\pi}\over T} = \left( {{GM}\over{4r^3}}\right)^{1/2}
  \ .\label{6.89}
\ee
In terms of $\Omega$ we can write down the explicit path of star
$a$,
\be
  x^1_a = r\cos\Omega t\ ,\qquad x^2_a = r\sin\Omega t\ ,\label{6.90}
\ee
and star $b$,
\be
  x^1_b = -r\cos\Omega t\ ,\qquad x^2_b =-r\sin\Omega t\ .\label{6.91}
\ee
The corresponding energy density is
\be
  T^{00}(t,\x) = M\delta(x^3)\left[\delta(x^1-r\cos\Omega t)
  \delta(x^2-r\sin\Omega t) + \delta(x^1+r\cos\Omega t)
  \delta(x^2+r\sin\Omega t)\right]\ .\label{6.92}
\ee
The profusion of delta functions allows us to integrate this 
straightforwardly to obtain the quadrupole moment from (6.83):
\bea
  q_{11} &=&  6Mr^2\cos^2\Omega t = 3Mr^2(1+\cos2\Omega t)\cr
  q_{22} &=&  6Mr^2\sin^2\Omega t = 3Mr^2(1-\cos2\Omega t)\cr
  q_{12} =q_{21} &=&  6Mr^2(\cos\Omega t)(\sin\Omega t) = 
  3Mr^2\sin2\Omega t\cr q_{i3} &=& 0\ .\label{6.93}
\eea
From this in turn it is easy to get the components of the metric
perturbation from (6.85):
\be
  \bh_{ij}(t,\x) = {{8GM}\over R}\Omega^2r^2\left(\matrix{
  -\cos2\Omega t_r & -\sin2\Omega t_r & 0\cr
  -\sin2\Omega t_r & \cos2\Omega t_r & 0 \cr
  0 & 0 & 0\cr}\right)\ .\label{6.94}
\ee
The remaining components of $\bh_\mn$ could be derived from demanding 
that the harmonic gauge
condition be satisfied.  (We have not imposed a subsidiary gauge 
condition, so we are still free to do so.)

It is natural at this point to talk about the energy emitted
via gravitational radiation.  Such a discussion, however, is
immediately beset by problems, both technical and philosophical.
As we have mentioned before, there is no true local measure of
the energy in the gravitational field.  Of course, in the weak
field limit, where we think of gravitation as being described 
by a symmetric tensor propagating on a fixed background metric,
we might hope to derive an energy-momentum tensor for the
fluctuations $h_\mn$, just as we would for electromagnetism or
any other field theory.  To some extent this is possible, but 
there are still difficulties.  As a result of these difficulties
there are a number of different proposals in the literature for
what we should use as the energy-momentum tensor for gravitation
in the weak field limit; all of them are different, but for 
the most part they give the same answers for physically 
well-posed questions such as the rate of energy emitted by a
binary system.

At a technical level, the difficulties begin to arise when we 
consider what form the energy-momentum tensor should take.
We have previously mentioned the energy-momentum tensors for
electromagnetism and scalar field theory, and they both
shared an important feature --- they were quadratic in the
relevant fields.  By hypothesis our approach to the weak field
limit has been to only keep terms which are linear in the 
metric perturbation.  Hence, in order to keep track of the
energy carried by the gravitational waves, we will have to
extend our calculations to at least second order in $h_\mn$.
In fact we have been cheating slightly all along.  In 
discussing the effects of gravitational waves on test particles,
and the generation of waves by a binary system, we have been
using the fact that test particles move along geodesics.  But
as we know, this is derived from the covariant conservation of
energy-momentum, $\nabla_\mu T^\mn=0$.  In the order to which
we have been working, however, we actually have $\p\mu T^\mn=0$,
which would imply that test particles move on straight lines
in the flat background metric.  This is a symptom of the
fundamental inconsistency of the weak field limit.  In practice,
the best that can be done is to solve the weak field equations
to some appropriate order, and then justify after the fact the
validity of the solution.

Keeping these issues in mind, let us consider Einstein's
equations (in vacuum) to second order, and see how the 
result can be interpreted in terms of an energy-momentum
tensor for the gravitational field.  If we write the metric
as $g_\mn = \eta_\mn + h_\mn$, then at first order we have
\be
  G^{(1)}_\mn[\eta+h] = 0\ ,\label{6.95}
\ee
where $G^{(1)}_\mn$ is Einstein's tensor expanded to first
order in $h_\mn$.  These equations determine $h_\mn$ up to
(unavoidable) gauge transformations, so in order to satisfy
the equations at second order we have to add a higher-order
perturbation, and write
\be
  g_\mn = \eta_\mn + h_\mn + h^{(2)}_\mn\ .\label{6.96}
\ee
The second-order version of Einstein's equations consists
of all terms either quadratic in $h_\mn$ or linear in 
$h^{(2)}_\mn$.  Since any cross terms would be of at least
third order, we have
\be
  G^{(1)}_\mn[\eta+h^{(2)}] +G^{(2)}_\mn[\eta+h] =0\ .
  \label{6.97}
\ee
Here, $G^{(2)}_\mn$ is the part of the Einstein tensor which
is of second order in the metric perturbation.  It can be
computed from the second-order Ricci tensor, which is given
by
\bea
  R^{(2)}_\mn &=&  {1\over 2}h^{\rho\sigma}\p\mu
  \p\nu h_{\rho\sigma} - h^{\rho\sigma}\p\rho\p{(\mu}
  h_{\nu)\sigma} +{1\over 4}(\p\mu h_{\rho\sigma})\p\nu
  h^{\rho\sigma} +(\partial^\sigma h^\rho{}_\nu)
  \p{[\sigma}h_{\rho]\mu} \cr
  &&\quad +{1\over 2}\p\sigma(h^{\rho\sigma}\p\rho h_\mn)
  -{1\over 4}(\p\rho h_\mn)\partial^\rho h - (\p\sigma
  h^{\rho\sigma} -{1\over 2}\partial^\rho h)\p{(\mu}
  h_{\nu)\rho}\ . \label{6.98}
\eea
We can cast (6.97) into the suggestive form
\be
  G^{(1)}_\mn[\eta+h^{(2)}] = 8\pi G t_\mn \ ,\label{6.99}
\ee
simply by defining
\be
  t_\mn =  -{1\over {8\pi G}}G^{(2)}_\mn[\eta+h]\ .\label{6.100}
\ee
The notation is of course meant to suggest that we think of
$t_\mn$ as an energy-momentum tensor, specifically that of
the gravitational field (at least in the weak field regime).
To make this claim seem plausible, note that the Bianchi 
identity for $G^{(1)}_\mn[\eta+h^{(2)}]$ implies that $t_\mn$
is conserved in the flat-space sense,
\be
  \p\mu t^\mn =0\ .\label{6.101}
\ee

Unfortunately there are some limitations on our interpretation
of $t_\mn$ as an energy-momentum tensor.  Of course it is
not a tensor at all in the full theory, but we are leaving that
aside by hypothesis.  More importantly, it is not invariant 
under gauge transformations (infinitesimal diffeomorphisms),
as you could check by direct calculation.  However, we can
construct global quantities which are invariant under certain
special kinds of gauge transformations (basically, those that
vanish sufficiently rapidly at infinity; see Wald).  These
include the total energy on a surface $\Sigma$ of constant
time,
\be
  E=\int_\Sigma t_{00}~d^3x\ ,\label{6.102}
\ee
and the total energy radiated through to infinity,
\be
  \Delta E = \int_S t_{0\mu} n^\mu ~d^2x~dt\ .\label{6.103}
\ee
Here, the integral is taken over a timelike surface $S$ made
of a spacelike two-sphere at infinity and some interval in time,
and $n^\mu$ is a unit spacelike vector normal to $S$.

Evaluating these formulas in terms of the quadrupole moment
of a radiating source involves a lengthy calculation which we
will not reproduce here.  Without further ado, the amount of
radiated energy can be written
\be
  \Delta E = \int P ~dt\ ,\label{6.104}
\ee
where the power $P$ is given by
\be
  P = {{G}\over {45}}\left[{{d^3 Q^{ij}}\over{dt^3}}
  {{d^3 Q_{ij}}\over{dt^3}}\right]_{t_r}\ ,\label{6.105}
\ee
and here $Q_{ij}$ is the traceless part of the quadrupole
moment,
\be
  Q_{ij} = q_{ij}-{1\over 3}\delta_{ij} \delta^{kl}
  q_{kl}\ .\label{6.106}
\ee

For the binary system represented by (6.93), the traceless
part of the quadrupole is 
\be
  Q_{ij} = Mr^2 \left(\matrix{(1+3\cos2\Omega t)&
  3\sin2\Omega t & 0\cr 3\sin2\Omega t & (1-3\cos2\Omega t) &0\cr
  0 & 0 & -2\cr}\right)\ ,\label{6.107}
\ee
and its third time derivative is therefore
\be
  {{d^3 Q_{ij}}\over{dt^3}}= 
  24 M r^2\Omega^3\left(\matrix{\sin2\Omega t &
  -\cos2\Omega t & 0\cr -\cos2\Omega t & -\sin2\Omega t & 0\cr
  0&0&0\cr}\right)\ .\label{6.108}
\ee
The power radiated by the binary is thus
\be
  P = {{2^7}\over 5}GM^2 r^4\Omega^6\ ,\label{6.109}
\ee
or, using expression (6.89) for the frequency,
\be
  P = {2\over 5}{{G^4 M^5}\over {r^5}}\ .\label{6.110}
\ee

Of course, this has actually been observed.  In 1974 Hulse and
Taylor discovered a binary system, PSR1913+16, in which both stars
are very small (so classical effects are negligible, or at least
under control) and one is a pulsar.  The period of the orbit is 
eight hours, extremely small by astrophysical standards.  The fact
that one of the stars is a pulsar provides a very accurate clock,
with respect to which the change in the period as the system loses
energy can be measured.  The result is consistent with the prediction
of general relativity for energy loss through gravitational radiation.
Hulse and Taylor were awarded the Nobel Prize in 1993 for their
efforts.

\eject

\thispagestyle{plain}

\setcounter{equation}{0}

\noindent{December 1997 \hfill {\sl Lecture Notes on General Relativity}
\hfill{Sean M.~Carroll}}

\vskip .2in

\setcounter{section}{6}
\section{The Schwarzschild Solution and Black Holes}

We now move from the domain of the weak-field limit to solutions of
the full nonlinear Einstein's equations.  With the possible exception
of Minkowski space, by far the most important such solution is that
discovered by Schwarzschild, which describes spherically symmetric
vacuum spacetimes.  Since we are in vacuum, Einstein's equations
become $R_\mn =0$.  Of course, if we have a proposed solution to
a set of differential equations such as this, it would suffice to
plug in the proposed solution in order to verify it; we would like
to do better, however.  In fact, we will sketch a proof of Birkhoff's
theorem, which states that the Schwarzschild solution is the {\it unique}
spherically symmetric solution to Einstein's equations in vacuum.
The procedure will be to first present some
non-rigorous arguments that any spherically symmetric metric (whether
or not it solves Einstein's equations) must take on a certain form,
and then work from there to more carefully derive the actual solution
in such a case.

``Spherically symmetric'' means ``having the same symmetries as a 
sphere.''  (In this section the word ``sphere'' means $S^2$, not
spheres of higher dimension.)
Since the object of interest to us is the metric on
a differentiable manifold, we are concerned with those metrics that
have such symmetries.  We know how to characterize symmetries of
the metric --- they are given by the existence of Killing vectors.
Furthermore, we know what the Killing vectors of $S^2$ are, and that 
there are three of them.  Therefore, a spherically symmetric manifold is 
one that has three Killing vector fields which are just like those
on $S^2$.  By ``just like'' we mean that the commutator of the
Killing vectors is the same in either case --- in fancier language,
that the algebra generated by the vectors is the same.  Something that
we didn't show, but is true, is that we can choose our three Killing
vectors on $S^2$ to be $(V^{(1)},V^{(2)},V^{(3)})$, such that
\bea
  [V^{(1)},V^{(2)}] &=&  V^{(3)}\cr
  [V^{(2)},V^{(3)}] &=&  V^{(1)}\cr [V^{(3)},V^{(1)}] &=&  V^{(2)}\ .
  \label{7.1}
\eea
The commutation relations are exactly those of SO(3), the group of 
rotations in three dimensions.  This is no coincidence, of course,
but we won't pursue this here.  All we need is that a spherically
symmetric manifold is one which possesses three Killing vector fields
with the above commutation relations.

Back in section three we mentioned Frobenius's Theorem, which states
that if you have a set of commuting vector fields then there exists a
set of coordinate
functions such that the vector fields are the partial derivatives
with respect to these functions.  In fact the theorem does not stop 
there, but goes on to say that if we have some vector fields which do
{\it not} commute, but whose commutator closes --- the commutator of
any two fields in the set is a linear combination of other fields in
the set --- then the integral curves of these vector fields ``fit
together'' to describe submanifolds of the manifold on which they are
all defined.  The dimensionality of the submanifold may be smaller 
than the number of vectors, or it could be equal, but obviously not
larger.  Vector fields which obey (7.1) will of course form 2-spheres.
Since the vector fields stretch throughout the space, every point will
be on exactly one of these spheres.  (Actually, it's almost every point
--- we will show below how it can fail to be absolutely every point.)
Thus, we say that a spherically symmetric manifold can be {\bf foliated}
into spheres.  

Let's consider some examples to bring this down to earth.  The simplest
example is flat three-dimensional Euclidean space.  If we pick an origin,
then $\R^3$ is clearly spherically symmetric with respect to rotations
around this origin.  
Under such rotations ({\it i.e.}, under the 
flow of the Killing vector fields) points move into each other, but
each point stays on an $S^2$ at a fixed distance from the origin.

\begin{figure}[h]
  \centerline{
  \psfig{figure=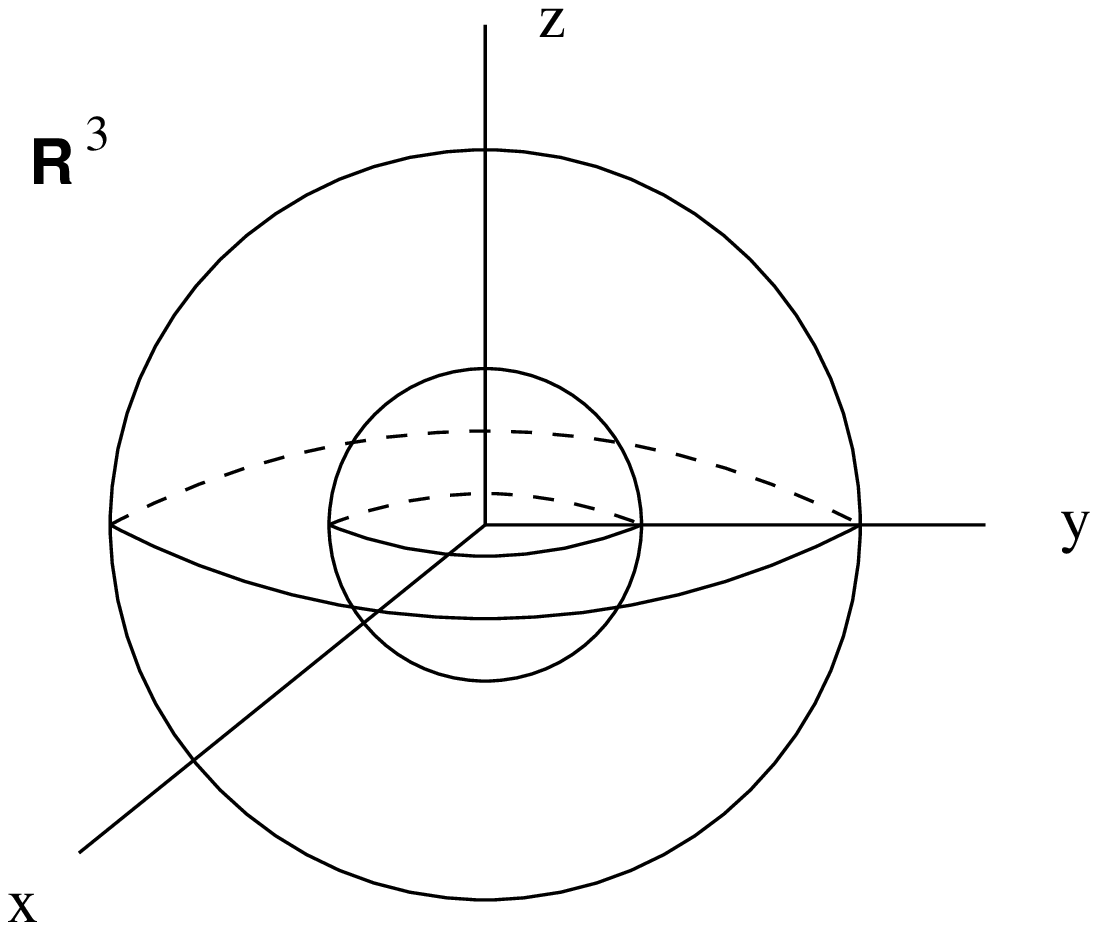,angle=0,height=6cm}}
\end{figure}

\noindent
It is these spheres which foliate $\R^3$.
Of course, they don't really foliate all of the space, since
the origin itself just stays put under rotations --- it doesn't move
around on some two-sphere.  But it should be clear that almost all of
the space is properly foliated, and this will turn out to be enough for
us.  

We can also have spherical symmetry without an ``origin'' to rotate
things around.  An example is provided by a ``wormhole'', with topology
$\R\times S^2$.  If we suppress a dimension and draw our two-spheres
as circles, such a space might look like this:

\eject

\begin{figure}[h]
  \centerline{
  \psfig{figure=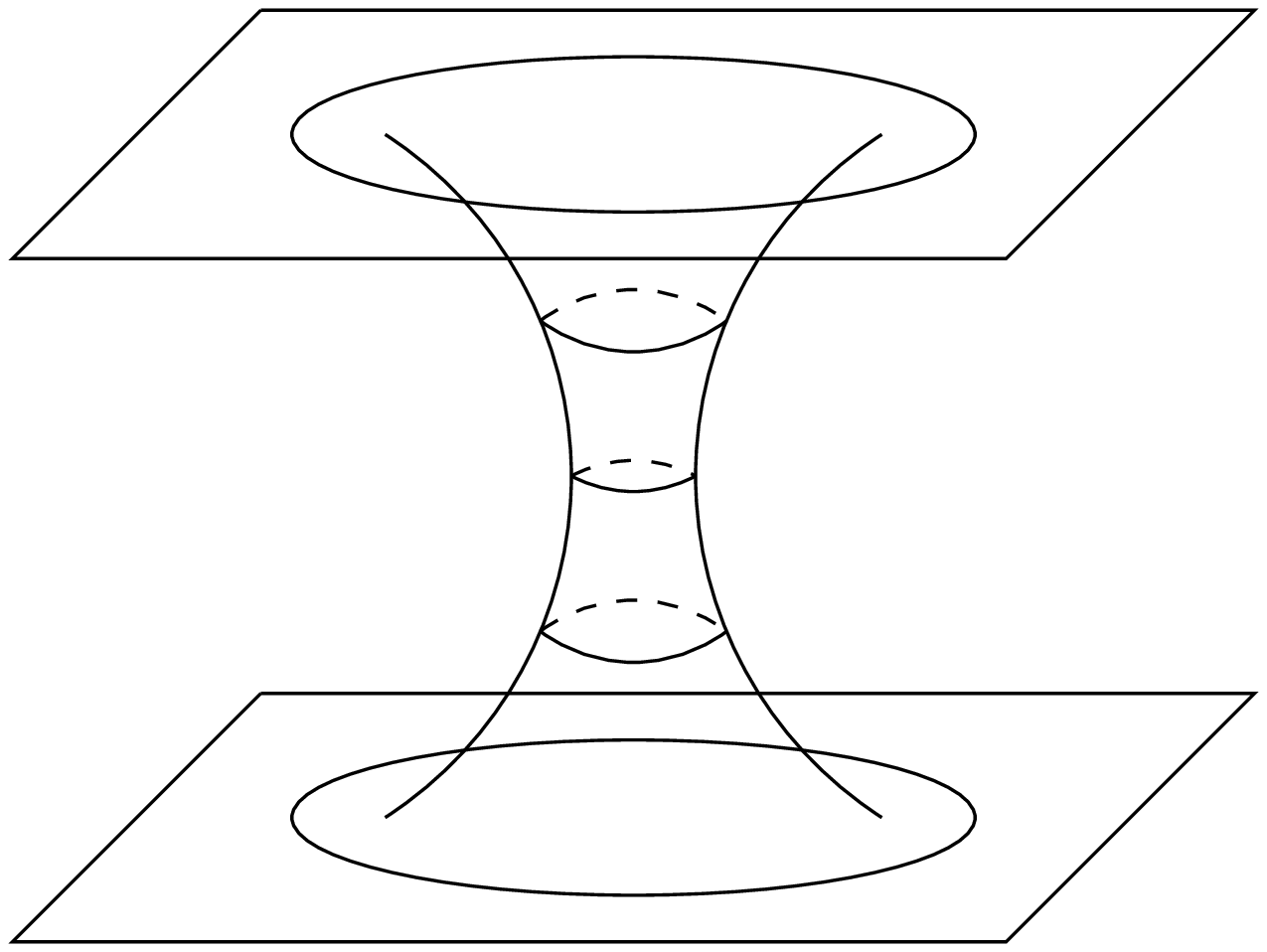,angle=0,height=6cm}}
\end{figure}

\noindent In this case the entire manifold can be foliated by two-spheres.

This foliated structure suggests that we put coordinates on our manifold
in a way which is adapted to the foliation.  By this we mean that, if
we have an $n$-dimensional manifold foliated by $m$-dimensional 
submanifolds, we can use a set of $m$ coordinate functions $u^i$ on
the submanifolds and a set of $n-m$ coordinate functions $v^I$ to tell
us which submanifold we are on.  (So $i$ runs from 1 to $m$, while
$I$ runs from 1 to $n-m$.)  Then the collection of $v$'s and $u$'s 
coordinatize the entire space.  If the submanifolds are maximally
symmetric spaces (as two-spheres are), then there is the following
powerful theorem: it is always possible to choose the $u$-coordinates
such that the metric on the entire manifold is of the form
\be
  ds^2 = g_\mn \d x^\mu \d x^\nu = g_{IJ}(v)\d v^I \d v^J
  +f(v)\gamma_{ij}(u)\d u^i \d u^j\ .\label{7.2}
\ee
Here $\gamma_{ij}(u)$ is the metric on the submanifold.
This theorem is saying two things at once: that there are no cross
terms $\d v^I \d u^j$, and that both $g_{IJ}(v)$ and $f(v)$ are functions
of the $v^I$ alone, independent of the $u^i$.  Proving the theorem is
a mess, but you are encouraged to look in chapter 13 of Weinberg.
Nevertheless, it is a perfectly sensible result.  Roughly speaking,
if $g_{IJ}$ or $f$ depended on the $u^i$ then the metric would change
as we moved in a single submanifold, which violates the assumption of
symmetry.  The unwanted cross terms, meanwhile, can be eliminated by
making sure that the tangent vectors $\partial/\partial v^I$ are
orthogonal to the submanifolds --- in other words, that 
we line up our submanifolds in the same way throughout the space.

We are now through with handwaving, and can commence some honest
calculation.
For the case at hand, our submanifolds are two-spheres, on which we
typically choose coordinates $(\theta,\phi)$ in which the metric takes
the form
\be
  d\Omega^2 = \d\theta^2 + \sin^2\theta\ \d\phi^2\ .\label{7.3}
\ee
Since we are interested in a four-dimensional spacetime, we need two
more coordinates, which we can call $a$ and $b$.  The theorem (7.2)
is then telling us that the metric on a spherically symmetric spacetime
can be put in the form
\be
  ds^2 = g_{aa}(a,b)\d a^2 + g_{ab}(a,b)(\d a\d b+\d b\d a)
  +g_{bb}(a,b)\d b^2 + r^2(a,b)d\Omega^2\ .\label{7.4}
\ee
Here $r(a,b)$ is some as-yet-undetermined function, to which we have
merely given a suggestive label.  There is nothing to stop us, 
however, from changing coordinates from $(a,b)$ to $(a,r)$, by
inverting $r(a,b)$.  (The one thing that could possibly stop us would
be if $r$ were a function of $a$ alone; in this case we could just
as easily switch to $(b,r)$, so we will not consider this situation 
separately.)  The metric is then
\be
  ds^2 = g_{aa}(a,r)\d a^2 + g_{ar}(a,r)(\d a\d r+\d r\d a)
  +g_{rr}(a,r)\d r^2 + r^2 d\Omega^2\ .\label{7.5}
\ee
Our next step is to find a function $t(a,r)$ such that, in the 
$(t,r)$ coordinate system, there are no cross terms $\d t\d r+\d r\d t$
in the metric.  Notice that
\be
  \d t = {{\partial t}\over{\partial a}}\d a + {{\partial t}\over
  {\partial r}}\d r \ ,\label{7.6}
\ee
so 
\be
  \d t^2 = \left({{\partial t}\over{\partial a}}\right)^2\d a^2
  + \left({{\partial t}\over{\partial a}}\right)\left({{\partial t}
  \over{\partial r}}\right)
  (\d a\d r+\d r\d a) + \left({{\partial t}\over{\partial r}}\right)^2
  \d r^2\ .\label{7.7}
\ee
We would like to replace the first three terms in the metric (7.5) by
\be
  m\d t^2 + n\d r^2\ ,\label{7.8}
\ee
for some functions $m$ and $n$.  This is equivalent to the
requirements
\be
  m\left({{\partial t}\over{\partial a}}\right)^2 = g_{aa}\ ,\label{7.9}
\ee
\be
  n+m\left({{\partial t}\over{\partial r}}\right)^2 = g_{rr}\ ,\label{7.10}
\ee
and
\be
  m\left({{\partial t}\over{\partial a}}\right)\left({{\partial t}\over
  {\partial r}}\right)=g_{ar}\ .\label{7.11}
\ee
We therefore have three equations for the three unknowns $t(a,r)$,
$m(a,r)$, and $n(a,r)$, just enough to determine them precisely (up
to initial conditions for $t$).  (Of course, they are ``determined''
in terms of the unknown functions $g_{aa}$, $g_{ar}$, and $g_{rr}$, so
in this sense they are still undetermined.)
We can therefore put our metric in the form
\be
  ds^2 = m(t,r)\d t^2 + n(t,r)\d r^2+ r^2 d\Omega^2\ .\label{7.12}
\ee

To this point the only difference between the two coordinates $t$ and
$r$ is that we have chosen $r$ to be the one which multiplies the
metric for the two-sphere.  This choice was motivated by what we know
about the metric for flat Minkowski space, which can be written
$ds^2 = -\d t^2 + \d r^2+ r^2 d\Omega^2$.  We know that the spacetime
under consideration is Lorentzian, so either $m$ or $n$ will have to
be negative.  Let us choose $m$, the coefficient of $\d t^2$, to be
negative.  This is not a choice we are simply allowed to make, and in
fact we will see later that it can go wrong, but we will assume it for
now.  The assumption is not completely unreasonable, since we know that 
Minkowski space is itself spherically symmetric, and will therefore be
described by (7.12).  With this choice we can trade in the functions
$m$ and $n$ for new functions $\alpha$ and $\beta$, such that
\be
  ds^2 = -e^{2\alpha(t,r)}\d t^2 + e^{2\beta(t,r)}\d r^2
  + r^2 d\Omega^2\ .\label{7.13}
\ee

This is the best we can do for a general metric in a spherically
symmetric spacetime.  The next step is to actually solve Einstein's
equations, which will allow us to determine explicitly the functions
$\alpha(t,r)$ and $\beta(t,r)$.  It is unfortunately necessary to
compute the Christoffel symbols for (7.13), from which we can get
the curvature tensor and thus the Ricci tensor.  If we use labels
$(0,1,2,3)$ for $(t,r,\theta,\phi)$ in the usual way, the Christoffel
symbols are given by
\bea
   &\Gamma^0_{00}=\p0\alpha\qquad\quad
  \Gamma^0_{01} =  \p1\alpha \qquad\quad
  \Gamma^0_{11} = e^{2(\beta-\alpha)}\p0\beta &\cr &
  \Gamma^1_{00} = e^{2(\alpha-\beta)}\p1\alpha\qquad
  \Gamma^1_{01} = \p0\beta \qquad\quad
  \Gamma^1_{11} = \p1\beta & \cr &
  \Gamma^2_{12} = {1\over r}\qquad
  \Gamma^1_{22} = - r e^{-2\beta}\qquad
  \Gamma^3_{13} = {1\over r} &\cr &
  \Gamma^1_{33} = -r e^{-2\beta}\sin^2\theta\qquad
  \Gamma^2_{33} = -\sin\theta \cos\theta \qquad
  \Gamma^3_{23} = {{\cos\theta}\over {\sin\theta}}\ .&\label{7.14}
\eea
(Anything not written down explicitly is meant to be zero, or related
to what is written by symmetries.)  From
these we get the following nonvanishing components of the Riemann
tensor:
\bea
  R^0{}_{101} &=&  e^{2(\beta-\alpha)}[\p0^2\beta +(\p0\beta)^2
  -\p0\alpha \p0\beta]+[\p1\alpha\p1\beta-\p1^2\alpha -(\p1\alpha)^2]\cr
  R^0{}_{202} &=&  -r e^{-2\beta}\p1\alpha \cr
  R^0{}_{303} &=&  -r e^{-2\beta}\sin^2\theta\ \p1\alpha \cr
  R^0{}_{212} &=&  -r e^{-2\alpha}\p0\beta \cr
  R^0{}_{313} &=&  -r e^{-2\alpha}\sin^2\theta\ \p0\beta \cr
  R^1{}_{212} &=&  r e^{-2\beta}\p1\beta \cr
  R^1{}_{313} &=&  r e^{-2\beta}\sin^2\theta\ \p1\beta \cr
  R^2{}_{323} &=&  (1-e^{-2\beta})\sin^2\theta\ .\label{7.15}
\eea
Taking the contraction as usual yields the Ricci tensor:
\bea
  R_{00} &=&  [\p0^2\beta +(\p0\beta)^2-\p0\alpha \p0\beta] + 
  e^{2(\alpha-\beta)}[\p1^2\alpha +(\p1\alpha)^2-\p1\alpha\p1\beta
  +{2\over{r}}\p1\alpha]\cr
  R_{11} &=&  -[\p1^2\alpha +(\p1\alpha)^2-\p1\alpha\p1\beta
  -{2\over{r}}\p1\beta] + e^{2(\beta-\alpha)}[\p0^2\beta +(\p0\beta)^2
  -\p0\alpha \p0\beta]\cr
  R_{01} &=&  {2\over{r}}\p0\beta \cr
  R_{22} &=&  e^{-2\beta}[r(\p1\beta-\p1\alpha)-1]+1\cr
  R_{33} &=&  R_{22}\sin^2\theta\ .\label{7.16}
\eea

Our job is to set $R_\mn=0$.  From $R_{01}=0$ we get
\be
  \p0\beta = 0\ .\label{7.17}
\ee
If we consider taking the time derivative of $R_{22}=0$ and using
$\p0\beta = 0$, we get
\be
  \p0\p1\alpha =0\ .\label{7.18}
\ee
We can therefore write
\bea
  \beta &=&  \beta(r)\cr
  \alpha &=&  f(r)+g(t)\ .\label{7.19}
\eea
The first term in the metric (7.13) is therefore $-e^{2f(r)}e^{2g(t)}
\d t^2$.  But we could always simply redefine our time coordinate by
replacing $\d t\rightarrow e^{-g(t)}\d t$; in other words, we are free
to choose $t$ such that $g(t)=0$, whence $\alpha(t,r)=f(r)$.  We therefore 
have
\be
  ds^2 = -e^{2\alpha(r)}\d t^2 + e^{\beta(r)}\d r^2
  + r^2 d\Omega^2\ .\label{7.20}
\ee
All of the metric components are independent of the coordinate $t$.
We have therefore proven a crucial result: {\it any spherically symmetric
vacuum metric possesses a timelike Killing vector.}

This property is so interesting that it gets its own name: a metric
which possesses a timelike Killing vector is called {\bf stationary}.
There is also a more restrictive property: a metric is called
{\bf static} if it possesses a timelike Killing vector which is 
orthogonal to a family of hypersurfaces.  (A hypersurface in an
$n$-dimensional manifold is simply an ($n-1$)-dimensional submanifold.)
The metric (7.20) is not
only stationary, but also static; the Killing vector field $\p0$ is
orthogonal to the surfaces $t=const$ (since there are no cross terms
such as $\d t\d r$ and so on).  Roughly speaking, a static metric is
one in which nothing is moving, while a stationary metric allows things
to move but only in a symmetric way.  For example, the static spherically
symmetric metric (7.20) will describe non-rotating stars or black holes,
while rotating systems (which keep rotating in the same way at all times)
will be described by stationary metrics.  It's hard to remember which
word goes with which concept, but the distinction between the two
concepts should be understandable.

Let's keep going with finding the solution.  Since both $R_{00}$ and
$R_{11}$ vanish, we can write
\be
  0=e^{2(\beta-\alpha)}R_{00} + R_{11} = {2\over r}(\p1\alpha+
  \p1\beta)\ ,\label{7.21}
\ee
which implies $\alpha = -\beta + {\rm~constant}$.  Once again, we can
get rid of the constant by scaling our coordinates, so we have
\be
  \alpha = -\beta\ .\label{7.22}
\ee
Next let us turn to $R_{22}=0$, which now reads
\be
  e^{2\alpha}(2r\p1\alpha+1)=1\ .\label{7.23}
\ee
This is completely equivalent to
\be
  \p1(r e^{2\alpha})=1\ .\label{7.24}
\ee
We can solve this to obtain
\be
  e^{2\alpha}=1+{\mu\over r}\ ,\label{7.25}
\ee
where $\mu$ is some undetermined constant.  With (7.22) and (7.25), 
our metric becomes
\be
  ds^2 = -\left(1+{\mu\over r}\right)\d t^2 + 
  \left(1+{\mu\over r}\right)^{-1}\d r^2
  + r^2 d\Omega^2\ .\label{7.26}
\ee
We now have no freedom left except for the single constant $\mu$, so
this form better solve the remaining equations $R_{00}=0$ and
$R_{11}=0$; it is straightforward to check that it does, for any
value of $\mu$.

The only thing left to do is to interpret the constant $\mu$ in
terms of some physical parameter.  The most important use of a
spherically symmetric vacuum solution is to represent the spacetime
outside a star or planet or whatnot.  In that case we would expect
to recover the weak field limit as $r\rightarrow\infty$.  In this 
limit, (7.26) implies
\bea
  g_{00}(r\rightarrow\infty) &=& -\left(1+{\mu\over r}\right)\ ,\cr
  g_{rr}(r\rightarrow\infty) &=& \left(1-{\mu\over r}\right)\ .
  \label{7.27}
\eea
The weak field limit, on the other hand, has
\bea
  g_{00} &=& -\left(1+2\Phi\right)\ ,\cr
  g_{rr} &=& \left(1-2\Phi\right)\ ,
  \label{7.28}
\eea
with the potential $\Phi=-GM/r$.  Therefore the metrics do agree in
this limit, if we set $\mu = -2GM$.

Our final result is the celebrated {\bf Schwarzschild metric},
\be
  ds^2 = -\left(1-{{2GM}\over r}\right)\d t^2 + 
  \left(1-{{2GM}\over r}\right)^{-1}\d r^2
  + r^2 d\Omega^2\ .\label{7.29}
\ee
This is true for any spherically symmetric vacuum solution to 
Einstein's equations; $M$ functions as a parameter, which we happen
to know can be interpreted as the conventional Newtonian mass that we
would measure by studying orbits at large distances from the
gravitating source.  Note that as $M\rightarrow 0$ we recover
Minkowski space, which is to be expected.  Note also that the metric
becomes progressively Minkowskian as we go to $r\rightarrow\infty$;
this property is known as {\bf asymptotic flatness}.

The fact that the Schwarzschild metric is not just a good solution,
but is the unique spherically symmetric vacuum solution, is known as
{\bf Birkhoff's theorem}.  It is interesting to note that the result
is a static metric.  We did not say anything about the source
except that it be spherically symmetric.  Specifically, we did not
demand that the source itself be static; it could be a collapsing
star, as long as the collapse were symmetric.  Therefore a process
such as a supernova explosion, which is basically spherical, would be
expected to generate very little gravitational radiation (in comparison
to the amount of energy released through other channels).  This is
the same result we would have obtained in electromagnetism, where the
electromagnetic fields around a spherical charge distribution do not
depend on the radial distribution of the charges.

Before exploring the behavior of test particles in the Schwarzschild
geometry, we should say something about singularities.  From the form
of (7.29), the metric coefficients become infinite at $r=0$ and
$r=2GM$ --- an apparent sign that something is going wrong.  The metric 
coefficients, of course, are 
coordinate-dependent quantities, and as such we should not make too
much of their values; it is certainly possible to have a ``coordinate
singularity'' which results from a breakdown of a specific coordinate
system rather than the underlying manifold.  An example occurs at
the origin of polar coordinates in the plane, where the metric 
$ds^2 = \d r^2 + r^2 \d \theta^2$ becomes degenerate and the component 
$g^{\theta\theta}=r^{-2}$ of the inverse metric blows up, even
though that point of the manifold is no different from any other.

What kind of coordinate-independent signal should
we look for as a warning that something about the geometry is out of
control?  This turns out to be a difficult question to answer, and
entire books have been written about the nature of singularities in
general relativity.  We won't go into this issue in detail, but
rather turn to one simple criterion for when something has gone wrong ---
when the curvature becomes infinite.  The curvature is measured by
the Riemann tensor, and it is hard to say when a tensor becomes infinite,
since its components are coordinate-dependent.  But from the curvature
we can construct various scalar quantities, and since scalars are
coordinate-independent it will be meaningful to say that they become
infinite.  This simplest such scalar is the Ricci scalar $R=g^\mn R_\mn$,
but we can also construct higher-order scalars such as $R^\mn R_\mn$,
$R^{\mn\rho\sigma}R_{\mn\rho\sigma}$, $R_{\mn\rho\sigma}
R^{\rho\sigma\lambda\tau} R_{\lambda\tau}{}^{\mn}$, and so on.  If any
of these scalars (not necessarily all of them) go to infinity as we
approach some point, we will regard that point as a singularity of the
curvature.  We should also check that the point is not ``infinitely
far away''; that is, that it can be reached by travelling a finite 
distance along a curve.

We therefore have a sufficient condition for a point to be considered
a singularity.  It is not a necessary condition, however, and it is
generally harder to show that a given point is nonsingular; for our 
purposes we will simply test to see if geodesics are well-behaved at 
the point in
question, and if so then we will consider the point nonsingular.
In the case of the Schwarzschild metric (7.29), direct 
calculation reveals that
\be
  R^{\mn\rho\sigma}R_{\mn\rho\sigma} = {{12 G^2 M^2}\over {r^6}}\ .
  \label{7.30}
\ee
This is enough to convince us that $r=0$ represents an honest 
singularity.  At the other trouble spot, $r=2GM$, you could check
and see that none of the curvature invariants blows up.  We therefore
begin to think that it is actually not singular, and we have simply
chosen a bad coordinate system.  The best thing to do is to transform
to more appropriate coordinates if possible.  We will soon see that
in this case it is in fact possible, and the surface $r=2GM$ is 
very well-behaved (although interesting) in the Schwarzschild metric.

Having worried a little about singularities, we should point out that
the behavior of Schwarzschild at $r\leq 2GM$ is of little day-to-day
consequence.  The solution we derived is valid only in vacuum, and
we expect it to hold outside a spherical body such as a star.  However,
in the case of the Sun we are dealing with a body which extends to a
radius of
\be
  R_\odot = 10^6 G M_\odot\ .\label{7.31}
\ee
Thus, $r=2GM_\odot$ is far inside the solar interior, where we do not
expect the Schwarzschild metric to imply.  In fact, realistic stellar
interior solutions are of the form
\be
  ds^2 = -\left(1-{{2Gm(r)}\over r}\right)\d t^2 + 
  \left(1-{{2Gm(r)}\over r}\right)^{-1}\d r^2
  + r^2 d\Omega^2\ .\label{7.32}
\ee
See Schutz for details.  Here $m(r)$ is a function of $r$ which goes
to zero faster than $r$ itself, so there are no singularities to 
deal with at all.  Nevertheless, there are objects for which the
full Schwarzschild metric is required --- black holes --- and therefore
we will let our imaginations roam far outside the solar system in this
section.

The first step we will take to understand this metric more fully is to 
consider the behavior of geodesics.  We need the nonzero
Christoffel symbols for Schwarzschild:
\bea
  &\Gamma^1_{00} = {{GM}\over {r^3}}(r-2GM)\qquad
  \Gamma^1_{11} = {{-GM}\over{r(r-2GM)}} \qquad
  \Gamma^0_{01} = {{GM}\over{r(r-2GM)}} \cr
  &\quad\qquad\Gamma^2_{12} = {1\over r}\qquad\quad
  \Gamma^1_{22} = - (r-2GM)\qquad\quad
  \Gamma^3_{13} = {1\over r}\cr
  &\quad\Gamma^1_{33} = -(r-2GM)\sin^2\theta\qquad
  \Gamma^2_{33} = -\sin\theta \cos\theta \qquad
  \Gamma^3_{23} = {{\cos\theta}\over {\sin\theta}}\ .\label{7.33}
\eea
The geodesic equation therefore turns into the following four
equations, where $\lambda$ is an affine parameter:
\be
  {{d^2 t}\over{d\lambda^2}} + {{2GM}\over{r(r-2GM)}}
  {{dr}\over{d\lambda}}{{dt}\over{d\lambda}} =0\ ,\label{7.34}
\ee
\bea  
  \lefteqn{{{d^2 r}\over{d\lambda^2}} + {{GM}\over {r^3}}(r-2GM)
  \left({{dt}\over{d\lambda}}\right)^2 - {{GM}\over{r(r-2GM)}}
  \left({{dr}\over{d\lambda}}\right)^2} \cr
  &&- (r-2GM)\left[
  \left({{d\theta}\over{d\lambda}}\right)^2+\sin^2\theta
  \left({{d\phi}\over{d\lambda}}\right)^2\right] = 0 \ ,\label{7.35}
\eea
\be
  {{d^2 \theta}\over{d\lambda^2}} + {2\over r}{{d\theta}\over{d\lambda}}
  {{dr}\over{d\lambda}} - \sin\theta \cos\theta 
  \left({{d\phi}\over{d\lambda}}\right)^2 = 0\ ,\label{7.36}
\ee
and
\be
  {{d^2 \phi}\over{d\lambda^2}} + {2\over r}{{d\phi}\over{d\lambda}}
  {{dr}\over{d\lambda}} + 2{{\cos\theta}\over {\sin\theta}}
  {{d\theta}\over{d\lambda}}{{d\phi}\over{d\lambda}} = 0\ .\label{7.37}
\ee
There does not seem to be much hope for simply solving this set of
coupled equations by inspection.  Fortunately our task is greatly
simplified by the high degree of symmetry of the Schwarzschild metric.
We know that there are four Killing vectors: three for the spherical
symmetry, and one for time translations.  Each of these will lead to
a constant of the motion for a free particle; if $K^\mu$ is a Killing
vector, we know that
\be
  K_\mu {{dx^\mu}\over{d\lambda}} = {\rm constant}\ .\label{7.38}
\ee
In addition, there is another constant of the motion that we always
have for geodesics; metric compatibility implies that along the path
the quantity
\be
  \epsilon = -g_\mn {{dx^\mu}\over{d\lambda}}{{dx^\nu}\over{d\lambda}} 
  \label{7.39}
\ee
is constant.
Of course, for a massive particle we typically choose $\lambda = \tau$,
and this relation simply becomes $\epsilon = -g_\mn U^\mu U^\nu=+1$.  For 
a massless particle we always have $\epsilon =0$.  We will also be 
concerned with spacelike geodesics (even though they do not correspond
to paths of particles), for which we will choose $\epsilon = -1$.

Rather than immediately writing out explicit expressions for the
four conserved quantities associated with Killing vectors, let's think
about what they are telling us.  Notice that the symmetries they
represent are also present in flat spacetime, where the conserved
quantities they lead to are very familiar.  Invariance under time
translations leads to conservation of energy, while invariance under
spatial rotations leads to conservation of the three components of
angular momentum.  Essentially the same applies to the Schwarzschild
metric.  We can think of the angular momentum as a three-vector with
a magnitude (one component) and direction (two components).  Conservation
of the direction of angular momentum means that the particle will move
in a plane.  We can choose this to be the equatorial plane of our 
coordinate system; if the particle is not in this plane, we can rotate
coordinates until it is.  Thus, the two Killing vectors which lead to
conservation of the direction of angular momentum imply
\be
  \theta = {\pi\over 2}\ .\label{7.40}
\ee
The two remaining Killing vectors correspond to energy and the
magnitude of angular momentum.  The energy arises from the timelike
Killing vector $K = \partial_t$, or
\be
  K_\mu = \left(-\left(1-{{2GM}\over r}\right),0,0,0\right)
  \ .\label{7.41}
\ee
The Killing vector whose conserved quantity is the magnitude of the
angular momentum is $L = \p\phi$, or
\be
  L_\mu = \left(0,0,0,r^2\sin^2\theta \right)\ .\label{7.42}
\ee
Since (7.40) implies that $\sin\theta = 1$ along the geodesics of
interest to us, the two conserved quantities are
\be
  \left(1-{{2GM}\over r}\right){{dt}\over{d\lambda}}=E\ ,\label{7.43}
\ee
and
\be
  r^2{{d\phi}\over{d\lambda}}=L\ .\label{7.44}
\ee
For massless particles these can be thought of as the energy and 
angular momentum; for massive particles they are the energy and
angular momentum per unit mass of the particle.  Note that the 
constancy of (7.44) is the GR equivalent of Kepler's second law
(equal areas are swept out in equal times).

Together these conserved quantities provide a convenient way to
understand the orbits of particles in the Schwarzschild geometry.
Let us expand the expression (7.39) for $\epsilon$ to obtain
\be
  -\left(1-{{2GM}\over r}\right)\left({{dt}\over{d\lambda}}\right)^2 +
  \left(1-{{2GM}\over r}\right)^{-1}\left({{dr}\over{d\lambda}}\right)^2
  +r^2\left({{d\phi}\over{d\lambda}}\right)^2 = -\epsilon\ .\label{7.45}
\ee
If we multiply this by $(1-2GM/r)$ and use our expressions for $E$ and
$L$, we obtain
\be
  -E^2+\left({{dr}\over{d\lambda}}\right)^2+\left(1-{{2GM}\over r}\right)
  \left({{L^2}\over{r^2}} +\epsilon\right) =0\ .\label{7.46}
\ee
This is certainly progress, since we have taken a messy system of
coupled equations and obtained a single equation for $r(\lambda)$.
It looks even nicer if we rewrite it as
\be
  {1\over 2}\left({{dr}\over{d\lambda}}\right)^2+ V(r) = 
  {1\over 2}E^2\ ,\label{7.47}
\ee
where
\be
  V(r) = {1\over 2}\epsilon - \epsilon{{GM}\over r} + 
  {{L^2}\over{2r^2}} - {{GML^2}\over {r^3}}\ .\label{7.48}
\ee
In (7.47) we have precisely the equation for a classical particle of unit 
mass and ``energy'' ${1\over 2}E^2$ moving in a one-dimensional potential
given by $V(r)$.  (The true energy per unit mass is $E$, but the
effective potential for the coordinate $r$ responds to ${1\over 2}E^2$.)

\eject

Of course, our physical situation is quite different from a classical
particle moving in one dimension.  The trajectories under consideration
are orbits around a star or other object:

\begin{figure}[h]
  \centerline{
  \psfig{figure=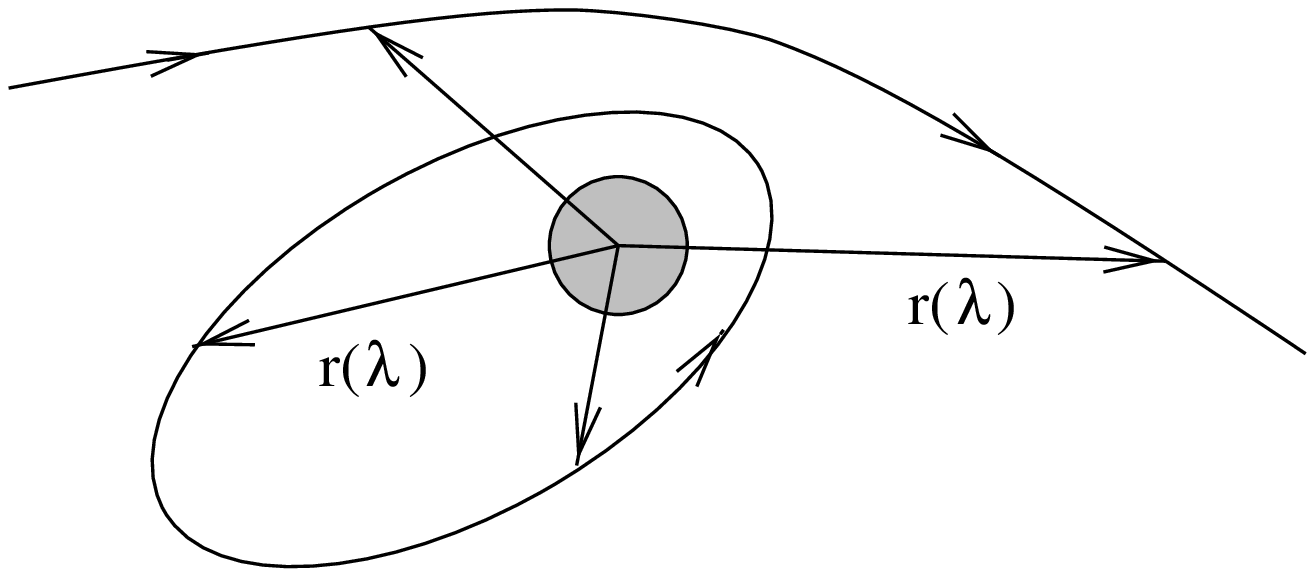,angle=0,height=4cm}}
\end{figure}

\noindent
The quantities of interest to us are not only $r(\lambda)$,
but also $t(\lambda)$ and $\phi(\lambda)$.  Nevertheless, we can go a
long way toward understanding all of the orbits by understanding their
radial behavior, and it is a great help to reduce this behavior to a 
problem we know how to solve.

A similar analysis of orbits in Newtonian gravity would have produced
a similar result; the general equation (7.47) would have been the
same, but the effective potential (7.48) would not have had the last
term.  (Note that this equation is not a power series in $1/r$, it is
exact.)  In the potential (7.48) the first term is just a constant, the
second term corresponds exactly to the Newtonian gravitational potential,
and the third term is a contribution from angular momentum which takes
the same form in Newtonian gravity and general relativity.  The last term,
the GR contribution, will turn out to make a great deal of difference,
especially at small $r$.

Let us examine the kinds of possible orbits, as illustrated in the
figures.  There are different curves $V(r)$ for different values
of $L$; for any one of these curves, the behavior of the orbit can be
judged by comparing the ${1\over 2}E^2$ to $V(r)$.  The general behavior of
the particle will be to move in the potential until it reaches a
``turning point'' where $V(r)={1\over 2}E^2$, where it will begin moving 
in the other direction.  Sometimes there may be no turning point to hit,
in which case the particle just keeps going.  In other cases the
particle may simply move in a circular orbit at radius $r_c= const$; this 
can happen if the potential is flat, $dV/dr=0$.  Differentiating
(7.48), we find that the circular orbits occur when
\be
  \epsilon GM r_c^2 - L^2 r_c +3GML^2\gamma =0\ ,\label{7.49}
\ee
where $\gamma=0$ in Newtonian gravity and $\gamma=1$ in general
relativity.  Circular orbits will be stable if they correspond to
a minimum of the potential, and unstable if they correspond to a
maximum.  Bound orbits which are not circular will oscillate
around the radius of the stable circular orbit.

\begin{figure}
  \centerline{
  \psfig{figure=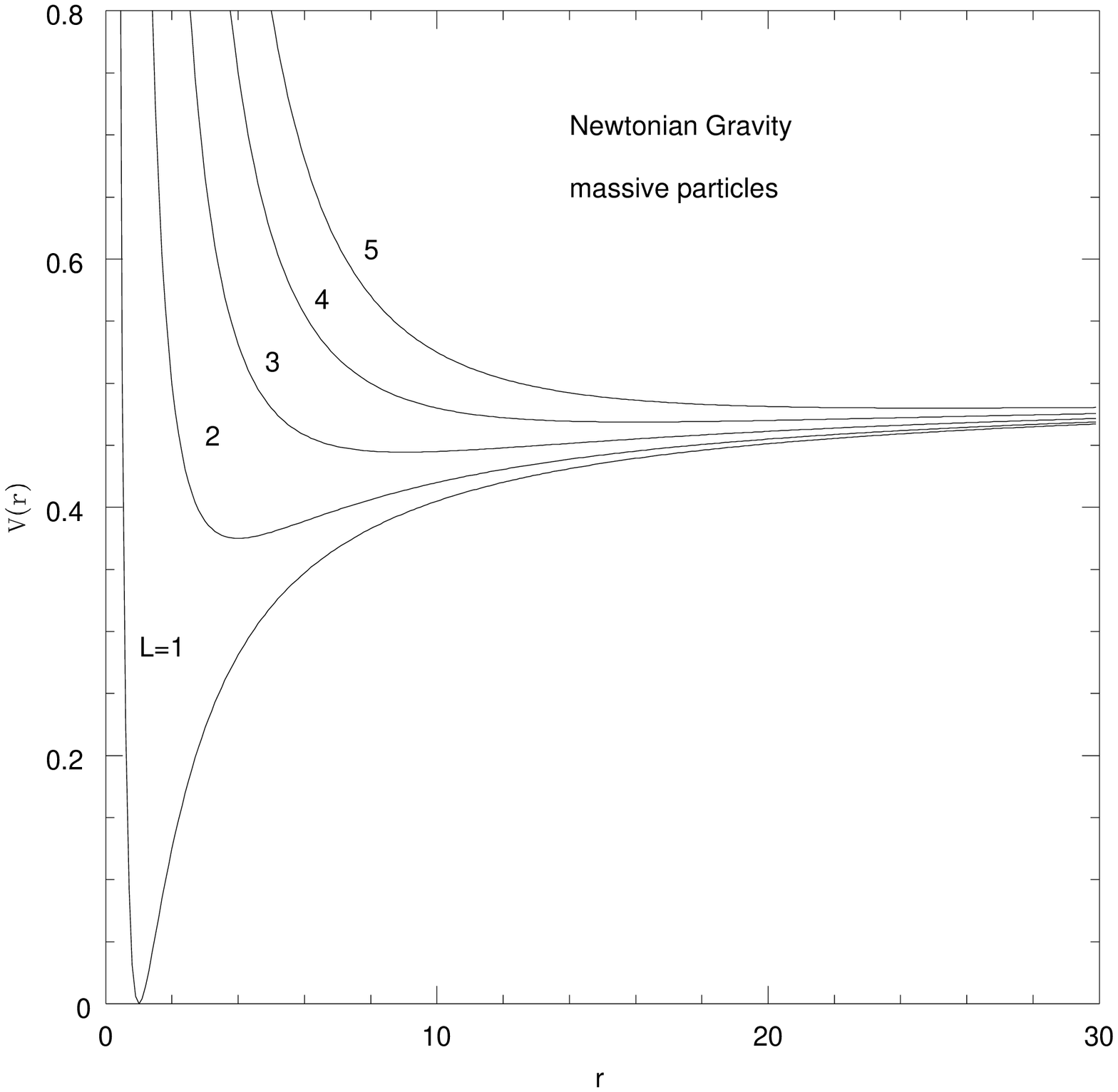,angle=0,height=11cm}}
  \vskip-2cm
\end{figure}

\begin{figure}
  \centerline{
  \psfig{figure=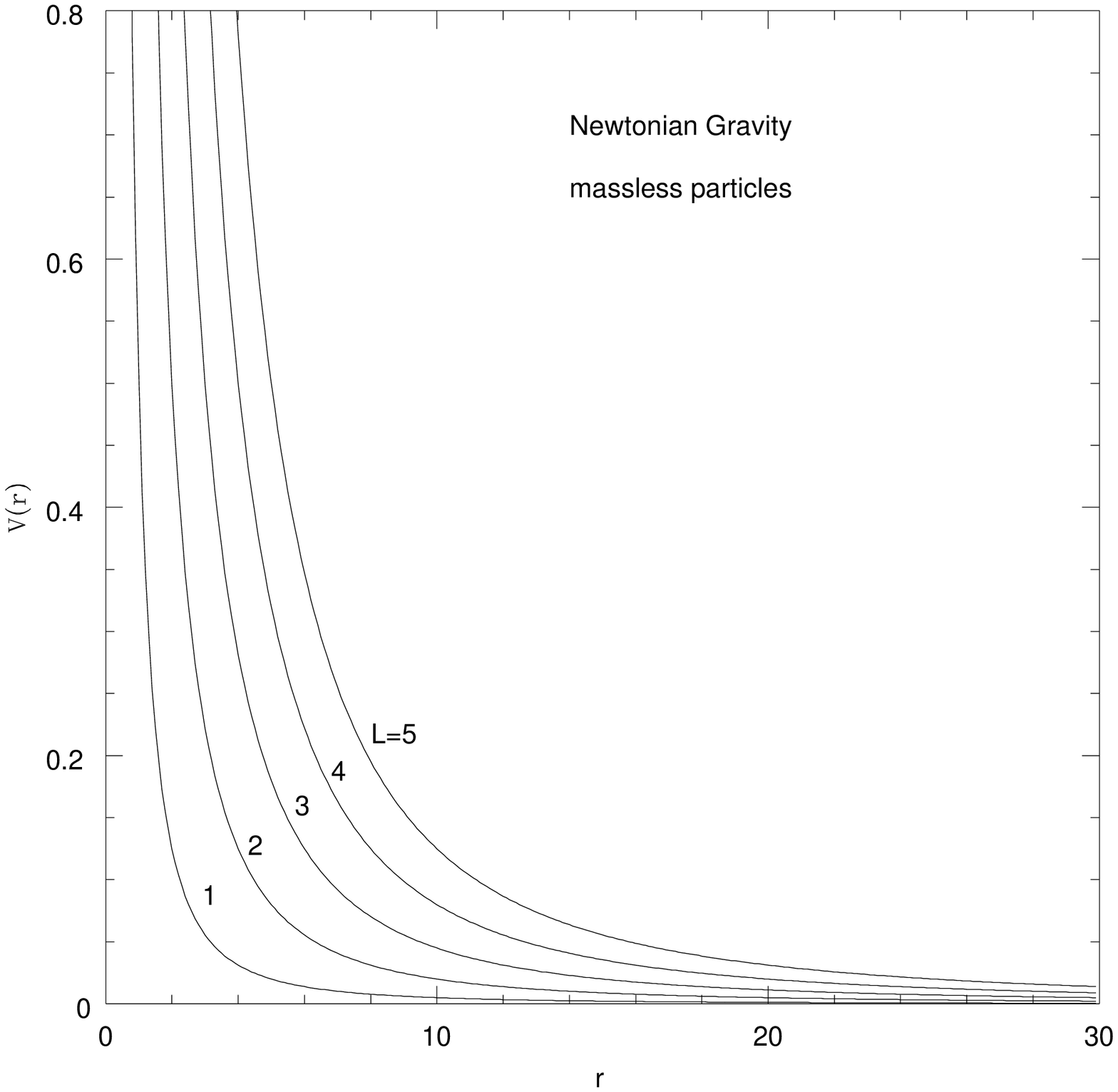,angle=0,height=11cm}}
  \vskip-2cm
\end{figure}

Turning to Newtonian gravity, we find that circular orbits appear at
\be
  r_c = {{L^2}\over{\epsilon GM}}\ .\label{7.50}
\ee
For massless particles $\epsilon=0$, and there are no circular orbits;
this is consistent with the figure, which illustrates that there are no
bound orbits of any sort.  Although it is somewhat obscured in this
coordinate system, massless particles actually move in a straight
line, since the Newtonian gravitational force on a massless particle is
zero.  (Of course the standing of massless particles in Newtonian theory
is somewhat problematic, but we will ignore that for now.)  In terms of
the effective potential, a photon with a given energy $E$ will come in 
from $r=\infty$ and gradually ``slow down'' (actually $dr/d\lambda$
will decrease, but the speed of light isn't changing) until it reaches
the turning point, when it will start moving away back to 
$r=\infty$.  The lower values of $L$, for which the photon will come
closer before it starts moving away, are simply those trajectories which
are initially aimed closer to the gravitating body.
For massive particles there will be stable circular orbits at the
radius (7.50), as well as bound orbits which oscillate around this
radius.  If the energy is greater than the asymptotic value $E=1$,
the orbits will be unbound, describing a particle that approaches the
star and then recedes.  We know that the orbits in Newton's theory are
conic sections --- bound orbits are either circles or ellipses, while
unbound ones are either parabolas or hyperbolas --- although we won't
show that here.

In general relativity the situation is different, but only for $r$ 
sufficiently small.  Since the difference resides in the term $-GML^2/r^3$,
as $r\rightarrow\infty$ the behaviors are identical in the two 
theories.  But as $r\rightarrow 0$ the potential goes to $-\infty$
rather than $+\infty$ as in the Newtonian case.  At $r=2GM$ the 
potential is always zero; inside this radius is the black hole, which
we will discuss more thoroughly later.   For massless particles
there is always a barrier (except for $L=0$, for which the potential
vanishes identically), but a sufficiently energetic photon will
nevertheless go over the barrier and be dragged inexorably down to
the center.  (Note that ``sufficiently energetic'' means ``in comparison
to its angular momentum'' --- in fact the frequency of the photon is
immaterial, only the direction in which it is pointing.)  At the top
of the barrier there are unstable circular orbits.
For $\epsilon=0$, $\gamma=1$, we can easily solve (7.49) to obtain
\be
  r_c = 3GM\ .\label{7.51}
\ee
This is borne out by the figure, which shows a maximum of $V(r)$ at
$r=3GM$ for every $L$.  This means that a photon can orbit forever in
a circle at this radius, but any 
perturbation will cause it to fly away either to $r=0$ or
$r=\infty$.

\begin{figure}
  \centerline{
  \psfig{figure=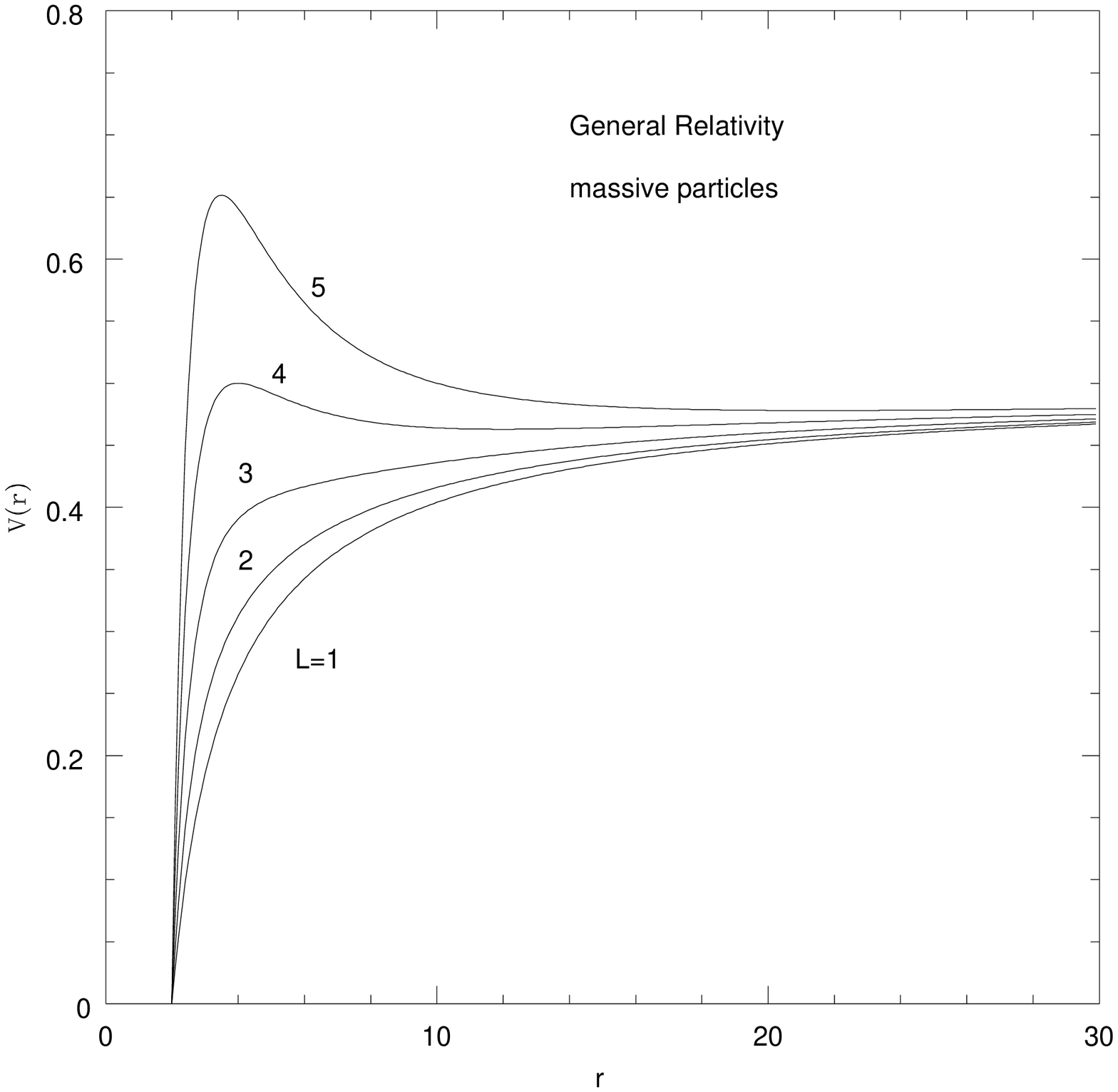,angle=0,height=11cm}}
  \vskip-2cm
\end{figure}

\begin{figure}
  \centerline{
  \psfig{figure=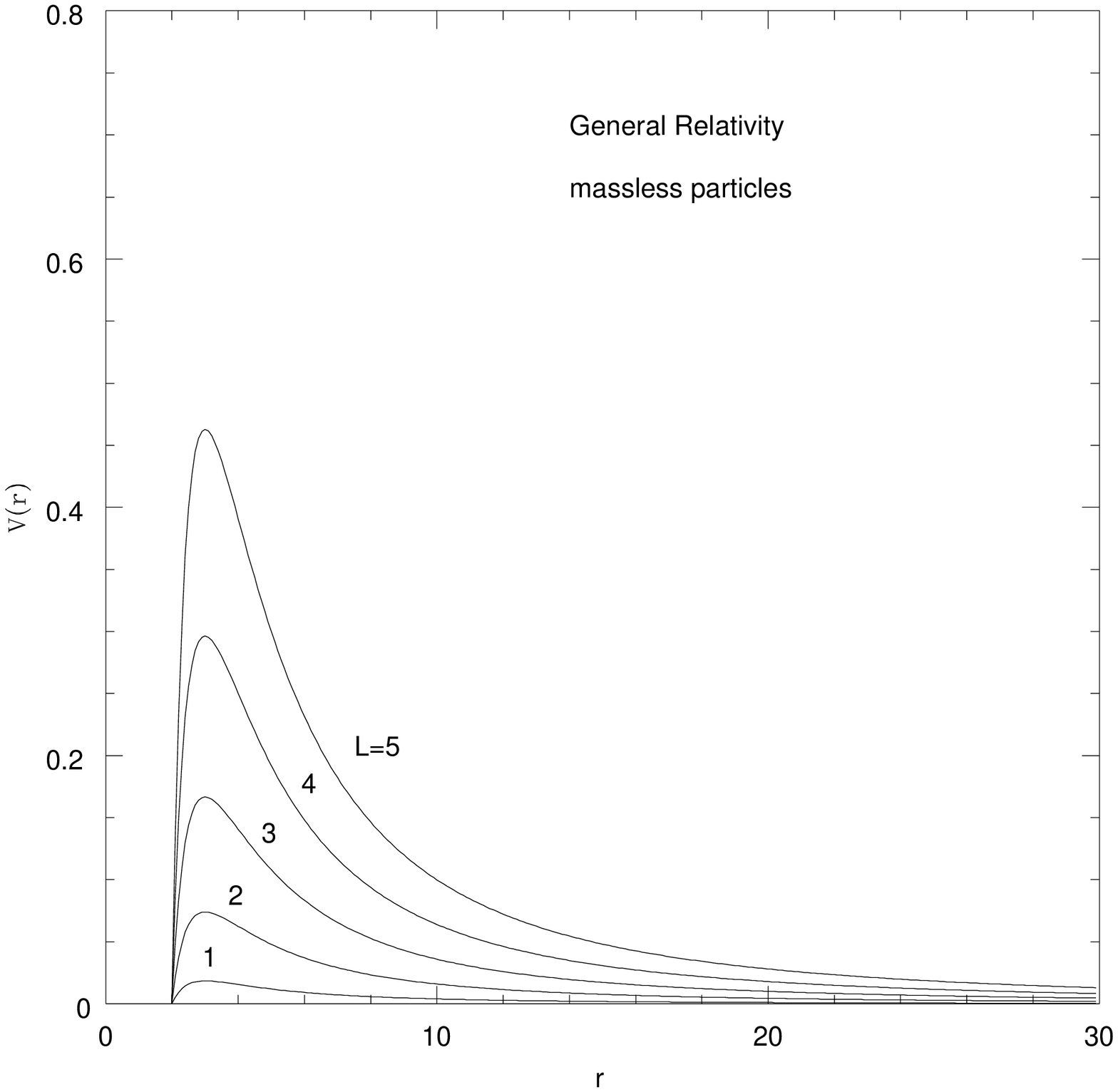,angle=0,height=11cm}}
  \vskip-2cm
\end{figure}

For massive particles there are once again different regimes depending
on the angular momentum.  The circular orbits are at
\be
  r_c = {{L^2\pm \sqrt{L^4 - 12 G^2 M^2 L^2}}\over{2GM}}\ .\label{7.52}
\ee
For large $L$ there will be two circular orbits, one stable and one 
unstable.  In the $L\rightarrow\infty$ limit their radii are given by
\be
  r_c = {{L^2\pm L^2(1 - 6 G^2 M^2/L^2)}\over{2GM}} =
  \left({{L^2}\over{GM}} ,\ 3GM\right)\ .\label{7.53}
\ee
In this limit the stable circular orbit becomes farther and farther 
away, while the unstable one approaches $3GM$, behavior which parallels 
the massless case.  As we decrease $L$ the two circular
orbits come closer together; they coincide when the discriminant in
(7.52) vanishes, at
\be
  L = \sqrt{12}GM\ ,\label{7.54}
\ee
for which
\be
  r_c = 6GM\ ,\label{7.55}
\ee
and disappear entirely for smaller $L$.  Thus $6GM$ is the smallest 
possible radius of a stable circular orbit in the 
Schwarzschild metric.  There are also unbound orbits, which come in 
from infinity and turn around, and bound but noncircular ones, which
oscillate around the stable circular radius.  Note that such
orbits, which would describe exact conic sections in
Newtonian gravity, will not do so in GR, although we would have to
solve the equation for $d\phi/dt$ to demonstrate it.  Finally, there are
orbits which come in from infinity and continue all the way in to 
$r=0$; this can happen either if the energy is higher than the barrier,
or for $L<\sqrt{12}GM$, when the barrier goes away entirely.

We have therefore found that the Schwarzschild solution possesses
stable circular orbits for $r>6GM$ and unstable circular orbits for
$3GM < r < 6GM$.  It's important to remember that these are only
the geodesics; there is nothing to stop an accelerating particle from
dipping below $r=3GM$ and emerging, as long as it stays beyond
$r=2GM$.

Most experimental tests of general relativity involve the motion of
test particles in the solar system, and hence geodesics
of the Schwarzschild metric; this is therefore a good place to
pause and consider these tests.  Einstein suggested three tests:
the deflection of light, the precession of perihelia, and gravitational
redshift.  The deflection of light is observable in the weak-field
limit, and therefore is not really a good test of the exact
form of the Schwarzschild geometry.  Observations of this deflection
have been performed during eclipses of the Sun, with results which
agree with the GR prediction (although it's not an especially clean
experiment).  The precession of perihelia reflects
the fact that noncircular orbits are not closed ellipses; to a good
approximation they are ellipses which precess, describing a flower pattern.

\begin{figure}
  \centerline{
  \psfig{figure=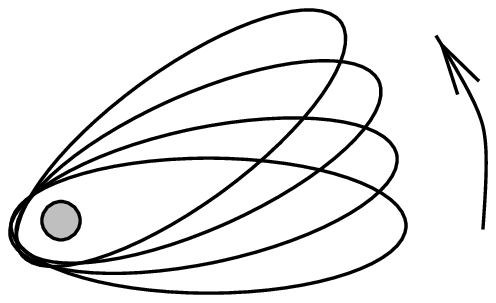,angle=0,height=4cm}}
\end{figure}

Using our geodesic equations, we could solve for 
$d\phi/d\lambda$ as a power series in the eccentricity $e$ of the
orbit, and from that obtain the apsidal frequency $\omega_a$,
defined as $2\pi$ divided by the time it takes for the ellipse to
precess once around.
For details you can look in Weinberg; the answer is
\be
  \omega_a={{3(GM)^{3/2}}\over {c^2(1-e^2)r^{5/2}}}\ ,\label{7.56}
\ee
where we have restored the $c$ to make it easier to compare with
observation.
(It is a good exercise to derive this yourself to lowest nonvanishing
order, in which case the $e^2$ is missing.)  Historically the
precession of Mercury was the first test of GR.  For Mercury the 
relevant numbers are
\bea
  {{GM_\odot}\over{c^2}}&=& 1.48\times 10^5 {\rm ~cm}\ ,\cr
  a &=&  5.55\times 10^{12}{\rm ~cm}\ ,\label{7.57}
\eea
and of course $c=3.00\times 10^{10}$~cm/sec.  This gives $\omega_a
= 2.35\times 10^{-14}$~sec$^{-1}$.  In other words, the major axis
of Mercury's orbit precesses at a rate of $42.9$~arcsecs every 100
years.  The observed value is $5601$~arcsecs/100~yrs.  However,
much of that is due to the precession of equinoxes in our geocentric
coordinate system; $5025$~arcsecs/100~yrs, to be precise.  The
gravitational perturbations of the other planets contribute an
additional $532$~arcsecs/100~yrs, leaving $43$~arcsecs/100~yrs
to be explained by GR, which it does quite well.

The gravitational redshift, as we have seen, is another effect
which is present in the weak field limit, and in fact will be predicted
by any theory of gravity which obeys the Principle of Equivalence.
However, this only applies to small enough regions of spacetime; over
larger distances, the exact amount of redshift will depend on the
metric, and thus on the theory under question.  It is therefore
worth computing the redshift in the Schwarzschild geometry.  We 
consider two observers who are not moving on geodesics, but are stuck
at fixed spatial coordinate values $(r_1,\theta_1,\phi_1)$ and
$(r_2,\theta_2,\phi_2)$.  According to (7.45), the proper time of
observer $i$ will be related to the coordinate time $t$ by
\be
  {{d\tau_i}\over{dt}} = \left(1-{{2GM}\over{r_i}}\right)^{1/2}\ .
  \label{7.58}
\ee
Suppose that the observer ${\cal O}_1$ emits a light pulse which travels 
to the observer ${\cal O}_2$, such that ${\cal O}_1$ measures the time
between two successive crests of the light wave to be $\Delta\tau_1$.
Each crest follows the same path to ${\cal O}_2$, except that they
are separated by a coordinate time
\be
  \Delta t= \left(1-{{2GM}\over{r_1}}\right)^{-1/2}\Delta \tau_1\ .
  \label{7.59}
\ee
This separation in coordinate time does not change along the photon
trajectories, but the second observer measures a time between successive
crests given by
\bea
  \Delta\tau_2 &=&  \left(1-{{2GM}\over{r_2}}\right)^{1/2}
  \Delta t\cr
  &=&  \left({{1-{{2GM}/{r_2}}}\over{1-{{2GM}/{r_1}}}}\right)^{1/2}
  \Delta\tau_1\ .\label{7.60}
\eea
Since these intervals $\Delta\tau_i$ measure the proper time between
two crests of an electromagnetic wave, the observed frequencies will be
related by 
\bea
  {{\omega_2}\over{\omega_1}}&=& {{\Delta\tau_1}\over
  {\Delta\tau_2}}\cr
  &=& \left({{1-{{2GM}/{r_1}}}\over{1-{{2GM}/{r_2}}}}\right)^{1/2}
  \ .\label{7.61}
\eea
This is an exact result for the frequency shift; in the limit $r>>2GM$
we have
\bea
  {{\omega_2}\over{\omega_1}}&=& 1-{{GM}\over{r_1}}+
  {{GM}\over{r_2}}\cr
  &=& 1+\Phi_1-\Phi_2\ .\label{7.62}
\eea
This tells us that the frequency goes down as $\Phi$ increases, which
happens as we climb out of a gravitational field; thus, a redshift.
You can check that it agrees with our previous calculation based on
the equivalence principle.

Since Einstein's proposal of the three classic tests, further tests
of GR have been proposed.  The most famous is of course the binary
pulsar, discussed in the previous section.  Another is the gravitational
time delay, discovered by (and observed by) Shapiro.  This is just
the fact that the time elapsed along two different trajectories between
two events need not be the same.  It has been measured by reflecting
radar signals off of Venus and Mars, and once again is consistent with
the GR prediction.  One effect which has not yet been observed is
the Lense-Thirring, or frame-dragging effect.  
There has been a long-term effort devoted to a proposed
satellite, dubbed Gravity Probe B, which would involve extraordinarily
precise gyroscopes whose precession could be measured and the 
contribution from GR sorted out.  It has a ways to go before being
launched, however, and the survival of such projects is always
year-to-year. 

We now know something about the behavior of geodesics outside the 
troublesome radius $r=2GM$, which is the regime of interest for the
solar system and most other astrophysical situations.  We will next turn 
to the study of objects which are described by the Schwarzschild solution
even at radii smaller than $2GM$ --- black holes.  (We'll use the
term ``black hole'' for the moment, even though we haven't introduced
a precise meaning for such an object.)

One way of understanding a geometry is to explore its causal structure,
as defined by the light cones.  We therefore consider radial null curves, 
those for which $\theta$ and $\phi$ are constant and $ds^2=0$:
\be
  ds^2 = 0 = -\left(1-{{2GM}\over r}\right)\d t^2 
  +\left(1-{{2GM}\over r}\right)^{-1}\d r^2 \ ,\label{7.63}
\ee
from which we see that
\be
  {{dt}\over{dr}}=\pm \left(1-{{2GM}\over r}\right)^{-1}\ .\label{7.64}
\ee
This of course measures the slope of the light cones on a spacetime diagram
of the $t$-$r$ plane.  For large $r$ the slope is $\pm 1$, as it would
be in flat space, while as we approach $r=2GM$  we get $dt/dr\rightarrow
\pm\infty$, and the light cones ``close up'':

\begin{figure}[h]
  \centerline{
  \psfig{figure=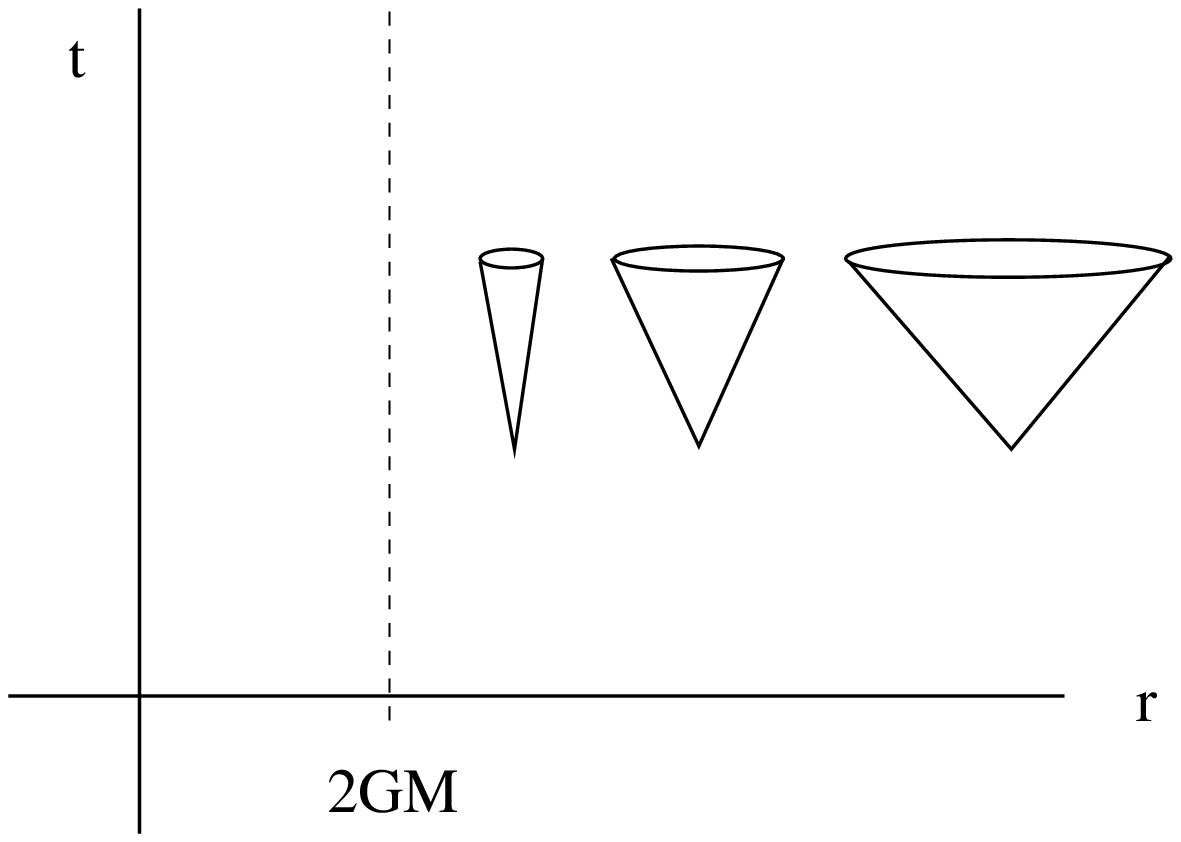,angle=0,height=5cm}}
\end{figure}

\noindent Thus a light ray which approaches $r=2GM$ never seems to
get there, at least in this coordinate system; instead it seems to
asymptote to this radius.

As we will see, this is an illusion, and the light ray (or a massive
particle) actually has no trouble reaching $r=2GM$.  But an observer
far away would never be able to tell.  If we stayed outside while
an intrepid observational general relativist dove into the black
hole, sending back signals all the time, we would simply see the
signals reach us more and more slowly.
\begin{figure}
  \centerline{
  \psfig{figure=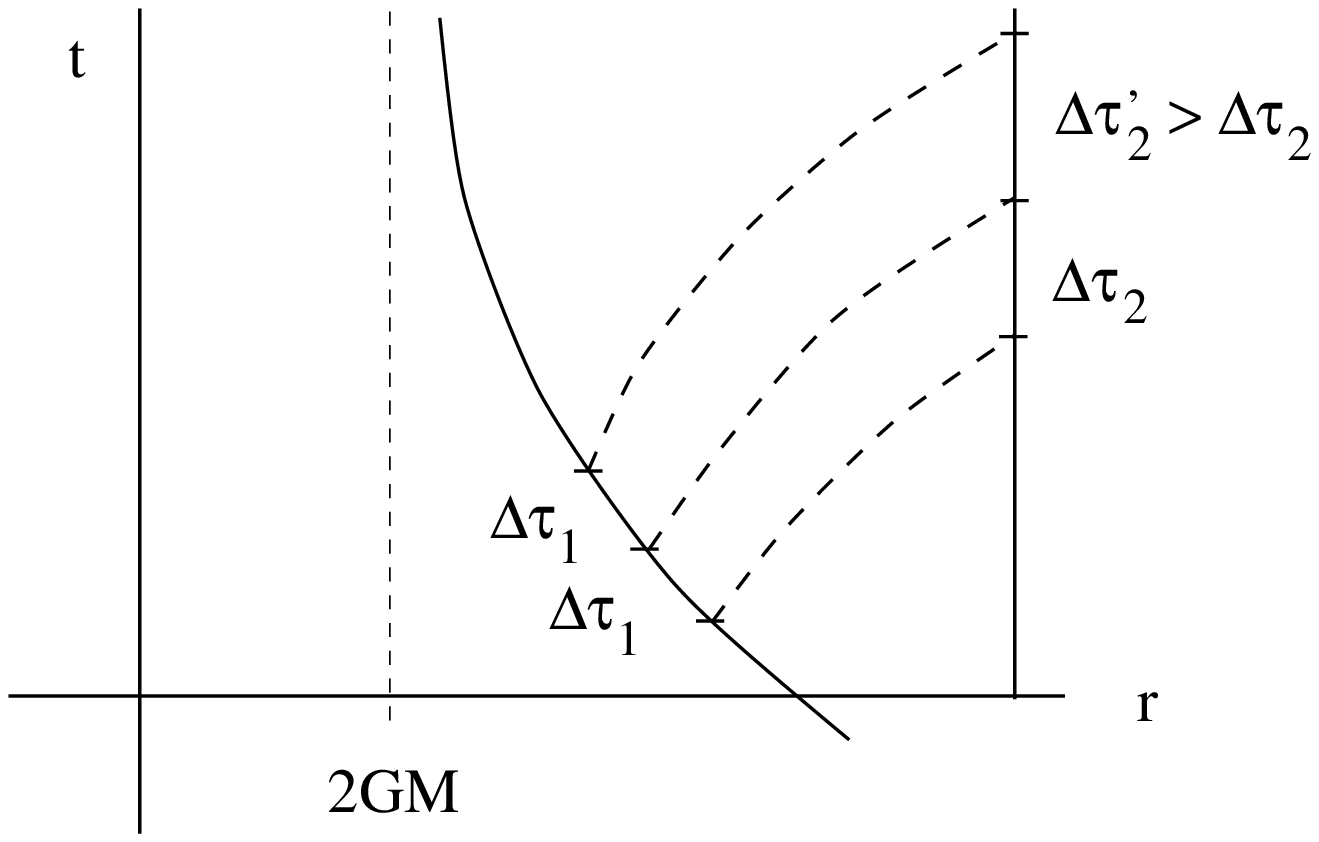,angle=0,height=6cm}}
\end{figure}
This should be clear from the pictures, and is confirmed
by our computation of $\Delta\tau_1/\Delta\tau_2$ when we discussed
the gravitational redshift (7.61).  As infalling astronauts
approach $r=2GM$, any fixed interval $\Delta\tau_1$ of their proper
time corresponds to a longer and longer interval $\Delta\tau_2$
from our point of view.  This continues forever; we would never see
the astronaut cross $r=2GM$, we would just see them move more and 
more slowly (and become redder and redder, almost as if they were
embarrassed to have done something as stupid as diving into a black
hole).

The fact that we never see the infalling astronauts reach $r=2GM$
is a meaningful statement, but the fact that their trajectory in
the $t$-$r$ plane never reaches there is not.  It is highly dependent
on our coordinate system, and we would like to ask a more 
coordinate-independent question (such as, do the astronauts reach
this radius in a finite amount of their proper time?).  The best
way to do this is to change coordinates to a system which is better
behaved at $r=2GM$.
There does exist a set of such coordinates, which we now set out to
find.  There is no way to ``derive'' a coordinate transformation, of
course, we just say what the new coordinates are and plug in the
formulas.  But we will develop these coordinates in several steps,
in hopes of making the choices seem somewhat motivated.

The problem with our current coordinates is that
$dt/dr\rightarrow \infty$ along radial null geodesics which approach
$r=2GM$; progress in the $r$ direction becomes slower and slower with
respect to the coordinate time $t$.  We can try to fix this problem
by replacing $t$ with a coordinate which ``moves more slowly'' along
null geodesics.  First notice that we can explicitly solve the
condition (7.64) characterizing radial null curves to obtain
\be
  t = \pm r^* +{\rm ~constant}\ ,\label{7.65}
\ee
where the {\bf tortoise coordinate} $r^*$ is defined by
\be
  r^* = r+2GM \ln\left({{r}\over{2GM}}-1\right)\ .\label{7.66}
\ee
(The tortoise coordinate is only sensibly related to $r$ when
$r\geq 2GM$, but beyond there our coordinates aren't very good
anyway.)  In terms of 
the tortoise coordinate the Schwarzschild metric becomes
\be
  ds^2 = \left(1-{{2GM}\over r}\right)\left(-\d t^2 
  +\d {r^*}^2 \right) + r^2 d\Omega^2\ ,\label{7.67}
\ee
where $r$ is thought of as a function of $r^*$.
This represents some progress, since the light cones now don't
seem to close up; furthermore, none of the metric coefficients becomes
infinite at $r=2GM$ (although both $g_{tt}$ and $g_{r^* r^*}$ become
zero).  The price we pay, however, is that the surface
of interest at $r=2GM$ has just been pushed to infinity.

\begin{figure}
  \centerline{
  \psfig{figure=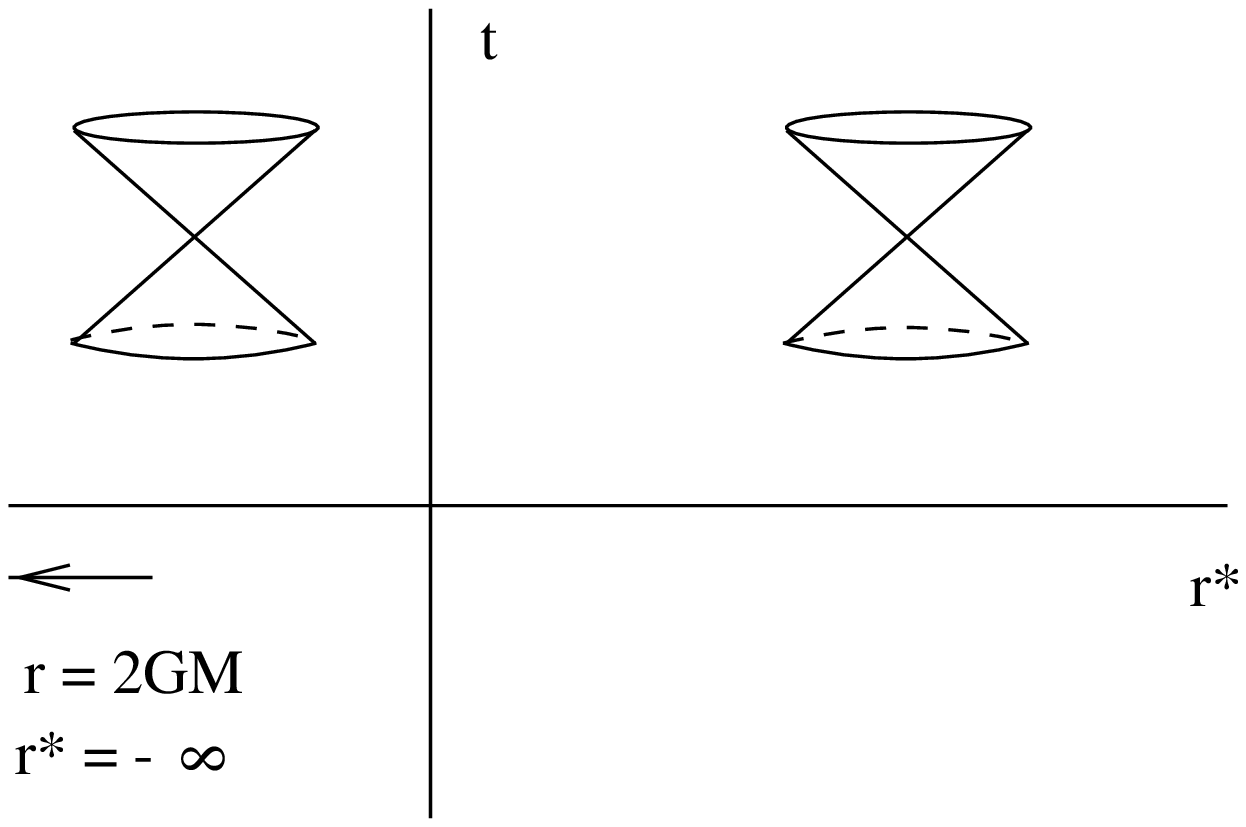,angle=0,height=5cm}}
\end{figure}

Our next move is to define coordinates which are naturally adapted
to the null geodesics.  If we let
\bea
  \tilde u &=&  t+r^*\cr \tilde v &=&  t-r^*\ ,\label{7.68}
\eea
then infalling radial null geodesics are characterized by
$\tilde u = $~constant, while the outgoing ones satisfy 
$\tilde v = $~constant.
Now consider going back to the original radial coordinate $r$,
but replacing the timelike coordinate $t$ with the new coordinate
$\tilde u$.  These are known as {\bf Eddington-Finkelstein coordinates}.
In terms of them the metric is
\be
  ds^2 = -\left(1-{{2GM}\over r}\right)\d{\tilde u}^2 +
  (\d\tilde u \d r + \d r\d\tilde u) + r^2 d\Omega^2\ .\label{7.69}
\ee
Here we see our first sign of real progress.  Even though the metric
coefficient $g_{\tilde u \tilde u}$ vanishes at $r=2GM$, there is
no real degeneracy; the determinant of the metric is
\be
  g = -r^4 \sin^2\theta\ ,\label{7.70}
\ee
which is perfectly regular at $r=2GM$.  Therefore the metric is
invertible, and we see once and for all that $r=2GM$ is simply a
coordinate singularity in our original $(t,r,\theta,\phi)$ system.
In the Eddington-Finkelstein coordinates the condition for radial
null curves is solved by
\be
  {{d\tilde u}\over {dr}} = \cases{ 0\ ,& {\rm (infalling)}\cr
  2\left(1-{{2GM}\over r}\right)^{-1}\ .& {\rm (outgoing)}\cr}
  \label{7.71}
\ee
We can therefore see what has happened: in this coordinate system 
the light cones remain well-behaved at $r=2GM$, and this surface
is at a finite coordinate value.  There is no problem in tracing
the paths of null or timelike particles past the surface.
On the other hand, something interesting is certainly going on.
Although the light cones don't close up, they do tilt over, such
that for $r< 2GM$ all future-directed paths are in the direction
of decreasing $r$.

\begin{figure}
  \centerline{
  \psfig{figure=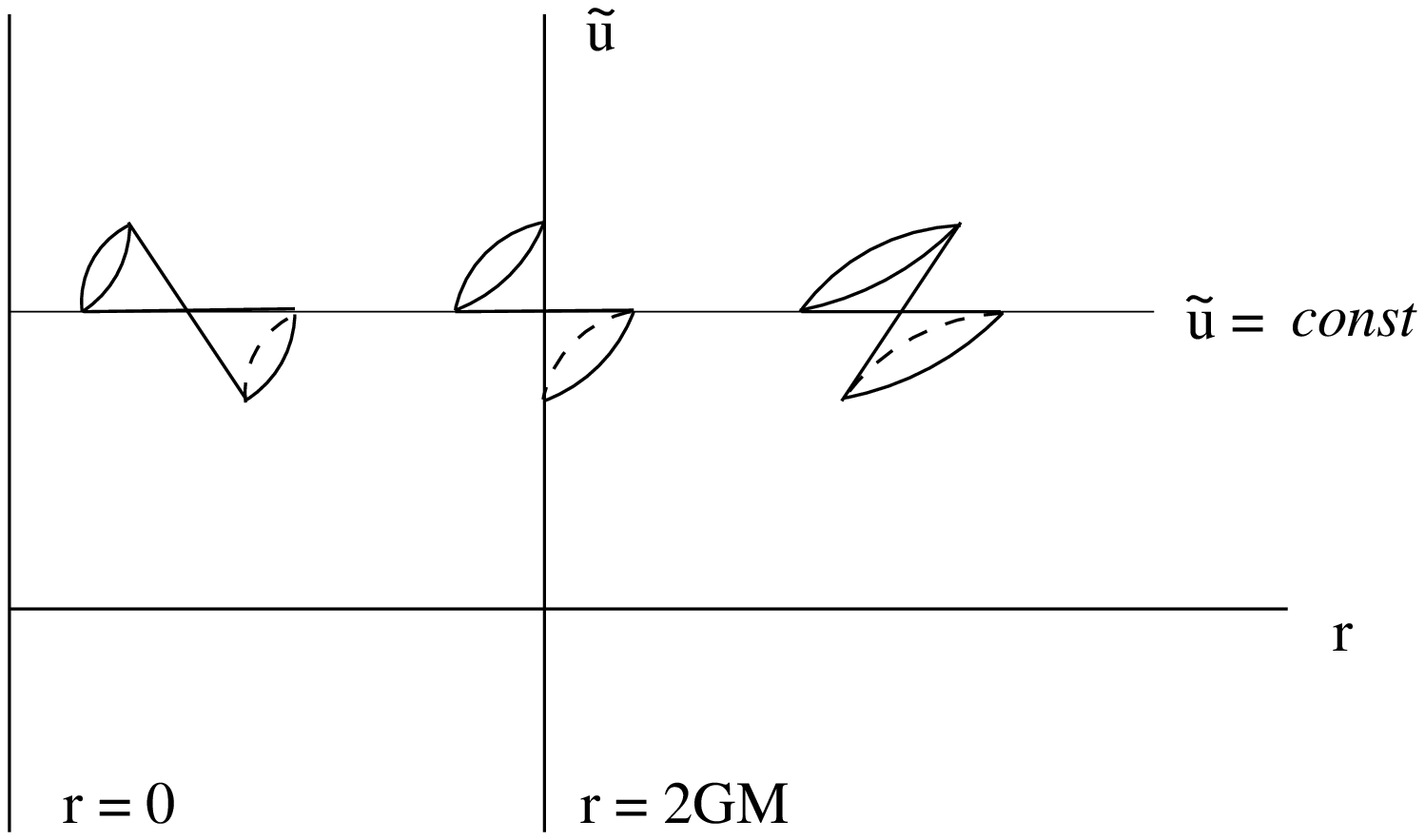,angle=0,height=5cm}}
\end{figure}

The surface $r=2GM$, while being locally perfectly regular, globally
functions as a point of no return --- once a test particle dips
below it, it can never come back.  For this reason $r=2GM$ is known
as the {\bf event horizon}; no event at $r\leq 2GM$ can influence any
other event at $r>2GM$.  Notice that the event horizon is a null surface, 
not a timelike one.  Notice also that since nothing can escape the
event horizon, it is impossible for us to ``see inside'' --- thus the
name {\bf black hole}.

Let's consider what we have done.  Acting under the suspicion that
our coordinates may not have been good for the entire manifold, we
have changed from our original coordinate $t$ to the new one $\tilde u$,
which has the nice property that if we decrease $r$ along a radial
curve null curve $\tilde u =$~constant, we go right through the
event horizon without any problems.  (Indeed, a local observer actually
making the trip would not necessarily know when the event horizon had
been crossed --- the local geometry is 
no different than anywhere else.)  We therefore
conclude that our suspicion was correct and our initial coordinate
system didn't do a good job of covering the entire manifold.  The region
$r\leq 2GM$ should certainly be included in our spacetime, since
physical particles can easily reach there and pass through.  However,
there is no guarantee that we are finished; perhaps there are other
directions in which we can extend our manifold.

In fact there are.  Notice that in the $(\tilde u, r)$ coordinate
system we can cross the event horizon on future-directed paths, but
not on past-directed ones.  This seems unreasonable, since we started
with a time-independent solution.  But we could have chosen $\tilde v$
instead of $\tilde u$, in which case the metric would have been
\be
  ds^2 = -\left(1-{{2GM}\over r}\right)\d{\tilde v}^2
  -(\d\tilde v\d r + \d r\d\tilde v) + r^2 d\Omega^2\ .\label{7.72}
\ee
Now we can once again pass through the event horizon, but this time
only along past-directed curves.

\begin{figure}
  \centerline{
  \psfig{figure=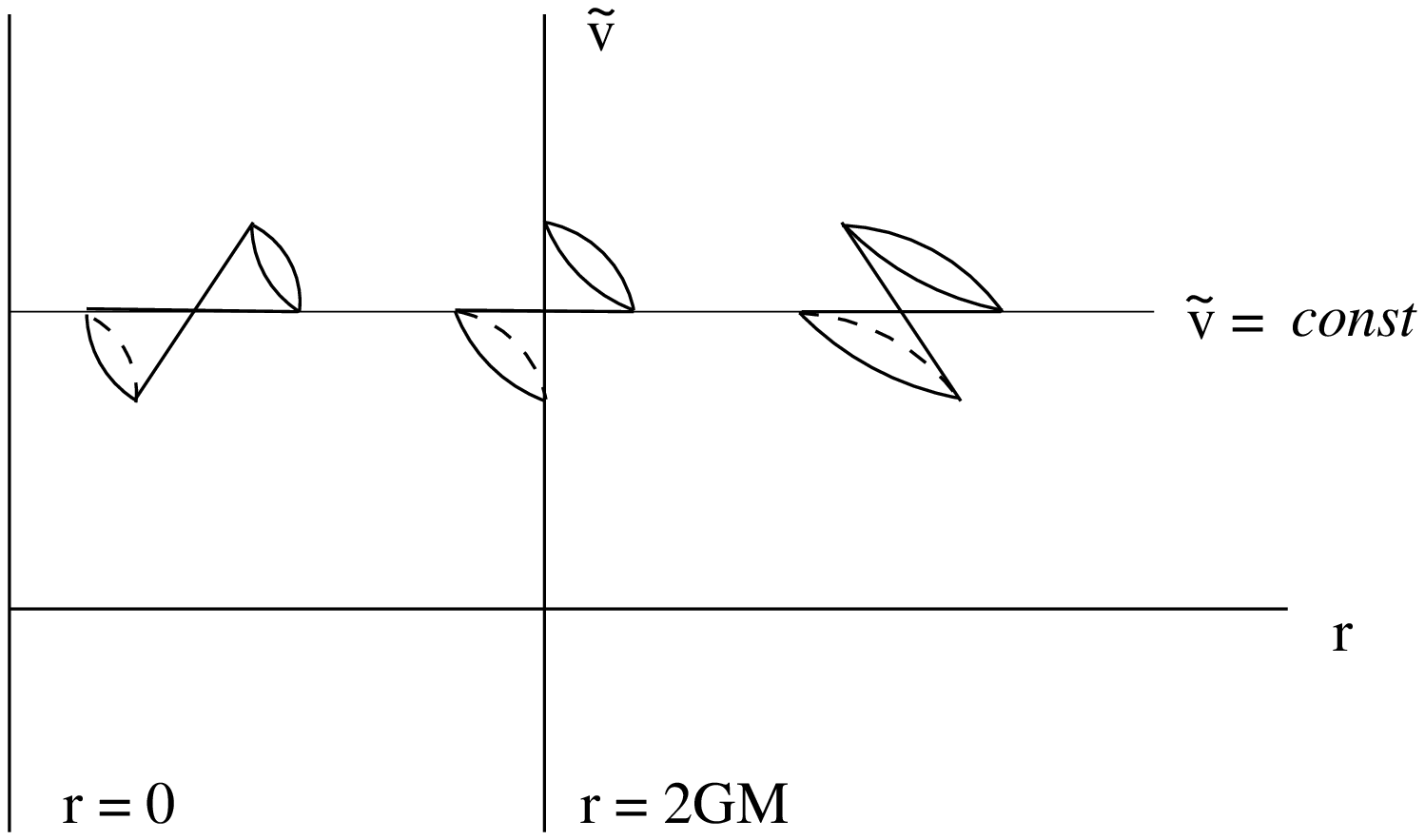,angle=0,height=5cm}}
\end{figure}

This is perhaps a surprise: we can consistently follow either 
future-directed or past-directed curves through $r=2GM$, but we
arrive at different places.  It was actually to be expected, since
from the definitions (7.68), if we keep $\tilde u$ constant and decrease
$r$ we must have $t\rightarrow +\infty$, while if we keep
$\tilde v$ constant and decrease $r$ we must have $t\rightarrow -\infty$.
(The tortoise coordinate $r^*$ goes to $-\infty$ as $r\rightarrow 2GM$.)
So we have extended spacetime in two different directions, one to
the future and one to the past.

The next step would be to follow spacelike geodesics to see if we 
would uncover still more regions.  The answer is yes, we would reach
yet another piece of the spacetime, but let's shortcut the process by
defining coordinates that are good all over.  A first guess might be to
use both $\tilde u$ and $\tilde v$ at once (in place of $t$ and $r$),
which leads to
\be
  ds^2 = {1\over 2}\left(1-{{2GM}\over r}\right)(\d{\tilde u}
  \d{\tilde v}+\d{\tilde v}\d{\tilde u}) +r^2 d\Omega^2\ ,\label{7.73}
\ee
with $r$ defined implicitly in terms of $\tilde u$ and $\tilde v$ by
\be
  {1\over 2}(\tilde u-\tilde v)=
  r+2GM \ln\left({{r}\over{2GM}}-1\right)\ .\label{7.74}
\ee
We have actually re-introduced the degeneracy with which we started
out; in these coordinates $r=2GM$ is ``infinitely far away'' (at
either $\tilde u=-\infty$ or $\tilde v=+\infty$).  The thing to do
is to change to coordinates which pull these points into finite
coordinate values; a good choice is
\bea
  u' &=&  e^{\tilde u/4GM} \cr v' &=&  e^{-\tilde v/4GM}\ ,
  \label{7.75}
\eea
which in terms of our original $(t,r)$ system is
\bea
  u' &=&  \left({{r}\over{2GM}}-1\right)^{1/2}e^{(r+t)/4GM} \cr 
  v' &=&  \left({{r}\over{2GM}}-1\right)^{1/2}e^{(r-t)/4GM}\ .
  \label{7.76}
\eea
In the $(u',v',\theta,\phi)$ system the Schwarzschild metric is
\be
  ds^2 =-{{16 G^3M^3}\over{r}}e^{-r/2GM}(\d u' \d v'+ \d v' \d u')
  +r^2 d\Omega^2\ .\label{7.77}
\ee
Finally the nonsingular nature of $r=2GM$ becomes completely manifest;
in this form none of the metric coefficients behave in any special way
at the event horizon.

Both $u'$ and $v'$ are null coordinates, in the sense that their 
partial derivatives $\partial/\partial u'$ and $\partial/\partial v'$
are null vectors.  There is nothing wrong with this, since the 
collection of four partial derivative vectors (two null and two
spacelike) in this system serve as a perfectly good basis for the
tangent space.  Nevertheless, we are somewhat more comfortable working
in a system where one coordinate is timelike and the rest are 
spacelike.  We therefore define
\bea
  u &=&  {1\over 2}(u'-v')\cr
  &=&  \left({{r}\over{2GM}}-1\right)^{1/2}e^{r/4GM}\cosh(t/4GM)
  \label{7.78}
\eea
and
\bea
  v &=&  {1\over 2}(u'+v')\cr
  &=&  \left({{r}\over{2GM}}-1\right)^{1/2}e^{r/4GM}\sinh(t/4GM)\ ,
  \label{7.79}
\eea
in terms of which the metric becomes
\be
  ds^2 ={{32 G^3M^3}\over{r}}e^{-r/2GM}(-\d v^2+\d u^2)
  +r^2 d\Omega^2\ ,\label{7.80}
\ee
where $r$ is defined implicitly from
\be
  (u^2-v^2)=
  \left({{r}\over{2GM}}-1\right)e^{r/2GM}\ .\label{7.81}
\ee
The coordinates $(v,u,\theta,\phi)$ are known as {\bf Kruskal
coordinates}, or sometimes Kruskal-Szekres coordinates.  Note that
$v$ is the timelike coordinate.

The Kruskal coordinates have a number of miraculous properties.
Like the $(t,r^*)$ coordinates, the radial null curves look like
they do in flat space:
\be
  v= \pm u + {\rm constant}\ .\label{7.82}
\ee
Unlike the $(t,r^*)$ coordinates, however, the event horizon $r=2GM$ 
is not infinitely far away; in fact it is defined by
\be
  v = \pm u\ ,\label{7.83}
\ee
consistent with it being a null surface.
More generally, we can consider the surfaces $r=$~constant.  From
(7.81) these satisfy
\be
  u^2-v^2 = {\rm ~constant}\ .\label{7.84}
\ee
Thus, they appear as hyperbolae in the $u$-$v$ plane.  Furthermore,
the surfaces of constant $t$ are given by
\be
  {v\over u} = \tanh(t/4GM)\ ,\label{7.85}
\ee
which defines straight lines through the origin with slope
$\tanh(t/4GM)$.  Note that as $t\rightarrow \pm\infty$ this
becomes the same as (7.83); therefore these surfaces are the
same as $r=2GM$.   

Now, our coordinates $(v,u)$ should be allowed to
range over every value they can take without hitting the real
singularity at $r=2GM$; the allowed region is therefore
$-\infty \leq u \leq \infty$ and $v^2 < u^2+1$.  We can now draw
a spacetime diagram in the $v$-$u$ plane (with $\theta$ and $\phi$
suppressed), known as a ``Kruskal diagram'', which represents the
entire spacetime corresponding to the Schwarzschild metric.

\begin{figure}[h]
  \centerline{
  \psfig{figure=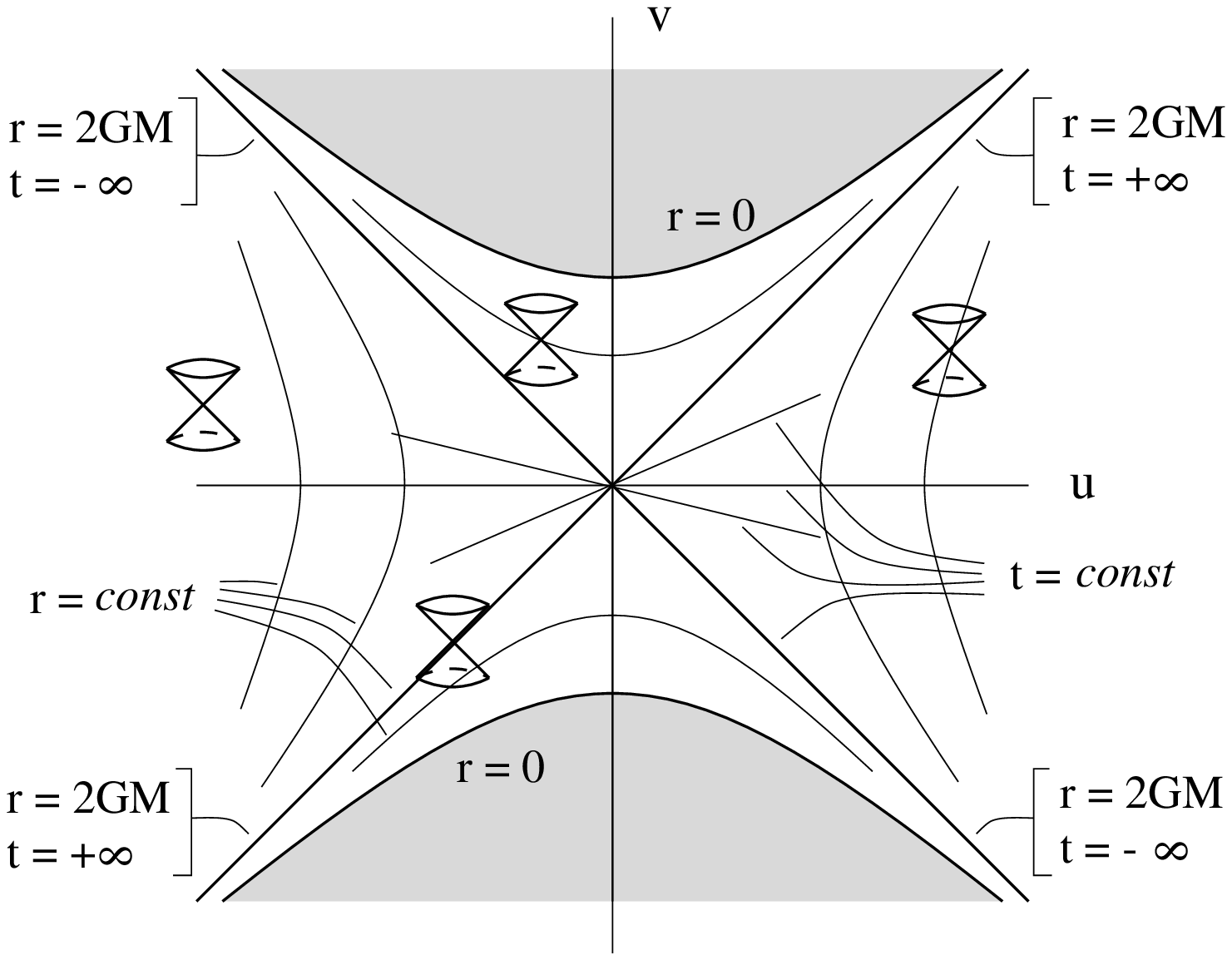,angle=0,height=12cm}}
\end{figure}

\noindent Each point on the diagram is a two-sphere.

\eject

Our original coordinates $(t,r)$ were only good for $r>2GM$, which is
only a part of the manifold portrayed on the Kruskal diagram.  It is
convenient to divide the diagram into four regions:

\begin{figure}[h]
  \centerline{
  \psfig{figure=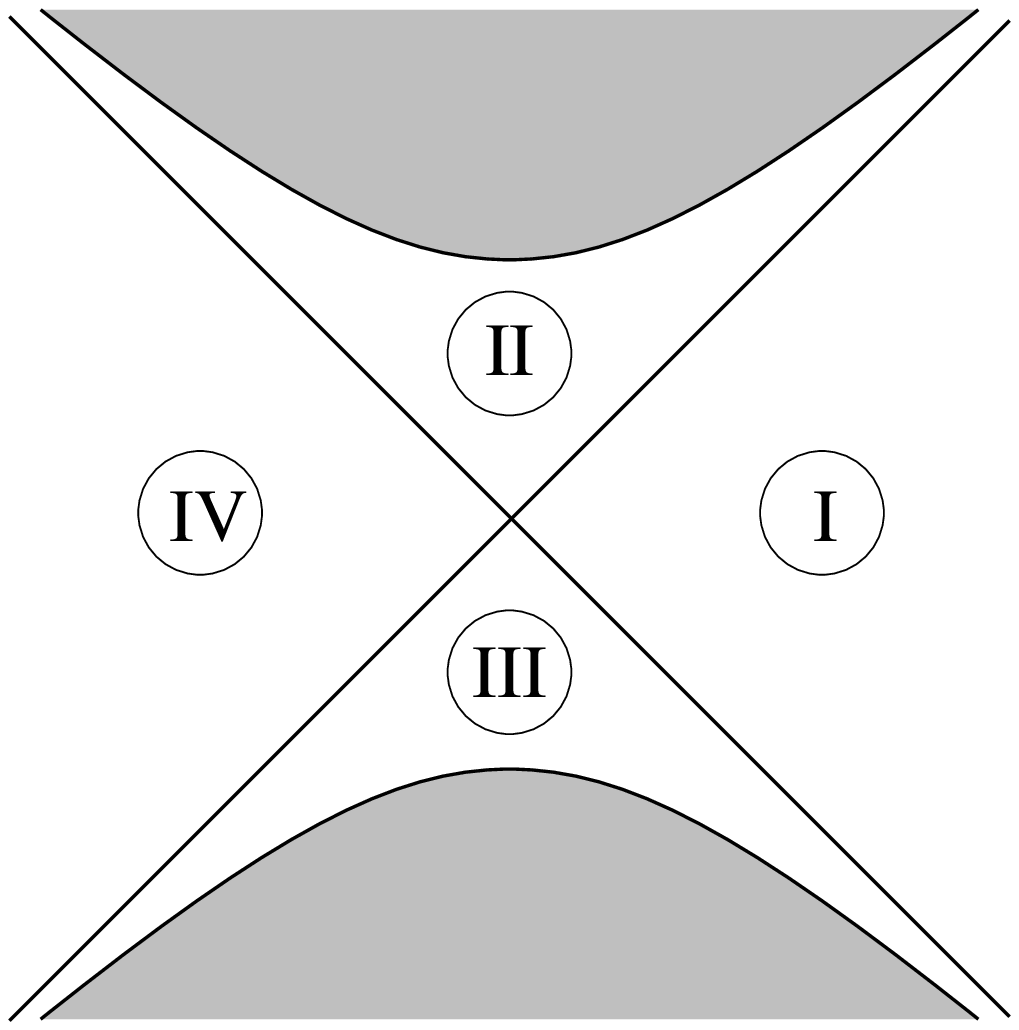,angle=0,height=5cm}}
\end{figure}

\noindent The region in which we started was region I; by following
future-directed null rays we reached region II, and by following
past-directed null rays we reached region III.  If we had explored
spacelike geodesics, we would have been led to region IV.
The definitions (7.78) and (7.79) which relate $(u,v)$ to $(t,r)$
are really only good in region I; in the other regions it is 
necessary to introduce appropriate minus signs to prevent the
coordinates from becoming imaginary.

Having extended the Schwarzschild geometry as far as it will go,
we have described a remarkable spacetime.  Region II, of course,
is what we think of as the black hole.  Once anything travels from
region I into II, it can never return.  In fact, every future-directed
path in region II ends up hitting the singularity at $r=0$; once you
enter the event horizon, you are utterly doomed.  This is worth
stressing; not only can you not escape back to region I, you cannot
even stop yourself from moving in the direction of decreasing $r$,
since this is simply the timelike direction.  (This could
have been seen in our original coordinate system; for $r<2GM$, $t$
becomes spacelike and $r$ becomes timelike.)  Thus you can no more
stop moving toward the singularity than you can stop getting older.
Since proper time is maximized along a geodesic, you will live the
longest if you don't struggle, but just relax as you approach the
singularity.  Not that you will have long to relax.  (Nor that the
voyage will be very relaxing; as you approach the singularity the
tidal forces become infinite.  As you fall toward the singularity
your feet and head will be pulled apart from each other, while 
your torso is squeezed to infinitesimal thinness.  The grisly
demise of an astrophysicist falling into a black hole is detailed
in Misner, Thorne, and Wheeler, section 32.6.  Note that they use
orthonormal frames [not that it makes the trip any more enjoyable].)

Regions III and IV might be somewhat unexpected.  Region III is simply
the time-reverse of region II, a part of spacetime from which things
can escape to us, while we can never get there.  It can be thought
of as a ``white hole.''  There is a singularity in the past, out of which 
the universe appears to spring.  The boundary of region III is sometimes
called the past event horizon, while the boundary of region II is called
the future event horizon.  Region IV, meanwhile, cannot be reached
from our region I either forward or backward in time (nor can 
anybody from over there
reach us).  It is another asymptotically flat region of spacetime,
a mirror image of ours.  It can be thought of as being connected to
region I by a ``wormhole,'' a neck-like configuration joining two
distinct regions.  Consider slicing up the Kruskal diagram into spacelike
surfaces of constant $v$:

\begin{figure}[h]
  \centerline{
  \psfig{figure=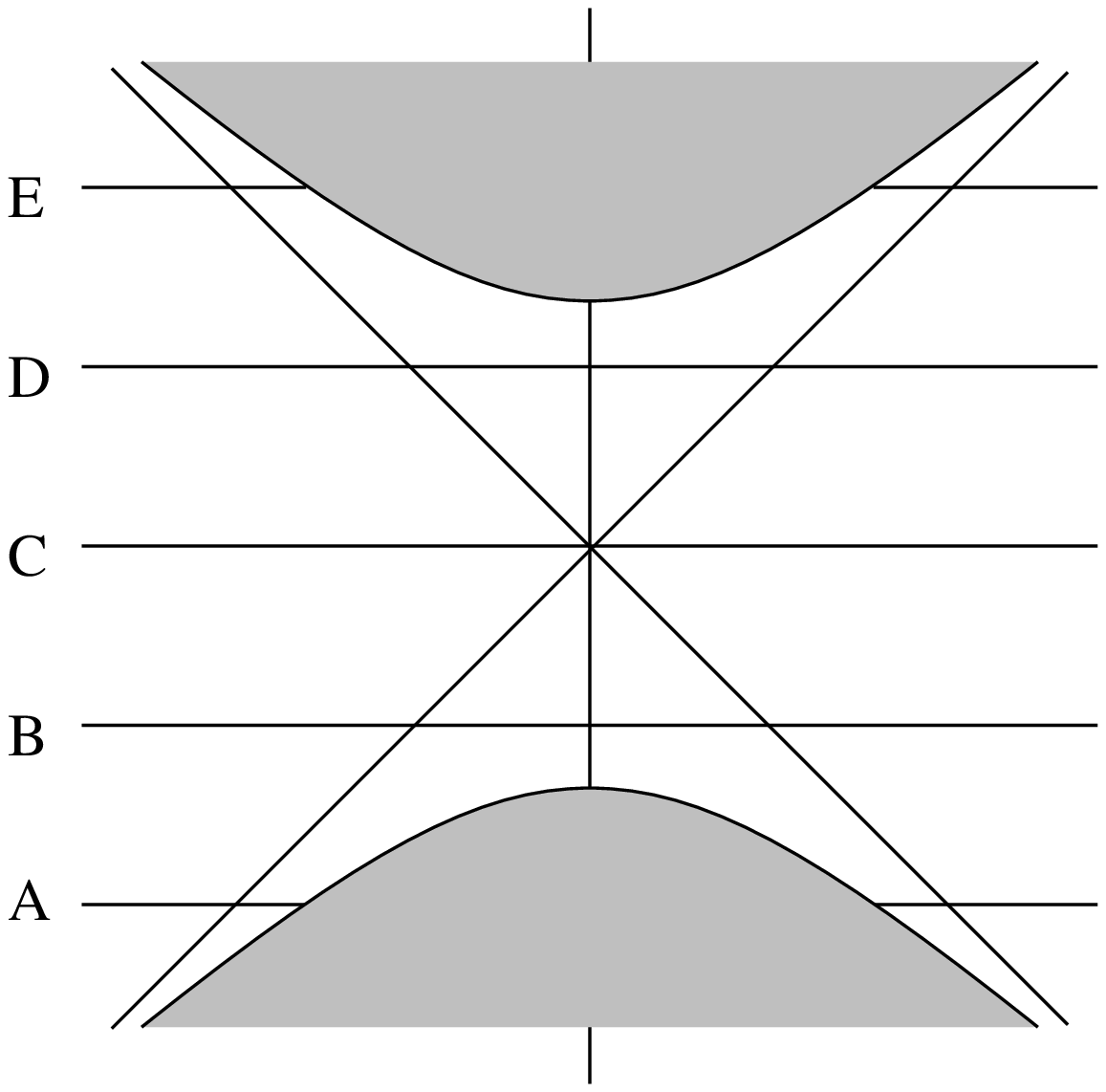,angle=0,height=6cm}}
\end{figure}

\noindent Now we can draw pictures of each slice, restoring one of 
the angular coordinates for clarity:

\begin{figure}[h]
  \centerline{
  \psfig{figure=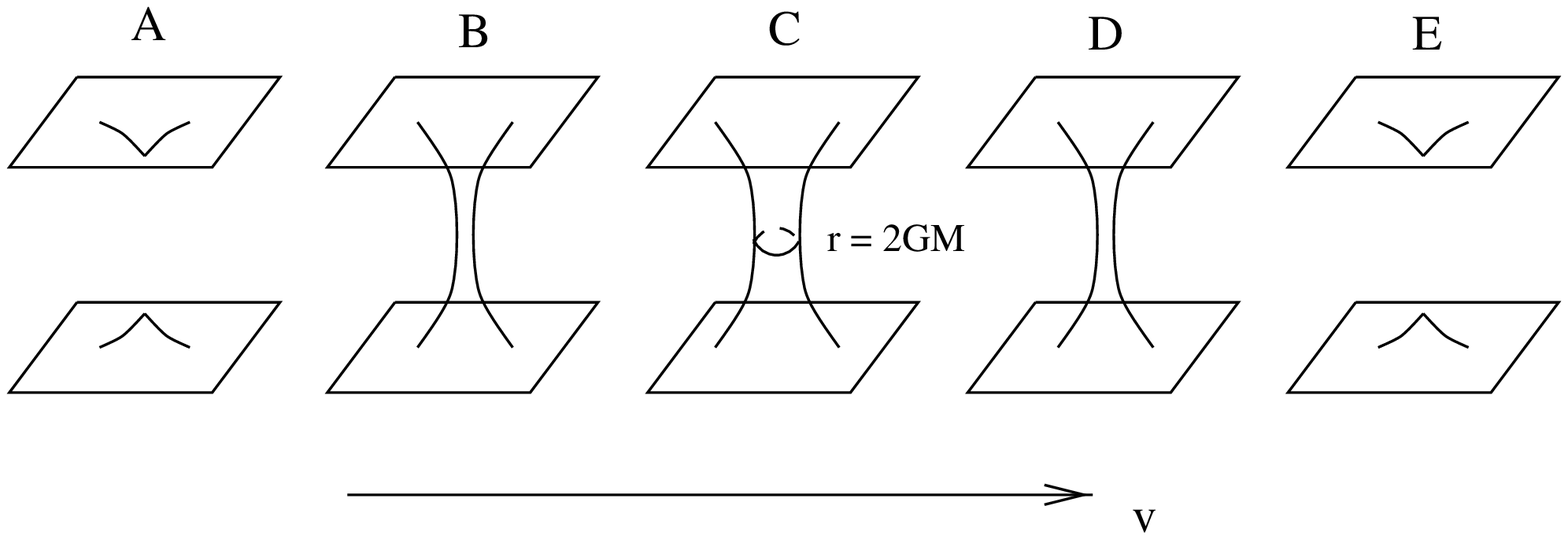,angle=0,height=6cm}}
\end{figure}

\noindent
So the Schwarzschild geometry really describes two asymptotically
flat regions which reach toward each other, join together via a
wormhole for a while, and then disconnect.  But the wormhole
closes up too quickly for any timelike observer to cross it from one
region into the next.

It might seem somewhat implausible, this story about two separate
spacetimes reaching toward each other for a while and then letting
go.  In fact, it is not expected to happen in the real world, since
the Schwarzschild metric does not accurately model the entire 
universe.  Remember that it is only valid in vacuum, for example
outside a star.  If the star has a radius larger than $2GM$, we
need never worry about any event horizons at all.  But we believe
that there are stars which collapse under their own gravitational
pull, shrinking down to below $r=2GM$ and further into a singularity,
resulting in a black hole.  There is no need for a white hole, however,
because the past of such a spacetime looks nothing like that of the
full Schwarzschild solution.  Roughly, a Kruskal-like diagram for
stellar collapse would look like the following:

\begin{figure}[h]
  \centerline{
  \psfig{figure=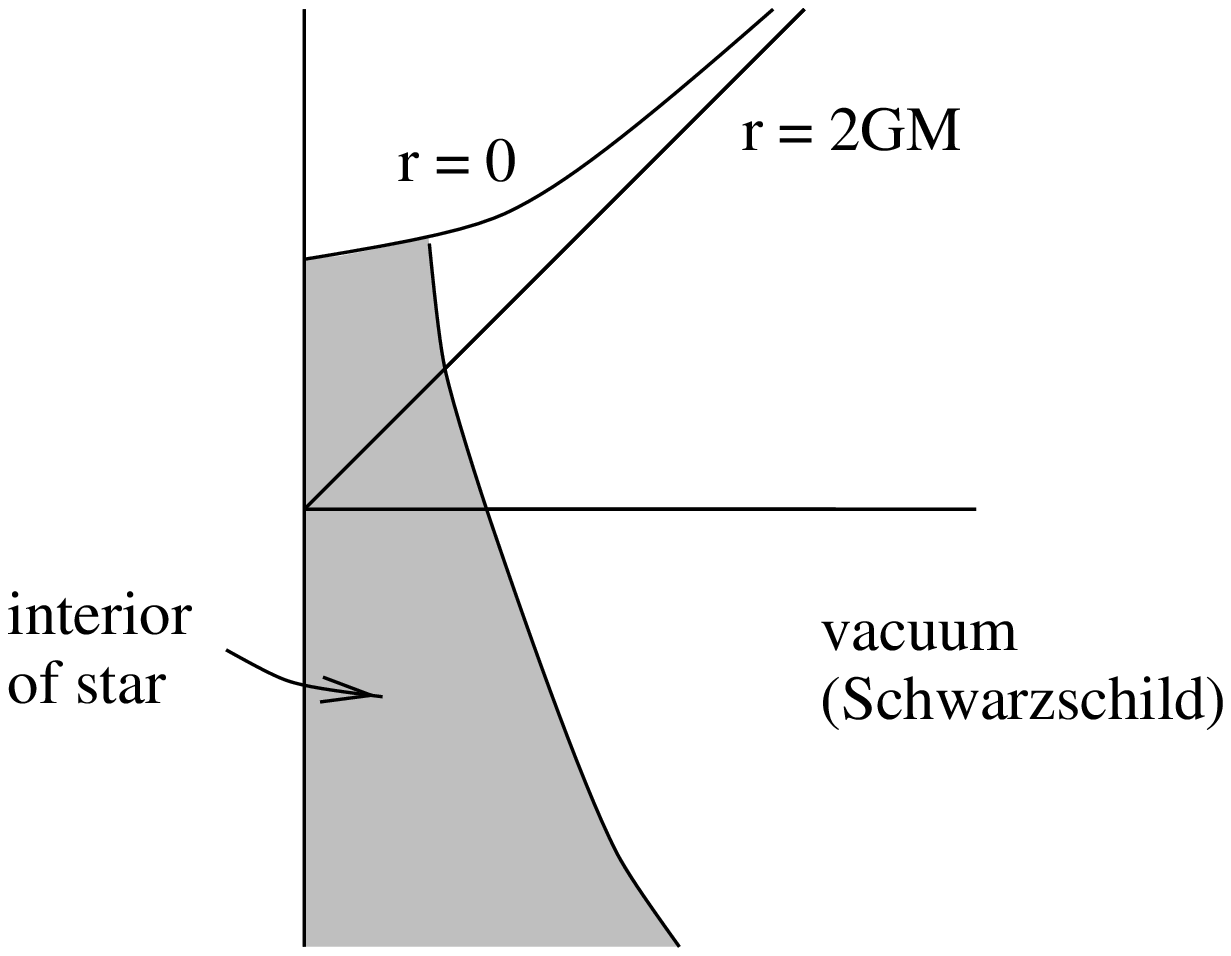,angle=0,height=7cm}}
\end{figure}

\noindent  The shaded region is not described by Schwarzschild, so
there is no need to fret about white holes and wormholes. 

While we are on the subject, we can say something about the formation
of astrophysical black holes from massive stars.  The life of a star
is a constant struggle between the inward pull of gravity and the
outward push of pressure.  When the star is burning nuclear fuel at
its core, the pressure comes from the heat produced by this burning.
(We should put ``burning'' in quotes, since nuclear fusion is unrelated
to oxidation.)  When the fuel is used up, the temperature declines and
the star begins to shrink as gravity starts winning the struggle.
Eventually this process is stopped when the electrons are pushed so
close together that they resist further compression simply on the
basis of the Pauli exclusion principle (no two fermions can be in the
same state).  The resulting object is called a {\bf white dwarf}.
If the mass is sufficiently high, however, even the electron 
degeneracy pressure is not enough, and the electrons will combine
with the protons in a dramatic phase transition.  The result is a
{\bf neutron star}, which consists of almost entirely neutrons (although
the insides of neutron stars are not understood terribly well).
Since the conditions at the center of a neutron star are very different
from those on earth, we do not have a perfect understanding of the
equation of state.  Nevertheless, we believe that a
sufficiently massive neutron star will itself be unable to resist the
pull of gravity, and will continue to collapse.  Since a fluid of
neutrons is the densest material of which we can presently conceive,
it is believed that the inevitable outcome of such a collapse is
a black hole.

The process is summarized in the following diagram of radius vs.
mass:

\begin{figure}[h]
  \centerline{
  \psfig{figure=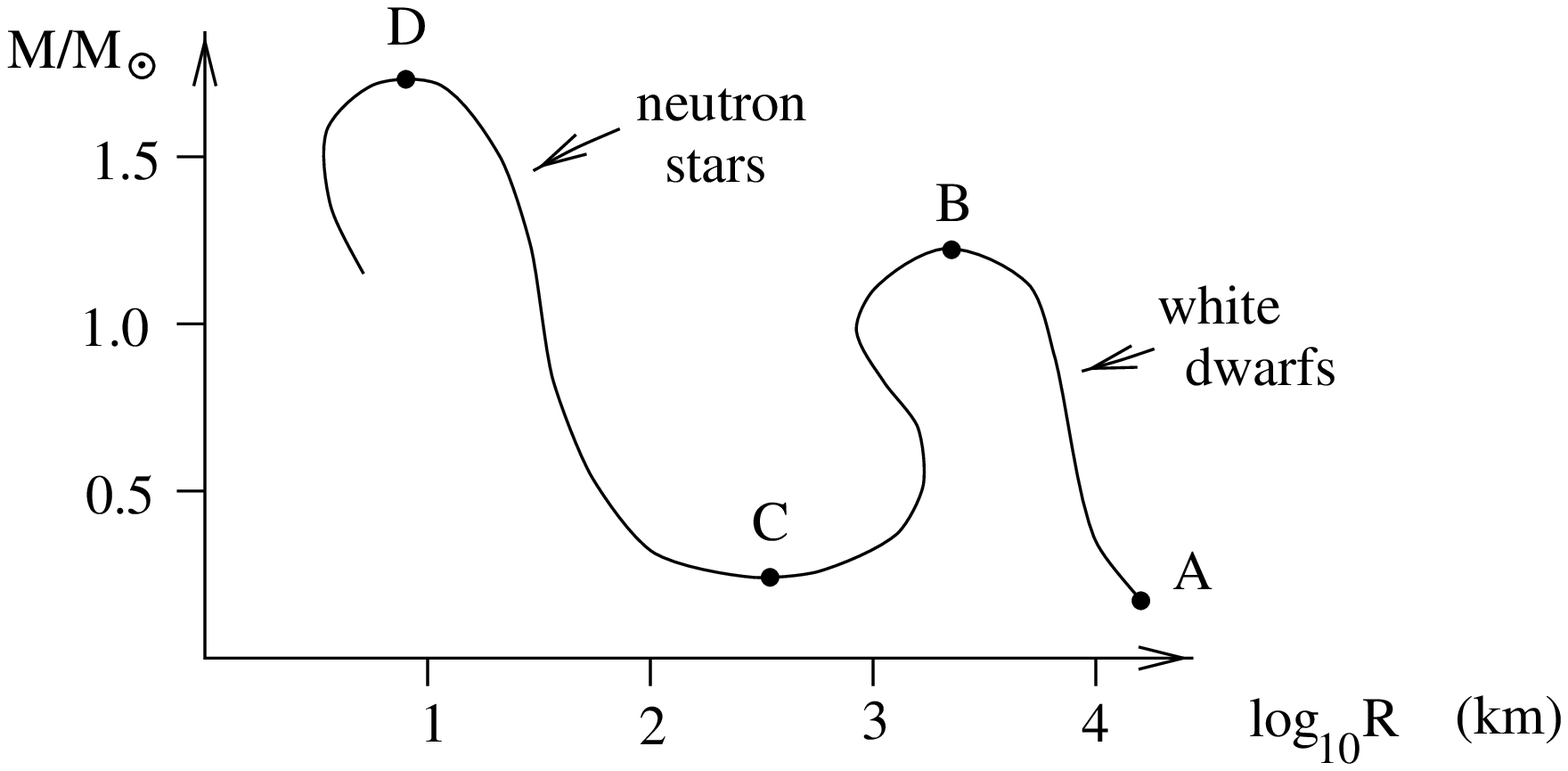,angle=0,height=7cm}}
\end{figure}

\noindent The point of the diagram is that, for any given mass $M$,
the star will decrease in radius until it hits the line.  White
dwarfs are found between points $A$ and $B$, and neutron stars 
between points $C$ and $D$.  Point $B$ is at a height of somewhat less
than 1.4 solar masses; the height of $D$ is less certain, but probably
less than 2 solar masses.  The process of collapse is complicated, and
during the evolution the star can lose or gain mass, so the endpoint
of any given star is hard to predict.  Nevertheless white dwarfs are
all over the place, neutron stars are not uncommon, and there are a
number of systems which are strongly believed to contain black holes.
(Of course, you can't directly see the black hole.  What you can see
is radiation from matter accreting onto the hole, which heats up as
it gets closer and emits radiation.)

We have seen that the Kruskal coordinate system provides a very
useful representation of the Schwarzschild geometry.  Before moving
on to other types of black holes, we will introduce one more
way of thinking about this spacetime, the Penrose (or Carter-Penrose,
or conformal) diagram.  The idea is to do a conformal transformation
which brings the entire manifold onto a compact region such that we
can fit the spacetime on a piece of paper.

Let's begin with Minkowski space, to see how the technique works.
The metric in polar coordinates is
\be
  ds^2 = -\d t^2 + \d r^2 + r^2 d\Omega^2\ .\label{7.86}
\ee
Nothing unusual will happen to the $\theta, \phi$ coordinates, but
we will want to keep careful track of the ranges of the other two
coordinates.  In this case of course we have 
\bea
  & -\infty < t < +\infty&\cr& 0 \leq r < +\infty\ .&
  \label{7.87}
\eea
Technically the worldline $r=0$ represents a coordinate singularity
and should be covered by a different patch, but we all know what is
going on so we'll just act like $r=0$ is well-behaved.

Our task is made somewhat easier if we switch to null coordinates:
\bea
  u &=&  {1\over 2}(t+r)\cr v &=&  {1\over 2}(t-r)\ ,
  \label{7.88}
\eea
with corresponding ranges given by
\bea
  &-\infty < u < +\infty &\cr &-\infty < v < +\infty &\cr
  & v \leq u\ .&\label{7.89}
\eea
\begin{figure}[hbt]
  \centerline{
  \psfig{figure=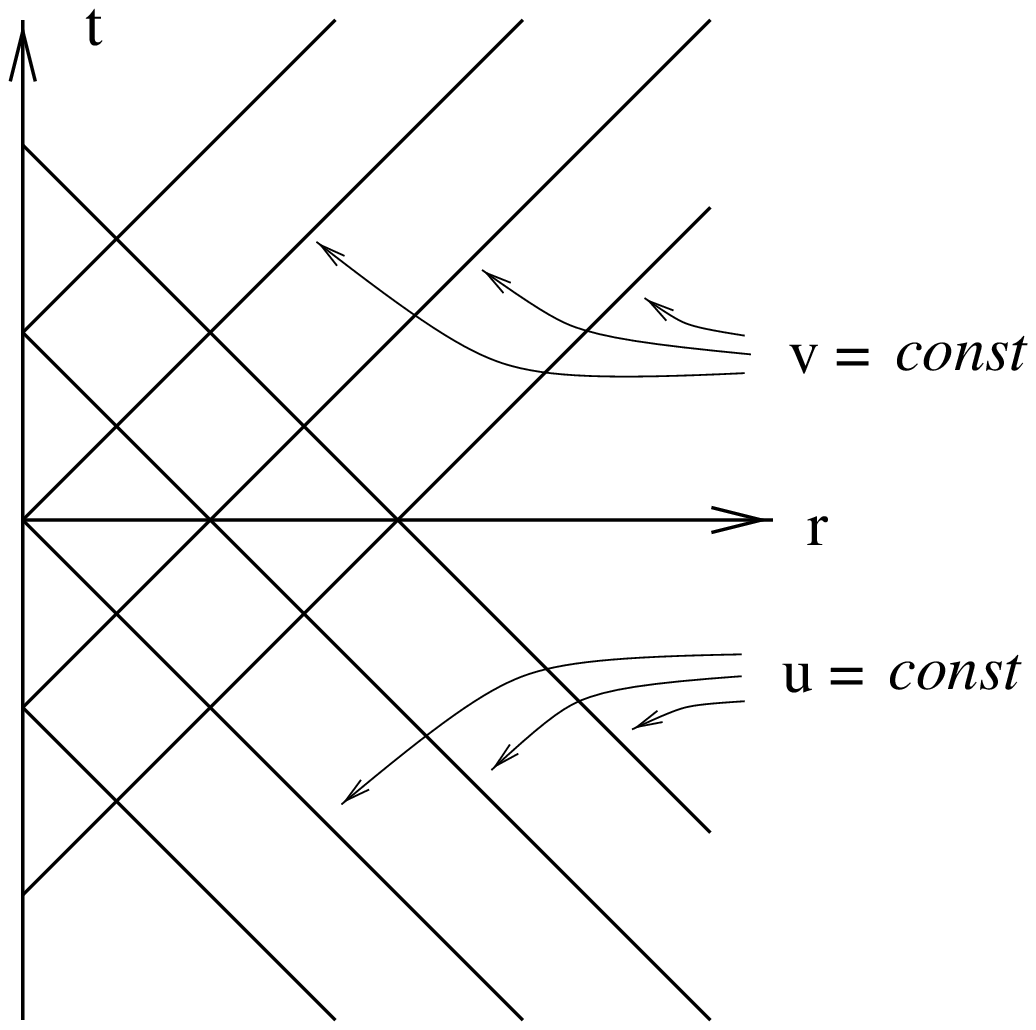,angle=0,height=7cm}}
\end{figure}
These ranges are as portrayed in the figure, on which each
point represents a 2-sphere of radius $r=u-v$.  The metric in
these coordinates is given by
\be
  ds^2 = -2(\d u\d v + \d v\d u) +(u-v)^2 d\Omega^2\ .\label{7.90}
\ee

We now want to change to coordinates in which ``infinity'' takes
on a finite coordinate value.  A good choice is
\bea
  U &=&  \arctan u\cr V &=&  \arctan v\ .\label{7.91}
\eea

\begin{figure}[ht]
  \centerline{
  \psfig{figure=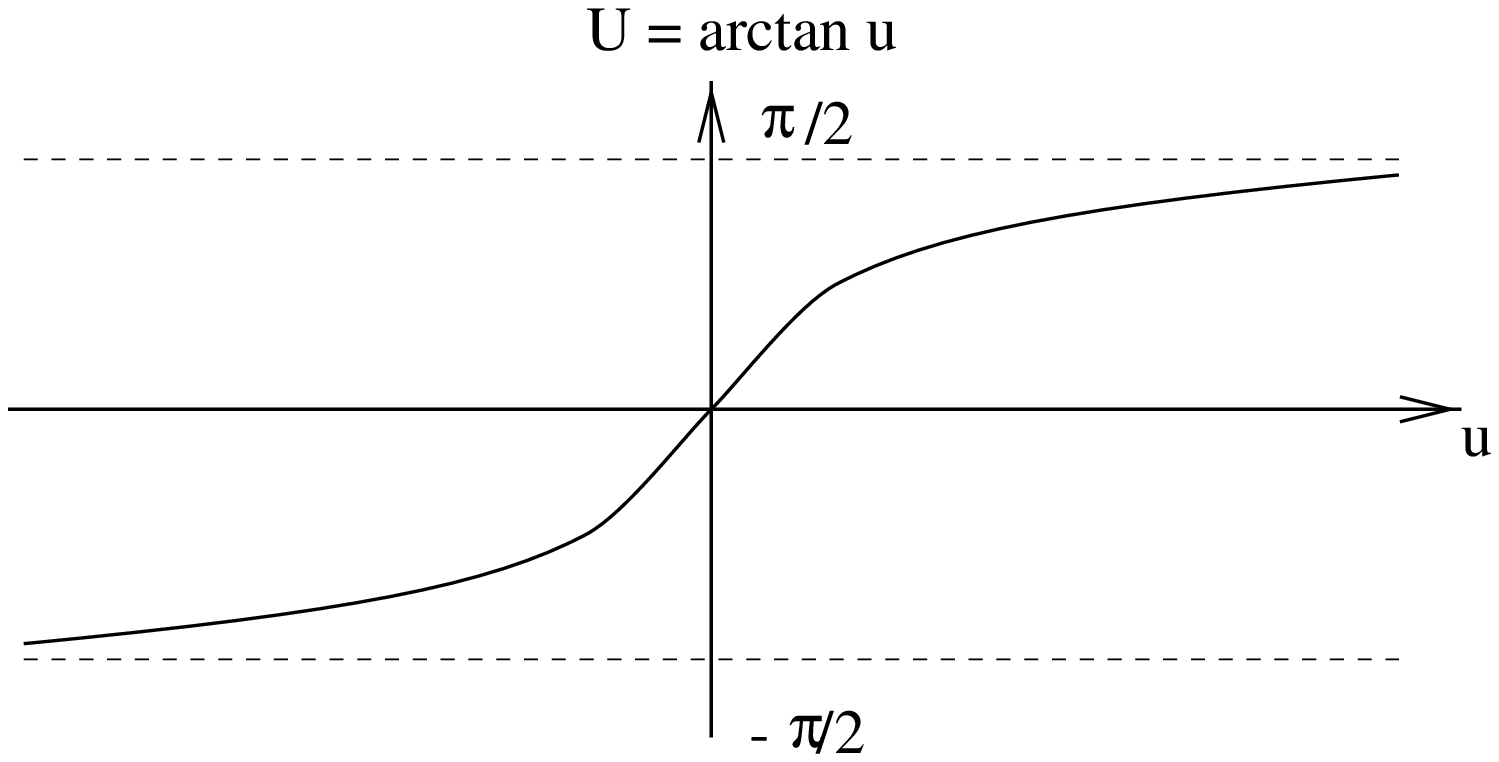,angle=0,height=5cm}}
\end{figure}

\noindent The ranges are now
\bea
  &-\pi/2  < U < +\pi/2&\cr &-\pi/2  < V < +\pi/2&\cr
  & V \leq U\ .& \label{7.92}
\eea
To get the metric, use
\be
  \d U = {{\d u}\over{1+u^2}}\ ,
  \label{7.93}
\ee
and 
\be
  \cos(\arctan{u}) = {{1}\over{\sqrt{1+u^2}}}\ ,\label{7.94}
\ee
and likewise for $v$.  We are led to
\be
  \d u\d v + \d v\d u = {{1}\over{\cos^2U \cos^2V}}
  (\d U\d V + \d V\d U)\ .\label{7.95}
\ee
Meanwhile,
\bea
  (u-v)^2 &=&  (\tan U - \tan V)^2\cr
  &=& {{1}\over{\cos^2U \cos^2V}}(\sin U\cos V- \cos U \sin V)^2\cr
  &=& {{1}\over{\cos^2U \cos^2V}}\sin^2(U-V)\ .\label{7.96}
\eea
Therefore, the Minkowski metric in these coordinates is
\be
  ds^2 = {{1}\over{\cos^2U \cos^2V}}\left[ -2(\d U\d V + \d V\d U)
  +\sin^2(U-V)d\Omega^2\right]\ .\label{7.97}
\ee

This has a certain appeal, since the metric appears as a fairly
simple expression multiplied by an overall factor.  We can make it
even better by transforming back to a timelike coordinate $\eta$
and a spacelike (radial) coordinate $\chi$, via
\bea
  \eta &=&  U+V\cr \chi &=&  U-V\ ,\label{7.98}
\eea
with ranges
\bea
   & -\pi< \eta < +\pi & \cr &  0  \leq \chi < +\pi\ . & 
  \label{7.99}
\eea
Now the metric is
\be
  ds^2 = \omega^{-2}\left(-\d \eta^2 + \d \chi^2 +\sin^2\chi\ 
  d\Omega^2\right)\ ,\label{7.100}
\ee
where
\bea
  \omega &=&  \cos U \cos V\cr &=&  {1\over 2}
  (\cos\eta +\cos\chi)\ .\label{7.101}
\eea

The Minkowski metric may therefore be thought of as related by a 
conformal transformation to the ``unphysical'' metric
\bea
  d\bar{s}^2 &=&  \omega^2 ds^2\cr
  &=& -\d \eta^2 + \d \chi^2 +\sin^2\chi\ d\Omega^2\ .\label{7.102}
\eea
This describes the manifold $\R\times S^3$, where the 3-sphere is
maximally symmetric and static.  There is curvature in this metric,
and it is not a solution to the vacuum Einstein's equations.  
This shouldn't bother us, since
it is unphysical; the true physical metric, obtained by a conformal
transformation, is simply flat spacetime.  In fact this metric is
that of the ``Einstein static universe,'' a static (but unstable)
solution to Einstein's equations with a perfect fluid and a cosmological
constant.  Of course, the full range of coordinates on $\R\times S^3$
would usually be $-\infty < \eta < +\infty$, $0\leq\chi \leq\pi$,
while Minkowski space is mapped into the subspace defined by (7.99).
The entire $\R\times S^3$ can be drawn as a cylinder, in which each
circle is a three-sphere, as shown on the next page.

\eject

\begin{figure}
  \centerline{
  \psfig{figure=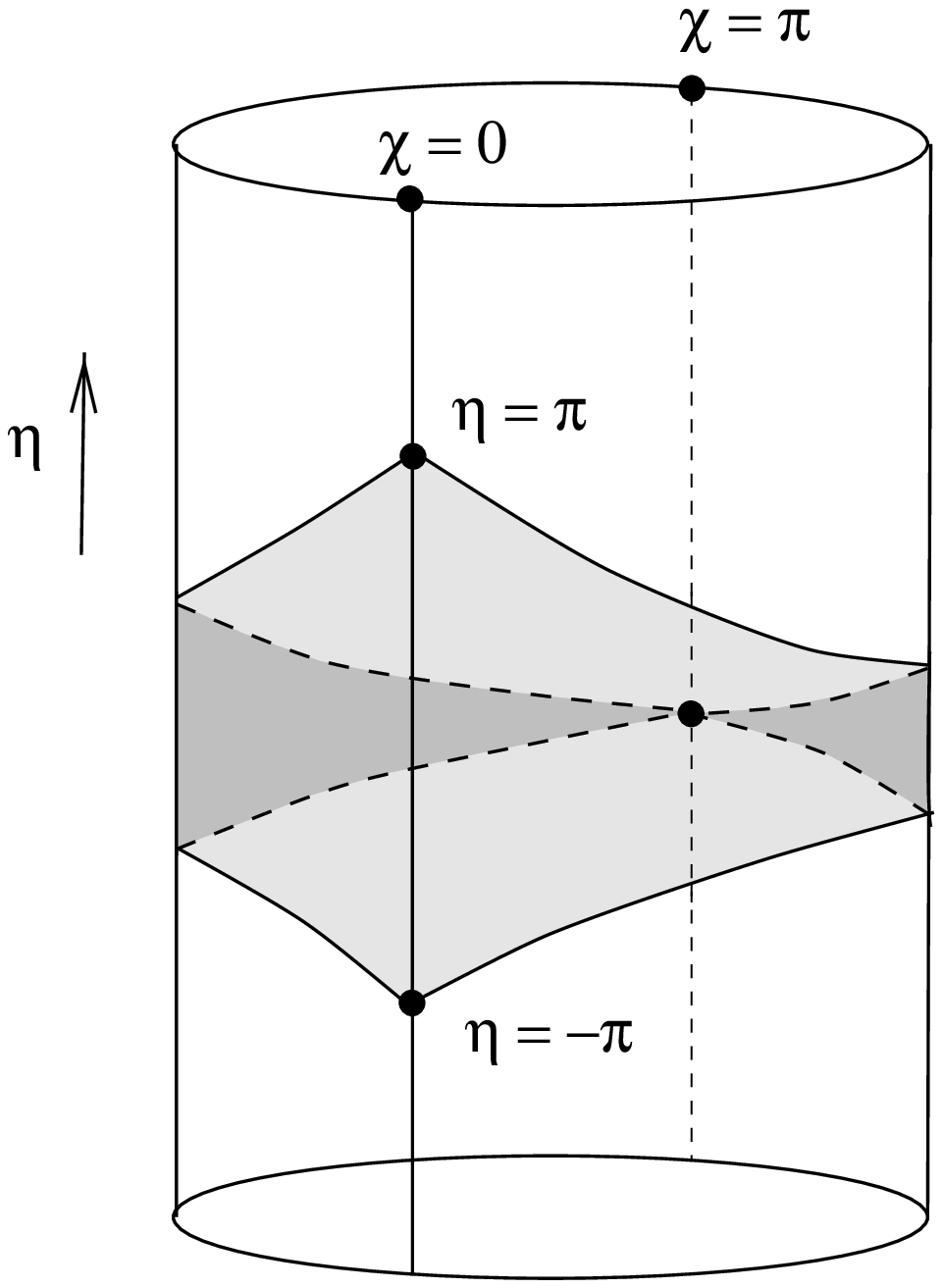,angle=0,height=8cm}}
\end{figure}

\noindent The shaded region represents Minkowski space.  Note that 
each point $(\eta,\chi)$ on this cylinder is half of a two-sphere, where 
the other half is the point $(\eta,-\chi)$.  We can unroll the 
shaded region to portray Minkowski space as a triangle, as shown
in the figure.
\begin{figure}[ht]
  \centerline{
  \psfig{figure=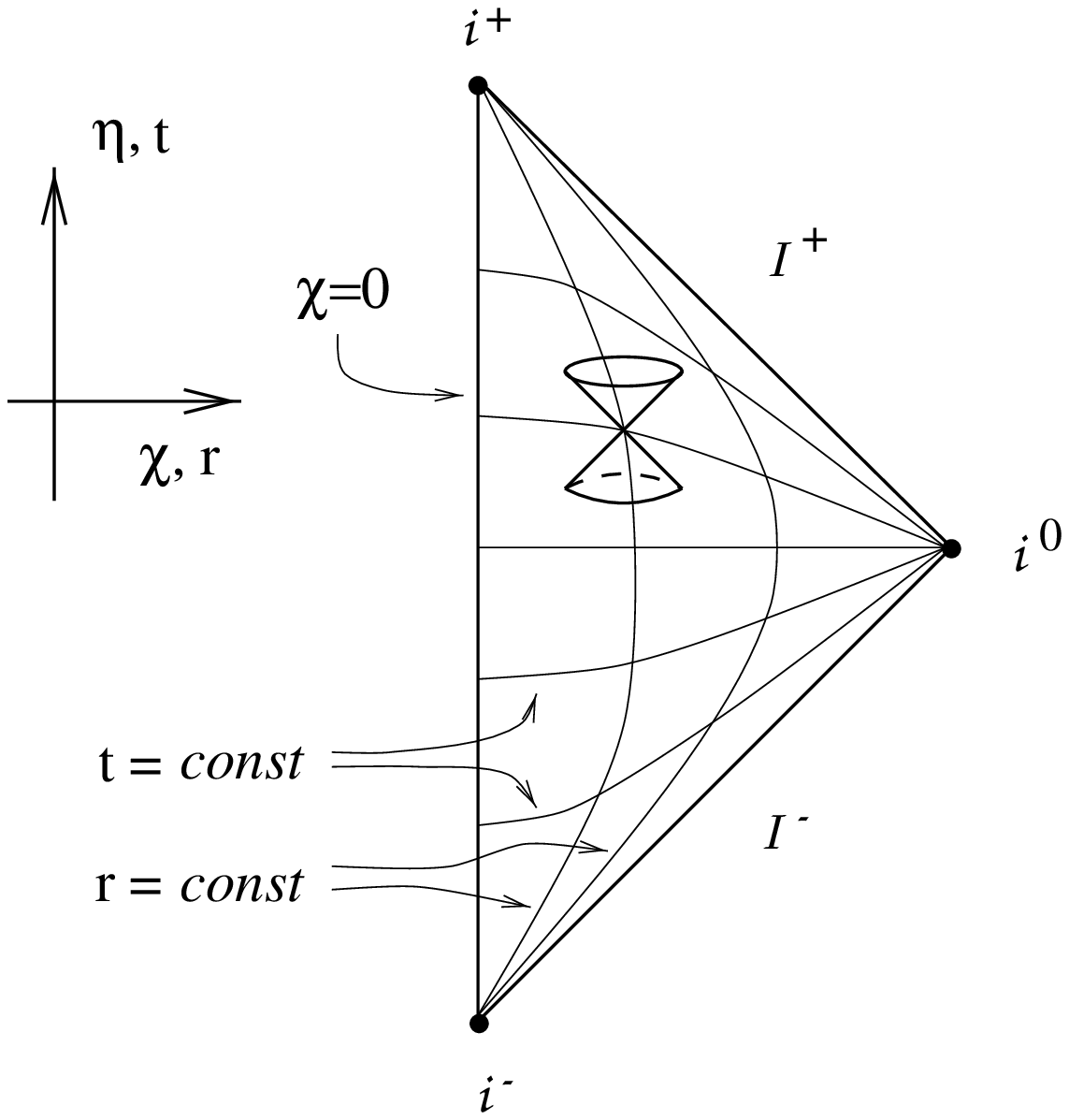,angle=0,height=7cm}}
\end{figure}
The is the {\bf Penrose diagram}.  Each point represents a 
two-sphere.

In fact Minkowski space is only the {\it interior} of the above
diagram (including $\chi=0$); the boundaries are not part of the
original spacetime.  Together they are referred to as {\bf conformal
infinity}.  The structure of the Penrose diagram allows us to subdivide 
conformal infinity into
a few different regions:
\[
  \begin{array}{rcl}
  i^+ &=&  {\rm future~timelike~infinity~} (\eta=\pi\ ,\ \chi=0)\cr
  i^0 &=&  {\rm spatial~infinity~} (\eta=0\ ,\ \chi=\pi)\cr
  i^- &=&  {\rm past~timelike~infinity~} (\eta=-\pi\ ,\ \chi=0)\cr
  {\cal I}^+ &=& {\rm future~null~infinity~} (\eta=\pi-\chi\ ,\ 0<\chi<\pi)\cr
  {\cal I}^- &=& {\rm past~null~infinity~} (\eta=-\pi+\chi\ ,\ 0<\chi<\pi)
  \end{array}
\]
(${\cal I}^+$ and ${\cal I}^-$ are pronounced as ``scri-plus'' and
``scri-minus'', respectively.)  Note that $i^+$, $i^0$, and $i^-$ are
actually {\it points}, since $\chi=0$ and $\chi=\pi$ are the north and
south poles of $S^3$.  Meanwhile ${\cal I}^+$ and ${\cal I}^-$ are
actually null surfaces, with the topology of $\R\times S^2$.

There are a number of important features of the Penrose diagram for
Minkowski spacetime.  The points $i^+$, and $i^-$ can be thought of
as the limits of spacelike surfaces whose normals are timelike;
conversely, $i^0$ can be thought of as the limit of timelike surfaces
whose normals are spacelike.  Radial null geodesics are at 
$\pm 45^\circ$ in the diagram.
All timelike geodesics begin at $i^-$
and end at $i^+$; all null geodesics begin at ${\cal I}^-$ and end
at ${\cal I}^+$; all spacelike geodesics both begin and end at $i^0$.
On the other hand, there can be non-geodesic timelike curves that 
end at null infinity (if they become ``asymptotically null'').

It is nice to be able to fit all of Minkowski space on a small piece
of paper, but we don't really learn much that we didn't already
know.  Penrose diagrams are more useful when we want to represent
slightly more interesting spacetimes, such as those for black holes.
The original use of Penrose diagrams was to compare spacetimes to
Minkowski space ``at infinity'' --- the rigorous definition of
``asymptotically flat'' is basically that a spacetime has a conformal
infinity just like Minkowski space.  We will not pursue these issues
in detail, but instead turn directly to analysis of the Penrose
diagram for a Schwarzschild black hole.

We will not go through the necessary manipulations
in detail, since they parallel the Minkowski case with
considerable additional algebraic complexity.  We would
start with the null version of the Kruskal coordinates, in
which the metric takes the form
\be
  ds^2 =-{{16 G^3M^3}\over{r}}e^{-r/2GM}(\d u' \d v'+ \d v' \d u')
  +r^2 d\Omega^2\ ,  \label{7.103}
\ee
where $r$ is defined implicitly via
\be
  u'v' = \left({{r}\over{2GM}}-1\right)e^{r/2GM}\ .
  \label{7.104}
\ee
Then essentially the same transformation as was used in
flat spacetime suffices to bring infinity into finite
coordinate values:
\bea
  u'' &=&  \arctan\left({{u'}\over{\sqrt{2GM}}}\right)\cr
  v'' &=&  \arctan\left({{v'}\over{\sqrt{2GM}}}\right)\ ,
  \label{7.105}
\eea
with ranges
\[\begin{array}{c}
  -\pi/2 < u'' < +\pi/2\cr -\pi/2 < v'' < +\pi/2\cr
  -\pi < u'' + v'' < \pi\ .
\end{array}\]
The $(u'',v'')$ part of the metric (that is, at constant angular
coordinates) is now conformally related to Minkowski space.
In the new coordinates the singularities
at $r=0$ are straight lines that stretch from timelike
infinity in one asymptotic region to timelike infinity in the 
other.  The Penrose diagram for the maximally extended 
Schwarzschild solution thus looks like this:

\begin{figure}[h]
  \centerline{
  \psfig{figure=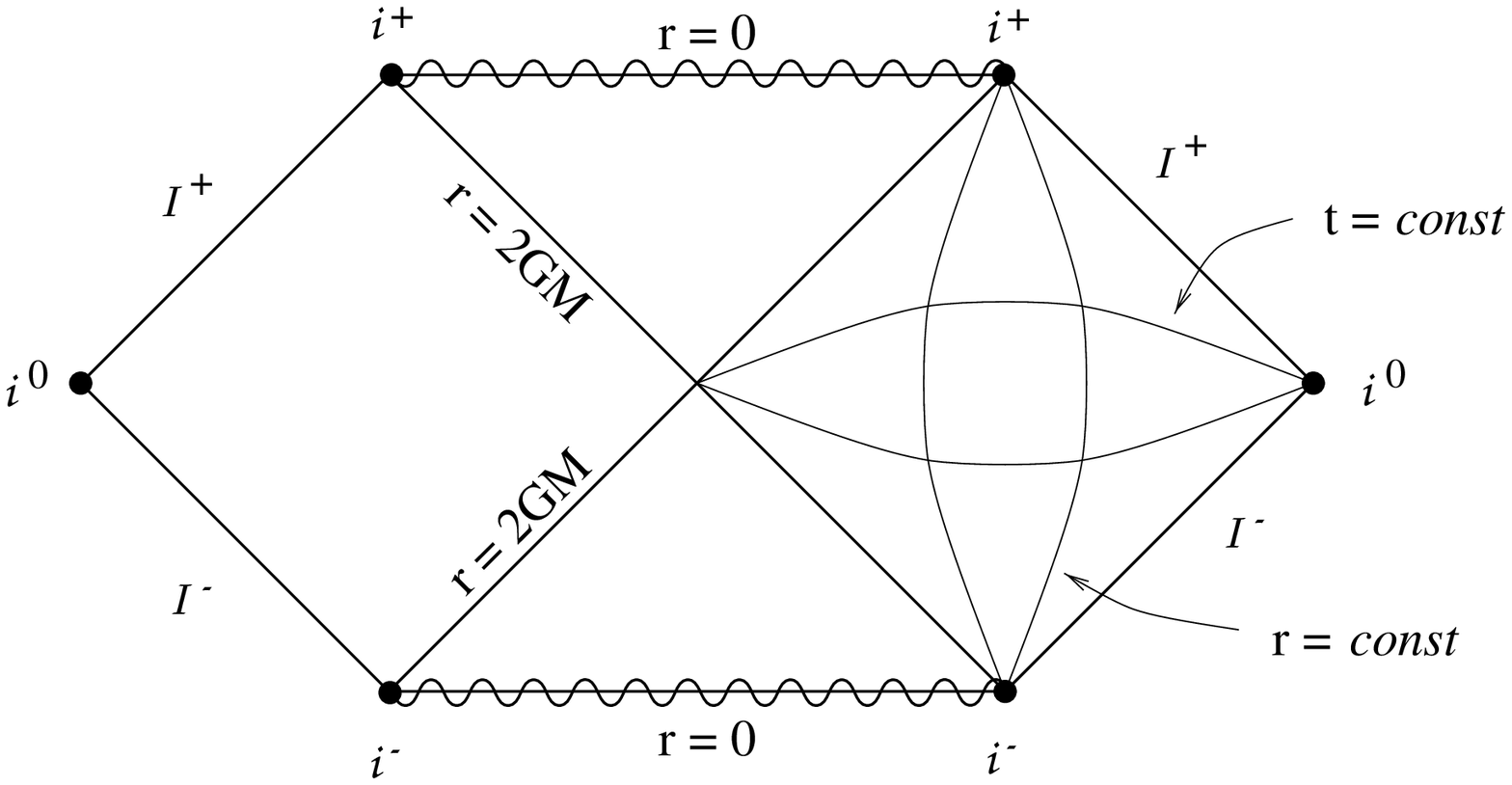,angle=0,height=6cm}}
\end{figure}

\noindent  The only real subtlety about this diagram is
the necessity to understand that $i^+$ and $i^-$ are distinct
from $r=0$ (there are plenty of timelike paths that do not
hit the singularity).  Notice also that the structure of
conformal infinity is just like that of Minkowski space,
consistent with the claim that Schwarzschild is asymptotically
flat.  Also, the Penrose diagram for a collapsing star that
forms a black hole is what you might expect, as shown on the
next page.

\begin{figure}
  \centerline{
  \psfig{figure=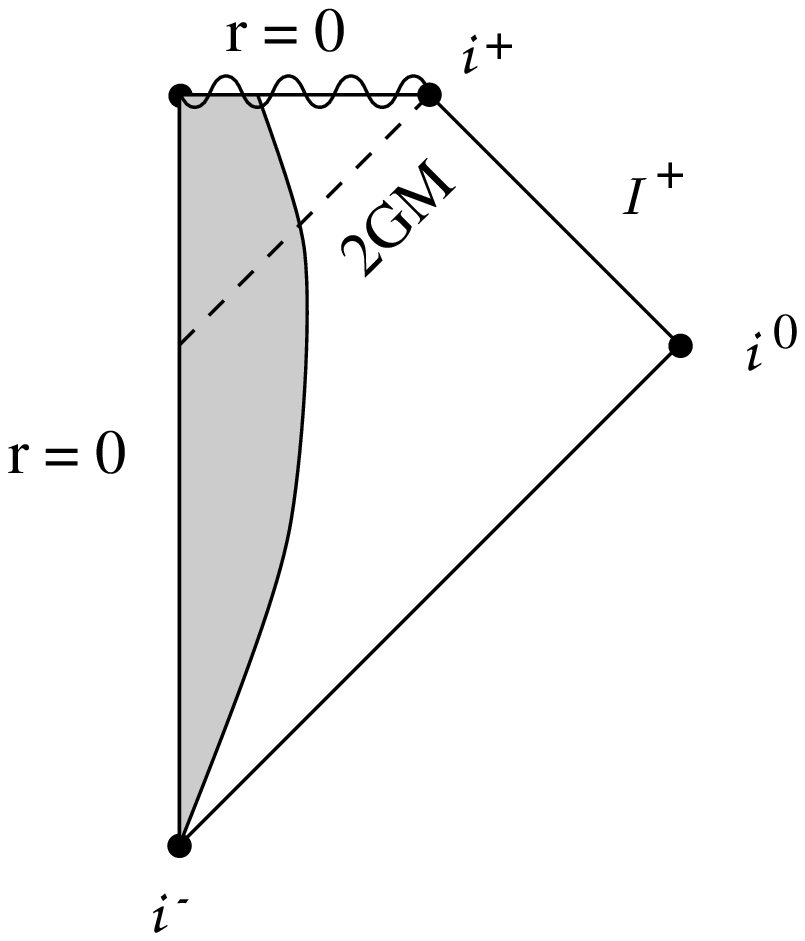,angle=0,height=6cm}}
\end{figure}

Once again the Penrose diagrams for these spacetimes don't
really tell us anything we didn't already know; their usefulness
will become evident when we consider more
general black holes.  In principle there
could be a wide variety of types of black holes, depending on
the process by which they were formed.  Surprisingly, however,
this turns out not to be the case; no matter how a black
hole is formed, it settles down (fairly quickly) into a
state which is characterized only by the mass, charge, and
angular momentum.  This property, which must be demonstrated
individually for the various types of fields which one might
imagine go into the construction of the hole, is often
stated as {\bf ``black holes have no hair."}  You can 
demonstrate, for example, that a hole which is formed from
an initially inhomogeneous collapse ``shakes off'' any
lumpiness by emitting gravitational radiation.  This is an
example of a ``no-hair theorem.''  If we are interested in
the form of the black hole after it has settled down, we thus
need only to concern ourselves with charged and rotating
holes.  In both cases there exist exact
solutions for the metric, which we can examine closely.

But first let's take a brief detour to the world of black
hole evaporation.
It is strange to think of a black hole ``evaporating,'' but in
the real world black holes aren't truly black --- they radiate
energy as if they were a blackbody of temperature $T=\hbar/8\pi
kGM$, where $M$ is the mass of the hole and $k$ is Boltzmann's
constant.  The derivation of this effect, known as {\bf Hawking
radiation}, involves the use of quantum field theory in curved
spacetime and is way outside our scope right now.  The informal
idea is nevertheless understandable.  
\begin{figure}[b]
  \centerline{
  \psfig{figure=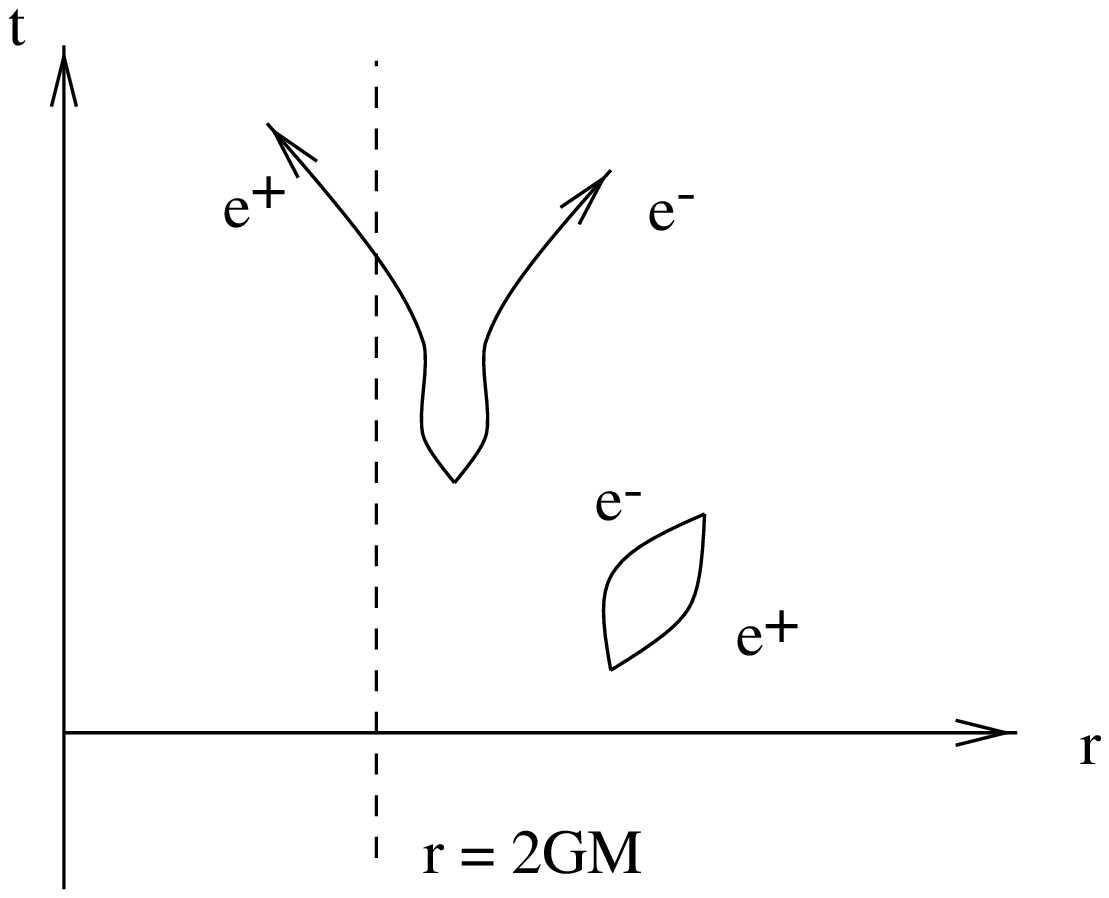,angle=0,height=6cm}}
\end{figure}
In quantum field theory
there are ``vacuum fluctuations'' --- the spontaneous creation and
annihilation of particle/antiparticle pairs in empty space.  
These fluctuations are precisely analogous to the zero-point
fluctuations of a simple harmonic oscillator.  Normally such
fluctuations are impossible to detect, since they average out
to give zero total energy (although nobody knows why; that's
the cosmological constant problem).  In the presence of an event
horizon, though, occasionally one member of a virtual pair will
fall into the black hole while its partner escapes to infinity.
The particle that reaches infinity will have to have a positive
energy, but the total energy is conserved; therefore the black
hole has to lose mass.  (If you like you can think of the particle
that falls in as having a negative mass.)  
We see the escaping particles as Hawking radiation.
It's not a very big effect, and the temperature goes down as the
mass goes up, so for black holes of mass comparable to the sun it
is completely negligible.  Still, in principle the black hole could
lose all of its mass to Hawking radiation, and shrink to nothing in
the process.  The relevant Penrose diagram might look like this:

\begin{figure}[h]
  \centerline{
  \psfig{figure=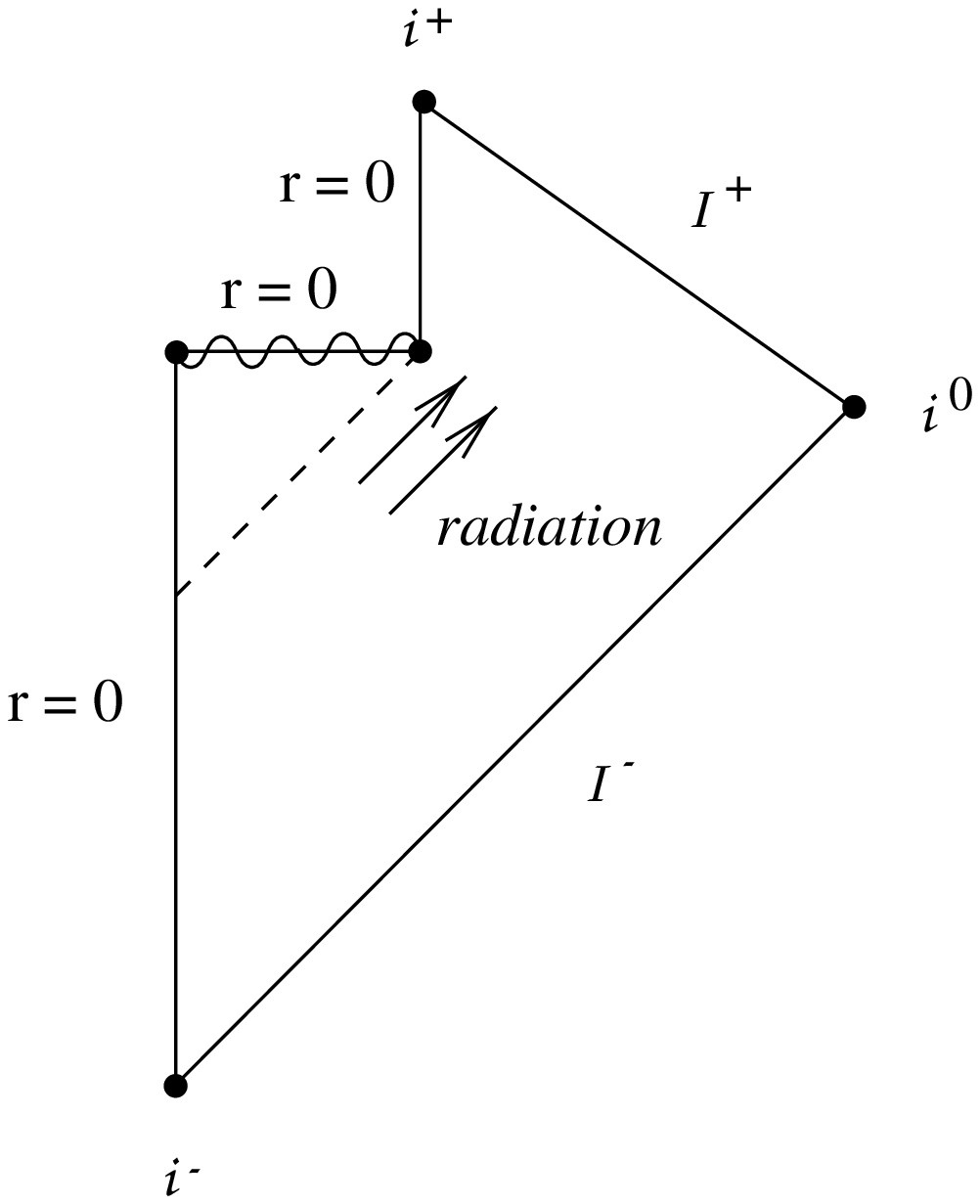,angle=0,height=7cm}}
\end{figure}

On the other hand, it might not.  The problem with this diagram is
that ``information is lost'' --- if we draw a spacelike surface
to the past of the singularity and evolve it into the future, part
of it ends up crashing into the singularity and being destroyed.
As a result the radiation itself contains less information than
the information that was originally in the spacetime.  (This is
the worse than a lack of hair on the black hole.  It's one thing
to think that information has been trapped inside the event horizon,
but it is more worrisome to think that it has disappeared entirely.)
 But such a process
violates the conservation of information that is implicit in both
general relativity and quantum field theory, the two theories that
led to the prediction.  This paradox is considered a big deal these
days, and there are a number of efforts to understand how the 
information can somehow be retrieved.  A currently popular explanation
relies on string theory, and basically says that black holes have a lot
of hair, in the form of virtual stringy states living near the event
horizon.  I hope you will not be disappointed to hear that we won't
look at this very closely; but you should know what the problem is
and that it is an area of active research these days.

With that out of our system, we now turn
to electrically charged black holes.  These
seem at first like reasonable enough objects, since there
is certainly nothing to stop us from throwing some net
charge into a previously uncharged black hole.  In an
astrophysical situation, however, the total amount of charge
is expected to be very small, especially when compared with
the mass (in terms of the relative gravitational effects).
Nevertheless, charged black holes provide a useful testing
ground for various thought experiments, so they are worth
our consideration.

In this
case the full spherical symmetry of the problem is still
present; we know therefore that we can write the metric as
\be
  ds^2 = -e^{2\alpha(r,t)}\d t^2 + e^{2\beta(r,t)}\d r^2 + 
  r^2 d\Omega^2\ .\label{7.106}
\ee
Now, however, we are no longer in vacuum, since the hole
will have a nonzero electromagnetic field, which in turn
acts as a source of energy-momentum.  The energy-momentum
tensor for electromagnetism is given by
\be
  T_\mn = {1\over{4\pi}}(F_{\mu\rho}F_\nu{}^\rho
  -{1\over 4}g_\mn F_{\rho\sigma}F^{\rho\sigma})\ ,\label{7.107}
\ee
where $F_\mn$ is the electromagnetic field strength tensor.
Since we have spherical symmetry, the most general field
strength tensor will have components
\bea
  F_{tr} &=&  f(r,t) = -F_{rt}\cr
  F_{\theta\phi} &=&  g(r,t)\sin\theta = -F_{\phi\theta}\ ,
  \label{7.108}
\eea
where $f(r,t)$ and $g(r,t)$ are some functions to be determined
by the field equations, and components not written are zero.  
$F_{tr}$ corresponds to a radial electric field, while $F_{\theta\phi}$
corresponds to a radial magnetic field.  (For those of you wondering
about the $\sin\theta$, recall that the thing which should be independent
of $\theta$ and $\phi$ is the radial component of the magnetic field,
$B^r = \epsilon^{01\mu\nu}F_{\mu\nu}$.  For a spherically symmetric metric,
$\epsilon^{\rho\sigma\mu\nu}={1\over\g}\tilde\epsilon^{\rho\sigma\mu\nu}$ 
is proportional to $(\sin\theta)^{-1}$, so we want a factor of $\sin\theta$
in $F_{\theta\phi}$.)  The field equations in this case are
both Einstein's equations and Maxwell's equations:
\bea
  g^\mn \nabla_\mu F_{\nu\sigma} &=& 0\cr
  \nabla_{[\mu}F_{\nu\rho]} &=& 0\ .\label{7.109}
\eea
The two 
sets are coupled together, since the electromagnetic field
strength tensor enters Einstein's equations through the
energy-momentum tensor, while the metric enters explicitly into
Maxwell's equations.

The difficulties are not insurmountable, however, and a
procedure similar to the one we followed for the vacuum
case leads to a solution for the charged case as well.  We
will not go through the steps explicitly, but merely quote
the final answer.  The solution is known as the {\bf
Reissner-Nordstr{\o}m metric}, and is given by 
\be
  ds^2 = -\Delta \d t^2 + \Delta^{-1}\d r^2 +
  r^2d\Omega^2\ ,\label{7.110}
\ee
where
\be
  \Delta = 1-{{2GM}\over r}+{{G(p^2+q^2)}\over{r^2}}\ .
  \label{7.111}
\ee
In this expression, $M$ is once again interpreted as 
the mass of the hole; $q$ is the total electric charge, and
$p$ is the total magnetic charge.  Isolated magnetic charges (monopoles)
have never been observed in nature, but that doesn't stop
us from writing down the metric that they would produce if
they did exist.  There are good theoretical reasons to think
that monopoles exist, but are extremely rare.  (Of course, there is
also the possibility that a black hole could have magnetic charge
even if there aren't any monopoles.)  In fact the
electric and magnetic charges enter the metric in the same
way, so we are not introducing any additional complications by keeping
$p$ in our expressions.
The electromagnetic fields associated with this 
solution are given by
\bea
  F_{tr} &=& - {q\over{r^2}}\cr
  F_{\theta\phi} &=&  p\sin\theta\ .\label{7.112}
\eea
Conservatives are welcome to set $p=0$ if they like.

The structure of singularities and event horizons is more
complicated in this metric than it was in Schwarzschild,
due to the extra term in the function $\Delta(r)$ (which can
be thought of as measuring ``how much the light cones tip over'').  
One thing remains the same: at $r=0$ there is a true curvature 
singularity (as could be checked by computing the curvature
scalar $R_{\mn\rho\sigma}R^{\mn\rho\sigma}$).  
Meanwhile,
the equivalent of $r=2GM$ will be the radius where $\Delta$
vanishes.  This will occur at
\be
  r_\pm = GM\pm \sqrt{G^2M^2 - G(p^2+q^2)}\ .\label{7.113}
\ee
This might constitute two, one, or zero solutions,
depending on the relative values of $GM^2$ and
$p^2+q^2$.  We therefore consider each case separately.
\begin{figure}
  \centerline{
  \psfig{figure=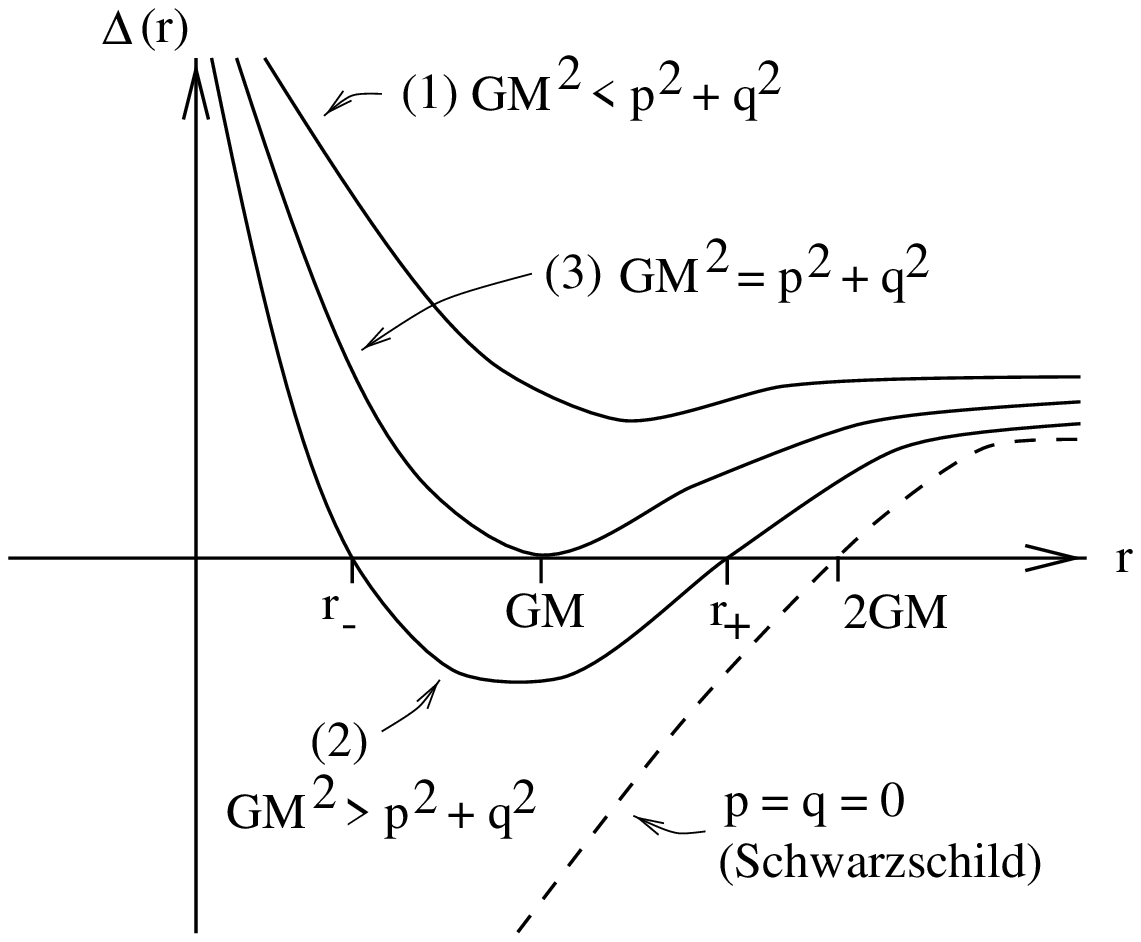,angle=0,height=7cm}}
\end{figure}

\vbox{\vskip .5cm}
\noindent {\it Case One} --- $GM^2<p^2+q^2$

In this case the coefficient $\Delta$ is always positive (never zero),
and the metric is completely regular in the ($t,r,\theta,\phi$)
coordinates all the way down to $r=0$.  The coordinate $t$ is
always timelike, and $r$ is always spacelike.  But there still
is the singularity at $r=0$, which is now a timelike line.
Since there is no event horizon, there is no obstruction
to an observer travelling to the singularity and returning
to report on what was observed.  This is known as a 
{\bf naked singularity}, one which  is not shielded by an
horizon.  A careful analysis of the geodesics reveals, however,
that the singularity is ``repulsive'' --- timelike geodesics
never intersect $r=0$, instead they approach and then reverse 
course and move away.  (Null geodesics can reach the singularity,
as can non-geodesic timelike curves.)

As $r\rightarrow\infty$ the solution approaches flat spacetime,
and as we have just seen the causal structure is ``normal'' everywhere.
The Penrose diagram will therefore be just like that of Minkowski
space, except that now $r=0$ is a singularity.

\begin{figure}[h]
  \centerline{
  \psfig{figure=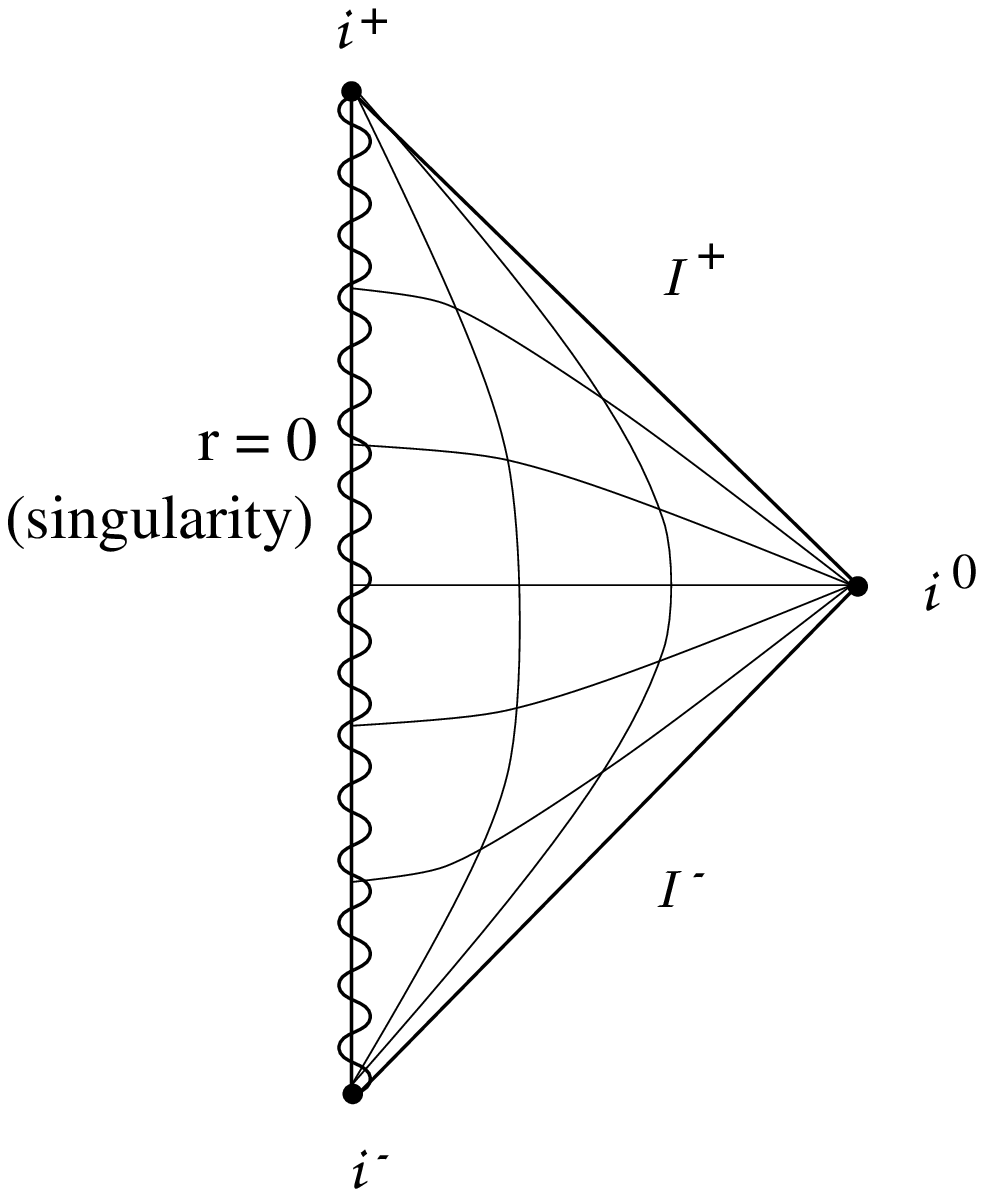,angle=0,height=7cm}}
\end{figure}

\noindent  The nakedness of the singularity offends our sense of
decency, as well as the {\bf cosmic censorship conjecture}, which
roughly states that the gravitational collapse of physical
matter configurations will never produce a naked singularity.
(Of course, it's just a conjecture, and it may not be right; there
are some claims from numerical simulations that collapse of
spindle-like configurations can lead to naked singularities.)
In fact, we should not ever expect to find a black hole with
$GM^2<p^2+q^2$ as the result of gravitational collapse.  Roughly
speaking, this condition states that the total energy of the hole
is less than the contribution to the energy from the electromagnetic
fields alone --- that is, the mass of the matter which carried the
charge would have had to be negative.  This solution is therefore
generally considered to be unphysical.  Notice also that there
are not good Cauchy surfaces (spacelike slices for which every
inextendible timelike line intersects them) in this spacetime, since
timelike lines can begin and end at the singularity.

\noindent {\it Case Two} --- $GM^2>p^2+q^2$

This is the situation which we expect to apply in real gravitational
collapse; the energy in the electromagnetic field is less than the
total energy.  In this case the metric coefficient $\Delta(r)$ is
positive at large $r$ and small $r$, and negative inside the two
vanishing points $r_\pm = GM\pm \sqrt{G^2M^2 - G(p^2+q^2)}$.  The 
metric has coordinate singularities at both $r_+$ and $r_-$; in
both cases these could be removed by a change of coordinates as we
did with Schwarzschild.

\begin{figure}[p]
  \centerline{
  \psfig{figure=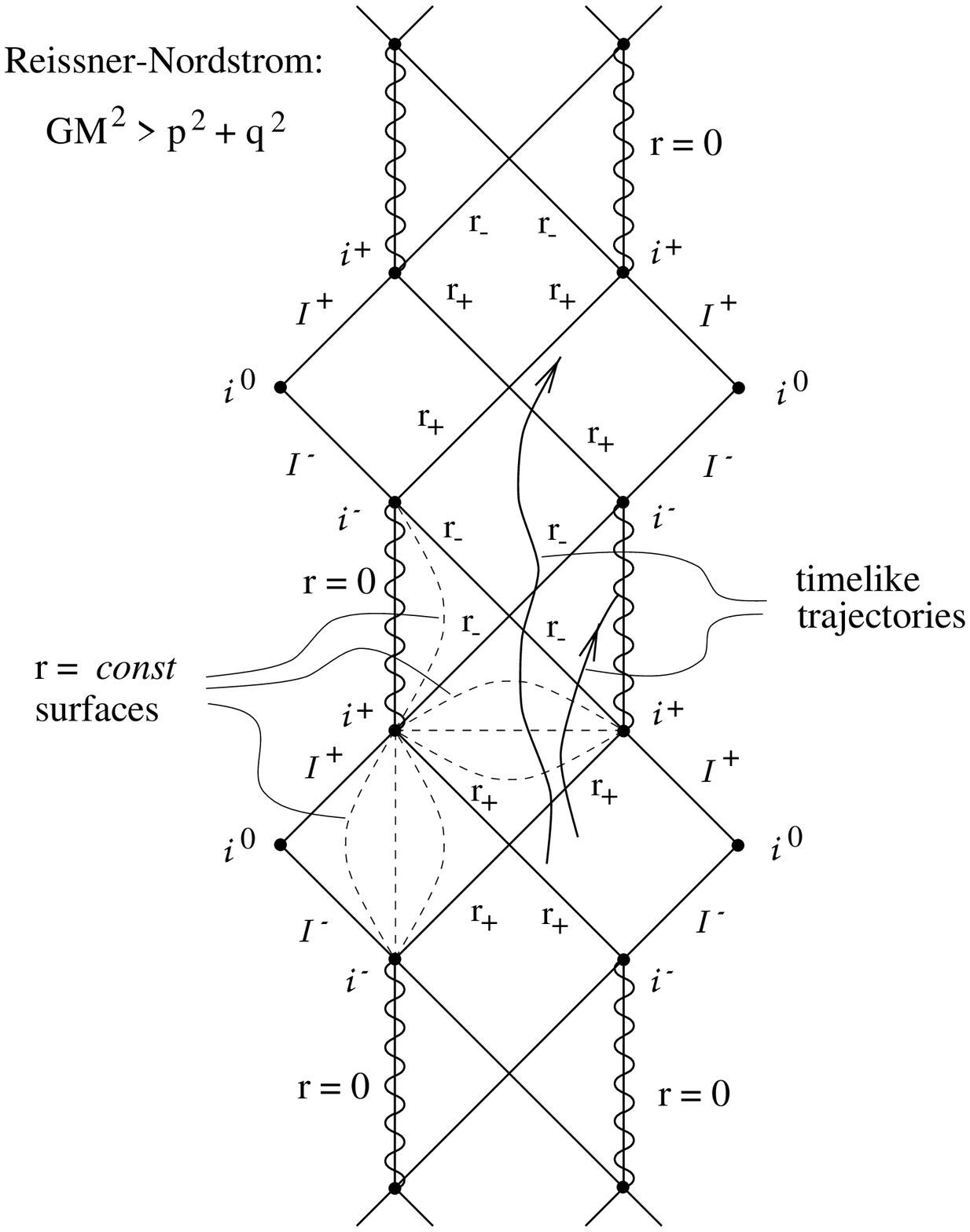,angle=0,height=18cm}}
\end{figure}

The surfaces defined by $r=r_\pm$ are both null, and in fact they
are event horizons (in a sense we will make precise in a moment).
The singularity at $r=0$ is a timelike line (not a spacelike
surface as in Schwarzschild).  If you are an observer falling into the 
black hole from far away, $r_+$ is just like $2GM$ in the Schwarzschild
metric; at this radius $r$ switches from being a spacelike coordinate
to a timelike coordinate, and you necessarily move in the
direction of decreasing $r$.  Witnesses outside the black hole also
see the same phenomena that they would outside an uncharged hole ---
the infalling observer is seen to move more and more slowly, and
is increasingly redshifted.

But the inevitable fall from $r_+$ to ever-decreasing radii only
lasts until you reach the null surface $r=r_-$, where $r$
switches back to being a spacelike coordinate and the motion in
the direction of decreasing $r$ can be arrested.  Therefore you
do not have to hit the singularity at $r=0$; this is to be expected,
since $r=0$ is a timelike line (and therefore not necessarily in your
future).  In fact you can choose either to continue on to $r=0$, or
begin to move in the direction of increasing $r$ back through the
null surface at $r=r_-$.  Then $r$ will once again be a timelike 
coordinate, but with reversed orientation; you are forced to move
in the direction of {\it increasing} $r$.  You will eventually be
spit out past $r=r_+$ once more, which is like emerging from a
white hole into the rest of the universe.  From here you can choose
to go back into the black hole --- this time, a different hole than
the one you entered in the first place --- and repeat the voyage
as many times as you like.  This little story corresponds to the
accompanying Penrose diagram, which of course can be derived more
rigorously by choosing appropriate coordinates and analytically
extending the Reissner-Nordstr{\o}m metric as far as it will go.

How much of this is science, as opposed to science fiction?
Probably not much.  If you think about the world as seen from
an observer inside the black hole who is about to cross the event
horizon at $r_-$, you will notice that they can look back in time
to see the entire history of the external (asymptotically flat)
universe, at least as seen from the black hole.  But they see this
(infinitely long) history in a finite amount of their proper time ---
thus, any  signal that gets to them as they approach $r_-$ is
infinitely blueshifted.  Therefore it is reasonable to believe
(although I know of no proof) that any non-spherically symmetric
perturbation that comes into a Reissner-Nordstr{\o}m black hole
will violently disturb the geometry we have described.  It's hard to 
say what the actual geometry will look like, but there is no very
good reason to believe that it must contain an infinite number of
asymptotically flat regions connecting to each other via
various wormholes.

\noindent {\it Case Three} --- $GM^2=p^2+q^2$

This case is known as the {\bf extreme} Reissner-Nordstr{\o}m
solution (or simply ``extremal black hole'').  
The mass is exactly balanced in some sense by the charge ---
you can construct exact solutions consisting of several extremal
black holes which remain stationary with respect to each other
for all time.  On the one hand the extremal hole is an amusing theoretical 
toy; these solutions are often examined in studies of the information
loss paradox, and the role of black holes in quantum gravity.
On the other hand it appears very unstable, since adding just a
little bit of matter will bring it to Case Two.

\begin{figure}[ht]
  \centerline{
  \psfig{figure=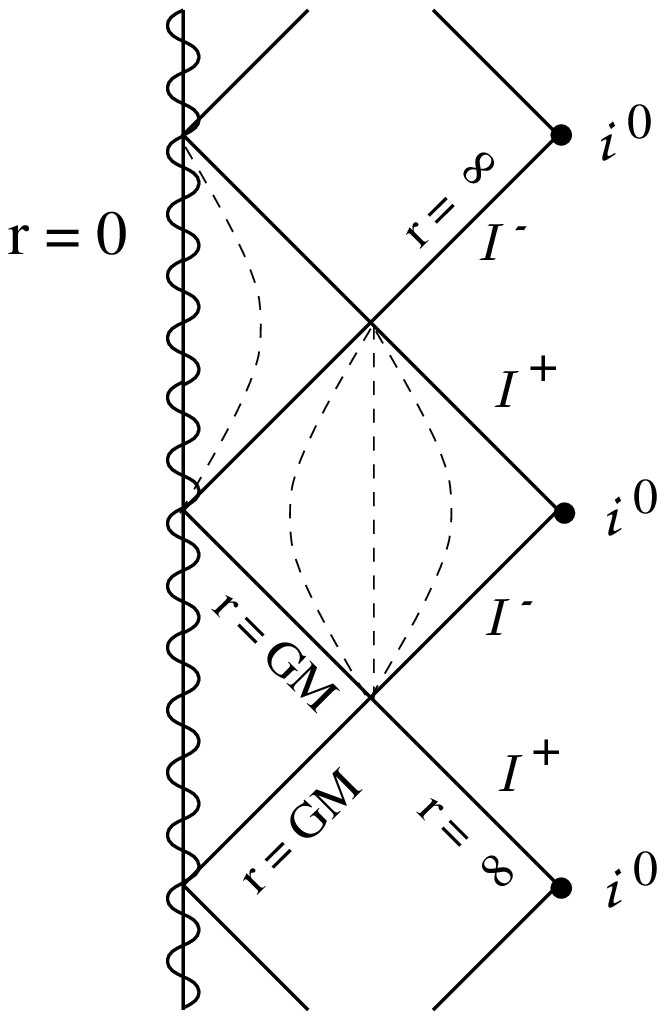,angle=0,height=10cm}}
\end{figure}

The extremal black holes have $\Delta(r)=0$ at a single radius,
$r=GM$.  This does represent an event horizon, but the $r$
coordinate is never timelike; it becomes null at $r=GM$, but is
spacelike on either side.  The singularity at $r=0$ is a 
timelike line, as in the other cases.  So for this black hole
you can again avoid the singularity and continue to move to the
future to extra copies of the asymptotically flat region, but
the singularity is always ``to the left.''  The Penrose diagram
is as shown.  

We could of course go into a good deal more detail about the
charged solutions, but let's instead move on to spinning 
black holes.  It is much more difficult to find
the exact solution for the metric in this case, since we have
given up on spherical symmetry.  To begin with all that is
present is axial symmetry (around the axis of rotation), but we can
also ask for stationary solutions (a timelike Killing vector).
Although the Schwarzschild and Reissner-Nordstr{\o}m solutions were
discovered soon after general relativity was invented, the solution
for a rotating black hole was found by Kerr only in 1963.  His
result, the {\bf Kerr metric}, is given by the following mess:
\be
  ds^2 = -\d t^2 + {{\rho^2}\over \Delta}\d r^2 +\rho^2\d\theta^2
  +(r^2+a^2)\sin^2\theta\,\d\phi^2 +{{2GMr}\over{\rho^2}}
  (a\sin^2\theta\,\d\phi - \d t)^2\ ,\label{7.114}
\ee
where
\be
  \Delta(r) = r^2 - 2GMr +a^2\ ,\label{7.115}
\ee
and
\be
  \rho^2(r,\theta) = r^2+a^2\cos^2\theta\ .\label{7.116}
\ee
Here $a$ measures the rotation of the hole and $M$ is the 
mass.  It is straightforward to include electric and magnetic charges
$q$ and $p$, simply by replacing $2GMr$ with $2GMr-(q^2+p^2)/G$; the result
is the {\bf Kerr-Newman metric}.  All of the interesting phenomena
persist in the absence of charges, so we will set $q=p=0$ from 
now on.

The coordinates $(t,r,\theta,\phi)$ are known as {\bf 
Boyer-Lindquist coordinates}.  It is straightforward to check that
as $a\rightarrow 0$ they reduce to Schwarzschild coordinates.  If
we keep $a$ fixed and let $M\rightarrow 0$, however, we recover
flat spacetime but not in ordinary polar coordinates.  The metric
becomes
\be
  ds^2 = -\d t^2 + {{(r^2+a^2\cos^2\theta)^2}\over (r^2+a^2)}\d r^2 
  +(r^2+a^2\cos^2\theta)^2\d\theta^2
  +(r^2+a^2)\sin^2\theta\,\d\phi^2\ ,\label{7.117}
\ee
and we recognize the spatial part of this as flat space in
ellipsoidal coordinates.

\begin{figure}
  \centerline{
  \psfig{figure=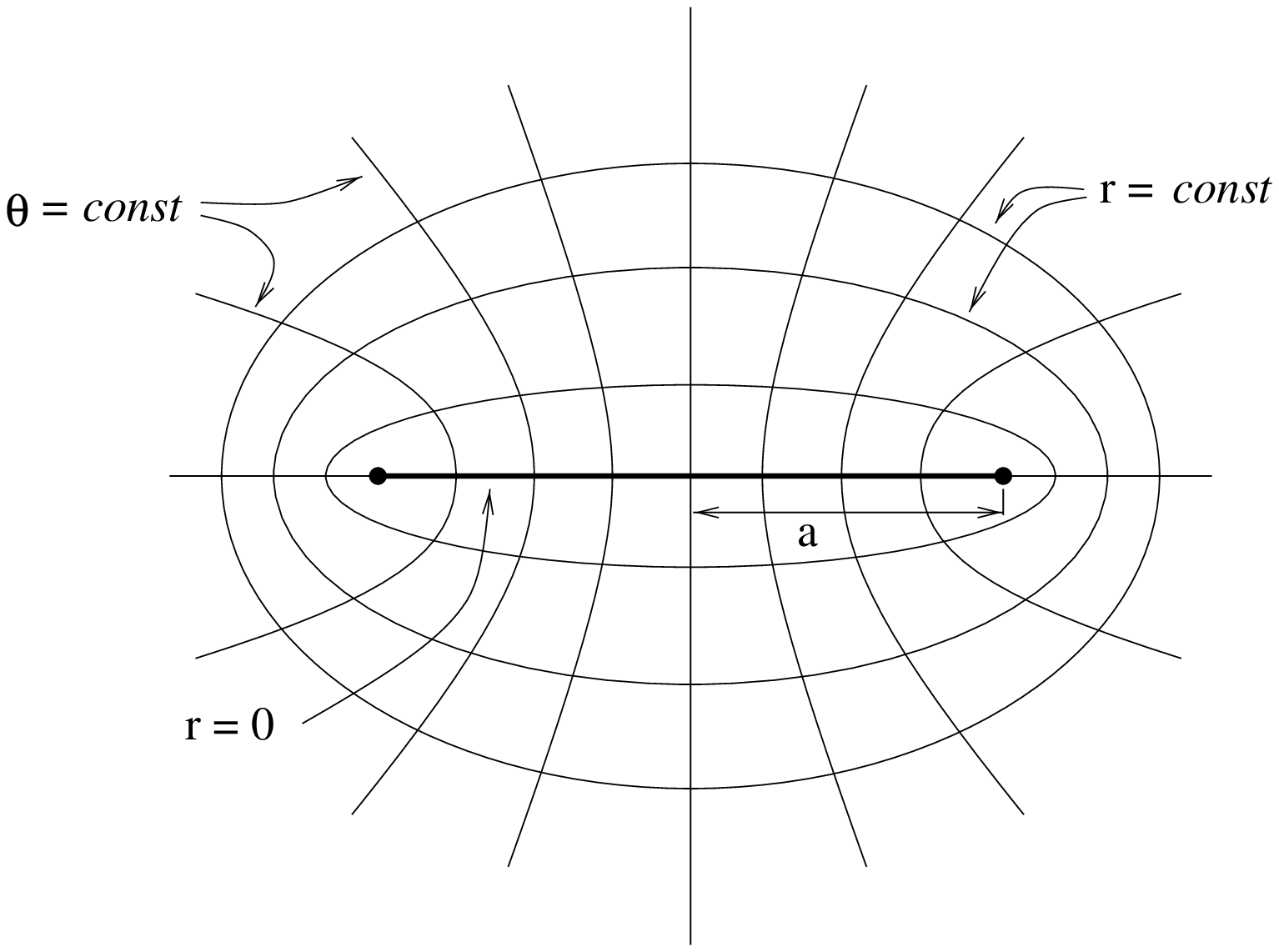,angle=0,height=6cm}}
\end{figure}

\noindent They are related to Cartesian coordinates in Euclidean
3-space by
\bea
  x&=& (r^2+a^2)^{1/2}\sin\theta\,\cos(\phi)\cr
  y&=& (r^2+a^2)^{1/2}\sin\theta\,\sin(\phi)\cr
  z&=& r\cos\theta\ .\label{7.118}
\eea

There are two Killing vectors of the metric (7.114), both of 
which are manifest; since the metric coefficients are independent
of $t$ and $\phi$, both $\zeta^\mu=\p{t}$ and $\eta^\mu=\p\phi$ are Killing 
vectors.  Of course $\eta^\mu$ expresses the axial symmetry of the solution.
The vector $\zeta^\mu$ is not orthogonal to $t=$~constant hypersurfaces,
and in fact is not orthogonal to any hypersurfaces at all; 
hence this metric is stationary, but not static.  (It's not changing
with time, but it is spinning.)  

What is more, the Kerr metric 
also possesses something called a {\bf Killing tensor}.  This is
any symmetric $(0,n)$ tensor $\xi_{\mu_1\cdots\mu_n}$ which
satisfies
\be
  \nabla_{(\sigma}\xi_{\mu_1\cdots\mu_n)}=0\ .\label{7.119}
\ee
Simple examples of Killing tensors are the metric itself, and
symmetrized tensor products of Killing vectors.  Just as a Killing
vector implies a constant of geodesic motion, if there exists a
Killing tensor then along a geodesic we will have
\be
  \xi_{\mu_1\cdots\mu_n}{{dx^{\mu_1}}\over{d\lambda}}\cdots
  {{dx^{\mu_n}}\over{d\lambda}} = {\rm constant}\ .\label{7.120}
\ee
(Unlike Killing vectors, higher-rank Killing tensors do not
correspond to symmetries of the metric.)
In the Kerr geometry we can define the $(0,2)$ tensor
\be
  \xi_\mn = 2\rho^2 l_{(\mu}n_{\nu)} + r^2 g_\mn\ .\label{7.121}
\ee
In this expression the two vectors $l$ and $n$ are given (with indices
raised) by
\bea
  l^\mu &=&  {1\over\Delta}\left(r^2+a^2, \Delta, 0, a\right)\cr
  n^\mu &=&  {1\over{2\rho^2}}\left(r^2+a^2, -\Delta, 0, a\right)\ .
  \label{7.122}
\eea
Both vectors are null and satisfy
\be
  l^\mu l_\mu =0\ ,\quad n^\mu n_\mu =0\ ,\quad l^\mu n_\mu =-1\ .
  \label{7.123}
\ee
(For what it is worth, they are the ``special null vectors'' of the
Petrov classification for this spacetime.)  With these definitions,
you can check for yourself that $\xi_\mn$ is a Killing tensor.

Let's think about the structure of the full Kerr solution.  Singularities
seem to appear at both $\Delta=0$ and $\rho=0$; let's turn our
attention first to $\Delta=0$.  As in the Reissner-Nordstr{\o}m 
solution there are three possibilities: $G^2M^2>a^2$, $G^2M^2=a^2$, and 
$G^2M^2<a^2$.  The last case features a naked singularity, and the
extremal case $G^2M^2=a^2$ is unstable, just as in Reissner-Nordstr{\o}m.
Since these cases are of less physical interest, and time is short,
we will concentrate on $G^2M^2>a^2$.  Then there are two radii at
which $\Delta$ vanishes, given by
\be
  r_\pm = GM\pm\sqrt{G^2M^2 - a^2}\ .\label{7.124}
\ee
Both radii are null surfaces which will turn out to be event 
horizons.  The analysis of these surfaces proceeds in close analogy
with the Reissner-Nordstr{\o}m case; it is straightforward to find
coordinates which extend through the horizons.

Besides the event horizons at $r_\pm$, the Kerr solution also
features an additional surface of interest.  Recall that in the
spherically symmetric solutions, the ``timelike'' Killing vector
$\zeta^\mu=\p{t}$ actually became null on the (outer) event 
horizon, and spacelike inside.  Checking to see where the 
analogous thing happens for Kerr, we compute
\be
  \zeta^\mu\zeta_\mu = -{1\over{\rho^2}}(\Delta-a^2\sin^2\theta)\ .
  \label{7.125}
\ee
This does not vanish at the outer event horizon; in fact, at $r=r_+$
(where $\Delta=0$), we have
\be
  \zeta^\mu\zeta_\mu={{a^2}\over{\rho^2}}\sin^2\theta \geq 0\ .
  \label{7.126}
\ee
So the Killing vector is already spacelike at the outer horizon,
except at the north and south poles ($\theta=0$) where it is null.
The locus of points where $\zeta^\mu\zeta_\mu =0$ is known as the
{\bf Killing horizon}, and is given by
\be
  (r-GM)^2 = G^2M^2 - a^2\cos^2\theta\ ,\label{7.127}
\ee
while the outer event horizon is given by 
\be
  (r_+-GM)^2 = G^2M^2 - a^2\ .\label{7.128}
\ee
There is thus a region in between these two surfaces, known as
the {\bf ergosphere}.  Inside the ergosphere, you must move in
the direction of the rotation of the black hole (the $\phi$ direction); 
however, you can still towards or away from the event horizon 
(and there is no trouble exiting the ergosphere).
It is evidently a place where interesting
things can happen even before you cross the horizon; more details
on this later.

\begin{figure}[t]
  \centerline{
  \psfig{figure=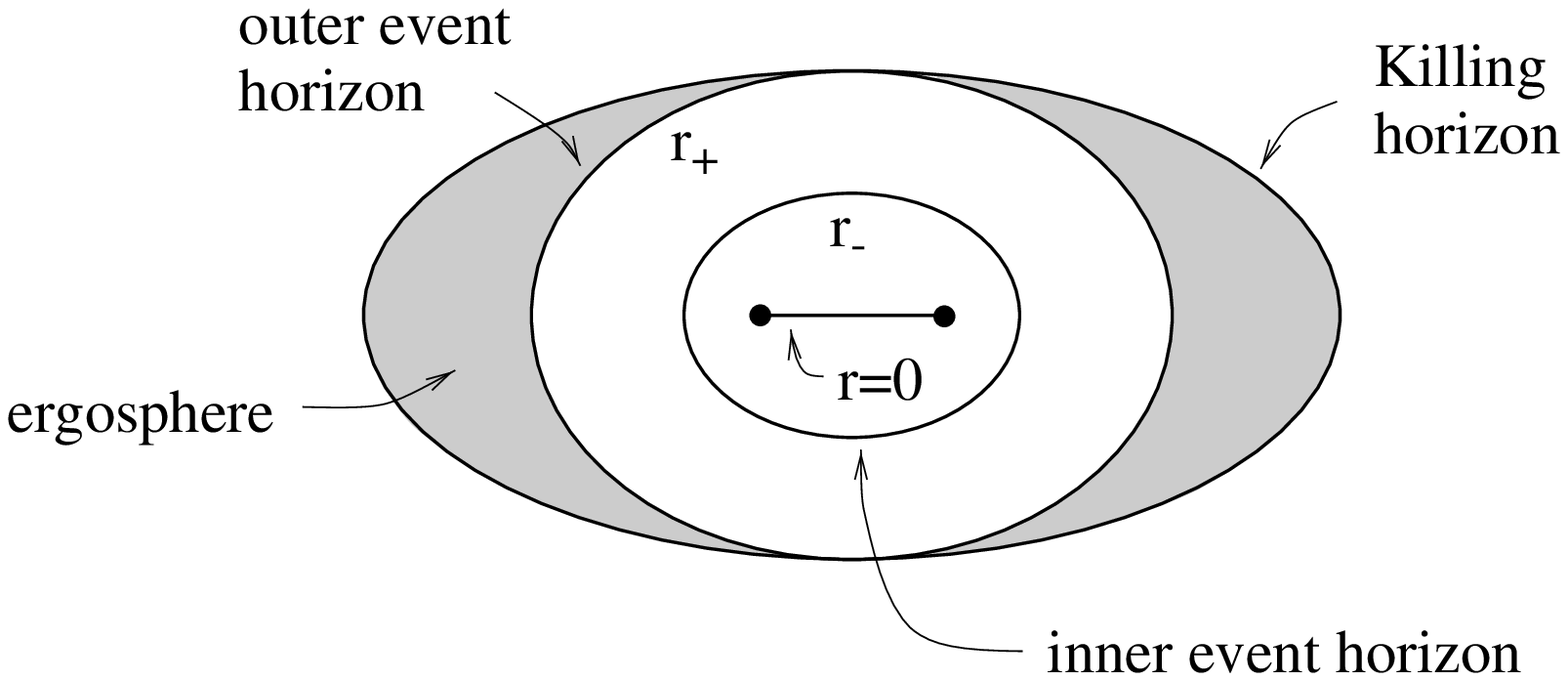,angle=0,height=5cm}}
\end{figure}

Before rushing to draw Penrose diagrams, we need to understand the
nature of the true curvature singularity; this does not occur at
$r=0$ in this spacetime, but rather at $\rho=0$.  Since $\rho^2
= r^2+a^2\cos^2\theta$ is the sum of two manifestly nonnegative
quantities, it can only vanish when both quantities are zero, or
\be
  r=0\ ,\qquad \theta = {\pi\over 2}\ .\label{7.129}
\ee
This seems like a funny result, but remember that $r=0$ is not
a point in space, but a disk; the set of points $r=0$, $\theta=\pi/2$
is actually the {\it ring} at the edge of this disk.  The rotation
has ``softened'' the Schwarzschild singularity, spreading it out
over a ring.

What happens if you go inside the ring?  A careful analytic
continuation (which we will not perform) would reveal that you
exit to another asymptotically flat spacetime, but not an identical
copy of the one you came from.  The new spacetime is described
by the Kerr metric with $r<0$.  As a result, $\Delta$ never vanishes
and there are no horizons.  The Penrose diagram is much like that for
Reissner-Nordstr{\o}m, except now you can pass through the singularity.

\begin{figure}[p]
  \centerline{
  \psfig{figure=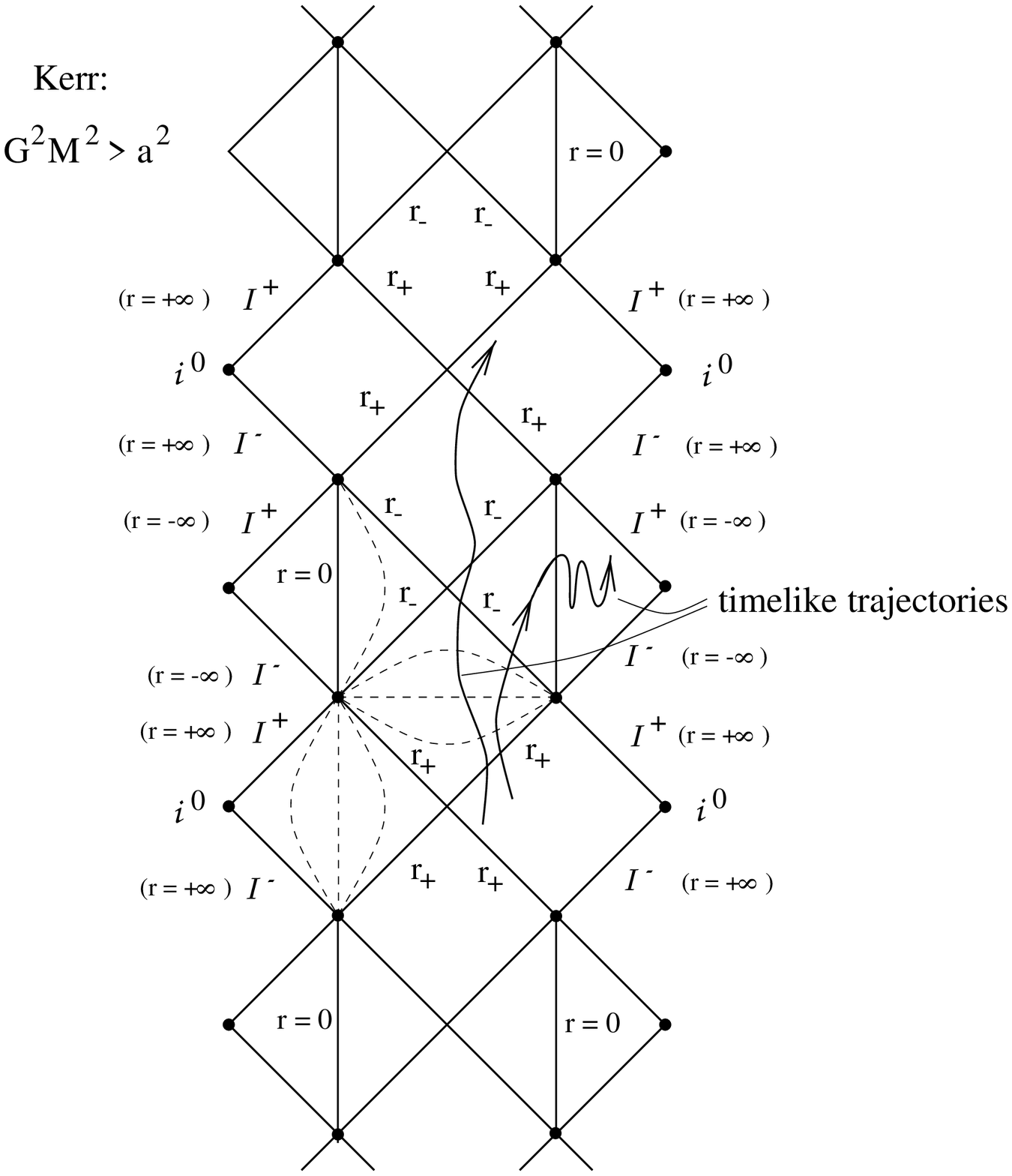,angle=0,height=18cm}}
\end{figure}

Not only do we have the usual strangeness of these distinct 
asymptotically flat regions connected to ours through the black
hole, but the region near the ring singularity has additional
pathologies: closed timelike curves.  If you consider trajectories
which wind around in $\phi$ while keeping $\theta$ and $t$
constant and $r$ a small negative value, the line element along such
a path is
\be
  ds^2 = a^2\left(1+{{2GM}\over r}\right)\d\phi^2\ ,\label{7.130}
\ee
which is negative for small negative $r$.  Since these paths are
closed, they are obviously CTC's.  You can therefore meet yourself
in the past, with all that entails.

Of course, everything we say about the analytic extension of Kerr
is subject to the same caveats we mentioned for Schwarzschild and
Reissner-Nordstr{\o}m; it is unlikely that realistic gravitational
collapse leads to these bizarre spacetimes.  It is nevertheless always
useful to have exact solutions.  Furthermore, for the Kerr metric
there are strange things happening even if we stay outside the
event horizon, to which we now turn.

We begin by considering more carefully the angular velocity of the
hole.  Obviously the conventional definition of angular velocity
will have to be modified somewhat before we can apply it to something
as abstract as the metric of spacetime.  Let us consider the fate
of a photon which is emitted in the $\phi$ direction at some
radius $r$ in the equatorial plane ($\theta=\pi/2$) of a Kerr black
hole.  The instant it is emitted its momentum has no components in
the $r$ or $\theta$ direction, and therefore the condition that it
be null is
\be
  ds^2 = 0 = g_{tt}\d t^2 + g_{t\phi}(\d t\d\phi+\d\phi \d t)
  +g_{\phi\phi}\d\phi^2\ .\label{7.131}
\ee
This can be immediately solved to obtain
\be
  {{d\phi}\over{dt}} = -{{g_{t\phi}}\over{g_{\phi\phi}}}
  \pm\sqrt{\left({{g_{t\phi}}\over{g_{\phi\phi}}}\right)^2
  -{{g_{tt}}\over{g_{\phi\phi}}}}\ .\label{7.132}
\ee
If we evaluate this quantity on the Killing horizon of the Kerr
metric, we have $g_{tt}=0$, and the two solutions are
\be
  {{d\phi}\over{dt}}=0\ ,\qquad {{d\phi}\over{dt}}={{2a}\over
  {(2GM)^2+a^2}}\ .\label{7.133}
\ee
The nonzero solution has the same sign as $a$; we interpret this
as the photon moving around the hole in the same direction as the
hole's rotation.  The zero solution means that the photon directed
against the hole's rotation doesn't move at all in this coordinate
system.  (This isn't a full solution to the photon's trajectory, just
the statement that its instantaneous velocity is zero.)  This is 
an example of the ``dragging of inertial frames'' mentioned earlier.  
The point of this exercise is to note that
massive particles, which must move more slowly than photons, are
necessarily dragged along with the hole's rotation once they are 
inside the Killing horizon.  This dragging continues as we approach
the outer event horizon at $r_+$; we can define the angular velocity
of the event horizon itself, $\Omega_H$, to be the minimum angular
velocity of a particle at the horizon.  Directly from (7.132) we
find that
\be
  \Omega_H = \left({{d\phi}\over{dt}}\right)_-(r_+)
  = {a\over{r_+^2+a^2}}\ .\label{7.134}
\ee

Now let's turn to geodesic motion, which we know will be
simplified by considering the conserved quantities associated with
the Killing vectors $\zeta^\mu=\p{t}$ and $\eta^\mu=\p\phi$.
For the purposes at hand we can restrict our attention to massive
particles, for which we can work with the four-momentum
\be
  p^\mu = m {{dx^\mu}\over{d\tau}}\ ,\label{7.135}
\ee
where $m$ is the rest mass of the particle.  Then we can take as
our two conserved quantities the actual energy and angular momentum
of the particle,
\be
  E=-\zeta_\mu p^\mu = m\left(1-{{2GMr}\over {\rho^2}}\right)
  {{dt}\over{d\tau}}
  +{{2mGMar}\over{\rho^2}}\sin^2\theta\, {{d\phi}\over{d\tau}}\label{7.136}
\ee
and
\be
  L=\eta_\mu p^\mu=-{{2mGMar}\over{\rho^2}}\sin^2\theta\, {{dt}\over{d\tau}}
  +{{m(r^2+a^2)^2 - m\Delta a^2\sin^2\theta}\over{\rho^2}}\sin^2\theta\,
  {{d\phi}\over{d\tau}}\ .\label{7.137}
\ee
(These differ from our previous definitions for the conserved
quantities, where $E$ and $L$ were taken to be the energy and angular
momentum {\it per unit mass}.  They are conserved either way, of course.)

The minus sign in the definition of $E$ is there because at infinity both
$\zeta^\mu$ and $p^\mu$ are timelike, so their inner product is negative,
but we want the energy to be positive.  Inside the ergosphere, however,
$\zeta^\mu$ becomes spacelike; we can therefore imagine particles for
which
\be
  E = -\zeta_\mu p^\mu < 0\ .\label{7.138}
\ee
The extent to which this bothers us is ameliorated somewhat by the
realization that {\it all} particles outside the Killing horizon
must have positive energies; therefore a particle inside the
ergosphere with negative energy must either remain on a geodesic
inside the Killing horizon, or be accelerated until its energy is
positive if it is to escape.

Still, this realization leads to a way to extract energy from a 
rotating black hole; the method is known as the {\bf Penrose process}.
The idea is simple; starting from outside the ergosphere, you arm
yourself with a large rock and leap toward the black hole.  If we
call the four-momentum of the (you + rock) system $p^{(0)\mu}$, then
the energy $E^{(0)}=-\zeta_\mu p^{(0)\mu}$ is certainly positive,
and conserved as you move along your geodesic.  Once you enter the
ergosphere, you hurl the rock with all your might, in a very
specific way.  If we call your momentum $p^{(1)\mu}$ and that of
the rock $p^{(2)\mu}$, then at the instant you throw it we have
conservation of momentum just as in special relativity:
\be
  p^{(0)\mu}=p^{(1)\mu}+p^{(2)\mu}\ .\label{7.139}
\ee
Contracting with the Killing vector $\zeta_\mu$ gives
\be
  E^{(0)}=E^{(1)}+E^{(2)}\ .\label{7.140}
\ee
But, if we imagine that you are arbitrarily strong (and accurate),
you can arrange your throw such that $E^{(2)}<0$, as per (7.158).
\begin{figure}
  \centerline{
  \psfig{figure=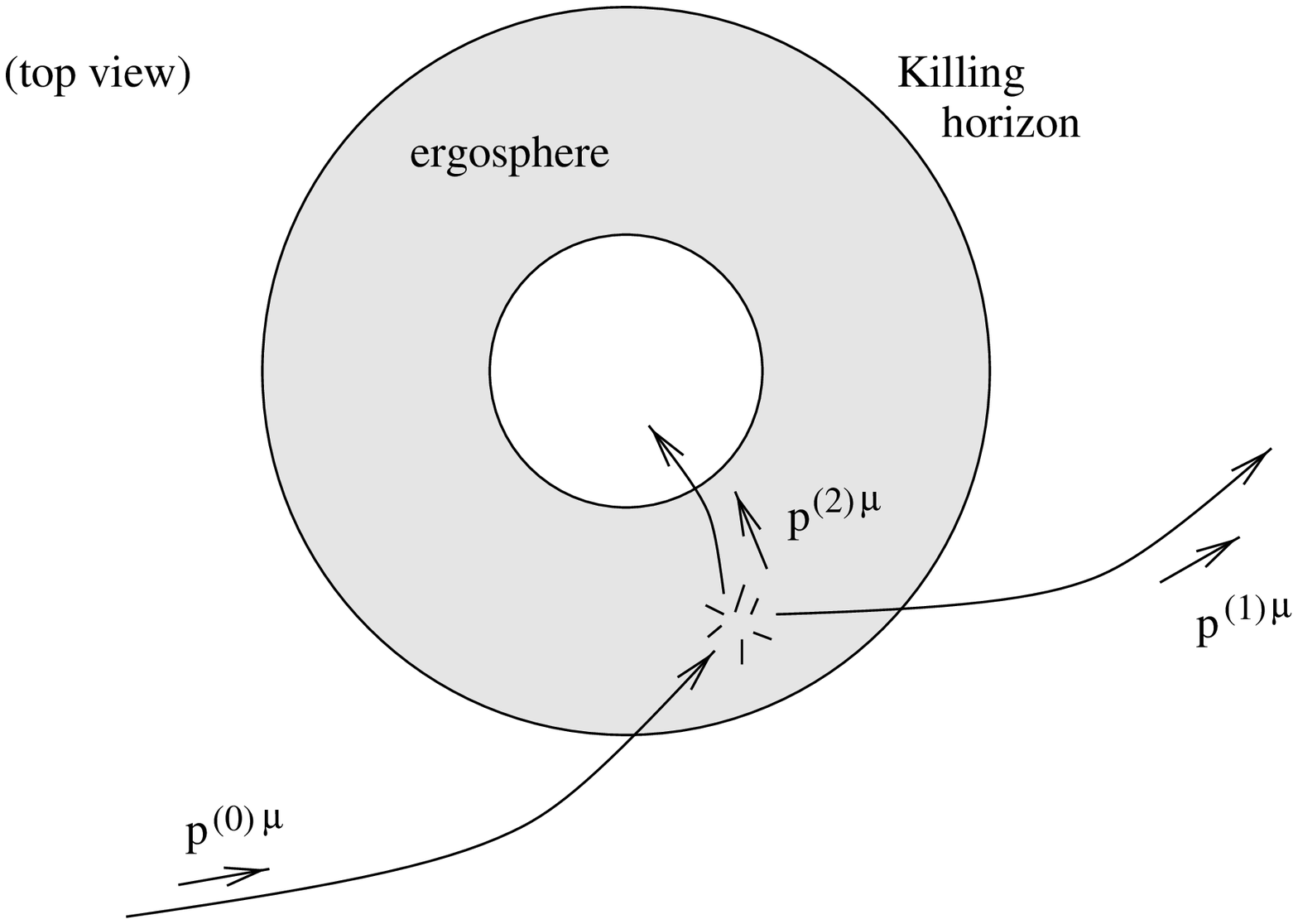,angle=0,height=6cm}}
\end{figure}
Furthermore, Penrose was able to show that you can arrange the
initial trajectory and the throw such that afterwards you follow a
geodesic trajectory back outside the Killing horizon into the
external universe.  Since your energy is conserved along the way,
at the end we will have
\be
  E^{(1)}>E^{(0)}\ .\label{7.141}
\ee
Thus, you have emerged with {\it more} energy than you entered
with.

There is no such thing as a free lunch; the energy you gained came
from somewhere, and that somewhere is the black hole.
In fact, the Penrose process extracts energy from the
rotating black hole by decreasing its angular momentum;
you have to throw the rock against the hole's rotation
to get the trick to work.  To see this more precisely,
define a new Killing vector
\be
  \chi^\mu = \zeta^\mu + \Omega_H\eta^\mu\ .\label{7.142}
\ee
On the outer horizon $\chi^\mu$ is null and tangent to
the horizon.  (This can be seen from $\zeta^\mu = \p{t}$,
$\eta^\mu=\p\phi$, and the definition (7.134) of $\Omega_H$.)
The statement that the particle with momentum $p^{(2)\mu}$
crosses the event horizon ``moving forwards in time'' is simply
\be
  p^{(2)\mu}\chi_\mu <0\ .\label{7.143}
\ee
Plugging in the definitions of $E$ and $L$, we see that
this condition is equivalent to
\be
  L^{(2)} < {{E^{(2)}}\over{\Omega_H}}\ .\label{7.144}
\ee
Since we have arranged $E^{(2)}$ to be negative, and $\Omega_H$
is positive, we see that the particle must have a negative
angular momentum --- it is moving against the hole's rotation.
Once you have escaped the ergosphere and the rock has
fallen inside the event horizon, the mass and angular momentum
of the hole are what they used to be plus the negative
contributions of the rock:
\bea
  \delta M &=&  E^{(2)}\cr \delta J &=&  L^{(2)}.
  \label{7.145}
\eea
Here we have introduced the notation $J$ for the angular
momentum of the black hole; it is given by
\be
  J=Ma\ .\label{7.146}
\ee
We won't justify this, but you can look in Wald for an
explanation.  Then (7.144) becomes a limit on how much you
can decrease the angular momentum:
\be
  \delta J < {{\delta M}\over {\Omega_H}}\ .\label{7.147}
\ee
If we exactly reach this limit, as the rock we throw in
becomes more and more null, we have the ``ideal'' process,
in which $\delta J=\delta M/\Omega_H$.

We will now use these ideas to prove a powerful result:
although you can use the Penrose process to extract energy
from the black hole, you can never decrease the area of the
event horizon.  For a Kerr metric, one can go through a
straightforward computation (projecting the metric and
volume element and so on) to compute the area of the event
horizon:
\be
  A = 4\pi(r_+^2 + a^2)\ .\label{7.148}
\ee
To show that this doesn't decrease, it is most convenient
to work instead in terms of the {\bf irreducible mass} of
the black hole, defined by
\bea
  M_{\rm irr}^2 &=&  {{A}\over {16\pi G^2}}\cr
  &=& {{1\over {4G^2}}}(r_+^2 + a^2)\cr
  &=& {1\over {2}}\left(M^2 + \sqrt{M^4-(Ma/G)^2}\right)\cr
  &=& {1\over {2}}\left(M^2 + \sqrt{M^4-(J/G)^2}\right)\ .
  \label{7.149}
\eea
We can differentiate to obtain, after a bit of work,
\be
  \delta M_{\rm irr} = {{a}\over{4G\sqrt{G^2M^2-a^2}M_{\rm irr}}}
  (\Omega_H^{-1}\delta M - \delta J)\ .\label{7.150}
\ee
(I think I have the factors of $G$ right, but it wouldn't
hurt to check.)  Then our limit (7.147) becomes
\be
  \delta M_{\rm irr} >0\ .\label{7.151}
\ee
The irreducible mass can never be reduced; hence the name.
It follows that the maximum amount of energy we can extract
from a black hole before we slow its rotation to zero is
\be
  M-M_{\rm irr} = M-{1\over\sqrt2}
  \left(M^2 + \sqrt{M^4-(J/G)^2}\right)^{1/2}\ .\label{7.152}
\ee
The result of this complete extraction is a Schwarzschild
black hole of mass $M_{\rm irr}$.
It turns out that the best we can do is to start with an
extreme Kerr black hole; then we can get out approximately
$29\%$ of its total energy.  

The irreducibility of $M_{\rm irr}$ leads immediately to
the fact that the area $A$ can never decrease.  From (7.149)
and (7.150) we have
\be
  \delta A = 8\pi G{a\over{\Omega_H\sqrt{G^2M^2-a^2}}}
  (\delta M - \Omega_H\delta_J)\ ,\label{7.153}
\ee
which can be recast as
\be
  \delta M = {{\kappa}\over{8\pi G}}\delta A +\Omega_H
  \delta J\ ,\label{7.154}
\ee
where we have introduced
\be
  \kappa = {{\sqrt{G^2M^2-a^2}}\over{2GM(GM+
  \sqrt{G^2M^2-a^2})}}				\ .\label{7.155}
\ee
The quantity $\kappa$ is known as the {\bf surface gravity}
of the black hole.

It was equations like (7.154) that first started people 
thinking about the relationship between black holes and
thermodynamics.  Consider the first law of thermodynamics,
\be
  dU = TdS + {\rm work~ terms}\ .\label{7.156}
\ee    
It is natural to think of the term $\Omega_H\delta J$ as
``work'' that we do on the black hole by throwing rocks into
it.  Then the thermodynamic analogy begins to take shape
if we think of identifying the area $A$ as the entropy
$S$, and the surface gravity $\kappa$ as $8\pi G$ times the
temperature $T$.  In fact, in the context of classical 
general relativity the analogy is essentially perfect.
The ``zeroth'' law of thermodynamics states that in 
thermal equilibrium the temperature is constant throughout
the system; the analogous statement for black holes is that
stationary black holes have constant surface gravity on
the entire horizon (true).  As we have seen, the first 
law (7.156) is equivalent to (7.154).  The second law,
that entropy never decreases, is simply the statement that
the area of the horizon never decreases.  Finally, the third
law is that it is impossible to achieve $T=0$ in any
physical process, which should imply that it is impossible
to achieve $\kappa=0$ in any physical process.  It turns
out that $\kappa=0$ corresponds to the extremal black
holes (either in Kerr or Reissner-Nordstr{\o}m) --- where
the naked singularities would appear.  Somehow, then,
the third law is related to cosmic censorship.

The missing piece is that {\it real} thermodynamic bodies
don't just sit there; they give off blackbody radiation
with a spectrum that depends on their temperature.  Black
holes, it was thought before Hawking discovered his radiation, 
don't do that, since they're truly black.  Historically,
Bekenstein came up with the idea that black holes should
really be honest black bodies, including the radiation at
the appropriate temperature.  This annoyed Hawking, who set
out to prove him wrong, and ended up proving that there
would be radiation after all.  So the thermodynamic analogy
is even better than we had any right to expect --- although
it is safe to say that nobody really knows why.

\eject

\thispagestyle{plain}
 
\setcounter{equation}{0}

\noindent{December 1997 \hfill {\sl Lecture Notes on General Relativity}
\hfill{Sean M.~Carroll}}

\vskip .2in

\setcounter{section}{7}
\section{Cosmology}

Contemporary cosmological models are based on the idea that the
universe is pretty much the same everywhere --- a stance sometimes
known as the {\bf Copernican principle}.  On the face of it, such
a claim seems preposterous; the center of the sun, for example, 
bears little resemblance to the desolate cold of interstellar
space.  But we take the Copernican principle to only apply on
the very largest scales, where local variations in density are
averaged over.  Its validity on such scales is 
manifested in a number of different observations, such as number counts
of galaxies and observations of diffuse X-ray and $\gamma$-ray
backgrounds, but is most clear in the $3^\circ$ microwave background
radiation.  Although we now know that the microwave background is not
perfectly smooth (and nobody ever expected that it was), the 
deviations from regularity are on the order of $10^{-5}$ or less,
certainly an adequate basis for an approximate description of 
spacetime on large scales.

The Copernican principle is related to two more mathematically
precise properties that a manifold might have: isotropy and homogeneity.
{\bf Isotropy} applies at some specific point in the space, and
states that the space looks the same no matter what direction you
look in.  More formally, a manifold $M$ is isotropic around a point
$p$ if, for any two vectors $V$ and $W$ in $T_pM$, there is an
isometry of $M$ such that the pushforward of $W$ under the isometry
is parallel with $V$ (not pushed forward).
It is isotropy which is indicated by the observations of the
microwave background.

{\bf Homogeneity} is the statement that the metric is the same
throughout the space.  In other words, given any two points $p$ and
$q$ in $M$, there is an isometry which takes $p$ into $q$.
Note that there is no necessary relationship between homogeneity
and isotropy; a manifold can be homogeneous but nowhere isotropic
(such as $\R\times S^2$ in the usual metric), or it can be isotropic
around a point without being homogeneous (such as a cone, which is
isotropic around its vertex but certainly not homogeneous).  On the
other hand, if a space is isotropic {\it everywhere} then it is
homogeneous.  (Likewise if it is isotropic around one point and
also homogeneous, it will be isotropic around every point.)
Since there is ample observational evidence for 
isotropy, and the Copernican principle would have us believe that we
are not the center of the universe and therefore observers elsewhere
should also observe isotropy, we will henceforth assume both
homogeneity and isotropy.

There is one catch.  When we look at distant galaxies, they appear
to be receding from us; the universe is apparently not static, but
changing with time.  Therefore we begin construction of cosmological
models with the idea that the universe is homogeneous and isotropic
in space, but not in time.  In general relativity this translates into
the statement that the universe can be foliated into spacelike slices
such that each slice is homogeneous and isotropic.  
We therefore consider our spacetime to be $\R\times\Sigma$, where
$\R$ represents the time direction and $\Sigma$ is a homogeneous and
isotropic three-manifold.  The usefulness of homogeneity and 
isotropy is that they imply that $\Sigma$ must be a maximally
symmetric space.  (Think of isotropy as invariance under rotations,
and homogeneity as invariance under translations.  Then homogeneity
and isotropy together imply that a space has its maximum possible
number of Killing vectors.)  Therefore
we can take our metric to be of the form
\be
  ds^2 = -dt^2 + a^2(t)\gamma_{ij}(u)\d u^i\d u^j\ .\label{8.1}
\ee
Here $t$ is the timelike coordinate, and $(u^1, u^2, u^3)$ are the
coordinates on $\Sigma$; $\gamma_{ij}$ is the maximally symmetric
metric on $\Sigma$.  This formula is a special case of (7.2), which we
used to derive the Schwarzschild metric, except we have scaled $t$
such that $g_{tt}=-1$.  The function $a(t)$ is known as the 
{\bf scale factor}, and it tells us ``how big'' the spacelike
slice $\Sigma$ is at the moment $t$.  The coordinates used here,
in which the metric is free of cross terms $\d t\,\d u^i$ and the
spacelike components are proportional to a single function of $t$, are 
known as {\bf comoving coordinates}, and an observer who stays at constant
$u^i$ is also called ``comoving''.  Only a comoving observer will
think that the universe looks isotropic; in fact on Earth we are
not quite comoving, and as a result we see a dipole anisotropy in
the cosmic microwave background as a result of the conventional
Doppler effect.

Our interest is therefore in maximally symmetric Euclidean three-metrics
$\gamma_{ij}$.  We know that maximally symmetric metrics obey
\be
  ^{(3)}R_{ijkl} = k(\gamma_{ik}\gamma_{jl}
  -\gamma_{il}\gamma_{jk})\ ,\label{8.2}
\ee
where $k$ is some constant, and we put a superscript $^{(3)}$ on
the Riemann tensor to remind us that it is associated with the
three-metric $\gamma_{ij}$, not the metric of the entire spacetime.
The Ricci tensor is then
\be
  ^{(3)}R_{jl} = 2k\gamma_{jl}\ .\label{8.3}
\ee
If the space is to be maximally symmetric, then it will certainly
be spherically symmetric.  We already know something about spherically
symmetric spaces from our exploration of the Schwarzschild
solution; the metric can be put in the form
\be
  d\sigma^2 = \gamma_{ij}\d u^i\,\d u^j =
  e^{2\beta(r)}\d r^2 + r^2(\d\theta^2 +
  \sin^2\theta\,\d\phi^2)\ .\label{8.4}
\ee
The components of the Ricci tensor for such a metric can be obtained
from (7.16), the Ricci tensor for a spherically symmetric spacetime,
by setting $\alpha=0$ and $\p0\beta=0$, which gives
\bea
  ^{(3)}R_{11} &=&  {2\over r}\p1\beta\cr
  ^{(3)}R_{22} &=&  e^{-2\beta}(r\p1 \beta-1)+1\cr
  ^{(3)}R_{33} &=&  [e^{-2\beta}(r\p1 \beta-1)+1]\sin^2\theta\, \ .
  \label{8.5}
\eea
We set these proportional to the metric using (8.3), and can solve
for $\beta(r)$:
\be
  \beta = -{1\over 2}\ln(1-kr^2)\ .\label{8.6}
\ee
This gives us the following metric on spacetime:
\be
  ds^2 = -dt^2 + a^2(t)\left[{{\d r^2}\over{1-kr^2}}
  + r^2(\d\theta^2 +\sin^2\theta\,\d\phi^2)\right]\ .\label{8.7}
\ee
This is the {\bf Robertson-Walker metric}.  We have not yet
made use of Einstein's equations; those will determine the behavior
of the scale factor $a(t)$.

Note that the substitutions
\bea
  k&\rightarrow{k\over{|k|}}\cr
  r &\rightarrow \sqrt{|k|}\, r\cr
  a &\rightarrow{a\over{\sqrt{|k|}}}\label{8.8}
\eea
leave (8.7) invariant.  Therefore the only relevant parameter
is $k/|k|$, and there are three cases of interest: $k=-1$,
$k=0$, and $k=+1$.  The $k=-1$ case corresponds to constant
negative curvature on $\Sigma$, and is called {\bf open}; the
$k=0$ case corresponds to no curvature on $\Sigma$, and is called 
{\bf flat}; the $k=+1$ case corresponds to positive curvature on $\Sigma$, 
and is called {\bf closed}.

Let us examine each of these possibilities.  For the flat case
$k=0$ the metric on $\Sigma$ is
\bea
  d\sigma^2 &=&  \d r^2 + r^2 d\Omega^2\cr
  &=&  \d x^2 + \d y^2 +\d z^2\ ,\label{8.9}
\eea
which is simply flat Euclidean space.  Globally, it could describe
$\R^3$ or a more complicated manifold, such as the three-torus
$S^1\times S^1 \times S^1$.  For the closed case $k=+1$ we can
define $r=\sin\chi$ to write the metric on $\Sigma$ as
\be
  d\sigma^2 = \d \chi^2 + \sin^2\chi\, d\Omega^2\ ,\label{8.10}
\ee
which is the metric of a three-sphere.  In this case the only
possible global structure is actually the three-sphere (except for
the non-orientable manifold $\R$P$^3$).  Finally in the open $k=-1$
case we can set $r=\sinh\psi$ to obtain
\be
  d\sigma^2 = \d\psi^2 + \sinh^2\psi\, d\Omega^2\ .\label{8.11}
\ee
This is the metric for a three-dimensional space of constant
negative curvature; it is hard to visualize, but think of the
saddle example we spoke of in Section Three.  Globally such a space could
extend forever (which is the origin of the word ``open''), but it
could also describe a non-simply-connected compact space (so ``open''
is really not the most accurate description).

With the metric in hand, we can set about computing the connection
coefficients and curvature tensor.  Setting $\dot a\equiv da/dt$,
the Christoffel symbols are given by
\bea
  \Gamma^0_{11} &=&  {{a\dot a}\over{1-kr^2}}\qquad 
  \Gamma^0_{22} = a\dot a r^2
  \qquad \Gamma^0_{33} = a\dot a r^2\sin^2\theta\cr
  \Gamma^1_{01} &=&  \Gamma^1_{10} = \Gamma^2_{02}= \Gamma^2_{20}
  = \Gamma^3_{03}= \Gamma^3_{30}={{\dot a}\over a}\cr
  \Gamma^1_{22} &=&   -r(1-kr^2) \qquad 
  \Gamma^1_{33} =  -r(1-kr^2)\sin^2\theta\cr
  \Gamma^2_{12} &=&  \Gamma^2_{21} = \Gamma^3_{13} = \Gamma^3_{31} = 
  {1\over r}\cr
  \Gamma^2_{33} &=&  -\sin\theta\,\cos\theta \qquad
  \Gamma^3_{23} = \Gamma^3_{32} = \cot\theta\ .\label{8.12}
\eea
The nonzero components of the Ricci tensor are
\bea
  R_{00} &=&  -3{{\ddot a}\over a}\cr
  R_{11} &=&  {{a\ddot a + 2\dot a^2 +2k}\over{1-kr^2}}\cr
  R_{22} &=&  r^2(a\ddot a+ 2\dot a^2 +2k)\cr
  R_{33} &=&  r^2(a\ddot a+ 2\dot a^2 +2k)\sin^2\theta \ ,\label{8.13}
\eea
and the Ricci scalar is then
\be
  R = {{6}\over{a^2}}(a\ddot a+ \dot a^2 +k)\ .\label{8.14}
\ee

The universe is not empty, so we are not interested in vacuum
solutions to Einstein's equations.  We will choose to model the
matter and energy in the universe by a perfect fluid.  We discussed
perfect fluids in Section One, where they were defined as fluids
which are isotropic in their rest frame.  The energy-momentum tensor
for a perfect fluid can be written
\be
  T_{\mn} = (p+\rho)U_\mu U_\nu + pg_\mn\ ,\label{8.15}
\ee
where $\rho$ and $p$ are the energy density and pressure (respectively)
as measured in the rest frame, and $U^\mu$ is the four-velocity of
the fluid.  It is clear that, if a fluid which is isotropic in some
frame leads to a metric which is isotropic in some frame, the two
frames will coincide; that is, the fluid will be at rest in comoving
coordinates.  The four-velocity is then
\be
  U^\mu = (1,0,0,0)\ ,\label{8.16}
\ee
and the energy-momentum tensor is
\be
  T_\mn = \left(\matrix{\rho &0&0&0\cr 0& & & \cr
  0& & g_{ij} p& \cr 0& & & \cr}\right)\ .\label{8.17}
\ee
With one index raised this takes the more convenient form
\be
  T^\mu{}_\nu = {\rm diag}(-\rho,p,p,p)\ .\label{8.18}
\ee
Note that the trace is given by
\be
  T = T^\mu{}_\mu = -\rho +3p\ .\label{8.19}
\ee

Before plugging in to Einstein's equations, it is educational to
consider the zero component of the conservation of energy equation:
\bea
  0 &=&  \nabla_\mu T^\mu{}_0\cr
  &=&  \p\mu T^\mu{}_0 +\Gamma^\mu_{\mu 0}T^0{}_0
  -\Gamma^\lambda_{\mu 0}T^\mu{}_\lambda\cr
  &=&  -\p0\rho -3{{\dot a}\over a}(\rho+p)\ .
  \label{8.20}
\eea
To make progress it is necessary to choose an {\bf equation of
state}, a relationship between $\rho$ and $p$.  Essentially all of
the perfect fluids relevant to cosmology obey the simple equation
of state
\be
  p=w\rho\ ,\label{8.21}
\ee
where $w$ is a constant independent of time.  The conservation
of energy equation becomes
\be
  {{\dot \rho}\over\rho} = -3(1+w){{\dot a}\over{a}}\ ,\label{8.22}
\ee
which can be integrated to obtain
\be
  \rho \propto a^{-3(1+w)}\ .\label{8.23}
\ee

The two most popular examples of cosmological fluids are
known as {\bf dust} and {\bf radiation}.  Dust is collisionless,
nonrelativistic matter, which obeys $w=0$.  Examples include 
ordinary stars and galaxies, for which the pressure is 
negligible in comparison with the energy density.  Dust is also
known as ``matter'', and universes whose energy density is mostly
due to dust are
known as {\bf matter-dominated}.  The energy density in matter
falls off as 
\be
  \rho\propto a^{-3}\ .\label{8.24}
\ee
This is simply interpreted
as the decrease in the number density of particles as the universe
expands.  (For dust the energy density is dominated by the rest
energy, which is proportional to the number density.)  ``Radiation''
may be used to describe either actual electromagnetic radiation, or
massive particles moving at relative velocities sufficiently close to 
the speed of light that they become indistinguishable from photons (at
least as far as their equation of state is concerned).
Although radiation is a perfect fluid and thus has an energy-momentum
tensor given by (8.15), we also know that $T_\mn$ can be expressed in
terms of the field strength as
\be
  T^\mn = {1\over{4\pi}}(F^{\mu\lambda}F^\nu{}_\lambda
  -{1\over 4}g^{\mu\nu} F^{\lambda\sigma}F_{\lambda\sigma})\ .
  \label{8.25}
\ee
The trace of this is given by
\be
  T^\mu{}_\mu = {1\over{4\pi}}\left[F^{\mu\lambda}F_{\mu\lambda}
  -{1\over 4}(4)F^{\lambda\sigma}F_{\lambda\sigma}\right] = 0\ .\label{8.26}
\ee
But this must also equal (8.19), so the equation of state is
\be
  p = {1\over 3}\rho\ .\label{8.27}
\ee
A universe in which most of the energy density is in the form of
radiation is known as {\bf radiation-dominated}.  The energy
density in radiation falls off as
\be
  \rho \propto a^{-4}\ .\label{8.28}
\ee
Thus, the energy density in radiation falls off slightly faster
than that in matter; this is because the number density of photons
decreases in the same way as the number density of nonrelativistic
particles, but individual photons also lose energy as $a^{-1}$
as they redshift, as we will see later.  (Likewise, massive but
relativistic particles will lose energy as they ``slow down'' in
comoving coordinates.)  We believe that today the
energy density of the universe is dominated by matter, with
$\rho_{\rm mat}/\rho_{\rm rad}\sim10^6$.  However, in the past 
the universe was much smaller, and the energy density in radiation
would have dominated at very early times.

There is one other form of energy-momentum that is sometimes 
considered, namely that of the vacuum itself.  Introducing energy
into the vacuum is equivalent to introducing a cosmological constant.
Einstein's equations with a cosmological constant are
\be
  G_\mn = 8\pi GT_\mn -\Lambda g_\mn \ ,\label{8.29}
\ee
which is clearly the same form as the equations with no cosmological
constant but an energy-momentum tensor for the vacuum,
\be
  T^{\rm (vac)}_\mn = -{{\Lambda}\over{8\pi G}}g_\mn\ .\label{8.30}
\ee
This has the form of a perfect fluid with
\be
  \rho = -p = {{\Lambda}\over{8\pi G}}\ .\label{8.31}
\ee
We therefore have $w=-1$, and 
the energy density is independent of $a$, which is what
we would expect for the energy density of the vacuum.  Since the
energy density in matter and radiation decreases as the universe
expands, if there is a nonzero vacuum energy it tends to win out over
the long term (as long as the universe doesn't start contracting).
If this happens, we say that the universe becomes {\bf
vacuum-dominated}.

We now turn to Einstein's equations.  Recall that they can be
written in the form (4.45):
\be
  R_\mn = 8\pi G\left(T_\mn - {1\over 2}g_\mn T\right)\ .
  \label{8.32}
\ee
The $\mn = 00$ equation is
\be
  -3{{\ddot a}\over a}=4\pi G(\rho+3p)\ ,\label{8.33}
\ee
and the $\mn = ij$ equations give
\be
  {{\ddot a}\over a} + 2\left({{\dot a}\over a}\right)^2
  +2{k\over{a^2}}= 4\pi G(\rho-p)\ .\label{8.34}
\ee
(There is only one distinct equation from $\mn = ij$, due to
isotropy.)  We can use (8.33) to eliminate second derivatives in
(8.34), and do a little cleaning up to obtain 
\be
  {{\ddot a}\over a}=-{{4\pi G}\over 3}(\rho+3p)\ ,\label{8.35}
\ee
and
\be
  \left({{\dot a}\over a}\right)^2={{8\pi G}\over 3}\rho
  -{{k}\over a^2}\ .\label{8.36}
\ee
Together these are known as the {\bf Friedmann equations},
and metrics of the form (8.7) which obey these equations define
Friedmann-Robertson-Walker (FRW) universes.

There is a bunch of terminology which is associated with the
cosmological parameters, and we will just introduce the basics
here.  The rate of expansion is characterized by the {\bf Hubble
parameter},
\be
  H ={{\dot a}\over a}\ .\label{8.37}
\ee
The value of the Hubble parameter at the present epoch is the
Hubble constant, $H_0$.  There is currently a great deal of 
controversy about what its actual value is, with measurements
falling in the range of 40 to 90 km/sec/Mpc.  (``Mpc'' stands for
``megaparsec'', which is $3\times 10^{24}$~cm.)  Note that we
have to divide $\dot a$ by $a$ to get a measurable quantity, since
the overall scale of $a$ is irrelevant.  There is also the
{\bf deceleration parameter},
\be
  q = -{{a\ddot a}\over {\dot a^2}}\ ,\label{8.38}
\ee
which measures the rate of change of the rate of expansion.

Another useful quantity is the {\bf density parameter},
\bea
  \Omega &=&  {{8\pi G}\over {3H^2}}\rho\cr
  &=& {\rho\over{\rho_{\rm crit}}}\ ,\label{8.39}
\eea
where the {\bf critical density} is defined by
\be
  \rho_{\rm crit} = {{3H^2}\over{8\pi G}}\ .\label{8.40}
\ee
This quantity (which will generally change with time) is called
the ``critical'' density because the Friedmann equation (8.36)
can be written
\be
  \Omega-1={{k}\over {H^2 a^2}} \ .\label{8.41}
\ee
The sign of $k$ is therefore determined by whether $\Omega$ is
greater than, equal to, or less than one.  We have
\[
  \matrix{\rho<\rho_{\rm crit} & \leftrightarrow & \Omega < 1 &
  \leftrightarrow & k=-1 & \leftrightarrow & {\rm open}\cr
  \rho=\rho_{\rm crit} & \leftrightarrow & \Omega = 1 &
  \leftrightarrow & k=0 & \leftrightarrow & {\rm flat}\cr
  \rho>\rho_{\rm crit} & \leftrightarrow & \Omega > 1 &		
  \leftrightarrow & k=+1 & \leftrightarrow & {\rm closed}\ .\cr}
\]
The density parameter, then, tells us which of the three 
Robertson-Walker geometries describes our universe.  Determining
it observationally is an area of intense investigation.

It is possible to solve the Friedmann equations exactly in 
various simple cases, but it is often more useful to know
the qualitative behavior of various possibilities.  Let us for
the moment set $\Lambda=0$, and consider the behavior of universes
filled with fluids of positive energy ($\rho > 0$) and nonnegative
pressure ($p\geq 0$).  Then by (8.35) we must have $\ddot a<0$.
Since we know from observations of distant galaxies that 
the universe is expanding ($\dot a>0$), 
this means that the universe is ``decelerating.'' This is what
we should expect, since the gravitational attraction of the matter
in the universe works against the expansion.  The fact that
the universe can only decelerate means that it must have been
expanding even faster in the past; if we trace the evolution 
backwards in time, we necessarily reach a singularity at 
$a=0$.  Notice that if $\ddot a$ were exactly zero, $a(t)$
would be a straight line, and the age of the universe would be
$H_0^{-1}$.  Since $\ddot a$ is actually negative, the universe
must be somewhat younger than that.

\begin{figure}
  \centerline{
  \psfig{figure=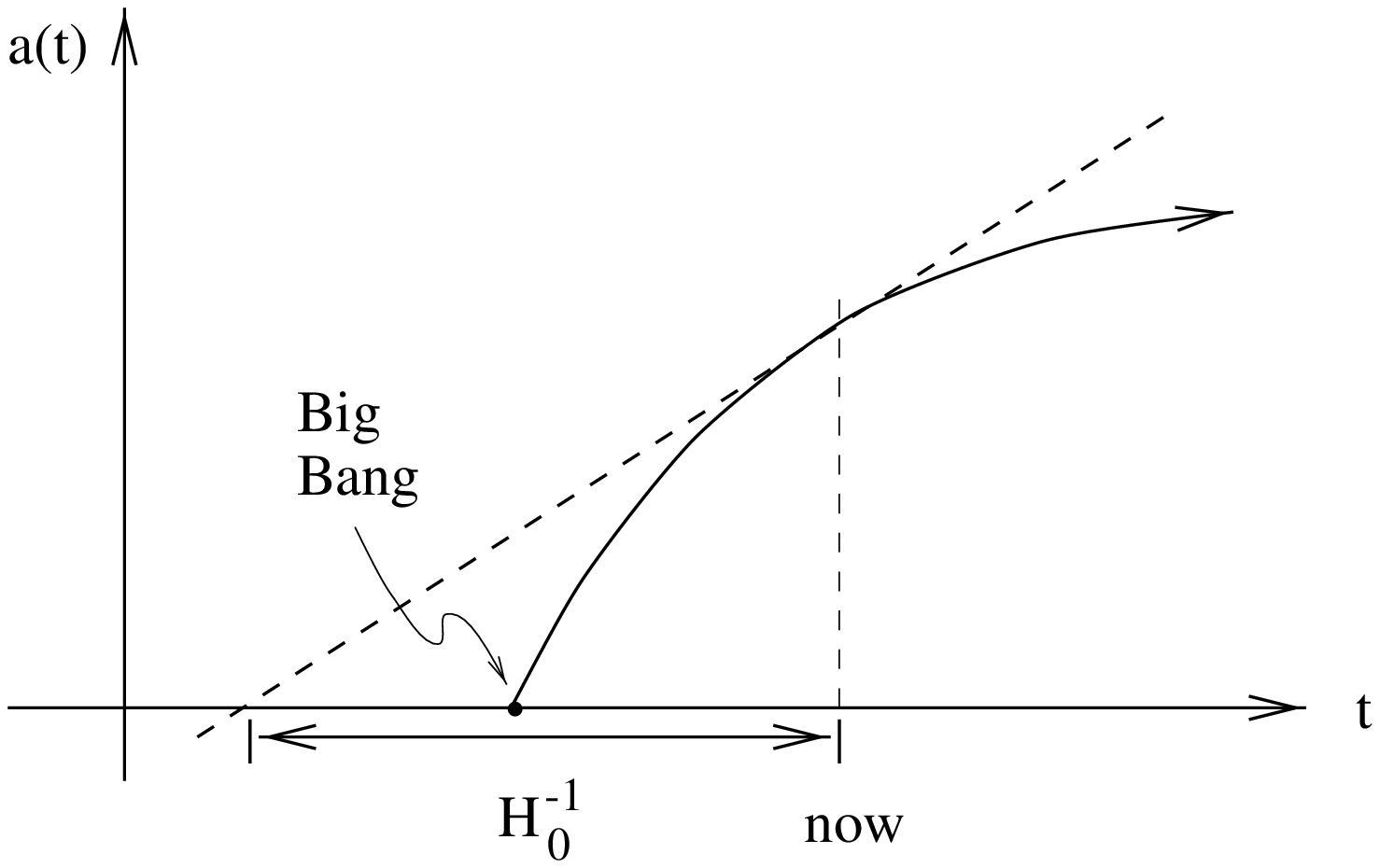,angle=0,height=6cm}}
\end{figure}

This singularity at $a=0$ is the {\bf Big Bang}.
It represents the creation of the universe from a singular state,
not explosion of matter into a pre-existing spacetime.  It might be
hoped that the perfect symmetry of our FRW universes was responsible
for this singularity, but in fact it's not true; the singularity
theorems predict that any universe with $\rho>0$ and $p\geq 0$ must
have begun at a singularity.  Of course
the energy density becomes arbitrarily high as $a\rightarrow 0$,
and we don't expect classical general relativity to be an
accurate description of nature in this regime; hopefully a 
consistent theory of quantum gravity will be able to fix things up.

The future evolution is different for different values of $k$.
For the open and flat cases, $k\leq 0$, (8.36) implies
\be
  \dot a^2 = {{8\pi G}\over 3}\rho a^2 + |k|\ .\label{8.42}
\ee
The right hand side is {\it strictly} positive (since we are
assuming $\rho>0$), so $\dot a$ never passes through zero.  Since
we know that today $\dot a>0$, it must be positive for all time.
Thus, the open and flat universes expand forever --- they are
temporally as well as spatially open.  (Please keep
in mind what assumptions go into this --- namely, that there
is a nonzero positive energy density.  Negative energy density
universes do not have to expand forever, even if they are ``open''.)

How fast do these universes keep expanding?  Consider the
quantity $\rho a^3$ (which is constant in matter-dominated
universes).  By the conservation of energy equation (8.20) we have
\bea
  {{d}\over {dt}}(\rho a^3) &=& 
  a^3\left(\dot\rho + 3\rho{{\dot a}\over a}\right)\cr
  &=&  -3p a^2\dot a\ .\label{8.43}
\eea
The right hand side is either zero or negative; therefore
\be
  {{d}\over {dt}}(\rho a^3)\leq 0\ .\label{8.44}
\ee
This implies in turn that $\rho a^2$ must go to zero in an
ever-expanding universe, where $a\rightarrow\infty$.  Thus (8.42)
tells us that
\be
  \dot a^2\rightarrow |k|\ .\label{8.45}
\ee
(Remember that this is true for $k\leq 0$.)  Thus, for $k=-1$
the expansion approaches the limiting value $\dot a\rightarrow 1$,
while for $k=0$ the universe keeps expanding, but more and more slowly.

For the closed universes ($k=+1$), (8.36) becomes
\be
  \dot a^2 = {{8\pi G}\over 3}\rho a^2 -1\ .\label{8.46}
\ee
The argument that $\rho a^2\rightarrow 0$ as $a\rightarrow\infty$
still applies; but in that case (8.46) would become negative, which
can't happen.  Therefore the universe does not expand indefinitely;
$a$ possesses an upper bound $a_{\rm max}$.  As $a$ approaches
$a_{\rm max}$, (8.35) implies
\be
  \ddot a \rightarrow -{{4\pi G}\over 3}(\rho +3p)a_{\rm max} <0
  \ .\label{8.47}
\ee
Thus $\ddot a$ is finite and negative at this point, so $a$ reaches
$a_{\rm max}$ and starts decreasing, whereupon (since $\ddot a <0$)
it will inevitably continue to contract to zero --- the Big Crunch.
Thus, the closed universes (again, under our assumptions of
positive $\rho$ and nonnegative $p$) are closed in time as well
as space.

\begin{figure}[h]
  \centerline{
  \psfig{figure=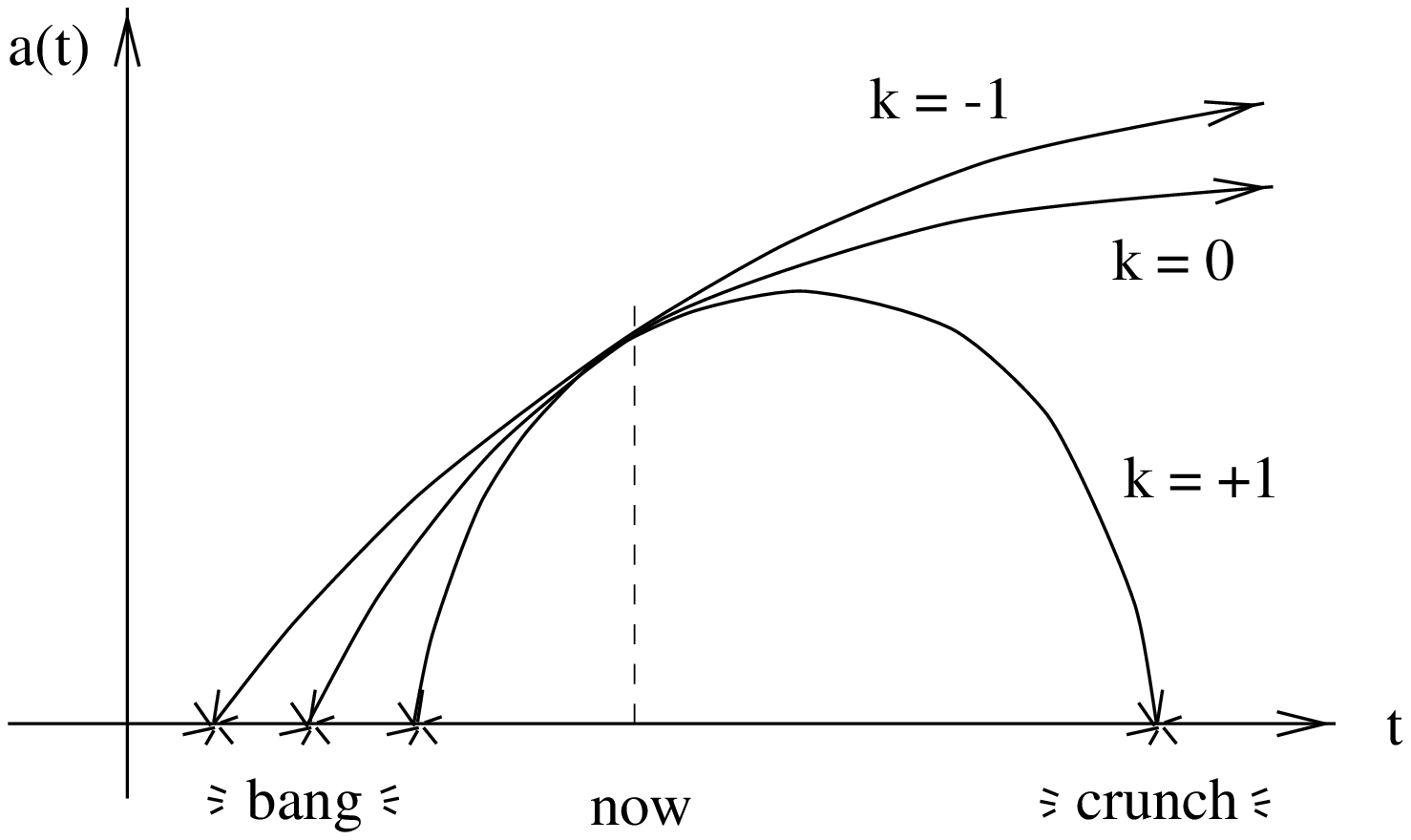,angle=0,height=6cm}}
\end{figure}

We will now list some of the exact solutions corresponding to 
only one type of energy density.
For dust-only universes ($p=0$), it is convenient to define
a {\bf development angle} $\phi(t)$, rather than using $t$ as
a parameter directly.  The solutions are then, for open
universes,
\be
  \cases{a={C\over 2}(\cosh\phi-1)\cr
  t={C\over 2}(\sinh\phi-\phi)\cr}\qquad (k=-1)\ ,\label{8.48}
\ee
for flat universes,
\be
  a = \left({{9C}\over 4}\right)^{1/3} t^{2/3}\qquad (k=0)\ ,
  \label{8.49}
\ee
and for closed universes,
\be
  \cases{a={C\over 2}(1-\cos\phi)\cr
  t={C\over 2}(\phi-\sin\phi)\cr}\qquad (k=+1)\ ,\label{8.50}
\ee
where we have defined
\be
  C={{8\pi G}\over 3}\rho a^3 = {\rm ~constant}\ .\label{8.51}
\ee
For universes filled with nothing but radiation, $p={1\over 3}\rho$,
we have once again open universes,
\be
  a=\sqrt{C'}\left[\left(1+{t\over{\sqrt{C'}}}\right)^2-1\right]^{1/2}
  \qquad (k=-1)\ ,\label{8.52}
\ee
flat universes,
\be
  a=(4C')^{1/4} t^{1/2}\qquad (k=0)\ ,\label{8.53}
\ee
and closed universes,
\be
  a=\sqrt{C'}\left[1-\left(1-{t\over{\sqrt{C'}}}\right)^2\right]^{1/2}
  \qquad (k=+1)\ ,\label{8.54}
\ee
where this time we defined
\be
  C'={{8\pi G}\over 3}\rho a^4 = {\rm ~constant}\ .\label{8.55}
\ee
You can check for yourselves that these exact solutions have 
the properties we argued would hold in general.

For universes which are empty save for the cosmological constant,
either $\rho$ or $p$ will be negative, in violation of the 
assumptions we used earlier to derive the general behavior of
$a(t)$.  In this case the connection between open/closed and 
expands forever/recollapses is lost.  We begin by considering
$\Lambda<0$.  In this case $\Omega$
is negative, and from (8.41) this can only happen if $k=-1$.
The solution in this case is
\be
  a = \sqrt{{-3}\over\Lambda}\sin\left(\sqrt{{-\Lambda}\over 3} \, t
  \right)\ .\label{8.56}
\ee
There is also an open ($k=-1$) solution for $\Lambda>0$, given by
\be
  a = \sqrt{{3}\over\Lambda}\sinh\left(\sqrt{{\Lambda}\over 3} \, t
  \right)\ .\label{8.57}
\ee
A flat vacuum-dominated universe must have $\Lambda>0$, and the
solution is
\be
  a\propto \exp\left(\pm\sqrt{{\Lambda}\over 3} \, t
  \right)\ ,\label{8.58}
\ee
while the closed universe must also have $\Lambda>0$, and satisfies
\be
  a= \sqrt{{3}\over\Lambda}\cosh\left(\sqrt{{\Lambda}\over 3} \, t
  \right)\ .\label{8.59}
\ee
These solutions are a little misleading.  In fact the three
solutions for $\Lambda>0$ --- (8.57), (8.58), and (8.59) ---
all represent the same spacetime, just in different coordinates.
This spacetime, known as {\bf de~Sitter space}, is actually
maximally symmetric as a spacetime.  (See Hawking and Ellis for
details.)  The $\Lambda<0$ solution (8.56) is
also maximally symmetric, and is known as {\bf anti-de~Sitter space}.

It is clear that we would like to observationally determine a 
number of quantities to decide which of the FRW models 
corresponds to our universe.  Obviously we would like to determine
$H_0$, since that is related to the age of the universe.  (For a
purely matter-dominated, $k=0$ universe, (8.49) implies that the
age is $2/(3H_0)$.  Other possibilities would predict similar 
relations.)  We would also like to know $\Omega$, which determines
$k$ through (8.41).  Given the definition (8.39) of $\Omega$,
this means we want to know both $H_0$ and $\rho_0$.   Unfortunately
both quantities are hard to measure accurately, especially $\rho$.
But notice that the deceleration parameter $q$ can be related
to $\Omega$ using (8.35):
\bea
  q &=&  -{{a\ddot a}\over {\dot a^2}}\cr
  &=&  -H^{-2}{{\ddot a}\over a}\cr
  &=& {{4\pi G}\over {3H^2}}(\rho+3p)\cr
  &=& {{4\pi G}\over {3H^2}}\rho(1+3w)\cr
  &=& {{1+3w}\over 2}\Omega\ .\label{8.60}
\eea
Therefore, if we think we know what $w$ is ({\it i.e.}, what kind
of stuff the universe is made of), we can determine $\Omega$ by
measuring $q$.  (Unfortunately we are not completely confident that
we know $w$, and $q$ is itself hard to measure.  But people are
trying.)

To understand how these quantities might conceivably be measured,
let's consider geo\-desic motion in an FRW universe.  There are a
number of spacelike Killing vectors, but no timelike Killing vector
to give us a notion of conserved energy.  There is, however, a
Killing tensor.  If $U^\mu=(1,0,0,0)$ is the four-velocity of
comoving observers, then the tensor
\be
  K_\mn = a^2(g_\mn + U_\mu U_\nu)\label{8.61}
\ee
satisfies $\nabla_{(\sigma}K_{\mn)}=0$ (as you can check), and is
therefore a Killing tensor.  This means that if a particle has
four-velocity $V^\mu = dx^\mu/d\lambda$, the quantity
\be
  K^2 = K_\mn V^\mu V^\nu = a^2[V_\mu V^\mu + (U_\mu V^\mu)^2]\label{8.62}
\ee
will be a constant along geodesics.  Let's think about this, first
for massive particles.  Then we will have $V_\mu V^\mu =-1$, or
\be
  (V^0)^2 = 1+|\vec V|^2\ ,\label{8.63}
\ee
where $|\vec V|^2 = g_{ij}V^iV^j$.  So (8.61) implies
\be
  |\vec V| = {{K}\over a}\ .\label{8.64}
\ee
The particle therefore ``slows down'' with respect to the 
comoving coordinates as the universe expands.  In fact this is an
actual slowing down, in the sense that a gas of particles with
initially high relative velocities will cool down as the universe
expands.

A similar thing happens to null geodesics.  In this case 
$V_\mu V^\mu =0$, and (8.62) implies 
\be
  U_\mu V^\mu = {{K}\over{a}}\ .\label{8.65}
\ee
But the frequency of the photon as measured by a comoving
observer is $\omega=-U_\mu V^\mu$.  The frequency of the photon
emitted with frequency $\omega_1$ will therefore be observed with
a lower frequency $\omega_0$ as the universe expands:
\be
  {{\omega_0}\over{\omega_1}} = {{a_1}\over {a_0}}\ .\label{8.66}
\ee
Cosmologists like to speak of this in terms of the {\bf redshift}
$z$ between the two events, defined by the fractional change in
wavelength:
\bea
  z &=&  {{\lambda_0-\lambda_1}\over{\lambda_1}}\cr
  &=& {{a_0}\over {a_1}}-1\ .\label{8.67}
\eea
Notice that this redshift is not the same as the conventional
Doppler effect; it is the expansion of space, not the relative
velocities of the observer and emitter, which leads to the
redshift.

The redshift is something we can measure; we know the rest-frame
wavelengths of various spectral lines in the radiation from
distant galaxies, so we can tell how much their wavelengths have
changed along the path from time $t_1$ when they were emitted to
time $t_0$ when they were observed.  We therefore know the
ratio of the scale factors at these two times.  But we don't know
the times themselves; the photons are not clever enough to tell
us how much coordinate time has elapsed on their journey.  We have
to work harder to extract this information.

Roughly speaking, since a photon moves at the speed of light its
travel time should simply be its distance.  But what is the
``distance'' of a far away galaxy in an expanding universe?
The comoving distance is not especially useful, since it is not
measurable, and furthermore because the galaxies need not be
comoving in general.  Instead we can define the {\bf luminosity
distance} as
\be
  d^2_L = {{L}\over{4\pi F}}\ ,\label{8.68}
\ee
where $L$ is the absolute luminosity of the source and $F$ is 
the flux measured by the observer (the energy per unit time per
unit area of some detector).  The definition comes from the 
fact that in flat space, for a source at distance $d$ the flux
over the luminosity is just one over the area of a sphere centered
around the source, $F/L=1/A(d)=1/4\pi d^2$.  In an FRW universe,
however, the flux will be diluted.  Conservation of photons
tells us that the total number of photons emitted by
the source will eventually pass through a sphere at comoving
distance $r$ from the emitter.  Such a sphere is at a physical
distance $d=a_0r$, where $a_0$ is the scale factor when the photons
are observed.  But the flux is diluted by two additional effects:
the individual photons redshift by a factor $(1+z)$, and the photons
hit the sphere less frequently, since two photons emitted a time
$\delta t$ apart will be measured at a time $(1+z)\delta t$ apart.
Therefore we will have
\be
  {F\over L} = {1\over{4\pi a_0^2 r^2 (1+z)^2}}\ ,\label{8.69}
\ee
or
\be
  d_L = a_0 r (1+z)\ .\label{8.70}
\ee
The luminosity distance $d_L$ is something we might hope to 
measure, since there are some astrophysical sources whose
absolute luminosities are known (``standard candles'').  But $r$
is not observable, so we have to remove that from our equation.
On a null geodesic (chosen to be radial for convenience) we have
\be
  0 =ds^2 = -\d t^2 + {{a^2}\over {1-kr^2}}\d r^2\ ,\label{8.71}
\ee
or
\be
  \int_{t_1}^{t_0} {{dt}\over{a(t)}} 
  = \int_{0}^{r} {{dr}\over{(1-kr^2)^{1/2}}}\ .
  \label{8.72}
\ee
For galaxies not too far away, we can expand the scale factor in
a Taylor series about its present value:
\be
  a(t_1)=a_0 + (\dot a)_0(t_1-t_0) +{1\over 2}(\ddot a)_0
  (t_1-t_0)^2 + \ldots \ .\label{8.73}
\ee
We can then expand both sides of (8.72) to find
\be
  r= a_0^{-1}\left[(t_0-t_1)+{1\over 2}H_0(t_0-t_1)^2 +\ldots
  \right]\ .
  \label{8.74}
\ee
Now remembering (8.67), the expansion (8.73) is the same as
\be
  {1\over {1+z}} = 1+H_0(t_1-t_0) -{1\over 2}q_0 H_0^2
  (t_1-t_0)^2 + \ldots \ .\label{8.75}
\ee
For small $H_0(t_1-t_0)$ this can be inverted to yield
\be
  t_0-t_1 = H_0^{-1}\left[z-\left(1+{{q_0}\over 2}\right)
  z^2 + \ldots\right]\ .\label{8.76}
\ee
Substituting this back again into (8.74) gives
\be
  r = {1\over{a_0 H_0}}\left[z-{1\over 2}\left(1+q_0\right)
  z^2 + \ldots\right]\ .\label{8.77}
\ee
Finally, using this in (8.70) yields {\bf Hubble's Law}:
\be
  d_L = H_0^{-1}\left[ z+{1\over 2}(1-q_0)z^2 + \ldots\right]\ .
  \label{8.78}
\ee
Therefore, measurement of the luminosity distances and redshifts
of a sufficient number of galaxies allows us to determine $H_0$
and $q_0$, and therefore takes us a long way to deciding what kind
of FRW universe we live in.  The observations themselves are
extremely difficult, and the values of these parameters in the
real world are still hotly contested.  Over the next decade or so
a variety of new strategies and more precise application of
old strategies could very well answer these questions once and for
all.

\end{document}